\documentclass[11pt,fleqn,twoside]{book}

\usepackage[top=1.5cm,bottom=1.5cm,inner=2cm,outer=3cm,headsep=8pt,paperwidth=17cm, paperheight=24cm]{geometry} % Page margins
\usepackage{changepage}
\usepackage{indentfirst}

\usepackage{graphicx} % Required for including pictures
\graphicspath{{images/}} % Specifies the directory where pictures are stored
\usepackage{afterpage}

\usepackage{lipsum} % Inserts dummy text

\usepackage{tikz} % Required for dng custom shapes
\usetikzlibrary{positioning}
\tikzset{%
  every neuron/.style={
    circle,
    draw,
    minimum size=1cm
  },
  neuron missing/.style={
    draw=none, 
    scale=4,
    text height=0.333cm,
    execute at begin node=\color{black}$\vdots$
  },
}

\usepackage[greek, english]{babel} % English language/hyphenation

\usepackage{enumitem} % Customize lists
\setlist{nolistsep} % Reduce spacing between bullet points and numbered lists

\usepackage{booktabs} % Required for nicer horizontal rules in tables

\usepackage{xcolor} % Required for specifying colors by name
\definecolor{orange}{HTML}{cc5200} %{ff6c00} %{ff6600}
\definecolor{gray}{HTML}{303030}
\definecolor{yellow}{HTML}{f0be52}
\definecolor{lightorange}{HTML}{f19e58}
\definecolor{darkorange}{HTML}{993d00}

\usepackage{physics}
\usepackage{subfigure}
\usepackage{qcircuit}
\usepackage{multirow}
\usepackage{booktabs,amsfonts,dcolumn}
\usepackage{textpos}
\usepackage[skins]{tcolorbox}
\usepackage[ruled,vlined,linesnumbered]{algorithm2e}
\usepackage{mathdots}
\usepackage{wrapfig}
\usepackage{multicol}
\usepackage{sidecap}
\sidecaptionvpos{figure}{c}
\usepackage[switch, modulo]{lineno}
\newcommand{\F}{\mathcal{F}}

\usepackage{caption}
\usepackage{color}
\DeclareCaptionFont{orange}{\color{orange}}
\DeclareCaptionFont{orange2}{\color{darkorange}}
\DeclareCaptionFont{blue}{\color{blue}}
\usepackage[draft]{minted}
\usepackage{avant} % Use the Avantgarde font for headings

\usepackage{microtype} % Slightly tweak font spacing for aesthetics
\usepackage{inputenc} % Required for including letters with accents
\usepackage[T1]{fontenc} % Use 8-bit encoding that has 256 glyphs

\usepackage{csquotes}
\usepackage[style=alphabetic,citestyle=alphabetic,sortcites=true, autopunct=true,autolang=hyphen,hyperref=true,abbreviate=true,backref=false, backend=biber,defernumbers=true, doi=true, url=false, giveninits=true]{biblatex}
\defbibheading{bibempty}{}
\DeclareFieldFormat[article]{title}{\textit{#1}}
\DeclareFieldFormat[article]{journaltitle}{#1}
\DeclareFieldFormat[article]{issn}{}
\DeclareFieldFormat[article]{editor}{}
\DeclareFieldFormat[inproceedings]{title}{\textit{#1}}
\DeclareFieldFormat[inproceedings]{booktitle}{#1}
\DeclareFieldFormat[inproceedings]{date}{(#1)}
\DeclareFieldFormat[incollection]{date}{(#1)}
\DeclareFieldFormat[misc]{date}{(#1)}
\DeclareFieldFormat[online]{date}{(#1)}
\DeclareFieldFormat[incollection]{title}{\textit{#1}}
\DeclareFieldFormat{citetitle}{\textit{#1}}
\renewbibmacro{in:}{}
\AtEveryBibitem{%
  \ifentrytype{article}{
  	\clearfield{issn}
  	\clearfield{editedby}
  	\clearfield{issue}
  	\clearfield{pages}
  	\clearfield{month}
    }{}
  \ifentrytype{inproceedings}{
  	\clearfield{editor}
  	\clearfield{publisher}
  	\clearfield{month}
    }{}
  \ifentrytype{incollection}{
  	\clearfield{month}
    }{}
}

\usepackage{acro}
\DeclareAcronym{nn}{
  short=NN,
  long=Neural Networks,
  first-style = long-short,
}

\DeclareAcronym{pdf}{
  short=PDF,
  long=Parton Distribution Functions,
  first-style = long-short,
}

\DeclareAcronym{svc}{
  short=SVC,
  long=Support Vector Classifiers,
  first-style = long-short,
}

\DeclareAcronym{ml}{
  short=ML,
  long=Machine Learning,
  first-style = long-short,
}

\DeclareAcronym{cml}{
  short=CML,
  long=Classical Machine Learning,
  first-style = long-short,
}

\DeclareAcronym{nisq}{
  short=NISQ,
  long=Noisy Intermediate-Scale Quantum,
    first-style = long-short,
}

\DeclareAcronym{qae}{
  short=QAE,
  long=Quantum Amplitude Estimation,
    first-style = long-short,
}

\DeclareAcronym{qgan}{
  short=qGAN,
  long=quantum Generative Adversarial Networks,
    first-style = long-short,
}

\DeclareAcronym{qpu}{
  short=QPU,
  long=Quantum Processing Unit,
    first-style = long-short,
}

\DeclareAcronym{qpe}{
  short=QPE,
  long=Quantum Phase Estimation,
    first-style = long-short,
}

\DeclareAcronym{bch}{
  short=BCH,
  long=Baker-Campbell-Haussdorf,
    first-style = long-short,
}

\DeclareAcronym{sgd}{
  short=SGD,
  long=Stochastic Gradient Descent,
    first-style = long-short,
}

\DeclareAcronym{uat}{
  short=UAT,
  long=Universal Approximation Theorem,
    first-style = long-short,
}

\DeclareAcronym{ffnn}{
  short=FfNN,
  long=Feedforward Neural Networks,
    first-style = long-short,
}

\DeclareAcronym{bm}{
  short=BM,
  long=Boltzmann Machines,
    first-style = long-short,
}

\DeclareAcronym{gpu}{
  short=GPU,
  long=Graphical Processing Units,
    first-style = long-short,
}

\DeclareAcronym{cpu}{
  short=CPU,
  long=Central Processing Units,
    first-style = long-short,
}

\DeclareAcronym{jit}{
  short=JIT,
  long=Just-In-Time,
    first-style = long-short,
}

\DeclareAcronym{qaoa}{
  short=QAOA,
  long=Quantum Approximate Optimization Algorithm,
    first-style = long-short,
}

\DeclareAcronym{vqe}{
  short=VQE,
  long=Variational Quantum Eigensolver,
    first-style = long-short,
}

\DeclareAcronym{bp}{
  short=BP,
  long=Barren Plateaus,
    first-style = long-short,
}

\DeclareAcronym{pca}{
  short=PCA,
  long=Principal Component Analysis,
    first-style = long-short,
}

\DeclareAcronym{qft}{
  short=QFT,
  long=Quantum Fourier Transform,
    first-style = long-short,
}

\DeclareAcronym{vqa}{
  short=VQA,
  long=Variational Quantum Algorithms,
    first-style = long-short,
}

\DeclareAcronym{ng}{
  short=NG,
  long=Natural Gradient,
    first-style = long-short,
}

\DeclareAcronym{iqae}{
  short=IQAE,
  long=Iterative Quantum Amplitude Estimation,
    first-style = long-short,
}

\DeclareAcronym{lhc}{
  short=LHC,
  long=Large Hadron Collider,
    first-style = long-short,
}

\DeclareAcronym{fpga}{
  short=FPGA,
  long=Field Programmable Gate Array,
    first-style = long-short,
} 

\DeclareAcronym{qml}{
  short=QML,
  long= Quantum Machine Learning,
    first-style = long-short,
}

\DeclareAcronym{qram}{
  short=QRAM,
  long= Quantum Random Access Memory,
    first-style = long-short,
}

\DeclareAcronym{hep}{
  short=HEP,
  long= High Energy Physics,
    first-style = long-short,
}

\DeclareAcronym{tn}{
  short=TN,
  long= Tensor Networks,
    first-style = long-short,
}

\usepackage{calc} % For simpler calculation - used for spacing the index letter headings correctly
\usepackage{makeidx} % Required to make an index
\makeindex % Tells LaTeX to create the files required for indexing

\usepackage{titletoc} % Required for manipulating the table of contents

\contentsmargin{0cm} % Removes the default margin

\titlecontents{part}[0cm]
{\addvspace{20pt}\centering\large\bfseries}
{}
{}
{}

\titlecontents{chapter}[1.25cm] % Indentation
{\addvspace{12pt}\large%\sffamily
\bfseries} % Spacing and font options for chapters
{\color{orange!80}\contentslabel[\Large\thecontentslabel]{1.25cm}\color{orange}} % Chapter number
{\color{orange}}  
{\color{orange!80}\normalsize\;\titlerule*[.5pc]{.}\;\thecontentspage} % Page number

\titlecontents{section}[2.25cm] % Indentation
{\addvspace{3pt}%\sffamily
\bfseries} % Spacing and font options for sections
{\contentslabel[\thecontentslabel]{1.25cm}} % Section number
{}
{\hfill\color{black}\thecontentspage} % Page number
[]

\titlecontents{subsection}[3.25cm] % Indentation
{\addvspace{1pt}%\sffamily
\small} % Spacing and font options for subsections
{\contentslabel[\thecontentslabel]{1.75cm}} % Subsection number
{}
{\ \titlerule*[.5pc]{.}\;\thecontentspage} % Page number
[]

\titlecontents{figure}[0em]
{\addvspace{-5pt}%\sffamily
}
{\thecontentslabel\hspace*{1em}}
{}
{\ \titlerule*[.5pc]{.}\;\thecontentspage}
[]

\titlecontents{table}[0em]
{\addvspace{-5pt}%\sffamily
}
{\thecontentslabel\hspace*{1em}}
{}
{\ \titlerule*[.5pc]{.}\;\thecontentspage}
[]

\titlecontents{lchapter}[0em] % Indenting
{\addvspace{15pt}\large%\sffamily
\bfseries} % Spacing and font options for chapters
{\color{black}\contentslabel[\Large\thecontentslabel]{1.25cm}\color{black}} % Chapter number
{}  
{\color{black}\normalsize%\sffamily
\bfseries\;\titlerule*[.5pc]{.}\;\thecontentspage} % Page number

\titlecontents{lsection}[0em] % Indenting
{%\sffamily
\small} % Spacing and font options for sections
{\contentslabel[\thecontentslabel]{1.25cm}} % Section number
{}
{}

\titlecontents{lsubsection}[.5em] % Indentation
{\normalfont\footnotesize%\sffamily
} % Font settings
{}
{}
{}

\usepackage{fancyhdr} % Required for header and footer configuration

\fancypagestyle{plain}{\fancyhead{}} % Style for when a plain pagestyle is specified

\makeatletter
\renewcommand{\cleardoublepage}{
\clearpage\ifodd\c@page\else
\hbox{}
\vspace*{\fill}
\thispagestyle{empty}
\newpage
\fi}

\usepackage{amsmath,amsfonts,amssymb,amsthm} % For math equations, theorems, symbols, etc

\newtheoremstyle{orangenumbox}% % Theorem style name
{0pt}% Space above
{0pt}% Space below
{\normalfont}% % Body font
{}% Indent amount
{\small\bf%\sffamily
\color{orange}}% % Theorem head font
{\;}% Punctuation after theorem head
{0.25em}% Space after theorem head
{\small%\sffamily
\color{orange}\thmname{#1}\nobreakspace\thmnumber{\@ifnotempty{#1}{}\@upn{#2}}% Theorem text (e.g. Theorem 2.1)
\thmnote{\nobreakspace\the\thm@notefont%\sffamily
\bfseries\color{black}---\nobreakspace#3.}} % Optional theorem note
\newtheoremstyle{orangebox}% % Theorem style name
{0pt}% Space above
{0pt}% Space below
{\normalfont}% % Body font
{}% Indent amount
{\small\bf%\sffamily
\color{orange}}% % Theorem head font
{\;}% Punctuation after theorem head
{0.25em}% Space after theorem head
{\small%\sffamily
\color{orange}\thmname{#1}\nobreakspace{\@ifnotempty{#1}{}\@upn{#2}}% Theorem text (e.g. Theorem 2.1)
\thmnote{\nobreakspace\the\thm@notefont%\sffamily
\bfseries\color{black}---\nobreakspace#3.}} % Optional theorem note
\newtheoremstyle{blacknumex}% Theorem style name
{5pt}% Space above
{5pt}% Space below
{\normalfont}% Body font
{} % Indent amount
{\small\bf%\sffamily
}% Theorem head font
{\;}% Punctuation after theorem head
{0.25em}% Space after theorem head
{\small%\sffamily
{\tiny\ensuremath{\blacksquare}}\nobreakspace\thmname{#1}\nobreakspace\thmnumber{\@ifnotempty{#1}{}\@upn{#2}}% Theorem text (e.g. Theorem 2.1)
\thmnote{\nobreakspace\the\thm@notefont%\sffamily
\bfseries---\nobreakspace#3.}}% Optional theorem note

\newtheoremstyle{blacknumbox} % Theorem style name
{0pt}% Space above
{0pt}% Space below
{\normalfont}% Body font
{}% Indent amount
{\small\bf%\sffamily
}% Theorem head font
{\;}% Punctuation after theorem head
{0.25em}% Space after theorem head
{\small%\sffamily
\thmname{#1}\nobreakspace\thmnumber{\@ifnotempty{#1}{}\@upn{#2}}% Theorem text (e.g. Theorem 2.1)
\thmnote{\nobreakspace\the\thm@notefont%\sffamily
\bfseries---\nobreakspace#3.}}% Optional theorem note

\newtheoremstyle{orangenum}% % Theorem style name
{5pt}% Space above
{5pt}% Space below
{\normalfont}% % Body font
{}% Indent amount
{\small\bf%\sffamily
\color{orange}}% % Theorem head font
{\;}% Punctuation after theorem head
{0.25em}% Space after theorem head
{\small%\sffamily
\color{orange}\thmname{#1}\nobreakspace\thmnumber{\@ifnotempty{#1}{}\@upn{#2}}% Theorem text (e.g. Theorem 2.1)
\thmnote{\nobreakspace\the\thm@notefont%\sffamily
\bfseries\color{black}---\nobreakspace#3.}} % Optional theorem note
\makeatother

\newcounter{dummy} 
\numberwithin{dummy}{section}
\theoremstyle{orangenumbox}
\newtheorem{theoremeT}[dummy]{Theorem}

\newtheorem{exerciseT}{Exercise}[chapter]
\theoremstyle{blacknumex}
\newtheorem{exampleT}{Example}[chapter]
\theoremstyle{blacknumbox}

\newtheorem{definitionT}{Definition}[section]
\newtheorem{corollaryT}[dummy]{Corollary}
\theoremstyle{orangenum}
\newtheorem{lemma}[dummy]{Lemma}

\RequirePackage[framemethod=default]{mdframed} % Required for creating the theorem, definition, exercise and corollary boxes

\newmdenv[skipabove=7pt,
skipbelow=7pt,
backgroundcolor=black!5,
linecolor=orange,
innerleftmargin=5pt,
innerrightmargin=5pt,
innertopmargin=5pt,
leftmargin=0cm,
rightmargin=0cm,
innerbottommargin=5pt]{tBox}

\newmdenv[skipabove=7pt,
skipbelow=7pt,
rightline=false,
leftline=true,
topline=false,
bottomline=false,
backgroundcolor=orange!10,
linecolor=orange,
innerleftmargin=5pt,
innerrightmargin=5pt,
innertopmargin=5pt,
innerbottommargin=5pt,
leftmargin=0cm,
rightmargin=0cm,
linewidth=4pt]{eBox}	

\newmdenv[skipabove=7pt,
skipbelow=7pt,
rightline=false,
leftline=true,
topline=false,
bottomline=false,
linecolor=orange,
innerleftmargin=5pt,
innerrightmargin=5pt,
innertopmargin=0pt,
leftmargin=0cm,
rightmargin=0cm,
linewidth=4pt,
innerbottommargin=0pt]{dBox}	

\newmdenv[skipabove=7pt,
skipbelow=7pt,
rightline=false,
leftline=true,
topline=false,
bottomline=false,
linecolor=gray,
backgroundcolor=black!5,
innerleftmargin=5pt,
innerrightmargin=5pt,
innertopmargin=5pt,
leftmargin=0cm,
rightmargin=0cm,linewidth=4pt,
innerbottommargin=5pt]{cBox}

\newenvironment{theorem}{\begin{tBox}\begin{theoremeT}}{\end{theoremeT}\end{tBox}}
				  
\newenvironment{definition}{\begin{dBox}\begin{definitionT}}{\end{definitionT}\end{dBox}}

\makeatletter
\renewcommand{\@seccntformat}[1]{{\textcolor{orange}{\csname the#1\endcsname}\hspace{1em}}}                    
\renewcommand{\section}{\@startsection{section}{1}{\z@}
{-4ex \@plus -1ex \@minus .4ex}
{1ex \@plus.2ex }
{\normalfont\large%\sffamily
\bfseries}}
\renewcommand{\subsection}{\@startsection {subsection}{2}{\z@}
{-3ex \@plus -0.1ex \@minus -.4ex}
{0.5ex \@plus.2ex }
{\normalfont%\sffamily
\bfseries}}
\renewcommand{\subsubsection}{\@startsection {subsubsection}{3}{\z@}
{-2ex \@plus -0.1ex \@minus -.2ex}
{.2ex \@plus.2ex }
{\normalfont\small%\sffamily
\bfseries}}                        
\renewcommand\paragraph{\@startsection{paragraph}{4}{\z@}
{-2ex \@plus-.2ex \@minus .2ex}
{.1ex}
{\normalfont\small%\sffamily
\bfseries}}

\newcommand{\@mypartnumtocformat}[2]{%
\setlength\fboxsep{0pt}%
\noindent\colorbox{orange!20}{\strut\parbox[c][.7cm]{\ecart}{\color{black}\Large%\sffamily
\bfseries\centering#1}}\hskip\esp\colorbox{orange!20}{\strut\parbox[c][.7cm]{\linewidth-\ecart-\esp}{\Large%\sffamily
\centering#2}}}%
\newcommand{\@myparttocformat}[1]{%
\setlength\fboxsep{0pt}%
\noindent\colorbox{orange!20}{\strut\parbox[c][.7cm]{\linewidth}{\Large%\sffamily
\centering#1}}}%
\newlength\esp
\newlength\ecart
\def\@part[#1]#2{%
\ifnum \c@secnumdepth >-2\relax%
\refstepcounter{part}%
\addcontentsline{toc}{part}{\texorpdfstring{\protect\@mypartnumtocformat{\thepart}{#1}}{\partname~\thepart\ ---\ #1}}
\else%
\addcontentsline{toc}{part}{\texorpdfstring{\protect\@myparttocformat{#1}}{#1}}%
\fi%
\startcontents%
\markboth{}{}%
{\thispagestyle{empty}%
\begin{tikzpicture}[remember picture,overlay]%
\node at (current page.north west){\begin{tikzpicture}[remember picture,overlay]%	
\fill[orange!70](0cm,0cm) rectangle (\paperwidth,-\paperheight);
\node[anchor=north] at (3cm,-3.25cm){\color{black}\fontsize{100}{40}%\sffamily
\bfseries\@Roman\c@part}; 
\node[anchor=south east] at (\paperwidth-1cm,-\paperheight+1cm){\parbox[t][][t]{9cm}{
\printcontents{l}{0}{\setcounter{tocdepth}{1}}%
}};
\node[anchor=north east] at (\paperwidth-1.5cm,-3.25cm){\parbox[t][][t]{9cm}{\strut\raggedleft\color{black}\fontsize{30}{30}
\bfseries#2}};
\end{tikzpicture}};
\end{tikzpicture}}%
\@endpart}
\def\@spart#1{%
\startcontents%
\phantomsection
{\thispagestyle{empty}%
\begin{tikzpicture}[remember picture,overlay]%
\node at (current page.north west){\begin{tikzpicture}[remember picture,overlay]%	
\fill[orange!70](0cm,0cm) rectangle (\paperwidth,-\paperheight);
\node[anchor=north east] at (\paperwidth-1.5cm,-3.25cm){\parbox[t][][t]{15cm}{\strut\raggedleft\color{black}\fontsize{30}{30}%\sffamily
\bfseries#1}};
\end{tikzpicture}};
\end{tikzpicture}}
\addcontentsline{toc}{part}{\texorpdfstring{%
\setlength\fboxsep{0pt}%
\noindent\protect\colorbox{orange!20}{\strut\protect\parbox[c][.7cm]{\linewidth}{\Large%\sffamily
\protect\centering #1\quad\mbox{}}}}{#1}}%
\@endpart}
\def\@endpart{\vfil\newpage
\if@twoside
\if@openright
\null
\thispagestyle{empty}%
\newpage
\fi
\fi
\if@tempswa
\twocolumn
\fi}

\newif\ifusechapterimage
\usechapterimagetrue
\newcommand{\thechapterimage}{}%
\newcommand{\chapterimage}[1]{\ifusechapterimage\renewcommand{\thechapterimage}{#1}\fi}%
\def\@makechapterhead#1{%
{\parindent \z@ \raggedright \normalfont
\ifnum \c@secnumdepth >\m@ne
\if@mainmatter
\begin{tikzpicture}[remember picture,overlay]
\node at (current page.north west)
{\begin{tikzpicture}[remember picture,overlay]
\node[anchor=north west,inner sep=0pt] at (0,0) {\ifusechapterimage\includegraphics[width=\paperwidth]{\thechapterimage}\fi};
\draw[anchor=west] (\Gm@lmargin,-7.5cm) node [line width=2pt,rounded corners=15pt,draw=orange,fill=white,fill opacity=0.8,inner sep=15pt]{\strut\makebox[22cm]{}};
\draw[anchor=west] (\Gm@lmargin+.3cm,-7.6cm) node {\huge %\sffamily
\bfseries\color{black}\thechapter. #1\strut};
\end{tikzpicture}
};
\end{tikzpicture}
\else
\begin{tikzpicture}[remember picture,overlay]
\node at (current page.north west)
{\begin{tikzpicture}[remember picture,overlay]
\node[anchor=north west,inner sep=0pt] at (0,0) {\ifusechapterimage\includegraphics[width=\paperwidth]{\thechapterimage}\fi};
\draw[anchor=west] (\Gm@lmargin,-7.5cm) node [line width=2pt,rounded corners=15pt,draw=orange,fill=white,fill opacity=0.8,inner sep=15pt]{\strut\makebox[22cm]{}};
\draw[anchor=west] (\Gm@lmargin+.3cm,-7.6cm) node {\huge %\sffamily
\bfseries\color{black}#1\strut};
\end{tikzpicture}};
\end{tikzpicture}
\fi\fi\par\vspace*{230\p@}}}

\def\@makeschapterhead#1{%
\begin{tikzpicture}[remember picture,overlay]
\node at (current page.north west)
{\begin{tikzpicture}[remember picture,overlay]
\node[anchor=north west,inner sep=0pt] at (0,0) {\ifusechapterimage\includegraphics[width=\paperwidth]{\thechapterimage}\fi};
\draw[anchor=west] (\Gm@lmargin,-7.5cm) node [line width=2pt,rounded corners=15pt,draw=orange,fill=white,fill opacity=0.8,inner sep=15pt]{\strut\makebox[22cm]{}};
\draw[anchor=west] (\Gm@lmargin+.3cm,-7.6cm) node {\huge %\sffamily
\bfseries\color{black}#1\strut};
\end{tikzpicture}};
\end{tikzpicture}
\par\vspace*{230\p@}}
\makeatother

\usepackage{hyperref}
\hypersetup{hidelinks,colorlinks=false,breaklinks=true,urlcolor= orange,bookmarksopen=false,pdftitle={Title},pdfauthor={Author}}
\usepackage{bookmark}
\bookmarksetup{
open,
numbered,
addtohook={%
\ifnum\bookmarkget{level}=0 % chapter
\bookmarksetup{bold}%
\fi
\ifnum\bookmarkget{level}=-1 % part
\bookmarksetup{color=orange,bold}%
\fi
}
}

\input{_minted-ms/default.pygstyle}
\input{_minted-ms/default-pyg-prefix.pygstyle}

\begin{document}
\frontmatter

\begingroup
\thispagestyle{empty}
\begin{tikzpicture}[remember picture,overlay]
\coordinate [below=10cm] (midpoint) at (current page.north);
\node at (current page.north west)
{\begin{tikzpicture}[remember picture,overlay]
\node[anchor=north west,inner sep=0pt] at (0,0) {\includegraphics[height=\paperheight]{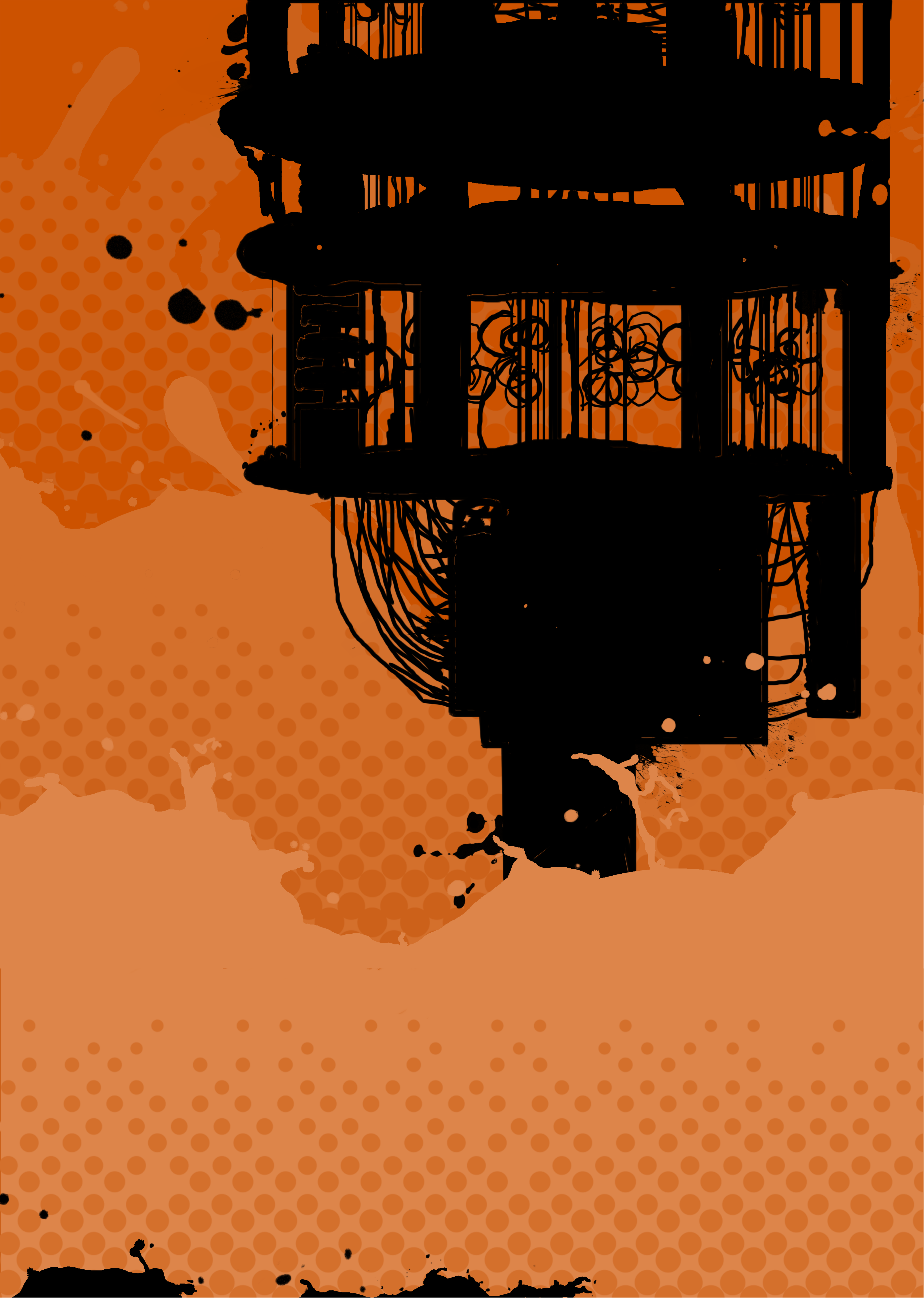}}; % Background image
\draw[anchor=north west] (midpoint) node at (0, -17.5) [fill=blue,fill opacity=0,text opacity=1,inner sep=0cm]{\Huge\centering\bfseries\itshape\parbox[c][][t]{\paperwidth}{\centering Algorithmic Strategies \\ for seizing Quantum Computing\\[15pt]}}; % Book title
\draw[anchor=north west] (midpoint) node at (0, -20) [fill=blue,fill opacity=0,text opacity=1,inner sep=0cm]{\Large\centering\bfseries\parbox[c][][t]{\paperwidth}{\centering PhD Thesis \\ Supervised by José I. Latorre and Artur Garcia-Saez\\[15pt]}}; % subtitle
\draw[anchor=north west] (midpoint) node at (.5, -12) [fill=blue,fill opacity=0,text opacity=1,inner sep=1cm]{\Huge\centering\bfseries\parbox[c][][t]{\paperwidth}{\flushleft Adrián \\ Pérez-Salinas\\[15pt]}  % Author name
};
%\draw[anchor=north west] (midpoint) node at (4.8, -20.5) [fill=blue,fill opacity=0,text opacity=1,inner sep=1cm]{\Huge\centering\bfseries\parbox[c][][t]{\paperwidth}{\centering \includegraphics[width=6cm]{logo_ub} \\ \includegraphics[width=6cm]{bsc_logo}} }; % logo 
\end{tikzpicture}};
\end{tikzpicture}
\endgroup

\cleardoublepage

\begingroup
\thispagestyle{empty}

~\vskip3cm
\begin{center}

{\Large Tesi doctoral}

\vspace{1cm}

{\Huge \bf Algorithmic Strategies for seizing Quantum Computing}
\end{center}

\vspace{2cm}

{\huge Autor: \hfill Adrián Pérez Salinas}

\vspace{1cm}

{\huge Director: \hfill José Ignacio Latorre Sentís}

{\huge Director: \hfill Artur García Sáez}

\vspace{3cm}

\begin{figure}[h!]
\centering

\includegraphics[width = .6\linewidth]{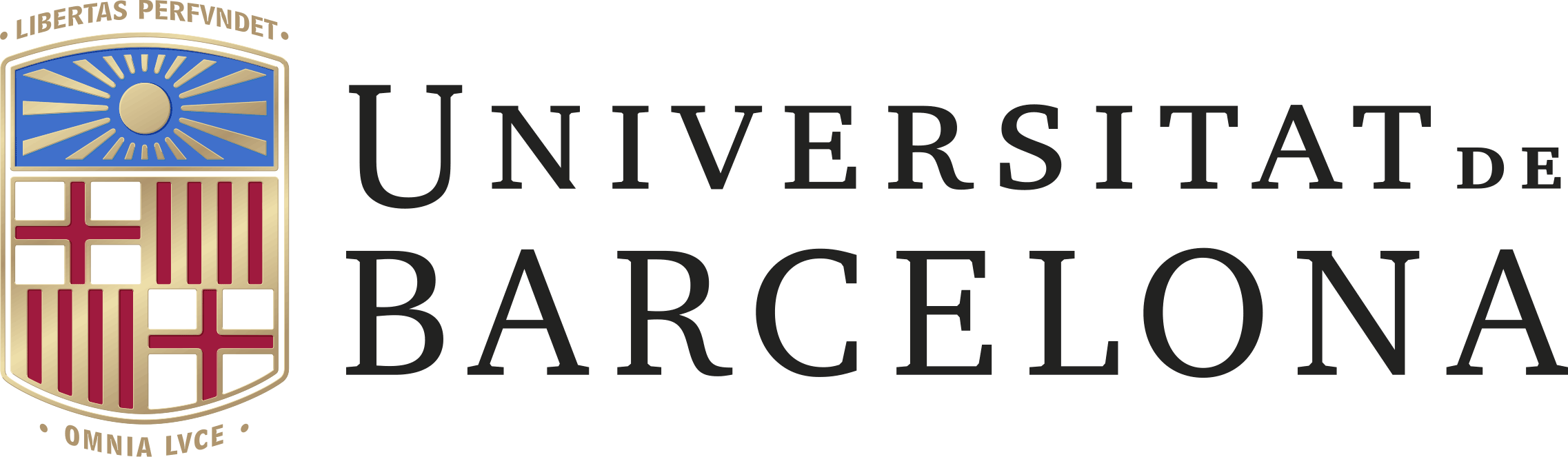}
\end{figure}

\endgroup
\cleardoublepage

\begingroup
\thispagestyle{empty}
\begin{center}

~\vskip2cm

{\Huge \bf Algorithmic Strategies for seizing Quantum Computing}

\vspace{2cm}

{\Large Programa de doctorat en Física}
\end{center}
\vspace{2cm}

{\huge Autor: \hfill Adrián Pérez Salinas}

\vspace{1cm}

{\huge Director: \hfill José Ignacio Latorre Sentís}

{\huge Director: \hfill Artur García Sáez}

\vspace{1cm}

{\huge Tutor: \hfill Joan Soto Riera}

\vspace{1cm}

\begin{center}
{\huge Barcelona, setembre 2021}
\end{center}

\vfill

\begin{figure}[h!]
\centering

\includegraphics[width = .6\linewidth]{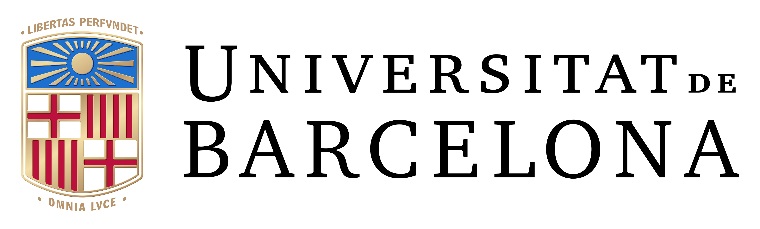}
\end{figure}

%\begin{tikzpicture}[remember picture,overlay]
%\node at (current page.north west)
%{\begin{tikzpicture}[remember picture,overlay]
%\node[anchor=north west,inner sep=0pt] at (9,-17) {\includegraphics[width=6cm]{/home/adrianps/Desktop/firma.png}};
%\end{tikzpicture}};
%\end{tikzpicture}

\endgroup

%color, add !30!black,
% New algorithms for few-qubits quantum computers
% New algorithms for near-term quantum computers
% Algorithmic strategies for exploiting Quantum Computing

%----------------------------------------------------------------------------------------
%	COPYRIGHT PAGE
%----------------------------------------------------------------------------------------

\newpage
\begin{adjustwidth}{0cm}{1cm}
~\vfill

{\bf Cover picture}: original work by María Esteban Ruiz: \href{https://instagram.com/maria_estebanco?utm_medium=copy_link}{@maria\_estebanco}\\

{\bf Chapters images}, credits to original sources:\\
{\sl Contents}: 

library\_mistress in {\tt Flickr} : \href{https://www.flickr.com/photos/library_mistress/525385511/in/photostream/}{\sl Alicia Martin: Biografias - Cascade of books}\\
{\sl Agradecimientos}: 

Unknown author : \href{https://pxhere.com/es/photo/920274}{\sl Ave del Paraíso}\\
{\sl List of publications}: 

Myself : {\sl My own signature}\\
{\sl Resumen}: 

Linnaea Mallette: \href{https://www.publicdomainpictures.net/en/view-image.php?image=275255&picture=porthole}{\sl Porthole, public domain picture}\\
{\sl Introduction}: 

Sara Díaz : {\sl Photo from Park Güell}\\
{\sl Quantum and Classical ML}:

Ramón y Cajal: \href{https://www.sharpbrains.com/wp-content/uploads/2008/02/cajal-chick-cerebellum.jpg}{\sl Drawing of the cells of the chick cerebellum from "Estructura de los centros nerviosos de las aves"}\\
{\sl Data re-uploading strategy for QML}: 

Maurits Cornelius Escher : {\sl Ascending and Descending}\\
{\sl Unary strategy}: 

Wikimedia commons : \href{https://commons.wikimedia.org/wiki/File:Fishing_net_IMGP8396.jpg}{\sl Fishing net IMGP8396.jpg}\\
{\sl Conclusions}:

NovoaR in {\tt Flickr}: \href{https://www.flickr.com/photos/novoar/6113973566/in/photostream/}{\sl Ronsel}\\
{\sl Appendices A and B}: 

Wikipedia - {\sl Trencadís}\\
{\sl Bibliography}: 

Erik Desmazières - {\sl La Bibliothèque de Babel}

\thispagestyle{empty}

\vspace{1cm}

\noindent Copyright \copyright\ 2021 Adrián Pérez-Salinas

License \href{http://creativecommons.org/licenses/by-nc-sa/3.0/}{CC BY-NC-SA 3.0}  \\ % Copyright notice

%\noindent \textsc{Published by Publisher}\\ % Publisher

%\noindent \textsc{book-website.com}\\ % URL

\noindent \LaTeX template {\sl The Legrand Orange Book}, version {\tt 2.1.1 (14/2/16)}, by Mathias Legrand, with modifications \\

\noindent Quantum circuits were designed with {\sl Q-circuit}, source can be found at \href{https://www.ctan.org/pkg/qcircuit}{\tt CTAN} \\

\noindent \textit{First printing, November 2021} % Printing/edition date
\end{adjustwidth}

\newpage

\begin{adjustwidth}{1cm}{0cm}

~\vfill

\begin{flushright}
{\sl A una estrella despistada}
\end{flushright}

~\vfill

{\centering 

{\bf Abstract}

}

Quantum computing is a nascent technology with prospects to have a huge impact in the world. Its current status, however, only counts on small and noisy quantum computers whose performance is limited. In this thesis, two different strategies are explored to take advantage of inherently quantum properties and propose recipes to seize quantum computing since its advent. First, the re-uploading strategy is a variational algorithm related to machine learning. It consists in introducing data several times along a computation accompanied by tunable parameters. This process permits the circuit to learn and mimic any behavior. This capability emerges naturally from the quantum properties of the circuit. Second, the unary strategy aims to reduce the density of information stored in a quantum circuit to increase its resilience against noise. This trade-off between performance and robustness brings an advantage for noisy devices, where small but meaningful quantum speed-ups can be found.

~\vfill

\end{adjustwidth}

\cleardoublepage

\chapterimage{chapter_contents.pdf} % Table of contents heading image

\pagestyle{empty} % No headers

\tableofcontents % Print the table of contents itself

\cleardoublepage % Forces the first chapter to start on an odd page so it's on the right

\pagestyle{fancy} % Print headers again

%auto-ignore
\chapterimage{chapter_agradecimientos.pdf}
\chapter*{Agradecimientos}
\setlength{\columnsep}{.5cm}
\begin{multicols}{2}
\textgreek{Σὰ βγεῖς στὸν πηγαιμὸ γιὰ τὴν Ἰθάκη,\\
νὰ εὔχεσαι νἆναι μακρὺς ὁ δρόμος,}\\
{\sl Cuando emprendas tu viaje a Ítaca\\
pide que el camino sea largo}

\hfill Konstantinos Petrou Kavafis

{\sl Shaking dust from my shoes \\
There's a road ahead \\
And there's no way back home \\
Ohh but I have to say\\
Leaving home ain't easy} 

\hfill Freddie Mercury
\end{multicols}

A pesar de que esta tesis la firma una persona, lo cierto es que poner el punto y final a este proyecto de libro no hubiera sido posible sin la ayuda de mucha gente, de todos los que pusieron un adoquín más en este camino que llamamos la vida. En mi caso, el camino tiene una parada de avituallamiento en la que me harán entrega de un papelito que me da el derecho a hacerme llamar doctor. Creo que es un papel que no vale nada, nada comparado con las ampollas que me ha levantado el sendero en los pies. Lo importante no es el final, es el viaje, y sobre todo las paradas que se hacen en el camino para compartir el té con alguien que en ese momento deja de ser un extraño. 

Las dos personas que más influencia han tenido en esta tesis han sido \href{https://open.spotify.com/track/2YkIDPL5lGhRhomCq4S2RO?si=24ff9390e2a444b5}{José Ignacio} y \href{https://open.spotify.com/track/6EPRKhUOdiFSQwGBRBbvsZ?si=0d3a3f6b84764afc}{Artur}. Ellos tenían el mapa del camino que había que seguir mientras yo sólo veía maleza. Gracias a sus consejos y sus indicaciones ha sido posible llegar hasta el final de este proyecto. Durante todo el doctorado han sido ellos los que han peleado por hacer posible que yo continuara en la brecha, a pesar de todos los obstáculos que han ido apareciendo en el camino. Ha sido un viaje lleno de momentos difíciles, la pandemia que nos confinó a todos en casa, los desplazamientos intercontinentales de José Ignacio, las nuevas y exigentes responsabilidades de Artur en el BSC, la financiación perdida y recuperada a falta de año y medio... Pero al final y casi de milagro todo ha ido saliendo bien. Muy atrás quedarán ya las discusiones armados con una tiza, pero estas fueron el sustrato que me permitió crecer después. Por ello les estaré a los dos eternamente agradecido. 

Ha habido otras muchas personas que me han ayudado a considerarme investigador. Quiero hacer una mención especial para Stefano Carrazza, con quien entendí que se podían hacer cosas más allá de los muros del grupo, y otra para Alba Cervera, a la que considero la hermana mayor que todo lo sabe y que nunca tuve, y de quien aprendí una barbaridad durante el poco tiempo que compartimos en la facultad. Quiero también acordarme de las muchas horas en común que pasé, sobre todo al principio, con Carlos, con Diego y con \href{https://open.spotify.com/track/2Wi5ubKr8zSk8L3CLemyS4?si=a178ae3aa5ef48c3}{Sergi}, y de todo lo que conseguimos juntos. Empezamos algunos, Elies, Josep, y con el tiempo conocí a nuevos compañeros, a Axel, a Bruna. De todos he aprendido y me he llevado cosas buenas.
Gracias también al grupo experimental, con Pol y David a la cabeza, por conseguir que nuestra imaginación despreocupada tenga un sitio por donde correr. Ir al laboratorio era como volver al pueblo a visitar a los primos lejanos, siempre un motivo de alegría. Thanks also to all collaborators I had during these years Juan, Abdullah, Stavros, Tarun, and everyone with whom I ever talked about quantum computing. Mille grazie a Lorenzo e Nicole per avermi insegnato ad insegnare.

Quiero agradecer también a los miembros de mi tribunal de tesis: Juanjo García-Ripoll, Jordi Tura i Brugués, Leandro Aolita, y también Ramón Muñoz-Tapia, Sofyan Iblisdir y Bruno Julià-Díaz por darme la bienvenida al gremio de la ciencia.

Gracias también a Jorge Cortada y a Jordi Planagumà por el primer impulso que necesitábamos para correr. Después, el suelo se derrumbó bajo nuestros pies, pero la inercia nos permitió encontrar un asidero del que colgarse. 

En el mundo de la computación cuántica no todo es investigación y academia, sino también inversiones delicadas y contactos. Por eso quiero agradecer a Parfait Atchadé, y también a los Qapitanes Sergio, Miquel y Guille por enseñarme que hay vida más allá de la ciencia. Gracias a vosotros ahora puedo mirar en otras direcciones. \\

Esta tesis doctoral culmina una etapa que comenzó hace casi tres años con un salto al vacío, un cambiar una ciudad por otra sin mucho más que la vida empaquetada como buenamente se pudo en una maleta. El agradecimiento más especial de todos es para \href{https://open.spotify.com/track/2ZtMNYog671T0UFfp0hhWq?si=3593a7f0eb2b4b1f}{Sara}, para quien se lió la manta a la cabeza y \href{https://open.spotify.com/track/5EskUV4Rg1VMNVPB32xUme?si=16924a4c346946be}{se montó en el tren} a mi lado aquella mañana de enero \href{https://open.spotify.com/track/2A8vMLgDHPEYfuAUmsJWvL?si=c28d2e41685748d9}{rumbo a una aventura desconocida}. Las cosas habrían sido muy diferentes \href{https://open.spotify.com/track/0d1q8eih1Ay5X8pGpiOJZm?si=0588698e61bd4160}{sintigo}, tú lo sabes bien. Han sido muchos días juntos en casa (muy juntos, \href{https://open.spotify.com/track/4QiTep2nCeTnBpnpyEjDPx?si=1732ad09f5db48b0}{recuerda los meses del COVID}), que nos han transformado en personas diferentes \href{https://open.spotify.com/track/7ffa3hboccVAKl1eCvIcyf?si=b06946cf3f2142dd}{una y otra y otra vez más}. Pase lo que pase, nunca podré agradecer lo suficiente la generosidad que supuso dejarlo todo atrás para acompañarme. Creo que, después de todo, las cosas han salido razonablemente bien, \href{https://open.spotify.com/track/5T8EDUDqKcs6OSOwEsfqG7?si=85a70ad3f29c466a}{pero todavía nos queda por vivir}.

Aunque se vean sólo las flores, hay que tener presente las ramas que las soportan. Gracias también a mis \href{https://open.spotify.com/track/47tWQbVn9bJZfH83QiDebk?si=2973d829c08c4c6e}{padres} porque aceptaron el viaje con un nudo en la garganta que nunca dejaron de reconocer, y porque me animaron a perseguir aquello que yo deseara. Y todo ello sin tener en cuenta los veintitantos años de cobijo anteriores que hicieron de mí quien hoy soy. Ha sido mucho tiempo de crecer, no juntos pero sí a su sombra, y aunque ahora toca enfrentarse a la intemperie es indescriptible la sensación de tener \href{https://open.spotify.com/track/39lrkYQZmewSOh4qMC6ApB?si=5736caa3c20d458f}{una casa a la que volver} en cualquier momento. Me queda pendiente un pase privado de la defensa de tesis sólo para ellos, para que me pregunten todo lo que quieran durante el tiempo que quieran. Una manera de rememorar otros tiempos de \href{https://www.youtube.com/watch?v=qv8S7ov4jbY}{espectáculo}. Les debo toda una vida. 

Aunque se recojan sólo los frutos, hay que tener presente las raíces que los alimentan. Esta tesis sólo fue posible gracias a cuatro vidas de trabajo, y sólo mucho tiempo después se ha podido apreciar. Sé que \href{https://open.spotify.com/track/7tzwxKhc52cWNFIc8Wj4sF?si=4982224a15bf4567}{dos de mis abuelos harán el enorme esfuerzo de recorrer medio país para presenciar el final de esta etapa}, como han hecho tantas veces antes, más cerca de casa, y estarán toda la mañana sentados mirándome, sin entender gran cosa, pero qué más da si está allí, tan guapo, tan seguro de sí mismo, tan haciendo lo que mejor sabe hacer. \href{https://open.spotify.com/track/1d90YckCzyZtoZ63XZ10D9?si=bc43505c6d9f413c}{Uno que me vio comenzar no me verá terminar}, o sí, si desde San Isidro los curas le ponen línea directa. Y mi \href{https://open.spotify.com/track/4zAtmTMEz1wagYtBe4e1SC?si=e0df500dc8b746ff}{estrellita} \href{https://open.spotify.com/track/4dBouHKaC5cJ2moBGkLXGX?si=98fac5efe87c4794}{despistada} nunca pudo entender de qué iba todo esto. Por la inmensa desgracia que supone estar pero no estar, esta tesis es sobre todo para ella. Gracias también a la rama del árbol de la \href{https://open.spotify.com/track/2AkmdLbVKS1steeZdy8H1l?si=3da32b1efeb942fb}{tía Nines}, para quienes la vida siempre ha ido \href{https://open.spotify.com/track/1NWIKFHA2gzqPleYlOQSHV?si=a9e5f9aefeb846ad}{partido a partido}. A Eriste, a Arán y a Leo, me hubiera encantado veros crecer. 

No puedo dejar de pensar mientras escribo en que me encantaría que me interrumpiese mi \href{https://open.spotify.com/track/4FjNOWxwlUgf30G9fMggpN?si=24baacfe184d4b2b}{hermano pequeño} para proponerme un \href{https://open.spotify.com/track/5OQgpnpZuZR0ovsEKUKBGe?si=3d7c45c5a5834550}{tentempié}, una partida a la \href{https://open.spotify.com/track/6qx6uKOJAOM587aZp4WlCk?si=214b7ec6ba2e429f}{consola}, o para preguntarme cómo se puede resolver tal integral o tal ejercicio. Creo que nunca he llegado a entender cuánto he llegado a \href{https://open.spotify.com/track/59bqZ237MW9CvFRLqu6Dgg?si=67dc350164564eed}{echarlo de menos}. \href{https://open.spotify.com/track/5n6RDaGFSN88oRWuGtYAIN?si=5f6e89f5563741ab}{La lástima es que se nos haya acabado eso de crecer juntos, y sólo nos queden algunos días sueltos.} Alguna vez tenía que llegar al final. \href{https://open.spotify.com/track/2bilsO0PSHRpsasfczHM01?si=9ad86803c1ed45b4}{Ojalá que nunca se deje ganar y siempre quede una revancha pendiente}. Quiero también pedirle perdón por haberlo dejado solo en \href{https://open.spotify.com/track/07eJs1PYa9KqhKnTtdcifa?si=346c584a2c434033}{un momento tan difícil en casa}. Sé que no me guarda rencor. Por supuesto, gracias también a \href{https://open.spotify.com/track/15tHagkk8z306XkyOHqiip?si=115b83acca714bba}{María} por estos últimos años, y por una portada tan especial para esta tesis. 

Gracias también a aquellos con los que siempre he podido contar sin compartir sangre. A todos y cada uno de los \href{https://open.spotify.com/track/5Fronw7eoL0huzCEYSZe80?si=cae294ca130249ca}{Vragasbundos}, a \href{https://open.spotify.com/track/4NU86l8sRt51lyxGvOGobY?si=fd8f146b688b4c9d}{Pedro}, a Sergio, a Helen, a Flow, a María, a Luisal, a Rodri, y especialmente a \href{https://open.spotify.com/track/19YMl1jI5e4s5GsfCiLMrl?si=18da1b8d855a46e9}{Diego}. Pasar estos años sabiendo que no era el único en una situación incierta ha significado mucho para mí. %Gracias también a \href{https://open.spotify.com/track/59CdyMTI878gUG87xaDcxF?si=acccd00781344a35}{Sarita}, porque a pesar de todo, seguimos. 
Gracias también a la gente de la universidad, a \href{https://open.spotify.com/track/2voFAPC3l9ZGEk0Y3Os1oH?si=f147621ee673466e}{Marina}, \href{https://open.spotify.com/track/4su3t6UIkZjXVc6W67U835?si=4c23428cabf14376}{Martín} y \href{https://open.spotify.com/track/7BXTCiQvCcfZyFtQ3hXQdh?si=80b47c0fdd3b47f0}{Teresa} en Madrid, y un recuerdo especial para Isa; a los que me he ido encontrando por el camino y siguen cerca, a \href{https://open.spotify.com/track/1kfrnPViuzKdNwmH21ehLg?si=61b14264ce864b12}{So}\href{https://open.spotify.com/track/4M4r04FdrMGRAmz6opGT0K?si=7b59b6e7234e4e7f}{fía}, a \href{https://open.spotify.com/track/3apjDg8RP6wLBSSuEcgreX?si=273230f6b5fc4b4b}{Elena}; a los que forman parte de los recuerdos de mi infancia, especialmente a \href{https://open.spotify.com/track/5Dq5lyTpl2xTdjjo0XtEkj?si=4c03beb68efb4ff7}{Ana}.

Los profesores que me han ayudado a ser como soy también merecen un espacio en estos agradecimientos. Gracias a todos los que he conocido durante muchos años. Gracias a \href{https://open.spotify.com/track/02wUyGp3YvXaeLt7kB60gn?si=c7c2c33341e94a4e}{Sergio Montañez} (yo también escribo mi tesis a ritmo de rock, yo he elegido a \href{https://open.spotify.com/track/37Tmv4NnfQeb0ZgUC4fOJj?si=aeb1901ad5584afe}{Mark Knopfler}) y a \href{https://open.spotify.com/track/0eVborSuxUeSg0meWYd9dZ?si=5762451081804465}{Ana Gutiérrez} por mostrarme el mundo de la Física. Gracias también a \href{https://open.spotify.com/track/3FornkeoA9iDwAHs8qCVyi?si=9ad13876d1394d87}{Juanflo}, han sido muchísimos años, y a \href{https://open.spotify.com/track/3btuYJrDm961RsTvEzQtJB?si=8ea004f5de3d4702}{Chemi}, por enseñarme música y por hacerme crecer como persona. 

Ser foraster mai no és fàcil. Cal aprendre un idioma, cal aprendre els costums de la nova ciutat. Però això sempre es possible si hi ha una família que t'acull. Moltes gràcies a tots els \href{https://open.spotify.com/track/6ivzLSDw82LpxXfToyb0He?si=b322d89c74024de0}{Warriors}, a O'Kantz, a Uri, a Laia, a Marc, a Pol, a Raquel, a Max, a Simona i Pere, a Ramón, a Pedret, a Luca, a Jeremy, a Miguel, a Sara, a Belinda, a Eric, a Joan, a Fosy; a Anna i Mar (Warriors honorífiques); i molt especialment a \href{https://open.spotify.com/track/1ejtrweBsVWgQVbuGXaTFj?si=e7517a9d0b1041db}{José-Sensei} per fer la nostra estada a L'Hospitalet de Llobregat molt més agradable. Moltes gràcies també a la gent del \href{https://open.spotify.com/track/03Ntkzzjkz7nFJldcPbL90?si=d13479c755054e51}{Freiburg}, a Valentín, a Sergio, a Raúl i a Guille, per crear un lloc on sentir-se com a casa. Gracias también a \href{https://open.spotify.com/track/4UjqoyfmarLo3Ub3LE19eC?si=ef4a448b58804dc2}{Alex, el escurialense}, por dejarme ser su aprendiz de forastero en Barcelona y en la ciencia.

Por último, este trabajo ha sido de los que pudieron continuar mientras el mundo paraba en mitad de una pandemia mundial provocada por el \href{https://open.spotify.com/track/25BbdnniImBaMpGlfNCPA2?si=a0a1e57fefbf4722}{COVID-19}. A todos los sanitarios que se dejaron la piel, y en algunos casos la vida, gracias por cerrar la herida que sufrimos todos.

\cleardoublepage

%auto-ignore
\chapterimage{chapter_publications.pdf}
\chapter*{List of publications}

The re-uploading strategy, Chapter~\ref{ch:reuploading}, of this thesis is based on references \cite{perezsalinas_data_2020, perezsalinas_proton_2021, perezsalinas_qubit_2021, dutta_realization_2021}. Corresponding software can be found in references \cite{github_data, github_qubit}. 

The unary strategy, Chapter~\ref{ch:unary}, of this thesis is based on reference \cite{ramos_unary_2021}. Corresponding software can be found in reference \cite{github_unary}. 

A main work during the last two years of this thesis is the development of the software for quantum circuits {\tt Qibo}, from reference \cite{qibo}, with corresponding code in \cite{qibo_code}. {\tt Qibo} is an open-source software to write and execute quantum circuits both on simulation and on actual quantum machines. The software has a variety of useful tools and is constantly growing to provide solutions to new challenges.

Additionally, reference \cite{perezsalinas_tangle_2020} was published during the development of this thesis, but it is finally not included.

\vspace{1cm}

{\bf Articles}
\printbibliography[heading=bibempty,keyword={own},type=article]

{\bf Preprints}
\printbibliography[heading=bibempty,keyword={own},keyword={preprint},type=misc]

{\bf Software}
\defbibfilter{software}{
  type=software or
  type=online
}
\printbibliography[heading=bibempty,keyword={own},keyword={software},filter=software]

\cleardoublepage

%auto-ignore
\chapterimage{chapter_resumen.pdf}
\chapter*{Resumen}

\subsection*{Castellano}
La computación cuántica es una tecnología emergente con potencial para resolver problemas hoy impracticables. Para ello son necesarios ordenadores capaces de mantener sistemas cuánticos y controlarlos con precisión. Sin embargo, construir estos ordenadores es complejo y a corto plazo sólo habrá ordenadores pequeños afectados por el ruido y sujetos a ruido (NISQ). Para aprovechar los ordenadores NISQ se exploran algoritmos que requieran pocos recursos cuánticos mientras proporcionan soluciones aproximadas a los problemas que enfrentan. 

En esta tesis se estudian dos propuestas para algoritmos NISQ: {\sl re-uploading} y {\sl unary}. Cada estrategia busca tomar ventaja de diferentes características de la computación cuántica para superar diferentes obstáculos. Ambas estrategias son generales y aplicables en diversos escenarios. 

En primer lugar, {\sl re-uploading} está diseñado como un puente entre la computación cuántica y el aprendizaje automático (Machine Learning). Aunque no es el primer intento de aplicar la cuántica al aprendizaje automático, {\sl re-uploading} tiene ciertas características que lo distinguen de otros métodos. En concreto, {\sl re-uploading} consiste en introducir datos en un algoritmo cuántico en diferentes puntos a lo largo del proceso. Junto a los datos se utilizan también parámetros optimizables clásicamente que permiten al circuito aprender cualquier comportamiento. Los resultados mejoran cuantas más veces se introducen los datos. El {\sl re-uploading} cuenta con teoremas matemáticos que sustentan sus capacidades, y ha sido comprobado con éxito en diferentes situaciones tanto simuladas como experimentales.

La segunda estrategia algorítmica es {\sl unary}. Consiste en describir los problemas utilizando sólo parte del espacio de computación disponible dentro del ordenador. Así, las capacidades computacionales del ordenador no son óptimas, pero a cambio las operaciones necesarias para una cierta tarea se simplifican. Los resultados obtenidos son resistentes al ruido, y mantienen su significado, y se produce una compensación entre eficiencia y resistencia a errores. Los ordenadores NISQ se ven beneficiados de esta situación para problemas pequeños. En esta tesis, {\sl unary} se utiliza para resolver un problema típico de finanzas, incluso obteniendo ventajas cuánticas en un problema aplicable al mundo real. 

Con esta tesis se espera contribuir al crecimiento de los algoritmos disponibles para ordenadores cuánticos NISQ y allanar el camino para las tecnologías venideras. 

\subsection*{English}

Quantum computing is an emergent technology with prospects to solve problems nowadays intractable. For this purpose it is a requirement to build computers capable to store and control quantum systems without losing their quantum properties. However, these computers are hard to achieve, and in the near term there will only be Noisy Intermediate-Scale Quantum (NISQ) computers with limited performance. In order to seize quantum computing during the NISQ era,  algorithms with low resource demands and capable to return approximate solutions are explored.

This thesis presents two different algorithmic strategies aiming to contribute to the plethora of algorithms available for NISQ devices, namely {\sl re-uploading} and {\sl strategy}. Each strategy takes advantage of different features of quantum computing, namely the superposition and the density of the Hilbert space in {\sl re-uploading}, and entanglement among different partitions of the system in {\sl unary}, to overcome a variety of obstacles. In both cases, the strategies are general and can be applied in a range of scenarios. Some examples are also provided in this thesis.

First, the {\sl re-uploading} is designed as a meeting point between quantum computing and machine learning. Machine learning is a set of techniques to build computer programs capable to learn how to solve a problem through experience, without being explicitly programmed for it. Even though the {\sl re-uploading} is not the first attempt to join quantum computers and machine learning, this approach has certain properties that make it different from other methods. 

In particular, the {\sl re-uploading} approach consists in introducing data into a classical algorithms in different stages along the process. This is a main difference with respect to standard methods, where data is uploaded at the beginning of the procedure. In the {\sl re-uploading}, data is accompanied by tunable classical parameters that are optimized by a classical method according to a cost function defining the problem. The joint action of data and tunable parameters grant the quantum algorithm a great flexibility to learn a given behavior from sampling target data. The more re-uploadings are used, the better results can be obtained. 

In this thesis, {\sl re-uploading} is presented by means of a set of theoretical results supporting its capabilities, and simulations and experiments to benchmark its performance in a variety of problems.

The second algorithmic strategy is {\sl unary}. This strategy describes a problem making use of only a small part of the available computational space. Thus, the computational capabilites of the computer are not optimal. In exchange, the operations required to execute a certain task become simpler. As a consequence, the retrieved results are more resilient to noise and decoherence, and meaningful. Therefore, a trade-off between efficiency and resillience against noise arises. NISQ computers benefit from this circumstance, especially in the case of small problems, where even quantum advantage and advantage over standard algorithms can be achieved. 

In this thesis, {\sl unary} is used to solve a typical problem in finance called option pricing, which is of interest for real world applications. Options are contracts to buy the right to buy/sell a given asset at certain time and price. The holder of the option will only exercise this right in case of profit. Option pricing concists in estimating this profit by handling stochastic evolution models.

This thesis aims to contribute to the growing number of algorithms available for NISQ computers and pave the way towards new quantum technologies.

\cleardoublepage

\mainmatter
\setcounter{page}{1}

%auto-ignore
\chapterimage{chapter_introduction.pdf}
\chapter{Introduction}
\begin{adjustwidth}{4cm}{0cm}
{\sl Nature isn’t classical, dammit, and if you want to make a simulation of Nature,
you’d better make it quantum mechanical, and by golly it’s a wonderful problem
because it doesn’t look so easy.}

\hfill Richard Feynman\\
\end{adjustwidth}

The scientific and innovation community is nowadays immersed in the advent of a new technological paradigm that promises to change the world as it is known. The emergence of quantum technologies will likely have a great impact in many different areas. In particular, a great revolution is expected to occur in the field of computing, as firstly proposed by Feynman \cite{Feynman_simulating_1982, preskill_feynman_2021}. Quantum computing is the main field of study of the present thesis. However, it is also valuable to highlight other emerging quantum technologies such as communication \cite{Gisin_communication_2007}, cryptography \cite{Bennett_cryptography_1992} or materials \cite{quantummaterials_2016}.

The first inspiration to come to the idea of quantum computers was the hardness to simulate natural phenomena. Classical computers were, and still are, extremely valuable to deal with the description of the world the human beings live in, but they fail at describing the Nature using the laws of quantum mechanics. The exponentially large size of the quantum Hilbert space is in fact responsible for this claim. %\cite{manin_computable_1980}. 
The ideal solution to this problem was first sighted by Feynman \cite{Feynman_simulating_1982}, that is, to build a computer following the same rules as Nature to describe its quantum behavior, that is, a quantum computer. 

In addition to simulating quantum mechanics, there are other relevant consequences that emerge from this new paradigm of computation. Classical computers are formally defined with the concept of a Turing Machine \cite{Turing_computable_1938}. Quantum computers cannot be a Turing Machine, but a different kind, commonly known as a Quantum Turing Machine, where all classical components are substituted by their more-sophisticated quantum counterparts. Quantum computing was gradually extended until its range of applicability reached fields outside the pure simulation of Physics, for instance in the seminal works from Refs. \cite{deutsch_quantum_1985, bernstein_quantum_1997, shor, grover, brassard_amplitude_estimation_2002}, where some problems are treated whose resolution using quantum devices is more efficient than using classical ones.  This case is commonly known as a quantum advantage. Quantum advantages can be exponential, the most desired but rarest kind, for instance in case of Shor's integer-factoring algorithm \cite{shor}, and polynomial as in Grover's search algorithm \cite{grover}. The main difference between both cases steeps in two different complexity classes available in the field of quantum computing \cite{arora_computational_2009,Vazirani_quantum_2002}

%The main difference between both cases steeps in two different complexity classes available in the field of quantum computing. In a nutshell, classical problems of class $P$ can be solved in polynomial time $\mathcal{O}(n^a)$, being $n$ the size of the problem, and problems $NP$ can only be guessed and verified in polynomial time, but not solved. It is believed that $P \in NP$, and $P \neq NP$ \cite{arora_computational_2009}. The quantum analogues to $P$ and $NP$ are known as $BQP$ and $QMA$ respectively \cite{Vazirani_quantum_2002}. Shor's algorithm shows that integer factorization belongs to $NP$ and $BQP$, that is, it is exponentially more efficient to solve the problem using a quantum than a classical computer. Grover's algorithm shows that the search problem is $NP$ since it is similar to solve it in both cases, in terms of complexity. Nevertheless a polynomial advantage is obtained, what can be useful in some cases. 

In addition to the theoretical advantages of quantum computing when dealing with a variety of problems, classical computers are slowly approaching their physical limits. Moore's law \cite{moore_cramming_1965} roughly predicts that the components of classical computer will reduce their size and energy requirements at a constant rate. This rate will necessarily come to an end in the fabrication of components as it is nowadays performed when the limit of quantum mechanics is reached. This phenomenon has already started \cite{moore_end}. 

Quantum computing is different from classical computing from first principles. 
There exist two main inherently quantum properties that differentiate both paradigms~\cite{nielsen_chuang_2010}, namely entanglement and superposition. Both are born from the definition of the quantum bit - qubit - and the conjunction of several qubits. 

One qubit is conceived as the quantum counterpart of a classical bit. If bits can take values $0$ and $1$, then qubits are
\begin{equation}
\ket{\psi} = \alpha\ket 0 + \beta \ket 1, 
\end{equation}
with $\vert \alpha \vert^2 + \vert \beta \vert^2 = 1$. The contemporary coexistence of both states is known as superposition, that is, quantum states are linear combinations of several 
well-defined states. A qubit only maintains its superposition state if it remains unobserved. At the moment a measurement is performed, the quantum state collapses to one of its well-defined states $\ket 0 / \ket 1$ with probability $\vert \alpha\vert^2 / \vert \beta\vert^2$ and the superposition is lost.

If several qubits are set together, it is possible to obtain a quantum state as
\begin{equation}
\ket\psi = \sum_{i=0}^{2^n - 1} \alpha_i \ket i, 
\end{equation}
with $\sum_{i=0}^{2^n - 1} \vert \alpha_i \vert^2 = 1$ and $\ket i$ is the combination of $\ket 0$ and $\ket 1$ that corresponds to the binary representation of $i$. It is clear that there are $2^n$ available complex coefficients, while a naive product conjunction of single-qubit states would return only $2n$ degrees of freedom. Entanglement is responsible for the emergent 
properties among different quantum systems to give rise to joint systems much larger than the sum of 
isolated parties. In particular, the dimensionality of the Hilbert space grows exponentially with the number of qubits. It is also important to note that quantum computing makes use of exponentially high-dimensional spaces, but these spaces are
dense as well. In contradistinction, classical computing is built upon discrete possibilities of strings 
composed of $0$'s and $1$'s.

Despite the theoretical advantages presented above and the aforementio-ned results, there are prominent inconvenients to be overcome before the existence of quantum computers is a reality. Those inconvenients are mainly related to quantum decoherence preventing quantum systems with many entangled particles to show quantum behavior and the inherent difficulty of experimental accurate control over these systems \cite{Unruh_coherence_1995,landauer_quantum_1995,Haroche_quantum_1996}. Even though different proposals already appeared to solve or at least reduce the impact of decoherence and reach a completely operational fault-tolerant quantum computing \cite{shor_scheme_1995, shor_fault_1996, Steane_error_1996, aharonov_fault_2008, gottesman_stabilizer_1997}, a physical realisation of such devices is still far from being possible. 

The current scenario corresponds to the \ac{nisq} era \cite{preskill_quantum_2018}, that is a moment in time when available quantum computers have moderate numbers of qubits, around hundreds of them, and the logical operations available to apply on the qubits are not completely accurate. In addition, quantum states cannot be indefinitely maintained along time and the purely quantum properties are steadily lost during the execution of a quantum algorithm. The two greatest achievements until the present time are a double attainment of the so-called quantum supremacy, that is, using a quantum computer to solve a problem more efficiently and with better performance than any classical computer \cite{google_supremacy_2019, zhong_quantum_2020}. In both cases, the problems solved are picked to favor the quantum implementation, and they are of no particular interest except for the experiment itself. However, the technological improvement needed to actually accomplish this goal must be highlighted. On the other hand, computers in the \ac{nisq} era are not expected to change the world by themselves, but rather to be an intermediate step towards a new generation of devices. 

One of the most prominent fields of research for \ac{nisq} devices are \ac{vqa} \cite{cerezo_variational_2020, bharti_noisy_2021}. This family of algorithms are constructed as hybrid models combining quantum and classical resources. The chosen models are usually quantum circuits with fixed architectures but adjustable classical parameters. The circuits are trained using a classical optimizer in such a way that an optimal configuration of parameters suffices to reach approximate solutions to a problem of interest. Variational approaches aim to use classical resources to mitigate possible hardware imperfections and limit the demand of quantum requirements. 

Another burning question is how to construct algorithms for quantum computers that are resilient enough to noise as to retrieve meaningful outcomes from the device. An extended, not unique, approach consists in spreading the information contained within a quantum system into a much larger one. The information is now encoded as a global property of the quantum system, much harder to destroy and with a chance to be recovered from a partial decay of the system. Specific implementations differ among different examples \cite{Steane_error_1996, Kitaev_topological_2003}. 

In this thesis a work for both a \ac{vqa} and a deterministic noise-resilient algorithm are presented. In the first case, a \ac{vqa} framework is developed to present a general strategy capable to address a wide variety of \acf{ml} problems. The strategy is here referred as the {\sl re-uploading strategy}. For the noise-resilient algorithm, an example of a real-world problem related to financial calculations is explored to demonstrate quantum advantage already feasible on current quantum computers. It is here called {\sl unary strategy}.

\subsubsection{Re-uploading strategy}
The re-uploading strategy described along Ch.~\ref{ch:reuploading} is a very general framework to bring the fields of quantum computing and \ac{ml} together. \ac{ml} comprisses all algorithms that can learn how to solve particular problems from sampling data, without being explicitly designed for it. Re-uploading is not the first approach to attempt this path, see Ch.~\ref{ch:MLbib} for a brief review on different subjects on this topic. However, re-uploading comes up with a novel idea: data serving as input to any \ac{ml} algorithm is introduced several times into a quantum circuit. The re-uploading strategy belongs to the class of hybrid quantum-classical algorithm.

The idea behind the repeated injection of data into the quantum circuit is to explore all the available Hilbert space and take advantage of it. As mentioned before, the Hilbert space available for performing computations is not only highly dimensional, but also dense, in contradistinction to the classical scheme of computation. To accomplish this goal, the data is interspersed with series of tunable parameters. These parameters, when optimized, drive the behavior of the quantum circuit to approximately what is required to solve a given problem. This behavior is learnt by sampling a training data as in all \ac{ml} algorithms.

The key ingredient that makes the re-uploading strategy useful for \ac{ml} is the natural emergence of non-linear properties. Non-linearities are needed in \ac{ml} to make models capable to approximate arbitrary functions, and then capable to learn and mimic the properties of any data. In classical models for \ac{ml}, non-linearities are artificially introduced, unlike in the re-uploading strategy here presented, where non-linearities arise naturally from the quantum properties of the circuits. 

In this thesis, several aspects of re-uploading are treated. First, theoretical support is given as a justification to use the method with general purposes. Although theoretical support is made explicit for particular cases, general arguments are given to glimpse universality for broader situations. Then, several applications of the re-uploading strategy for different problems are developed with satisfying results to benchmark its performance in different scenarios. The implementations have been implemented on noiseless and noisy simulations of quantum systems, and on actual quantum devices, where the performance degrades in average as the noise increases. The examples of this thesis include regression of test functions, classification of data and extracting physical results from experimental data using the re-uploading strategy.

\subsubsection{Unary strategy}
The unary strategy aims to explore the possibility to reduce the overall performance of quantum algorithms in exchange to gain robustness against noise. In the present work, this is accomplished by reducing the Hilbert space used along the computation. This way, the information in the quantum state does not spread over all the available space, and thus decoherence and noise do not destroy the quantum state entirely. The unary strategy is used in this work to solve a problem of quantitative finances called option pricing. This is a real-world problem with realistic applications. 

The algorithm is built on the unary representation of quantum states. The Hilbert space considered takes into account only those states in the computational basis where there is only one qubit in the $\ket 1$ state, and all others qubit are $\ket 0$. On the one hand, this reduces significantly the amount of information that can be stored into the quantum state, from $e^{\mathcal{O}(n)}$ to $\mathcal{O}(n)$. On the other hand, this reduction simplifies the circuit needed to execute a given algorithm, which translates into a mitigation of potential errors. In addition, the unary algorithm always resides in the restricted area of the Hilbert space, and thus any measurement must reflect this fact. This triggers a native post-selection mechanism that permits to mitigate errors. 

The financial problem to be solved is known as option pricing. The holder of an option gets the right to buy/sell a given asset at a certain price and date. This right is only exercised if it grants some economical benefit. The problem of option pricing consists in estimating the expected pay-off of this option by  running a stochastic model of price evolution.

Even though the theoretical and asymptotical performance of the unary algorithm is low with respect to the standard representations. The aim of unary strategy is not to compete against other efficient methods, but rather to show that it is possible to obtain a trade-off between the quantum advantage obtained and the resilience against noise. In the current \ac{nisq} era, robustness brings advantage since more theoretically efficient approaches retrieve meaningless final results. In addition, a slight quantum advantage is possible even in the less efficient unary scheme thanks to the \acf{qae} recipe. Furthermore, practical applications of the option pricing problem make it useful in the first \ac{nisq} era. 

\subsubsection{Structure of this thesis}

Chapter~\ref{ch:MLbib} covers a brief overview of the status of both classical and quantum \ac{ml}. This chapter gives, first, an overall context of the current situation, and second, all background serving as preliminary contents for next chapters.

In Chapter~\ref{ch:reuploading}, the entire re-uploading strategy is covered as described above. This chapter is composed by different sections treating distinct subjects. Technical appendices to this chapter can be found in App.~\ref{app:reuploading}. The content of this chapter is based on the works in Refs. \cite{perezsalinas_data_2020, perezsalinas_proton_2021, perezsalinas_qubit_2021, dutta_realization_2021, github_data, github_qubit}.

In Chapter~\ref{ch:unary}, the unary strategy and its implementation for financial problems is explained. Technical appendices to this chapter can be found in App.~\ref{app:unary}. This chapter is based on the works in Refs. \cite{ramos_unary_2021,github_unary}

%auto-ignore
\chapterimage{chapter_ml.pdf}

\chapter[Quantum and Classical Machine Learning]{Quantum and Classical ML}\label{ch:MLbib}
\begin{adjustwidth}{4cm}{0cm}
{\sl The world isn't getting any easier. With all these new inventions I believe that people are hurried more and pushed more... The hurried way is not the right way; you need time for everything - time to work, time to play, time to rest.
}

\hfill Hedy Lamarr\\
\end{adjustwidth}

\acf{ml} is nowadays one of the most important fields of computation, being ubiquitious both in research and industry. In recent years, it has gained a strong presence mainly due to the improvement of computing techniques and the increase of available data, both aspects supported by the emergent surge of technological capabilities. \ac{ml} has been used to develop  algorithms capable to solve complex tasks in an automatic manner. For instance, a classic problem of \ac{ml} is to automatically recognize handwritten digits~\cite{deng_mnist_2012}. Current capabilities allow to solve much more complex problems, being the most prominent playing chess~\cite{chess}, Go~\cite{go} or solving the protein folding problem \cite{alpha_fold}. The scientific research also benefits from the development of \ac{ml}~\cite{carleo_ml_2019}

\ac{ml} is a broad field including all different algorithms and techniques with the capability to improve automatically by collecting experience and sampling data.~\cite{mitchell_machine_1997, michalski_machine_2013, russell_artificial_2010}. These algorithms can learn in a general sense, and are prepared to carry specific tasks without being specifically programmed with this purpose. 

The main three steps needed to carry a given \ac{ml} algorithm are essentially:
\begin{itemize}
\item[\bf 1] {\bf Model design:} the architecture of the \ac{ml} model is created. These models usually have some more or less fixed structure with tunable parameters. Those parameters can reach the number of millions in some complex cases. The chosen architecture can be in principle designed from scratch. Nevertheless, there already exists a catalogue of pre-defined models whose competitive performances have been proven under broad conditions. In most cases, the pre-defined architectures can be adjusted to match the needs of the problem to solve. 
\item[\bf 2] {\bf Training:} once the model is obtained, it is compulsory to tune the parameters in such a way that the algorithm is capable to solve a given task. This is done by learning from some training dataset. The action of the model is to receive some input $(x)$ and return some output $(y) = f(x)$. Ideally, the output $y$ provides the solution to the problem of interest. The training is done by improving the obtained solution for the values of $x$ provided by the dataset. In case the dataset also includes the expected solution, the training is performed by comparing output and solution for the same point and minimizing those differences. Otherwise, some more imaginative methods are needed. 
\item[\bf 3] {\bf Generalization:} the final aim of any \ac{ml} technique is to generalize the method, that is, being able to provide good results even for data previously unseen. This step lies at the core of the \ac{ml} strategies. This step is checked by making the algorithm act on a different dataset known as the test set. The output of this test set should be correct and similar to the one of the training set. Otherwise, the traning must be repeated to ensure generalization. There exist techniques for achieving this goal.
\end{itemize}

There exist mainly three different approaches to tackle a problem using \ac{ml} techniques, depending on the features of the datasets to deal with:
\begin{itemize}
\item {\bf Supervised learning:} In supervised learning the dataset contains couples of both input and output. The goal of the algorithm is to mimic the general behavior mapping input to output, both for the training and for unseen data. Common applications of this approach are regression and classification~\cite{nielsen_neural_2015}.
\item {\bf Unsupervised learning:} In this case the dataset only has input, and there is no possible reference for the possible output of the algorithm. The training of the algorithm is accomplished by comparing outputs of different inputs and understanding the similarities. A celebrated application of unsupervised learning is clustering~\cite{duda_pattern_2012,EstivillCastro_clustering_2002}.
\item {\bf Reinforcement learning:} The main relationship for this problem is the one between an agent (the model) and the environment. The agent must obtain the best possible cumulative reward through a given process by combining strategies of inmediate gains and further exploring~\cite{Schmidhuber_ageneral_1998, Kaelbling_reinforcement_1996}.
\end{itemize}

The aim of this section is not to provide an exhaustive review of \ac{ml} techniques. Instead, it is a bibliographical shallow survey of those techniques of both classical and quantum \ac{ml} related to future content in this thesis,  in particular in Ch.~\ref{ch:reuploading}. 

\section{Classical Machine Learning}
In this section general concepts of \ac{cml} are covered as a technical introduction to \ac{ml}.
\subsection{Neural Networks}\label{ssec:nn}

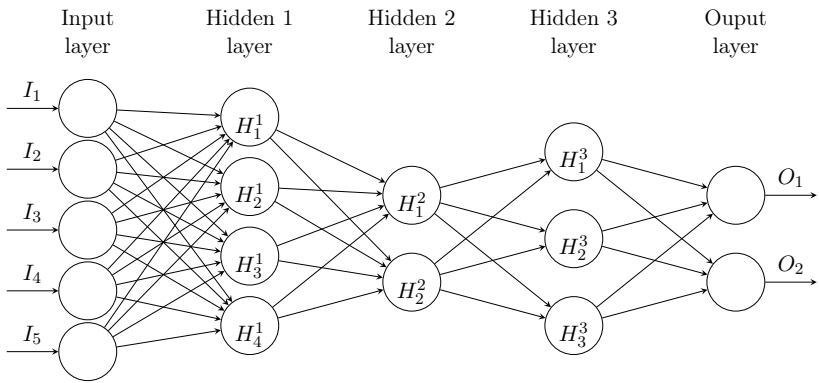
\begin{figure}[t!]
\centering
\resizebox{.9\linewidth}{!}{
\begin{tikzpicture}[x=1.4cm, y=1.5cm, >=stealth]
\foreach \m/\l [count=\y] in {1,2,3,4,5}
  \node [every neuron/.try, neuron \m/.try] (input-\m) at (0,2.2-\y*0.7) {};
  
\foreach \m [count=\y] in {1,2,3,4}
  \node [every neuron/.try, neuron \m/.try ] (hidden1-\m) at (2,2.2-\y*0.8) {};

\foreach \m [count=\y] in {1,2}
  \node [every neuron/.try, neuron \m/.try ] (hidden2-\m) at (4,1.5-\y) {};
  
  \foreach \m [count=\y] in {1,2,3}
  \node [every neuron/.try, neuron \m/.try ] (hidden3-\m) at (6,2-\y) {};

\foreach \m [count=\y] in {1,2}
  \node [every neuron/.try, neuron \m/.try ] (output-\m) at (8,1.5-\y) {};

\foreach \l [count=\i] in {1,2,3,4,5}
  \draw [<-] (input-\i) -- ++(-1,0)
    node [above, midway] {$I_\l$};

\foreach \l [count=\i] in {1,2,3,4}
  \node [above] at (hidden1-\i.south) {$H^{1}_\l$};
  
\foreach \l [count=\i] in {1,2}
  \node [above] at (hidden2-\i.south) {$H^{2}_\l$};
  
\foreach \l [count=\i] in {1,2,3}
  \node [above] at (hidden3-\i.south) {$H^{3}_\l$};

\foreach \l [count=\i] in {1,2}
  \draw [->] (output-\i) -- ++(1,0)
    node [above, midway] {$O_\l$};

\foreach \i in {1,...,5}
  \foreach \j in {1,...,4}
    \draw [->] (input-\i) -- (hidden1-\j);
    
\foreach \i in {1,...,4}
  \foreach \j in {1,...,2}
    \draw [->] (hidden1-\i) -- (hidden2-\j);
    
\foreach \i in {1,...,2}
  \foreach \j in {1,...,3}
    \draw [->] (hidden2-\i) -- (hidden3-\j);

\foreach \i in {1,...,3}
  \foreach \j in {1,...,2}
    \draw [->] (hidden3-\i) -- (output-\j);

\foreach \l [count=\x from 0] in {Input, Hidden 1, Hidden 2, Hidden 3, Ouput}
  \node [align=center, above] at (\x*2,2) {\l \\ layer};

\end{tikzpicture}}
\caption{\ac{ffnn}. Left layer is the input, while right layer is the output. Data is processed from left to right. In the same layer, several processings are taken in parallel, one in each neuron. Different steps are then carried in different layers.}
\label{fig:ffnn}
\end{figure}

\acf{cml} is mostly dominated by the presence of \ac{nn}. They are a successful family of models capable to solve a great variety of tasks with high performance. \ac{nn}s are inspired in animal brains. All \ac{nn}s are composed by unit cells, neurons, that process data. Each neuron receives some input and returns some output. Input data is processed in each neuron depending on some fixed behavior, commonly known as the propagation function, and tunable parameters. In a \ac{nn}, neurons are connected to other neurons following a specific architecture. The connections between neurons can also be tunable.  

The main strength of a \ac{nn} resides on the emergent properties appearing from the correlations between all different neurons. Data is processed in several parallel and / or subsequent steps to give the chance to the \ac{nn} to disclose the most important joint properties of data. Therefore, a \ac{nn} is a method capable to find the most interesting pieces of the dataset to solve the problem of interest. 

\ac{ffnn} are the most general model for \ac{nn}s and their basic architecture~\cite{hoffmann_nn_1992, nielsen_neural_2015, Kaelbling_reinforcement_1996}. This architecture is versatile and provide good performances for a wide variety of problems. 
Standard \ac{ffnn}s can be simply defined as a series of layers where neurons are connected between consecutive layers, see Fig.~\ref{fig:ffnn}. Each neuron receives some input $\vec x$ coming from the previous layer and produces an output $y$, which is forwarded to the next layer. The general behaviour of each neuron in the \ac{ffnn} is
\begin{equation}
y = \sigma\left(\vec w \cdot \vec x + b \right),
\end{equation}
where $\vec w \cdot \vec x = \sum_j{w_j \cdot x_j}$, $\vec w$ is the weight vector connecting neurons, $b$ is the bias and $\sigma(\cdot)$ is known as the activation function. This function can be chosen among different options, and it is not required that all neurons have the same one. It is important to mention that this scheme of weights and biases plays a key role to grant the \ac{nn} enough flexibility as to solve different problems by adjusting its parameters.

The explicit computation performed in every neuron is then as follows. The role of the input layer is to introduce data into the \ac{nn}, possibly with some pre-processing performed within the neuron. In general, this processing is void and the action of the input layer is just an identity map. However, depending on the problems it can be convenient to apply some refinement of raw data. In the case of the hidden layers, the processing is given by the function
\begin{equation}
y^l_j = \sigma\left(\sum_{k = 0}^{n_{l - 1}} w^l_{j, k} y^{l - 1}_{k} + b^{l}_j\right);
\end{equation}
where $l$ is the index of the hidden layer and $j, k$ run over all neurons of the corresponding layer. In terms of indices, the input layer can be labelled with $l = 0$. Notice that this definition is recursive and the activation function of each neuron has some other activation functions, possibly different, together with the corresponding weights and biases. Therefore, each layer adds a new step in the complexity of the final output provided by the \ac{nn}. 

It is worth mentioning that there are theoretical results supporting the general use of the \ac{ffnn} model. The first result guaranteeing that a \ac{ffnn} can represent any continuous function, that is most functions of interest, was obtained in Ref.~\cite{cybenko_approximation_1989}. In this preliminar but fundamental result, the \ac{ffnn} is restricted to be a single-hidden-layer \ac{nn}, and the $\sigma(\cdot)$ function is restricted to be a sigmoid (hence the symbol). A sigmoid function satisfies the property
\begin{equation}
\sigma(x) \rightarrow \left\lbrace \begin{matrix}
1 & \textrm{as} & t \rightarrow \infty \\ 
0 & \textrm{as} & t \rightarrow - \infty \\ 
\end{matrix} \right. ,
\end{equation}
and the most celebrated example is $\sigma(x) = (1 + e^{-x})^{-1}$. Further results extended the role of the $\sigma(\cdot)$ function to be any non-constant non-bounded continuous function~\cite{hornik_approximation_1991}. The results supporting universality for multi-layer \ac{nn}s were achieved later~\cite{leshno_multilayer_1993}.

Apart from the basic \ac{ffnn} model, there exists a great variety of \ac{nn}s whose peak performances are achieved under different conditions. Recursive \ac{nn}s apply the same connections recursively over a structured input. They are for instances broadly used for Natural Language Processing \cite{Hammer2005}. Recurrent \ac{nn}s connect layers with themselves and are useful for handwriting or speech recognition \cite{Graves2009, li2015constructing}. Autoencoders efficiently encode high-dimensional into small parameter spaces data~\cite{Kramer1991}, and they are applied in problems such as face-recognition and face-generation~\cite{Kingma_autoencoders_2019}. Convolutional \ac{nn}s apply filters to input data and are commonly used in image processing~\cite{Valueva2020}. \ac{bm} have capabilities to learn probability distributions over sets of inputs~\cite{ackley_learning_1985, du_neural_2019}.

\subsubsection{Training the \ac{nn}}
The aim of any \ac{nn}, in particular the \ac{ffnn} model, is to learn from input data to return input capable to solve a given problem. For the simple assumption of a supervised learning problem, a given sample of outputs is provided by the problem itself, namely $\vec y_o$. The action of the \ac{nn} can be in general described as a function
\begin{equation}
NN(\vec x; W, B), 
\end{equation}
where $W = \{ w^l_{j,k} \}$ is the set of all weights and $B = \{ b^l_j\}$ is the set of all biases. It is straightforward to see that the aim is to make $\vec y \approx NN$. In order to train the \ac{nn} it is required to measure and minimize the differences between these two quantities. This is usually accomplished by defining a cost function $\chi^2(W, B)$ such that the approximation is better as $\chi^2(W, B)$ decreases. A common example of this quantity is
\begin{equation}
\chi^2(W, B) = \frac{1}{2}\mathop{\rm Average}_{\{\vec x, \vec y\}} \left( \vec y - NN(\vec x; W, B) \right)^2,
\end{equation}
although other possibilities can also be considered~\cite{nielsen_neural_2015}.

The next step consists in finding an optimal set of parameters $(W, B)$ such that
\begin{equation}
(W, B)_{\rm opt} = \mathop{\rm argmin}_{\{W, B\}} \chi^2(W, B).
\end{equation}
This operation can be achieved in many different ways. In general, this problem is passed to an optimization program returning an instance of the optimal parameters, the corresponding value of $\chi^2(W, B)$ and other informations depending on the method to optimize. Optimizing a multi-variable function as in this scenario is in general a NP-hard problem~\cite{pardalos_non-convex_2017, jain_non-convex_2017}, unless the landscape of $\chi^2(W, B)$ is convex, as in this case~\cite{nielsen_neural_2015}. There exists a large variety of optimization methods capable to solve different problems, for instance those based on gradients, such as \ac{sgd}~\cite{nielsen_neural_2015}, quasi-Newton methods~\cite{bfgs, l-bfgs, Nash_tnc_1984}, conjugate-direction~\cite{powell_1964}, simplex sampling~\cite{nelder_mead_1965} or genetic strategies~\cite{cma}. The performance of each method strongly depends on the characteristics of the function to optimize, and in general it is not possible to know which method is more convenient for a particular problem prior to carrying the optimization.

The family of \ac{nn}s have found in \ac{sgd} a method returning high quality solutions for many instances of problems of interest. The gradient-descent piece consists in updating the sets of parameters $(W, B)$ along many iterations by
\begin{eqnarray}
W^{l} = W^{l - 1} - \eta_W \frac{\partial \chi^2(W, B)}{\partial W} \\
B^{l} = B^{l - 1} - \eta_B \frac{\partial \chi^2(W, B)}{\partial B}, 
\end{eqnarray}
where $\eta_{W, B}$ can change along the process following different recipes~\cite{kingma_adam_2017, Qian1999,nesterov_method_1983,zeiler_adadelta_2012,sutskever2013training}, and the derivative $\partial \chi^2 / \partial \{W, B\}$ can be computed exactly or approximately~\cite{spall_stochastic_2005,spall_spsa_1998}. Thus, any gradient-descent based method looks for a standard steepest descent. \ac{nn}s can use \ac{sgd} methods very efficiently thanks to two features, namely batch optimization and backpropagation~\cite{nielsen_neural_2015}. 

Batch optimization consists in estimating the gradient over only a subset of the training set instead of averaging all the possible values of $\{\vec x\}$. In every iteration, the choice of the training data subset is different. Batch optimization brings two advantages respect to the standard gradient-descent strategy. First, there is an improvement in the efficiency since the number of function evaluations required per iteration is reduced. After few iterations all the available values of input $\{ \vec x\}$ have been used and thus participate in the optimization process, and thus the final result is statistically identical to a standard procedure. Second, the alternation between among subsets of $\{x\}$ permits to circumvent local minima. In case a given subset encounters a local minima for a given configuration, this scenario disappears in the next iteration. It is then more likely to reach nearly-optimal minima. 

Backpropagation techniques lie at the core of optimization in \ac{nn}s~\cite{nielsen_neural_2015}. This method allows to compute the gradient values with respect to all different weights and biases efficiently. The results from layer $i + i$ are recovered for layer $i$ successively, and the global amount of operations required is largely diminished. The backpropagation algorithm has as main computing rules
\begin{eqnarray}
\frac{\partial \chi^2(W, B)}{\partial b^{i}_j} &=& \delta^i_j \\
\frac{\partial \chi^2(W, B)}{\partial w^{i}_{j, k}} &=& y^{i - 1}_k \delta^i_j,
\end{eqnarray}
with
\begin{eqnarray}
\delta^{L}_j &=& (O_j - y^{L}_j) \sigma^\prime(z^L_j) \\
\delta^{l}_j &=& \sum_k w^{l + 1}_{j, k} \delta^{l + 1}_k \sigma^\prime(z^l_j).
\end{eqnarray}
Backpropagation is efficiently implemented because the number of operations needed to obtain an estimate of the derivative is small. In addition, it is entirely based on linear algebra and matrix-vector multiplication. The hardware progress accomplished in the last years focuses on fast and efficient implementation of these operations, in particular through \ac{gpu}~\cite{cuda}.

\begin{wrapfigure}{O}{.5\linewidth}
    \includegraphics[width=\linewidth]{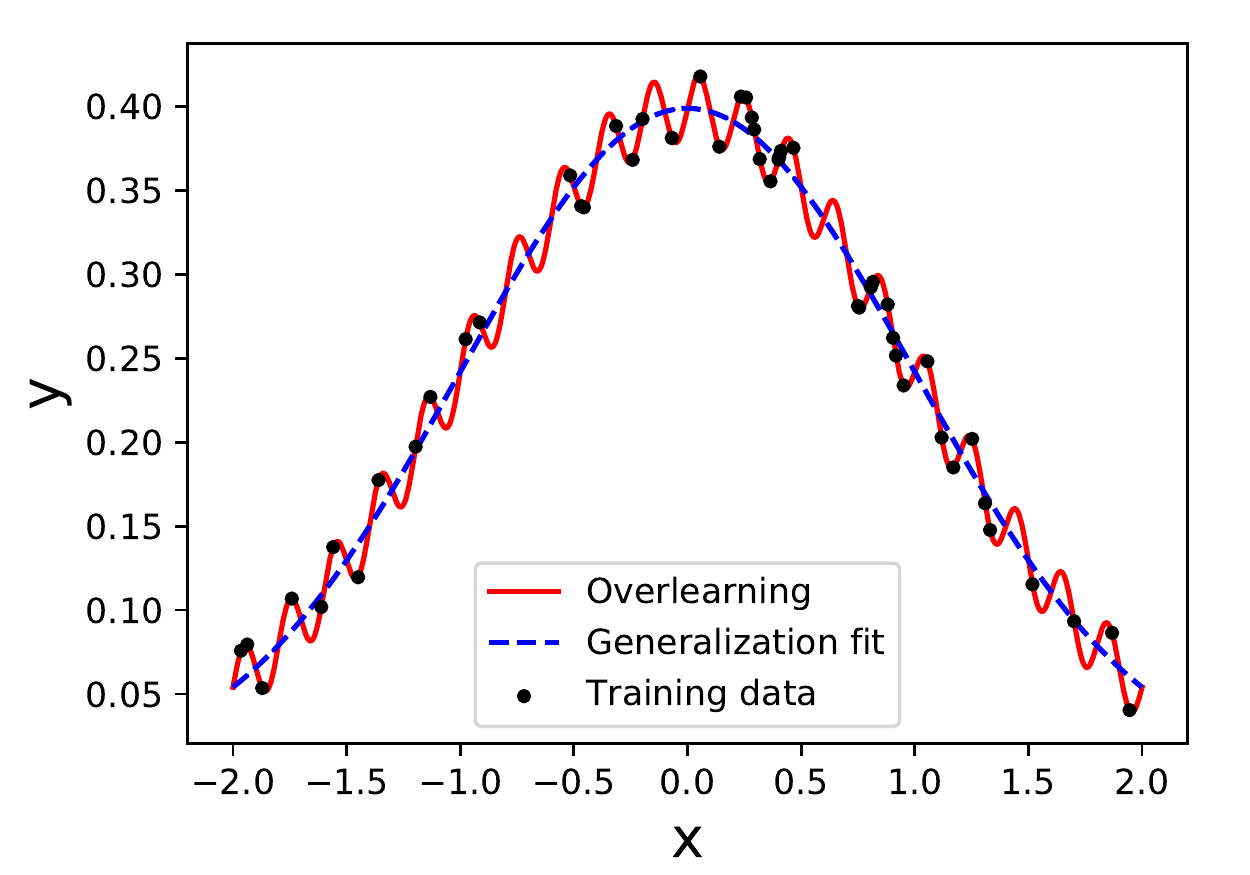} 
    \caption{Graphical description of the overlearning phenomenon. The model is capable to fit the training data (black dots) extremely well (solid red line). However, the general behavior of the training data (dashed blue line) is lost in the process and not captured by the fitting model.}
    \label{fig:overlearning}
\end{wrapfigure}
An extremely successful training process for \ac{nn}s with large number of parameters may lead to an overlearning phenomenon, see Fig.~\ref{fig:overlearning}.
Overlearning appears when the training data offers a great complexity. A model with a large enough number of parameters can represent all tiny details present in the training data, when properly trained. However, it is important to have in mind that the scope of \ac{ml} in general is not to fit a given training data, but to generalize the properties of the training data to provide competitive solutions for unseen datasets. Thus, overfitting must be avoided. There are several techniques with this purpose, like controlling the number of parameters $(W, B)$, adding a normalization term $\lambda (|W|^2 + |B|^2)$ to the cost function, or controlling the performance on the test set to pick the best configuration with respect to this metric, probably not the optimal one according to $\chi^2(W, B)$. 

\subsection{Support Vector Classifier}
\ac{svc} are an alternative method to \ac{nn}s for supervised and unsupervised learning~\cite{Cortes_svc_1995, benhur_svc_2002}. In these models, the classification of data is performed by means of a support vector, hence the name, capable to distinguish different classes. For instance, a couple $\{\vec x, y\}$, with $y = \pm 1$, is classified by means of the hyperplane
\begin{equation}
\vec w \cdot \vec x - b = 0, 
\end{equation}
where the classes $ y = \pm 1$ correspond to both sides of the plane, ideally with some margin if data is linearly separable. In case it is not possible, the optimal boundary can be found by optimizing the cost function
\begin{equation}
\chi^2(\vec w, b) = \frac{1}{2}\mathop{\rm Average}_{\{\vec x, \vec y\}} \left(\max(0, 1 - y (\vec w \cdot \vec x - b)\right) + \lambda |\vec w |^2, 
\end{equation}
where $\lambda$ determines a trade-off between increasing the margin-size and retaining all samples in the correct side of the space. This parameter is also related to the overlearning processes of Sec.~\ref{ssec:nn}.

There exists an alternative method to linearly separate non-separable data, known as the kernel trick~\cite{press_numerical_1986}. In this case, the data is embedded into some non-linear function $\varphi(\vec x)$ and a kernel function 
\begin{equation}\label{eq:kernel}
K_{ij} = k(\vec x_i, \vec x_j) = \varphi(\vec x_i) \cdot \varphi(\vec x_j).
\end{equation}
Subject to this transformation, weights and bias are transformed accordingly. Tha main purpose of this kernel trick is to find a mapping from input data $\vec x$ to other embedded space where classification can be done linearly.

\section{Quantum Machine Learning}
Quantum computers have some properties that make them suitable, in principle, to solve problems related to \ac{ml} more efficiently or accurately than the standard classical counterparts. In particular, a quantum computer with $n$ qubits can store up to $\mathcal{O}(2^n)$ real numbers in its inner quantum state, while $n$ bits are only capable to store $n$ binary variables. In recent years, a new surge of \ac{qml} algorithms has emerged dealing with many different problems. Some of these examples are reviewed in this section, see for instance Refs.~\cite{cerezo_variational_2020,bharti_noisy_2021} for condensed overviews. Complete reviews in \ac{qml} can be read in Refs.~\cite{dunjko_machine_2017,PerdomoOrtiz_opportunities_2018}.

This section does not include any quantum-inspired algorithm designed to be executed on a classical computer in spite of 
their potential utility. Quantum-inspired methods, in particular the celebrated \ac{tn}s~
\cite{Vidal_mps_2004,Vidal_entanglement_2007,verstraete_tn_2008,Vidal_mera_2008, 
orus_tn_2014,orus_tn2_2014,biamonte_tensor_2017}, take advantage of efficient representations of quantum states to 
approximate them with high accuracies. \ac{tn}s were originally conceived to store quantum states, but they present high 
levels of flexibility and capabilities to carry arbitrary data structures. Some methods apply the general \ac{tn} structure 
to solve \ac{ml} problems, both in the field of quantum physics~\cite{torlai_quantum_2020} and general problems 
\cite{Cheng_supervised_2021, convy_mutual_2021, martyn_entanglement_2020, wang_anomaly_2020, stoudenmire_supervised_2017, 
reyes_multiscale_2020}. General \ac{ml} strategies used to solve problems of quantum physics are neither considered in this 
section~\cite{huang_provably_2021}.

\ac{qml} has not accomplished yet an efficient and scalable manner to introduce arbitrary data into a quantum circuit. This lack of uploading methods constitutes a problem when looking for quantum advantages, specially for exponential speed-ups. The reason is that embedding data in a quantum state exploiting the complete storage capability requires the specification of exponentially many terms. In case this translates into an exponential number of operations, a bottleneck appears preventing any remarkable quantum speed-up. This problem is expected to be overcome by a \ac{qram} \cite{Giovannetti_qram_2008}, that is, a quantum operation whose action is
\begin{equation}
U(x)\ket{0} = \ket{\psi(x)}, 
\end{equation}
where $x$ is the input data and $\ket{\psi(x)}$ is a quantum state where the input data is encoded in some convenient manner. \ac{qram}s aim to condensate the loading of large amounts of data into a small number of quantum operations. Nevertheless, no experimental implementation of a \ac{qram} has been achieved yet.

\subsection[Supervised learning in QML]{Supervised learning in \ac{qml}}\label{ssec:supervised_qml}

Many \ac{qml} algorithms developed so far tackle the problem of supervised learning, both for classification or regression of data. Strategies can include hybrid quantum-classical schemes for optimization, but most of them follow the same scheme of embedding classical data into the quantum circuits and look for the optimal measurement dividing them, as it was carried by classical \ac{svc}s.  

To extend classical kernels to the realm of quantum computing it is compulsory to define a quantum operation $V(\vec x)$ taking as input the data of interest and performing the operation
\begin{equation}
\ket{\psi(\vec x)} = V(\vec x)\ket 0,
\end{equation}
where the choice of $\ket 0$ as initial state is arbitrary since any other quantum state could by chosen and the corresponding transformation could be absorbed in $V(\vec x)$. In the case where the data of interest is quantum, there is no special needs to embed it in a quantum circuit.

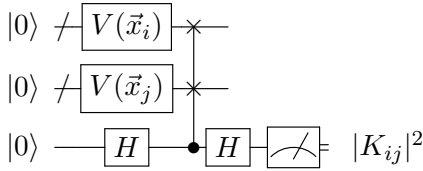
\begin{wrapfigure}{O}{.5\linewidth}
\hspace{1cm} \Qcircuit @R=0.5em @C=0.3em{
\lstick{\ket{0}} & \qw / & \qw & \gate{V(\vec x_i)} & \qw & \qswap & \qw \\
\lstick{\ket{0}} & \qw / & \qw & \gate{V(\vec x_j)} & \qw & \qswap & \qw \\
\lstick{\ket{0}} & \qw  & \qw & \gate{H} & \qw & \ctrl{-2} & \gate{H} & \qw & \meter & \cw & \rstick{\vert K_{ij}\vert ^2} 
}
\caption{Quantum circuit for computing the kernel of two input data. $V(\vec x)$ embeds data into the quantum circuit. The ancilla qubit performs a standard swap test to measure the value of the kernel for the $\vec x$ instances of interest.}
\label{fig:kernel_circuit}
\end{wrapfigure}
The kernel function can be directly measured as 
\begin{equation}
K_{ij} = \braket{\psi(\vec x_i)}{\psi(\vec x_j)},
\end{equation}
where the final $K$ matrix is hermitian with $K_{ii} = 1$. Ideally, $\vert K_{ij}\vert \sim 0$ if $\vec x_i$ and $\vec x_j$ differ between them. Moreover, the properties of this kernel are a cornerstone in the success of any supervised learning task. See Fig.~\ref{fig:kernel_circuit} for the circuit representation of a simple kernel.

The existence of this kernel allows to perform supervised learning both for classification and regression. During the first period of \ac{qml}, these methods were used to develop standard tools of \ac{ml} using quantum circuits with no variational parameter, such as a standard \ac{svc}~\cite{rebentrost_qsvm_2014} and \ac{pca}~\cite{Lloyd_qpca_2014}. Recent works show that quantum kernel methods can only achieve quantum advantage if an appropriate kernel is computed more efficiently using quantum means than classical computation~\cite{kübler_inductive_2021}. In fact, a theoretical quantum speed-up on supervised learning has been already accomplished~\cite{liu_rigorous_2020}. The main feature of such algorithm is the dataset to be classified, specifically chosen to be mapped to the discrete-logarithm problem. The discrete logarithm belongs to the BQP class and can be solved with a quantum computer exponentially faster than using a classical one~\cite{shor}.

Two problems arise when implementing kernel methods as described above on actual quantum computers. First, finding embeddings $V(\vec x)$ leading to kernel methods with competitive performances is far from trivial. The Hilbert space reaches $2^n$ dimensions, where $n$ is the number of qubits. The enormous dimensionality of the Hilbert space opens up the space to embed input data in such a way that different instances lie far from each other in the HIlbert space. To develop the mapping, it is in general a requirement to include deep circuits with many entangling gates and high connectivity between qubits. The literature does not provide embeddings for almost any problem. Second, deep and complex circuits do not perform properly on nowadays quantum computers due to noise and decoherence. Thus, it is not expected that experimental implementations of kernel methods provide competitive results in the near-term. In fact, the discrete logarithm example~\cite{liu_rigorous_2020} requires a \ac{qpe} step, including a \ac{qft} ~\cite{nielsen_chuang_2010}, which is out of scope for any state-of-the-art quantum device. 

It is also worth it mentioning a general algorithm to be used, among others, in \ac{qml}. The HHL algorithm \cite{hhl} is a linear algebra algorithm to invert matrices exponentially more efficient than any classical algorithm under certain conditions. These conditions are a best-case scenario and are usually not completely fulfilled. Inverting matrices has plenty of applications in the field of \ac{ml}, in particular for \ac{svc}s \cite{rebentrost_qsvm_2014} or bayesian learning \cite{Zhao_bayesian_2019}. However, the hardware demands of this algorithm and derivatives cannot be yet satisfied by current machines. 

In order to bring \ac{qml} closer to the \ac{nisq}, several approaches emerged to include \ac{vqa}s in kernel recipes to exploit the capabilities of these methods. \ac{vqa}s constitute the largest family of quantum algorithms at the present time. The usefulness of such algorithms is that they are expected to provide approximate solutions to specific problems without a great size or quality of the quantum computers, that is during the \ac{nisq} era. \ac{vqa}s are based on hybrid quantum-classical schemes. The quantum part is composed by a circuit with a fixed structure and gates depending on classical parameters. The exact operation performed by the circuit depends on the set of parameters serving as input. The classical part is a classical optimizer looking for an optimal set of circuit parameters such that a given function of the measurement is minimized. Canonic examples of a \ac{vqa} are the \ac{vqe}~\cite{Peruzzo_vqe_2014} and \ac{qaoa} methods~\cite{farhi_qaoa_2014}. From a mathematical perspective, a \ac{vqa} can be seen as a circuit $U(\theta)$ performing the operation 
\begin{equation}
\ket{\psi(\theta)}=U(\theta)\ket 0.
\end{equation}
This $U(\theta)$ is commonly known as the {\sl Ansatz} of the circuit and comprehends both the architecture and the distribution of classical parameters. Then, an observable $M$ is measured for the output state $\ket{\psi(\theta)}$, and some cost function encoding the problem and depending both on $M$ and $\theta$, $\chi_M^2(\theta)$ drives the search for the optimal configuration of parameters as
\begin{equation}
\theta_{\rm opt} = \mathop{\rm argmin}_{\theta} \chi_M^2(\theta).
\end{equation}
This kind of algorithms is only successful if two conditions are matched. First, $U(\theta)$ must be flexible enough as to obtain an accurate approximation to the lowest possible value of $\chi^2_M(\theta)$. Finding an Ansatz with the properties to approximate a given state is no trivial problem~\cite{sim_expressibility_2019,nakaji_expressibility_2021}. Nevertheless, the existence Solovay-Kitaev theorem guarantees accurate approximations with a manageable depth of the Ansatz~\cite{christopher_solovay_2006, nielsen_chuang_2010}. Second, the classical optimizer must be capable to find a competitive configuration of optimal parameters. Related to this topic, it has been recently demonstrated that training any \ac{vqa} is a NP problem~\cite{bittel_training_2021}, and further difficulties, such as \ac{bp} to be commented later, appear~\cite{McClean_bp_2018}. As an attempt to avoid these problems, adiabatic strategies have been added to the existing variational ones \cite{garciasaez_aavqe_2018, schiffer_adiabatic_2021}.

Mixing the framework of \ac{vqa}s and kernel methods requires to define the global operation \begin{equation}
\ket{\psi(\vec x, \theta)} = U(\theta) V(\vec x) \ket 0,
\end{equation}
and the output of the quantum circuit can be defined as 
\begin{equation}\label{eq:var_kernel}
K_{ij} = \bra{\psi(\vec x_i, \theta)} M \ket{\psi(\vec x_j, \theta)}, 
\end{equation}
where $M$ is an observable encoding some property of interest. It is possible to compute the kernel by comparing the states corresponding to labels $i, j$. However, it is also common to retrieve information only of one index, say $i$, by directly measuring that state. 
\begin{wrapfigure}{O}{.4\linewidth}
\hspace{.4cm} \Qcircuit @R=0.5em @C=0.3em{
\lstick{\ket{0}} & \qw / & \gate{V(\vec x)} & \qw & \gate{U(\theta)} & \qw & \meter & \cw & \rstick{\langle M \rangle}
}
\caption{Quantum circuit for measuring the expected value of a given data $\vec x$, as in Eq.~\eqref{eq:var_kernel}, with $\langle M \rangle = \bra{\psi(\vec x, \theta)}M \ket{\psi(\vec x, \theta)}$. If no comparison between input data $\vec x_i, \vec x_j$ is desired, no ancilla qubit is required.}
\label{fig:measure_circuit}
\end{wrapfigure}
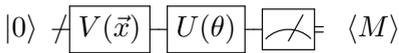
For example, taking $\ket{\psi(\vec x_i, \theta)}$ as a single-qubit state it is possible to relate $K_{ii}$ to some function $f(\vec x_i) \in [-1, 1]$ by selecting $M = Z$. This approach is usually taken both in classification~\cite{schuld_circuit_2020, lloyd_embeddings_2020} and regression~\cite{mitarai_quantum_2018}. See Fig.~\ref{fig:measure_circuit} for a graphical description of such circuits.

The canonical example of kernel methods both with and without \ac{vqa} pieces is described in Ref.~\cite{havlicek_supervised_2019}. In this work, a simple yet useful embedding map to classify 2-dimensional input data instances into two different classes is implemented on a superconducting device. The classes are defined based on measurement outcomes. The standard kernel method show a clear separation between classes. The classifier using a \ac{vqa} reaches a success of $\sim 95\%$. 

Notice that the operator $U^\dagger(\theta)M U(\theta)$, where $\theta$ stands for a set of tunable parameters, is equivalent to finding the optimal measurement to distinguish a certain property~\cite{schuld_supervised_2021}. This observation links with the optimal separation of quantum states~\cite{helstrom_quantum_76}. Two quantum states are completely distinguishable if they are orthogonal, and the certainty to discriminate them decreases as the relative overlap increases. The computational cost of obtaining the optimal measurement strategy grows exponentially with the number of qubits. Thus, variational methods aim to look for approximately optimal measurements with reduced computational costs. 

In order to improve the performance of these variational methods, two main actions can be carried. The first one is to include variational parameters $\phi$ into the embedding layer, $V(\vec x, \phi)$ to gain expressibility in the kernel method. The second approach is to introduce the data $\vec x$ redundantly into the circuit. A motivation is to multiply the available information in the circuit while circumventing the no-cloning theorem \cite{wootters_single_1982}. This theorem prevents the copy of quantum states. The idea of redundancy was included in theoretical~\cite{GilVidal_redundancy_2020}, numerical \cite{mitarai_quantum_2018} and experimental works~\cite{havlicek_supervised_2019}.

A conjunction of both redundancy and variational parameters in the embedding step was developed in terms of the {\sl re-uploading} strategy, greatly detailed in Ch.~\ref{ch:reuploading}~\cite{perezsalinas_data_2020}. In the re-uploading and similar strategies, the embedding and optimal measurement steps are fused into one combined step. The addition of flexible embedding schemes allow to find a quantum circuit capable to separate data in a nearly optimal manner. In addition, a measurement maximizing distinguishability is also accomplished. This strategy has already shown its universal representation capability~\cite{perezsalinas_qubit_2021, schuld_effect_2021}.

\subsubsection{Training a \ac{vqa}}\label{sssec:training_vqa}

Several properties of quantum computers must be considered when \ac{qml} algorithms rely on findign an optimal configuration of parameters must be found. Many references delegate this search to the classical optimizer without providing any other instruction to help with the procedure. It would be straightforward to think that the process followed by any classical optimizer is equivalent both for quantum and classical computation. It is in fact expected that \ac{vqa} can outperform classical methods during the \ac{nisq} era~\cite{Harrow_gradient_2021} under certain conditions. This is, however, not true in most cases. 

The landscape defined by any quantum variational method is affected by sampling uncertainties, what difficults the attainment of accurate measurements. The only possible strategy to retrieve information from a quantum circuit is by measuring observables. A measurement of interest is repeated $N$ to return an estimate of the expected value of such observable as
\begin{equation}
\langle \bar M \rangle = \bra{\psi} M \ket{\psi}_{N},
\end{equation}
where the approximation to the exact expected value $\langle M \rangle$ measured with an infinite number of shots is bounded by 
\begin{equation}
\vert \langle\bar M\rangle - \langle M \rangle \vert \sim \mathcal{O}(N^{-1/2}).
\end{equation}
This phenomenon is inherently quantum and cannot be avoided due to the nature of quantum computers. It can be only mitigated considering more shots or performing techniques such as \acf{qae}~\cite{brassard_amplitude_estimation_2002}. Many modern classical optimization algorithms are designed to seize all the numeric precision computers can provide~\cite{scipy, cma_package}, in particular for gradient-based methods~\cite{l-bfgs, bfgs, spall_spsa_1998, nielsen_chuang_2010}. The fine-tuning is required to achieve the best possible final results. The reason for this phenomenon is that many algorithms take advantage of approximation methods whose peak performances are matched at a particular and small computing precision. In case these conditions are not fulfilled, the performance of the recipes degrades severely. 

A variety of methods have been developed to deal with the problem of computing gradients in \ac{vqa}s. First, there exist recipes to analytically evaluate the gradient of a given quantum circuit with respect to some parameter~\cite{mitarai_quantum_2018, schuld_evaluating_2019}. The evaluation is exact up to measurement uncertainty. The key observation to design a gradient evaluator for quantum circuits depending on $\vec \theta$ is known as the {\sl parameter shift rule}. Gradient components with respect to some parameters are evaluated as a combination of two circuit evaluations as
\begin{equation}
\frac{\partial f}{\partial \theta_i} = r \left(f(\vec \theta + s \hat e_i) - f(\vec \theta - s \hat e_i) \right), 
\end{equation}
where $s = \pi / 4r$ and $r$ is the absolute value of the eigenvalues of the unitary gate where $\theta_i$ comes into, annd $\hat e_i$ is the unitary vector in the direction $i$, that is only the $i$-th component of $\vec\theta$ is modified. Notice that this observation may looks similar to an standard finite-differences method. However, it does not depend on any approximation and is, in fact, exact. This way, only a sampling uncertainty error comes into the calculation. Further works have extended these results to consider also measurement phenomena~\cite{sweke_stochastic_2020}, and have generalized the parameter shift rule to compute not only first, but higher order derivatives~\cite{hubregtsen_singlecomponent_2021}.

Measurement uncertainty is not the only source of inaccuracies appearing when optimizing \ac{vqa}s. The presence of noise and decoherence also contribute to erratic evaluations of functions on a quantum circuit. There already exist attempts to use the possible noise occurring in the quantum devices to estimate gradients more accurately \cite{Meyer_variational_2021}. 

Another important inconvenient encountered for quantum circuit optimi-zation is related to the landscape defined by the parameters and the cost function to be minimized. Nowadays it is not clear what is the shape of the parameters landscape, and thus it is difficult to actually design an optimization method suiting these properties. 

A first approach to minimize an objective function taking into account the geometry of the landscape is provided by the \ac{ng} algorithm. This algorithm is inherited from classical computing~\cite{Amari_natural_1998}, and consists in evaluating gradients by weighting them with the Fisher information matrix~\cite{fisher_mathematical_1922,Savage_fisher_1976}. The aim of doing such trick is to gain information about the landscape for optimizing a given function since the Fisher information gives insights on the local geometry structure of the landscape, and it is then more likely to obtain competitive results. Since the first proposal of \ac{ng} for quantum circuits~\cite{Stokes_quantum_2020}, several recipes have been published to take profit of this feature~\cite{meyer_fisher_2021,beckey_variational_2020,gacon_simultaneous_2021,Cerezo_fisher_2021}. The main reason preventing a systematic application of \ac{ng} to the field of \ac{vqa} is the large amount of measurements needed per iteration. 

A prominent problem for all the family of \ac{vqa}s is known as the \ac{bp}s. This phenomenon is recognized when the average value of the derivatives of a given circuit vanishes exponentially as the system size increases~\cite{McClean_bp_2018}. The direct consequence is that the landscape becomes essentially flat and it is then extremely difficult to find a parameter configuration with low values of the cost function of interest. The phenomenon of \ac{bp}s appear in many instances of \ac{vqa}s~\cite{BravoPrieto_scaling_2020, Huembeli_characterizing_2021, bravoprieto_variational_2020}. The immediate consequence of the existence of\ac{bp}s is that an exponentially large number of measurements is needed to resolve an average derivative. \ac{bp}s compromise the performance of gradient-free optimization methods as well~\cite{arrasmith_effect_2020}.

Two different sources give rise to \ac{bp}s. First, the mapping of a set of classical algorithms to an exponentially large Hilbert space induces large separations between close sets, that is, the probability of two objects to be close is exponentially small. The appearance of \ac{bp}s is closely related with the definition of the cost function of interest~\cite{Uvarov_bp_2021,Uvarov_variational_2020, Cerezo_bp_2021}. In fact, local cost functions reduce the presence of \ac{bp}. The expressibility of the Ansätze has also a relationship with this source of error \cite{sim_expressibility_2019,nakaji_expressibility_2021, holmes_connecting_2021}. The landscape of \ac{bp}s due to cost function can be understood as a large flat space with a narrow deep well in some point. It is extremely difficult to find the optimal point since most of the parameter space does not provide any information about it. The second source of error is the noise of the quantum circuit~\cite{wang_noiseinduced_2021}. Random errors make different outputs indistinguishable, and thus the landscape is in average flattened. 

\ac{bp}s are nowadays the main problem to be overcome for achieving quantum advantage in any \ac{vqa}-dependent procedure. In case the \ac{bp} issue cannot be solved, it will be exponentially hard to train a quantum circuit. Current methods cannot deal with a problem of such difficulties, and quantum hardware is not competitive enough yet to execute this kind of algorithms accurately. However, it has been shown for small circuits that finding an optimal set of parameters is feasible. The improvement of all different parts of \ac{vqa}s will translate into the exploitation of \ac{nisq} devices. 

\subsection[Other approaches in QML]{Other approaches in \ac{qml}}

This chapter is mainly focused on supervised learning due to its connection to the novel work presented in subsequent chapters. Nevertheless, other approaches in \ac{qml} have also experimented large improvements in recent years. In this section, some major contributions to this field are covered in order to give a broad perspective of the status of the field. 

\subsubsection{Quantum Boltzmann Machines}
\ac{bm}s are a model for \ac{ml} coming from its classical approach. Even though the model is well understood from a theoretical and mathematical point of view, it is not broadly used. The main reason is that the training of such model is not efficient in the classical case~\cite{Ackley_bm_1985}. This property may change with the advent of quantum computers since the training methods differ and could take advantage of the special features of \ac{bm}s.

The idea of \ac{bm}s come from statistical physics. There exists a system with a total energy, and the probability of finding a particular state of the system when meauring depends on this energy. In a nutshell, a \ac{bm} is composed by neurons of two kinds: some neurons are visible and some are hidden. The neurons can take continuous or discrete values depending on the exact choice, and these neurons are connected by a two-local hamiltonian
\begin{equation}
H(x;a, b) = \sum_{\{i, j\}} a_{ij} x_i x_j + \sum_i b_i x_i,
\end{equation}
where the weights $a_{i, j}$ and $b_i$ are supposed to control the behavior of the model. In this hamiltonian there is no distinction between visible $(v)$ and hidden $(h)$ neurons. The \ac{bm} is then sampled to obtain a specific configuration of the visible neurons. The probability of obtaining a given visible result $v$ is
\begin{equation}
P(v) = Z^{-1} \sum_{h} e^{-H((v, h); a, b)},
\end{equation}
where $Z$ is the partition function taking all possible outcomes into account for normalization. 

The main application of \ac{bm}s, both in classical and quantum computers, is to learn a probability distribution. The distribution can be taught as a training dataset to mimic, or can be modeled by looking at the relationships between different inputs as in unsupervised or reinforcement learning schemes. 

\ac{bm}s have been recently explored both for circuit and adiabatic quantum computing. In the case of circuits, examples considering \ac{vqa}s \cite{Zoufal_bm_2021} and time evolution \cite{shingu_bm_2020} were developed with appreciable success. In the case of quantum annealing, the computer is used to train the model adiabatically \cite{Dixit_training_2021}, leading to a much more efficient optimization technique. 

\subsubsection{Autoencoders}
Autoencoders are general recipes to transform raw data into much more compressed versions of it to store and send it more efficiently while minimizing the loss of information. In the most extreme case where the only information of interest is a label, a very large and sophisticated data instance can be compressed to just an integer variable. In any case, autoencoders must carry a compressor and an uncompressor to recover the original data. 

Quantum computing has developed several autoencoding algorithms during recent years \cite{Romero_autoencoders_2017, Wan_generalisation_2017, verdon_universal_2018}. In addition to theoretical proposals, experimental implementations have been already achieved \cite{Pepper_autoencoder_2019}. It is worth it highlighting the work from Ref.~\cite{BravoPrieto_autoencoders_2021}, which uses the data re-uploading strategy presented in Ch.~\ref{ch:reuploading} to build an autoencoder. 

\subsubsection{Generative models}

The aim of generative models, both for \ac{cml} and \ac{qml} is to learn a probability distribution and sample synthetic data from this learning ~\cite{Du_expressive_2020, verdon_quantum_2019, Liu_differentiable_2018}. An interesting feature of generative models in \ac{qml} is that models receive as input some white noise that creates similar but different outcomes. Quantum computers are capable to generate purely random states that are beneficial in this subject. In addition, it has been already shown that Born machines are capable to give rise to generative models more efficiently than any classical method \cite{Coyle_born_2020}. 

The model of \ac{qgan} is the generative proposal for quantum computing, in this case using \ac{vqa}s. The overview of this approach is that two networks, a generator and a discriminator, compite against each other. Only the discriminator must be previously trained. This way, the generator learns to produce data very close to the real one, and the discriminator knows to distinguish accurately \cite{qGAN-lloyd2018}. Several examples of this approach have recently emerged \cite{qGAN-zoufal2019, romero_variational_2019}.

\subsubsection{Quantum neural networks}
In quantum neural networks, every processing unit from the classical models is substituted by a qubit, and the connections among neurons are translated into quantum gates. For instance, standard \ac{ffnn}-like quantum circuits \cite{altaisky_quantum_2001, farhi_classification_2018, Torrontegui_unitary_2019} and other models with different purposes and architectures \cite{Cong_quantum_2019, franken_explorations_2020} and convenient trainability properties \cite{zhang_trainability_2020, pesah_absence_2020}.

\subsubsection{Reinforcement learning}
In reinforcement learning for \ac{qml}, the main expectation is that quantum phenomena, namely superposition and entanglement, can provide quantum speed-up with respect to classical methods \cite{Dong_quantum_2008, Dunjko_quantum_2016, Dunjko_advances_2017}. Preliminary reinforcement learning algorithms have been already tested \cite{Jerbi_quantum_2021}.

In the quantum approach for reinforcement learning, agent and environment have access to different Hilbert spaces and interchange information via unitary maps. The quantum mapping opens up a larger range of possibilities as compared to the classical counterpart since there is much more freedom in the former. 

Reinforcement learning has been used both as an application of \ac{vqa}s and quantum annealing. In the case of variational circuits, subsequent improvements have been accomplished increasing the efficiency and hardness of the problem to solve \cite{Chen_variational_2020, lockwood_reinforcement_2020, lockwood_atari_2021}. For the case of annealing, the main quantum advantage can be achieved by optimization of complex systems, in particular for \ac{bm} \cite{crawford_reinforcement_2019}. Experimental implementations have also appeared recently \cite{Lamata_superconducting_2017, cardenaslopez_multiqubit_2018, Yu_reconstruction_2019}.

\chapterimage{chapter_reuploading.pdf}

\newcommand{\relu}{{\rm ReLU}}
\newcommand{\poly}{{\rm poly}}
\newcommand{\step}{{\rm step}}

\chapter[Data re-uploading strategy for QML]{Data re-uploading strategy for \ac{qml}}\label{ch:reuploading}

\begin{adjustwidth}{4cm}{0cm}
{\sl Las cosas que parecen duras tienen una elasticidad... \\ ~\hfill ...una elasticidad retardada.}

\hfill Julio Cortázar\\
\end{adjustwidth}

% For introduction?
One of the main features of quantum computing relies on the high density of the computational space.
Classical computing is built upon discrete frameworks of strings 
composed of $0$'s and $1$'s, in contradistinction to quantum computing where all superpositions of a number of states are, in principle, available.
Thus, every quantum computational system conforms a dense space where an infinite number of states
can be described. The main focus now are the capabilities of the smallest possible quantum system. A single qubit is defined 
by the computational eigenstates $\ket 0$ and $\ket 1$. As mentioned, this qubit can in principle be any state of an 
infinite plethora of superpositions of both eigenstates. This observation permits proposing
algorithms where the focus is not on the different states carried by a quantum computation, but rather on 
the coefficients that define a given quantum state. In fact, a qubit can only store one bit of 
information (0, 1, alternatively) in their states of the computational basis, but there is room for two natural numbers (or one 
complex) in its internal degrees of freedom. An example of a state fulfilling this condition would be

\begin{equation}\label{eq:qubit_state}
\ket\psi = \sqrt{1 - f^2} \ket 0 + f e^{i\phi}\ket 1, 
\end{equation}
with $f \in [0, 1]$ and $\phi\in [0, 2\pi)$. In this example, the encoded complex number is $f e^{i\phi}$

It is straightforward and efficient to drive one qubit to the state $\ket\psi$ by means of some single-qubit
rotation. For example, starting from the standard state $\ket 0$, $\ket\psi = R_z(\phi) R_y(2 
\arcsin(f)) \ket 0$. It is also direct to extend this recipe to store functions $f(x), \phi(x)$ instead of fixed values in a 
quantum circuit, only if the functional forms are known. Therefore, there is room for any complex function $z(x) \in 
\mathbb{C}; \vert z(x)\vert \leq 1$ in a single-qubit circuit. Thus, storing mathematical functions
can be achieved by passing the dependency on $x$ to the parameters defining the
operations in the circuit. 

In case the quantum system to manipulate is larger, the complexity of 
the algorithm to tune all the coefficients of such state grows exponentially with the number of qubits. 
In addition to naive approaches, some other sophisticated techniques exist to deal with similar problems, like signal 
processing or qubitization~\cite{low_optimal_2017, low_hamiltonian_2019, huang_power_2021}.

The situation becomes much more complicated when the function to store in the quantum system is known only through samples, that is, its functional form remains veiled. In this case, it is simply not possible to load well-defined functions to the circuit. This scenario becomes of interest in the field of \ac{ml}. The re-uploading strategy here presented is a \ac{ml} framework for learning functions from samples, that is solving a problem generalizing data, using a quantum computer assisted by a classical optimizer. % Puede dársele una vuelta

Re-uploading makes use of several gates 
depending on independent variables or data $x$ and some tunable parameters. The gates are standard single-qubit rotations on the Bloch sphere. These gates are applied sequentially, that is, 
the independent variable is {\sl re-uploaded} throughout the circuit. The more re-uploadings of $x$ come 
into the circuit, the more flexibility the circuit has to produce a determined output state. The dependency in $x$ within the gates is linear and kept simple, so that the scheme can adapt to any kind of output state. It also avoids the emergence of biases since no feature of the desired function is introduced in the gates. The tunable 
parameters present in the different gates have the power to arbitarily shape the final output state of a 
quantum circuit given a sufficient number of re-uploadings. To find the optimal parameters, classical 
optimization methods are utilized to force the output state to match some conditions encoded within the definition of a loss 
function. With these ingredients it is possible to claim that single-qubit systems are capable to store in 
its degrees of freedom functions only learnt from samples.

The re-uploading strategy permits to circumvent a fundamental limitation existing in quantum computing, 
namely the no-cloning theorem~\cite{wootters_single_1982}. Quantum no-cloning theorem prevents a quantum 
state to be copied into some other quantum register. This limits the available processing of quantum data 
to either modify it only once or to have several copies of such data to process it in different steps. In 
contradistinction, classical devices can copy data. For instance, a \ac{nn} takes the same 
input as many times as desired when processing it. The implicit solution considered by the re-uploading 
strategy consists therefore in using a classical device to copy data and introduce it repeatedly in the 
quantum circuit. In fact, using this line of thought it is possible to find an equivalence between \ac{nn}s 
and this re-uploading scheme, see Sec.~\ref{ssec:theorems}.

This re-uploading technique introduces changes in the overall structure of \acf{qml} algorithms as well. As seen 
in Ch.~\ref{ch:MLbib}, many of these algorithms are performed in three steps, namely uploading data, 
processing data and measurement. Re-uploading combines the upload and processing steps into one stage.
This avoids several difficulties concerning these issues. Re-uploading tecnhniques do not make use 
of sophisticated embedding schemes nor large quantum systems to transfer classical data into the quantum 
ciruits. The linear encoding used in this scheme is capable to 
capture correlations among different features of the sampling data. Due to the classical optimization 
in the recipe, this step is accomplished with no prior knowledge on the dataset the quantum algorithm deals with. 

The linear encoding gives also rise to highly non-trivial functions naturally due to the quantum nature of the algorithm. 
One single-qubit operation is capable to introduce a rotation in the Bloch sphere, where the only complexity in the operation depends on the parameter describing such rotation, in this case a linear dependency. This does not supply enough computational power to deal with non-trivial problems. 
However, a consecutive application of non-commuting single-qubit operations triggers the appearence of non-linear terms. These terms are the reason why the re-uploading strategy is a flexible and general model to address \ac{qml} problems.

The re-uploading technique emphasizes the capabilities of quantum circuits that make use of a small 
number of quantum resources. They are of most importance to the field of quantum computing, in particular 
for the first phase of quantum computing or \ac{nisq} era~\cite{preskill_quantum_2018}. In fact, algorithms 
that need few qubits may be proven relevant even though they do not attempt any quantum improvement, since 
they may be useful pieces of larger and more advantageous circuits. Considering circuits with few quantum 
resources as the building blocks for larger circuits avoids a drawback to arise as well. Even though it is 
possible to store a complex function in a single-qubit circuit, retrieving that information from the 
quantum state is costly and requires a large amount of measurements, for instance performing full tomography methods~\cite{dariano_tomography_2003}. Indeed, the Holevo bound~\cite{holevo_bounds_1973} limits the accesible information when a quantum system of $n$ qubits is 
measured to only $n$ classical bits at a time. 

The linear encoding used in the re-uploading scheme is another advantage when using this strategy
as a piece of larger circuits. Due to this encoding it is possible to include in the upload-processing
step data stored in some quantum register. This is feasible using controlled gates whose control qubits 
are the data register, and the target qubits are the processing ones. To do so, a map between the 
original scheme and the controlled rotations must be included into the quantum circuit. 

This chapter is based on the articles from 
Refs.~\cite{perezsalinas_data_2020,perezsalinas_qubit_2021, 
perezsalinas_proton_2021,dutta_realization_2021}. It is structured as follows. First, the 
theoretical aspects and comparison of re-uploading with \ac{cml}
approaches are detailed in Sec.~\ref{sec:reuploading_theorems}. Then, numerical 
benchmarks for some test functions both in classical simulators and experimental 
devices can be found in Sec.~\ref{sec:benchmark}. Two applications based on this 
scheme with superficial modifications are then explained in the following sections. Sec.~\ref{sec:qlassifier} 
contains a quantum classifier accomplished in classical simulations, and Sec.~\ref{sec:exp_qlassifier} describes the experimental implementation of this classifier on
an 
ion-trapped quantum device. Sec.~\ref{sec:qpdf} is devoted to a machine learning 
approach to \ac{hep} application to determine the content of protons 
from experimental data obtained at \ac{lhc}.

\section{Theoretical support}\label{sec:reuploading_theorems}

There exists a fundamental question in the field of \ac{ml} of whether a given model can represent any function. In this case, the model is a quantum circuit following the re-uploading strategy, and
any possible functionality must be encoded
within the degrees of freedom of the output state. If that is the case, it is important to find the sequence of gates required to accomplish this goal. This section is devoted to answer these questions.

In classical cases, this problem was solved by a series of theorems establishing that a given function can be re-expressed as a 
linear combina-tions of other specific functions. The most fundamental family of theoretical results in this field is the harmonic anaylisis and Fourier series~\cite{dirichlet_fourier_1829, riemann_fourier_1867}. Those results demonstrate that a great range of functions can be re-expressed as sums of trigonometric functions with fixed frequencies. In classical machine learning, the \ac{uat} proves that a \ac{nn} with a unique intermediate hidden layer can converge to approximate 
any continuous function~\cite{cybenko_approximation_1989, hornik_approximation_1991}. In both approaches, it is important to notice that each step 
of the process, namely neurons for \ac{uat} or terms for Fourier series, is fed with the original data of the problem. The query 
complexity of the process increases linearly with the number of steps, namely the degree of approximation.

Data re-uploading strategies are conceived as a quantum analogous of the well-known classical model of \ac{nn}s. 
In the case of feed-forwarding \ac{nn}s, data is entered in the network in such a way that it is processed by subsequent layers of neurons. The key observation is that the original data is introduced as many times as neurons in the first hidden layer, and then the result of one layer is introduced several times in the next one until the output layer is reached. Strictly speaking, data is re-uploaded onto the \ac{nn}. In case \ac{nn} were affected by some sort of no-cloning theorem, they could not work as they do. Thus, a circuit is designed such that its architecture allows data to be introduced several times. This observation is critical to develop theoretical support  for the expressibility of quantum circuits in order to support progress 
in \ac{qml}~\cite{perezsalinas_data_2020, perezsalinas_proton_2021, bravoprieto_quantum_2020, 
mitarai_quantum_2018, zhu_training_2019, lloyd_embeddings_2020, liu_rigorous_2020, rebentrost_qsvm_2014, 
lloyd_quantum_2013, schuld_effect_2021, 
goto_universal_2020, sim_expressibility_2019, nakaji_expressibility_2021}. 

\begin{figure}[t]
\centering
\subfigure[Classical \ac{nn}]{
\includegraphics[height=.4\linewidth]{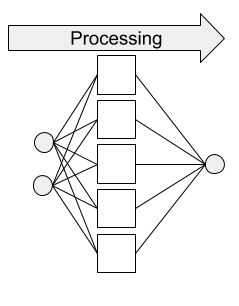}}
\subfigure[Re-uploading Quantum Circuit]{
\includegraphics[height=.4\linewidth]{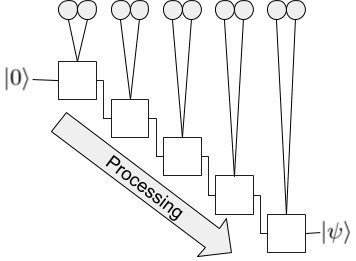}}
\caption{Simplified working schemes of a classical \ac{nn} and a single-qubit re-uploading scheme. In the \ac{nn}, the processing layer receives information from every neuron in the input layer, and the processing is done in parallel. In contradistinction, the single-qubit circuit receives input from the previous processing unit and the data (introduced classically). It processes everything sequentially. The output of both models are flexible enough to accomodate many constraints.}
\label{fig:reuploading_scheme}
\end{figure}

An architecture where data can be re-uploaded and processed along the computation is designed. Fig.~\ref{fig:reuploading_scheme} shows a comparison between a single-layer classical \ac{nn} model and the quantum circuit here proposed.
In the case of \ac{nn}, data is re-uploaded many times in one step, once per neuron, and processed in parallel. Then, all partial processings are collected into a final output neuron. This model can approximate any continuous function in the output neuron, as stated in Th.~\ref{th:UAT}, \cite{cybenko_approximation_1989, hornik_approximation_1991}. In the case of the quantum circuit, data points are introduced in each processing unit, in this case a single-qubit operation. As it is a single-qubit system, there is no room for parallel processing, and then the calculation is done sequentially, since every gate receives as input the result from the last one.

In this chapter, it is shown that quantum and classical models here discussed are formally equivalent for the same number of processing units. Two independent proofs that any bounded complex function can be approximated in a convergent way by a single-qubit quantum 
circuit are presented. Thus, it constitutes a {\bf single-qubit approximant}. This demonstrates the precise representation 
power of a single-qubit circuit, which increases as more layers are added. As for classical models, query complexity, achieved through re-uploading of data, is 
attached to accuracy. The first proof makes contact with harmonic analysis. This is a natural step as single-qubit gates 
are expandable in Fourier series that can be arranged to fit existing theorems. The second method is analogous to the \ac{uat} 
using a translation into quantum circuits. In both cases, the quantum theorems inherit the applicability and characteristics 
of their classical counterparts. 

In this section, the problem is first defined in Sec.~\ref{ssec:setup}. The two theorems constituting the core of the work are presented in Sec.~\ref{ssec:theorems}, and demonstrated in Sec.~\ref{ssec:proofs_univ_theorems}. Conclusions are commented in Sec.~\ref{ssec:conclusions_th}.

\subsection{Set-up of the problem}\label{ssec:setup}

The most general representation of a single-qubit quantum state stores a single complex number, as stated in Eq.~\eqref{eq:qubit_state}, explicitly
with $f, \phi$ real numbers and $f \in [0, 1]$, $\phi \in [0, 2\pi)$. The aim is to encode a complex function within the values ($f, \phi$) by defining them as $ f:\mathbb{R}^m \rightarrow [0, 1]$ and $\phi: \mathbb{R}^m \rightarrow [0, 2\pi)$. The functional forms of $f(x)$ and $\phi(x)$ are unknown, otherwise solving this problem is trivial. To do so,
the circuit $\mathcal U^{(k)}_{f, \phi}(x)$ is designed in such a way that its output state approximates the desired complex function as
\begin{equation}\label{eq:aprox}
    \bra{1}\mathcal{U}^{(k)}_{f, \phi}\ket{0} \sim f(x) e^{i \phi(x)},
\end{equation}
where $k$ stands for the number of uploadings of data, also referred to as query complexity.
Note that building an approximation to a bounded complex function suffices to address any bounded complex function by shifting and re-scaling the target function to another one that fits in the model. In addition, approximating a complex function includes the capability of fitting real-valued functions by either setting $\phi(x) = 0$ or relating the real-valued function to the modulus of other complex functions. The latter approach is lesser demanding since a degree of freedom is set free. Initial $\ket 0$ and comparing $\bra 1$ can be chosen arbitrarily without loss of generality, since any state can be transformed into any other by slightly modifying the $\mathcal{U}^{(k)}_{f, \phi}$ operation. The general operation is described in a specific way.
\begin{definition} \label{def:prod_general}
The $k$-th approximating circuit is defined as 
\begin{equation} \label{eq:prod_general}
    \mathcal{U}^{(k)}_{f, \phi}(x, \Theta) = \prod_{i = 1}^k U(x, \vec\theta_i),
\end{equation}
where $U(x, \vec\theta)$ is a fundamental gate depending on $x$ and a set of parameters $\vec\theta$, with $\Theta = \{\vec\theta_1, \ldots, \vec\theta_k\}$. 
\end{definition}

This general construction allows to obtain great performances in \ac{qml} problems for a wide variety of fundamental gates $U(x, \vec\theta)$, as it is seen in later examples. However, there are two particular choices, to be defined later, that provide mathematically provable universality. The expected behavior of this operation $\mathcal{U}^{(k)}_{f, \phi}$ is that the approximation from Eq.~\eqref{eq:aprox} will improve as the number $k$ increases, that is, as the independent variable is re-uploaded multiple times and the query complexity increases. The appropriate choice of the parameters $\Theta$ enables a systematic approximation of any functionality. The optimal configuration will depend on $f(x)$ and $\phi(x)$.

\begin{figure}[t!]
\begin{adjustwidth}{-2cm}{-1cm}
\centering
\subfigure[$x$-independent rotations\label{fig:bloch_a}]{\includegraphics[width=.3\linewidth]{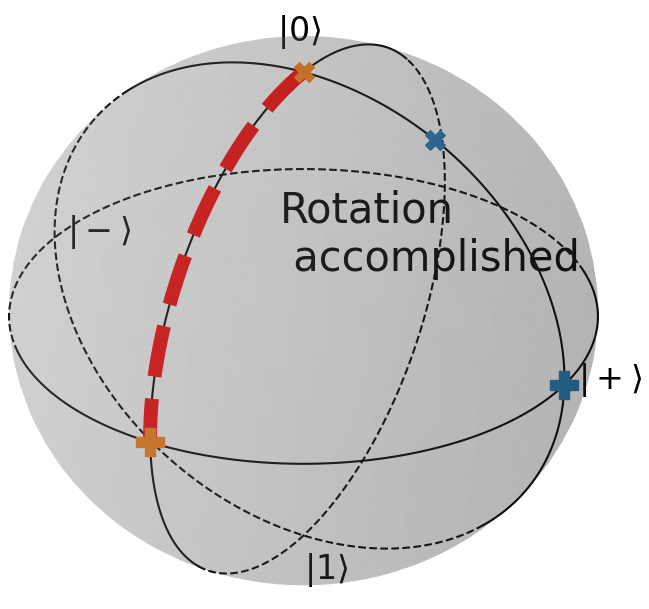} \includegraphics[width=.3\linewidth]{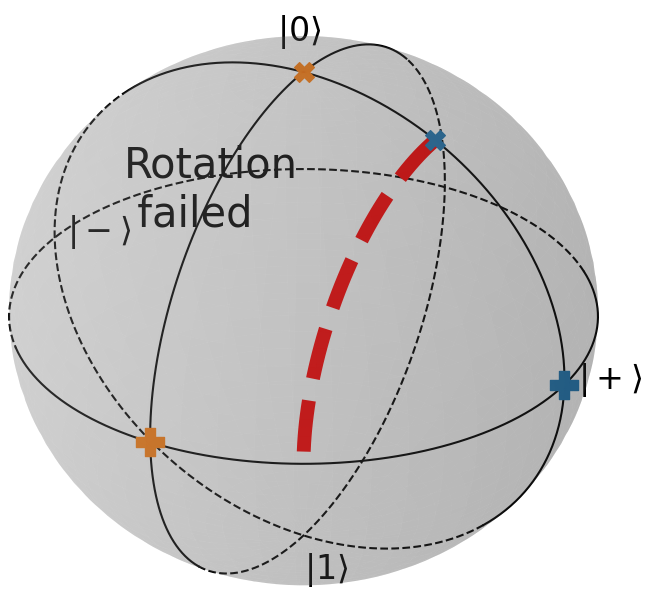}}
\hfill
\subfigure[$x$-dependent rotations\label{fig:bloch_b}]{\includegraphics[width=.3\linewidth]{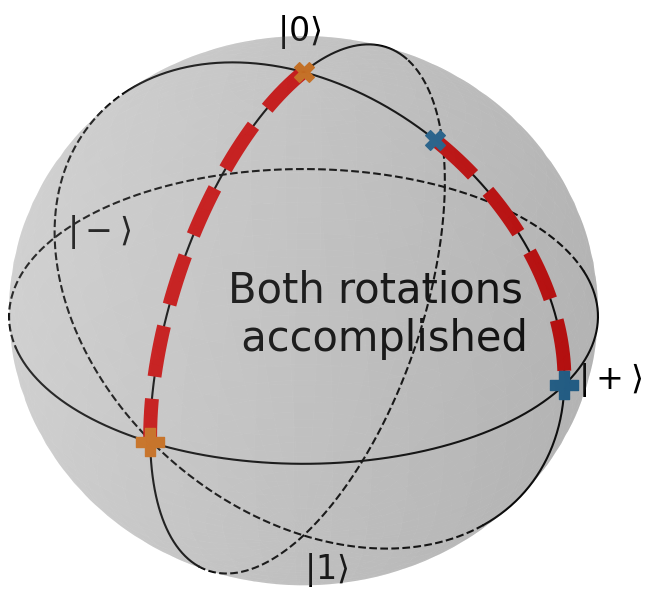}}
\end{adjustwidth}
\caption{Geometrical interpretation of standard single-qubit operations (a) and the re-uploading strategy (b). The initial states, $X$ crosses around the $\ket 0$ state, must move towards two different target states, $+$ crosses in the equatorial plane $\langle Z\rangle = 0$. Standard operations are combination of rotations, what leads to another rotation. Thus, it is not possible to achieve a tailored rotation for both initial states. On the contrary, many $x$-dependent rotations allows to obtain overall flexible rotations.}
\label{fig:bloch}
\end{figure}

It is possible to interpret the operation for the re-uploading strategy in Def.~\ref{def:prod_general} for a single-qubit system by making use of the geometrical arguments picturing the Bloch sphere. Two initial states on the Bloch sphere are considered. Those states must be rotated by the same operation $U(x, \vec\theta)$ to another arbitrary pair of states. In the case where rotations do not depend on $x$ is first considered, many different operations can be applied, but all of them can be fused into an overall one. The optimal rotation for one point does not fit another one, and thus accomodating several data is not possible, see Fig.~\ref{fig:bloch_a} for an illustration. On the other hand, making the operations $x$-dependent provides the possibility to transport different points along different paths, see Fig.~\ref{fig:bloch_b}. As more layers are applied, more independent rotations are included into the circuit, and the conjunction of many simple operations permits to obtain systems of increasing flexibility.

It is also worth to mention that the re-uploading strategy allows to increase the distinguishability between two data points. General strategies of \ac{qml} look for circuits that provides optimal measurement schemes for some given datasets, see Sec.~\ref{ssec:supervised_qml}. In that case the performance of the classifier is upper-bounded by the distinguishability obtained by the embedding scheme when uploading data into the quantum circuit. Re-uploading circumvents this limitation. This becomes particularly useful when two close points $x_1$, $x_2$ have very different properties, for example in the vicinity of the border between two classes in a supervised learning problem. 

In general, the set of parameters for a given gate
$\vec\theta_i$ is composed of a set of angles.
The quest for the optimal set of parameters $\Theta$ is driven by optimizing a particular loss function $\mathcal{L}(\Theta; f, \phi, x)$. This loss function must be designed in such a way that Eq.~\eqref{eq:aprox} becomes an equality as $\mathcal{L} \rightarrow 0$. The optimal parameters are then
\begin{equation}
    \Theta_{\rm opt} = {\rm argmin}_\Theta \mathcal{L}(\Theta; f, \phi, x).
\end{equation} 
The presence of classical optimization methods makes this scheme belong to the family of variational algorithms.

\subsubsection{Optimization techniques}
In this chapter, optimizers are taken as black boxes. Finding the optimal configuration corresponding to a minimum (or equivalently a maximum) of a given cost function is a extremely complicated, in fact a NP, problem. As the number of parameters increases, the search space grows exponentially, and so does the difficulty of the problem. Of course, this also depends on the particular loss function that is to be solved. Purely convex functions are usually easier to optimize since all movements lead to the global minimum. Take for instance the function $f(x) = \vert x \vert^2$. In this case, the obvious solution is $x = 0$. However, this scenario is certainly not common in general optimization problems.

One of the hardest features of the optimization procedures is that there is no possible way, apart from exhaustive search, to ensure that a given configuration $\Theta_{\rm opt}$ provided by an optimizer is the best possible one, that is the global minimum. One can know if this is a local minimum, that is whether there are or not better points in the vicinity of that solution, but all remaining space is unexplored. This becomes particularly difficult when the loss function is full of local minima, for example if $f(x)$ is some combination of trigonometric functions. In this case it is very likely to get trapped in a local minimum, while this local minimum will not probably be the global one. 

\ac{nn}s are commonly trained using the Stochastic Gradient Descent~(SGD) \cite{nielsen_neural_2015} with standardized acceptable results, see Sec.~\ref{ssec:nn}. This is a consequence of a previous knowledge on the landscape of the loss functions. Unlike \ac{nn}s, quantum algorithms do not have any particular optimization method with high performance in wide varieties of problems. In the case of the \ac{qml} problems presented along this chapter, even the landscape is unknown. Therefore, the optimization strategy to follow to obtain the best possible results is far from being trivial. 

A first attempt on developing direct SGD was intended following known recipes for estimating gradients on quantum circuits, see Sec.~\ref{ssec:supervised_qml}, \cite{hubregtsen_singlecomponent_2021, sweke_stochastic_2020, schuld_evaluating_2019}. This was tried in the seminal quantum classifier paper~\cite{perezsalinas_data_2020}. However, the results obtained in this scenario did not rise to the challenge when compared against other methods. In summary, final results did get stuck in local minima far away from results obtained with other methods. 

There are two optimization methods that have worked along the examples presented in this chapter, namely the quasi-Newton {\tt L-BFGS(-B)} method \cite{l-bfgs}, and the genetic {\tt CMA} algorithm \cite{cma}. 

The {\tt L-BFGS(-B)} method constructs in every iteration an approximation of the Hessian matrix taking as input information the function evaluation in the current iteration and previous ones, up to an adjustable limit. On the one hand, this method can explore the vicinity of a given point by computing first and second order derivatives. On the other hand, the algorithm saves memories of the previous steps, and then small increases of the landscapes can be overcome using informetion collected before. This idea follows from the inertial thought implemented in other optimizers \cite{kingma_adam_2017, nielsen_neural_2015}. Thus, the search for optimal points is a downhill descent in average. These properties make the algorithm resilient to get stuck in shallow local minima. In this work, the method is implemented as given by {\tt scipy}~\cite{scipy}.

The {\tt L-BFGS(-B)} method is sometimes used in \ac{nn}s when the training dataset is small. This is understood in terms of local minima. Landscapes of \ac{nn}s, as in the quantum models observed in this section, are full of local minima. SGD overcomes them by using different batches of training data and descending in different directions so that the final average is a good result. If there is no data available, some other method less sensitive to local minima is needed. 

The second method commonly used through this chapter is the genetic algorithm {\tt CMA}. Genetic algorithms are inspired by biological evolution and natural selection. An initial population is generated, and from those individuals only the most adapted ones are chosen to propagate their information to the next generation, which is created by adding changes to their parents. This way, the landscape is explored and only the individuals with smaller values of the cost function survive, contributing to the overall improvement of the population. This algorithm, which is not gradient-based, permits exploring vast landscapes where the derivative information is missing or useless. In exchange, large numbers of function evaluations are usually required to go depp into the landscape and find actual global minima, specially due to the \ac{bp} phenomenon, see Sec.~\ref{sssec:training_vqa} \cite{McClean_bp_2018}. Nevertheless, genetic algorithms can extract the interesting combination of parameters very efficiently. The particular {\tt CMA} strategy chosen stands for the specific manner to create new generations. It generates new individuals as gaussian perturbations of the parents \cite{cma}. In this chapter, the algorithm is implemented as provided by Ref.~\cite{cma_package}.

It is worth mentioning that optimization is carried mostly on quantum simulators in all examples here presented, unless stated otherwise. This eliminates a further difficulty that optimization of quantum circuits have. In the future, specific optimization methods for quantum computing will be needed to address these problems efficiently. The reason is that quantum computng has two main sources of errors that degrade the retrievement of values for loss functions: noise and decoherence from the circuit, and sampling uncertainty from the measurements. In principle, optimization techniques can help to mitigate systematic errors, but statistical ones pose a challenge on the attainment of acceptable results. 

In terms of noise and decoherence, the functions evaluations are distorted when measured, and thus it is not possible to accurately determine the function values. Thus, the optimizer receives corrupted information and is is simply not possible to drive the search for optimal parameters efficiently. As the quality of quantum computers increases towards fault tolerant computing, these errors will be slowly corrected and mitigated. 

Sampling uncertainty adds further unaccuracies to the measured values obtained. This becomes important when decoherence is neglictible since it becomes the main source of errors. The error can be only reduced by increasing the number of measurements, but it is not possible to overcome it. The effect of sampling uncertainties attacks the core of many optimization methods that rely on high precision computing. This dependency must be circumvented to progress towards efficient optimizers, see Sec. \ref{sssec:training_vqa} \cite{sweke_stochastic_2020}.

\subsection{Two theorems on universality}\label{ssec:theorems}
The structure of the algorithm previously presented is completed with the design of the single-qubit gates $U$ aforementioned in Definition~\ref{def:prod_general}. In the following, two sets of single-qubit gates are used to construct quantum circuits that represent arbitrary complex functions. Each set is based on known results from the theory of function approximations, namely Fourier series~\cite{dirichlet_fourier_1829, riemann_fourier_1867} and \ac{uat}~\cite{cybenko_approximation_1989, hornik_approximation_1991}, respectively. The range of applicability of these theorems for quantum circuits and the conditions for universality are thus inherited from their classical counterparts.

\paragraph*{Non-linearities}\label{paragraph:non-linear}

Before coming into technical details for demonstrating universality, it is important to highlight a requirement of $\mathcal{U}^{(k)}_{f, \phi}$, that is the emergence of non-linearities. Non-linearities are needed in all methods looking for universal representability. They appear as an essential ingredient in all classical theorems. For instance, the Fourier series~\cite{dirichlet_fourier_1829, riemann_fourier_1867} are built upon trigonometric functions, and \ac{uat}~\cite{cybenko_approximation_1989, hornik_approximation_1991} explicitly require non-linear functional forms. While in the classical case non-linearities are introduced artificially, the quantum case makes them appear taking advantage of inherent properties of quantum operations. For this purpose, the fundamental gates $U(x, \vec\theta)$ cannot commute between them for any combination of $(x, \theta)$, except for the trivial case where two gates are equal. Due to the mathematical structure of the $SU(2)$ group containing all possible single-qubit operators, non-linear terms emerge naturally. This can be observed via the \ac{bch} formula~\cite{van-brunt_explicit_2018, eichler_new_1968, hall_bakercampbellhausdorff_2015}.

\begin{definition}{\ac{bch} formula}

Let $e^M = e^Y e^Z$ be, for $M, Y, Z$ matrices and possibly $[Y, Z] \neq 0$. The \ac{bch} formula gives the solution to the matrix $M$ in terms of a formal series not necessarily convergent whose first terms are
\begin{equation}
M = Y + Z + \frac{1}{2}[Y, Z] + \frac{1}{12} \left([Y,[Y,Z]] + [Z,[Z,Y]] \right) +  \ldots
\end{equation}
\end{definition}

The quantum operations $R_z(\alpha), R_y(\varphi)$, corresponding to exponentiation for $ = -i \alpha / 2 \sigma_z, Y = -i \varphi/2 \sigma_y$, are considerd. The overall operation of interest is $U = e^{-i \varphi / 2 \sigma_y}e^{-i \alpha/2 \sigma_z}$. Following the \ac{bch} formula, this final operation can be written as
\begin{equation}
U = e^{-i M};\qquad M = \frac{\varphi}{2} \sigma_Y +\frac{\alpha}{2} \sigma_z + \frac{\alpha \varphi}{4} \sigma_X -\frac{\alpha\varphi}{24}\left(\varphi \sigma_Y + \alpha \sigma_Y \right).
\end{equation}

In this particular case, the series comes to an end because of the commutation relationships between the Pauli matrices $\sigma_Y, \sigma_Z$. However, even in this simple case, it is possible to detect polynomial terms up to degree 3 $(\alpha^2\varphi, \alpha\varphi^2)$, while the original operations depend only on variables with polynomial degree 1. 

This formulation gives some insight on how to design global universal operations \cite{perezsalinas_data_2020}. Rotations around at least two axis are required to make non-linear quantities appear. However, it is also remarkable that a rotation around a third axis contributes with more parameters, but it is not strictly needed since this third axis emerges naturally with the first non-commuting term. 

The single-qubit classifier is constructed as a series of gates which are in general $SU(2)$ matrices. There exist many possible decompositions of an $SU(2)$ rotational matrix. In particular, 

\begin{equation}
U(\beta, \varphi, \alpha) = e^{\frac{-i\beta}{2} \sigma_z} e^{\frac{-i\varphi}{2} \sigma_y} e^{\frac{-i\alpha}{2} \sigma_z}.
\end{equation} 

According to Def.~\ref{def:prod_general}, the overall operation is generated as a series of simple rotations. Thus, with no loss of generality, $\beta = 0$, since this parameter can always be absorbed by the $\alpha$ parameter of the following layer. Using the $SU(2)$ decomposition law, the above parametrization can be written in a single exponential

\begin{equation}
U(\beta = 0, \varphi, \alpha)=e^{-i \vec{\omega}(\varphi, \alpha)\cdot\vec{\sigma}},
\end{equation}
with $\vec{\omega}(\varphi, \alpha)=\left(\omega_{1}(\varphi, \alpha),\omega_{2}(\varphi, \alpha),\omega_{3}(\varphi, \alpha)\right)$ and
\begin{align}
\omega_{1}(\varphi, \alpha)&= d \  \mathcal{N}\sin\left(-\alpha/2\right)\sin\left(\varphi/2\right),\\
\omega_{2}(\varphi, \alpha)&= d \  \mathcal{N}\cos\left(\alpha/2\right)\sin\left(\varphi/2\right) , \\ 
\omega_{3}(\varphi, \alpha)&= d \ \mathcal{N}\sin\left(\alpha/2\right)\cos\left(\varphi/2\right),
\end{align}
where $\mathcal{N}=\left(\sqrt{1-\cos^2d} \right)^{-1}$ and $\cos d=\cos\left(\alpha/2\right)\cos\left(\varphi/2\right)$.

The re-uploading scheme codifies an independent variable $x$ into the parameters of the $U$ gate as $\alpha(x), \varphi(x)$, where the exact dependency is to be defined yet. Thus 
\begin{equation}
\mathcal{U}^{(k)}_{f, \phi}(x, \Theta) = \prod_{i = 1}^k e^{-i \vec{\omega}(\varphi(x), \alpha(x))\cdot\vec{\sigma}}.
\end{equation}

Applying the \ac{bch} formula to the above equation, the overall expression of the global operation is
\begin{equation}
\mathcal{U}^{(k)}_{f, \phi}(x, \Theta) = \exp\left(-i \sum_{i=1}^k \vec{\omega}(\varphi_i(x), \alpha_i(x)) + \mathcal{O}_{\rm corr} \right),
\end{equation}
where $\mathcal{O}_{\rm corr}$ involves higher order terms of Pauli matrices due to the property $[\sigma_i, \sigma_j] = 2 i \varepsilon_{ijk} \sigma_k$, where $\varepsilon_{ijk}$ is the Levi-Civitta symbol. 

All $\vec \omega$ terms are trigonometric function, and thus unconstant, bounded and continuous, as required by \ac{uat}~\cite{hornik_approximation_1991}. Hence, the sum of all of them must fulfill the same properties.
\begin{equation}
\sum_{i=1}^k \vec{\omega}(\varphi_i(x), \alpha_i(x)) = \vec{\eta}(x).
\end{equation}

There are still remaining terms $\mathcal{O}_{\rm corr}$ of the \ac{bch} expansion. Instead of applying such expansion, it is possible to use again the SU(2) group composition law to obtain the analytical formula of $\mathcal{U}^{(k)}_{f, \phi}(x, \Theta)= e^{i\vec{\xi}\left(x\right)\cdot\vec{\sigma}}$, where $\vec{\xi}(x)$ will be an inextricably trigonometric function of $x$. The $\mathcal{O}_{corr}$ terms are proportional to $\vec{\sigma}$ matrices, so $\mathcal{O}_{corr} = \vec{\varrho}(x)\cdot\vec{\sigma}$ for some function $\vec{\varrho}(x)$. Then,
\begin{equation}
\mathcal{U}^{(k)}_{f, \phi}(x, \Theta) = e^{i\vec{\xi}(x)\cdot\vec{\sigma}} = e^{i\vec{\eta}(x)\cdot\vec{\sigma} + i\vec{\varrho}(x)\cdot\vec{\sigma}}.
\end{equation}
Thus, $\mathcal{O}_{corr}$ terms can be absorbed in $\vec{\eta}(\vec{x})$. As the resulting function is a combination of trigonometric function, which satisfies the constraints for Th. \ref{th:UAT}, it is expected to have enough flexibility as to represent a huge variety of functions.

Notice that the arguments on the appearances of higher-order terms may provide insights and lines of thought towards the attainment of universality, but they do not suffice to demonstrate it. Up to this point there is no further knowledge of the inner structure of $\vec\xi(x)$. The many combined trigonometric functions can contribute to cancel terms with each other so that the final result does not provide universality, even though it has all the required ingredients. Therefore, further development is needed to prove universality.

\subsubsection*{Theorems} 

\begin{figure}[p!]
\centering
{\large \bf FOURIER SERIES}\\

\begin{theorem}\label{th:fourier}
{\bf Fourier series}\\ \cite{dirichlet_fourier_1829, riemann_fourier_1867, carleson_fourier_1966}

Let $z$ be any function $z: \mathbb{R} \rightarrow \mathbb{C}$ with a finite number of finite discontinuities integrable within an interval $[a, b] \in \mathbb{R}$ of length $P$. Then 
\begin{equation}\label{eq:fourier_exp_complex}
    z_N(x) = \sum_{n = -N}^N c_n e^{i \frac{2\pi n x}{P}},
\end{equation}
where
\begin{equation}
    c_n = \frac{1}{P} \int_P z(x) e^{-i\frac{2\pi n x}{P}} dx,
\end{equation}
approximates $z(x)$ as
\begin{equation}
    \lim_{N\rightarrow\infty} z_N(x) = z(x).
\end{equation}
\end{theorem}

\begin{definition}\label{def:fourier_gate}
Let the fundamental Fourier gate $U^\mathcal F$ be
\begin{equation}\label{eq:unitary_f}
 U^{\mathcal{F}}(x; \underbrace{\omega, \alpha, \beta, \varphi, \lambda}_{\vec\theta}) = R_z\left(\alpha + \beta\right) R_y(2\lambda)R_z\left(\alpha - \beta\right) R_z(2\omega x) R_y(2\varphi),
\end{equation}
with $\alpha, \beta, \varphi, \lambda, \omega \in \mathbb{R}$.
\end{definition}

\begin{theorem}\label{th:q_fourier}
{\bf Quantum Fourier series} 

Let $f,\phi$ be any pair of functions $f: \mathbb{R} \rightarrow [0, 1]$ and $\phi: \mathbb{R} \rightarrow [0, 2\pi)$ , such that $z(x) = f(x)e^{i\phi(x)}$ is a complex function with a finite number of finite discontinuities integrable within an interval $[a, b] \in \mathbb{R}$ of length $P$.
Then, there exists a set of parameters $\lbrace\vec\theta_1, \vec\theta_2, \ldots, \vec\theta_N \rbrace$ such that
\begin{equation}
     \bra 1 \prod_{i=1}^N U^\mathcal{F}(x, \vec\theta_i) \ket 0 = z_N(x), 
\end{equation}
with $z_N(x)$ the $N$-terms Fourier series. 

Proof in Sec.~\ref{sssec:dem_fourier}.
\end{theorem}
\end{figure}

\begin{figure}
\centering
{\large \bf UAT}

\begin{theorem}\label{th:UAT}
{\bf Universal Approximation Theorem}\\ \cite{cybenko_approximation_1989, hornik_approximation_1991, leshno_multilayer_1993}

Let $I_m$ denote the $m$-dimensional cube $[0, 1]^m$. The space of continuous functions on $I_m$ is denoted by $C(I_m)$, and  $|\cdot|$ denotes the uniform norm of any function in $C(I_m)$. 
Let $\sigma:\mathbb R \rightarrow \mathbb{R}$ be any non-constant bounded continuous function. 
Given a function $f \in C(I_m)$ there exists an integer $N$ and a function 
\begin{equation}\label{eq:UAT}
    G(\vec x) = \sum_{n=1}^N \alpha_n \sigma(\vec w_n \cdot \vec x + b_n),
\end{equation}
such that 
\begin{equation}
    |G(\vec x) - f(\vec x)| < \varepsilon, \qquad \forall \vec x \in I_m,
\end{equation}
for $\vec w_n\in \mathbb{R}^m$ and $b_n, \alpha_n \in \mathbb{R}$ for any $\varepsilon > 0$.
\end{theorem}
\begin{definition}\label{def:uat_gate}
Let the fundamental \ac{uat} gate $U^{\rm UAT}$ be
\begin{equation}\label{eq:unitary_uat}
 U^{\rm UAT}(\vec x; \underbrace{\vec \omega, \alpha, \varphi}_{\vec\theta}) = R_y(2\varphi)R_z(2\vec\omega\cdot\vec x + 2\alpha), 
\end{equation}
with $\lbrace \vec\omega, \alpha, \varphi \rbrace \in \lbrace \mathbb{R}^m, \mathbb{R}, \mathbb{R}\rbrace$.
\end{definition}

\begin{theorem}\label{th:q_UAT}
{\bf Quantum \ac{uat}} \\

Let $f,\phi$ be any pair of functions $f: I_m \rightarrow [0, 1]$ and $\phi: I_m \rightarrow [0, 2\pi)$ , such that $z(\vec x) = f(\vec x)e^{i\phi(\vec x)}$ is a complex continuous function on $I_m$, with $I_m = [0, 1]^m$. Then there is an integer $N$ and a set of parameters $\lbrace\vec\theta_1, \vec\theta_2, \ldots, \vec\theta_N \rbrace$ such that
\begin{equation}
    \left\vert f(\vec x)e^{i\phi(\vec x)} -  \bra 1 \prod_{i=1}^N U^{\rm UAT}(\vec x, \vec\theta_i) \ket{0} \right\vert < \epsilon,
\end{equation}
for any $\epsilon > 0$. 

Proof in Sec.~\ref{sssec:dem_uat}.
\end{theorem}
\end{figure}

Two different methods are explored to link the universality of the quantum circuit here presented with standard mathematical representability theorems, namely Fourier series and \ac{uat}. In both cases, first the mathematical theorems are stated. Then, a quantum gate is defined such that the repeated application of this gate allows to extend those mathematical results to a quantum circuit. 

Fourier series as a constructive method permits expressing a great range of target functions defined within an interval as a sum of a set of known functions, see Th.~\ref{th:fourier}. If a circuit as explicited in Def.~\ref{def:prod_general} is created by repeating the gate $U^\mathcal{F}$from Def.~\ref{def:fourier_gate}, the output state is compatible with a Fourier series, see Th.~\ref{th:q_fourier}. Intuitively, $\alpha, \beta, \varphi, \lambda$ are related to the coefficients of a single Fourier step, while $\omega$ may be identified as the corresponding frequency.
The relationship between these parameters and the original Fourier coefficients is explicitly shown in Sec.~\ref{sssec:dem_fourier}.

When the building blocks are the $U^\mathcal{F}(x, \vec\theta_i)$ defined in Eq.~\eqref{eq:unitary_f}, the unitary operation as defined in Eq.\eqref{eq:prod_general} generates a total unitary gate that outputs a $N$-term Fourier series when applied to an initial state $\ket 0$. Taking $\ket 0$ as the initial state implies no loss of generality, since $\ket 0$ can be transformed into any other initial state by adjusting the first $U^\mathcal{F}$. The Fourier series behavior is only achieved if all $\lbrace\vec\theta_i\rbrace$ take specific values leading to a final result that exactly matches the Fourier coefficients. However, since this procedure relies on quantum-classical variational methods, optimal parameters are searched by means of a classical optimizer. This freedom gives room to configurations surpassing the performance of the standard Fourier series, especially for shallow circuits. However, the recipe to construct the Fourier series by performing well-defined calculations is instead lost. In addition, the Fourier series is obtained only in the last step. Intermediate steps have the functional forms of a Fourier series, but not its values. For details on the proof of this theorem the reader is referred to Sec~\ref{sssec:dem_fourier}.

The \ac{uat}, Th.~\ref{th:UAT} demonstrates that any continuous function of a $m$-dimensional variable can be uniformly approximated as a sum of a specific set of functions with adjustable parameters. The first formulation restricted the functions to be sigmoidal functions~\cite{cybenko_approximation_1989}. Later works extended the result to any non-constant bounded continuous function~\cite{hornik_approximation_1991}. This theorem is directly applied to \ac{nn}s containing one hidden layer. 

This theorem is an existence theorem, and thus it does not specify how many terms from Eq.~\eqref{eq:UAT} are needed to achieve an accuracy $\varepsilon$ nor what values the parameters must take. 
Note that although the \ac{uat} supports approximations for real functions, it can be immediately applied to complex functions by substituting the real-valued function $\sigma(\cdot)$ with some complex-valued function. In particular, it works if $\sigma(\cdot) \rightarrow e^{i (\cdot)}$. A proof is shown in Appendix~\ref{app:real_to_complex_uat}.

In the case of quantum circuits as designed in Def.~\ref{def:prod_general}, a quantum analogous of the UAT is available if the repeated gates is the one from Def.~\ref{def:uat_gate}. Intuitively, $\vec\omega$ and $\alpha$ are equivalent to the weights and bias in a \ac{nn}, while $\varphi$ plays the role of the coefficient. The output state, although different to the standard \ac{uat}, fulfills the same requirements to ensure universality, see Th.~\ref{th:q_UAT}. Both classical and quantum theorems are analogous, and one can arrive at its proof by following the steps developed in Ref.~\cite{cybenko_approximation_1989}. All theorems supporting the original formulation of the \ac{uat} also hold for the quantum version. For more details on the demonstration of the quantum \ac{uat}, the reader is referred to Sec.~\ref{sssec:dem_uat}.

The quantum universality theorems here proposed inherit the range of applicability, advantages and limitations of their classical counterparts. The Fourier approach is guaranteed to work for all integrable functions with a finite number of finite discontinuities. This range of functions includes --but is not limited to-- continuous functions. The \ac{uat} only gives support to continuous functions, which is useful from a practical perspective, but less robust than the Fourier series.

The Fourier theorem holds for functions depending on a single variable. However, the extension to multi-dimensional spaces is complicated and requires a space of parameters whose size increases exponentially with the number of dimensions~\cite{riemann_fourier_1867}. However, in the \ac{uat} case the use of multi-variable $\vec x$ arises naturally by adjusting the dimension of the weights.

\subsection{Proofs of universality theorems}\label{ssec:proofs_univ_theorems}

This section is devoted to the proofs of Theorems \ref{th:q_fourier} and \ref{th:q_UAT} supporting universality for quantum circuits. The reader interested in final results may skip this subsection without any regret. 

\subsubsection{Demonstration for the quantum Fourier series}\label{sssec:dem_fourier}

The quantum circuit proposed in Theorem \ref{th:q_fourier} fulfills the requirement that every new gate plays the role of a new step in the original Fourier series. The proof is based on an inductive procedure  and can be then decomposed in two steps. First, it is shown that the first gate of the circuit is equivalent to the $0$-th constant Fourier term. Then, if there are $N$ gates in a row forming a $N$-term Fourier series, adding a new gate provides a $(N+1)$-terms Fourier series if previous values are modified.

Let the fundamental gate $U^\mathcal{F}(x, \vec\theta)$ defined in Eq. \eqref{eq:unitary_f} be

\begin{adjustwidth}{-1cm}{-2cm}
\small
\begin{multline}
 U^\F(x; \vec\theta) = U^\mathcal{F}(x; \omega, \alpha, \beta, \varphi, \lambda) = R_z\left(\alpha + \beta\right) R_y(2\lambda)R_z\left(\alpha - \beta\right) R_z(2\omega x) R_y(2\varphi) = \\ = 
    \begin{pmatrix}
    \cos\lambda \cos\varphi e^{i \alpha} e^{i\omega x} - \sin\lambda \sin\varphi e^{i \beta} e^{- i \omega x} & 
    -\cos\lambda \sin\varphi e^{i \alpha} e^{i\omega x} - \sin\lambda \cos\varphi e^{i \beta} e^{- i \omega x} \\
    \sin\lambda \cos\varphi e^{-i \beta} e^{i\omega x} + \cos\lambda \sin\varphi e^{-i \alpha} e^{- i \omega x} & 
    -\sin\lambda \sin\varphi e^{-i \beta} e^{i\omega x} + \cos\lambda \cos\varphi e^{-i \alpha} e^{- i \omega x} \\
    \end{pmatrix},
\end{multline}
\end{adjustwidth}

It is possible to recast the above choice of fundamental gate using the following redefinition of parameters,
\begin{eqnarray}
    a_+  & = &   \cos\lambda \cos\varphi e^{i \alpha}, \\
    a_-  & = &  -\sin\lambda \sin\varphi e^{i \beta}, \\
    b_+ & = &  -\cos\lambda \sin\varphi e^{i \alpha}, \\
    b_- & = & - \sin\lambda \cos\varphi e^{i \beta}.
\end{eqnarray}

A more compact representation of the fundamental gate follows

\bigskip
\begin{lemma}

The fundamental gate can be expressed as
\begin{equation}\label{eq:u_f_2}
\small
    U^{\mathcal{F}}(x; \omega, \alpha, \beta, \varphi, \lambda) =
    \begin{pmatrix}
    a_+ e^{i\omega x} + a_- e^{- i \omega x} & 
    b_+ e^{i\omega x} + b_- e^{- i \omega x} \\
    -b_-^{*} e^{i\omega x} - b_+^{*} e^{- i \omega x} & 
    a_-^{*} e^{i\omega x} + a_+^{*} e^{- i \omega x} \\
    \end{pmatrix},
\end{equation}

as can be verified by simple substitution from Definition \ref{def:fourier_gate}. 

\end{lemma} 
Note that this expression corresponds to a unitary matrix, due to the relations involved in the definition of the coefficients $a_\pm$ and $b_\pm$.
Note also that a unitary matrix has three degrees of freedom, which are here fixed by 5 parameters. An intuition behind the role of these parameters is that $\alpha, \beta, \varphi, \lambda$ are related to the coefficients of one Fourier step, that is $a_\pm, b_\pm$, while $\omega$ can be identified with the corresponding frequency.

A total circuit can be constructed by multiplying $k$ fundamental gates to obtain $\mathcal{U}^{(k)}_{f, \phi}$ as in Definition \ref{def:prod_general}. Starting with this composite gate, the proof for the main Fourier approximation theorem~\ref{th:q_fourier} is feasible.

\begin{proof}
The proof of this constructive theorem consists in making contact with harmonic analysis and proceeds by induction.

{\bf i)} The first circuit consists only of one fundamental gate, chosen with frequency $\omega=0$, that is
\begin{equation}
    U_0^\mathcal{F} = 
    \begin{pmatrix}
    A_0 &  B_0 \\
    -B_0^{*} & A_0^{*} 
    \end{pmatrix},
\end{equation}
This, indeed corresponds to the first constant term of Fourier series.

{\bf ii)} It is assumed that the $N$-th approximant circuit takes the form for a Fourier series, but not its value constraints.
\begin{equation}
\small
    \prod_{i=0}^N U_i^\mathcal{F} = \begin{pmatrix}
    \sum_{n = -N}^N A_n e^{i \Omega_n x} & \sum_{n = -N}^N B_n e^{i \Omega_n  x} \\
    -\sum_{n = -N}^N B_n^ {*} e^{-i \Omega_n  x} & \sum_{n = -N}^N A_n^ {*} e^{-i \Omega_n  x}
    \end{pmatrix}.
\end{equation}
where the frequencies are  $\Omega_n$ are free, and to be fixed later. The  result of adding a new fundamental gate my left multiplication. corresponds to 
\begin{equation}\label{eq:uN+1}
\small
\prod_{i=0}^{N+1} U_i^\mathcal{F} =
\begin{pmatrix}
    \sum_{n = -N-1}^{N + 1} \tilde{A}_n e^{i \tilde \Omega_n x} & \sum_{n = -N-1}^{N + 1} \tilde{B}_n e^{i \tilde \Omega_n x} \\
    -\sum_{n = -N-1}^{N + 1} \tilde{B}_n^{*} e^{-i \tilde \Omega_n x} & \sum_{n = -N-1}^{N + 1} \tilde{A}_n^{*} e^{-i \tilde \Omega_n x}
    \end{pmatrix}
\end{equation}{}
where the new coefficients $\tilde\Omega_n$ need to be fixed and frequencies in terms of the old ones $\Omega_n$ and the new single gate frequency $\omega$ added to the circuit. It is easy to see that the addition of a gate changes the frequency in one unit, that is, $\tilde\Omega_n=\Omega_n \pm \omega$. Then, the general structure of the series can be adapted to  a Fourier expansion by choosing
\begin{equation}
    \Omega_n=(2 n+1)\frac{\pi}{2}.
\end{equation}

After fixing the values that the frequencies must take, it is straightforward to re-arrange terms in the matrix and reach
\begin{eqnarray}\label{eq:induction}
\tilde{A}_0 & = & A_0 a_- - B_0^{*} b_- \\
\tilde{A}_{\pm n} & = & A_{\pm n} a_- - B_{\mp n}^{*} b_- + A_{\pm (n-1)} a_+ - B^{*}_{\mp (n-1)} b_+ \\
\tilde{A}_{\pm (N + 1)} & = & A_{\pm N} a_+ - B^{*}_{\mp N} b_+\\
\tilde{B}_0 & = & B_0 a_- + A^{*}_0 b_- \\ 
\tilde{B}_{\pm n} & = & B_{\pm n} a_- + A^{*}_{\mp n} b_- + B_{\pm (n-1)} a_+ + A^{*}_{\mp (n-1)} b_+ \\
\tilde{B}_{\pm (N + 1)} & = & A^{*}_{\mp N} a_+ + A^{*}_{\mp N} b_+
\end{eqnarray}
This provides the explicit connection between approximant circuits and Fourier expansions for the coefficients of the global unitary matrix.

\end{proof}

The above theorem is sufficient to prove that the output probability of a series of approximant circuits
can reproduce any functionality. Notice that the proof ensures that for any number of re-uploadings $N$ it is possible to arrange terms in such a way that the final output state is a Fourier series. However, the intermediate steps follow a Fourier-like expression but do not maintain the required values of coefficients and frequencies. The mathematical form is the same, but the values of different coefficients must be changed to match the Fourier ones only in the last step. 

\subsubsection{Demonstration for the quantum \ac{uat}}\label{sssec:dem_uat}

An alternative manner to design a single-qubit universal approximant is related to the equivalent \ac{uat} broadly used in \ac{nn}s ~\cite{cybenko_approximation_1989}. The idea is to start from a different fundamental gate.

Let the {\sl fundamental gate} $U^{\rm UAT}(\vec x; \vec\theta)$ defined in Eq. \eqref{eq:unitary_uat} be explicitly
\begin{adjustwidth}{-1cm}{-2cm}
\begin{multline}\label{eq:u_uat}
    U^{\rm UAT}(x; \vec \omega, \alpha, \varphi) = R_z(2\left(\vec \omega \cdot \vec x + \alpha\right)) R_y(2\varphi) = \\ =
    \begin{pmatrix}
        \cos(\varphi) e^{i(\vec \omega \cdot \vec x + \alpha)} & -\sin(\varphi) e^{i(\vec \omega \cdot \vec x + \alpha)} \\
        \sin(\varphi) e^{-i(\vec \omega \cdot \vec x + \alpha)} & \cos(\varphi) e^{-i(\vec \omega \cdot \vec x + \alpha)}
        \end{pmatrix},
\end{multline}
\end{adjustwidth}

A full circuit can be constructed by multiplying $k$ fundamental gates to obtain $\mathcal{U}^{(k), {\rm UAT}}_{f, \phi}$ as in Def. \ref{def:prod_general}. The quantum \ac{uat}~\ref{th:q_UAT} using this fundamental gate is now proven. 

\begin{proof}
The classical $U^{\rm UAT}$ is defined in Eq. \eqref{eq:u_uat}.
By direct inspection it is straigthforward to check that every entry in this matrix can be understood as one term of $\bar f_N$ in Eq. \eqref{eq:UAT}. From this definition the recursive rule that defines all steps is obtained. If 
\begin{eqnarray}
A_{N} = \bra 0 \prod_{n=1}^{N}U_n^{UAT} \ket 0  \\
B_{N} = \bra 1 \prod_{n=1}^{N}U_n^{UAT} \ket 0
\end{eqnarray}

then the updating rule is 

\begin{adjustwidth}{-2cm}{0cm}
\begin{eqnarray}
A_{N+1} & = &
A_N  \cos(\varphi_{N+1}) e^{i \vec \omega_{N+1} \cdot \vec x} e^{i \alpha_{N+1}} -
B_N  \sin(\varphi_{N+1}) e^{i \vec \omega_{N+1} \cdot \vec x} e^{i \alpha_{N+1}}    
 \\
B_{N+1} & = &
A_N \sin(\varphi_{N+1}) e^{-i \vec \omega_{N+1} \cdot \vec x}e^{\alpha_{N+1}} +
B_N \cos(\varphi_{N+1}) e^{-i \vec \omega_{N+1} \cdot \vec x}e^{i\alpha_{N+1}}     
\end{eqnarray}
\end{adjustwidth}

Having this updating rule in mind, it is possible to write 
\begin{equation}\label{eq:B_N}
B_N = \sum_{m = 0}^{2^{N - 1}} c_m(\varphi_1, \ldots, \varphi_N) e^{i \delta_m(\alpha_1, \ldots, \alpha_N)} e^{i \vec w_m(\vec\omega_1, \ldots, \vec\omega_N) \cdot \vec x}, 
\end{equation}
where the inner dependencies of $c_m$ are products of sines and cosines of $\varphi_n$, and those of $\delta_m$ and $\vec w_m$ are linear combinations of $\alpha_n$ and $\omega_n$.

The procedure follows now as in the proof of the \ac{uat} in Ref.~\cite{cybenko_approximation_1989}. $S$ is the set of functions of the form $B_N(\vec x)$, and $C^{\mathbb{C}}(I_m)$ the set of continuous complex-valued functions in $I_m$, defined as in Theorem \ref{th:UAT}. It is assumed that $S \subset C^{\mathbb{C}}(I_m)$, and $S \neq C^{\mathbb{C}}(I_n)$. Now, the Theorem \ref{th:hahn_banach} is applied, known as Hahn-Banach theorem. This theorem allows to state that there exists a linear functional $L$ acting on $C^{\mathbb{C}}(I_n)$ such that
\begin{equation}
L(S) = L(\bar{S}) = 0, \qquad L\neq 0.
\end{equation}
Notice that this theorem is applicable since there are no restriction in working only with real numbers. 

Theorem \ref{th:riesz}, known as Riesz representation theorem, is called now. The functional $L$ is 
\begin{equation}
L(h) = \int_{I_n} h(x) d\mu(x)
\end{equation}
for $\mu \in M(I_n)$ non-null and $\forall \, h \in  C^{\mathbb{C}}(I_n)$. In particular, 
\begin{equation}
L(h) = A_N(\vec x) d\mu(\vec x) = 0,
\end{equation}
and thus
\begin{equation}
\int_{I_n} e^{i\vec{v_m}(\omega_1, \ldots, \omega_N) \cdot \vec x} d\mu(\vec x) = 0.
\end{equation}
This is the usual Fourier transform of $\mu$. By calling Theorem \ref{th:lebesgue}, Lebesgue Bounded Convergence theorem, if the $\mathcal{FT}(\mu) = 0$, then $\mu = 0$, and a contradiction is encountered with the only made assumption. 

The measure of all half-planes being 0 implies that $\mu = 0$. $\vec w$ is fixed, and for a bounded measurabe function $h$, the linear functional is defined as
\begin{equation}
F(h) = \int_{I_n} h(\vec w \cdot \vec x) d\mu(x),
\end{equation}
which is bounded on $L^\infty(\mathbb{R})$ since $\mu$ is a finite signed measure. Let $h$ be an indicator of the half planes $h(u) = 1$ if $u\geq -b$ and $h(u) = 0$ otherwise, then
\begin{equation}
F(h) = \int_{I_n} h(\vec w \cdot \vec x) d\mu(x) =  \mu(\Pi_{\vec w, b}) + \mu(H_{\vec w, b}) = 0.
\end{equation}
By linearity, $F(h) = 0$ for any simple function, such as sum of indicator functions of intervales~\cite{ash_analysis_1972}. 

In particular, for the bounded measurable functions $s(u) = \sin(\vec w \cdot \vec x), c(u) = \cos(\vec w \cdot \vec x)$ 
\begin{equation}
F(c + is) = \int_{I_n} \exp{i \vec w \cdot \vec x} d\mu(\vec x) = 0.
\end{equation}
The Fourier Transform of this $F$ is null, thus $\mu = 0$.

\end{proof}

For the sake of completeness, the three theorems required for the proof are covered in App.~\ref{app:math_theorems}.

\subsubsection{Link to output of quantum circuits}
Last sections were devoted to prove that specific series of circuits return functionalities able to represent a wide range of functions. In this last step, previous results are related to the output of quantum circuits.

\begin{theorem}
The computational basis output of a single-qubit quantum circuit can provide a convergent approximattion to any desired function.
\end{theorem}  

\begin{proof}
The output of a $k$-th approximant circuit can be cast a an approximation expansion of an arbitrary function.
It is sufficient to initalize a register in the $\vert 0\rangle$ state and measure the output in the computational basis. It follows 
\begin{equation}
    \bra{1} \prod_{i=0}^{N} U_i\ket{0} =  z_N(x)
\end{equation}
where $z_N(x)$ can take different forms and $U$ can provide Fourier or \ac{uat} approximations.

If the fundamental gate is $U^\mathcal{F}$, then the output is the truncated Fourier series
\begin{equation}
    z_N(x) = \sum_{n=-N}^N B_n e^{i 2\pi n x},
\end{equation}
where $B_n$ are free complex coefficients. This result holds for single-variable functions.

If the fundamental gate is $U^{\rm UAT}$, then the output is a function
\begin{equation}
\small
    z_N(\vec x) = \sum_{m = 0}^{2^{N - 1}} c_m(\varphi_1, \ldots, \varphi_N) e^{i \delta_m(\alpha_1, \ldots, \alpha_N)} e^{i \vec w_m(\vec\omega_1, \ldots, \vec\omega_N) \cdot \vec x}, 
\end{equation}
according to Eq. \eqref{eq:B_N}. This result holds for single- and multi-variable functions.

According to Theorems \ref{th:fourier} and \ref{th:UAT}, both expressions can approximate any desired function.

\end{proof}

\subsection{Discussion}\label{ssec:conclusions_th}
The theorems here presented demonstrate formally that a single-qubit circuit has enough flexibility to store any complex function $z(x)$ in its inner degrees of freedom by sequentially applying single-qubit gates depending on the independent variable $x$ and tunable parameters. This result is useful when the functional form of $z(x)$ is unknown and can only be learnt by sampling data. The learning process is performed by a classical optimizer. 

This result guarantees that two different independent real functions can be represented by single-qubit circuits. It provides the highest possible degree of compression of data in such a small state since there is no more possible room in the Hilbert space. In addition, there are no fundamental limitations in the complexity or dimensionality of the functions that can be encoded into the circuit. 

Two different approaches were followed for the proof, leading to two different sets of single-qubit gates. First, there is a link between Fourier series and quantum circuits if a quantum gate with 5 tunable parameters is defined, only if $x$ is one-dimensional. A quantum circuit composed by $N$ gates provides an output state whose components can be written as a $N$-terms Fourier series. Unlike in the classical case, since the process is variational, this result ensures at least the Fourier series, but permits better approximations. This proof inherits the assumptions of Fourier theorems. For the second proof, a link with the \ac{uat} was found by applying gates depending on 3 parameters. Such circuit delivers a output state compatible with the formulation of \ac{uat}. This proof is inherited from the equivalent one for \ac{nn}s and supports only approximation for continuous functions. In exchange, multidimensional dependencies on $x$ are supported as well. 

These results can serve as a starting point for studying the expressibility of quantum systems beyond one qubit, see also \cite{schuld_effect_2021, sim_expressibility_2019, nakaji_expressibility_2021}. When more qubits are added, the exponential size of the Hilbert space triggered by the entanglement is likely to play an essential role to be fully understood yet.

The core of the proof relies on the non-linearities arising from the consecutive application of non-commuting gates, and the linear encoding of data as arguments for the different gates. As a consequence, rotations around only two axis are necessary and sufficient to achieve any kind of non-linearities. The linear encoding provides unbiased approximations. Therefore, although the theorems give support to specific designs of quantum gates, it is likely that some other gates constructed taking these two ingredients into account can also provide accurate approximations. This can be advantageous for some problems with challenging properties. 

These results will serve as full theoretical support to the examples provided in Secs.~\ref{sec:benchmark}, and as partial support to the examples provided in Secs.~\ref{sec:qlassifier} and~\ref{sec:qpdf}.

\section{Numerical benchmark}\label{sec:benchmark}

In this section, the practical performance of the theorems explained in Sec.~\ref{ssec:theorems} is explored by fitting test functions from sample data using the theoretical models here presented. Two different kinds of benchmarks for real and complex functions, respectively, are presented. These benchmarks collect results using both $U^\mathcal{F}$ and $U^{\rm UAT}$ gates from Defs.~\ref{def:fourier_gate} and~\ref{def:uat_gate}. 
Benchmarks are performed using first simulations that include no decoherence nor sampling uncertainty, and then passing the obtained results to a quantum device to test the experimental performance. Simulations with up to 6 layers are considered. 

The aim of this benchmark is to compare the results of quantum and classical methods. The classical Fourier representation can be obtained by following Theorem~\ref{th:fourier}. In the \ac{uat} case, the description from Theorem~\ref{th:UAT} is followed, with $\sigma(\cdot)$ being a cosine for real functions and $e^{i (\cdot)}$ for complex functions. The parameters are found by employing specific classical optimization methods. For the quantum \ac{uat} case, $H \ket 0 = \ket +$ is taken as the initial state. Note that the choice of initial state does not compromise the validity of any result since it is possible to transform any state into any other by adjusting the first layer of the approximation method. 

All simulations are performed using the framework {\tt QIBO}~\cite{qibo}. The code computing the numerical experiments as well as the final results can be found on {\tt GitHub}~\cite{github_qubit}. Benchmarks for real and complex functions are described in Sec.~\ref{ssec:benchmark_real} and Sec.~\ref{ssec:benchmark_complex}, respectively. Results are covered in Sec.~\ref{ssec:th_results}. Final remarks can be read in Sec.~\ref{ssec:conclusions_universal}.

\subsection{Benchmark for real functions}\label{ssec:benchmark_real}

For the first benchmark, a single-variable, real-valued function $-1 \leq f(x) \leq 1$ related to the observable $\langle Z \rangle \sim f(x)$ is considered. The quantum state to represent is then
\begin{equation}
    \ket{\psi(x)}_Z = \sqrt{\frac{1 + f(x)}{2}}\ket 0 + e^{i\phi}  \sqrt{\frac{1 - f(x)}{2}}\ket 1,
\end{equation}
where $\phi$ is a phase that in general may be $x$-dependent, but it is neglected at this stage. The $\chi^2$ function that drives the optimization is then
\begin{equation}
    \chi^2 = \frac{1}{M} \sum_{j=1}^M \left(\langle Z(x_j) \rangle - f(x_j) \right)^2,
\end{equation}
where $M$ is the total number of samples of $x$.

The $Z$ benchmark is first tested against four different functions of interest
\begin{eqnarray}\label{eq:functions}
    \relu(x) &=& \max(0, x),\\\label{eq:functions1}
    \tanh(a  x) & \; {\rm for} \; & a = 5, \\\label{eq:functions2}
    \step(x) &=& x / \vert x \vert;\quad 0\; {\rm if}\; x=0, \\\label{eq:functions3}
    \poly (x) &=& \vert 3 x^3 (1 - x^4)\vert. 
\end{eqnarray}
All functions are conveniently rescaled to fit the limits $-1 \leq f(x) \leq 1$. In all cases, $x \in [-1, 1]$. The $\relu(\cdot)$ and $\tanh(\cdot)$ functions are chosen given the central role they play in the field of \ac{ml}. $\step(\cdot)$ presents a discontinuity, which implies a challenge in the approximation. $\poly(\cdot)$ is chosen as it contains wavy features arising from  non-trigonometric functions. 

Next, the approach is tested against four functions of two variables in order to check how the quality of the approximations evolves as more dimensions are added to the problem. Those are known 2D functions named {\tt adjiman, brent, himmelblau, threehump}~\cite{2d_functions}. These functions are chosen as representatives of a variety of difficulties the algorithm needs to overcome.
In the 2D case, the functions are 
conveniently rescaled to fit the limits $-1 \leq f(x, y) \leq 1$ and $(x, y) \in [-5, 5]^2$. A definition of these functions can be found in Appendix~\ref{app:2D_benchmark}.

In this benchmark, both the \ac{uat} and Fourier quantum and classical methods are considered for the one-dimensional functions. However, 2D functions are only tested for \ac{uat} methods since theorem~\ref{th:q_fourier} from Sec.~\ref{sec:reuploading_theorems} does not support multidimensional Fourier series.

\subsection{Benchmark for complex functions}\label{ssec:benchmark_complex}

In order to test the performance of the presented algorithm for fitting complex functions, a tomography-like benchmark is proposed. Since complex functions have real and imaginary parts, one needs to measure at least two observables in the qubit space. In this case, the observables are $\langle X \rangle$ and $\langle Y \rangle$ for the real and imaginary parts, that is {$\langle X \rangle + i \langle Y \rangle \sim f(x) e^{i g(x)}$}. The quantum state that permits this identification is
\begin{equation}
    \ket{\psi(x)}_{XY}  = \sqrt{\frac{1 + \sqrt{1 - f(x)}}{2}} \ket 0+ e^{i g(x)}\sqrt{\frac{1 - \sqrt{1 - f(x)}}{2}} \ket 1. 
\end{equation}
for $0\leq f(x)\leq 1$It is then possible to construct a $\chi^2$ function as 
\begin{equation}
    \chi^2 = \frac{1}{M} \sum_{j=1}^M \left\vert \langle X(x) \rangle + i \langle Y(x) \rangle - f(x)e^{i g(x)} \right\vert^2.
\end{equation}

For the $X-Y$ benchmark the algorithm is tested against all possible combinations of real and imaginary parts of the functions defined in Eqs.~\eqref{eq:functions}--\eqref{eq:functions3}, conveniently renormalized to ensure that $\langle X \rangle^2 + \langle Y \rangle^2 \leq 1$.

\begin{figure}[b!]
\begin{adjustwidth}{-2cm}{-1cm}
    \centering
    \includegraphics[width=.9\linewidth]{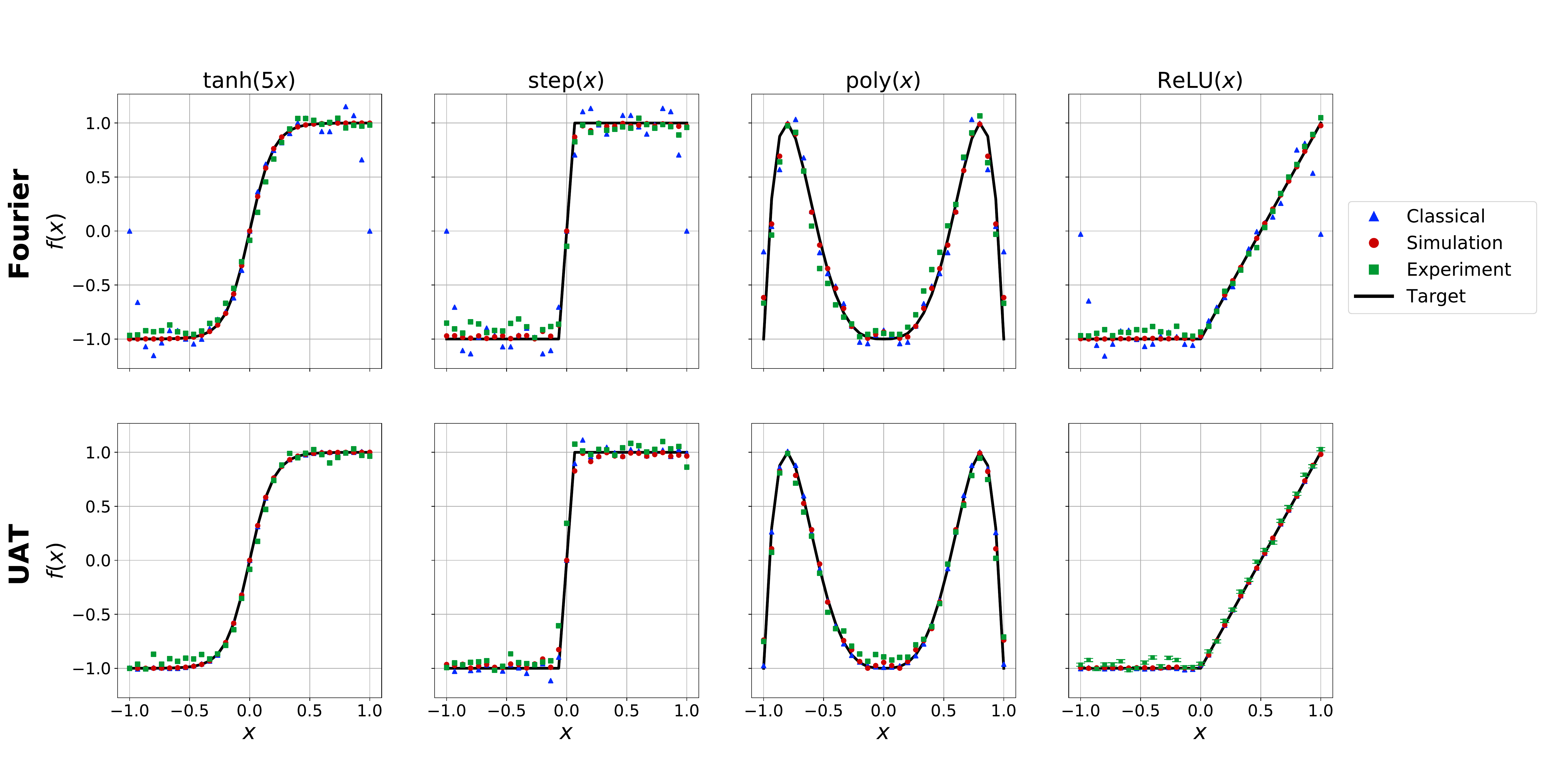}
\end{adjustwidth}
    \caption{Fits for four real-valued functions using the $Z$ benchmark with five layers. Blue triangles represent classical models, namely Fourier and \ac{uat}, while red dots represent its quantum counterparts computed using a classical simulator. Green squares are the experimental execution of the optimized quantum model using a superconducting qubit. The target function is plotted in black for comparison. The analysis for experimental errors is plotted for the $\relu$ function and the \ac{uat} model.}
    \label{fig:real_funs}

\begin{adjustwidth}{-2cm}{-1cm}
\centering
    \includegraphics[width=.9\linewidth]{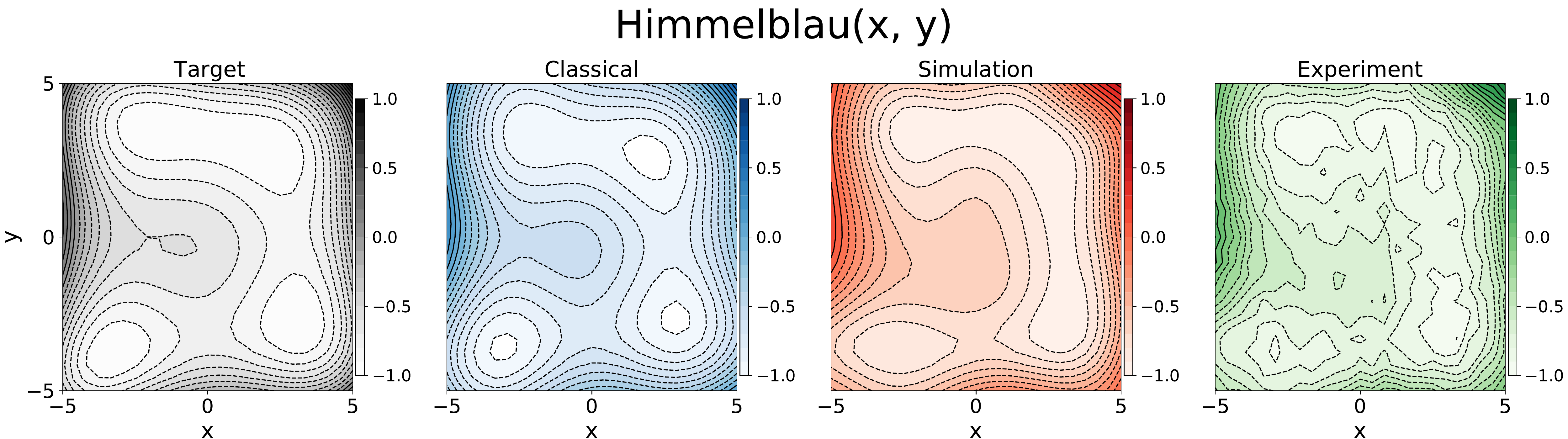}
\end{adjustwidth}
    \caption{Fits for the 2D function Himmelblau properly normalized using the $Z$ benchmark for five layers. The blue (classical) plot represents the classical \ac{uat} model, while the red (simulation) plot represents its quantum counterpart simulated. The green (experiment) plot is the experimental execution of the optimized quantum model. The target function is painted in black (target). In all drawings, the lines corresponds to the same levels in the $Z$ axis.}
    \label{fig:2d_himmelblau}
\end{figure}

\subsection{Results}\label{ssec:th_results}

\begin{figure}[b!]
\begin{adjustwidth}{-1cm}{-2cm}
    \centering
    \subfigure[\hspace{2mm} 1D functions\label{fig:chi2_real}]{\includegraphics[width=.48\linewidth]{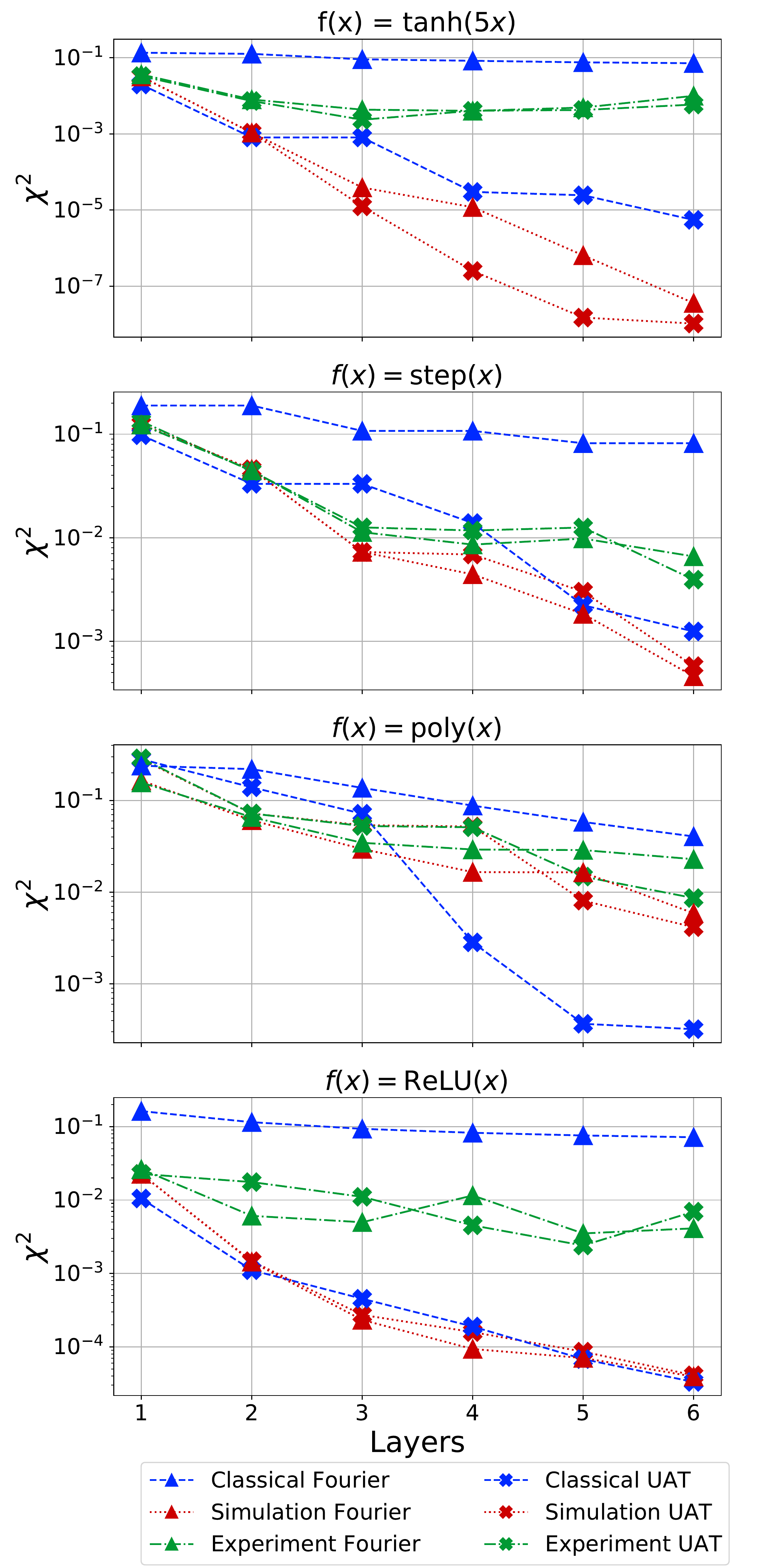}}
    \subfigure[\hspace{2mm} 2D functions\label{fig:2d_chi}]{\includegraphics[width=.48\linewidth]{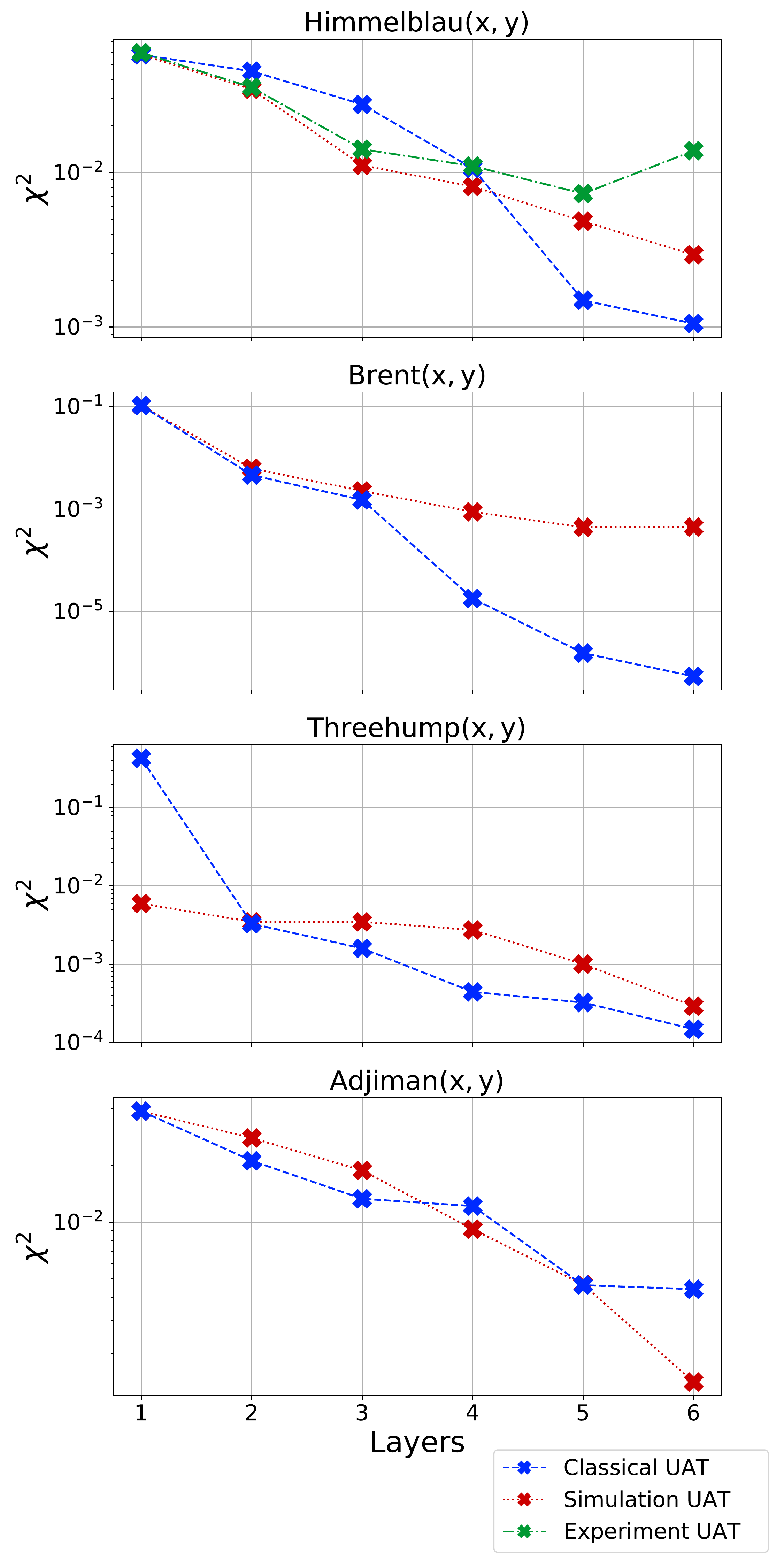}}
\end{adjustwidth}
    \caption{Values of $\chi^2$ for the $Z$ benchmark in all four test 1D and 2D functions using classical computation (blue scatter), classical simulation of the quantum algorithm (red scatter) and experimental implementation with a superconducting qubit (green scatter). Fourier models are depicted with triangles, while \ac{uat} models are represented by crosses. In the 2D case, only \ac{uat} models are considered.}
\end{figure}

In all results presented in this section there are three different final values. First, the Fourier and the \ac{uat} classical methods are used to approximate a target function. The Fourier method is obtained following the constructive recipe of Th.~\ref{th:fourier}. The \ac{uat} is applied using a single-hidden-layer \ac{nn}. Second, the same function is approximated using the quantum procedures defined in this work, simulating the wave function evolution with classical methods. In both cases, the best outcome obtained with different initial conditions used in the optimization step is retained. Finally, the parameters obtained using the simulation of the quantum procedure are translated to the experimental device to execute the circuit of interest in the actual superconducting machine. Details concerning the experimental implementation can be found in App.~\ref{app:exp_universal}. The theoretical optimal parameters may be, in principle, different than the experimental ones. Hence, an optimization performed directly on the experimental parameters could improve the final results \cite{future-work}.

\begin{figure}[b!]
    \centering
    \includegraphics[width=\linewidth]{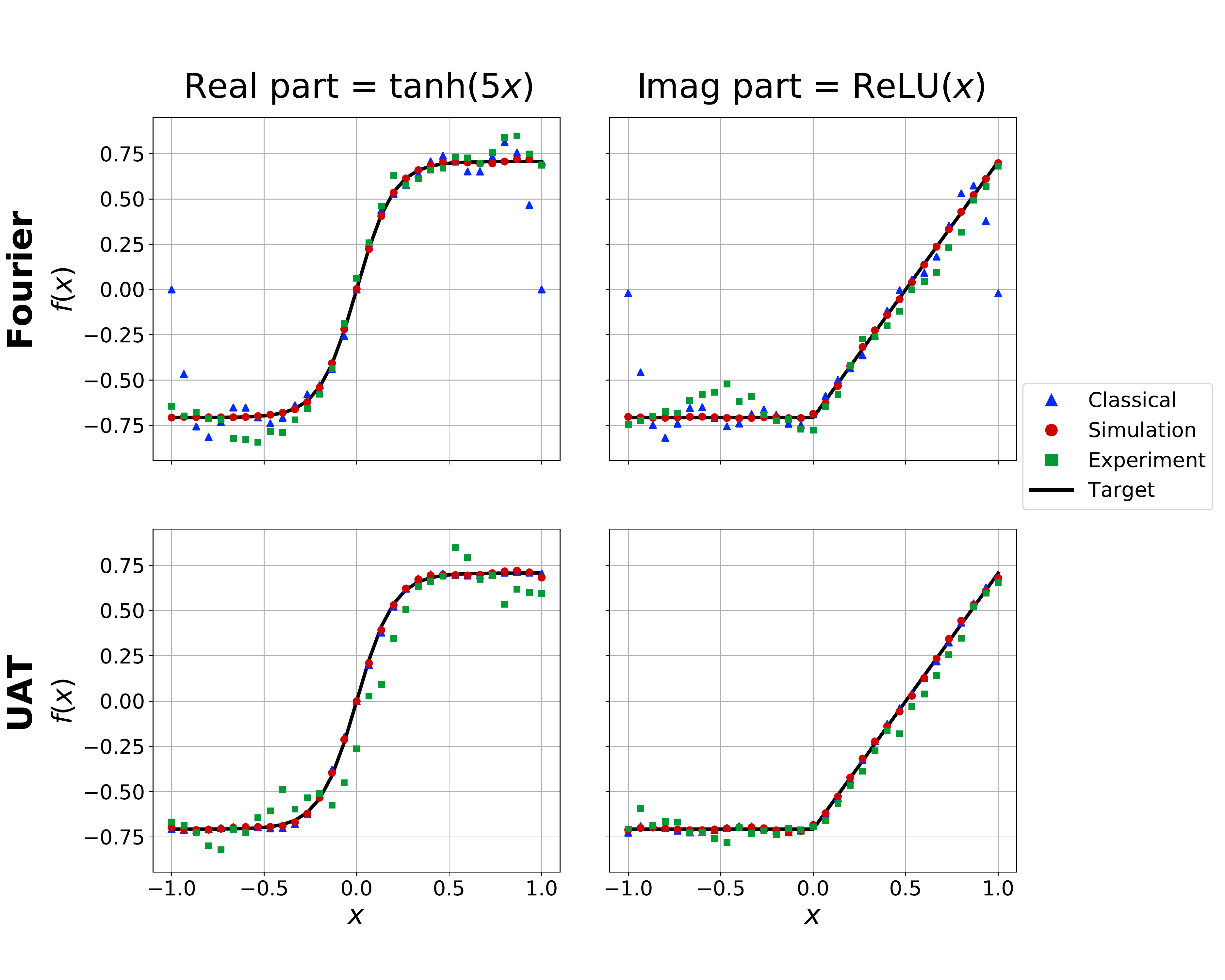}
    \caption{Fits for the complex function $f(x) = \tanh(5x) + i \relu(x)$ properly normalized using the $X-Y$ benchmark for five layers. Blue triangles represent a classical model, while red dots represent its quantum counterparts computed using a classical simulator. Green squares are the experimental execution of the optimized quantum model using a superconducting qubit. The target function is plotted in black for comparison.}
    \label{fig:complex_funs}
\end{figure}

The resulting fit for all four single-variable real-valued functions from Eqs.~\eqref{eq:functions}--\eqref{eq:functions3} is shows in Fig.~\ref{fig:real_funs}. In this case the $Z$ benchmark with 5 layers is considered. A classical approximation (blue), a quantum exact simulation (red) and its experimental implementation (green) are depicted. 
All methods follow the overall shape of the target function. Classical Fourier approximations return less accurate predictions on the value of $f(x)$ due to the periodic nature of the model. The quantum Fourier and both classical and quantum \ac{uat} models return better results for all values of $x$. This behaviour is observed in all benchmarks. The experimental results retain the qualitative properties of the exact models, although a loss in performance is visible. In addition, an analysis of experimental uncertainties is also depicted at the \ac{uat} $\relu$ plot from Fig.~\ref{fig:real_funs}.

\begin{figure}[b!]
\begin{adjustwidth}{-1cm}{-2cm}
	\centering
    \includegraphics[width=\linewidth]{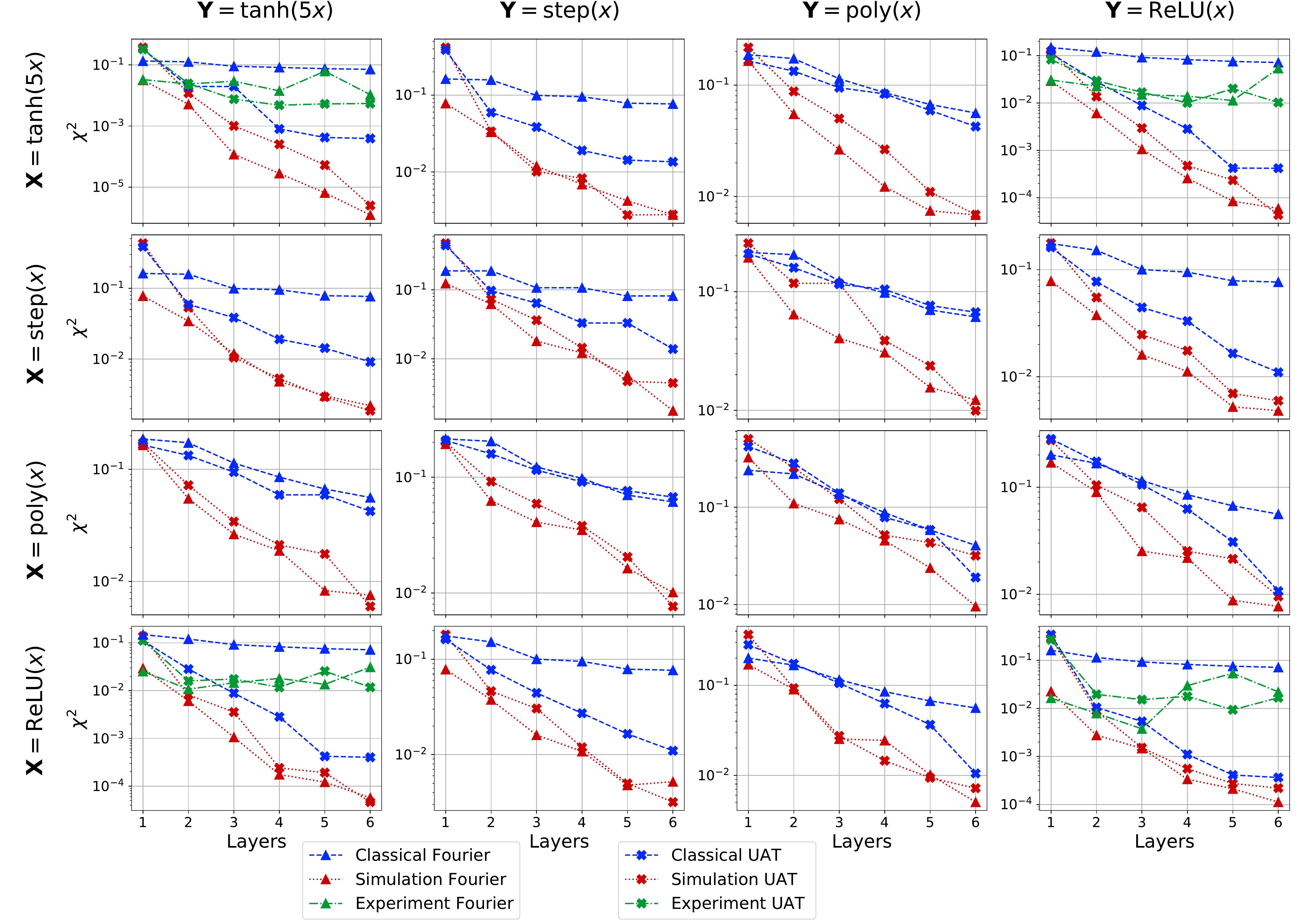}
\end{adjustwidth}
    \caption{Values of $\chi^2$ for the X-Y benchmark in all possible combinations for real and imaginary parts of the four test functions from Eqs.~\eqref{eq:functions} to~\eqref{eq:functions3} using classical computation (blue scatter), classical simulation of the quantum algorithm (red scatter), and experimental implementation with a superconducting qubit (green scatter). Fourier models are depicted with triangles, while \ac{uat} models are represented by crosses.}
    \label{fig:chi2_complex}
\end{figure}

Fig.~\ref{fig:chi2_real} shows a summary of the values of $\chi^2$ for classical and their analogous quantum simulated models and their experimental validation. In the case of classical and simulated quantum models a general trend towards better approximations --implying lower values of $\chi^2$-- is observed with an increasing numbers of layers.  

The simulated Fourier model performs better than its classical counterpart. This is due to the fact that a classical Fourier series does not contain tunable parameters, while its quantum version does.
However, the result from the classical Fourier series constitutes a lower bound for any approximation method based on optimization since at least the quality of the Fourier series is guaranteed. 

In the \ac{uat} case of Fig.~\ref{fig:chi2_real}, no approach returns significantly better results. The classical algorithm performs better in the $\poly(x)$ case, but the results with the simulated quantum method improve the classical ones in the $\tanh(5x)$ case. Both models present similar trends as the number of layers increases. 

Despite the fact that the Fourier model contains more parameters than the \ac{uat} model, the latter performs better as seen in Fig.~\ref{fig:chi2_real}. Therefore, the \ac{uat} method seems more appropriate for the functions used here. 

The experimental realization of the quantum approximation models suffers from circuit noise and sampling uncertainties, and therefore degrades the quantity $\chi^2$. This is more prominent as more layers are added to the model. As a direct consequence, the approximation of the quantum model to the target function loses accuracy. The inherent sampling uncertainty sets a lower bound in the value of $\chi^2$ obtained through experiments.

In general, Fig.~\ref{fig:chi2_real} supports the claim that every layer grants the model more flexibility, and thus enhances the capability of fitting the target function. This flexibility is given by the number of re-uploadings of the independent variable and not by the amount of parameters. In addition, having too many parameters likely hinders the optimization procedure.

Figure~\ref{fig:2d_himmelblau} depicts the approximations obtained for the ${\rm Himmelblau (x, y)}$ function comparing the target function and all different methods considered. Figure~\ref{fig:2d_chi} summarizes the values of $\chi^2$ for all 2D-functions taken into account in this work. 

All different executions capture the overall shape of the function, but some differences exist in the different plots. Classical simulations return values for $Z < -1, Z > 1$, and thus lead to three minima in this case. On the other hand, the quantum simulation cannot clearly distinguish those minima. The experimental execution presents sharp contours because of the inherent noise and sampling uncertainty.

The values of $\chi^2$ in Fig.~\ref{fig:2d_chi} measure the accuracy of the approximations. As before, is is seen that a larger number of layers provides better approximations to the target function. In agreement to the one-dimensional $Z$ benchmark, the scaling is similar for both quantum and classical methods.

A complex function in the $X-Y$ benchmark is depicted in Fig.~\ref{fig:complex_funs}. In that case, the $X$ measurement leads to $\tanh(5x)$ while the imaginary part contains $\relu(x)$. All the observations made for the $Z$ benchmark hold in this case. 

Fig.~\ref{fig:chi2_complex} shows values of $\chi^2$ for all possible combinations of real and imaginary parts using the functions described in Eqs.\eqref{eq:functions}--\eqref{eq:functions3}, being the real and imaginary parts. In this case, it is possible to see a common advantage for the quantum models. In particular, the functions $\tanh(5x)$ and $\relu(x)$ work better in any combination. This reflects the behaviour already observed in Fig.~\ref{fig:chi2_real}, where these functions present better performance than other functions considered. 

\subsection{Discussion}\label{ssec:conclusions_universal}
The results presented in this section are the numerical and experimental confirmations that the theoretical works presented in Sec.~\ref{sec:reuploading_theorems} have some utility in practice. With this purpose, numerical evidences on the flexibility and approximation capabilities of the quantum circuits are given.

The benchmarks were obtained using first simulations of quantum systems optimized by classical means for a set of test functions. As benchmarks, 1D and 2D real functions, and 1D complex functions are included. Final results are compared against classical counterparts. In general, it is possible to see that quantum and classical methods scale equivalently. This is the numerical confirmation that quantum procedures can perform similarly to classical ones, at least theoretically. 

A further step has been carried by implemented the found solutions in an actual quantum device built upon superconducting trasmon qubits. No further optimization is performed in this stage. Experimental results confirm the trend seen by simulation, and the finite coherence and imperfections of the qubit do not seem to impact results significantly for small numbers of gate. 

The obvious next step to tackle in this experiment is to actually carry the optimization procedure on the quantum device. However, the lack of accurate retrievement of values for the cost function limits the performance achievable in this direction. Future work will deal with problems of this kind \cite{future-work}.

\section{Re-uploading for a quantum classifier}\label{sec:qlassifier}

The re-uploading strategy can be used to address classification problems with minimal amounts of quantum resources. As it is seen in this section, even a single-qubit circuit is a system versatile enough as to build a universal quantum classifier upon it. The computational capabilities are obtained by combining quantum features and classical optimization subroutines. In fact, there exist no theoretical limits in the dimensionality of data that can be classified with this strategy. On the other hand, multiple class classification can be accomodated even in the smallest possible Hilbert space. The main reason why this is possible is the density of the Hilbert space the quantum processing takes place in. 

The single-qubit classifier here proposed does not attempt to address classically intractable problems, and the achievement of quantum advantage is out of scope. The focus of the work of this chapter is rather to illustrate that even the most minimalistic quantum systems provide large computational power. In this work, the aim is to distill the minimal amount of quantum resources required to solve a given supervised learning problem in practice. Quantum resources include qubits and gates. The problems to be solved in this section are not trivial, even though they can be expressed with simple datasets. 

This quantum classifier also extends the idea of re-uploading to multiple qubits and entangling architectures. There is currently no theoretical work supporting the universality of such circuits. However, extensive benchmarks were carried to compare the efficiency of this strategy as entanglement expands the superpositions carried along with the classification. 
The main result in this section is to show that there is a trade-off between the number of qubits and depth of the circuit. That is, fewer qubits may be used at the price of re-uploading data in several steps along the computation. In summary, similar results can be obtained for similar query complexities even in the case of different architectures. Thus, despite the lack of theoretical evidence, it is expected that quantum systems with several entangled qubits possess more capabilities to perform classification than the single-qubit systems. To what extent complex systems outperform simple ones is still unclear.

The power of single- and multi-qubit classifiers following the data re-uploading strategy is illustrated by a series of examples. All the examples here presented were computed by means of classical simulations. Later, this approach was experimentally demonstrated on a trapped-ions quantum computer, see Sec.~\ref{sec:exp_qlassifier}. First, it is attempted to distinguish between points in a plane divided in two classes. The classes are determined by a geometrical boundary. Then, the number of regions, that is classes, is extended on the plane to address more complicated datasets. As a last extension, the dimensionalty of the datasets is increased from plane (2D) to points in space (3D) and hyperspace (4D). A complete benchmark is done for every problem as addressed with quantum circuits with differentes properties, that is number of qubits and entanglement schemes. 

This section is structured as follows. First, the adaptations needed to transform the general re-uploading strategy to a single-qubit quantum classifier scheme including measurement strategy are explained in Sec.~\ref{ssec:qlassifier}. The extension from single- to multi-qubit is depicted in Sec.~\ref{ssec:multiq}. Results are reviewed in Sec.~\ref{ssec:results_qlassifier}. Sec.~\ref{ssec:conclusions_classifier} covers final comments.

The contents of this section are based on the work in Ref.~\cite{perezsalinas_data_2020}. This work constitutes the seminal paper of all the re-uploading strategy. It is important to mention that some technical details were polished as more comprehension of the scheme was acquired. These improvements, however, do not compromise the validity of the results here displayed, even though these implementations do not constitute an optimal one. 

\subsection{Quantum classifier}

In this section the different ingredients required to build the quantum classifier are described in detail, namely the quantum circuit, the measure-ment strategy, the cost functions and the general working principle for mixing all pieces in a consistent way.

\subsubsection{Quantum circuit}\label{ssec:qlassifier}

We take as the starting line of the quantum classifier the general scheme defined in Def.~\ref{def:prod_general}, and specify the quantum circuit used in terms of their processing gates. For this application of the re-uploading strategy, the gates take the role of Defs.~\ref{def:fourier_gate} and~\ref{def:uat_gate}, but they do not exactly match the ones previously proposed. Nevertheless, it mantains the most important properties, namely re-uploading and linear encoding of data, and non-commuting matrices interspersed. In the single-qubit classifier, data is introduced in simple rotations which are easy to characterize. The definition of each layer is 
\begin{equation}\label{eq:u_qlassifier}
U(\vec x; \vec \theta, \vec w) = U_3(\vec\theta + \vec w \circ \vec x),
\end{equation}
where $\vec\theta, \vec w, \vec x$ are 3D vectors and $\vec{w}\circ\vec{x}=\left(w^{1}x^{1},w^{2}x^{2},w^{3}x^{3}\right)$ is the Hadamard product of two vectors. $U_3$ is the most general single-qubit unitary gate defined as $U_3(\vec\theta) = R_z(\theta_1)R_y(\theta_2)R_z(\theta_3)$. In case the data points have dimension lesser than $3$, the rest of the components of $\vec x$ until reaching this dimensionality are set to $0$, and the corresponding terms in $\vec w$ are then irrelevant. 

It is also possible to enlarge the dimensionality of the input space in the following way. The definition of one layer can be extended to 
\begin{equation}\label{eq:multiple_dim}
\hat{U}\left(\vec{x}; \vec{\theta},\vec{w}\right)=U\left(\vec{x}^{(j)}; \vec{\theta}^{(j)},\vec{w}^{(j)}\right)\cdots U\left(\vec{x}^{(1)} \vec{\theta}^{(1)},\vec{w}^{(1)}\right),
\end{equation}
where each data point is divided into $j$ vectors of dimension three. In general, each unitary $U$ could absorb as many variables as degrees of freedom in an SU(2) unitary. Each set of variables act at a time, and all of them have been shown to the circuit after $j$ iterations. Then, the layer structure follows. The complexity of the circuit only increases linearly with the size of the input space. This recipe to enlarge dimensionality is not supported by the theorems from Sec.~\ref{sec:reuploading_theorems}. However, numerical results show that slight modifications on the scheme permit to keep the functioning principle of the quantum algorithm.

Further refinement and understanding of the re-uploading technique sheds light on some properties of the proposed circuit that could be more efficient only in the single-qubit case. Since one layer is defined as the sequence of gates $R_z R_y R_z$, when two layers are set together pairs of $R_z$ gates appear in consecuive positions. These gates do commute between them, and therefore the inner parameters of those gates can fuse to give rise to only one parameter, and non-linearities do not rise. This way, the overall depth of the circuit is reduced. As a consequence, this scheme includes some tunable parameters whose presence does not entail any improvement in the capability of the circuit. 

For the sake of completeness the complete circuit is made further explicit now. Following the recipe from Def.~\ref{def:prod_general}, the overall operation for the classifier is 
\begin{equation}\label{eq:def_classifier}
\mathcal{U}^{(k)}(\vec x; \Theta, W) = \prod_{i=1}^k \hat{U}(\vec x; \vec \theta_i, \vec w_i), 
\end{equation}
where the optimals parameters $\Theta = \lbrace\vec\theta_i\rbrace, W= \{\vec w_i\}$ are different for each layer $(i)$.  Then, the output state of this circuit depending on $\vec x$ will be
\begin{equation}
\ket{\psi(\vec x; \Theta, W)} = \mathcal{U}^{(k)}(\vec x; \Theta, W) \ket 0
\end{equation}

The quest for optimal configurations of $\Theta$ and $W$ is done by classically minimizing some loss function. This loss function $\chi^2$ will depend both on the parameters and on the dataset to be classified. The quantity $\chi^2$ is designed in such a way that $\chi^2 \rightarrow 0$ implies that the circuit can correctly guess all the classes from the training dataset.
However, it is worth to mention than an optimal set of parameters do not always imply a good performance on unseen test datasets, since classifiers must learn to generalize their training sets, see~\ref{ssec:nn}. Therefore, to measure the quality of the quantum classifier the accuracy $\mathcal A$, measured on the testset as

\begin{equation}
\mathcal{A}(\Theta, W) = \frac{\textrm{Correct guesses}}{\textrm{Number of samples}},
\end{equation}
is considered.
Thus, benchmark not only the adaptability of the quantum classifier, but also its generalization capability are benchmarked

\subsubsection{Measurement}\label{ssec:measurement}

The measurement strategy to retrieve useful information for the classifi-cation from the circuit is key to achieve versatile models. The final states $\ket{\psi(\vec x; \Theta, W)}$ are measured, and the results are used to compute the cost function $\chi^2$ that quantifies the error made in the classification of the training set. The minimization of this quantity in terms of the classical parameters of the circuit can be organized using any preferred supervised machine learning technique.

To design a successful measurement recipe find an optimal way to distinguish among output quantum states belonging to different classes must be found. Following the guiding principle of Ref.~\cite{helstrom_quantum_76} the optimal scenario to distinguish between two quantum states appears when the states are orthogonal. This is easily achieved for the single-qubit case for binary classifications. Classes $A$ and $B$ are attached to two orthogonal states, for instance $\ket 0$ ad $\ket 1$. The output state of the circuit is then measured, with a probability $P(0), P(1)$ of obtaining $\ket{0}, \ket 1$ respectively, with $P(0) + P(1) = 1$. A given datapoint is then guessed as $A$ if $P(0) > P(1)$, and as $B$ otherwise. A more flexible criterium is possible by introducing some boundary $\lambda$. In this case a point is guessed as $A$ if $P(0) > \lambda$. Notice that for $\lambda = 1/2$ the direct comparison is recovered. Thus, the presence of $\lambda$ can only improve the final result. An optimal $\lambda$ is chosen to get the best possible accuracy $\mathcal{A} $ on some unseen test set. 

It is convenient for understanding more complex datasets to picture the division of the Hilbert space into different areas, where each area corresponds to a different class. In this dichotomic classification, the Bloch sphere is divided in northern $\ket 0$ and southern $\ket 1$ hemispheres. The parameter $\lambda$ can move the boundary from the equator to some other convenient parallel. 

The assignment of classes to the output reading of a single qubit becomes a much more difficult problem when many classes are involved in the dataset. The Hilbert space spanned in a single qubit is two dimensional, and there only exist pairs of orthogonal states. Thus, it is convenient to develop some other strategies to distinguish between different classes when measuring the output states.

The geometrical vision provides insights to address this problem. One possible strategy is to divide the Bloch sphere through parallels. The quantity $P(0)$ is compared to three thresholds $0 \leq \lambda_1 \leq \lambda_2 \leq \lambda_3 \leq 1$. The value of $P(0)$ must then fall into one of these areas. A second more robust strategy consists in dividing the Bloch sphere into equivalent classes as separate as possible. This is obtained by computing the overlap between the output state and some label-state that targets a given class. The set of label-states is chosen in such a way that the members are maximally orthogonal among them. This strategy divides the Bloch sphere into different classes, where the boundary between two of them lies on the points where the relative fidelity of a given state with two different label states is equal.

\begin{figure}
\centering
\includegraphics[width=.4\linewidth]{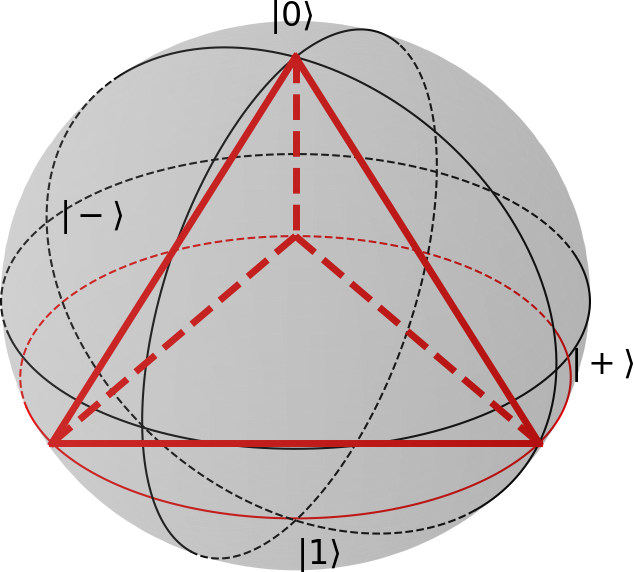} \hfill 
\includegraphics[width=.4\linewidth]{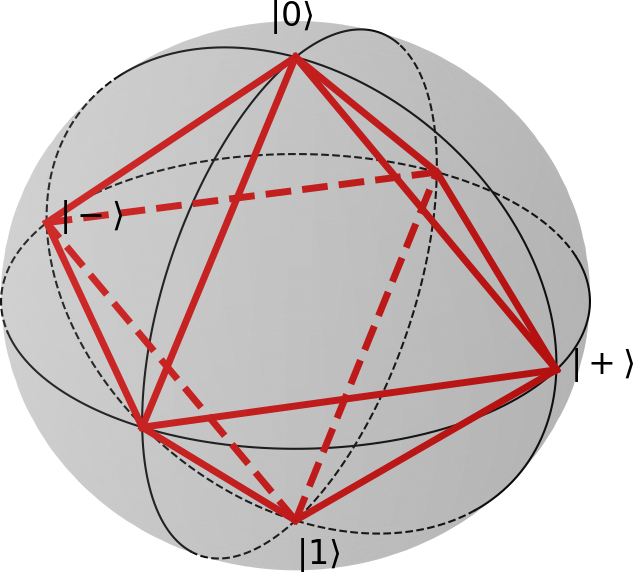}
\caption{Representation in the Bloch sphere of four and six maximally orthogonal points, corresponding to the vertices of a tetrahedron and an octahedron respectively. The single-qubit classifier will be trained to distribute the data points in one of these vertices, each one representing a class.}
\label{Fig:BlochSphere}
\end{figure}

Finding the optimal set of label-states is a problem on its own. It can be solved almost trivially in a small number of cases. For 2 classes, solutions are $\ket 0$ and $\ket 1$. For 3 classes, the label-states must form a regular triangle in any equatorial plane. For 4 classes, a tetrahedron is needed. This opens up the field of platonic solids~\cite{atiyah_polyhedra_2003}, that are the solutions to 4, 6, 8, 12 and 20 classes, corresponding to the number of verteices. For other numbers of classes, solutions must be found through some other method. Figure \ref{Fig:BlochSphere} shows the particular cases that can be applied to a classification task of four and six classes.  

In general, a good measurement strategy may need some prior computa-tional effort and refined tomography of the final state. For a single-qubit classifier, the tomography protocol will only require three measurements. 
There is an alternative experimental approach to avoid tomography. Comparing some output state with some other is equivalent to measure their relative fidelities~\cite{nielsen_chuang_2010}, that is
\begin{equation}\label{eq:fidelity}
\mathcal{F}_y(\vec x, \Theta, W) = \vert\braket{\phi_y}{\psi(\vec x; \Theta, W)}\vert^2
\end{equation}
This can be done by adding a unitary gate $U_y$ at the end of the circuit such that $V_y\ket 0 = \ket{\phi_y}$ and measuring the probability of getting $\ket 0$ as 
\begin{equation}
\mathcal F_y(\vec x, \Theta, W)= \vert\bra{0}V_y^\dagger \mathcal{U}(\vec x; \Theta, W) \ket{0}\vert^2
\end{equation}
In the case where more qubits are available, a swap test can be used as well~\cite{kang_swap_2019}.

\subsubsection{A fidelity cost function}
We propose a simple cost function motivated by the geometrical interpretation introduced above. The final goal of the quantum classifier is then to force the output states $\ket{\psi(\vec{x}; \Theta, W)}$ to be geometrically close to the corresponding label state. That implies that the average fidelity between the output states and the corresponding label states must be maximal in average. The following cost function carries out this task:
\begin{equation}\label{eq:fidelity_chi2}
\chi^2_{f}(\Theta, W) = \sum_{\{\vec x, y\}}^{M} \left(1- \mathcal{F}_{y}(\vec x, \Theta, W)^2\right), 
\end{equation}
where data and classes pairs $\{\vec x, y\}$ run over the training dataset.

\subsubsection{A weighted fidelity cost function}

The next step is to define a refined version of the previous fidelity cost function to be minimized. Maximixing the overlap between output states and the label states is equivalent to minimizing the overlap to other label states. Since the room in a single-qubit Hilbert space has dimension two, the latter overlap cannot be zero. A vector whose components are the relative fidelities among the label state is defined as, 
\begin{equation}
(Y_y)_i = \vert \braket{\phi_y}{\psi_i}\vert ^2.
\end{equation}
This quantity stores the expected fidelities for a successful classification. For example, given a four-class classification and using the vertices of a tetrahedron as label states (as shown in Figure \ref{Fig:BlochSphere}), the ideal output state would have complete overlap with the corresponding class, say $0$, and a relative fidelity $1/3$ for any other class. Thus, $\vec{Y_y} = (1, 1/3, 1/3, 1/3)$. 

One expects $Y_{s}(\vec{x}) = 1$, where $s$ is the correct class, and $Y_{r}(\vec{x}) = 1/3$ for the other $r$ classes. In general, $Y_{y}(\vec{x})$ can be written as a vector with one entry equal to 1, the one corresponding to the correct class, and the others containing the overlap between the correct class label-state and the other label-states.The expected fidelities are then
compared with the obtained fidelities for a given output state of a successful classification, $Y_{y}(\vec{x})$. 

With this definition, a cost function inspired by conventional cost functions in artificial \ac{nn}s can be constructed. By weighting the fidelities of the final state of the circuit with all label states, the {\sl weighted fidelity} cost function is
\begin{equation}
\chi^2_{wf}	(\Theta, W, \vec\alpha) = \frac{1}{2} \sum_{\{\vec x, y\}}^M \left(\sum_{j=1}^{\mathcal{C}}\left(\alpha_{j}\mathcal F_{j}(\vec{x};\Theta, W) - (Y_{y})_j\right)^2\right),
\label{eq:conventional_chi2}
\end{equation}
where $M$ is the total number of training points, $\mathcal{C}$ is the total number of classes, $(\vec{x}, y)$ are set of training data and classes and $\vec{\alpha}=(\alpha_{1},\cdots,\alpha_{\mathcal{C}})$ are introduced as \emph{class weights} to be optimized together with $\Theta$ and $W$ parameters. The weighted fidelity cost function needs more parameters than the fidelity one. Notice however that the parameters in $\vec\alpha$ have the capability to enlarge and decrease the size of the areas corresponding to the different classes by weighting the classes. Large values of $\alpha_y$ increase the probability to guess the class $y$. This is useful when datasets are unbalanced. \\

The main computational difference between $\chi^2_f$ and $\chi^2_{wf}$ lies in how many relative fidelities must be measured to compute the cost function. The $\chi_{wf}^2$ requires as many fidelity measurements as classes for every evaluation of the cost function in the optimization subroutine, while the $\chi_{f}^2$ needs just one. For few classes and one qubit, it is not a such a big difference since tomography is not extremely costly. The extra cost in terms of number of parameters and then hardness of the optimization affects only the classical part of the computation. The classical requirements are increased to reduce the overall contribution of quantum resources. This fact links with the \ac{nisq} line of thought. 

Besides the larger computational cost of $\chi^2_{wf}$ with respect to $\chi^2_f$, there is a qualititative difference betweeen both. The fidelity cost function retrieves information about how close an output state is from its corresponding label state. Thus, the only mechanism of the optimizer to obtain good results is to minimize this distance. Loosely speaking, $\chi^2_f$ looks for an optimal configuration by moving towards where the output state should be. On the contrary, $\chi^2_{wf}$ retrieves a more complete information, since the distance to all label states is measured in every interaction. There is then more insight when updating the parameters in every new evaluation. The geometrical interpretation is that the output state moves away from the wrong classes while approaching the right one. The differences between both cost functions induce that the $\chi^2_{wf}$ is expected to provide better results.

\subsubsection{Optimization and retrievement of results}

The cost function defines the two steps needed to carry classification task, namely the training and the retrievement of results. 

\begin{figure}[t!]
\begin{adjustwidth}{-2cm}{-1cm}
\begin{tcolorbox}[enhanced,width=1\linewidth, center, colback=white, colframe=orange]
\begin{center}

\Large \sl Quantum classifier: multi-variable $\vec x$ $\Rightarrow$ multi-class $\{ y \}$
\end{center}

\vskip5mm
\begin{flushleft}
{\bf 1. Optimization on the training dataset}\newline
\resizebox{.75\textwidth}{!}{\hskip1cm $\Qcircuit @R=0.5em @C=0.3em{
\lstick{\ket 0} & \qw & \gate{U(\vec x, \vec\theta_{1}, , \vec w_{1})} & \gate{U(\vec x, \vec\theta_{2}, \vec w_{2})} & \qw & \push{\cdots} & \gate{U(\vec x, \vec\theta_{k}, \vec w_{2})} &\qw & \meter & \ustick{\hspace{5mm} \Rightarrow}\cw & \control \cw\\
 &  & \control \cwx & \control \cwx\cw & \dstick{\Leftarrow}\cw & \cw & \control \cwx\cw & \gate{\rm Classical\; Optimizer} \cw & \dstick{\Leftarrow} \cw & \cw & \gate{\chi^2(\Theta, W)} \cwx \\
}$}
\begin{textblock}{10}(10, -0.3)
{\Large$
\Rightarrow \Theta, W$
}
\end{textblock}
\begin{textblock}{10}(9.5, -0.87)
{\Large$
\longrightarrow \vec{\mathcal{F}}(\vec x, \Theta, W)$
}
\end{textblock}

\vspace{5mm}

{\bf 2. Retrieving results from test dataset}\newline
\resizebox{.49\textwidth}{!}{\hskip1cm $\Qcircuit @R=0.5em @C=0.3em{
\lstick{\ket 0} & \qw & \gate{U(\vec x, \vec\theta_{1}, , \vec w_{1})} & \gate{U(\vec x, \vec\theta_{2}, \vec w_{2})} & \qw & \push{\cdots} & \gate{U(\vec x, \vec\theta_{k}, \vec w_{2})} &\qw & \meter &
}$}

\begin{textblock}{10}(6.4, -0.45)
{\Large$
\longrightarrow \vec{\mathcal{F}}(\vec x, \Theta, W) \Rightarrow {\rm guessed} \; y$
}
\end{textblock}
\end{flushleft}
\vskip2mm
\end{tcolorbox}
\end{adjustwidth}
\caption{Functioning scheme of the quantum classifier. In the first stage, a training dataset is used to find optimal parameters. The output states are measured to construct a cost function $\chi^2_f, \chi^2_{wf}$, to be minimized. Once this process is complete, second stage starts. In this step, the quantum circuit is executed using data points $vec x$ from the testset. Fidelities are measured, and from that information a class $y$ is guessed.}
\label{fig:opt_qlassifier}
\end{figure}
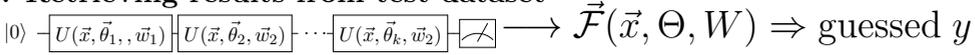

For optimizing the circuit, the well-known hybrid classical-quantum models can be used, see Fig.~\ref{fig:opt_qlassifier}. The quantum circuit is executed and measured several times. From those measurements the fidelity results from Eq.~\eqref{eq:fidelity} are retrieved. If the fidelity cost function $\chi^2_f$ is used, then is is only needed to measure the fidelity with respect to the label state corresponding to the class of the data point. If $\chi^2_{wf}$ is the chosen cost function, then all fidelities must be measured for each point. Then, those results are given to the classical optimizer to look for the optimal set of parameters $\Theta, W$.

When the set of optimal parameters has been obtained, it is time to run the quantum circuit with data points from the testset. At the end of the execution, measurements are made to obtain the relative fidelities between the output state and all label states. In this step it is compulsory to take all possibilities into account since no prior knowledge is available. From that information, it is possible to guess a class $y$, commonly by choosing the largest fidelity. The number of correctly guessed points will determine the accuracy $\mathcal{A}(\Theta, W)$.

\subsection{From single- to multi-qubit quantum classifier} \label{ssec:multiq}

Entanglement makes the dimensionality of the Hilbert space of composite quantum systems increase exponentially as more particles are added. This phenomenon lies at the core of the impossibility to simulate quantum mechanics by classical means. Therefore, quantum advantage can only be achieved by those quantum algorithms whose number of qubits is at least moderately large, only if they are entangled. In this case, the single-qubit quantum classifier cannot expect to overcome any classical method. 

\begin{figure}[b!]
\centering
\begin{adjustwidth}{-1cm}{-2cm}
\subfigure[\hspace{0.05cm} Ansatz with no entanglement]{
\resizebox{.45\linewidth}{!}{
\hspace{10mm}\Qcircuit @C=0.85em @R=.9em  @!R{
&\lstick{\ket 0} & \qw & \gate{U(1, 1)} & \gate{U(1, 2)} & \gate{U(1, 3)} & \qw & \cdots & & \gate{U(1, N)} & \qw \\
&\lstick{\ket 0} & \qw & \gate{U(2, 1)} & \gate{U(2, 2)} & \gate{U(2, 3)} & \qw & \cdots & & \gate{U(2, N)} & \qw \\
\vspace{0cm}
}
}} \hfill 
\subfigure[\hspace{0.05cm} Ansatz with entanglement]{
\resizebox{.45\linewidth}{!}{
\hspace{10mm}\Qcircuit @C=0.85em @R=.9em  @!R{
&\lstick{|0\rangle} & \qw & \gate{U(1, 1)} & \ctrl{1} & \gate{U(1, 2)} & \ctrl{1} & \qw & \cdots & & \ctrl{1} & \gate{U(1, N)} & \qw  \\
&\lstick{|0\rangle} & \qw & \gate{U(2, 1)} & \ctrl{-1} & \gate{U(2, 2)} & \ctrl{-1} & \qw & \cdots & & \ctrl{-1} & \gate{U(2, N)} & \qw \\
\vspace{0cm}
\gategroup{1}{4}{2}{5}{.7em}{--}
\gategroup{1}{6}{2}{7}{.7em}{--}
\gategroup{1}{12}{2}{12}{.7em}{--}
}
}}
\end{adjustwidth}
\caption{Two-qubit quantum classifier circuit without and with entanglement. Here, each layer includes a single-qubit operation with data re-uploading per plus a CZ gate if entanglement is considered. The last layer never carries an entanglement gate. For a fixed number of layers, the number of parameters to be optimized doubles the one needed for a single-qubit classifier. The depth is also doubled, up to physical realization of the CZ gate.}
\label{Fig:2qubit_circuit}

\centering
\begin{adjustwidth}{-1cm}{-2cm}
\subfigure[Ansatz with no entanglement]{
\resizebox{.45\linewidth}{!}{
\hspace{5mm}\Qcircuit @C=0.85em @R=.9em  @!R{
&\lstick{|0\rangle} & \qw & \gate{U(1, 1)} & \gate{U(1, 1)} & \gate{U(1, 2)} & \qw & \cdots & & \gate{U(1, N)} & \qw \\
&\lstick{|0\rangle} & \qw & \gate{U(1, 1)} & \gate{U(2, 1)} & \gate{U(2, 2)} & \qw & \cdots & & \gate{U(2, N)} & \qw \\
&\lstick{|0\rangle} & \qw & \gate{U(1, 1)} & \gate{U(3, 1)} & \gate{U(3, 2)} & \qw & \cdots & & \gate{U(3, N)} & \qw \\
&\lstick{|0\rangle} & \qw & \gate{U(1, 1)} & \gate{U(4, 1)} & \gate{U(4, 2)} & \qw & \cdots & & \gate{U(4, N)} & \qw \\
\vspace{0cm}
}
}}\hfill 
\subfigure[Ansatz with entanglement]{
\resizebox{.45\linewidth}{!}{
\hspace{5mm}\Qcircuit @C=0.85em @R=.9em  @!R{
&\lstick{|0\rangle} & \qw & \gate{U(1, 1)} & \ctrl{1} & \gate{U(1, 2)} & \qw & \ctrl{3} & \qw & \cdots & & \ctrl{1} & \gate{U(1, N)} & \qw  \\
&\lstick{|0\rangle} & \qw & \gate{U(2, 1)} & \ctrl{-1} & \gate{U(2, 2)} & \ctrl{1} & \qw & \qw & \cdots & & \ctrl{-1} & \gate{U(2, N)} & \qw \\
&\lstick{|0\rangle} & \qw & \gate{U(3, 1)} & \ctrl{1} & \gate{U(3, 2)} & \ctrl{-1} & \qw &  \qw & \cdots & & \ctrl{1} & \gate{U(3, N)} & \qw \\
&\lstick{|0\rangle} & \qw & \gate{U(4, 1)} & \ctrl{-1} & \gate{U(4, 2)} & \qw & \ctrl{-3} & \qw & \cdots & & \ctrl{-1} & \gate{U(4, N)} & \qw \\
\vspace{0cm}
\gategroup{1}{4}{4}{5}{.7em}{--}
\gategroup{1}{6}{4}{8}{.7em}{--}
\gategroup{1}{13}{4}{13}{.7em}{--}
}
}}
\end{adjustwidth}
\caption{Four-qubit quantum classifier circuit without and with entanglement. Here, each layer includes a single-qubit operation with data re-uploading per qubit plus two CZ gates if entanglement is considered. The order of CZ gates alternates in each layer between (1)-(2) and (3)-(4) qubits and (2)-(3) and (1)-(4) qubits. The last layer never carries an entanglement gate. For a fixed number of layers, the number of parameters to be optimized quadruples the one needed for a single-qubit classifier. The depth is only doubled, up to physical realization of the CZ gate.}
\label{Fig:4qubit_circuit}
\end{figure}

On the other hand, the classical analog to the single-qubit classifier here presented is a single-hidden-layer \ac{nn}, as stated in Theorem~\ref{th:q_UAT}. Nevertheless, using this architecture is rare. The amount of neurons needed to achieve good performance and the inefficiency of training methods prevents an extended use of this \ac{nn}. Instead, other kinds of \ac{nn}s were developed to circumvent this limitation. One of the most celebrated models is the deep \ac{ffnn}, see Sec.~\ref{ssec:nn}. In this case, many neurons are distributed through several layers, and each layer is connected to the previous and next ones, except for the extrema. Connections between two consecutive layers can happen in principle among all possible pairs of neurons. In this case, data is processed in several steps, and more complex features and correlations of the training data are captured. In addition, the training of such models can be done more efficiently using techniques such as back-propagation~\cite{rumelhart_backprop_1986, goodfellow_deep_2016, nielsen_neural_2015}.

The natural step to take is to extend the single-qubit formalism to a framework which is more similar to the deep-learning approach of classical methods. The generalization should carry entanglement as a crucial resource for obtaining quantum advantage. The expectation of the model is to improve the overall performance of the quantum classifier as more qubits are added, aiming to mimic the features of hidden layers in deep \ac{nn}s. Ideally, incorporating entanglement among qubits could lead to a reduction in the number of re-uploadings needed to solve some classification problem. However, even though the relationship between a single-qubit quantum classifier and a single-hidden-layer \ac{nn} is well understood, see Fig.~\ref{fig:reuploading_scheme} and properly supported by theoretical results, see Th.~\ref{th:q_UAT}, the interpretation of multi-qubit classifiers in terms of deep-learning scheme is far from straightforward. 

In contradistinction to the single-qubit quantum classifier, the extension to many qubits and addition of entanglement does not have yet theoretical results supporting its applicability and universality. However, it is expected that these circuits have at least the same flexibility as single-qubit circuits. The work in Ref.~\cite{schuld_effect_2021} deals with the representation capabilities of single-depth circuits when distributed along several qubits. Thus, this work does not deal with the possible correlations among subsequent re-uploadings of data on different qubits. 

As in multi-layer \ac{nn}s, the design of multi-qubit quantum classifiers is heuristic. In the classical case there is no recipe to know the optimal number of layers and neurons, and neurons per layer, to solve some problem with the best possible performance. In general, it depends strongly on the problem. In the quantum case, several properties of the circuits must be decided. For instance, how to upload data onto the circuit, namely whether all features are included in all gates and repeated, or distributed along larger parts of the circuit. Another purely quantum extra degree of freedom is the addition of entanglement. Entanglement is in general an open problem when finding Ansätze for parameterized quantum circuits~\cite{nakaji_expressibility_2021, sim_expressibility_2019}

Figures~\ref{Fig:2qubit_circuit} and~\ref{Fig:4qubit_circuit} show the explicit circuits proposed for classification in this work. In these models, layers are a set of parallel single-qubit operations occurring simultaneously. Entanglement is introduced by means of CZ gates between layers of rotations. In the case of circuits with entanglement, those gates are absorbed in the definition of layer. For the two-qubit circuit, the CZ entangling gate is applied after each set of rotations, except for the last one. For the four-qubit classifier, two CZ gates are applied after each rotation set interspersed between qubits (1)-(2) and (3)-(4); and (2)-(3) and (1)-(4) qubits, see Fig.~\ref{Fig:4qubit_circuit} for a graphical description. In both images, the gate $U(i, j)$ corresponds to the gates from Eq.~\eqref{eq:u_qlassifier} and~\eqref{eq:multiple_dim} depending on the dimension of the training data, applied on $i$-th qubit, $j$-th layer.

Each rotational layer is composed by the same gates as in the single-qubit case. Thus, the the number of parameters needed to define the circuit is multiplied times the number of qubits present. The depth of the circuit, however, is only doubled with respect to the single-qubit one, up to the physical realization of entangling gates. 

\subsubsection{Measurement strategy and cost function for a multi-qubit classifier}

In the single-qubit classifier, the only available measurement strategy consists on comparing the output quantum state of the circuit with the label-states representing the different classes. This can be done using performing simple tomography procedures or directly measuring relative fidelities. However, when more qubits are considered, this protocol becomes rapidly outdated. First, the number of dimension of the Hilbert space defined by the quantum system increases exponentially, and thus the possibilities of finding maximally orthogonal sets label-states become much more diverse. Second, tomography protocols suffer from exponentially larger costs in terms of number of measurements as the size of the quantum system increases. To overcome these barriers, two different measurement strategies for the multi-qubit classifiers are proposed. 

First, the single-qubit measurement is generalized in a natural way. The final output state is compared with one chosen label-state from the computational basis. This, however, becomes unrealizable for a large number of qubits. Due to the exponential increase of the dimensionality of Hilbert spaces, the number of orthogonal states becomes quickly much larger than the number of classes provided by the dataset. With this method, it is only possible to retrieve information from an exponentially small subspace of the Hilbert space. In particular for the first steps of the optimization, where the output state is created following random parameters, this measurement protocol only captures insufficient and random oddments.

The second strategy consists in measuring only one qubit and assign different classes depending on the result. Notice that this qubit cannot be longer described as a pure state $\ket{\psi}$, but rather as a density matrix $\rho$. This permits to compress the information of large Hilbert spaces into smaller subspaces. This strategy follows similar to the single-qubit one. This approach aims to join ideas of binary multi-qubit classifiers~\cite{farhi_classification_2018} and the possibility of multi-class classification by introducing thresholds and single-qubit label states, see Section~\ref{ssec:measurement}.

Another piece that should be adapted to accomodate multi-qubit measurements is the definition of the cost function. A different cost function for each measurement strategy is attached. The new cost functions are inspired in the previous $\chi^2_f, \chi^2_{wf}$ from Eqs.~\eqref{eq:fidelity_chi2} and~\eqref{eq:conventional_chi2}. 

For the first strategy, the fidelity cost function $\chi^2_f$ is used. Its generalization to more qubits is straightforward. However, the orthogonal states used for a multi-qubit classifier are taken as the computational basis states. A more sophisticated set of states could be considered to improve the performance of this method.

For the second strategy, the weighted fidelity cost function $\chi^2_{wf}$ is used. As stated above, only one qubit is considered, thus
\begin{equation}
\mathcal F_{y,q}(\vec{x}; \Theta, W) = \bra{\phi_y}\rho_{q}(\vec{x}; \Theta, W)\ket{\phi_y},
\label{eq:fidelity_multiq}
\end{equation}
where $\rho_{q}$ is the reduced density matrix of the qubit to be measured.
Then, the weighted fidelity cost function can be adapted as
\begin{equation}
\chi_{wf}^2(\Theta, W, \alpha) = \frac{1}{2} \sum_{\{\vec x, y\}}\sum_{j=1}^{\mathcal{C}}\left(\sum_{q=1}^{Q}\left(\alpha_{j,q}F_{c,q}(\vec{x}; \Theta, W) - (Y_y)_{j}\right)^2\right),
\label{eq:conventional_chi2_multiq}
\end{equation}
where an average is computed over all $Q$ qubits that form the classifier. Notice that the $\alpha$ parameters evolved from being a vector to a $\mathcal C \times Q$ matrix. Eventually, the number of measured qubits and effectively the number of optimizable parameters can be reduced.

\subsection{Numerical benchmark of the quantum classifier}\label{ssec:results_qlassifier}
In this section the performance of the quantum classifier previously described is numerically benchmarked against several distinct classification problems. The results here present demonstrate numerically that this method is capable to successfully solve multi-class classification problems for multi-dimensional data. A single-qubit classifier suffices, but multi-qubit circuits reach comparable final results with less processing layers. In summary, the flexibility of the quantum classifier depends mainly on the query complexity of the algorithm, that is, how many times the data is re-uploaded into the circuit. The complete code can be viewed in Ref. \cite{github_data}.

The benchmark is carried by constructing several different classifiers with different numbers of layers. This makes possible to control the query complexity of a given circuit. Then, the models are trained to obtain the optimal parameters $\{\Theta, W\}$ for each layer, plus $\{\alpha\}$ when applicable. The cost functions used to drive the optimization are $\chi^2_f, \chi^2_{wf}$ from Eqs.~\eqref{eq:fidelity_chi2} and~\eqref{eq:conventional_chi2} for the single-qubit classifiers, and its multi-qubit analogues, see Eq.~\eqref{eq:conventional_chi2_multiq}, for multi-qubit classification. 

The datasets to classify for the benchmark are composed by random points in the space $[-1, 1]^{d}$ whose classes are defined by geometrical means. The data points are always the same, while the assignment of classes differs depending on the problem. After the training is complete, the classifier is tested against an unseen testset one order of magnitude larger than the training set. The reason to proceed in this way is to check the generalization capabilities of the quantum classifier without facing a too costly optimization step. 

Classifications with 1, 2 and 4 qubits were carried, with and without entanglement for the multi-qubit cases, for the two cost fuctions defined above. The number of layers tested in every case are $L = \{1,2,3,4,5,6,8,10\}$. The problems tackled run from simple two-dimensional binary classification to multi-dimensional and multi-class classifications. Non-convex classes, which are considered difficult for classification, are also covered. 

\subsubsection{Binary classification of a circle}
This example is the easiest one considered.The dataset is a random set of data points $\{\vec x\} \in [-1, 1]^2$, that is $\vec x = (x_1, x_2); -1\leq x_i \leq 1$ . The classification task is to determine whether these points satisfy the condition $x_1^2 + x_2^2 < r^2$ for some radius $r$. From a geometrical perspective, this is equivalent to infer from a point whether it is inside a circle of radius $r$ or not. The value of $r$ was chosen in such a way that the area of the circle is half the total area of the feature space, in this case $r = \sqrt{2 / \pi}$. This way, a random classifier obtains an accuracy of $50\%$. The training dataset is composed of 200 random points, while the testset has 4000 previously unseen points to densely populate the feature space.

\begin{table}[t]
\centering
\scriptsize
\begin{tabular}{c|c|cc|c|cc|cc}
  & \multicolumn{3}{c|}{$\chi^{2}_{f}$} & \multicolumn{5}{c}{$\chi^{2}_{wf}$} \\
 \hline
Qubits & 1 & \multicolumn{2}{c|}{2 } & 1 & \multicolumn{2}{c|}{2} & \multicolumn{2}{c}{4 }   \\
Layers  & & No Ent. & Ent. & & No Ent. & Ent. & No Ent. & Ent. \\ 
 \hline
 1 & 0.50 & 0.75 & -- & 0.50 & 0.76 & -- & 0.76 & --  \\
 2 & 0.85 & 0.80 & 0.73 & 0.94 & 0.96 & 0.96 & 0.96 & 0.96  \\
 3 & 0.85 & 0.81 & 0.93 & 0.94 & 0.97 & 0.95 & 0.97 & 0.96  \\
 4 & 0.90 & 0.87 & 0.87 & 0.94 & 0.97 & 0.96 & 0.97 & 0.96  \\
 5 & 0.89 & 0.90 & 0.93 & 0.96 & 0.96 & 0.96 & 0.96 & 0.96  \\
 6 & 0.92 & 0.92 & 0.90 & 0.95 & 0.96 & 0.96 & 0.96 & 0.96  \\
 8 & 0.93 & 0.93 & 0.96 & 0.97 & 0.95 & 0.97 & 0.95 & 0.96  \\
10 & 0.95 & 0.94 & 0.96 & 0.96 & 0.96 & 0.96 & 0.96 & 0.97  \\
\end{tabular}
\caption{Results of the single- and multi-qubit classifiers with data re-uploading for the circle problem. Numbers indicate the success rate, that is correctly guessed samples over total number of samples. Words ``Ent." and ``No Ent." refer to considering entanglement between qubits or not, respectively. Minimization was in this case done using the L-BFGS-B algorithm. For this problem, both cost functions $\chi^2_f$, $\chi^2_{wf}$ lead to high success rates, although the weighted fidelity one achieves it with lesser numbers of layers. The multi-qubit classifier increases this success rate, although entanglement does not produce any appreciable effect.}
\label{tab:results_circle}
\end{table}

The results of the classification for the circle problem are summarized in Tab.~\ref{tab:results_circle}. First, it is worth noticing that the $\chi^2_{wf}$ cost function delivers much better results than the simpler $\chi^2_f$, in particular for classifiers with few layers. With $\chi^2_{wf}$, the single qubit classifier achieves over 90\% of success with only two layers, 12 parameters. The two- and four-qubit classifiers reach that threshold with two layers as well, that is 22 and 42 parameters. In addition, the introduction of entanglement does not change the final result in any case. The results show a saturation in the success rate, so that adding more layers leads to no further improvement. The flexibility achieved with few layers is enough to capture all the available data, and the fine tuning required here to improve the results needs of further strategies in the training, such as increasing and densifying the training dataset around the border between classes. 

\begin{figure}[t]
\centering
\begin{adjustwidth}{-2cm}{-1cm}
\subfigure[\hspace{0.05cm} 1 layer]{\includegraphics[width=0.24\linewidth]{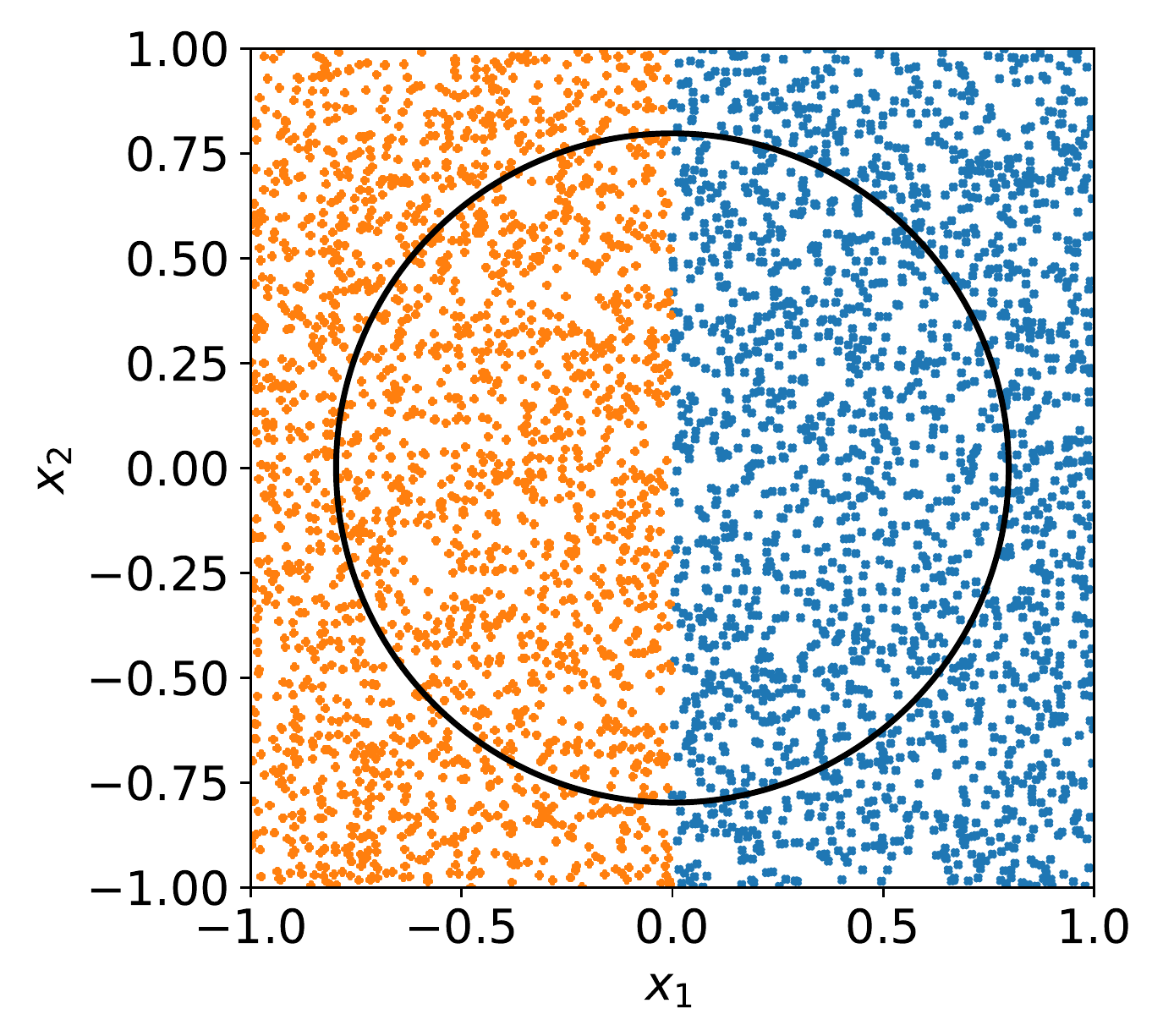}}
\subfigure[\hspace{0.05cm} 2 layers]{\includegraphics[width=0.24\linewidth]{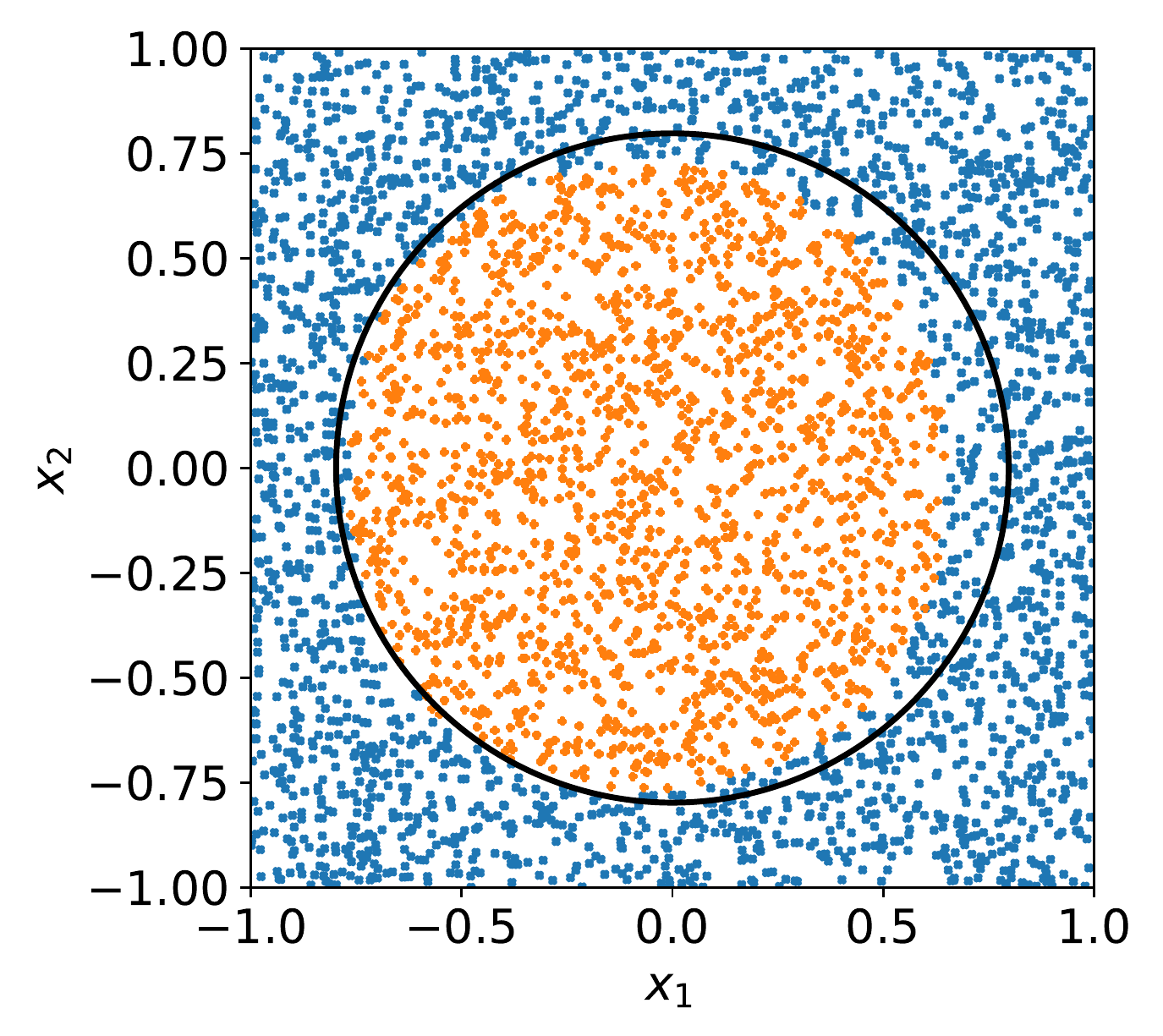}}
\subfigure[\hspace{0.05cm} 4 layers]{\includegraphics[width=0.24\linewidth]{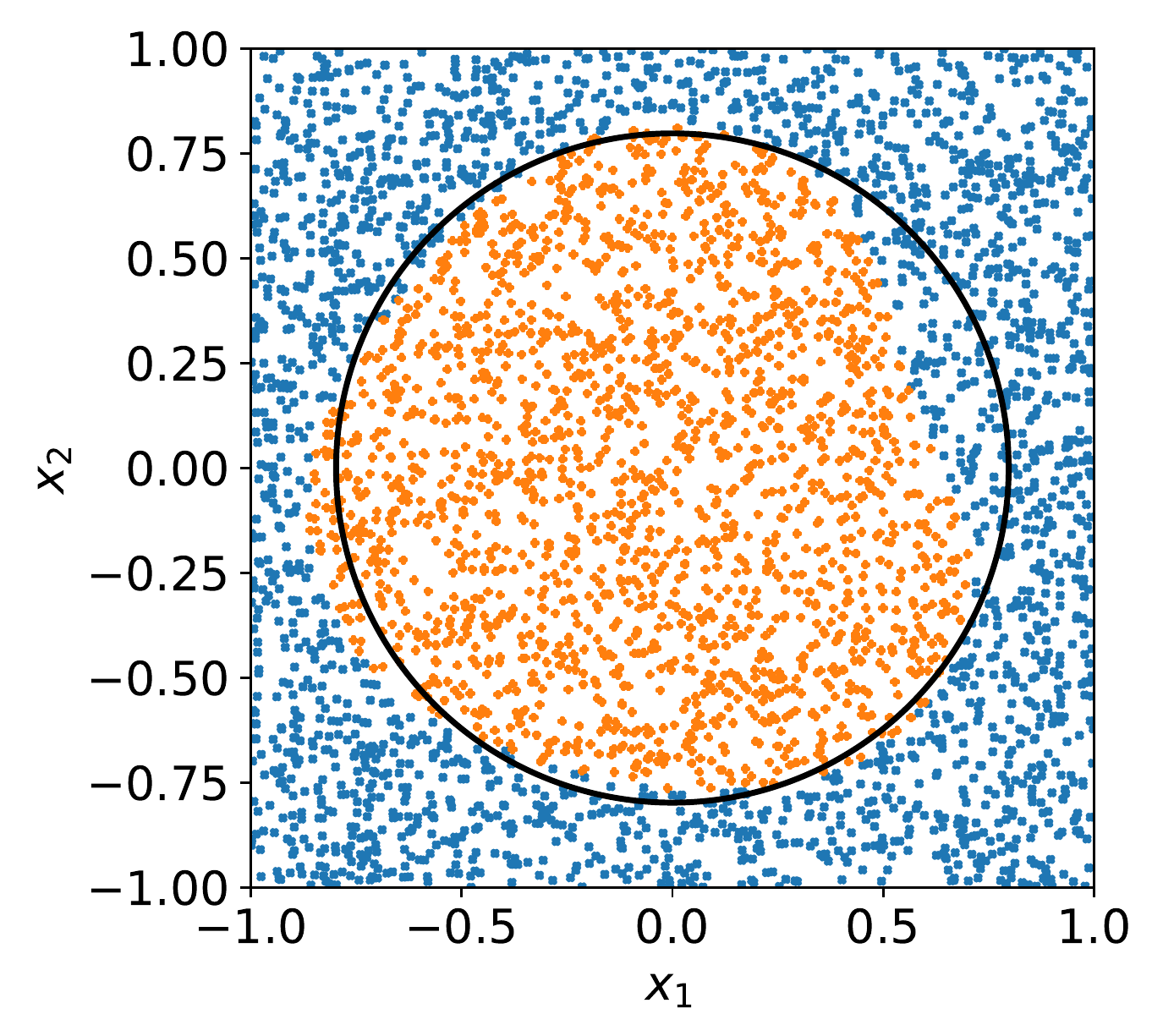}}
\subfigure[\hspace{0.05cm} 8 layers]{\includegraphics[width=0.24\linewidth]{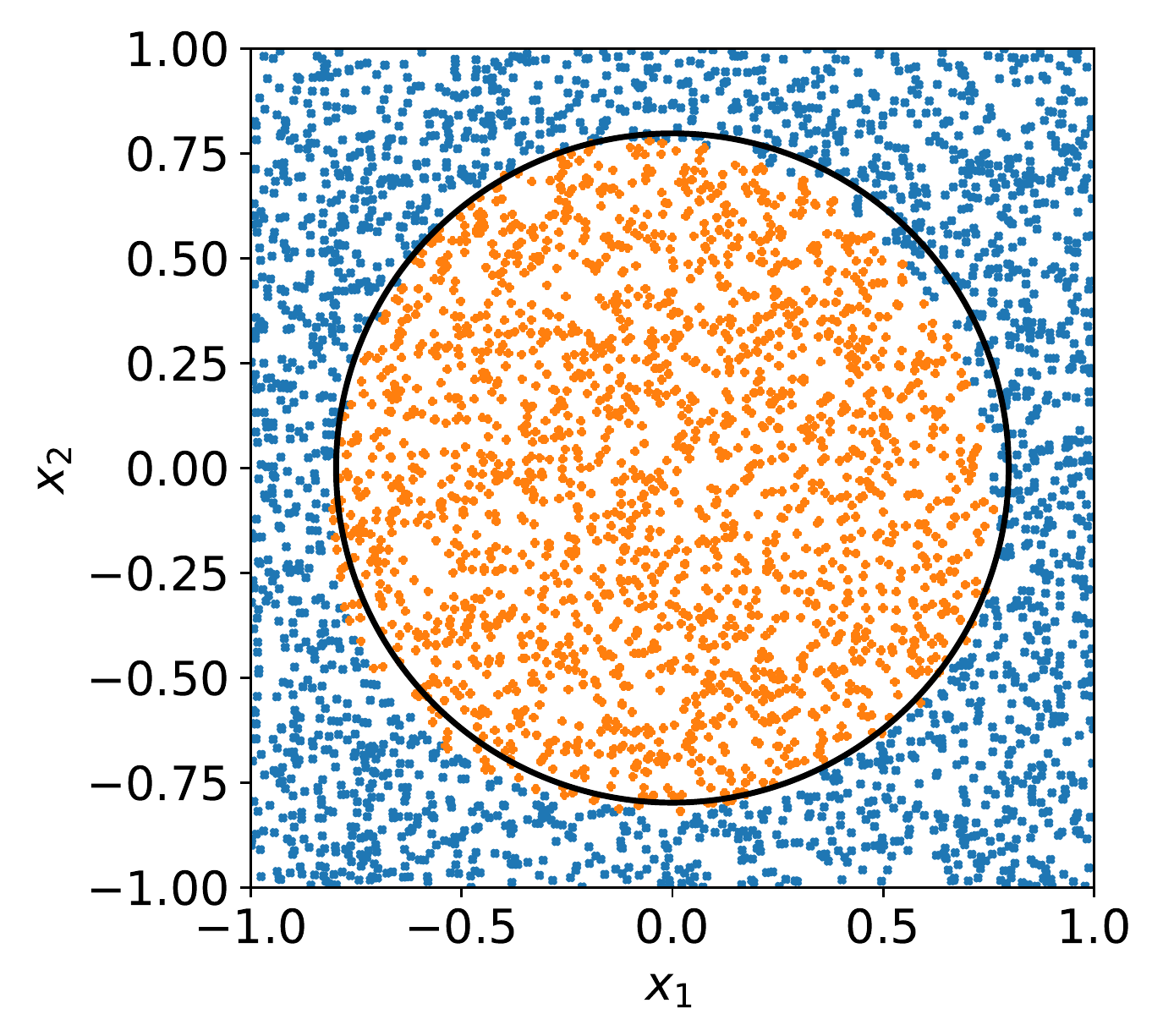}}

\end{adjustwidth}
\caption{Results of the circle classification obtained with a single-qubit classifier with different number of layers using the L-BFGS-B minimizer and $\chi^2_{wf}$ cost function. One layer returns a random classifier, while the circular shape is already captured with two layers. More layers refine the previous result.}
\label{Fig:1circle_evol}
\end{figure}

In Fig.~\ref{Fig:1circle_evol} it is depicted the evolution of the single-qubit classifier for the circle problem as more layers (1, 2, 4, 8) are added. The first layer cannot provide any information since the classification is essentially random. However, adding a second layer is enough for the classifier to broadly capture the general properties of the classifier. When this comprehension is complete, subsequent layers lead to further refinements of the classification. 

The characterization of a closed curve is not a trivial problem in classical Machine Learning. Single-layer \ac{nn}s work in a linear regime, that is, the approximations to any curve must be done by superposition of many linear functions. The quantum classifier is constructed using rotations as building-blocks. Thus, the classification of a circle seems a natural function to classify, in the same sense as a linear dataset is easily understood by a \ac{nn}. Thus, the quantum classifier must be tested in more complex scenarios to properly benchmark its capabilities. 

\subsubsection{Multi-class classification: 3 circles}
Multi-class classification is now addressed for the first time. Here it is shown that a single-qubit classifier is capable to solve this problem. The same 2D plane as for the circle problem is divided in four different regions of different shapes and sizes. In this case, three classes correspond to three different circular sectors with different centers and radii. The fourth class is the remaining space among circles. This dataset is referred to as the {\sl 3 circles} problem. This dataset is non-linear and conceptually difficult to solve for a classical \ac{nn}. 

The summary of results for this multi-class problem is depicted in Tab.~\ref{tab:results_3circles}. The single-qubit classifiers surpasses the 90\% threshold with 10 layers, 54 parameters. In this problem, the difference between cost functions $\chi^2_f$ and $\chi^2_{wf}$ is smaller than in the circle problem. In addition, the classifier saturates at success rates of $\sim 91\%$. The introduction of several qubits and entanglement makes possible to reach the saturation regime with less parameters, specially for the weighted fidelity function.

In light of these results it is possible to observe that the performance of the classifier does not only depend on the number of parameters, but also on the minimization process and the presence of local minima. Notice that success rates do not always improve with the number of layers and parameters.  

\begin{table}[t!]
\centering
\scriptsize
\begin{tabular}{c|c|cc|c|cc|cc}
  & \multicolumn{3}{c|}{$\chi^{2}_{f}$} & \multicolumn{5}{c}{$\chi^{2}_{wf}$} \\
 \hline
Qubits & 1 & \multicolumn{2}{c|}{2 } & 1 & \multicolumn{2}{c|}{2} & \multicolumn{2}{c}{4 }   \\
Layers  & & No Ent. & Ent. & & No Ent. & Ent. & No Ent. & Ent. \\ 
 \hline
 1 & 0.73 & 0.56 & -- & 0.75 & 0.81 & -- & 0.88 & -- \\
 2 & 0.79 & 0.77 & 0.78 & 0.76 & 0.90 & 0.83 & 0.90 & 0.89 \\
 3 & 0.79 & 0.76 & 0.75 & 0.78 & 0.88 & 0.89 & 0.90 & 0.89 \\
 4 & 0.84 & 0.80 & 0.80 & 0.86 & 0.84 & 0.91 & 0.90 & 0.90 \\
 5 & 0.87 & 0.84 & 0.81 & 0.88 & 0.87 & 0.89 & 0.88 & 0.92 \\
 6 & 0.90 & 0.88 & 0.86 & 0.85 & 0.88 & 0.89 & 0.89 & 0.90 \\
 8 & 0.89 & 0.85 & 0.89 & 0.89 & 0.91 & 0.90 & 0.88 & 0.91 \\
10 & 0.91 & 0.86 & 0.90 & 0.92 & 0.90 & 0.91 & 0.87 & 0.91 \\
\end{tabular}
\caption{Success rates of the single- and multi-qubit classifiers with data re-uploading for the 3-circles problem. Words ``Ent." and ``No Ent." refer to considering entanglement between qubits or not, respectively. The L-BFGS-B minimization method with the weighted fidelity and fidelity cost functions is used. Weighted fidelity cost function presents better results than the fidelity cost function. The multi-qubit classifier reaches 0.90 success rate with a lower number of layers than the single-qubit classifier. The introduction of entanglement slightly increases the success rate respect the non-entangled circuit.}
\label{tab:results_3circles}
\end{table}

\begin{figure}[t!]
\begin{adjustwidth}{-2cm}{-1cm}
\centering
\subfigure[\hspace{0.05cm} 1 layer]{\includegraphics[width=0.24\linewidth]{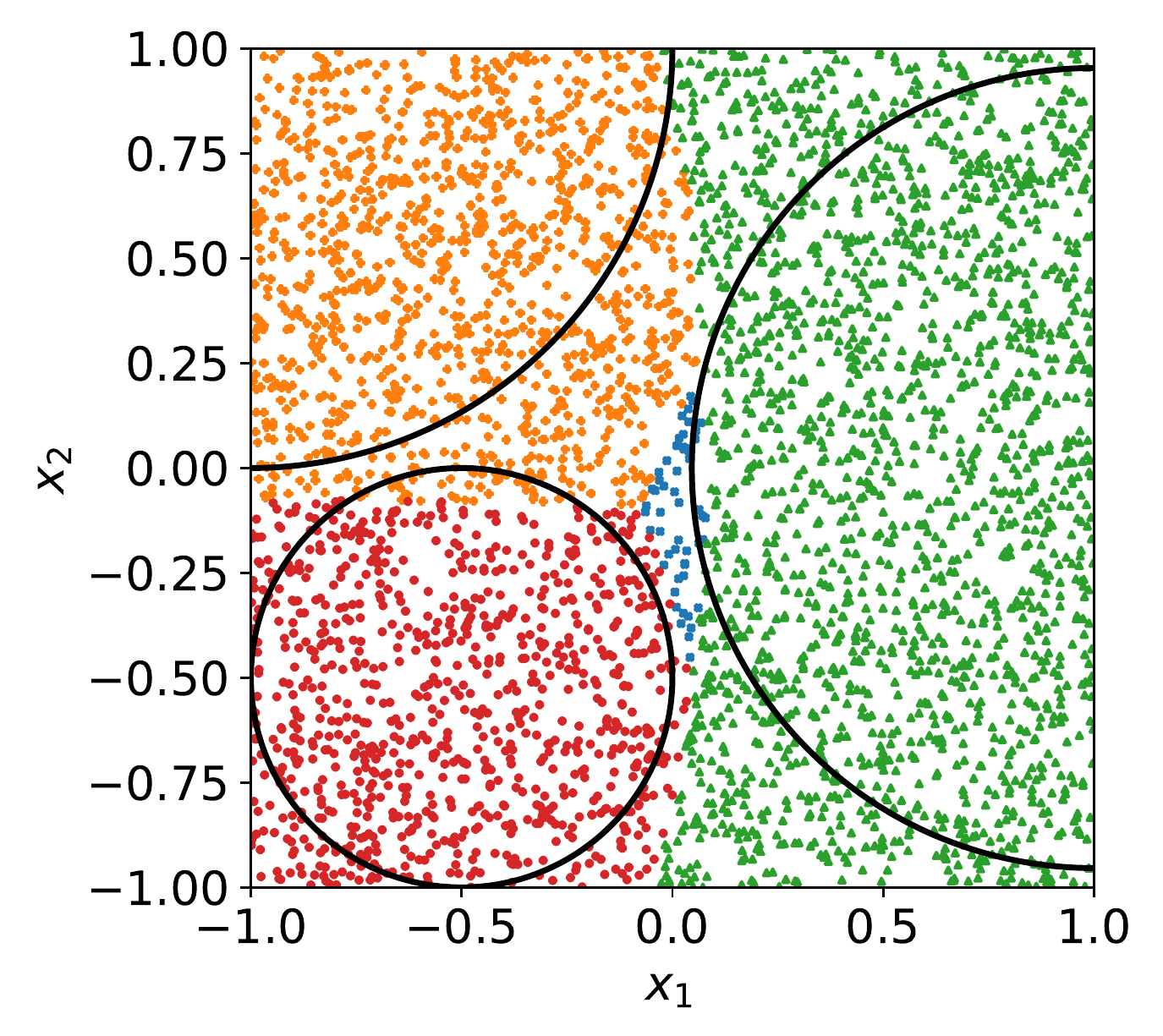}}
\subfigure[\hspace{0.05cm} 3 layers]{\includegraphics[width=0.24\linewidth]{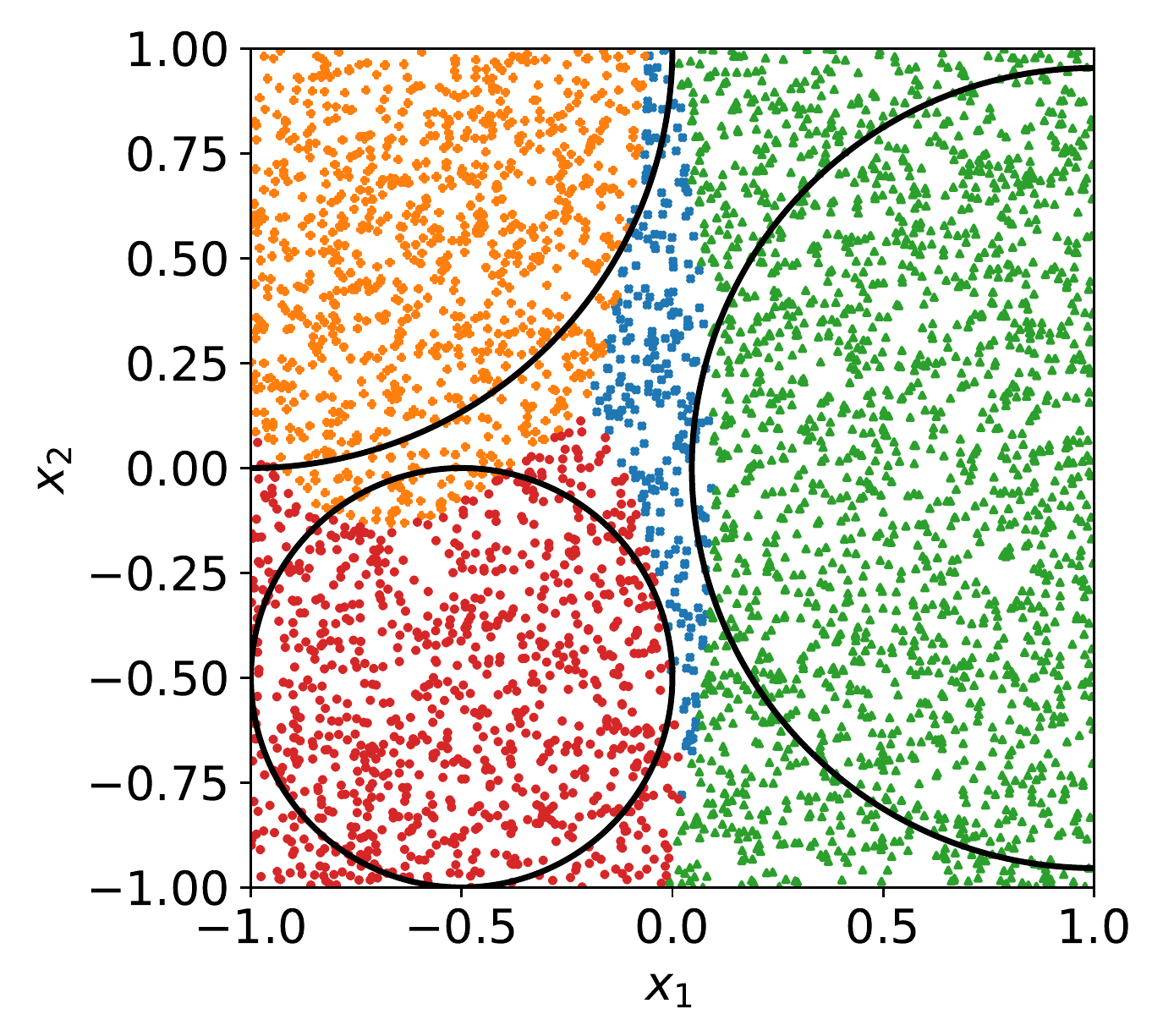}}
\subfigure[\hspace{0.05cm} 4 layers]{\includegraphics[width=0.24\linewidth]{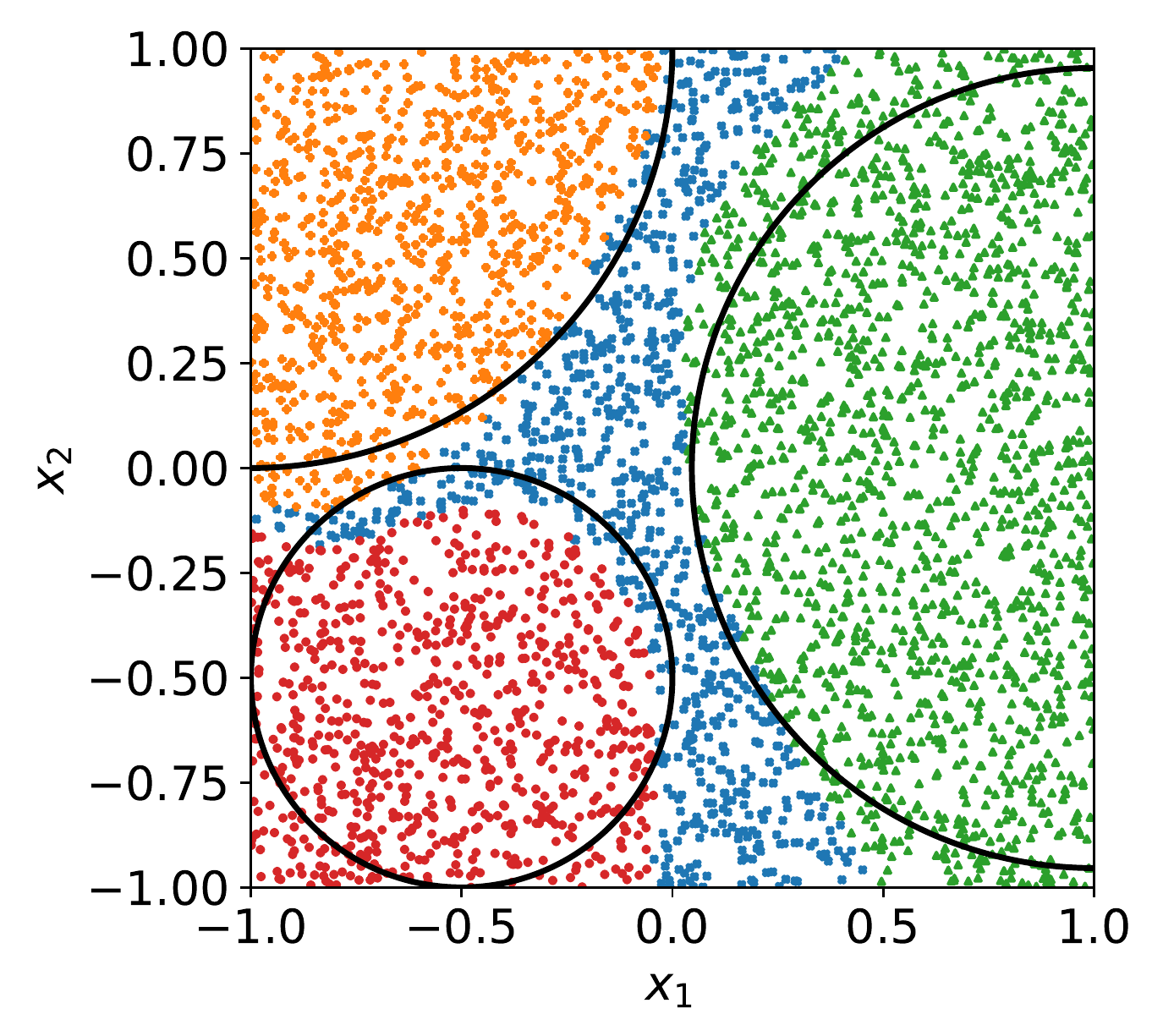}}
\subfigure[\hspace{0.05cm} 10 layers]{\includegraphics[width=0.24\linewidth]{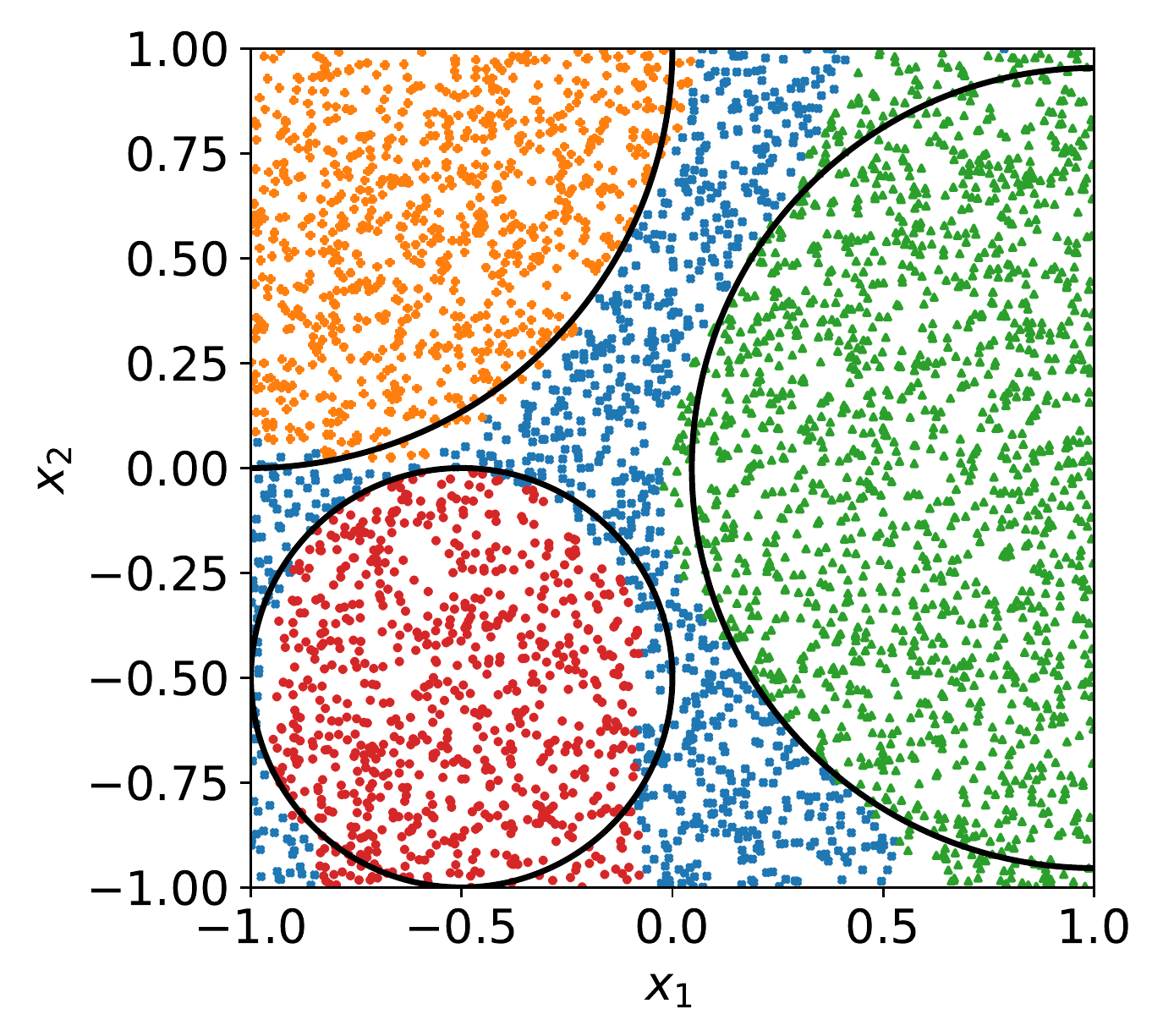}}
\end{adjustwidth}
\caption{Results of the 3 circles classification obtained with a single-qubit classifier with different number of layers using the L-BFGS-B minimizer and $\chi^2_{wf}$ cost function. With one layer, the classifier intuits the four regions although the central one is difficult to tackle. With more layers, this region is clearer for the classifier and the circular regions are adjusted.}
\label{Fig:3circles_evol}
\end{figure}

\begin{figure}[t!]
\centering
\begin{adjustwidth}{-1cm}{-2cm}
\subfigure[\hspace{2mm} Guessed points \label{fig:worldmap_a}]{\includegraphics[height=.3\linewidth]{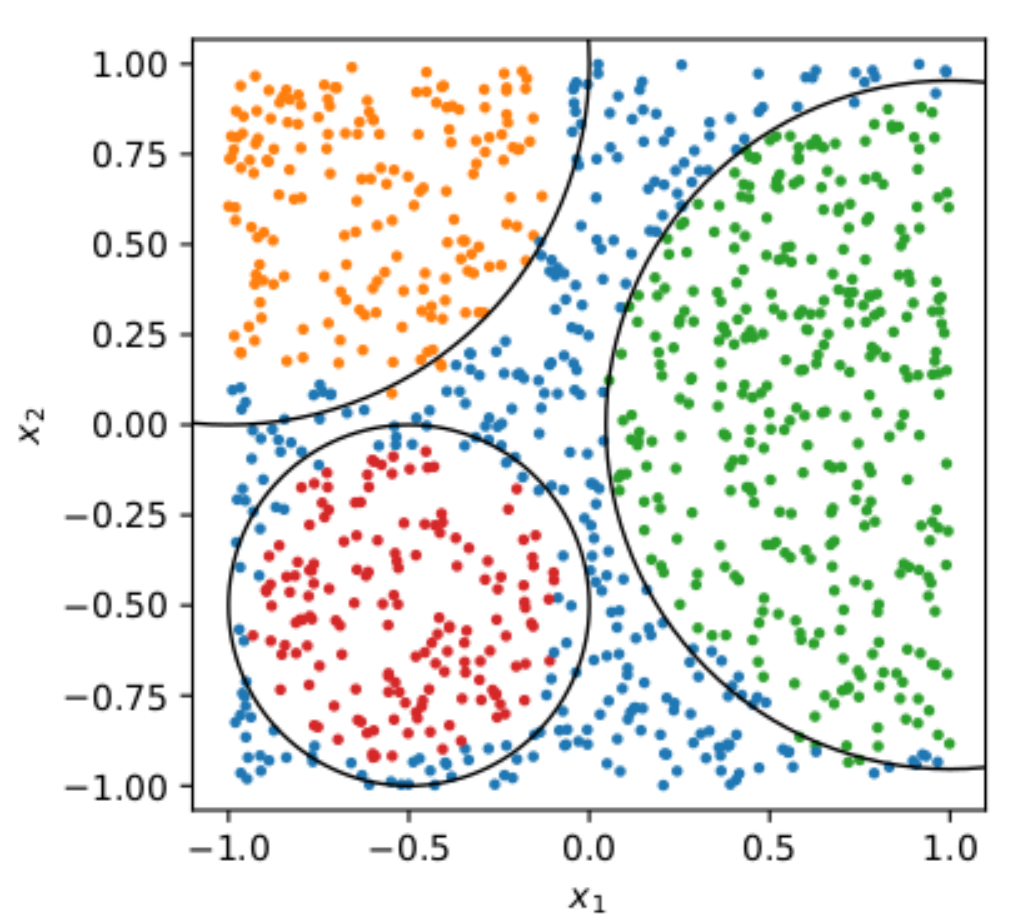}}
\subfigure[\hspace{2mm} Output states on the Bloch sphere \label{fig:worldmap_b}]{\includegraphics[height=.3\linewidth]{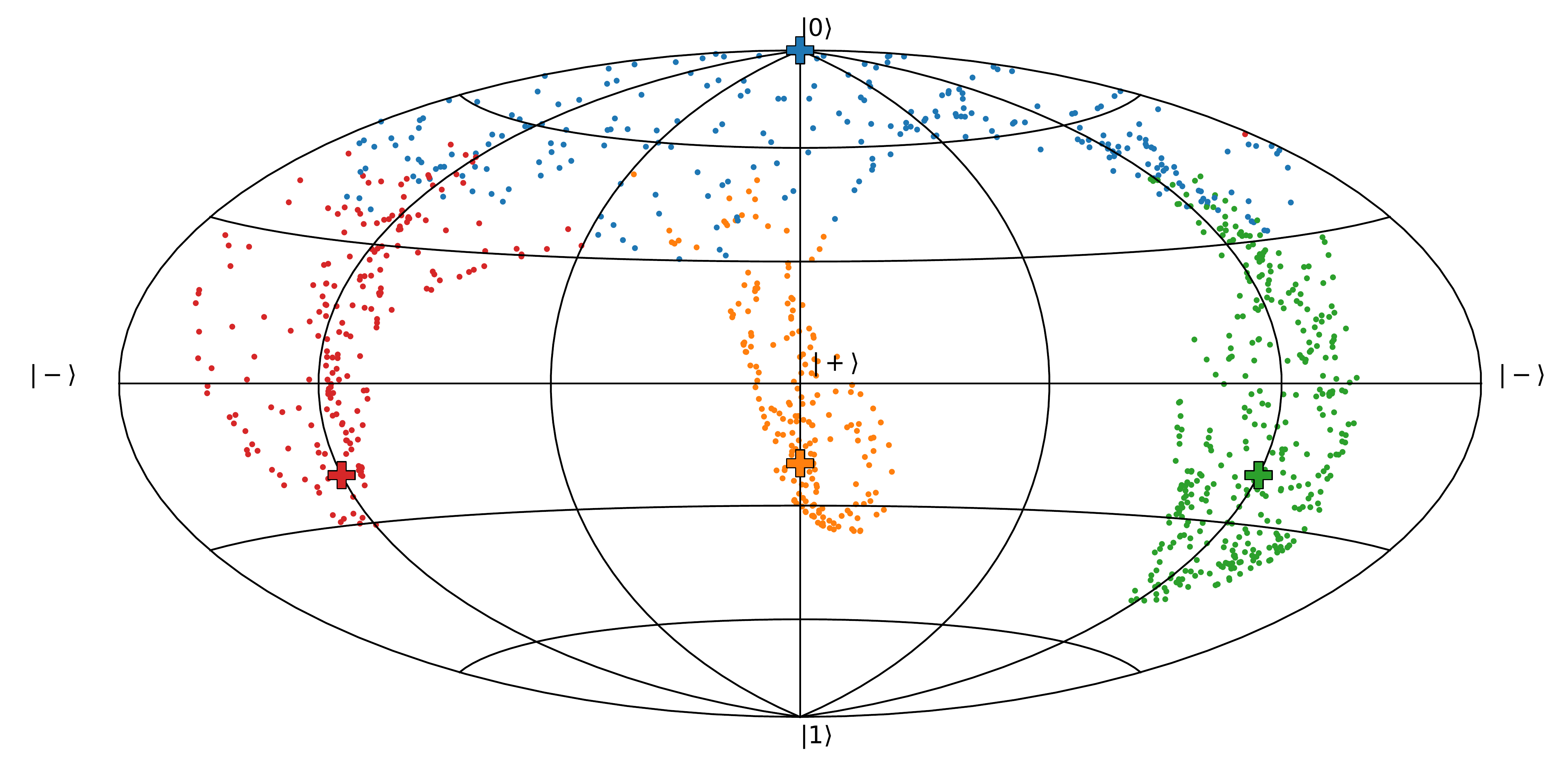}}
\end{adjustwidth}
\caption{Left: Results for a classification problem. The color corresponds to the classification guess. Bottom: output states of all data points projected on the Bloch sphere, where some states are printed for reference. Color corresponds to the actual class, and not to the classifier guess. Crosses stand for the label states of the different classes. Points belonging to the same class tend to gather around its label state.}
\label{fig:worldmap}
\end{figure}

As for the previous problem, the evolution of the final results as more layers are taken into account is observed. Fig.~\ref{Fig:3circles_evol} shows this data for an increasing number of layers. It is worth mentioning that in the very first attempt with one layer the classifier is capable to identify four different regions, that is four different classes, but the boundaries among them are only vaguely learnt. A significant change is observed from 3 to 4 layers. At this stage, the geometrical figures of the datasets are captured by the classifier. With 10 layers, further refinements are accomplished. 

The 3-circles problem is an illustrative example to see how the output state behaves for classifying data. In Fig.~\ref{fig:worldmap} we can see a classification of data, as extracted from the documentation {\tt Qibo} \cite{qibo_code}. The first plot, Fig.~\ref{fig:worldmap_a}, shows the guesses of the classifier (left) and what points are correctly guesses by the classifier (left), as in other examples here provided. In the second plot, Fig.~\ref{fig:worldmap_b}, the output states of all different points are projected in a Bloch sphere as in a world map. The states $\ket 0, \ket 1, \ket +, \ket -$ are printed for reference. The colors correspond to the actual classes, and not to the guesses of the classification, while crosses stand for the label states of each class.  It is straightforward to see that all points corresponding to the same class gather around the label states. 

\subsubsection{Classification of high-dimensional datasets: hypersphere}

The quantum classifier does not have any restriction in the dimension of the re-uploaded data. As mentioned in Sec.~\ref{ssec:qlassifier}, every gate can accomodate up to three features of data. In case the dimension is larger, then the re-uploading steps can be split into different steps, see Eq~\eqref{eq:multiple_dim}. Larger dimensionality can be managed by using more gates. 

\begin{table}[t!]
\centering
\scriptsize
\begin{tabular}{c|c|cc|c|cc|cc}
  & \multicolumn{3}{c|}{$\chi^{2}_{f}$} & \multicolumn{5}{c}{$\chi^{2}_{wf}$} \\
 \hline
Qubits & 1 & \multicolumn{2}{c|}{2 } & 1 & \multicolumn{2}{c|}{2} & \multicolumn{2}{c}{4 }   \\
Layers  & & No Ent. & Ent. & & No Ent. & Ent. & No Ent. & Ent. \\ 
 \hline
 1 & 0.87 & 0.87 & -- & 0.87 & 0.87 & -- & 0.90 & -- \\
 2 & 0.87 & 0.87 & 0.87 & 0.87 & 0.92 & 0.91 & 0.90 & 0.98 \\
 3 & 0.87 & 0.87 & 0.87 & 0.89 & 0.89 & 0.97 & -- & -- \\
 4 & 0.89 & 0.87 & 0.87 & 0.90 & 0.93 & 0.97 & -- & -- \\
 5 & 0.89 & 0.87 & 0.87 & 0.90 & 0.93 & 0.98 & -- & -- \\
 6 & 0.90 & 0.87 & 0.87 & 0.95 & 0.93 & 0.97 & -- & -- \\
 8 & 0.91 & 0.87 & 0.87 & 0.97 & 0.94 & 0.97 & -- & -- \\
10 & 0.90 & 0.87 & 0.87 & 0.96 & 0.96 & 0.97 & -- & -- \\
\end{tabular}
\caption{Success rates of the single- and multi-qubit classifiers with data re-uploading for the four-dimensional hypersphere problem. Words ``Ent." and ``No Ent." refer to considering entanglement between qubits or not, respectively. The L-BFGS-B minimization method with the weighted fidelity and fidelity cost functions is used. The fidelity cost function is likely to get stuck in local minima for the multi-qubit classifiers. $\chi^2_{wf}$ results are much better, peaking at 0.98 success rates with only two layers in the entangled four-qubit classifier. Unlike in other examples, the presence of entanglement significantly improves the performance.}
\label{tab:results_hypersphere}
\end{table}

With this idea more complicated classifications can be addressed. In particular, the classification of a 4D hypersphere is used as testbed. This problem is just an extension of the circle one, where the radius of the hypersphere changed to fill half the volume of the feature space. This time, the training set is composed of 1000 random points. 

Results are summarized in Tab.~\ref{tab:results_hypersphere}. A single-qubit classifier reaches its maximum success rate 97\% for 8 layers, 82 parameters, with the $\chi^2_{wf}$ cost function. Two-qubit classifiers peak at 5 layers (62 parameters) for the entangled case, and four-qubits classifiers have their maximum with 2 layers (82 parameters). Note that, in this case, entanglement provides better final results as compare to other problems. Four-qubit classifiers with more layers are not considered due to their training computational cost.

\subsubsection{Classification of non-convex datasets: tricrown}
\begin{figure}[t!]
\centering
\subfigure[\hspace{0.05cm} 1 layer ]{\includegraphics[width=0.24\linewidth]{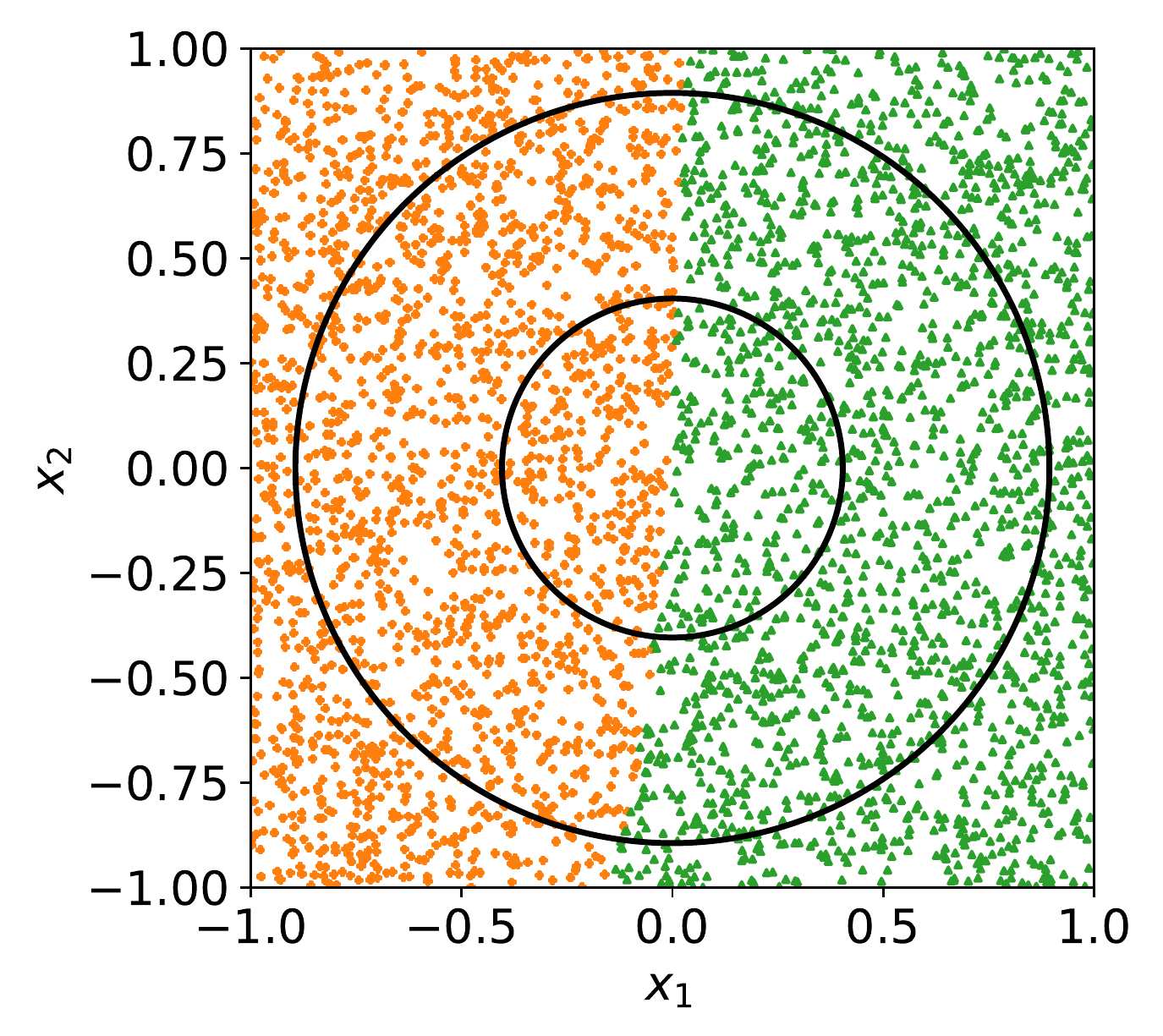}}
\subfigure[\hspace{0.05cm} 2 layers]{\includegraphics[width=0.24\linewidth]{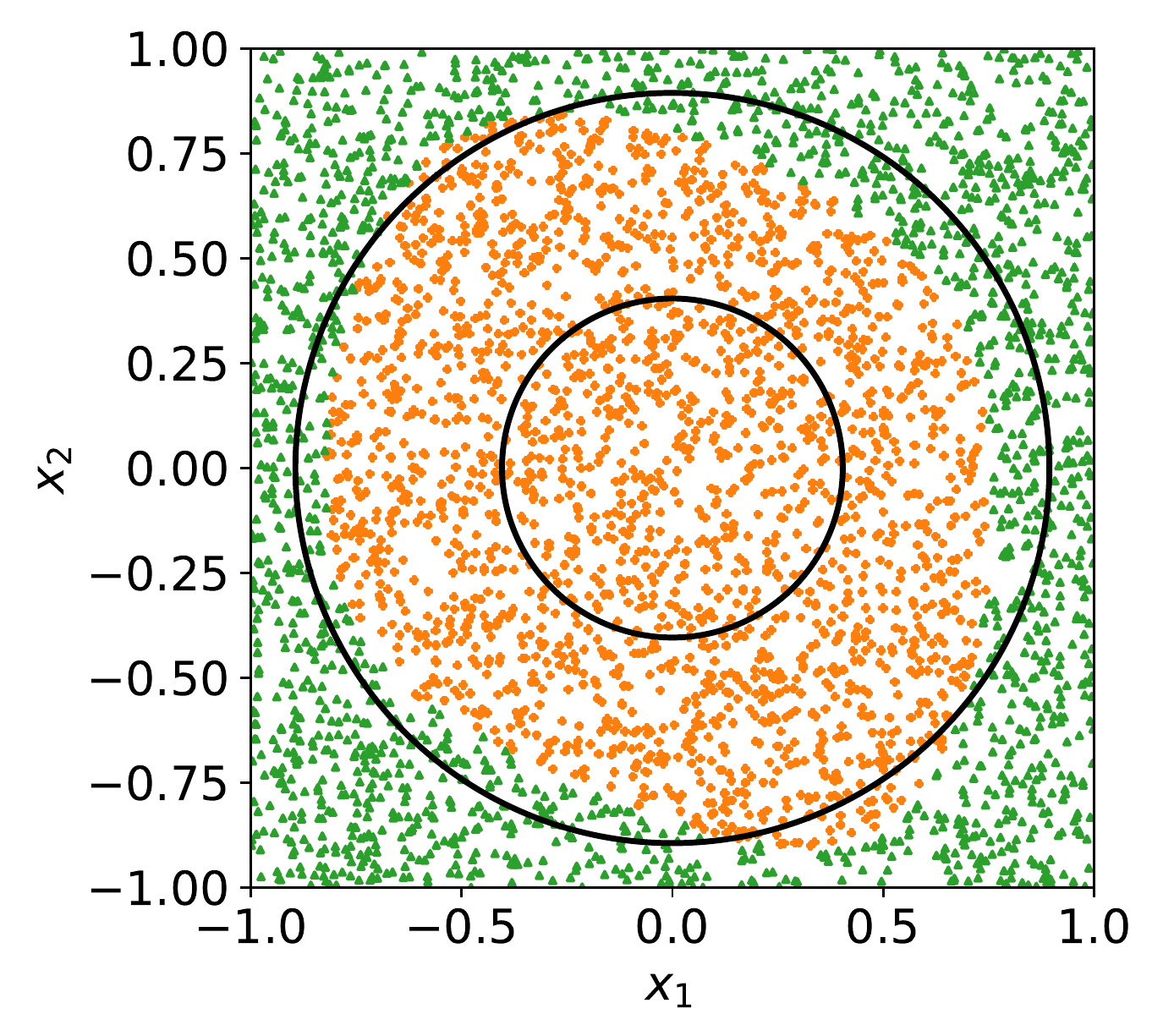}}
\subfigure[\hspace{0.05cm} 3 layers]{\includegraphics[width=0.24\linewidth]{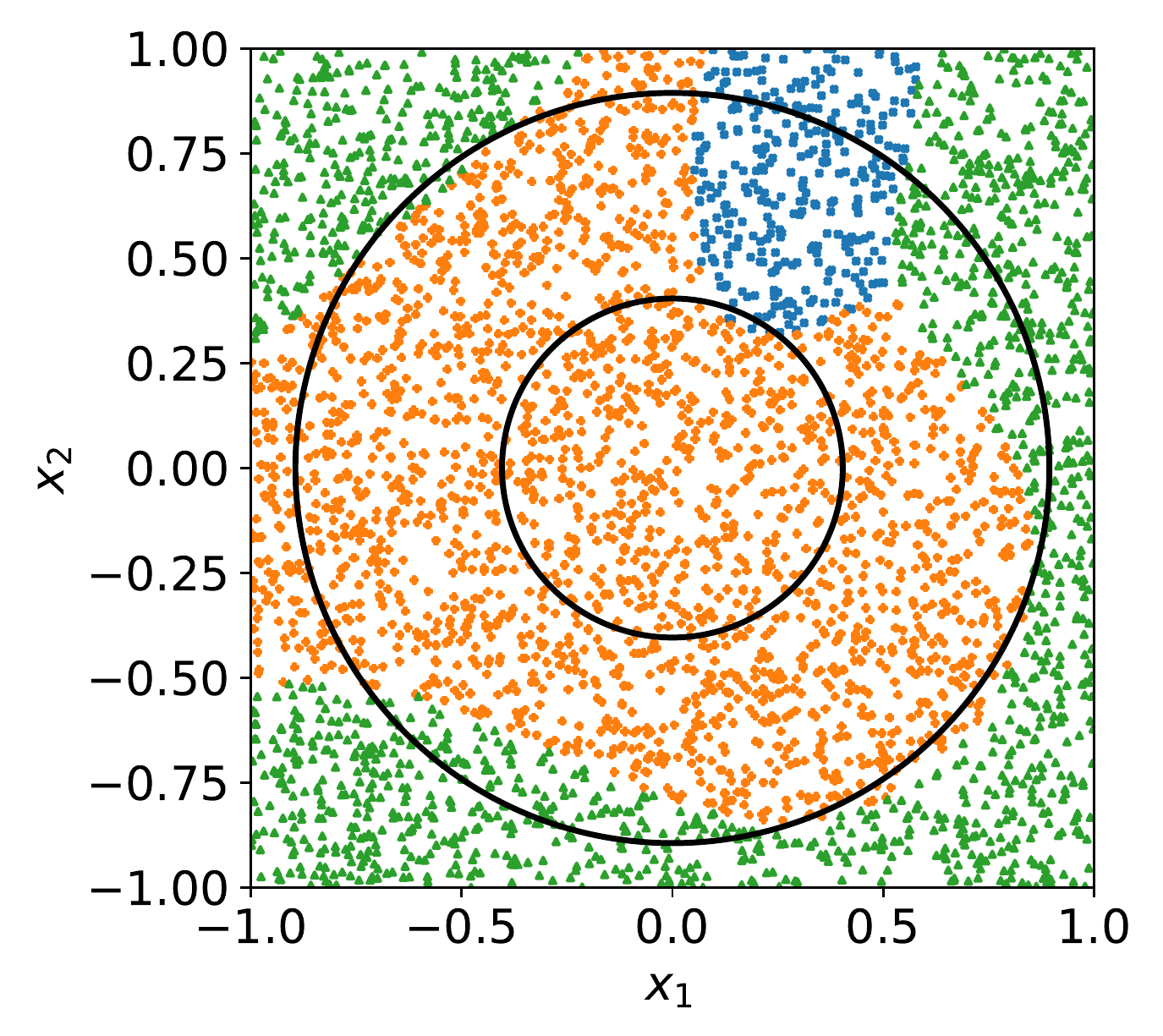}}
\subfigure[\hspace{0.05cm} 4 layers]{\includegraphics[width=0.24\linewidth]{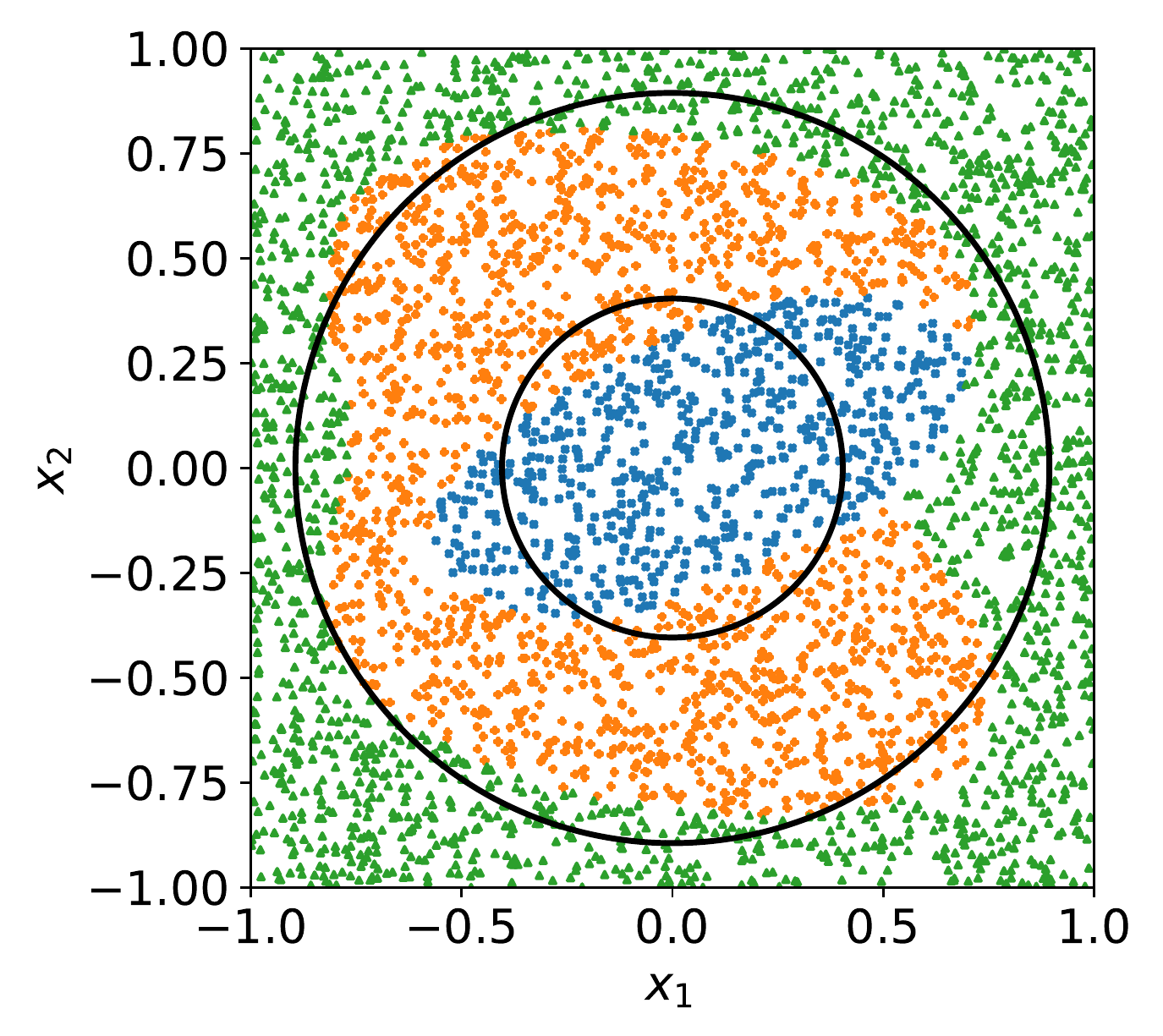}}

\subfigure[\hspace{0.05cm} 5 layers]{\includegraphics[width=0.24\linewidth]{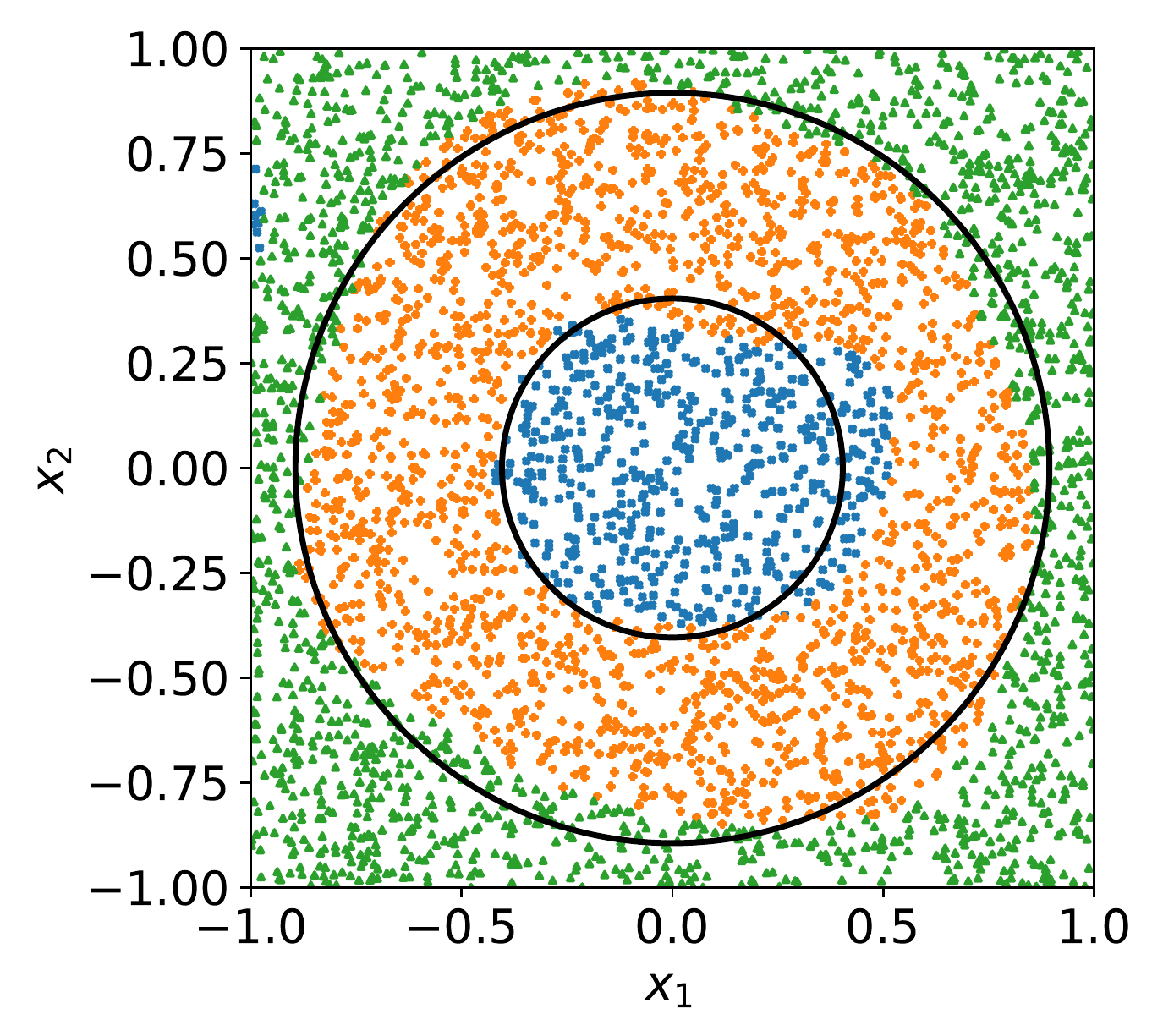}}
\subfigure[\hspace{0.05cm} 6 layers]{\includegraphics[width=0.24\linewidth]{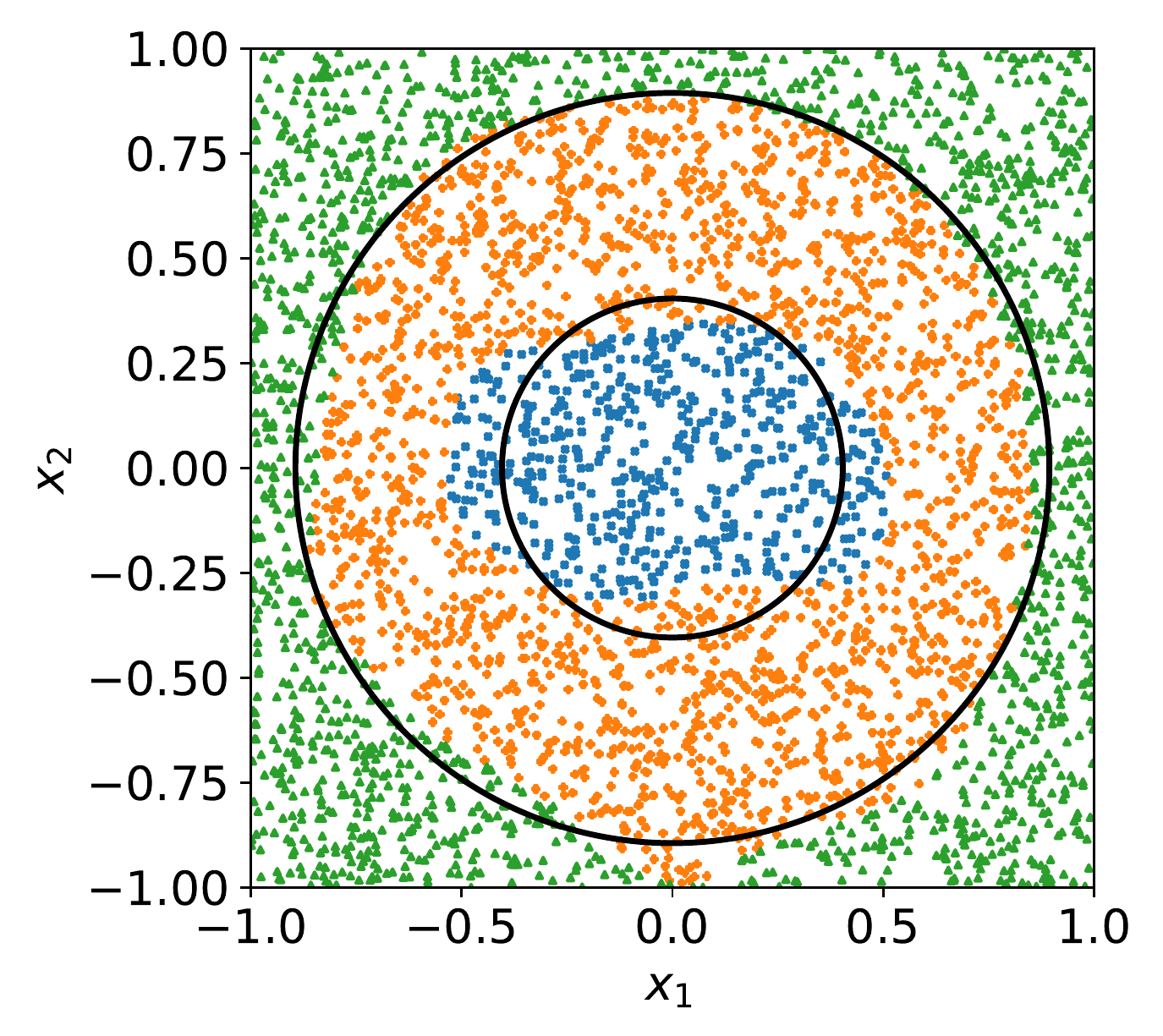}}
\subfigure[\hspace{0.05cm} 8 layers]{\includegraphics[width=0.24\linewidth]{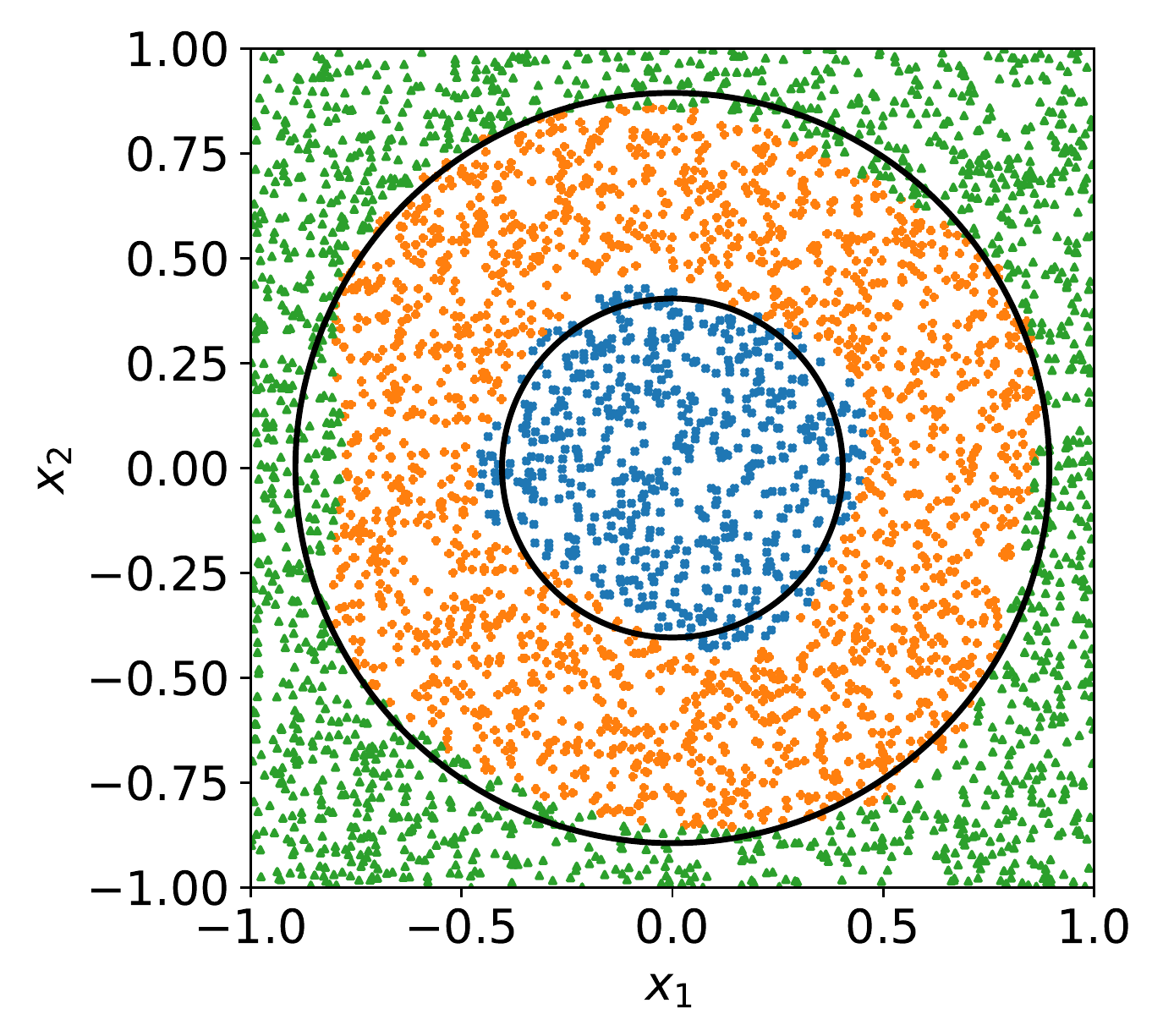}}
\subfigure[\hspace{0.05cm} 10 layers]{\includegraphics[width=0.24\linewidth]{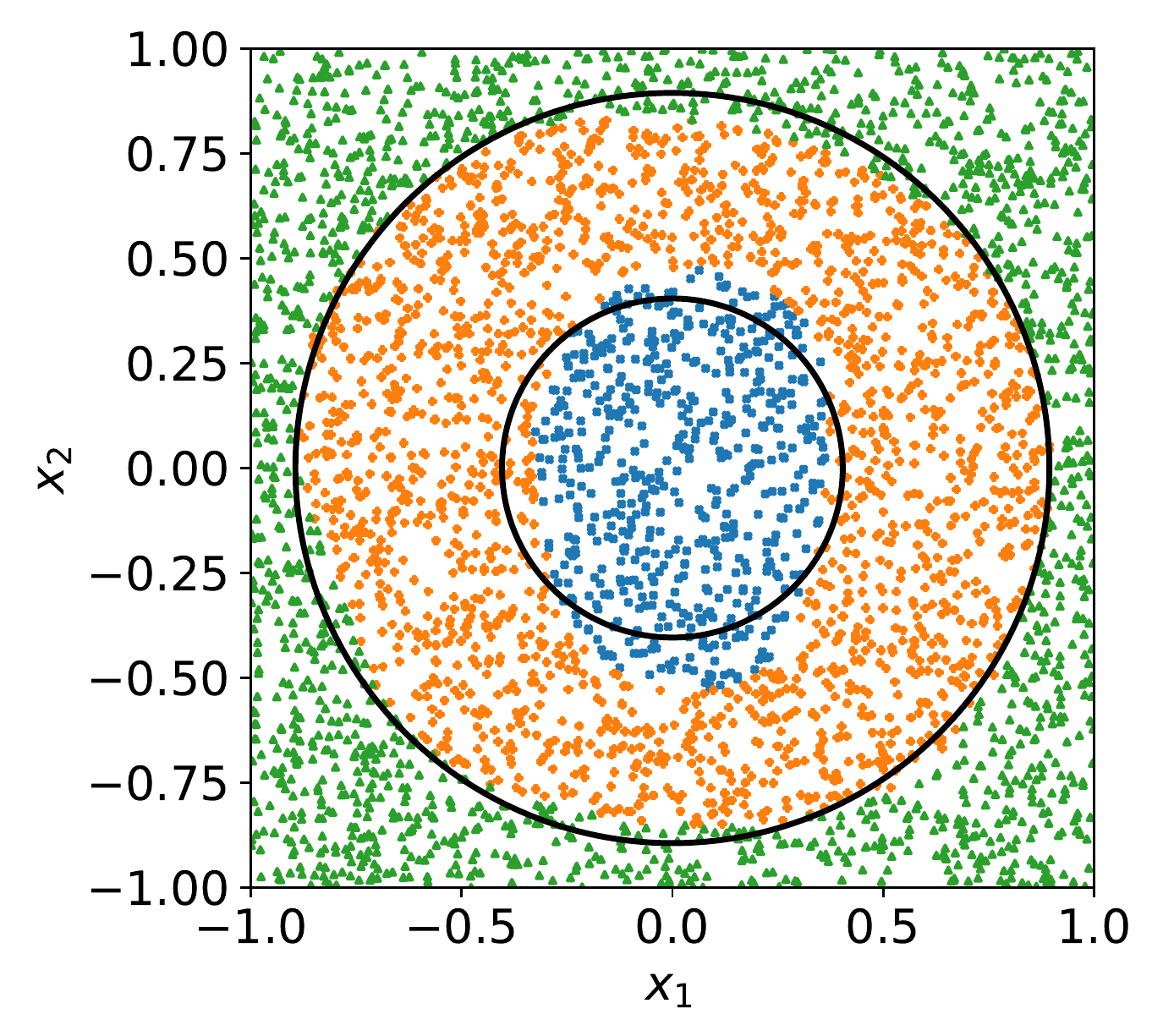}}
\caption{Results obtained with the single-qubit classifier for the tricrown problem, using the weighted fidelity cost function during the training. Notice that 2 layers can capture 2 classes reasonably well, while the inner one is forgotten. 4 layers locate all classes in their corresponding places, and 5 layers learn the general shape of the datset. Then, further layers add refinement to the classification.}
\label{Fig:3crown_evol}
\end{figure}

\begin{table}[t!]
\centering
\scriptsize
\begin{tabular}{c|c|cc|c|cc|cc}
  & \multicolumn{3}{c|}{$\chi^{2}_{f}$} & \multicolumn{5}{c}{$\chi^{2}_{wf}$} \\
 \hline
Qubits & 1 & \multicolumn{2}{c|}{2 } & 1 & \multicolumn{2}{c|}{2} & \multicolumn{2}{c}{4 }   \\
Layers  & & No Ent. & Ent. & & No Ent. & Ent. & No Ent. & Ent. \\ 
 \hline
 1 & 0.34 & 0.51 & -- & 0.43 & 0.77 & -- & 0.81 & -- \\
 2 & 0.57 & 0.63 & 0.59 & 0.76 & 0.79 & 0.82 & 0.87 & 0.96 \\
 3 & 0.80 & 0.68 & 0.65 & 0.68 & 0.94 & 0.95 & 0.92 & 0.94 \\
 4 & 0.84 & 0.78 & 0.89 & 0.79 & 0.93 & 0.96 & 0.93 & 0.96 \\
 5 & 0.92 & 0.86 & 0.82 & 0.88 & 0.96 & 0.96 & 0.96 & 0.95 \\
 6 & 0.93 & 0.91 & 0.93 & 0.91 & 0.93 & 0.96 & 0.97 & 0.96 \\
 8 & 0.90 & 0.89 & 0.90 & 0.92 & 0.94 & 0.95 & 0.95 & 0.94 \\
10 & 0.90 & 0.91 & 0.92 & 0.93 & 0.95 & 0.96 & 0.95 & 0.95 \\
\end{tabular}
\caption{Success rates of the single- and multi-qubit classifiers for the tricrown problem. Words ``Ent." and ``No Ent." refer to considering entanglement between qubits or not, respectively. The $\chi^2_{wf}$ cost function presents better success rates than $\chi^2_f$. The multi-qubit classifiers improve the results obtained with the single-qubit classifier but the using of entanglement does not introduce significant changes.}
\label{tab:results_3crown}
\end{table}

Non-convex patterns are usually difficult to classify for in Supervised Learning frameworks. In this example two concentric circles with different radii defining three different classes of similar sizes are studied. Thus, this is a non-convex and multi-class problem. 

Results for this tricrown problem are summarized in Tab.~\ref{tab:results_3crown}. A success rate of 93\% is achieved with 10 layers for the single-qubit classifier and cost function $\chi^2_{wf}$. Two-qubit classifiers peak at 94\% (3 layers), while four-qubit obtain 96\% (2 entangled layers). 

The results of different layers for single-qubit classifiers with different numbers of layers show the evolution in the performance of the classification, see Fig.~\ref{Fig:3crown_evol}. 4 layers are needed to learn the concentric patterns, and the fifth one establishes the topology of the different regions. From this point, more layers just refine the final results. 

\subsubsection{Other datasets: non-convex, sphere, squares, wavy lines}
The single- and multi-qubit classifiers are tested in more datasets to complete the benchmark. These extra datasets cover different kinds of training data and aim to show that the quantum classifier can adapt itself to large varieties of problems. From a qualitative point of view, the results here presented are just an extension of the ones seen before. Thus, the reader can skip these lines to Sec.~\ref{sssec:comp_cl_class} without any regrets. All tables and figures summarizing data are depicted at the end of the section. The datasets are similar to all previous datasets. Data points $\vec x \in [-1, 1]^d$, and the sizes are 200 / 4000 for the training / test sets. The only exception is the {\sl sphere} problem, where the sizes are 500 / 4000. The design of the problems was made in such a way that a random classifier guesses 1 / (\# classes) correctly. 

\begin{figure}[b!]
\begin{adjustwidth}{-1cm}{-2cm}
\centering
\subfigure[\hspace{0.05cm} Non-convex: $\chi^2_{wf}$, 1 qubit, 6 layers \label{Fig:non convex}]{\includegraphics[width=0.45\linewidth]{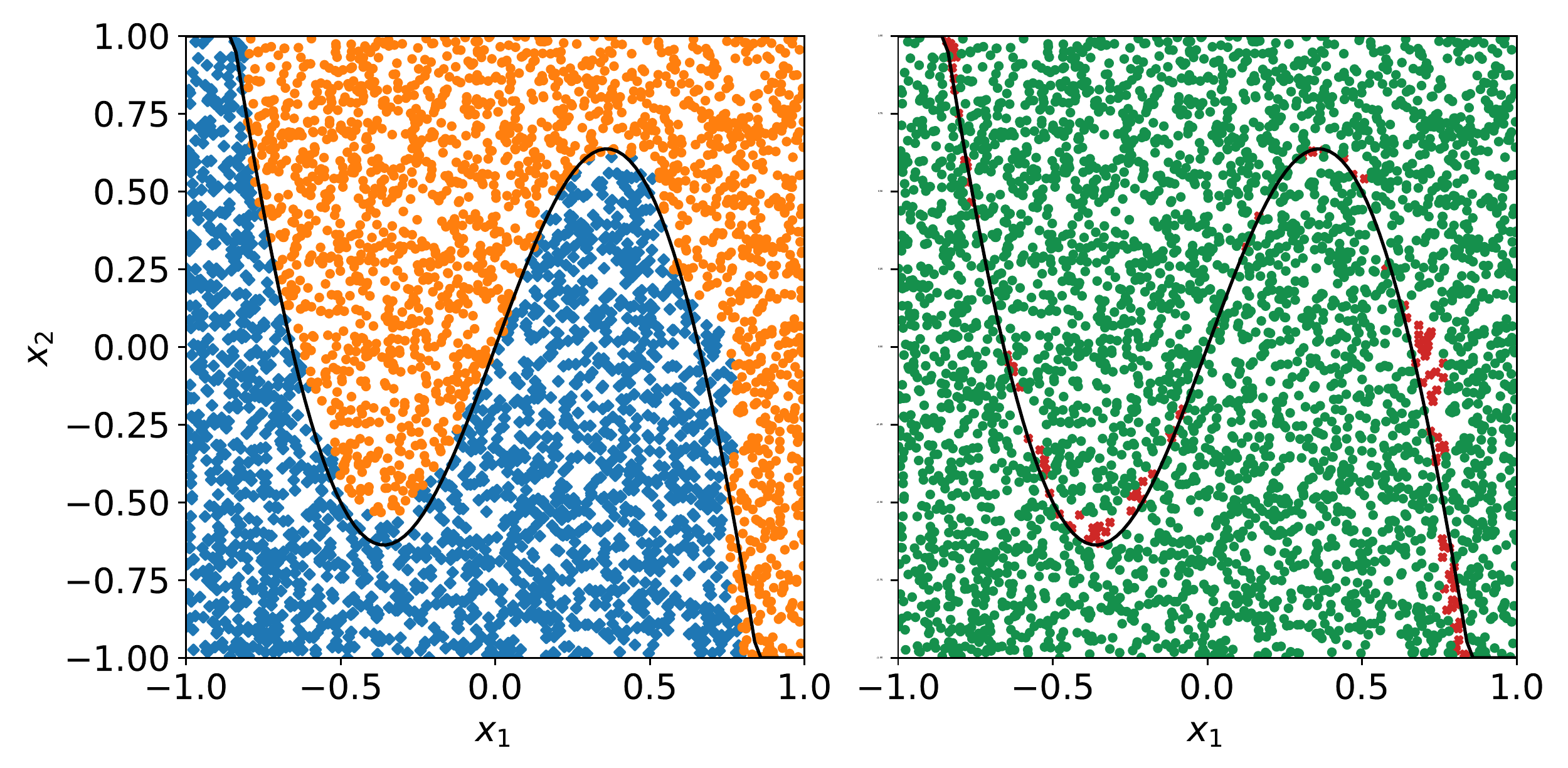}}\hfill
\subfigure[\hspace{0.05cm} Crown: $\chi^2_{wf}$, 2 qubits without entanglement, 4 layers \label{Fig:crown}]{\includegraphics[width=0.45\linewidth]{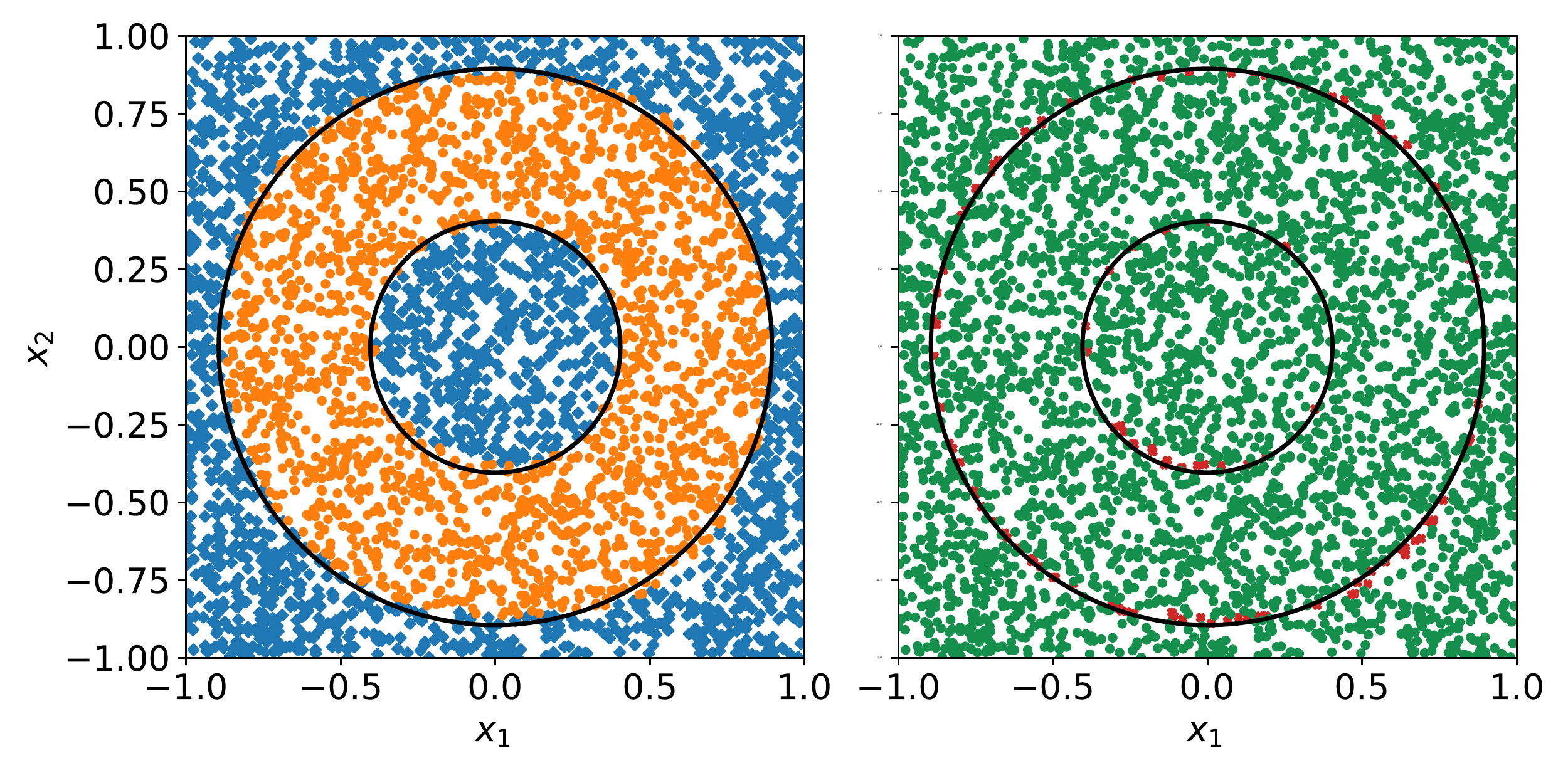}} \\
\subfigure[\hspace{0.05cm} Squares: $\chi^2_{f}$, 2 qubits without entanglement, 6 layers \label{Fig:squares}]{\includegraphics[width=0.45\linewidth]{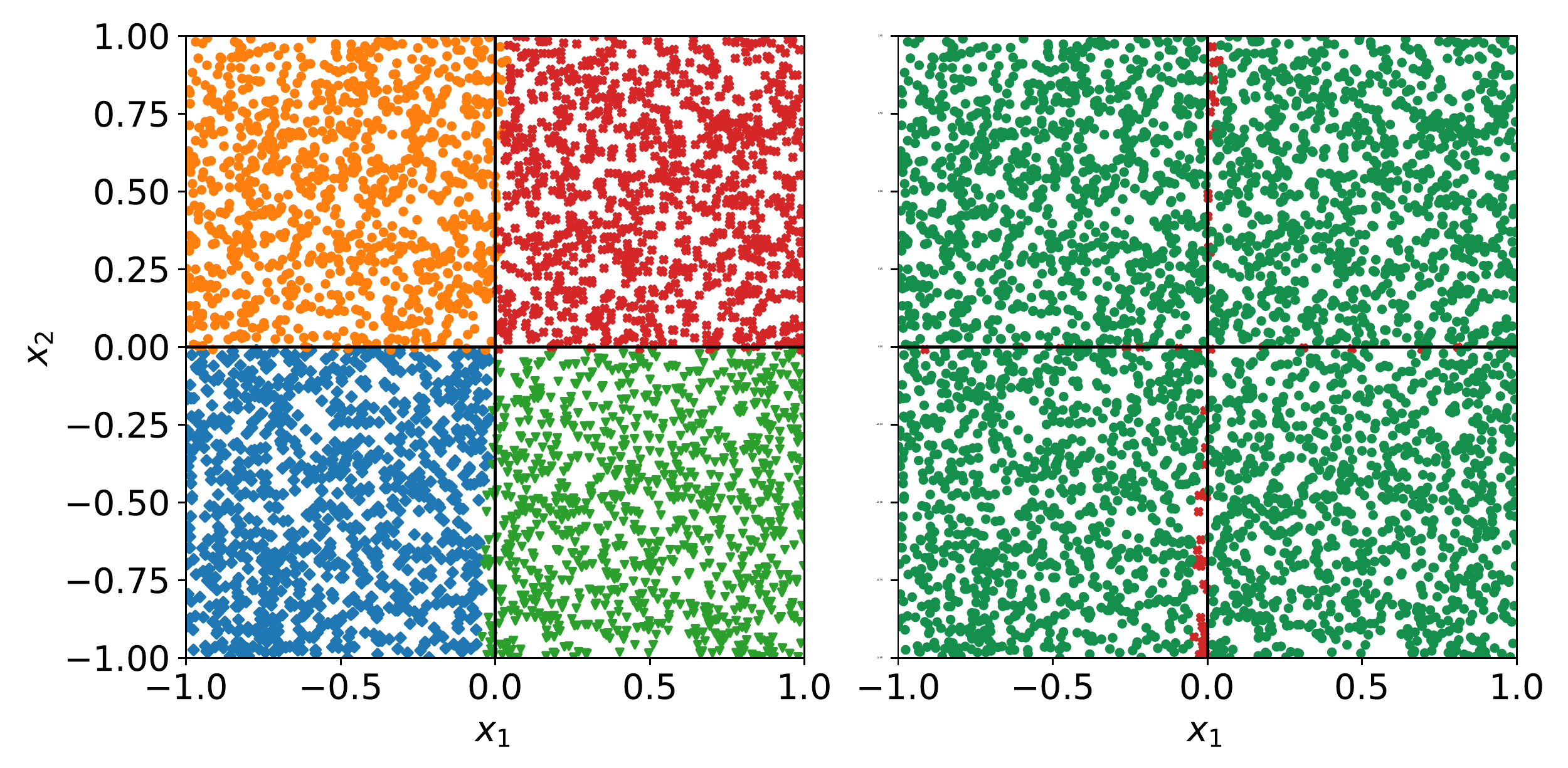}}\hfill
\subfigure[\hspace{0.05cm} Wavy lines: $\chi^2_{wf}$, 2 qubits with entanglement, 6 layers \label{Fig:wavy}]{\includegraphics[width=0.45\linewidth]{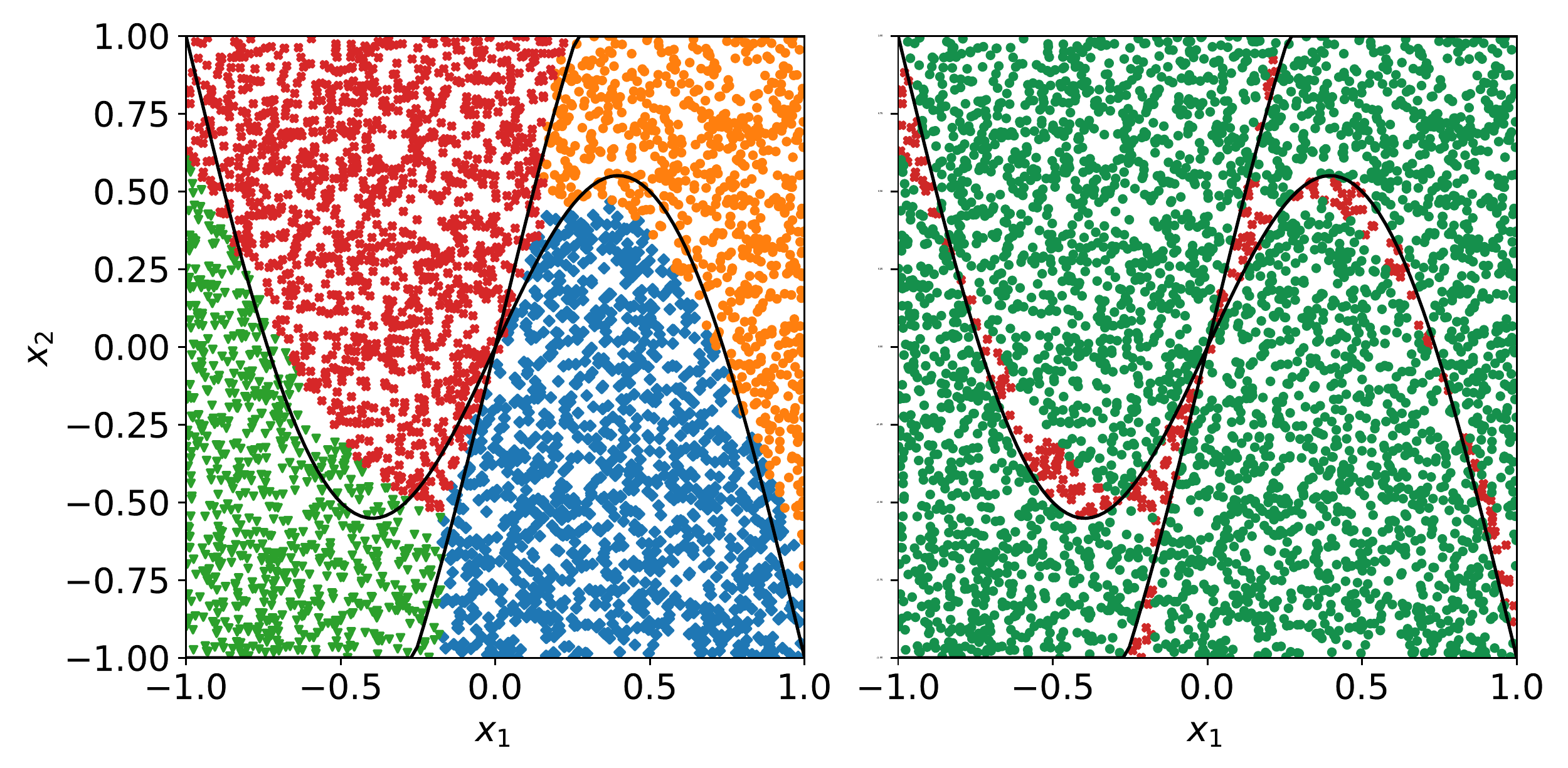}}
\end{adjustwidth}
\caption{Best results for 2D problems analyzed in this section. The problem and architecture is detailed in each caption. For similar results, the simpler architecture was chosen. Colors in the left part of each figure represent the different classes obtained from the classifier outputs, while the right images show correctly (green) and wrongly (red) classified points. Black solid lines define the problem boundaries. }
\label{Fig:Results_extra}
\end{figure}

The results described in the following reinforce the properties previously observed. The classifier starts in a first stage where the performance improves as more layers are added. Then, a stationary stage is reached. Larger number of qubits and entanglement advance the appearance of this stationary regime.

In the following, each different problem carries its own results table. The best instance for each problem in a 2D feature space is plotted in Fig.~\ref{Fig:Results_extra}. Notice in this figure that all missed points are located near the borders of the problem. This means that the classifier is properly understanding the properties of the dataset, but more training is needed to fine-tune the bordering regions.

\paragraph*{Non-convex problem} Classification problems where classes are mutually non-convex are considered difficult since the separation between both classes is hard to characterize. In this problem, both zones are separated through the line $x_2 = -2x_1 + 3 / 2 \sin(\pi x_1)$. With this boundary, there is no area so small that the classifier can achieve good results even if this area is neglected. 

\begin{table}[t!]
\centering
\scriptsize
\begin{tabular}{c|c|cc|c|cc|cc}
  & \multicolumn{3}{c|}{$\chi^{2}_{f}$} & \multicolumn{5}{c}{$\chi^{2}_{wf}$} \\
 \hline
Qubits & 1 & \multicolumn{2}{c|}{2 } & 1 & \multicolumn{2}{c|}{2} & \multicolumn{2}{c}{4 }   \\
Layers  & & No Ent. & Ent. & & No Ent. & Ent. & No Ent. & Ent. \\ 
 \hline
 1 & 0.49 & 0.55 & -- & 0.49 & 0.76 & -- & 0.76 & -- \\
 2 & 0.82 & 0.75 & 0.75 & 0.86 & 0.94 & 0.85 & 0.96 & 0.96 \\
 3 & 0.93 & 0.74 & 0.85 & 0.96 & 0.95 & 0.95 & 0.95 & 0.97 \\
 4 & 0.93 & 0.74 & 0.88 & 0.95 & 0.96 & 0.97 & 0.95 & 0.96 \\
 5 & 0.91 & 0.95 & 0.90 & 0.97 & 0.95 & 0.96 & 0.95 & 0.97 \\
 6 & 0.96 & 0.94 & 0.93 & 0.98 & 0.97 & 0.97 & 0.95 & 0.97 \\
 8 & 0.96 & 0.96 & 0.95 & 0.98 & 0.98 & 0.97 & 0.96 & 0.97 \\
10 & 0.95 & 0.92 & 0.95 & 0.96 & 0.96 & 0.96 & 0.96 & 0.97 \\
\end{tabular}
\caption{Results of the single- and multi-qubit classifiers with data re-uploading for the non-convex problem. Numbers indicate the success rate, i.e. the number of data points classified correctly over the total number of points. Words ``Ent." and ``No Ent." refer to considering entanglement between qubits or not respectively. The L-BFGS-B minimization method with the weighted fidelity and fidelity cost functions is use. Both cost functions lead to higher success rates, although the weighted fidelity cost function is better. It achieves the 0.98 success with two qubits, entanglement, and four layers.}
\label{tab:results_non_convex}
\end{table}

Tab.~\ref{tab:results_non_convex} summarizes the results for this problem. Best performance (98\%) is achieved with a single-qubit classifier of 6 layers (32 parameters) using the $\chi^2_{wf}$ cost function. The fidelity cost function $\chi^2_f$ gets 96\% in the same conditions, but peaks at 97\% for 2 entangled qubits with 8 layers (80 parameters). See this example in Fig.~\ref{Fig:non convex}. 

\paragraph*{Crown}

This one is a binary version of the tricrown dataset. The most interesting feature of this dataset is that one class is composed of two different disconnected regions. Thus, the classifier must find a way to understand disjoint regions as belonging to the same class. 

The $\chi^2_f$ function reaches its best result (94\%) for 2 entangled qubits and 6 layers (60 parameters). For $\chi^2_{wf}$, a 97\% accuracy is obtained for 2 unentangled qubits, 4 layers (40 parameters). See Tab.~\ref{tab:results_crown} for a summary, and Fig.~\ref{Fig:crown} for an example. 

\begin{table}[t!]
\centering
\scriptsize
\begin{tabular}{c|c|cc|c|cc|cc}
  & \multicolumn{3}{c|}{$\chi^{2}_{f}$} & \multicolumn{5}{c}{$\chi^{2}_{wf}$} \\
 \hline
Qubits & 1 & \multicolumn{2}{c|}{2 } & 1 & \multicolumn{2}{c|}{2} & \multicolumn{2}{c}{4 }   \\
Layers  & & No Ent. & Ent. & & No Ent. & Ent. & No Ent. & Ent. \\ 
 \hline
 1 & 0.44 & 0.50 & -- & 0.44 & 0.59 & -- & 0.66 & -- \\
 2 & 0.48 & 0.50 & 0.51 & 0.53 & 0.73 & 0.72 & 0.70 & 0.96 \\
 3 & 0.91 & 0.50 & 0.56 & 0.74 & 0.75 & 0.95 & 0.78 & 0.96 \\
 4 & 0.80 & 0.74 & 0.56 & 0.86 & 0.97 & 0.97 & 0.92 & 0.96 \\
 5 & 0.90 & 0.93 & 0.88 & 0.89 & 0.97 & 0.96 & 0.97 & 0.94 \\
 6 & 0.92 & 0.91 & 0.94 & 0.95 & 0.94 & 0.95 & 0.95 & 0.93 \\
 8 & 0.90 & 0.93 & 0.95 & 0.92 & 0.94 & 0.94 & 0.96 & 0.94 \\
10 & 0.90 & 0.92 & 0.91 & 0.92 & 0.95 & 0.93 & 0.96 & 0.93 \\
\end{tabular}
\caption{Results of the single- and multi-qubit classifiers with data re-uploading for the crown problem. Numbers indicate the success rate, i.e. the number of data points classified correctly over the total number of points. Words ``Ent." and ``No Ent." refer to considering entanglement between qubits or not respectively. The L-BFGS-B minimization method with the weighted fidelity and fidelity cost functions is use. As happens in other problems, the results obtained with the weighted fidelity cost function are better than the ones obtained with the fidelity cost function. The multi-qubit classifiers and the introduction of entanglement increase the success rates.}
\label{tab:results_crown}
\end{table}

\paragraph*{Sphere}

\begin{table}[t!]
\centering
\scriptsize
\begin{tabular}{c|c|cc|c|cc|cc}
  & \multicolumn{3}{c|}{$\chi^{2}_{f}$} & \multicolumn{5}{c}{$\chi^{2}_{wf}$} \\
 \hline
Qubits & 1 & \multicolumn{2}{c|}{2 } & 1 & \multicolumn{2}{c|}{2} & \multicolumn{2}{c}{4 }   \\
Layers  & & No Ent. & Ent. & & No Ent. & Ent. & No Ent. & Ent. \\ 
 \hline
 1 & 0.53 & 0.70 & -- & 0.53 & 0.70 & -- & 0.70 & --\\
 2 & 0.77 & 0.73 & 0.53 & 0.78 & 0.94 & 0.96 & 0.96 & 0.96 \\
 3 & 0.76 & 0.74 & 0.77 & 0.78 & 0.92 & 0.94 & 0.94 & 0.95 \\
 4 & 0.84 & 0.83 & 0.78 & 0.89 & 0.92 & 0.94 & 0.95 & 0.94 \\
 5 & 0.89 & 0.85 & 0.77 & 0.90 & 0.94 & 0.94 & 0.95 & 0.94 \\
 6 & 0.90 & 0.89 & 0.88 & 0.92 & 0.87 & 0.93 & 0.94 & 0.94 \\
 8 & 0.89 & 0.87 & 0.90 & 0.93 & 0.92 & 0.89 & 0.94 & 0.93 \\
10 & 0.93 & 0.91 & 0.90 & 0.93 & 0.94 & 0.92 & 0.92 & 0.92 \\
\end{tabular}
\caption{Results of the single- and multi-qubit classifiers with data re-uploading for the three-dimensional sphere problem. Numbers indicate the success rate, i.e. the number of data points classified correctly over the total number of points. Words ``Ent." and ``No Ent." refer to considering entanglement between qubits or not respectively. The L-BFGS-B minimization method with the weighted fidelity and fidelity cost functions is use. The weighted fidelity cost function is better than the fidelity cost function. There are no significant differences between the two-qubit and the four-qubit classifiers. Both are better than the single-qubit classifier and the introduction of entanglement does not increase the success rates.}
\label{tab:results_sphere}
\end{table}

This quantum classifier is able to classify multidimensional data, as shown with the four-dimensional hypersphere. A three-dimensional figure is also tested, a regular sphere of size half the feature space. 

For this problem, the fidelity cost function reaches its maximum, 93\%, with a single-qubit classifier of 10 layers (60 parameters). The same success is obtained with a two-qubit entangled classifier and 6 layers (72 parameters). With the weighted fidelity, this success rate grows up to 96\% for two- and four- qubit classifier of 2 layers (24 and 48 parameters respectively) with and without entanglement. 
All results are written in Table \ref{tab:results_sphere}.

\paragraph*{Squares}

This problem divides a 2D area into four quadrants with straight lines. This is one of the easiest problems for a \ac{nn}. By construction, \ac{nn}s can establish a separation between classes by using biases, and thus straight lines are immediate to understand. This problem aims to see how a quantum classifier performs against a \ac{nn} in the latter's field. 

The fidelity cost function reaches 99\% of success in a two-qubit classifier without entanglement and 6 layers (60 parameters). Any two-qubit result is comparable with the success rate of the classical models. Something similar can be found for the weighted fidelity. The maximum success, 96\%, is obtained with a two-qubit entangled classifier with 4 layers (40 parameters). The results are written in Table \ref{tab:results_squares} and the best performance is plotted in Figure \ref{Fig:squares}.

\begin{table}[h!]
\centering
\scriptsize
\begin{tabular}{c|c|cc|c|cc|cc}
  & \multicolumn{3}{c|}{$\chi^{2}_{f}$} & \multicolumn{5}{c}{$\chi^{2}_{wf}$} \\
 \hline
Qubits & 1 & \multicolumn{2}{c|}{2 } & 1 & \multicolumn{2}{c|}{2} & \multicolumn{2}{c}{4 }   \\
Layers  & & No Ent. & Ent. & & No Ent. & Ent. & No Ent. & Ent. \\ 
 \hline
 1 & 0.58 & 0.48 & -- & 0.70 & 0.92 & -- & 0.90 & -- \\
 2 & 0.76 & 0.96 & 0.97 & 0.74 & 0.91 & 0.94 & 0.95 & 0.95 \\
 3 & 0.90 & 0.96 & 0.98 & 0.90 & 0.94 & 0.95 & 0.95 & 0.95 \\
 4 & 0.89 & 0.98 & 0.96 & 0.88 & 0.94 & 0.95 & 0.95 & 0.95 \\
 5 & 0.91 & 0.97 & 0.98 & 0.89 & 0.94 & 0.94 & 0.95 & 0.94 \\
 6 & 0.92 & 0.99 & 0.94 & 0.93 & 0.94 & 0.94 & 0.94 & 0.94 \\
 8 & 0.93 & 0.98 & 0.94 & 0.93 & 0.94 & 0.95 & 0.95 & 0.94 \\
10 & 0.94 & 0.97 & 0.93 & 0.94 & 0.94 & 0.94 & 0.94 & 0.93 \\
\end{tabular}
\caption{Results of the single- and multi-qubit classifiers with re-uploading data for the squares problem. Numbers indicate the success rate. Words ``Ent." and ``No Ent." refer to considering entanglement between qubits or not respectively. The L-BFGS-B minimization method is used. In this problem, the fidelity cost function $\chi^2_f$ presents better results. It achieves the 0.99 success with the two-qubit classifier with six layers and no entanglement.}
\label{tab:results_squares}
\end{table}

\paragraph*{Wavy lines}This problem is the four-class version of the non-convex problem. Now the area is divided into four regions by two different functions. The borders' equations are $x_2 =  \sin(\pi x_1) \pm x_1$. The important feature of this problem is that there are some areas in the problem too small to be caught by the classifier. 

As can be seen in Figure \ref{Fig:wavy}, most of the failure points are in these small non-convex areas. The classifier would rather adjust the rest of the points instead of tuning those zones and losing everything else. The results for this problem are not as good as for other problems, but still 94\% for the fidelity cost function is obtained, two entangled qubits and 10 layers (200 parameters) and the weighted fidelity, four entangled qubits and 4 layers (80 parameters). 

\begin{table}[h!]
\centering
\scriptsize
\begin{tabular}{c|c|cc|c|cc|cc}
  & \multicolumn{3}{c|}{$\chi^{2}_{f}$} & \multicolumn{5}{c}{$\chi^{2}_{wf}$} \\
 \hline
Qubits & 1 & \multicolumn{2}{c|}{2 } & 1 & \multicolumn{2}{c|}{2} & \multicolumn{2}{c}{4 }   \\
Layers  & & No Ent. & Ent. & & No Ent. & Ent. & No Ent. & Ent. \\ 
 \hline
 1 & 0.70 & 0.52 & -- & 0.76 & 0.75 & -- & 0.88 & -- \\
 2 & 0.86 & 0.75 & 0.80 & 0.84 & 0.89 & 0.88 & 0.91 & 0.92 \\
 3 & 0.74 & 0.82 & 0.84 & 0.84 & 0.92 & 0.91 & 0.92 & 0.92 \\
 4 & 0.80 & 0.85 & 0.87 & 0.87 & 0.89 & 0.93 & 0.92 & 0.93 \\
 5 & 0.85 & 0.90 & 0.88 & 0.87 & 0.92 & 0.92 & 0.93 & 0.93 \\
 6 & 0.92 & 0.92 & 0.91 & 0.88 & 0.93 & 0.94 & 0.93 & 0.93 \\
 8 & 0.90 & 0.91 & 0.91 & 0.92 & 0.92 & 0.92 & 0.93 & 0.94 \\
10 & 0.92 & 0.91 & 0.93 & 0.90 & 0.93 & 0.93 & 0.93 & 0.93 \\
\end{tabular}
\caption{Results of the single- and multi-qubit classifiers with re-uploading data for the wavy lines problem. Numbers indicate the success rate. Words ``Ent." and ``No Ent." refer to considering entanglement between qubits or not respectively. The L-BFGS-B minimization method with the weighted fidelity and fidelity cost functions is used. Results with the weighted fidelity cost function and multi-qubit classifiers are vaguely better than other configurations. Entanglement does not change significantly the results.}
\label{tab:results_wavy}
\end{table}

\subsubsection{Comparison to classical classifiers}\label{sssec:comp_cl_class}

The field of Machine Learning has a great development for classical computers. Thus, the performance of the quantum classifiers against classical methods can be compared to check if this proposal can in some sense compete against well established methods. 

To do so, the standard ML library {\tt scikit-learn}~\cite{scikit-learn} is used to solve the same examples as in the quantum classifier. The classical models here used are simple ones in order to get fair comparisons, that is, quantum and classical models with similar complexities. The aim of this benchmark is not to review the capabilities of classical Machine Learning, which are known to be extensive, but rather to settle whether quantum or classical models work better for similar circumstances. 

\begin{table}[t!]
\centering
\begin{tabular}{c|cc|cc}
\multirow{2}{*}{Problem (\# classes)} & \multicolumn{2}{c|}{Classical classifiers} & \multicolumn{2}{c}{Quantum classifier} \\
 & \ac{nn} & \acs{svc} & $\chi_{f}^2$ & $\chi^2_{wf}$ \\ \hline
 Circle \hfill (2) & 0.96 & 0.96 & 0.96 & 0.97 \\
  Crown\hfill (2)& 0.96 & 0.82 & 0.95 & 0.97 \\
  Non-Convex\hfill (2)& 0.98 & 0.97 & 0.96 & 0.98 \\
 Sphere\hfill (2)& 0.93 & 0.91 & 0.93 & 0.96 \\
 Hypersphere\hfill (2)& 0.89 & 0.92 & 0.91 & 0.98 \\
 Tricrown\hfill (3)& 0.96 & 0.83 & 0.93 & 0.97 \\
 3 circles\hfill (4)& 0.94 & 0.92 & 0.91 & 0.91 \\
 Squares\hfill (4)& 0.99 & 0.95 & 0.99 & 0.95 \\
 Wavy Lines\hfill (4)& 0.98 & 0.89 & 0.93 & 0.94 \\ 
\end{tabular}
\caption{Comparison between single-qubit quantum classifier and two well-known classical classification techniques: \ac{nn}s with a single hidden layer composed of 100 neurons and \acf{svc}, both with the default parameters as defined in \texttt{scikit-learn} python package~\cite{scikit-learn}.
Results of the single-qubit quantum classifier are obtained with the fidelity and weighted fidelity cost functions, $\chi^2_{f}$ and $\chi^{2}_{wf}$ defined in Eq. \eqref{eq:fidelity_chi2} and Eq. \eqref{eq:conventional_chi2} respectively. This table shows the best success rate, being 1 the perfect classification, obtained after running ten times the \ac{nn} and \ac{svc} algorithms and the best results obtained with single-qubit classifiers up to 10 layers.}
\label{tab:classical_benchmark}
\end{table}

The results were obtained using two different methods. First, single-hidden-layers \ac{nn}s of 20 neurons with all activation functions available in {\tt scikit-learn} and a {\sl lbfgs} solver. The function used to this extent is {\tt sklearn.neural\_network.} {\tt MLPClassifier}. For a \ac{svc}, {\tt sklearn} {\tt .svm.SVC} with different kernels was used. Note that \ac{nn}s have a controllable number of tunable parameters that can be adjusted to match the number of the quantum classifier. For \ac{svc}s this is not possible and the complexity of the algorithm depends on the size of the training set.

A summary of results can be seen in Tab.~\ref{tab:classical_benchmark}. In classical models, only the best final result is retrieved for each problem and \ac{nn} or \ac{svc}. For quantum results, single-qubit results with different cost functions are depicted. In all problems it is possible to see that quantum and classical performance are at least comparable to classical methods. 

It is important to mention that even though final results of quantum and classical methods perform in a similar way, the effort required to achieve these results is very different. Quantum algorithms are not as efficient as classical ones. 

\subsection{Discussion}\label{ssec:conclusions_classifier}
The quantum classifier here proposed and based on the general re-uploading strategy has shown capabilities to successfully accomplish non-trivial supervised learning tasks, in an equivalent sense as simple \ac{nn}s can. The classification problems here addressed consists in learning complex geometrical figures in multidimensional spaces. 

The key ingredient for the quantum classifier, apart from the re-uploading scheme, is the measurement strategy. A set of label states is created so that each one corresponds to a different class. The quantum circuit is forced to deliver output states in average as close as possible to the corresponding label state, depending on the class of the training data. The identification of the class is then defined as the smallest distance between the output and the different label states. These target states are chosen in a maximally orthogonal way as a strategy to make them as distinguishable as possible. As a consequence, each label corresponds to a region in the Bloch sphere.

The optimization process is driven by two different cost functions. The simplest one, fidelity cost function, simply measures the fidelity between the output state and its corresponding label. A more sohpsticated one, weighted fidelity cost function, is inspired in \ac{nn}s. It measures the distance to all classes and compare it to the ideal case. 

The single-qubit classifier is extendable to multi-qubit architectures. This allows for the introduction of entanglement between the qubits. An entangling Ansatz is shown as a proof of concept, while exhaustive exploration of Ansätze is out of the scope for the moment. 

We benchmarked several quantum classifiers with different numbers of layers, qubits and entanglement mappings against classical classification methods. The test problems are data points embedded in 2D, 3D and 4D feature spaces, where each class is defined by means of geometrical figures. In all cases, the single-qubit classifier provides success rates of over 90\%. More qubits and entanglement can increase this success while reducing the effective number of layers. However, the number of calls to classical data remains approximately constant for different classifier with the same performance. In terms of cost functions, the weighted fidelity one helps to find better results, at the cost of increasing the number of trainable parameters. In addition, as more layers are considered, the probability to get stuck in local minima increases as well, as expected from an optimization problem involving large numbers of parameters. 

After publishing the original work on the quantum classifier \cite{perezsalinas_data_2020}, several publications have been developed based on this method. Reference \cite{mccaldin_reuploading_2021} carried an exhaustive study and nice representations of the re-uploading classifier. Reference~\cite{huembeli_characterizing_2021} develops an exploration on the landscape of loss functions generated by the quantum classifier. This model is also part of the documentation of some quantum computing packages~\cite{reuploading_pennylane, reuploading_qibo}. 

\section{Experimental quantum classifier}\label{sec:exp_qlassifier}
A surge of algorithms for \ac{qml} that theoretically or even numerically work has appeared during recent years. However, examples of successful implementations of some \ac{qml} recipe on an actual quantum processor are much scarcer. Quantum devices at the state of the art suffer the effects of noise and decoherence, thus the performance of their calculations is limited. In addition, different experimental platforms have different properties that make them more or less suitable for a particular algorithm, and it is then difficult to gain insights on the optimal experimental configuration to accomplish a given task. 

%Some extended platforms for quantum devices are photonics, superconducting materials and ion traps. There exists, for example, an experiment realizing a toy model for quantum Support Vector Machines \cite{rebentrost_qsvm_2014} on a photonic device \cite{cai_qsvm_2015}. Superconducting qubits are more related to variational algorithms, including those in the field of QML \cite{kandala_hardware_2017,havlicek_supervised_2019}. 

A technology that suits the data re-uploading strategy algorithm is the ion trap. The framework here described makes use of quantum systems that are sparse in qubits, but those qubits must be faithfully controlled to successfully accomplish the \ac{qml} task faced. Ion trap devices are capable to control small systems very accurately, even though the scalability of the machines is not as convenient as in other platforms like photonics or superconductors \cite{resch_quantum_2019, brown_single_2011, Wang_single_2021}. In recent years, ion-trap experiments have shown its worth and widened the range of feasible experiments \cite{figgatt_complete_2017, Nam_groundstate_2020, johri_nearest_2020, rudolph_generation_2020, hempel_quantum_2018}. Those results support the choice of ion traps for the implementation here presented as well. 

In this chapter, it is shown that the data re-uploading strategy can be implemented on an ion-trap quantum processing unit~(QPU). The experiment here performed consists in applying a simplified version of the quantum classifier, that is, a simplified circuit to solve the same problems. This simplified version is close to the formulation described in Th.~\ref{th:q_UAT}, since it provides the best results with the smallest number of parameters and depth of the circuit. The experimental approach here presented constitutes, to the best of our knowledge, the first successful implementation of a quantum classifier in a system that is very sparse in qubits. 

The main difference between running quantum experiments on classical simulators or actual quantum devices is that simulated methods do not always capture the complexity and features of the experiment faithfully. Elements as native gates or noise environment are completely device-dependent. In this work, the qubit is controlled in an optimal way, so that the inherent properties of the system are utilized to improve the overall performance of the classifier. 

In this section, the translation from simulation to experiment is treated in Sec.~\ref{ssec:classifier_experiment}. Results are reviewed in Sec.~\ref{ssec:results_exp_classifier}. Final remarks are collected in Sec.~\ref{ssec:conclusions_exp_classifier}.The experimental setup is detailed in App.~\ref{app:experimental_setup_classifier}. 

\subsection{Single-qubit quantum classifier on the experiment}\label{ssec:classifier_experiment}
In order to implement the quantum classifier on the ion-trap quantum device, some shallow modifications must be taken into account.

\subsubsection{Quantum circuits}

The circuits used to perform the classification tasks on the experiment are a simplified version of the simulated ones. The reason to simplify the algorithm is to obtain a quantum circuit with equivalent query complexity as in the original formulation, but less parameters and rotations. This is an attempt to reduce the overall number of operations required to carry the task. The circuit is again defined as a series of gates, as in Eq.~\eqref{eq:def_classifier}, but the building blocks, corresponding to Eqs.~\eqref{eq:u_qlassifier} changes. Two different single-qubit operations are proposed. 

The Ansatz A is directly inherited from Th.~\ref{th:q_UAT}, and it is defined as
\begin{equation}\label{eq:ansatz_a}
U_A(\vec x, \vec\theta, \vec w) = R_z(\theta_z)R_y(\vec w \cdot \vec x + \theta_y), 
\end{equation}
where $\vec x$ is the data point and $\theta_z, 	\theta_y, \vec w$ are tunable parameters. Note that this operation unlike the one in Eq.~\eqref{eq:u_qlassifier} compresses all the $x$-dependency in only one argument of the rotation, namely the angle around the $y$-axis. On the other hand, the $z$-rotation is only included to generate non-linearities. It is also worth mentioning that this Ansatz permits the data $\vec x$ to have any number of features. All data can be easily accomodated by increasing the dimensionality of $\vec w$.

Ansatz B is more similar to the definition in Eq.~\eqref{eq:u_qlassifier}. For simplicity, only 2D problems are addressed with this Ansatz. The gate is defined as
\begin{equation}
U_B(\vec x, \vec w, \vec \theta) R_z(w_2 x_2 + \theta_2) R_y(w_1 x_1 + \theta_1).
\end{equation}
In this case, the non-linearities are expected to arise faster than for Ansatz A since both features of $\vec x$ interact with each other. 

\subsubsection{Optimization technique}
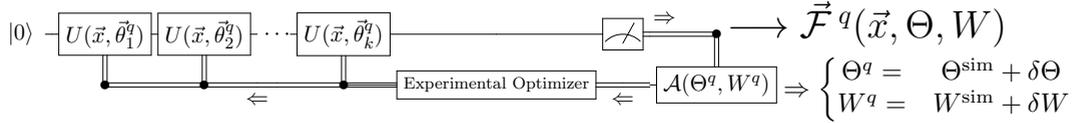
\begin{figure}[t!]
\begin{adjustwidth}{-2cm}{-1cm}
\begin{tcolorbox}[enhanced,width=1\linewidth, center, colback=white, colframe=orange]
\begin{center}

\Large \sl Quantum classifier: multi-variable $\vec x$ $\Rightarrow$ multi-class $\vec c$
\end{center}

\vskip5mm
\begin{flushleft}
{\bf 1. TRAINING USING CLASSICAL SIMULATION}\newline

\resizebox{.75\textwidth}{!}{\hskip1cm $\Qcircuit @R=0.5em @C=0.3em{
\lstick{\ket 0} & \qw & \gate{U(\vec x, \vec\theta^{\rm sim}_{1})} & \gate{U(\vec x, \vec\theta^{\rm sim}_{2})} & \qw & \push{\cdots} & \gate{U(\vec x, \vec\theta^{\rm sim}_{k})} &\qw & \meter & \ustick{\hspace{5mm} \Rightarrow}\cw & \control \cw\\
 &  & \control \cwx & \control \cwx\cw & \dstick{\Leftarrow}\cw & \cw & \control \cwx\cw & \gate{\rm Classical\; Optimizer} \cw & \dstick{\Leftarrow} \cw & \cw & \gate{\chi_f^2(\Theta^{\rm sim}, W^{\rm sim})} \cwx \\
}$}
\begin{textblock}{10}(10, -0.45)
{\Large$
\Rightarrow (\Theta^{\rm sim}, W^{\rm sim})$
}
\end{textblock}
\begin{textblock}{10}(9.2, -0.88)
{\Large$
\longrightarrow \vec{\mathcal{F}}^{\rm \; sim}(\vec x, \Theta, W)$
}
\end{textblock}
\vskip5mm

{\bf 2. TRAINING USING QUANTUM PROCESSING UNIT}\newline

\resizebox{.75\textwidth}{!}{\hskip1cm $\Qcircuit @R=0.5em @C=0.3em{
\lstick{\ket 0} & \qw & \gate{U(\vec x, \vec\theta^{q}_{1})} & \gate{U(\vec x, \vec\theta^{q}_{2})} & \qw & \push{\cdots} & \gate{U(\vec x, \vec\theta^{q}_{k})} &\qw & \meter & \ustick{\hspace{5mm} \Rightarrow}\cw & \control \cw\\
 & & \control \cwx & \control \cwx\cw & \dstick{\Leftarrow}\cw & \cw & \control \cwx\cw & \gate{\scriptsize \textrm{Experimental Optimizer}} \cw & \dstick{\Leftarrow} \cw & \cw & \gate{\mathcal{A}(\Theta^{q}, W^{q})} \cwx \\
}$}
\begin{textblock}{10}(9.9, -0.4)
{\small 
$\Rightarrow \left\lbrace\begin{matrix}\Theta^{q} =& \Theta^{\rm sim} + \delta\Theta \\ W^{q} =& W^{\rm sim} + \delta W\end{matrix} \right. $
}\end{textblock}
\begin{textblock}{10}(9.2, -.9)
{\Large $
\longrightarrow \vec{\mathcal{F}}^{\; q}(\vec x, \Theta, W)$
}
\end{textblock}
\end{flushleft}

\vskip2mm
\end{tcolorbox}
\end{adjustwidth}
\caption{Schematic description of the training algorithm used in this work. In a first step, data re-uploading is trained using a classical simulation. The simulated quantum circuit is composed of single-qubit gates $U$ depending on variational parameters $(\Theta, W)$ and the data points $\vec x$. The output state is measured to obtain the fidelity between the output state and the corresponding label state. This quantity encodes the probabilities that will serve to classify the given pattern into a category. A classical optimization is performed to obtain optimal values $\Theta^{\rm sim}, W^{\rm sim}$. This optimization is driven by the cost function $\chi_f^2$ evaluated on training data. In a second step, a further optimization is accomplished only using the quantum device, taking as starting point the values $\Theta^{\rm sim}, W^{\rm sim}$ and delivering a set $\Theta^{q}, W^q$ suiting better the experiment. The quantity to minimize is the accuracy $\mathcal A$ evaluated on the test dataset. The aim of the experimental optimization is to mitigate and even compensate possible systematic experimental errors.}\label{fig:exp_optimization}
\end{figure}

The working procedure used for the experimental quantum classifier follows the scheme depicted in Fig.~\ref{fig:opt_qlassifier} for the simulated one. The retrieving stage is similar, up to the technical details required to use an actual quantum machine, to be detailed later. However, the optimization stage itself is performed in two steps, a simulated and a experimental one, see Fig.~\ref{fig:exp_optimization} for a schematic description of the optimization procedure.

First, the training set is considered for optimization on a simulated framework, as in the original proposal. This stage returns a set of optimal parameters in the simulated scenario $\Theta^{\rm sim}, W^{\rm sim}$, attached to a given value of the cost function $\chi^2_f(\Theta^{\rm sim}, W^{\rm sim})$ and a theoretical accuracy $\mathcal{A}(\Theta^{\rm sim}, W^{\rm sim})$ computed on the test set. In this experiment only the fidelity cost function $\chi^2_f$ from Eq.~\eqref{eq:fidelity_chi2} is considered. 

The second step is then performed on the experiment. The parameters $\Theta^{\rm sim}, W^{\rm sim}$ provide a given accuracy when the quantum circuit is run on the experimental setup. It is likely that the imperfections of the experimental device limit the capabilities of the algorithm. In order to mitigate this effect, a new optimization step is performed on the quantum machine taking as starting point the previously obtained $\Theta^{\rm sim}, W^{\rm sim}$. This will ideally lead to some new parameters $\Theta^q = \Theta^{\rm sim} + \delta\Theta$ and equivalently for $W$. The new set of parameters is ideally capable to suit better the requirements and mitigate any systematic error of the experimental setup, and thus to obtain improved accuracies. 

The experimental optimization is carried by scanning the parameter space in the vicinity of $\Theta^{\rm sim}, W^{\rm sim}$ and retrieving the configuration $\Theta^{q}, W^q$ with best accuracy $\mathcal{A}^{q}$. This second step is available for quantum devices only if the loss function in the parameter space near the vicinity of the minimum is smooth and large deviations from the optimal parameters translate into small changes in the loss function. Unfortunately, this full scan is extremely time expensive for current machines, and then it is only possible to provide results optimized in these two steps in a small number of examples. 

\subsection{Results}\label{ssec:results_exp_classifier}
Results from the experimenta setup were obtained following the same problems as in the simulated classifier, and all descriptions hold in this approach. To reduce the computational cost of the experimental implementation, the training dataset is set to 250 points, and the test dataset to 1000.

\begin{figure}[t!]
     \begin{center}
        \subfigure[\hspace{1mm} QPU, 1 layer]{%
            \label{fig:exp_circle_a}
            \includegraphics[width=0.30\textwidth]{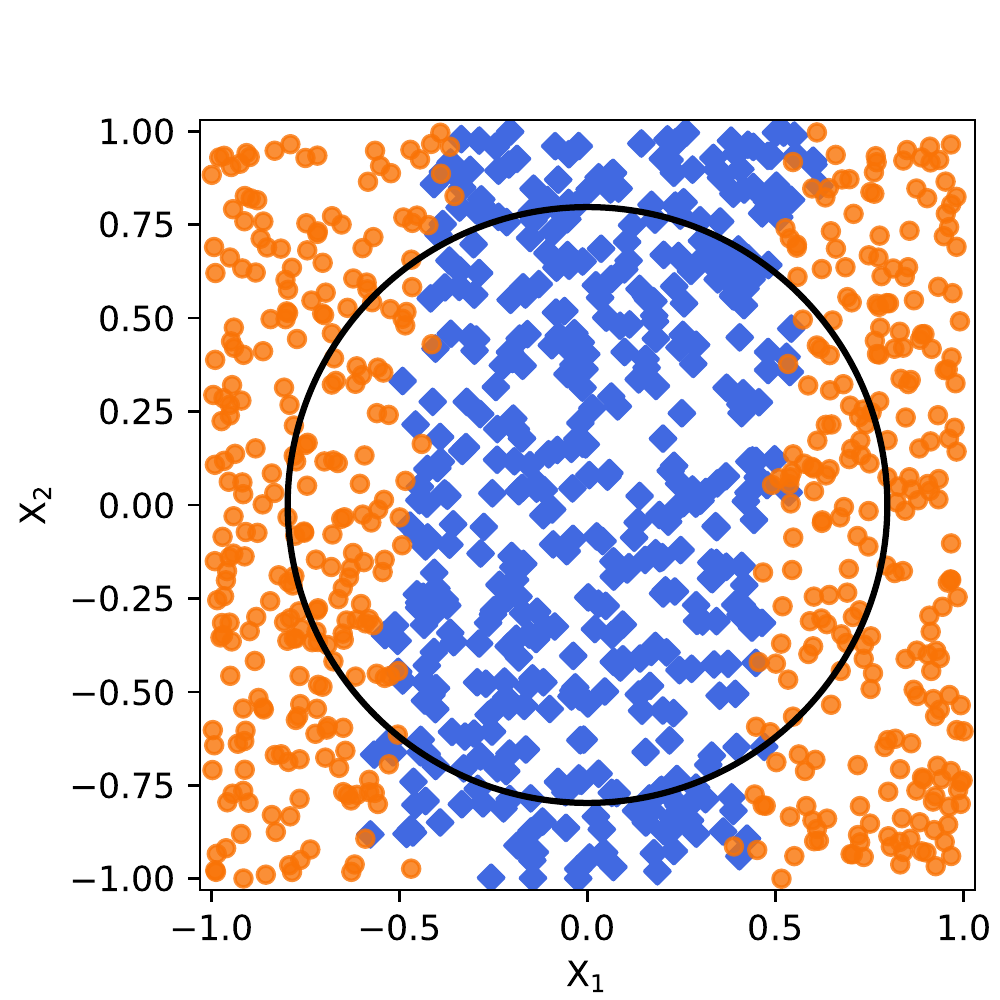}
        }%
        \subfigure[\hspace{1mm} QPU, 2 layers]{%
           \label{fig:exp_circle_b}
           \includegraphics[width=0.30\textwidth]{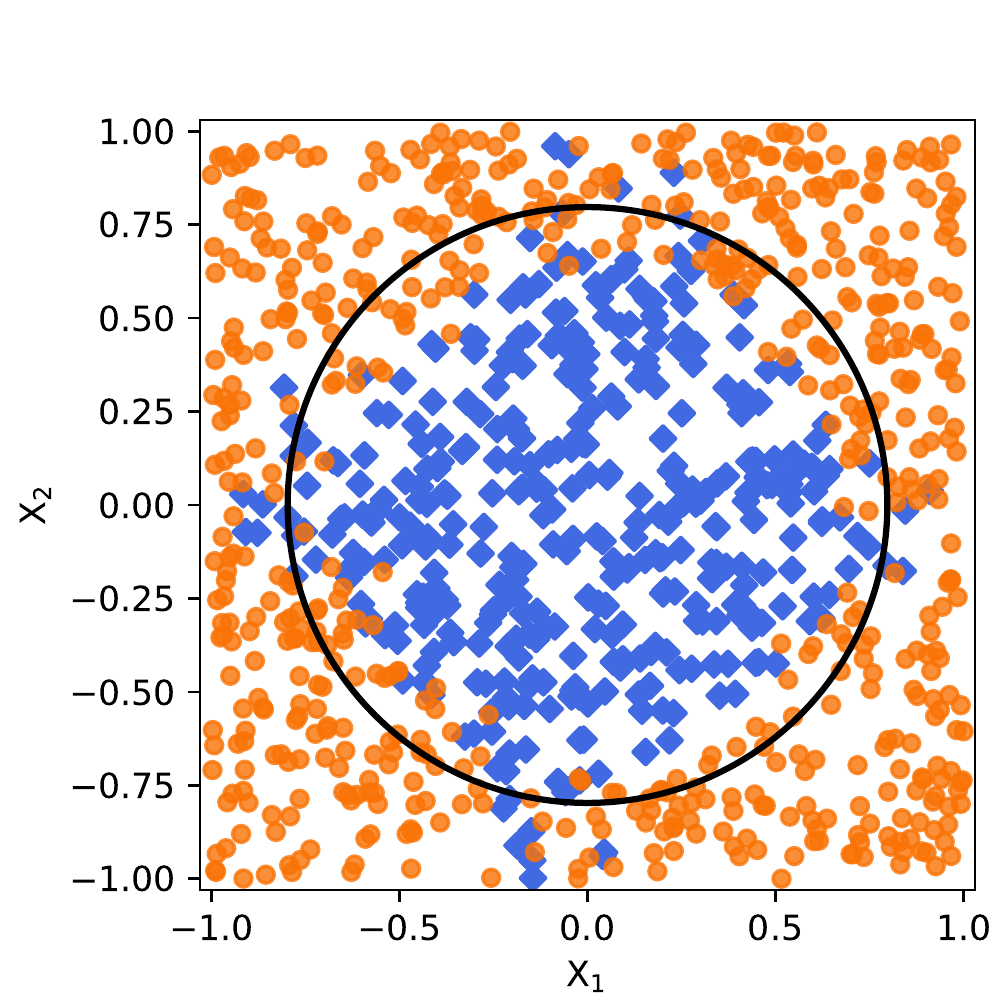}
        } 
        \subfigure[\hspace{1mm} QPU, 3 layers]{%
            \label{fig:exp_circle_c}
            \includegraphics[width=0.30\textwidth]{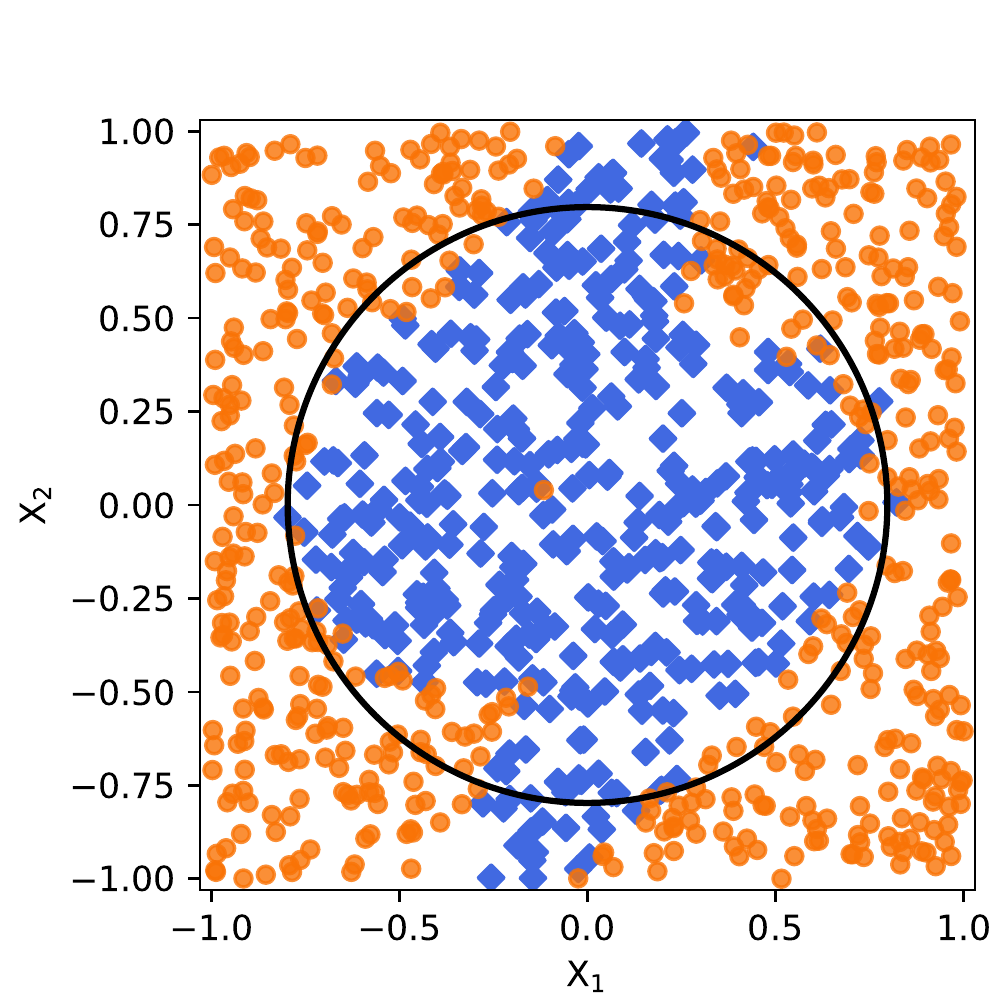}
        }\\%  ------- End of the first row ----------------------%
        \subfigure[\hspace{1mm} QPU, 4 layers; $\mathcal{A}_{q} = 93\pm{2}\%$]{
            \label{fig:exp_circle_d}
            \includegraphics[width=0.45\textwidth]{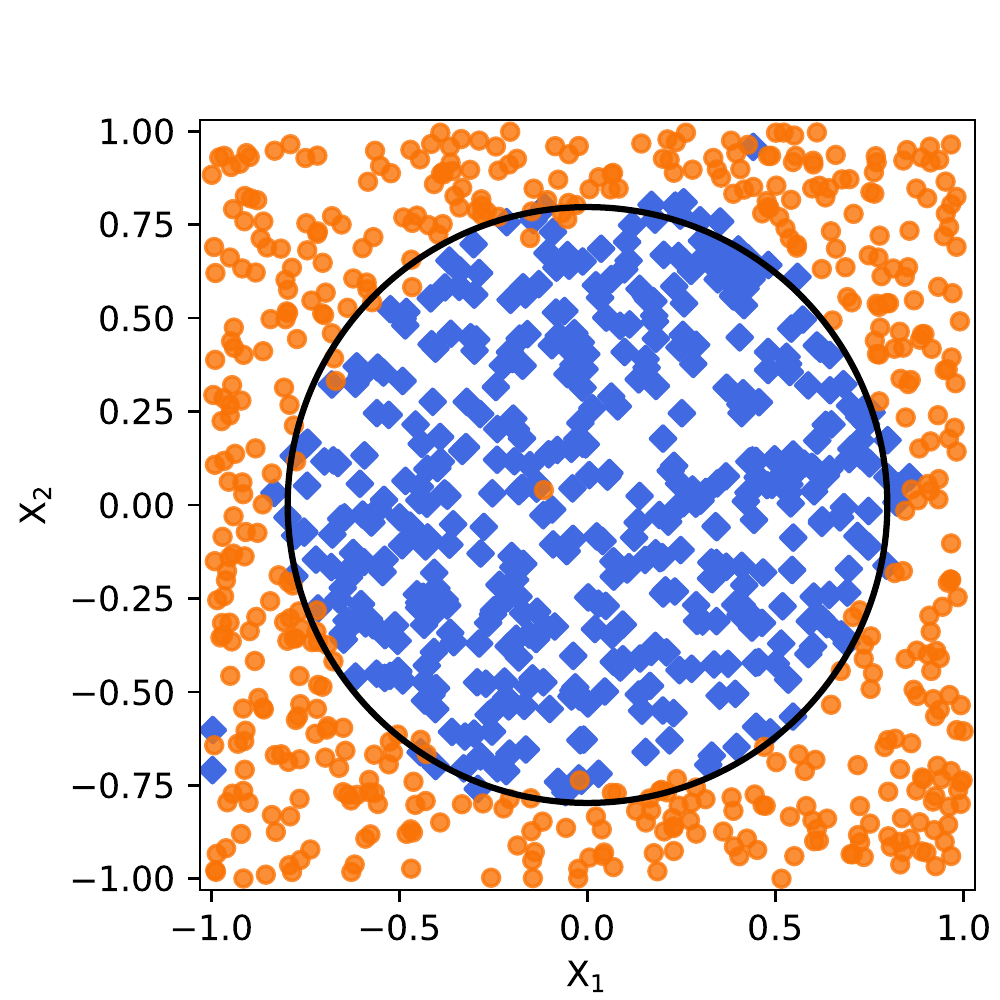}
        } \hfill
        \subfigure[\hspace{1mm} Cl. simulation, 4 layers; $\mathcal{A}_{\rm sim} = 97\%$]{
            \label{fig:exp_circle_e}
            \includegraphics[width=0.45\textwidth]{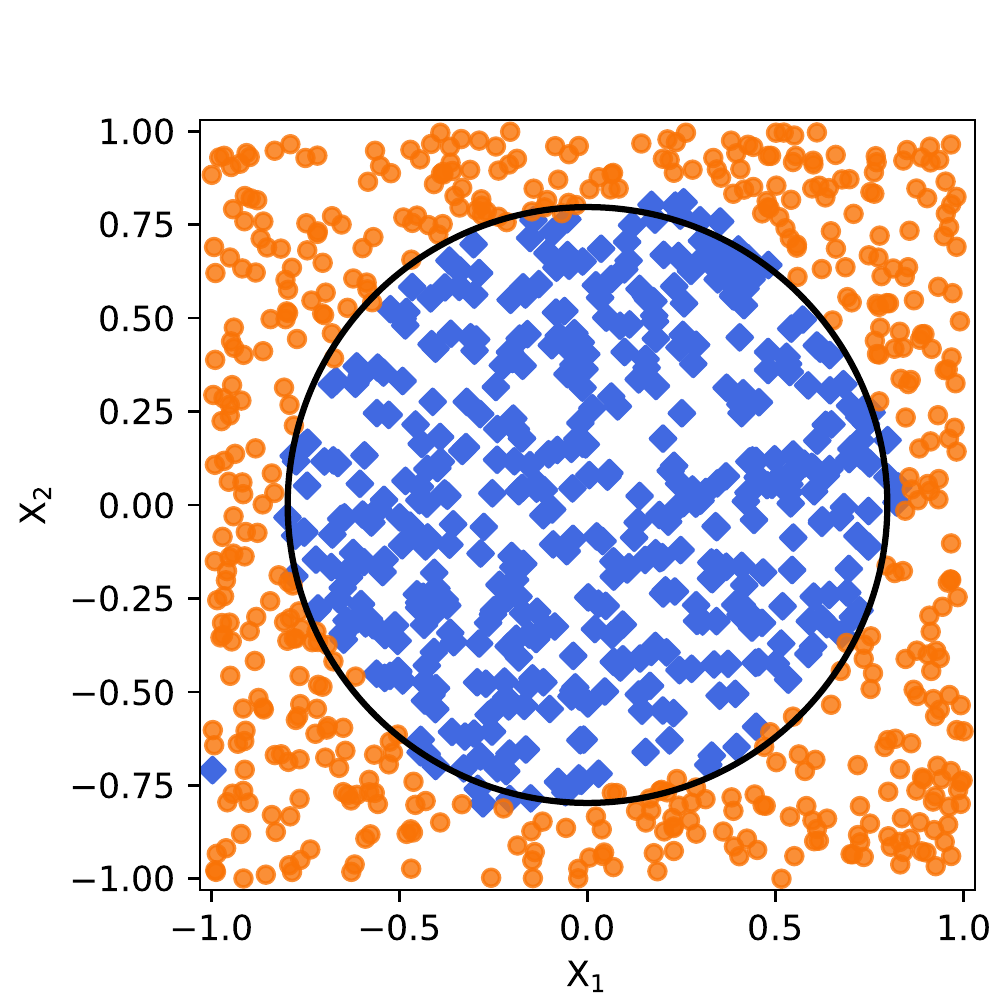}
        }
    \end{center}
\caption{Classifier test results. The ion trap based QPU classifier performed on $1000$ random data points tests depicted in blue for points within and orange outside the boundary separating the circular feature shown in solid line. The depth of the circuit is increased by one, starting from 1-layer in (a) to 4 in (d). The result of the $4$-layer QPU classifier (d) is compared with the same $4$-layer simulator (e). Notice that the border between classes in the experimental results is not as sharply defined as in the simulated classification. This difference is due to the uncertainty of the quantum measurements and systematic errors.
}
   \label{fig:exp_circle}
\end{figure}

We describe the results for the circle problem in a detailed way to compare the performance of the theoretical model and its experimental counterpart. Fig.~\ref{fig:exp_circle} shows experimental results for this classification problem for an increasing number of layers, and a comparison between the test data as classified by the QPU and an ideal classical qubit simulator. It is possible to see the improvement in the classification of results as more layers are added, up to 4 layers. In that case, the classification accuracy for the QPU is $\mathcal{A}^{q} = 93\pm{2}\%$, slightly lower than its classically simulated counterpart $\mathcal{A^{\rm sim}}= 97\%$. The error here refers to the standard deviation of $10$ repeated trials performed on the same dataset and it reflects the underlying systematic uncertainty leading to an uncertainty of the accuracy. This confirms experimentally the expected behavior of a re-uploading scheme, namely the classifier is more accurate as more layer are added to the quantum circuit.

It is worth noting a difference between simulated and experimental results. In Fig.~\ref{fig:exp_circle_e}, the guessed boundary between classes is sharply defined, even though it does not match exactly the theoretical boundary and the classification is slightly deformed. That is a consequence of simulation, the output state is described with arbitrary accuracy and thus the border between two zones in the Bloch sphere is perfectly defined. The results on the experimental data, \ref{fig:exp_circle_e}, show uncertainty in the determination of classes. For instance, in the higher part of the circuit, there is a cluster of points where different classes are interspersed. The origin of this phenomenon is the sampling uncertainty. All points near the boundary can be wrongly measured, leading to misclassification. In addition to the uncertainty region, some outliers are misclassified, see blue points at the border of the feature space and an orange one in the center. In the simulation scenario, the reason for misclassification is a defective training or model. 

\begin{figure}[h!]
\begin{adjustwidth}{-1cm}{-2cm}
    \begin{center}
        \subfigure[\hspace{1mm}Landscape of $\mathcal{A}_q$ around $\boldsymbol\theta_{\rm sim}(\star)$]{%
            \label{fig:supervised_a}
            \includegraphics[width=0.3\linewidth]{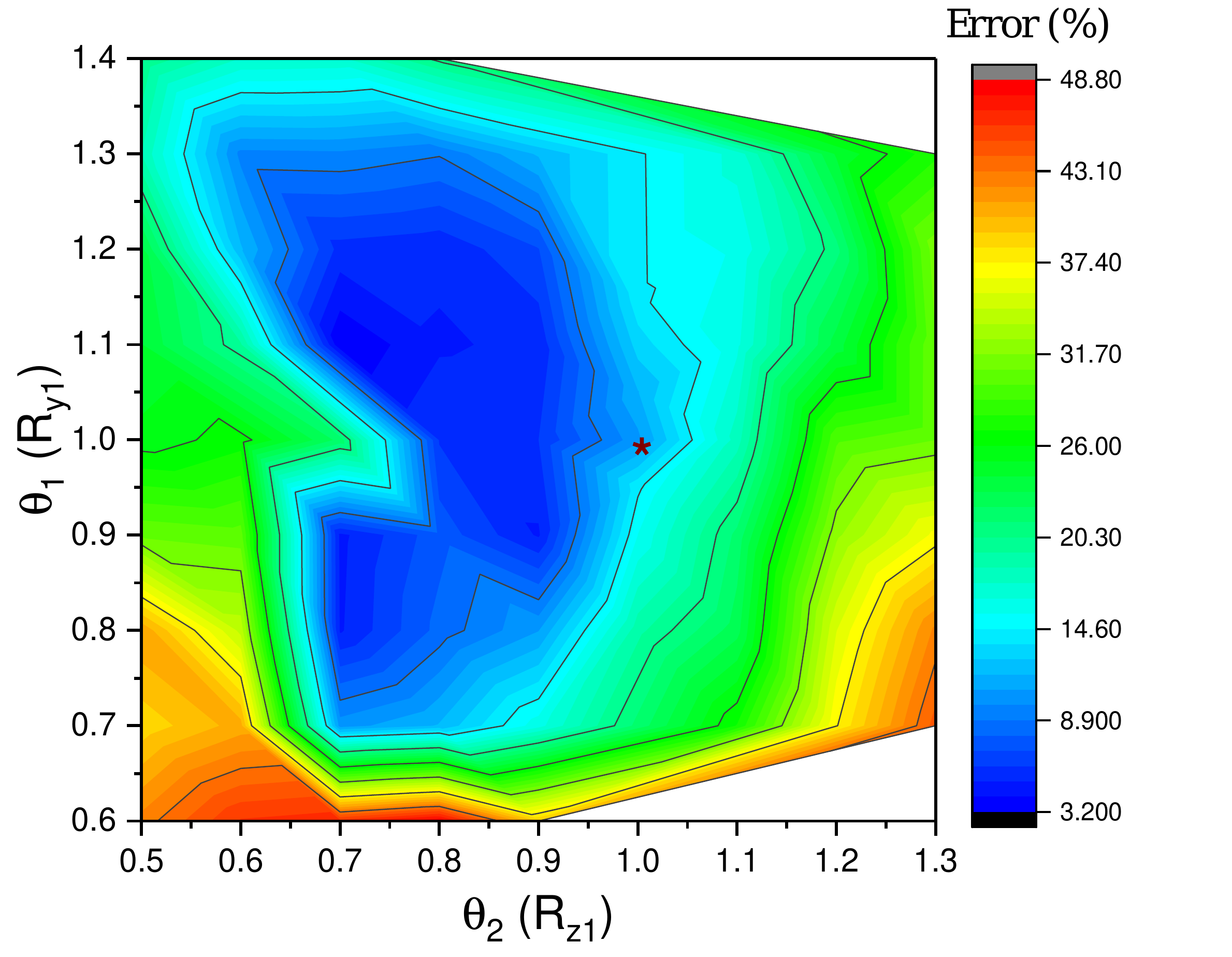}
            \includegraphics[width=0.3\linewidth]{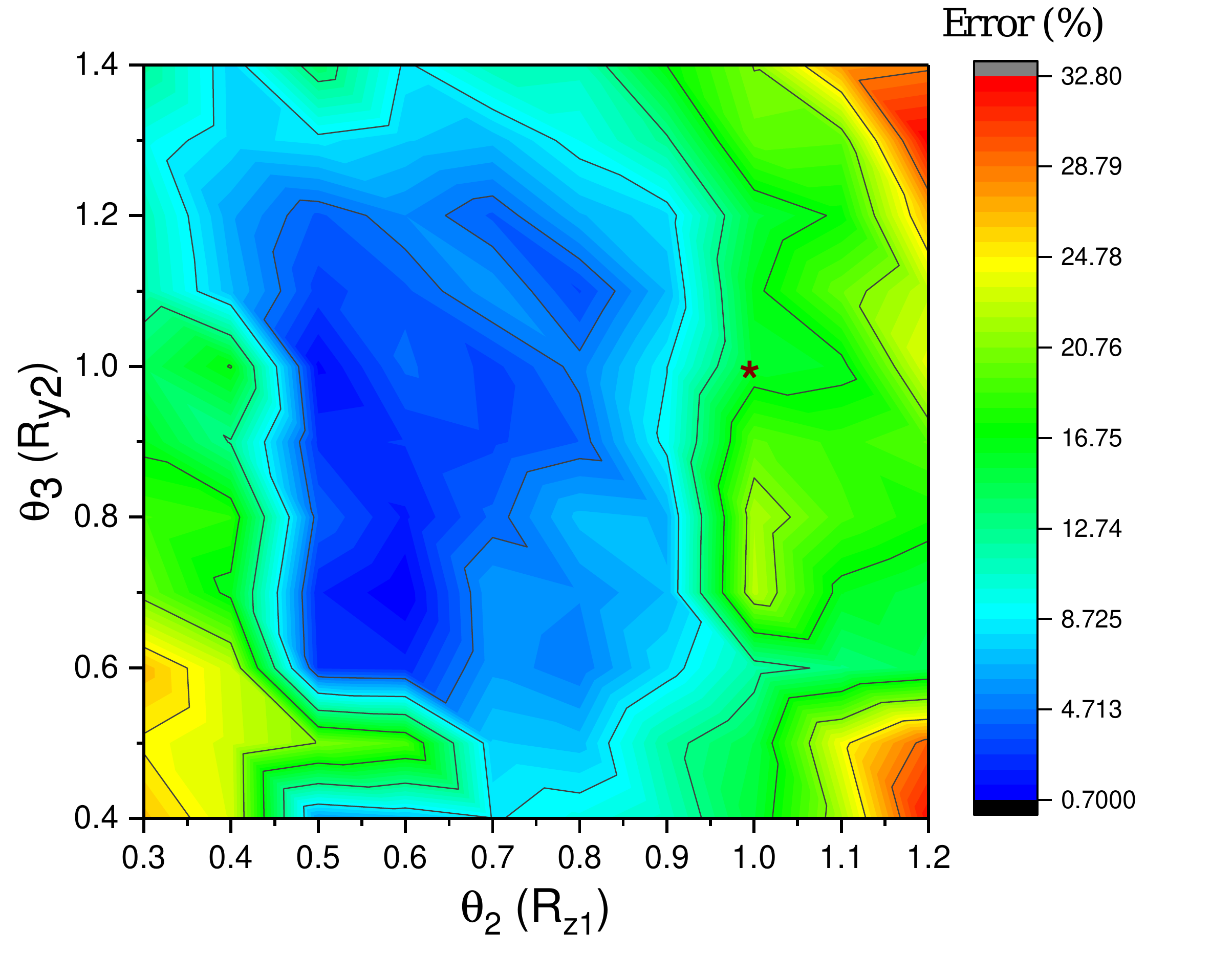}
        } \hfill 
        \subfigure[\hspace{1mm}Error in classification]{%
            \label{fig:supervised_c}
            \includegraphics[width=0.3\linewidth]{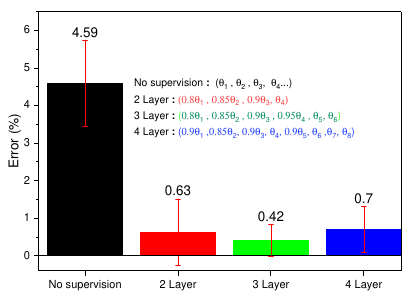}
       
        }%
\end{center}
\end{adjustwidth}
\caption{Experimental optimization: The ion trap based QPU classifier used for binary classification of data within or outside a bounded circle is trained by varying the training parameters ($\theta_1,~ \theta_2,~\theta_3)$ in the vicinity of the parameters obtained from the simulated training. The error surface plotted in color-coded surface plot in (a) shows the deviation of the optimal parameters from the trained minimum {\color{red} $(\star)$}. These plots are for 2-layer QPU and corresponds to Fig.~\ref{fig:exp_circle_b}. Similar training performed on the QPU leads to the betterment of the accuracy in the 3 and 4 layer QPU as shown in (b). Note that the improvement is more pronounced in the first two layers only.}%
\label{fig:supervised}
\end{figure}

\begin{figure}[t!]
\begin{adjustwidth}{-1cm}{-2cm}
     \begin{center}
        \subfigure[\hspace{1mm} { hypersphere} - QPU]{%
            \label{fig:hypersphere_qpu}
            \includegraphics[width=0.28\linewidth]{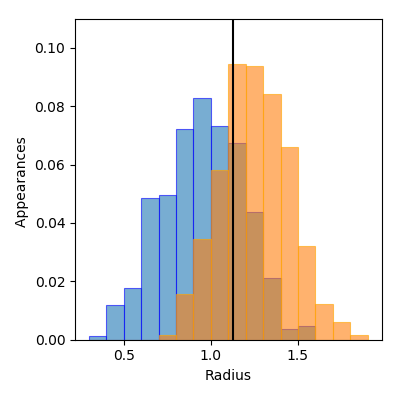}
        }%
        \subfigure[\hspace{1mm} { hypersphere} -  Simulation]{%
           \label{fig:hypersphere_sim}
           \includegraphics[width=0.28\linewidth]{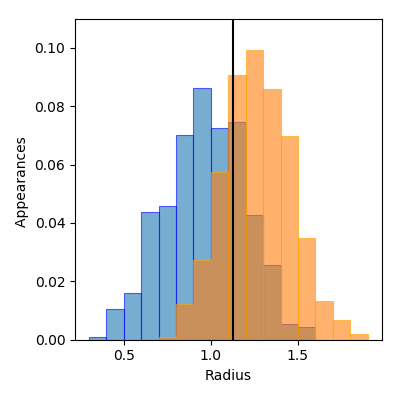}
        }
        \subfigure[\hspace{1mm} Experimental optimization]{
            \label{fig:hypersphereerror}
            \includegraphics[width=0.35\linewidth]{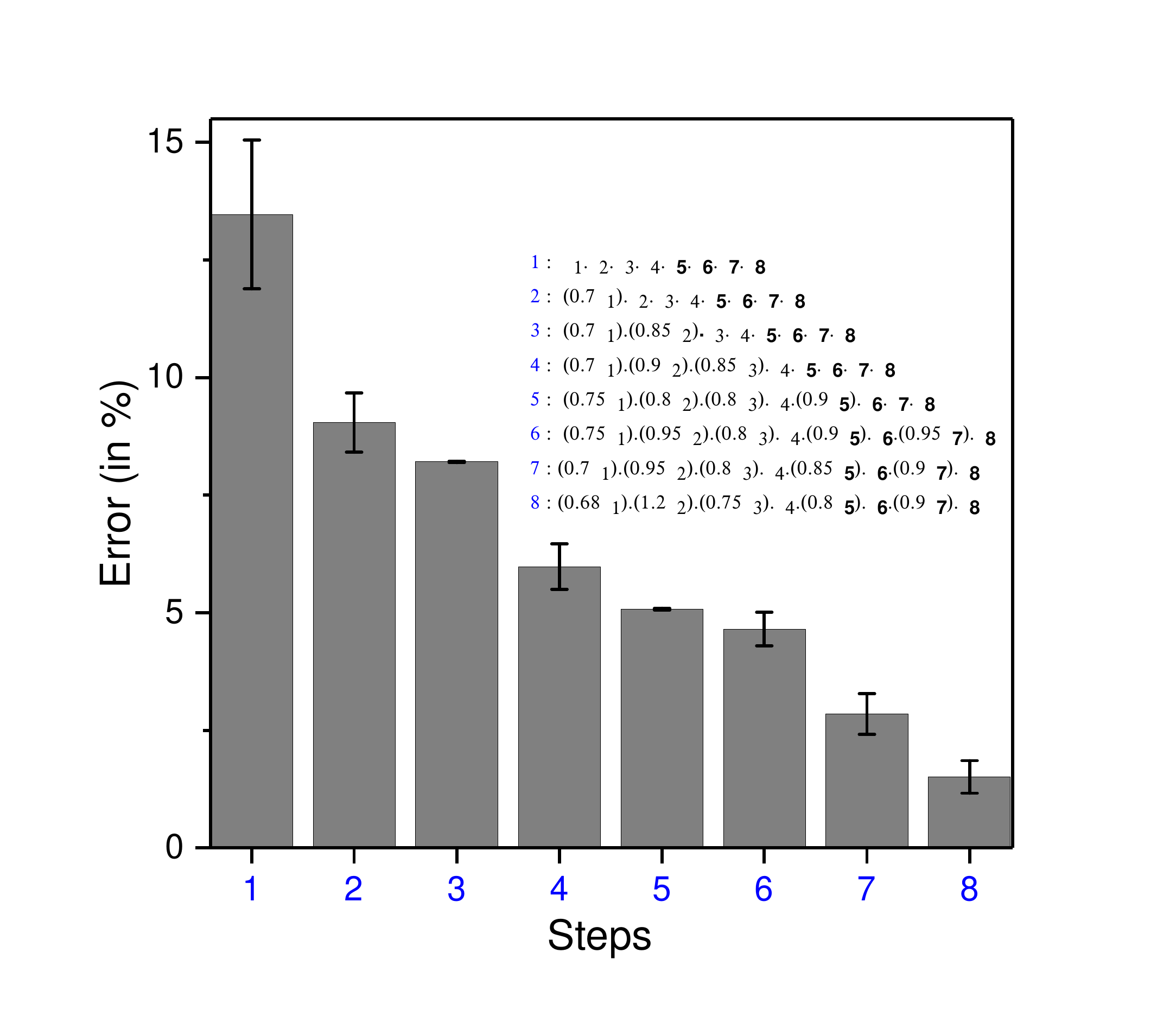}
        }%
    \end{center}
\end{adjustwidth}
\caption{Binary classification of the hypersphere dataset. The histograms represent the class association of points within a hyper shell denoted by blue (inside the hyper-sphere) and orange (outside the hyper-sphere) for QPU (a) and simulation (b). The overlap region shows the ambiguity in classifying the points within a certain radius. The accuracy of the QPU is improved by performing experimental optimization in a series of ten training steps. The reduction in the error with respect to the simulated results (b) is shown in (c).}
   \label{fig:hyper_4d}
\end{figure}

\begin{figure}[t!]
\begin{adjustwidth}{-2cm}{-1cm}
    \centering
    \subfigure[\scriptsize Non-convex - QPU\label{fig:fig3a}]{
    \includegraphics[width=.22\linewidth]{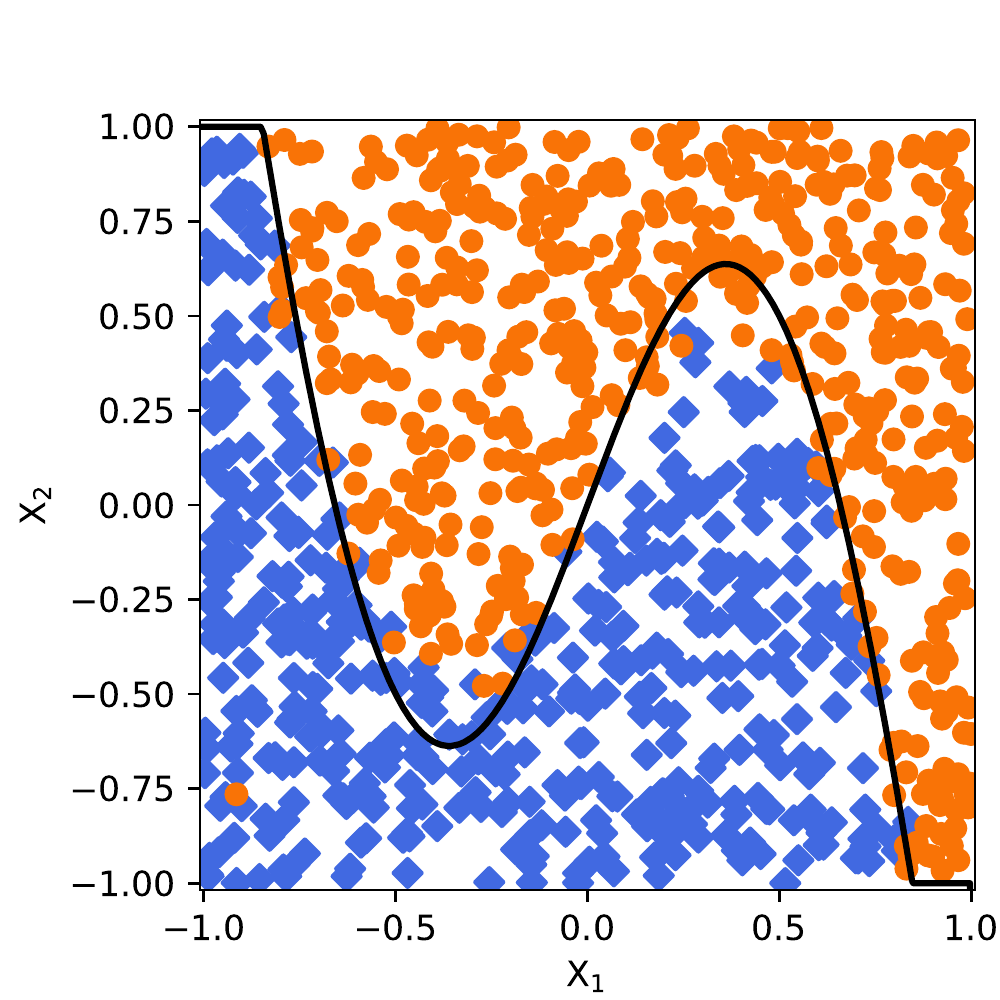}}
    \subfigure[\scriptsize Non-convex - Sim.\label{fig:fig3c}]{
    \includegraphics[width=.22\linewidth]{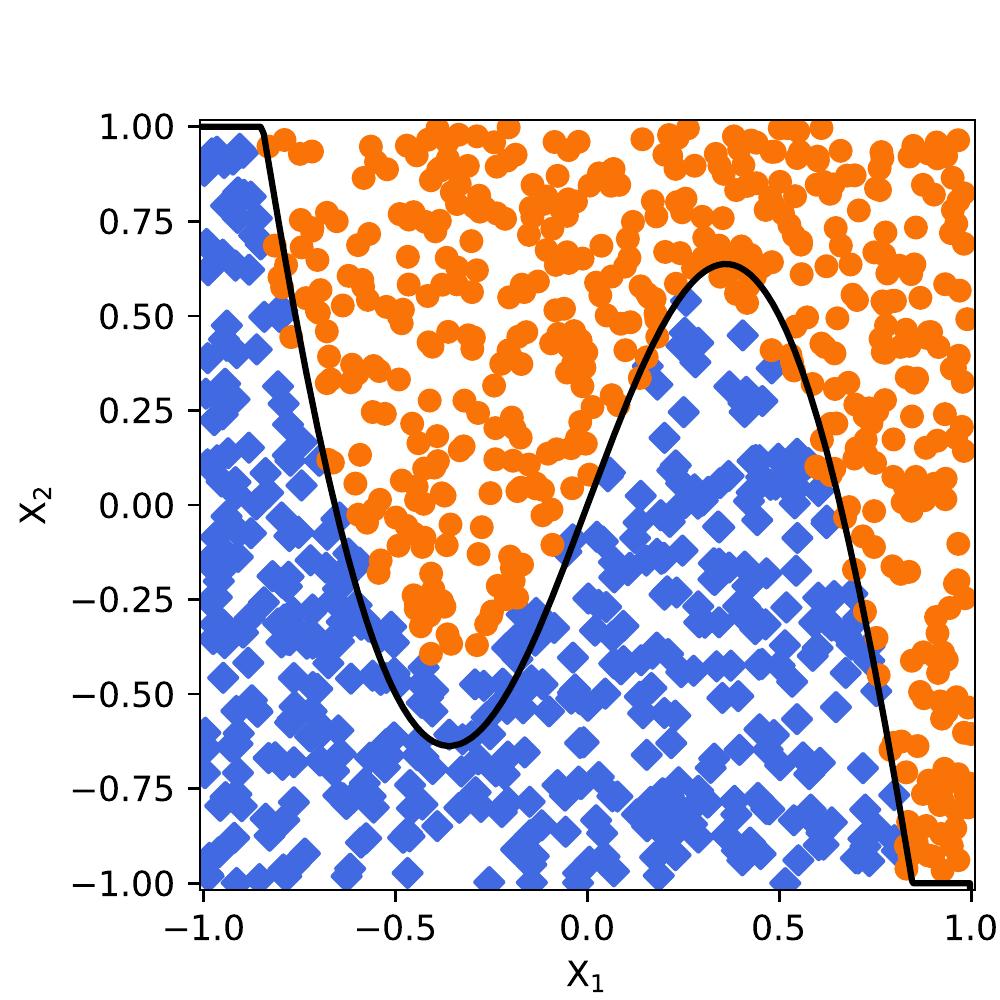}
    }
    \subfigure[\scriptsize Crown - QPU\label{fig:fig3b}]{
    \includegraphics[width=.22\linewidth]{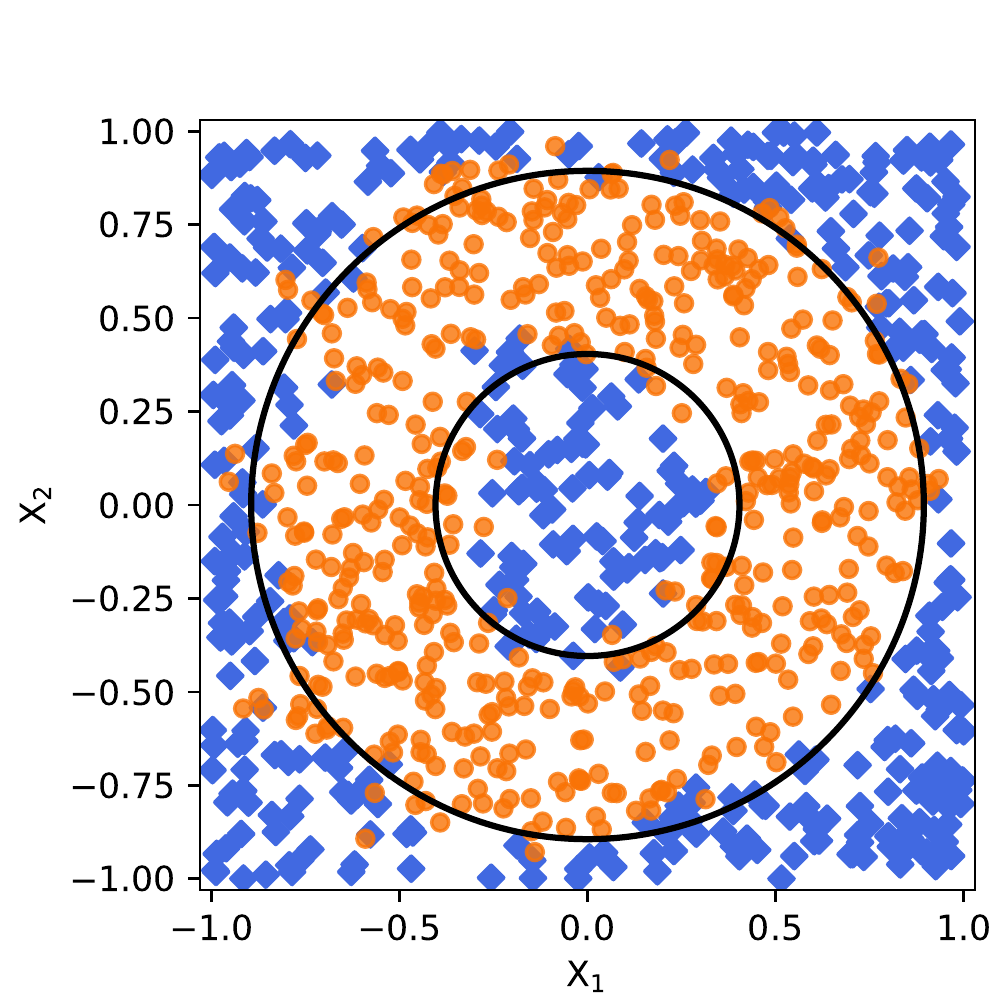}
    }
    \subfigure[\scriptsize Crown - Sim.\label{fig:fig3d}]{
    \includegraphics[width=.22\linewidth]{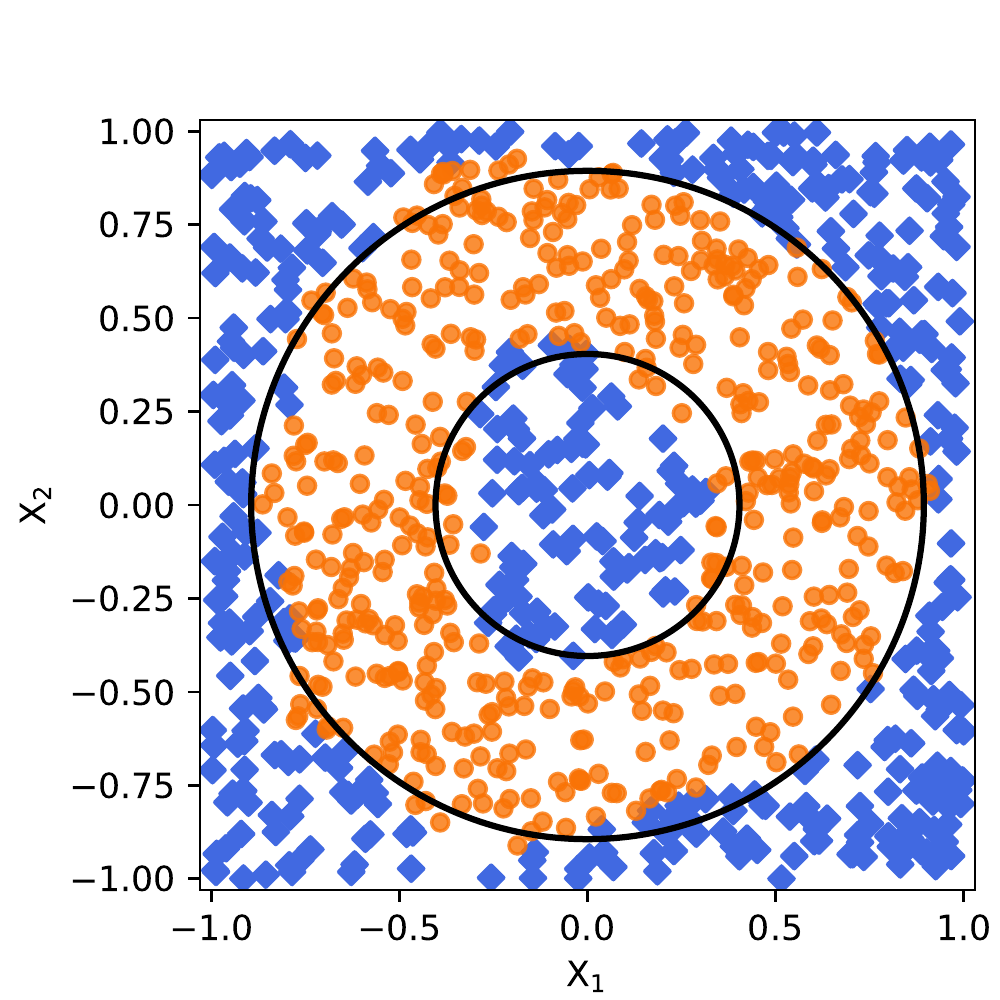}
    }
    \subfigure[\scriptsize Tricrown - QPU\label{fig:tricrown_QPU}]{
    \includegraphics[width=.22\linewidth]{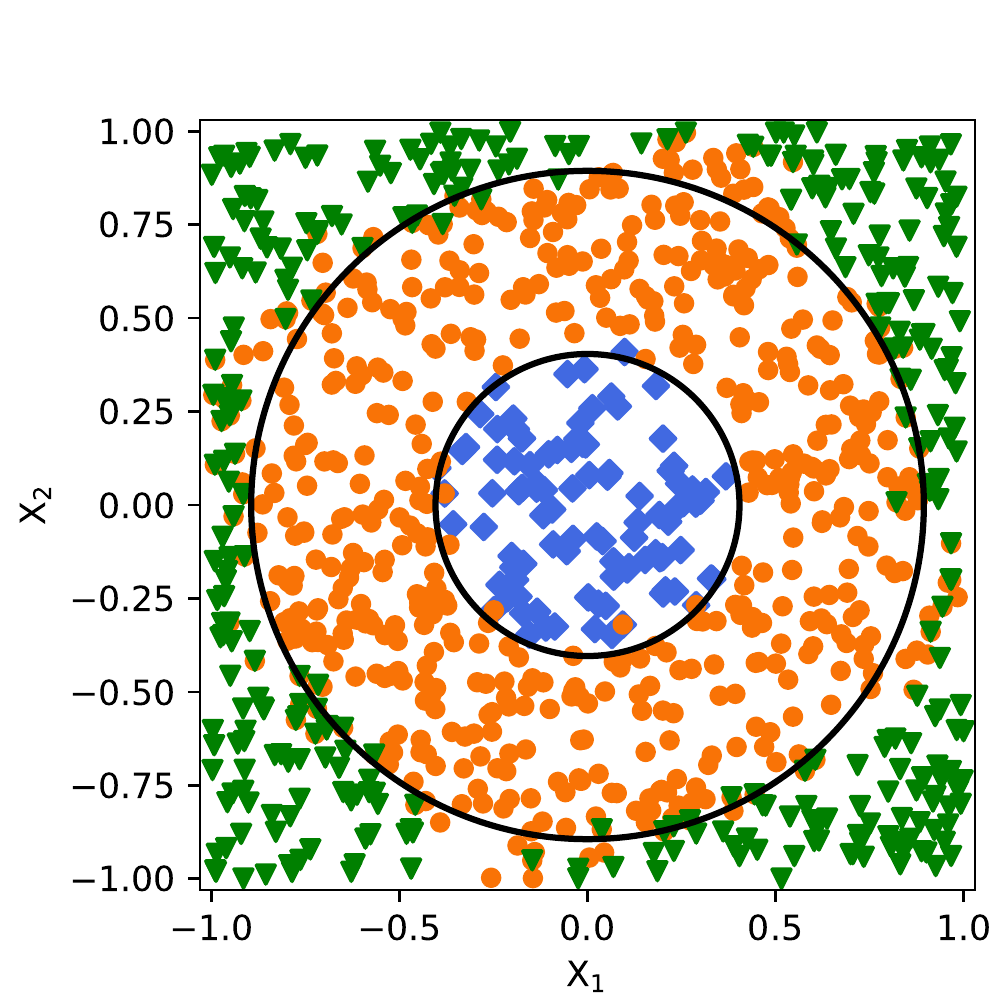}}
    \subfigure[\scriptsize Tricrown - Sim.\label{fig:tricrown_sim}]{
    \includegraphics[width=.22\linewidth]{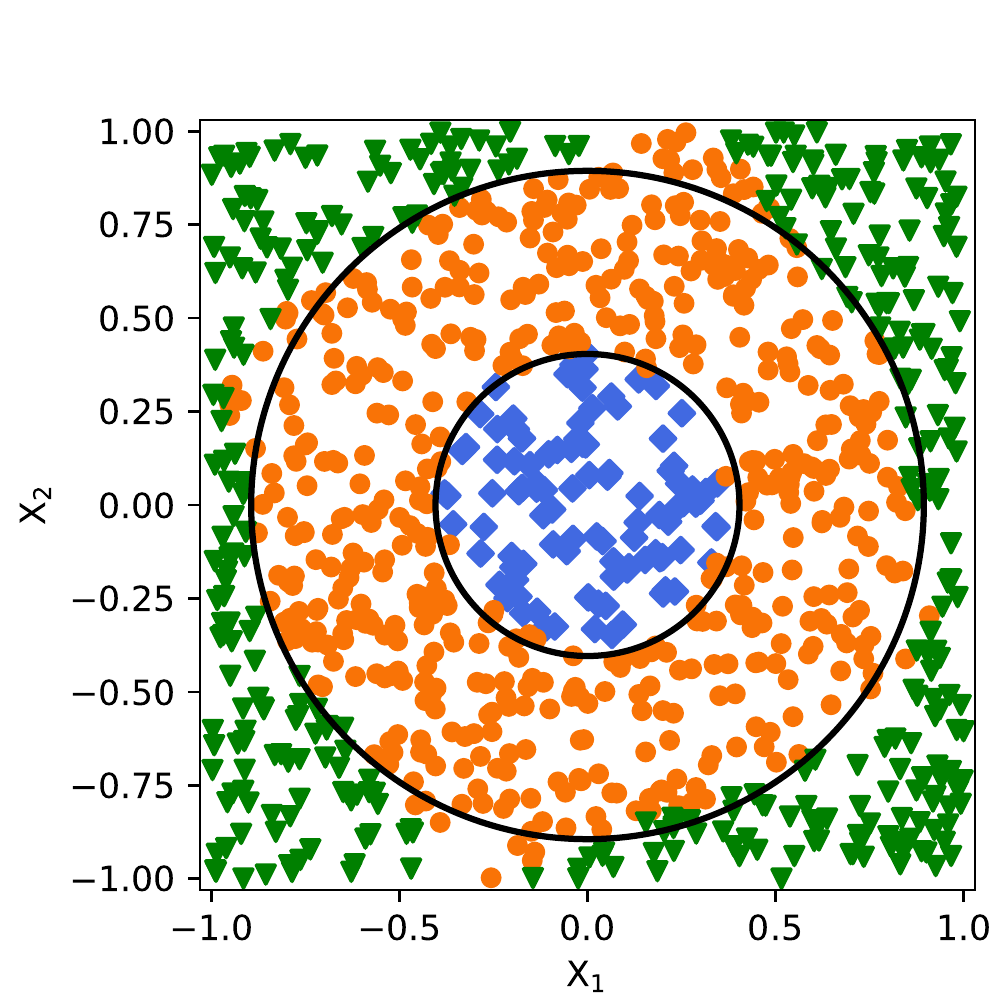}
    }
    \subfigure[\scriptsize 3 circles - QPU]{%
           \label{fig:multiclass_a}
           \includegraphics[width=0.22\linewidth]{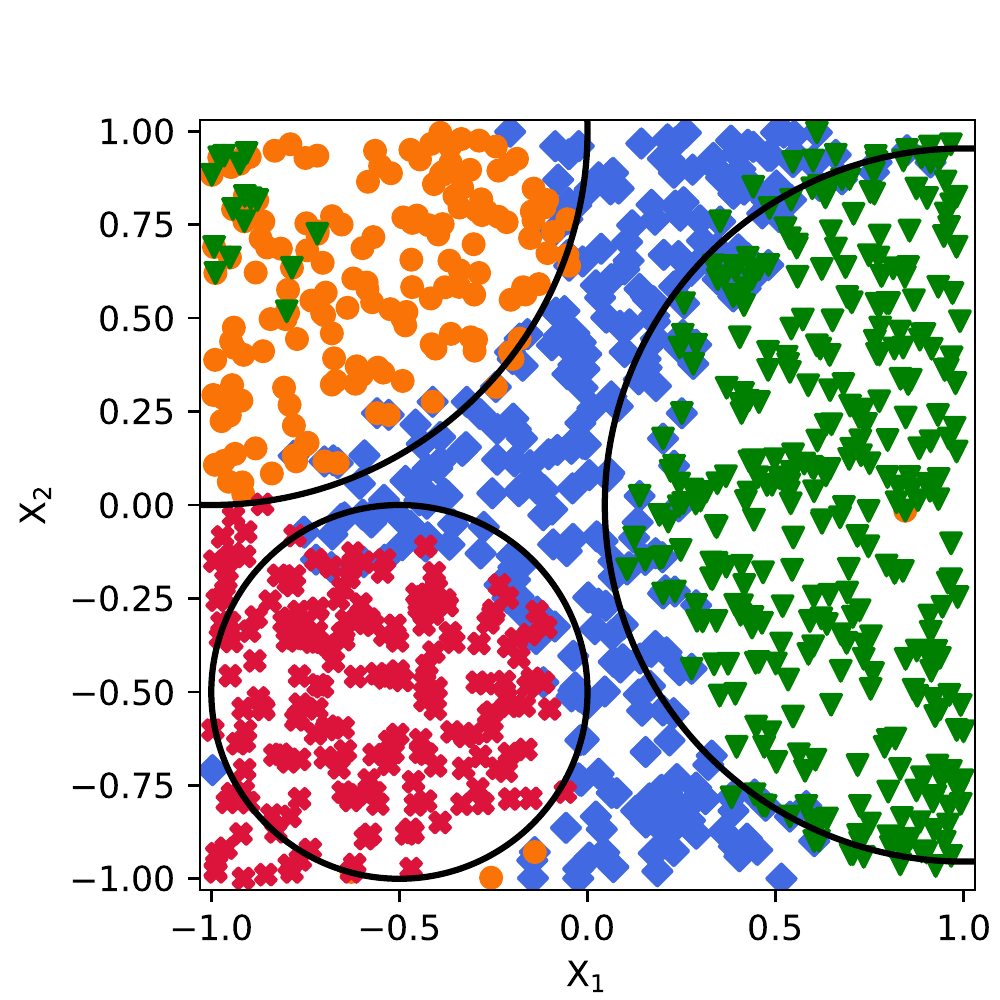}
        } 
    \subfigure[\scriptsize 3 circles - Sim.]{%
            \label{fig:multiclass_d}
            \includegraphics[width=0.22\linewidth]{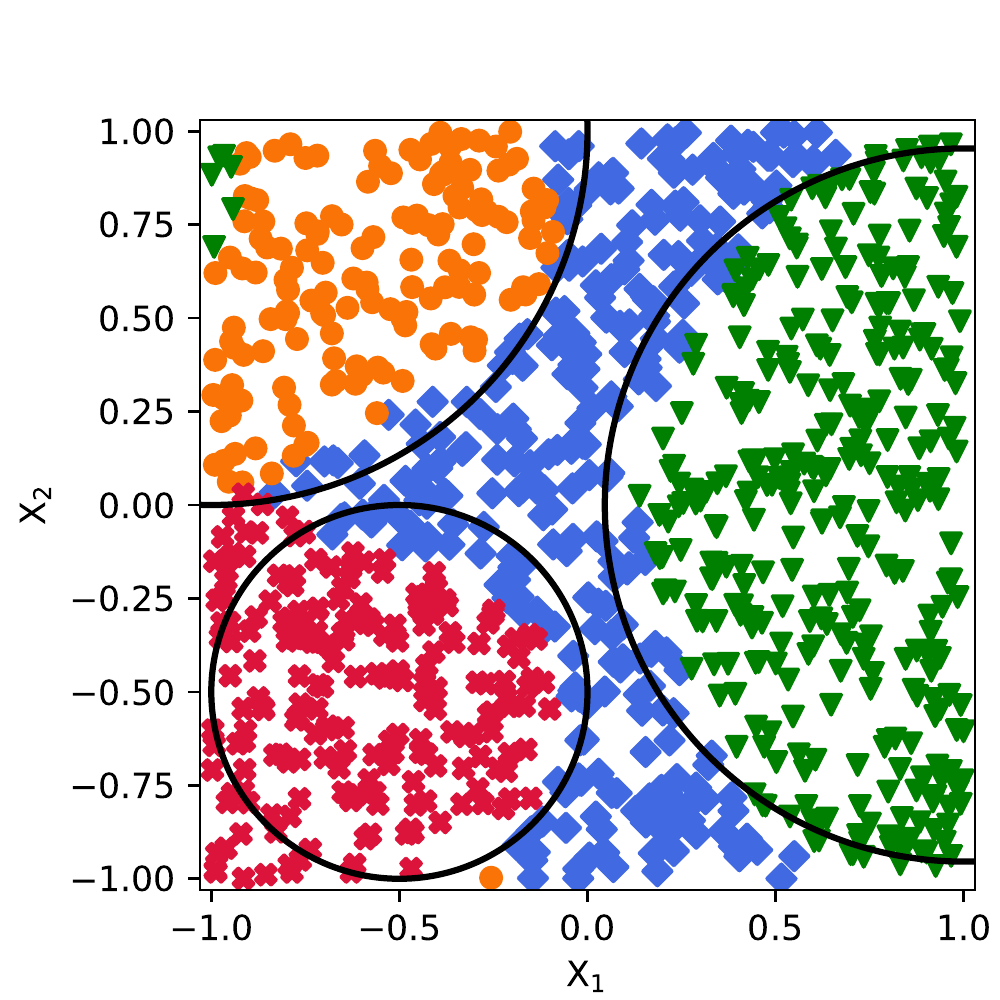}
        }
        \subfigure[\scriptsize squares - QPU]{
            \label{fig:multiclass_b}
             \includegraphics[width=0.22\linewidth]{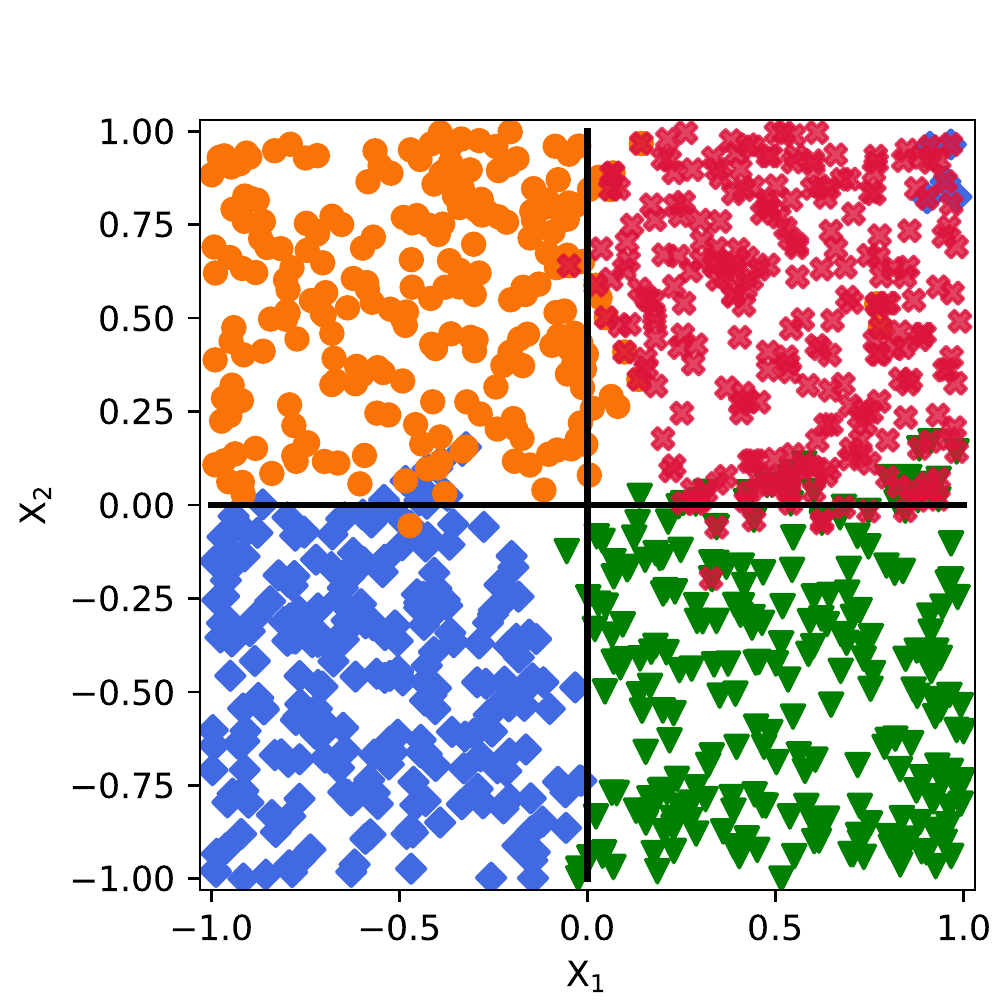}
        }
   \subfigure[\scriptsize squares - Sim.]{%
            \label{fig:multiclass_e}
            \includegraphics[width=0.22\linewidth]{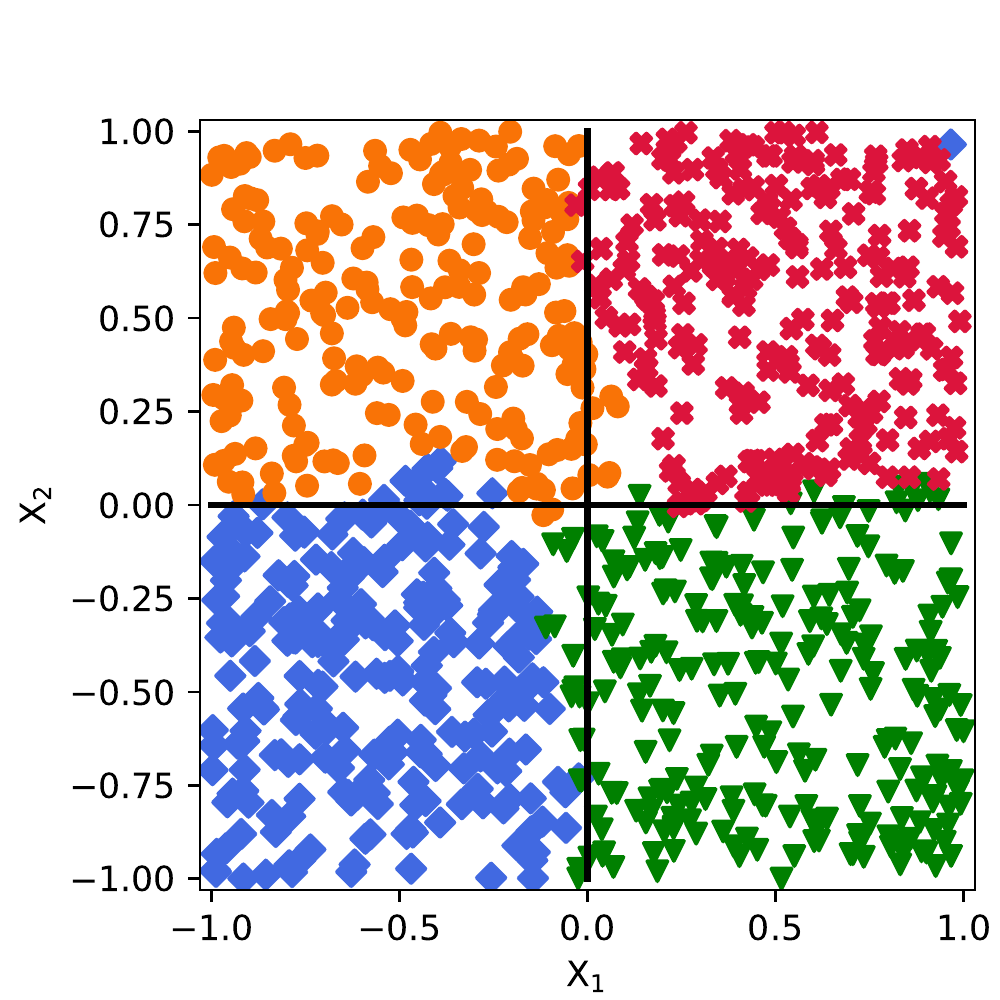}}
        \subfigure[\scriptsize wavy lines - QPU]{%
            \label{fig:multiclass_c}
            \includegraphics[width=0.22\linewidth]{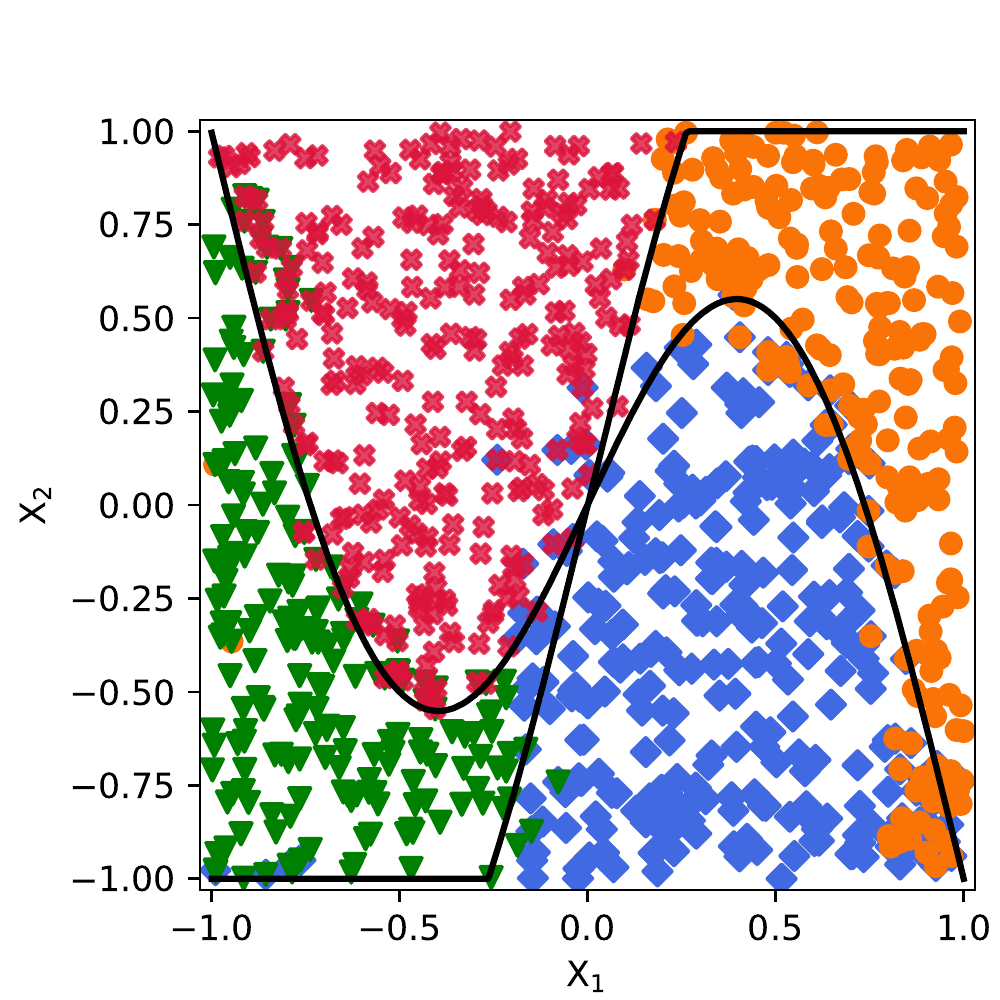}
        }
        \subfigure[\scriptsize wavy lines - Sim.]{%
          \label{fig:multiclass_f}
          \includegraphics[width=0.22\linewidth]{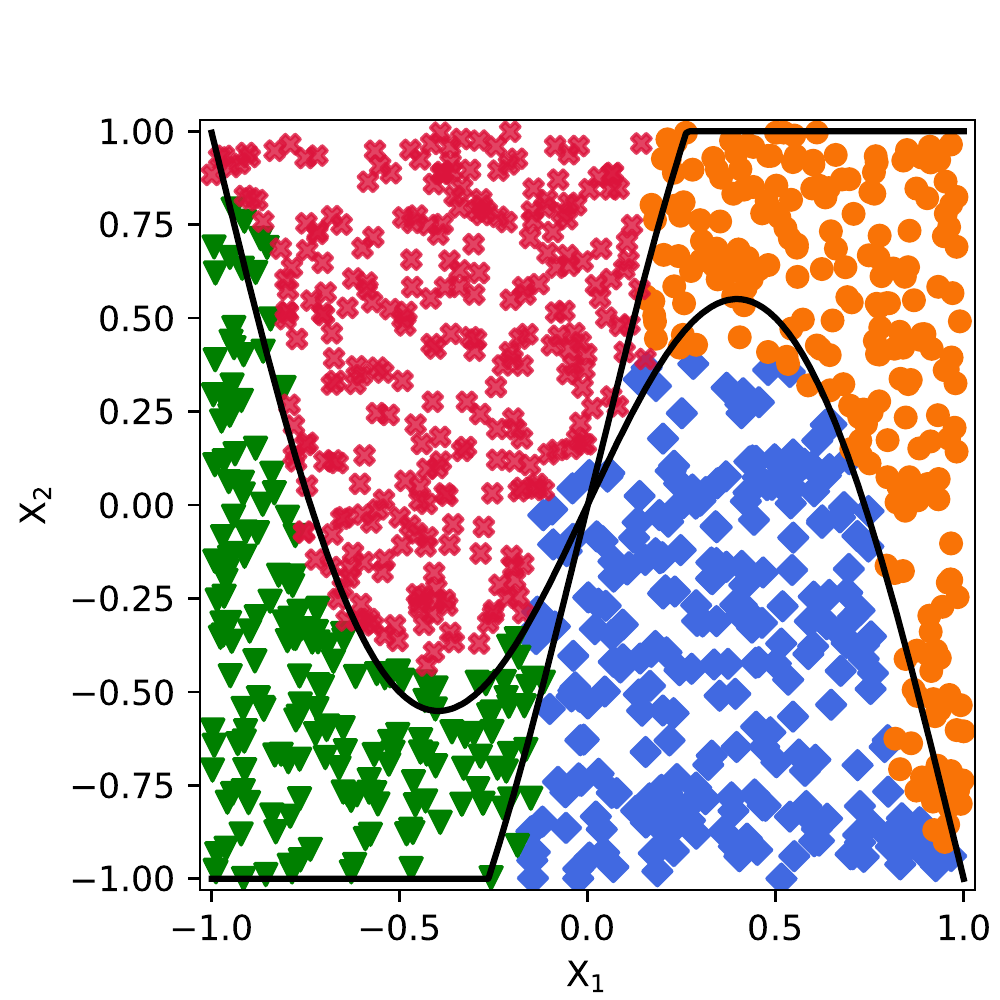}
        }
\end{adjustwidth}
    \caption{Classifier test results for other classification problems. The ion trap based QPU classifier performed on $1000$ random test data points. Colors and symbols stand for different classes, separated by the lines shown in solid black. The results are computed using 4 layers, both from QPU and simulation (Sim.). Notice that the border between classes in the experimental results is not as sharply defined as in the simulated classification. This difference is due to the uncertainty of the quantum measurements.}

    \label{fig:problems}
    
\end{figure}

Results for the experimental optimization step are also provided in this example. Figs.~\ref{fig:supervised_a} show the landscape of the accuracy for a specific subset of parameters in the vicinity of the optimum point as provided by the classical simulation. The smoothness of the landscape around the starting point $(\Theta^{\rm sim}, W^{\rm im})$ induces that only three parameters are actually contributing to significant changes in the cost function, and thus the computational cost of scanning for the optimal configuration is manageable. A deviation between the theoretical and experimental optimal parameters is appreciable. In addition to the landscape, the error in the classification is depicted in Fig.~\ref{fig:supervised_a}, where the experimental errors are estimated separatedly, see App.~\ref{app:experimental_setup_classifier}. After corrections, accuracies for the simulated and experimental setups are similar. The experimental optimization brings an improvement of nearly $5\%$ with respect to the initial parameters $(\Theta^{\rm sim}, W^{\rm sim})$.

The next result presented corresponds to the 4D-hypersphere problem, depicted in Fig.~\ref{fig:hyper_4d}. In this figure, the number of points guessed as inside and outside the hypersphere are printed in different colors and represented in a histogram. A black line corresponding to the boundary is also depicted. The $x$ axis corresponds to the radius of the data point. The overlap region around the boundary corresponds to the failure rate, and thus to that area where the classification is ambiguous. This representation is only feasible due to the spherical symmetry of the problem. 

Our results show that is is feasible to classify the data as well as the simulator does. As in the circle example, the parameter set inherited from classical simulation implies a high error rate. This error can be further reduced from $\sim 13\% $ to $\sim 2\% $ after the experimental optimization step is performed, see Fig.~\ref{fig:hypersphereerror}. 

Equivalent results for all problems presented in the theoretical model of the classifier were tested without experimental optimization. Results for 2D problems are depicted in Fig.~\ref{fig:problems}. A summary of results comparing \ac{qpu} and simulated methods is written in Tab.~\ref{tab:results_qpu}

\begin{table}[t!]
\centering
\begin{adjustwidth}{-1cm}{-2cm}
\resizebox{\linewidth}{!}{
\begin{tabular}{|c||c|c||c|c|c||c|}\hline
\multirow{2}{*}{Problem (\# classes)} & \multicolumn{2}{c||}{\bf Classical algorithms} & \multicolumn{3}{c||}{\bf Quantum re-uploading} & \multirow{2}{*}{Ansatz} \\
 & Neural Network & Support Vector Machine & Simulation & QPU(${\boldsymbol \theta}^{\rm sim})$ & QPU(${\boldsymbol \theta}^{\rm q})$ &\\ \hline\hline
 Circle \hfill (2) & 0.98 & 0.96  & 0.97 & 0.93 & 0.96& A\\
 Crown \hfill (2) & 0.71 & 0.82 & 0.92 & 0.87 && B \\  
 Non-Convex \hfill (2)& 0.98 & 0.79 & 0.95 & 0.92 && B\\
 Sphere \hfill (2) & 0.95 & 0.91 &0.74 & 0.66 && A\\
 Hypersphere \hfill (2) & 0.76 & 0.92 & 0.75 & 0.64 & 0.73& A\\ 
 Tricrown  \hfill (3) & 0.97 & 0.83 & 0.95 & 0.91 && A\\ 
 3 circles \hfill (4) & 0.93 & 0.92  & 0.90 & 0.85 && B\\
 Squares \hfill (4) & 0.99 & 0.95 & 0.97 & 0.93 && A\\ 
 Wavy Lines \hfill (4)& 0.99 & 0.89 & 0.94 & 0.90 && A\\ \hline

\end{tabular}}
\end{adjustwidth}
\caption{Comparison between single-qubit re-uploading quantum classifier and two well-known classical classification techniques, namely  single-hidden-layer neural networks and support vector machines. The experimental data and its simulated analogue is provided here with 4 Layer and 100 repetitions on the quantum part, and its equivalent in complexity for the neural network. The uncertainty of experimental data is $\pm 2\%$. The error refers to the standard deviation of $10$ repeated trials performed on the same dataset and it implies that underlying systematic uncertainty leads to an uncertainty of the accuracy. Only two cases have been further optimized using an exploration done only with the quantum device.}
\label{tab:results_qpu}
\end{table}

\subsection{Discussion}\label{ssec:conclusions_exp_classifier}
The experimental implementation of the quantum classifier on an ion trap QPU was accomplished. The key ingredient for this implementation is the high quality of the control achieved by ion trap platforms for small systems. This is, to the best of our knowledge, the first experimental attainment of supervised learning problems with a single-qubit quantum processing unit. The experiment shows an advantage on the number of physical gates required when compared to classical approaches. 

The experimental classifier is trained in two steps, a simulated and an experimental one, to enhance the performance of the algorithm. Gate-level fine tuning on the application of operations is also performed for further experiments. The experimental optimization ehances the accuracy up to $5-10\%$ depending on the problem, and final results are comparable to those obtained by the classical simulation of the quantum classifier and by classical methods, for models with similar numbers of parameters. 

As in Sec.~\ref{sec:benchmark}, this experiment suffers a lack of purely experimental optimization, in spite of the experimental optimization step. This stage is a refinement of parameters to obtain the same results provided by classical simulators while mitigating possible experimental noise. The difficulty of this task is not to be compared with a full optimization. Future works are expected to deal with this subject. 

This work is the second experimental confirmation that the data re-uploading strategy is a useful scheme towards implementing \ac{qml} algorithms in NISQ devices with few qubits. 

\section{Re-uploading for determining the proton content}\label{sec:qpdf}

It has been shown through this chapter that data re-uploading is a general strategy for applying \ac{qml} to classical data. Theoretical mathematical support that ensures the universality of this strategy is provided, and it is demonstrated by means of both numerical classical simulation and experiments on quantum processing devices that regression of functions and classification of data is possible on some testbeds. In this last section of the chapter a real-world machine learning problem is addressed with the data re-uploading strategy and show that this approach is general enough to be useful in a huge variety of scenarios. The problem here faced is related to \acf{hep}: determining the proton content from experimental data using \ac{ml} strategies. This problem is of most importance since the knowledge of hadron contents, in particular for protons, is crucial for \ac{hep} experiments, such as \ac{lhc}. The standard model predicts the behavior of the interactions between fundamental particles, but not between hadrons. Thus, the inner structure of hadrons must be well known in order to interpret experimental results properly. 

This is not the first attempt to develop works in the frontier between quantum computing and \ac{hep}. This field is highly demanding in terms of computational resources, and thus any advantage provided by quantum computing would be useful. Some examples are computation of helicities \cite{Bepari2021}, simulation of final-state radiation \cite{Nachman2021}, and studies on the description of hadronic structure \cite{li_partonic_2021, Alexandru2019, Lamm2020}.

Quantum Chromodynamics~(QCD) provide theoretical recipes to under-stand and compute all possible interactions between particles in the standard model with strong charge, namely quarks and gluons~\cite{halzen_quark_1985}. These particles compose the nucleons, namely protons and neutrons. Thus, from a theoretical perspective, it is possible to know the exact content of a nucleon from first principles. However, the complexity of the calculations needed to accomplish this task is enormous. One must take into account all different possibilities of interactions to occur, which is extremely hard since the number of particles is not conserved, but it rather increases during collision processes. In addition, the theory is highly non-perturbative, and thus one cannot approximate to some given order since the remaining contribution is never neglictible with respect to the considered one. In summary, theoretical computa-tions to describe nucleons are not feasible nowadays. 

The main framework to describe the non-perturbative structure of hadrons, protons in this case, are \ac{pdf} \cite{feynman_parton_1988, Forte:2020yip}. \ac{pdf}s are typically determined from regression models that compare a large amount of experimental data and theoretical predictions. A well established technique for obtaining \ac{pdf}s is the NNPDF methodology~\cite{Ball:2014uwa}, where regression models are implemented through \ac{nn}s.  

The approach here proposed to address the \ac{pdf} problem with \ac{qml} is to make use of data re-uploading strategy to extract \ac{pdf}s from experimental data, which is referred to as quantum \ac{pdf}~(qPDF). To accomplish this task, two steps were taken. First, the reference data is a set of \ac{pdf}s obtained by classical methods and a quantum circuit, the qPDF, able to mimic this behavior is designed. Second, the obtained simulated results are submitted to public quantum hardware to benchmark the current performance of experimental devices on this real-world problem. Finally, the quantum model substitute the \ac{nn} in NNPDF methodology to learn \ac{pdf} directly from experimental data.

There are several reasons to attempt qPDF recipes, mainly regarding the efficiency of the algorithms. Since this example is the only one presented in the chapter that potentially matches the requirements of real-world problems, any waving must be taken into account. The first one is the reduction in energy consumption required to perform computations. Secondly, it is shown in the results here presented that the number of parameters required to reach acceptable qPDF fits is in average lower than in modern classical approaches. Furthermore, since \ac{pdf} determination is an inherently quantum problem, the presence of entanglement may bring advantage to solve the regression task. Finally, quantum hardware can bring advantage in terms of running time as compared to the classical counterparts. The lower number of parameters implies that the number of operations needed to obtain comparable solutions is smaller. In addition, this model possesses an exact hardware representation, which is not possible in classical cases. 

The results presented in this section are conceived as a proof-of-concept for future implementations of qPDFs on quantum hardware. Neither the performance of quantum simulation nor the quality of quantum hardware at the present time suffice to implement the qPDF approach more efficiently than current classical methods used in modern \ac{pdf} determinations. 

This section is structured as follows. First, the definition for the qPDF model is described in Sec.~\ref{ssec:qcpdf}. Sec.~\ref{ssec:ansatze} is devoted to the exact quantum circuits here used in terms of gates, parameters and test performance. Experiments for the circuits in actual quantum devices are carried in Sec.~\ref{ssec:experiment_proton}. Sec.~\ref{ssec:lhc_data} describes how to use the qPDF framework to extract \ac{pdf}s from experimental data from \ac{lhc}. Final remarks are written in Sec.~\ref{ssec:coclusions_proton}.

\subsubsection{Workflow design}

\begin{wrapfigure}{L}{.5\linewidth}
\centering
  \includegraphics[width=\linewidth]{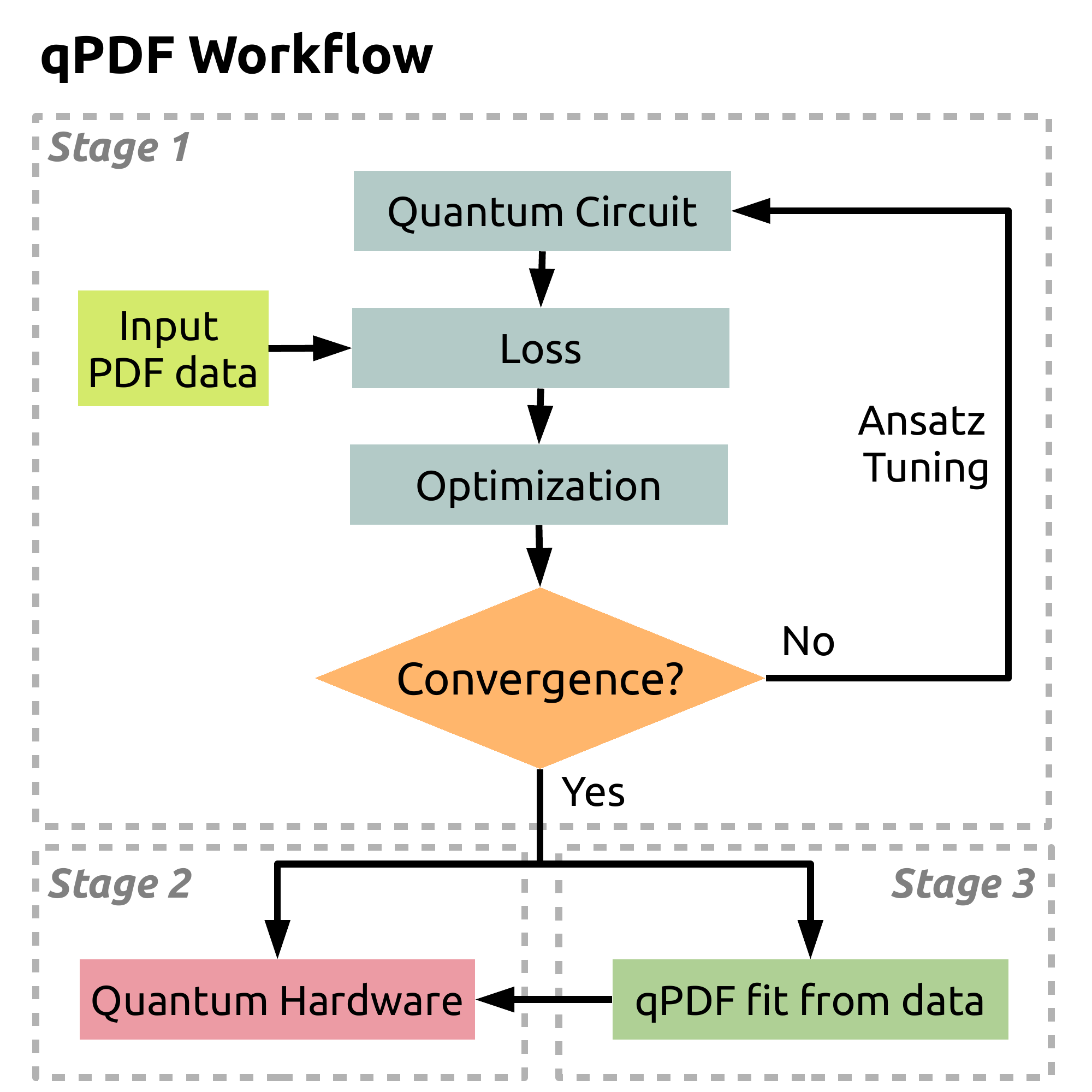}
  \caption{\label{fig:workflow}Schematic workflow for the implementation of
  qPDF. The initial input data is a classical model for \ac{pdf}. The Ansätze are modified until they are flexible enough to fit these data. Then, the same model is testes on an experimental quantum device. As a last step, the simulated circuit is introduced into a full \ac{pdf} procedure to compare classical and quantum performances.}
\end{wrapfigure}

The complete workflow of this project is composed of three different steps, schematically shown in Fig.~\ref{fig:workflow}. 
\begin{itemize}
\item[1] Design of the most adapted quantum circuit Ansatz for qPDF
\item[2] Feasibility study to deploy the qPDF architecture on real quantum devices
\item[3] Integration of the qPDF model into a global \ac{pdf} fitting framework taking the role of classical \ac{nn}s
\end{itemize}

In the first step several Ansätze for circuits were tested to choose the one with more flexibility to learn and generalize \ac{pdf}-like datasets. The re-uploading strategy pro-vides a general framework to design circuits capable to represent any data, but particular architecture is not determined by this approach. This stage is analogous to the hyper-optimization tuning performed in classical machine learning applications. However, since there are many possible architectures and no pre-defined initial Ansatz is assumed, empirical tests and refinement through trial and error is needed. The calculations are done using the exact classical simulation provided by {\tt QIBO}~\cite{qibo, qibo_code}, where both exact wavefunctions and expected values for hamiltonians are computed. For all tests, the model is trained to fit \ac{pdf} generated by classical means, in particular from the NNPDF3.1 set~\cite{Ball:2017nwa}. This serves to find the optimal quantum circuits defined in the next subsection. 

The second stage deals with the deployment of the qPDF model in actual quantum devices. In this case, both measurement and its uncertainty and noise models become relevant. This stage helps to determine the minimum number of measurements required to retrieve acceptable descriptions of \ac{pdf}s on quantum hardware. The experimental implementation of the qPDF model is done using the {\tt Qiskit} language~\cite{qiskit} from the {\tt OpenQASM}~\cite{cross2017open} codegenerated by {\tt Qibo}. 

Finally, we use the qPDF model in an actual functioning classical \ac{pdf} fitter based on experimental data, mostly from \ac{lhc} measurements. The quantum model is integrated in the NNPDF fitting framework {\tt n3fit}~\cite{Carrazza:2019mzf,Forte:2020yip} using the simulation tools. This implementation opens the possibility to perform quantum fits of \ac{pdf} on similar datasets for modern \ac{pdf} releases.

\subsection{Quantum circuits for \ac{pdf}}\label{ssec:qcpdf}

We define in the following the Ansätze for quantum circuits needed to accomplish the qPDF fitting. First, the \ac{pdf}s are defined as
\begin{equation}\label{eq:pdf}
f_i(x, Q_0), 
\end{equation}
where $x \in [0, 1]$ is the momentum fraction of the incoming hadron carried by the parton of flavour $i$ (quarks and gluon), at an energy scale $Q_0$. The normalization rule for \ac{pdf}s is
\begin{equation}
\sum_{i} \int_{x = 0}^{x=1} x f_i(x,Q_0) = 1.
\end{equation}
Following this definition, modifications to the data re-uploading are now added to accomodate it to \ac{pdf}. These modifications leave unaffected the most important properties of data re-uploading, namely query complexity and arising of non-linearities. 

To introduce the $x$ variable in the quantum circuit, the same structure as in the previous examples is followed
\begin{equation}\label{eq:quantumcircuit}
\ket{\psi(x, \Theta)} = \mathcal{U}^{(k)}(x, \Theta) \ket 0 = \prod_{j = 1}^k U(x, \vec\theta_i) \,\ket{0}, 
\end{equation}
where $\Theta$ are tunable parameters to be optimized through a loss function. The exact definition of the different layers is left for later since there are some features of the problem to be discussed before. 

The second key ingredient in the model is the way to retrieve information from the circuit. In this problem, the structure of the proton is determined through several different functions, as many as flavours in the protons. Thus, many different independent measurements must be designed to extract information. A $n$-qubit circuit is considered to run the quantum algorithm on, where each qubit corresponds to one flavour. The set of hamiltonians to build is then
\begin{equation}
Z_i = \bigotimes_{j = 0}^n Z^{\delta_{ij}},
\end{equation}
where $\delta_{ij}$ is the Kronecker delta. The choice of hamiltonians is heuristic, and measures the population of $\ket 0$ and $\ket 1$ states of a particular qubit. The values of qPDFs will be then related to the probability to measure one particular qubit in the excited state. The function
\begin{equation}\label{eq:z_function}
z_i(x, \Theta) = \bra{\psi(x, \Theta)}Z_i\ket{\psi(x, \Theta)},
\end{equation}
is taken as the outcome of the circuit. The next step is then to relate this outcome to the \ac{pdf} values. Every $z_i(x,\Theta)$ will be associated to only one parton $i$, and thus as many qubits as partons are needed. The qPDF model at a given $(x, Q_0)$ coordinates is then
\begin{equation}
qPDF(x, Q_0, \Theta) = \frac{1 - z_i(x, \Theta)}{1 + z_i(x, \Theta)}. 
\end{equation}
This choice, as well as the hamiltonian, is heuristic and supported by empirical results. It only allows the qPDFs to take positive values, although there is no upper bound in this quantity. This is no hard constraint since it is possible to drop the possitivity with simple re-scaling. Theoretical motivations can be drawn from the fact that \ac{pdf}s can be made non-negative~\cite{Candido:2020yat}, but their values can in principle grow to any real value, see
for instance the gluon \ac{pdf} in Fig.~\ref{fig:all_flavours}.

\subsection{Ansätze}\label{ssec:ansatze}
It is seen and discussed that the re-uploading of classical data in conjunction with tunable weights and biases following the scheme $w x + b$ permits to represent arbitrary functions, for example \ac{pdf}s, in single-qubit systems, see Theorems~\ref{th:q_fourier} and~\ref{th:q_UAT}. On the other hand, Ref.~\cite{schuld_effect_2021} shows that a Fourier approximation of an arbitrary function is possible even if the weights are fixed. Therefore, two different Ansätze were considered in this work, where the main difference between them is the presence or absence of tunable weights. 

The $x$ independent variable of the dataset is constrained to take values between $0$ and $1$. However, the representation of \ac{pdf} is usually given in a logarithmic scale in $x$ since variations in \ac{pdf} among orders of magnitude are prominent. This observation encourages using not only $x$ as the independent variable, but also $\log(x)$ to properly capture the behavior of the functions at small scales.

The first Ansatz, named Weighted Ansatz, is inherited from the works in Refs.~\cite{perezsalinas_data_2020} and later~\cite{perezsalinas_qubit_2021}. The $x$ variable is introduced following the weights and biases scheme. The single-qubit operation acting as building block of the Ansatz is
\begin{equation}\label{eq:weighted_ansatz}
U_w(\theta, x) = R_z(w_2 \log(x) + b_2) R_y(w_1 x + b_2).
\end{equation}
Note the presence of rotations around two different axis, $Y$ and $Z$, in the definition of this gate, to make non-linearities emerge from the quantum mechanical nature of the single-qubit operations. In addition, each axis introduces one scale, namely linear or logarithmic. 

The second Ansatz, Fourier Ansatz, limits the weights to fixed values~\cite{schuld_effect_2021}. Its single-qubit operation is
\begin{equation}
U_f(\theta, x) = R_y(b_2)R_z(w_2)R_y(-\pi/2 \log x) R_y(b_1)R_z(w_1)R_y(\pi x), 
\end{equation}
where again two axis are involved. The choice of the weights $\pi$ in the linear scale and $-\pi/2$ in the logarithmic scale depends on the dataset. For the \ac{pdf} set chosen to test the Ansätze, $x\in[10^{-4}, 1]$. This way, all the amplitude of the single-qubit gates is exploited. 

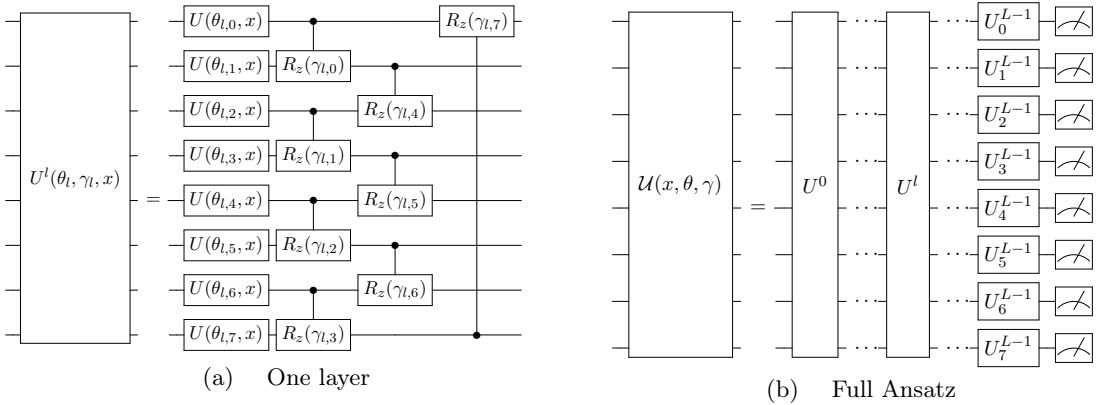
\begin{figure}[t]
  \centering
\begin{adjustwidth}{-1cm}{-2cm}
  \subfigure[\hspace{2mm} One layer]{
\resizebox{.505\linewidth}{!}{
\Qcircuit @R=.75em @C=.4em{
      &\qw & \multigate{7}{U^{l}(\theta_l, \gamma_l, x)} & \qw & \push{\hspace{5mm}} & \qw & \gate{U(\theta_{l,0}, x)} & \ctrl{1} & \qw & \gate{R_z(\gamma_{l,7})} & \qw \\
      &\qw & \ghost{U^{l}(\theta_l, \gamma_l, x)} & \qw & \push{\hspace{5mm}} & \qw & \gate{U(\theta_{l,1}, x)} & \gate{R_z(\gamma_{l,0})} & \ctrl{1} & \qw & \qw & \\
      &\qw & \ghost{U^{l}(\theta_l, \gamma_l, x)} & \qw & \push{\hspace{5mm}} & \qw & \gate{U(\theta_{l,2}, x)} & \ctrl{1} & \gate{R_z(\gamma_{l,4})} & \qw & \qw & \\
      &\qw & \ghost{U^{l}(\theta_l, \gamma_l, x)} & \qw & \push{\hspace{5mm}} & \qw & \gate{U(\theta_{l,3}, x)} & \gate{R_z(\gamma_{l,1})} & \ctrl{1} & \qw & \qw & \\
      &\qw & \ghost{U^{l}(\theta_l, \gamma_l, x)} & \qw & \push{ = \hspace{1.5mm}} & \qw & \gate{U(\theta_{l,4}, x)} & \ctrl{1} & \gate{R_z(\gamma_{l,5})} & \qw & \qw & \\
      &\qw & \ghost{U^{l}(\theta_l, \gamma_l, x)} & \qw & \push{\hspace{5mm}} & \qw & \gate{U(\theta_{l,5}, x)} & \gate{R_z(\gamma_{l,2})} & \ctrl{1} & \qw & \qw &  \\
      &\qw & \ghost{U^{l}(\theta_l, \gamma_l, x)} & \qw & \push{\hspace{5mm}} & \qw & \gate{U(\theta_{l,6}, x)} & \ctrl{1} & \gate{R_z(\gamma_{l,6})} & \qw & \qw &  \\
      &\qw & \ghost{U^{l}(\theta_l, \gamma_l, x)} & \qw & \push{\hspace{5mm}} & \qw & \gate{U(\theta_{l,7}, x)} & \gate{R_z(\gamma_{l,3})} & \qw & \ctrl{-7} & \qw & \\
      & & & & & & & & & & & & & & & & &
      }
}
  } ~\hfill
  \subfigure[\hspace{2mm} Full Ansatz]{
\resizebox{.45\linewidth}{!}{
   \Qcircuit @R=.46em @C=.4em{
      &\qw & \multigate{7}{\mathcal U(x, \theta, \gamma)} & \qw & \push{\hspace{5mm}} & \qw & \multigate{7}{U^0} & \qw & \push{\cdots} & \multigate{7}{U^l} & \qw & \push{\cdots} & \gate{U^{L-1}_0} & \qw & \meter \\
      &\qw & \ghost{\mathcal U(x, \theta, \gamma)} & \qw & \push{\hspace{5mm}} & \qw & \ghost{U^0} & \qw & \push{\cdots} & \ghost{U^l} & \qw & \push{\cdots} & \gate{U^{L-1}_1} & \qw & \meter \\
      &\qw & \ghost{\mathcal U(x, \theta, \gamma)} & \qw & \push{\hspace{5mm}} & \qw & \ghost{U^0} & \qw & \push{\cdots} & \ghost{U^l} & \qw & \push{\cdots} & \gate{U^{L-1}_2} & \qw & \meter \\
      &\qw & \ghost{\mathcal U(x, \theta, \gamma)} & \qw & \push{\hspace{5mm}} & \qw & \ghost{U^0} & \qw & \push{\cdots} & \ghost{U^l} & \qw & \push{\cdots} & \gate{U^{L-1}_3} & \qw & \meter \\
      &\qw & \ghost{\mathcal U(x, \theta, \gamma)} & \qw & \push{ = \hspace{1.5mm}} & \qw & \ghost{U^0} & \qw & \push{\cdots} & \ghost{U^l} & \qw & \push{\cdots} & \gate{U^{L-1}_4} & \qw & \meter \\
      &\qw & \ghost{\mathcal U(x, \theta, \gamma)} & \qw & \push{\hspace{5mm}} & \qw & \ghost{U^0} & \qw & \push{\cdots} & \ghost{U^l} & \qw & \push{\cdots} & \gate{U^{L-1}_5} & \qw & \meter \\
      &\qw & \ghost{\mathcal U(x, \theta, \gamma)} & \qw & \push{\hspace{5mm}} & \qw & \ghost{U^0} & \qw & \push{\cdots} & \ghost{U^l} & \qw & \push{\cdots} & \gate{U^{L-1}_6} & \qw & \meter \\
      &\qw & \ghost{\mathcal U(x, \theta, \gamma)} & \qw & \push{\hspace{5mm}} & \qw & \ghost{U^0} & \qw & \push{\cdots} & \ghost{U^l} & \qw & \push{\cdots} & \gate{U^{L-1}_7} & \qw & \meter \\
      & & & & & & & & & & & & & & & & &
      }
      }
  } 
\end{adjustwidth}  
  \caption{On the left, an example of one layer architecture is shown. On the
  right, the scheme of a full Ansatz circuit including 8 qubits and
  entangling gates. The $U^l(\theta_l, \gamma_l, x)$ from the left figure enters
  the full ansatz as $U^l$. Note that the last layer does not have any
  entangling gate.}
  \label{fig:ansatz}
  \end{figure}
  
The single-qubit operations are introduced into a more general structure to create several-qubits multi-layered global Ansätze to fit the \ac{pdf}s. The reason for this procedure is that it is expected to obtain better approximations as more layers are added to the circuit and the query complexity of the algorithm increases. The construction of layers is made in two steps. First, a layer of parallel single-qubit operations is applied to each qubit in the circuit. Second, a layer of entangling gates is added to the circuit. In this problem, all entangling gates are controlled $Z$ rotations depending on some parameter $R_z(\gamma)$. Entangling gates connect one qubit with next or previous one alternatively. Sections of rotations and entangling gates are interspersed along the circuit, except for the last iteration where only single-qubit gates are considered. Parameters for every gate are independent from all other parameters and optimized simultaneously. See Fig.~\ref{fig:ansatz} for a schematic description of the Ansatz structure.   
  
For the first stage of tuning Ansätze, simulation methods are used. The optimization procedure takes as loss function the standard Pearson's $\chi^2$\cite{pearson_chi_1900} to compare qPDF predctions against classical calculations of NNPDF3.1 NNLO \cite{Ball:2017nwa}. This dataset is composed by a central value $f_i$ and an uncertainty $\sigma_{f_i}$, both depending on $x$ and $Q_0$. For this exercise, a grid of $x$ values $x \in [10^{-4}, 1]$ and a fixed value $Q_0 = 1.65~GeV$ are considered. The dataset of interest contains 8 different flavours, namely quarks, antiquarks and the gluon: $i \in \{\bar{s}, \bar{u}, \bar{d}, g, d, u, s, c(\bar{c})\}$. 

\begin{table}[t]
\begin{adjustwidth}{-2cm}{-1cm}
\resizebox{.48\linewidth}{!}{
\begin{tabular}{
  |c | c!{\vrule width 1.6pt} c | c | }\hline
  \multicolumn{2}{|c!{\vrule width 1.6pt}}{Single-flavour fit} & \multicolumn{2}{c|}{Multi-flavour fit} \\ \hline
  Layers (Parameters) & $\chi^2$ & $\chi^2$ & Layers (Parameters)\\ \noalign{\hrule height .1em}
  1 (32) & \multicolumn{2}{c|}{28.6328} & 1 (32)\\ \hline
  2 (64)& 1.0234 & -- & -- \\ \hline
  3 (96)& 0.0388 & 0.1500  & 2 (72) \\ \hline
  4 (128)& 0.0212 & 0.0320 & 3 (112)\\ \hline
  5 (160)& 0.0158 & 0.0194 & 4 (152)\\ \hline
  6 (192)& 0.0155 & 0.0154  & 5 (192) \\ \hline
\end{tabular}} ~\hfill 
\resizebox{.48\linewidth}{!}{
\begin{tabular}{
  |c | c!{\vrule width 1.6pt} c | c | }\hline
  \multicolumn{2}{|c!{\vrule width 1.6pt}}{Single-flavour fit} & \multicolumn{2}{c|}{Multi-flavour fit} \\ \hline
  Layers (Parameters) & $\chi^2$ & $\chi^2$ & Layers (Parameters)\\ \noalign{\hrule height .1em}
 1 (32) & \multicolumn{2}{c|}{900.694} & 1 (32)\\ \hline
  2 (64)& 57.2672 & -- & -- \\ \hline
  3 (96)& 0.0410 & 47.4841  & 2 (72) \\ \hline
  4 (128)& 0.0232 & 0.0371 & 3 (112)\\ \hline
  5 (160)& 0.0165 & 0.0216 & 4 (152)\\ \hline
  6 (192)& 0.0156 & 0.0160  & 5 (192) \\ \hline
\end{tabular}}
\end{adjustwidth}
\caption{Comparison of $\chi^2$ values for the Weighted (left) and Fourier (right) Ansätze 
average of all single-flavour fits and the corresponding multi-flavour
fit.}
\label{tab:chi}

\centering 
    \begin{tabular}{| c | c | c | c | c |}\cline{2-5}
    \multicolumn{1}{c|}{} & \multicolumn{2}{c|}{Single-flavour} & \multicolumn{2}{c|}{Multi-flavour} \\ \cline{2-5}
    \multicolumn{1}{c|}{} & Weighted & Fourier & Weighted & Fourier \\ \hline
    Qubits $(q)$ & \multicolumn{2}{c|}{1 (per flavour)} & \multicolumn{2}{c|}{8} \\ \hline
    Layers $(l)$ & \multicolumn{2}{c|}{5} & \multicolumn{2}{c|}{5} \\ \hline
    \multirow{3}{*}{Parameters} & $2 \cdot l \cdot q$ weights & \multirow{2}{*}{$4 \cdot l \cdot q$} & $16 \cdot l$ weights & \multirow{2}{*}{$32 \cdot l$} \\
     & $2 \cdot l \cdot q$ biases &  & $16 \cdot l$ biases &  \\ \cline{2-5}
     & \multicolumn{2}{c|}{No entanglement} & \multicolumn{2}{c|}{$8 (l - 1)$ entangling} \\ \hline
    \end{tabular}
    \caption{Summary for the Ansätze chosen for this work. The preferred number of l
    ayers was chosen as a compromise between small $\chi^2$ and number of parameters.
    Results depicted in Tab.~\ref{tab:chi}
    determine that the multi-flavour Weighted Ansatz is the best candidate model.}
    \label{tab:summary_ansatz}
\end{table}

Values of Pearson's $\chi^2$ after full optimization are summarized in Tab.~\ref{tab:chi}, both for Weighted and Fourier Ansätze. In both cases, the left column shows an average fit for all flavours when optimized individually, while the right column shows the fit for all flavors simultaneously. Compa-risons are made between models with similar numbers of parameters. For the same number of layers, circuits with entanglement have more parameters since every entangling gate has one of them. Thus, unentangled circuits of $n$ layers with entangled ones with $n-1$ layers are compared until both numbers of parameters agree. The reason for this comparison is that entanglement is expected to improve the overall quality of this method. 

Standard classical optimization methods were used to find optimal parameters. The optimization procedure was carried in two steps. First, the {\tt CMA} genetic algorithm is used to find optimal solutions in the single-flavour scenario~\cite{cma, cma_package}. Then, this result is taken as starting point in the multi-flavour optimization carried through the {\tt L-BFGS-B} method \cite{l-bfgs} included in the library {\tt scipy} \cite{scipy}. This two-step optimization ensures that competitive results are obtained for the multi-flavour cases. 

\begin{figure}[t!]
\begin{adjustwidth}{-1cm}{-2cm}
\includegraphics[width=\linewidth]{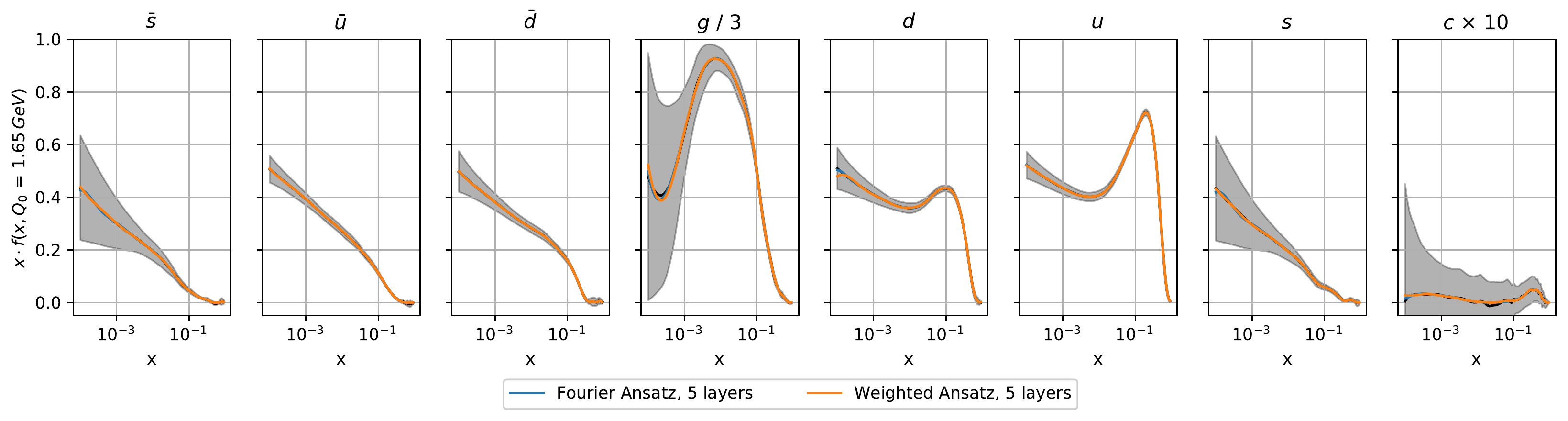}
\end{adjustwidth}
\caption{Multi-flavour qPDF fits using the Weighted Ansatz (orange curves) and
the Fourier Ansatz (blue curves) with 5 layers and 8
qubits. The mean value and $1\sigma$ uncertainty of the target \ac{pdf} data is shown
by means of a solid black line and a shaded grey area.}
\label{fig:all_flavours}
\end{figure}
\begin{SCfigure}
\centering
\caption{Comparison between single-flavour fits (left) and multi-flavour fits
(right) for the gluon, up and strange quarks \ac{pdf}s. For the single-flavour fits
the Weighted Ansatz (orange curves) and Fourier Ansatz (blue curves) are
composed by 1 qubits and 6 layers. On the other hand for the
multi-flavour fits, the Ansätze are composed by 8 qubits and 5 layers. The mean value and $1\sigma$ uncertainty of the target \ac{pdf} data is sshown
by means of a solid black line and a shaded grey area.
\label{fig:single_flavours}}
\includegraphics[width=.6\linewidth]{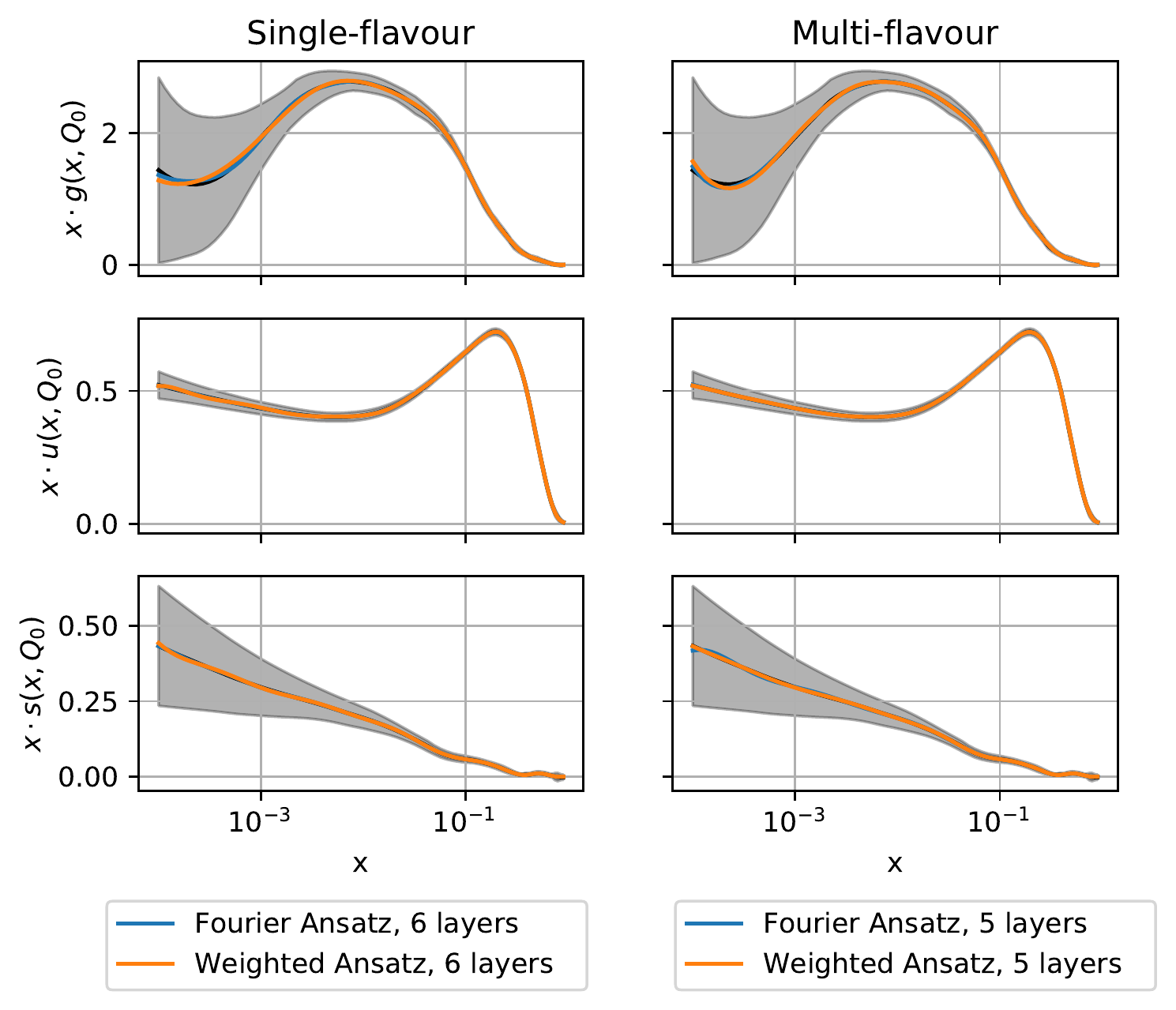}
\end{SCfigure}

Some results of the final configuration after full optimization are depicted in Figs.~\ref{fig:all_flavours} and~\ref{fig:single_flavours}. Fig.~\ref{fig:all_flavours} show the multi-flavour qPDF both for Weighted and Fourier Ansätze, for 5 layers, and 8 qubits, as compared to the classical results. In Fig.~\ref{fig:single_flavours}, a comparison of single- and multi-flavour optimizations focusing on certain flavours is detailed, for Weighted and Fourier Ansätze with similar numbers of parameters. Notice that in both cases qPDFs and classical data overlap with high accuracy

The results here presented permit to claim several interpretations. First, entanglement is not enough to provide good approximations by its own. Entanglement can in principle access to the correlations between different qubits, which in this case encode different flavours and qPDF. However, every layer introduces a new re-uploading of data and takes another step towards non-linearity, which is needed for representing arbitrary functions. Results from Tab.~\ref{tab:chi} show that entanglement, and possibly more parameters, help to obtain better results when comparing models with the same query complexity. For the same number of parameters, both number of layers and entanglement provide the same capabilities. Thus, both entanglement and query complexity contribute to the overall performance. Secondly, the goodness of the Weighted Ansatz respect to the Fourier one is unveiled. Built-in weights grant large flexibility, especially in cases with small numbers of layers. 

We retain the Weighted Ansatz with 5 layers as the final model, both for single- and multi-flavour scenarios. For the sake of comparison, equivalent Fourier Ansätze were chosen as well. The total amount of parameters is 192, which is a manageable number, see Tab. \ref{tab:summary_ansatz} for a detailes comparison. In addition, tests run on both Ansätze reveiled that the Weighted Ansatz is easier to train using gradient-based methods like {\tt L-BFGS-B}.

\subsection{Experimental configuration}\label{ssec:experiment_proton}
\begin{figure}[b!]
\centering
\begin{adjustwidth}{-1cm}{-2cm}
  \subfigure[\hspace{5mm} Single flavour fits in IBM Athens  \label{fig:SingleFlavorExperiment}]{\includegraphics[width=\linewidth]{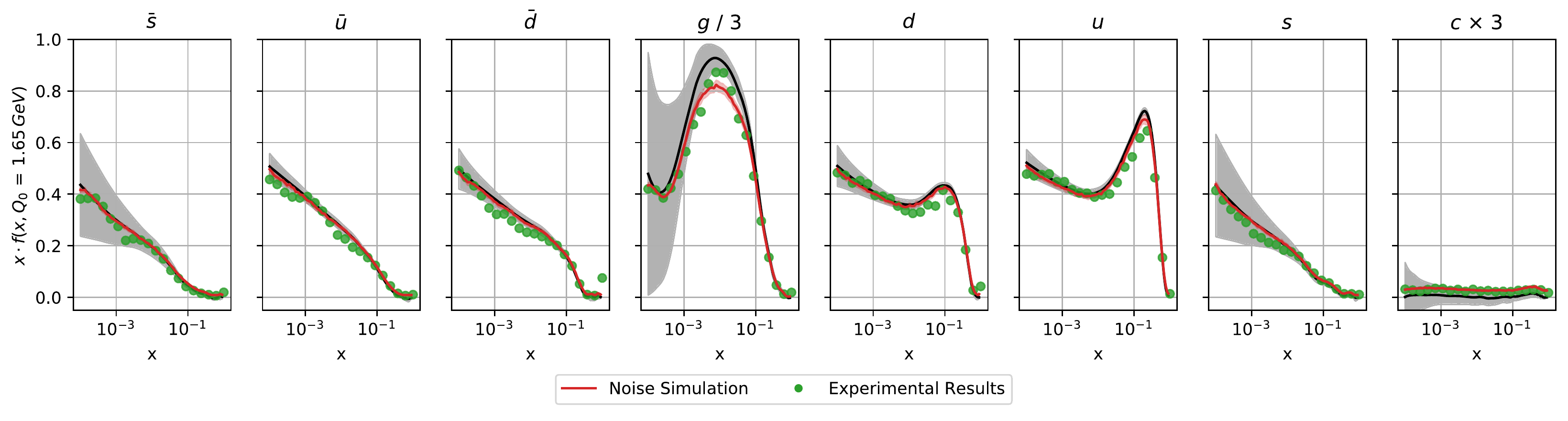}}
  \subfigure[\hspace{5mm} Multi-flavour fits in IBM Melbourne-like simulator \label{fig:Ideal_8000}]{\includegraphics[width=\linewidth]{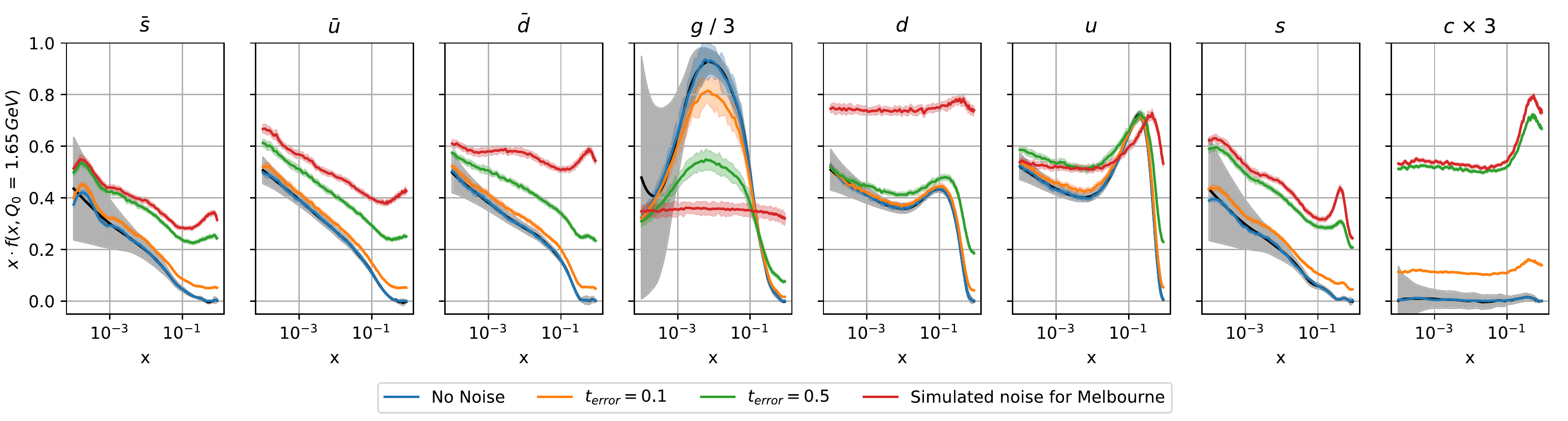}}
\end{adjustwidth}
\caption{a) Single-flavour fit for all flavours. The red lines represent
  the prediction of the qPDF model with simulated noise from the IBM Athens
  processor \cite{qiskit}. Green points are the results of running the circuit
  on the Athens quantum processor. The mean value and $1\sigma$ uncertainty of the
  target \ac{pdf} data is shown by means of a solid black line and a shaded grey area. b) Multi-flavour fit for all flavours. Blue lines are the mean and the blue shadowed area the
    $1\sigma$ uncertainty of the circuit measurement results for an
    ideal noise free quantum device. The red curve refers to simulated
    circuit measurements using the noise model for the IBM Melbourne
    processor \cite{qiskit}. Similarly, green and orange curves
    show simulation results with noise reduced by 50\% and 90\%
    respectively. The mean value and $1\sigma$ uncertainty of the
    target \ac{pdf} data is shown by means of a solid black line and a
    shaded grey area. In both cases the Weighted Ansatz, for 5
  layers and 8 qubits was used.}

\end{figure}

Low-depth quantum circuits are able to represent a full set of \ac{pdf}s when the calculations are carried by means of classical simulation. However, in the results presented up to this moment, no measurement uncertainty or noisy executions of quantum circuits have been taken into account. In this section it is explored how the capabilities are transferred the theoretical models to realistic quantum computers. First, the trained single-flavour model onto the IBM Athens quantum processor \cite{qiskit} is loaded to check how much the noise degrades the final results. The resilience of ths quantum model in the single-flavour scenario is expected to be larger than in the multi-flavour case because of the absence of entangling gates. Gate fidelities for two-qubit gates are around an order og magnitud worse than for single-qubit operations. 

For experimental results, each parton is evaluated at 20 values of $x$ logarithmically spaced in $x\in [10^{-4}, 1]$. The expecation value $z_i(x, \Theta)$ is evaluated for every $x$ with $2^{13}=8192$ shots. Then, each evaluation is repeated $5$ times to probe statistical averages and uncertainties in estimation. See Fig.~\ref{fig:SingleFlavorExperiment} for a comparison between experimental results in the IBM Athens quantum and its corresponding noisy simulation as provided by {\tt qiskit}. From those results it is possible to claim that single-flavor models with only one qubit perform properly on cloud-accessible quantum processors, and it is then possible to extrapolate this model to actual machines. In addition, the agreement between experimental results and simulation credits the simulation environment to properly simulate the real situation. 

The next step is to extend the experimental implementation to multi-flavor models. Quantum computers required for theses models must have several qubits and be able to execute two-qubit gates. It is expected that decoherence play an important role in this case. It is benchmarked by executing the optimized model on the {\tt qiskit} simulator for noisy computers, taking as noise model the one corresponding to the best possible configuration of the IBM Melbourne quantum processor. This processor was chosen since it is the only publicly available processor with enough number of qubits. The 8 qubit circuit is mapped onto Melbourne in a way that it matches the chip architecture and the entangling gates are directly applicable.

\begin{wrapfigure}{R}{.5\linewidth}
\centering
  \includegraphics[width=\linewidth]{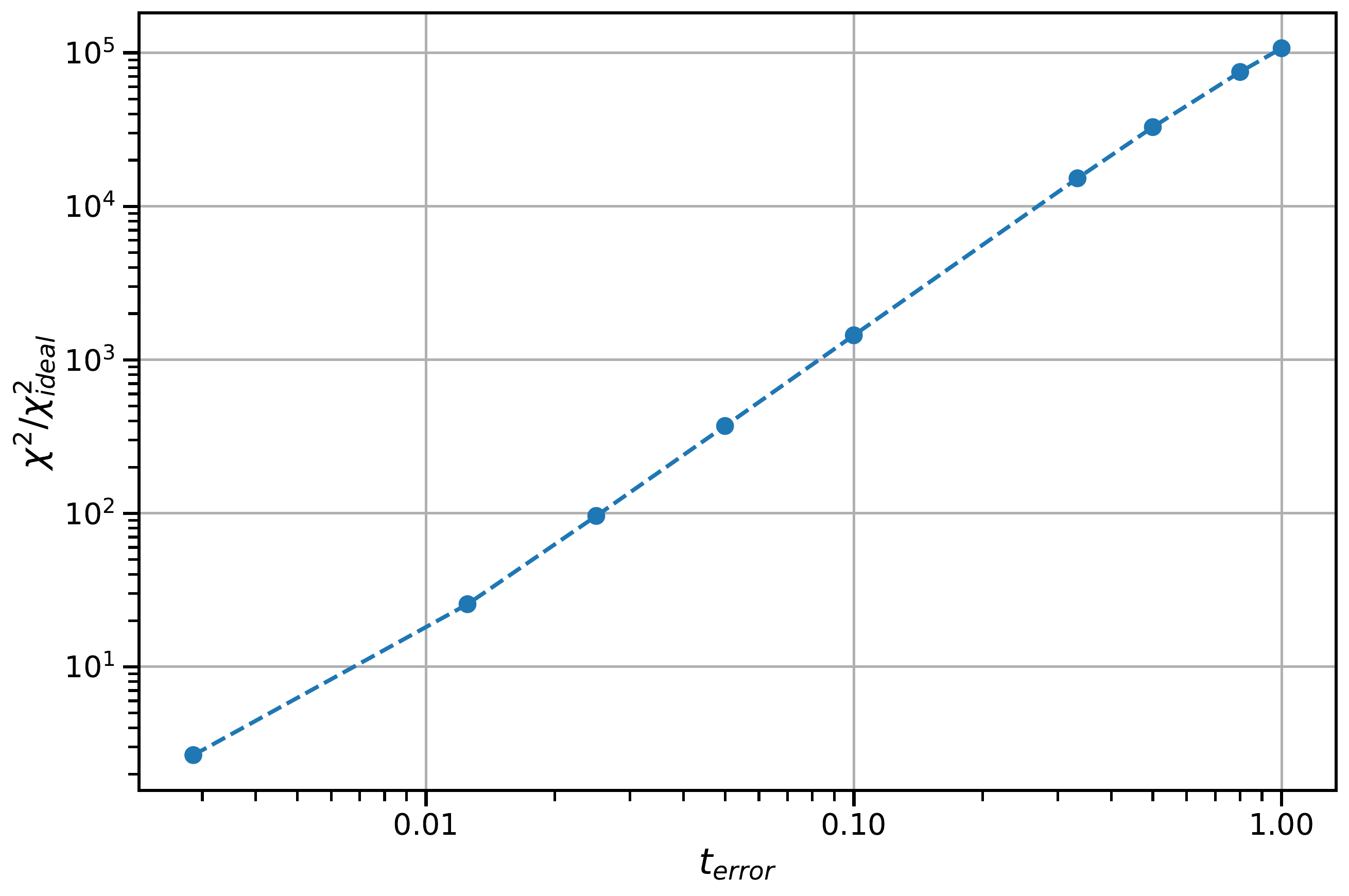}
  \caption{The error as a function of the error interpolation parameter $t_{\text{error}}$. The $y$-axis is given as the ratio between the error, $\chi^2$, and the error on an ideal quantum computer, $\chi_{\text{ideal}}^2$.}
  \label{fig:errorScaling}
\end{wrapfigure}The first step of the multi-flavor execution on a virtual quantum computer is to add measurement gates. In this case, the circuit must be run 8 times, one per flavor, to measure each qubit independently from the other ones. As in the single-flavor scenario, each qubit is measured 8192 times, a sufficient number to estimate the expected value of the hamiltonians of interest with low uncertainty, and thus reconstruct the \ac{pdf} accurately. It is immediately seen that the errors present in the IBM Melbourne device drastically deteriorate the performance of the pre-trained quantum model, see Fig.~\ref{fig:Ideal_8000}. To understand better the effect of noise, the noise environment is used to create simulators with the same noise structure as IBM Melbourne but smaller noise values. Thus, a noise model $\mathcal{N} = \mathcal{N}_{\rm Melbourne} t$ is available, where $\mathcal{N}_{\rm Melbourne}$ is the noise model provided by {\tt qiskit} and $t$ is an interpolation parameter. This way, the noise model can be linearly scaled down while maintaining all the characteristics, namely connections among qubits, single-, two-qubit and readout errors and thermalization. For $t = 0$, no noise is considered, while for $t = 1$, the full device is simulated. Results for $t = \{0, 0.1, 0.5, 1\}$ are depicted in Fig~\ref{fig:Ideal_8000}.

A summary of the obtained results is also depicted in Fig.~\ref{fig:errorScaling}. Here, the relation between the obtained $\chi^2$ and the ideal one is explored as a function of the error parameter $t$. The aim is to explore how robust must a quantum computer be in order fo return acceptable representations of qPDF. It is possible to see that a $\chi^2$ an order of magnitude larger than the ideal one is achieved at (extrapolated) value $t \sim 0.007$. 

The analysis shows that even though it is theoretically possible to fit \ac{pdf} with the qPDF model here proposed, as it was demonstrated by means of classical simulations, the noise and decoherence in state-of-the-art quantum devices are still to high to provide accurate computation frameworks. 
 
\subsection{qPDF determination from experimental data}\label{ssec:lhc_data}
In previous sections the process of finding a quantum circuit capable to capture the properties of physical \ac{pdf}s by mimicking classically known results is described. Furthermore, the possibility to extend theoretical models to experimental devices is also explored. In this section the last step of the workflow is addressed, see Fig.~\ref{fig:workflow}, andthe qPDF model is used to learn \ac{pdf}s from the only available dataset in reality, that is experimental measurements of physical observables, in this case physical cross sections measured at the \ac{lhc}. 

In this stage it is proven that the qPDF methodology has the potential to replace \ac{nn}s underlying at the core of the NNPDF family of proceedings to learn \ac{pdf}s from experimental data classically. Current quantum devices are far from supplying enough computational capability to tackle this problem in practice. However, classical simulation shows that the data re-uploading strategy can indeed replace \ac{nn}s as a universal function approximation of arbitrary functions, such as \ac{pdf}s, at least from a theoretical perspective.

This section describes the NNPDF methodology and the changes needed to incorporate the qPDF model to the general framework to perferm a full fit. The dataset used to fit \ac{pdf}s is the NNPDF3.1, including deep inelastic scattering and hadronic collider data. Finally, the obtained \ac{pdf}s and qPDF are compared showing that results are compatible and usable in realistic computation of physical observables. 

\subsubsection{The NNPDF methodology}

NNPDF methodology is based on two main ingredients. First, a Monte Carlo approach to synthetic generation of artificial measurements is required. Second, \ac{nn}s are used to model \ac{pdf}s. In the following some main aspects are outlined, and the reader is referred to a more in-depth review for further details \cite{Ball:2014uwa}.

First, data replicas must be generated. This procedure propagates experimental uncertainties through the \ac{pdf} fit by leveraging experimental uncertainties obtained from experiments. Synthetic copies of data are then created and indistinguishable from actual data. 

The \ac{pdf} fit is done following the functional form
\begin{equation}\label{eq:nnpdf}
f_i(x, Q_0) = x^{-\alpha_i} (1 - x)^{\beta_i}NN_i(x, Q_0),
\end{equation}
where $i$ is the parton of interest and $NN(\cdot)$ is the function provided by the \ac{nn}. The preprocessing factors $x^{-\alpha_i}, (1 - x)^{\beta_i}$ guarantee a correct behavior for $x \approx \{0, 1\}$, where there could be a lack of experimental data to properly constraint the \ac{nn}. This function cannot be directly compared to experimental data, it must be convoluted with the partonic cross rection to obtain physical predictions comparable to measurable observables, 
\begin{equation}\label{eq:convolution}
    P = \displaystyle\int \dd{x_{1}}\dd{x_{2}}  f^{i}_{1}(x_{1}, q^{2})f^{j}_{2}(x_{2}, q^{2})|M_{ij}(\{p_{n}\})|^{2}, 
\end{equation}
where $x_1, x_2$ are momentum fractions of two particles, and $\{i, j\}$ run over all possible partons. The quantity $M_{ij}$ is the matrix element for the given processes, computed analytically with other methods, and $\{p_n\}$ represent all momenta involved in the computation. Nevertheless, a numerical integration of this quantity would be impractible. Instead, theoretical predictions are approximated as products between the \ac{pdf} functions and a fastkernel table with the relevant information \cite{Ball:2010de,Bertone:2016lga}. 

The optimization process consists in minimizing the Pearson's $\chi^2$ defined as
\begin{equation}
    \chi^2 = \sum_{i,j}^{N_{\rm dat}} (D-P)_i \sigma_{ij}^{-1} (D-P)_j, \label{eq:chi2}
\end{equation}
where $D_{i}$ and $P_{i}$ are respectively the $i$-nth data point from the training set and
its theoretical prediction and $\sigma_{ij}$ is the experimental covariance matrix.

This procedure is then repeated for each synthetic replica. In all cases, only the experimental data changes and must be calculated in every step. The final \ac{pdf} is the average among all replicas and the error bands are given by enveloping $68\%$ of the replicas, that is the number associated to a $1\sigma$ width in a normal distribution.

The latest NNPDF methodology, NNPDF3.1, as described in Ref. \cite{Carrazza:2019mzf} is taken. The \ac{nn} module is replaced with the qPDF model. Since both {\tt qibo} and NNPDF3.1 are based on {\tt tensorflow}, integration can be accomplished without adapting the programming language. A number of changes must be done in order to make the integration complete. The dataset included in this fit correspond to that of NNPDF3.1, which is detailed in
Ref.~\cite{Ball:2017nwa} and
includes data from deep-inelastic scattering experiments, fixed-target data
and hadronic collider data from experiments at Tevatron and \ac{lhc}. Technical details can be found in App.~\ref{app:nnpdf}.

\subsubsection{Experimental qPDF results}

\begin{figure}[t]
\begin{adjustwidth}{-1cm}{-2cm}
    \centering
    \subfigure[\ Gluon \ac{pdf}.]{
        \includegraphics[width=0.31\linewidth]{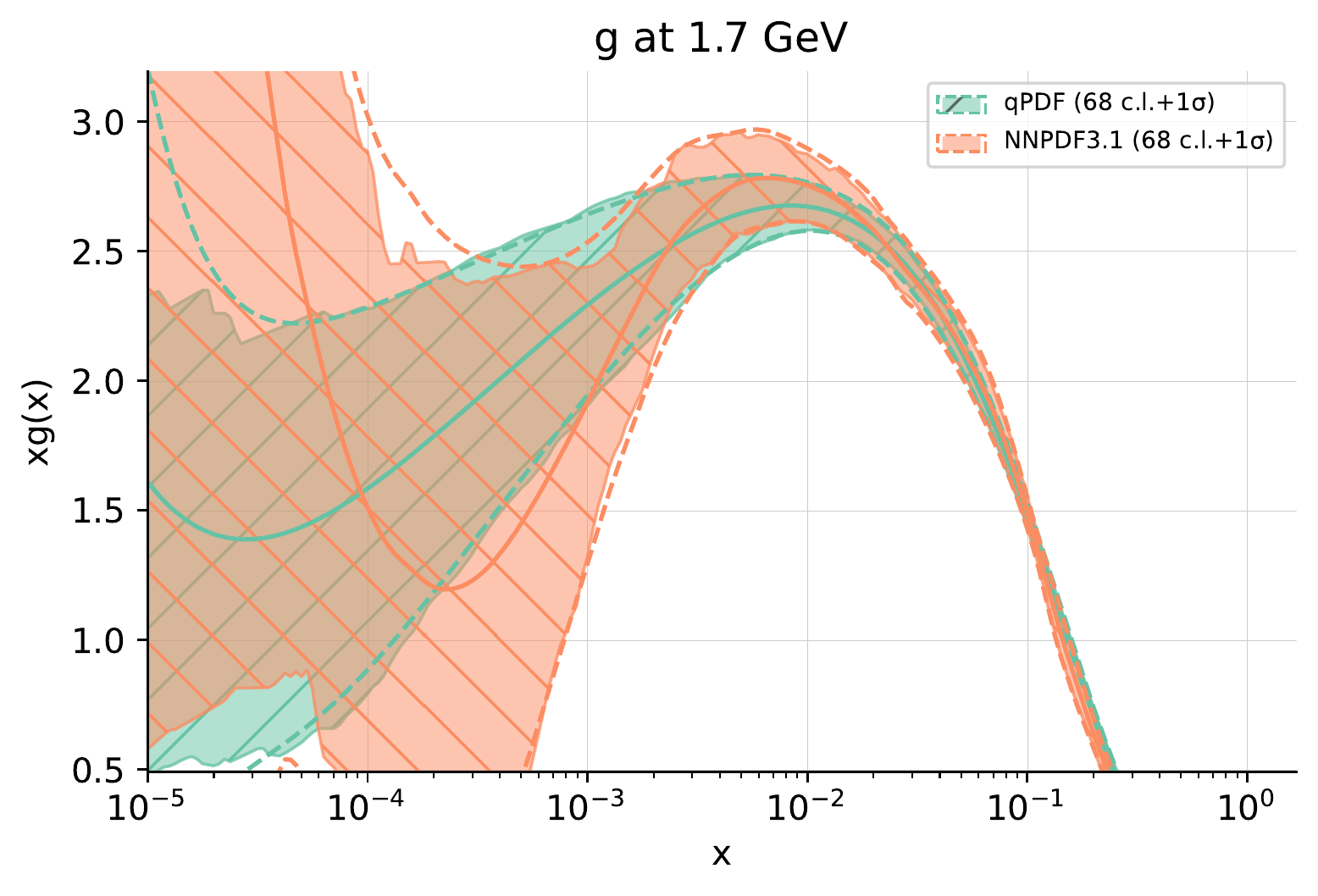}
    }
    \subfigure[\ u quark \ac{pdf}.]{
        \includegraphics[width=0.31\linewidth]{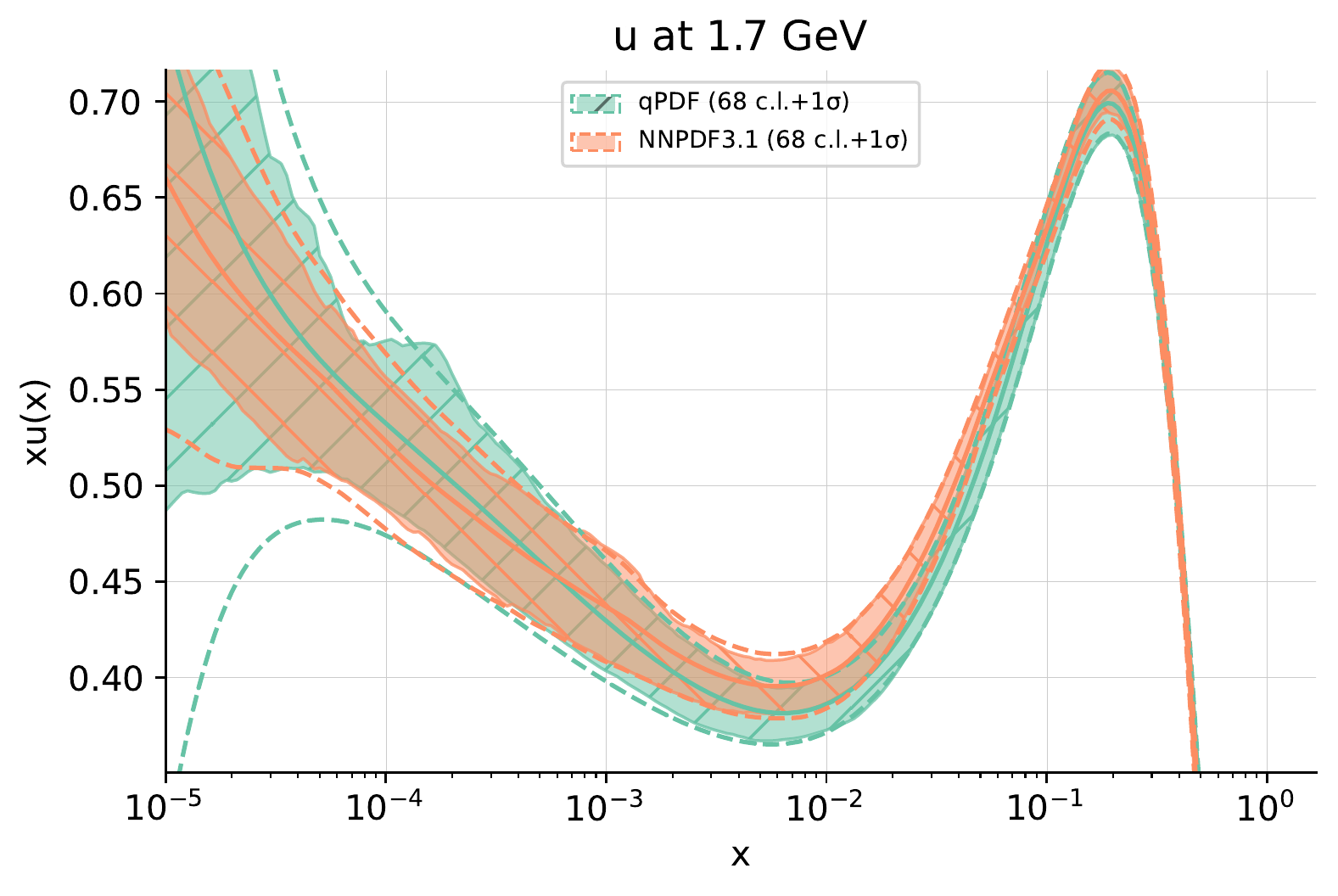}
    }
    \subfigure[\ s quark \ac{pdf}.]{
        \includegraphics[width=0.31\linewidth]{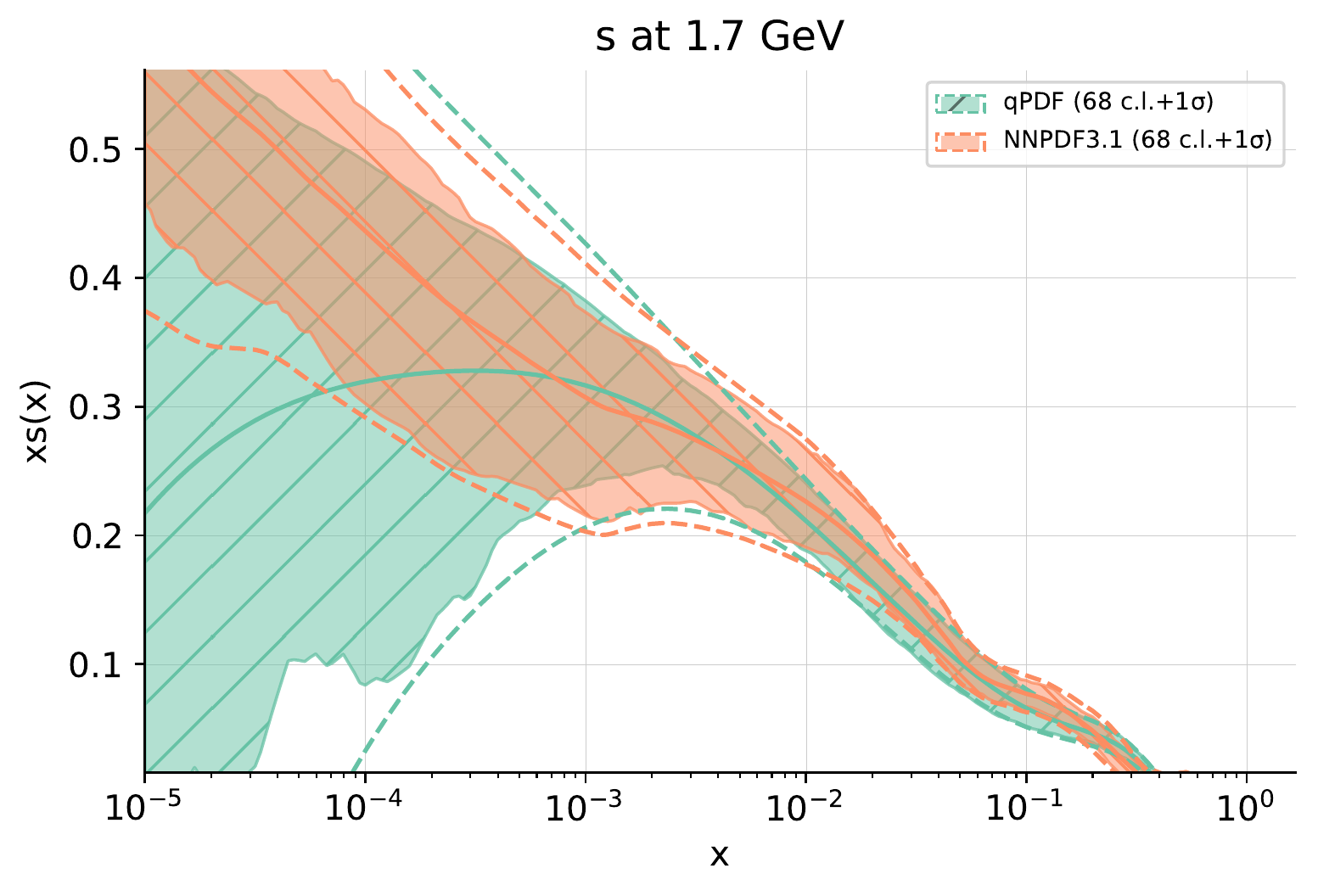}
    }
\end{adjustwidth}
    \caption{Fit results for the gluon and the $u$ and $s$ quarks. As previously seen in
    Fig.~\ref{fig:all_flavours}, qPDF is able to reproduce the features of NNPDF3.1. This is also true when
the fit performed by comparing to data and not by comparing directly to the goal function. The differences seen at
low-x can be attributed to the lack of data in that region.}\label{fig:fitperflavour}
\end{figure}

We compare the published reference \ac{pdf}s and their uncertainties to the qPDF results obtaining with this method to check that both results are compatible. Quarks $u$ and $d$, and gluon $g$ are depicted in Fig.~\ref{fig:fitperflavour}. For these flavors, the qPDF central result is within the $1-\sigma$ uncertainty bar of the reference classical \ac{pdf}, and both classical and quantum uncertainty bars have strong overlaps for the considered range. 

Phenomenological implications must be addressed to benchmark the performance of the qPDF model. Examples related to the most common Higgs production channels with $m_H=125$ GeV are summarized in Tab.~\ref{tab:higgs}. In this table, cross sections are depicted for different channels and compared to reference values \cite{Carrazza:2020gss,christopher_schwan_2020_3992765,Alwall:2014hca}. For all examples, reference values and qPDF results are compatible within uncertainty ranges. 

\begin{table}[t]
\centering
  \begin{tabular}{c||cc}
  \hline
  Channel & NNPDF3.1 NNLO  & qPDF \tabularnewline
  \hline
  \hline
  $ggH$ & $31.04 \pm 0.30$ pb & $31.71 \pm 0.51$ pb \tabularnewline
  \hline
  $t\bar{t}H$ & $0.446 \pm 0.003$ pb & $0.464 \pm 0.008$ pb \tabularnewline
  \hline
  $WH$ & $0.133 \pm 0.002$ pb & $0.135 \pm 0.002$ pb \tabularnewline
  \hline
  $ZH$ & $0.0181 \pm 0.0002$ pb & $0.0184 \pm 0.0002$ pb \tabularnewline
  \hline
  VBF & $2.55 \pm 0.03$ pb & $2.62 \pm 0.04$ pb \tabularnewline
  \hline
  \end{tabular}
  \caption{\label{tab:higgs}The cross-sections for Higgs production at 13 TeV in
  various channels at NLO using the settings described in the text. From top to
  bottom: gluon fusion, $t\bar{t}H$ production, $WH$ production, $ZH$ production
  and vector boson fusion. Standard Model Higgs boson with
  mass $m_H=125$ GeV is assumed.}
\end{table}

These results support the claim that the NNPDF methodology where \ac{nn}s are replaced with quantum circuits can be used for regression problems to unknown functional forms, in particular the \ac{pdf}s and inner structure of the proton. Classical state-of-the-art and quantum results are coherent from a phenomenological point of view. In addition, it is reasonable that the same level of accuracy from classical methods can be reached with adequate fine-tuning. 

\subsection{Discussion}\label{ssec:coclusions_proton}
It is possible to highlight certain advantages that the quantum model here proposed has in comparison to standard \ac{ml} methodology for fitting \ac{pdf}s. First, non-linearities emerging from quantum operations and entanglement help to reduce the number of parameters required to obtain a flexible \ac{pdf} representation, as compared to equivalent \ac{nn}s. Second, from a hardware perspective, the implementation of a qPDF fit on a quantum processor and the use of native gates as operations can accelerate the evaluation and training of \ac{pdf}s. In addition, it is expected that the energy consumption of quantum devices is smaller than of classical hardware based on hardware accelerators such as graphical processing units. 

However, the current status of quantum devices does not provide enough quality to implement this model. All typical difficulties from experimental quantum hardware appear in this problem, including noise, decoherence and measurement uncertainty. More accurate or even fault tolerant quantum computing will be necessary for succesfully carrying this method on quantum devices. 

On the other hand, the results here presented should be considered a proof-of-concept. Standard machine learning implementations are so optimized that no quantum simulation algorithm is competitive in terms of performance nor efficiency against classical methods. The advantage of qPDFs will come as the quantum hardware becomes more precise. 

The work here presented is a first attempt to join quantum machine learning into the field of \ac{pdf} determination. This approach can open a new surge of algorithms for the field of \ac{hep}, which can benefit from quantum computing.

\section{Conclusions}

Throughout this chapter the re-uploading strategy is developed since its inception to several examples of applications. All the contents here covered support the claim that data re-uploading is a general strategy to employ in \ac{qml} problems and to apply in the first generation of NISQ devices, including some already existing processors. 

Re-uploading strategy is a general approach whose main focus is to replace a \ac{nn} with some quantum circuit. The circuit can be as small as one qubit. The key ingredient is to upload data several times in subsequent operations applied to the circuit to permit multiple processing of data. As compared to \ac{nn}s, each operation receiving data is the quantum counterpart of a neuron. For single-qubit circuits, the comparison to single-hidden-layer \ac{nn}s is direct. In fact, it is mathematically proven that both models are formally equivalent. Even though the relationship between deep learning classical approaches and multi-qubit re-uploading circuits is not clear, it is expected that steps are taken in this direction in the future. 

Data is uploaded to the circuit following the linear mapping used in \ac{nn}s, that is as $\vec x \rightarrow \vec w \cdot \vec x + \theta$, with $\vec w$ the weight and $\theta$ the bias. This encoding allows to encode data in an unbiased manner, thus it is possible to carry supervised learning problems without any prior knowledge of the dataset to classify. 

The cornerstone of the re-uploading strategy is that non-linearities emerge naturally from the quantum properties of the circuits. By applying two gates around different axis, one of them depending linearly on $x$, non-linear terms arise as a consequence of the non-commutativity of quantum operations. Quantum circuits are capable to represent any functional form thanks to this inherently quantum property. 

Data re-uploading is also a strategy to circumvent the no-cloning theorem. In this approach, classical data is introduced several times in the circuit, but the copy of data is not performed using quantum, but classical resources. It is then possible to increase the number of calls from the circuit to the data, unlike in most \ac{qml} approaches where data is introduced at the beginning of the algorithm. 

The performance of the re-uploading strategy is directly related to the query complexity of the circuit, that is the number of re-uploading along the algorithm. In fact, this quantity remains approximately constant for similar performances, while the number of qubits or depth in circuits are more unstable. Entanglement plays a role in the performance, but the improvement attained is not as significant as expected. 

Present results give support to different \ac{qml} applications. The first one is regression. It is shown in Sec.~\ref{sec:benchmark} that the quantum algorithm is capable to learn arbitrary functional forms from sampling data. Sections~\ref{sec:qlassifier} and~\ref{sec:exp_qlassifier} address supervised classification tasks successfully. In both cases, both numerical and experimental benchmarks are performed. Also in both cases, the training step is mostly carried on classical simulations of quantum systems. The reason to proceed in this way is that optimization on quantum devices is much harder than on classical devices. Thus, actual quantum optimization is a task left for future extensions of the present work.

In Sec.~\ref{sec:qpdf}, a real-world problem is addressed with the re-uploading strategy. The problem is determining the proton content from experimental data of \ac{hep}. This application, although only classically simulated, presents a strong evidence that re-uploading approaches can actually deal with real problems. Current status of experimental devices does not endure running running this kind of algorithms due to the yet persistent quantum noise and decoherence.  

It is expected that future improvements in the device quality and optimization methods will help the data re-uploading strategy to enhance its range of applicability and performance, and it will hopefully be an interesting ingredient in the development of \ac{qml}.

\cleardoublepage

%auto-ignore
\chapterimage{chapter_unary.pdf}

\def\la{{\langle}}
\def\ra{{\rangle}}

\newcommand{\A}{\mathcal{A}}
\newcommand{\D}{\mathcal{D}}
\newcommand{\C}{\mathcal{C}}
\newcommand{\R}{\mathcal{R}}
\newcommand{\Q}{\mathcal{Q}}
\renewcommand{\S}{\mathcal{S}}

\chapter{Unary strategy for finance}\label{ch:unary}

\begin{adjustwidth}{4cm}{0cm}
{\sl Más vale el buen nombre que las muchas riquezas.}

\hfill Miguel de Cervantes\\
\end{adjustwidth}

Quantum computing posseses the inherent property of entanglement. Entanglement is the reason
for the exponential size of the Hilbert space with respect to the number of qubits. Entanglement is also a key feature in the speed up achieved in the most prominent quantum algorithms \cite{shor, grover}. However, there is another application available for entanglement, that is effectively distributing the information of a small quantum system across a larger one. The reason is that this procedure allows to store the information as a global and share property of many small quantum system, making it resilient against errors and decoherence. For example, this strategy lies at the core of quantum error correction codes \cite{shor_scheme_1995, cory_experimental_1998, gottesman_stabilizer_1997}. 

In this chapter an application of this line of thought is explored, looking for computing financial products using a near-term \ac{nisq} device. The main difference between the algorithm here proposed and other quantum algorithms, in particular for finance, is that the {\bf unary representation} is used, that is, the Hilbert space is restricted to those components of the computational basis with only one $\ket 1$ qubit, and $\ket 0$ for all other qubits. Previous works considering the unary representation can be found in Refs.~\cite{poulin_quantum_2018,babbush_encoding_2018,steudtner_estimating_2020}.

The unary basis has some advantages and inconvenients with respect to the standard binary computations. It is clear that the unary basis does not have the capabilities to store an exponential number of coefficients with respect to the number of qubits, but rather a linear one. Thus, a worse asymptotic scaling is retrieved in the unary case. However, this permits to simplify logical operations carried on those quantum states and leads to lesser gates to be applied on a given circuit. In addition, the unary representation brings a native post-selection strategy that results in error mitigation. It is likely that all algorithms to be executed \ac{nisq} must perform some error mitigation technique to obtain proper results. 

The properties of the unary basis make it useful in a near-term regime, both with respect to the size of the problem and the quality of the quantum computer. In case the size of the problem increases, the overhead of the standard binary representation with respect to the unary one is compensated by the asymptotic behavior. In terms of resilience against error, the unary basis can provide better results as well. Quantum advantage is feasible for small problems, and it is also possible to find interesting problems where this small size is useful.

The benefits of the unary representation are explored in the field of quantitative finances. This field is expected to be transformed with the bloom of the first generation of quantum computers, see Ref. \cite{qfinance-orus2019}. In recent years there has been a surge of methods and algorithms for solving financial problems using quantum computers \cite{crashes-orus2019, derivatives-martin2019, credit-egger2019} , in particular for hard optimization problems \cite{portfolio-kerenidis2019,optimization-rosenberg2016, portfolio-rebentrost2018, optimization-moll2018, optimization-lopez2015}.

The prominent problem of pricing financial derivatives is taken into consideration. Many computational obstacles of this problem can be overcome by quantum computing, in particular, the pricing of european options. Options are contracts that allow the holder to buy / sell some asset at a pre-established time at a future date. The problem is then to estimate if the price of the given option will increase or not with respect to the agreed value. The evolution of the asset price follows a stochastic process described by the Black-Scholes model \cite{black_blackscholes_1973}. Then, a payoff function, specified by contract as well, must be incorporated to this evolution to obtain the expected return of the option. The main method to classically perform this computation is the costly Monte Carlo simulation. 

Quantum algorithms have been already proposed to solve the option pricing problem more efficiently that their classical counterparts \cite{stamatopoulos_binary_2020, rebentrost_quantum_2018, woerner_quantum_2019}. The key ingredient to develop this algorithm is that quantum computers provide a quadratic speed-up in the number of evaluations required to obtain a given accuracy using a Monte Carlo simulation. This exploits the idea of \ac{qae} \cite{brassard_amplitude_estimation_2002, aaronson_counting_2020, montanaro_montecarlo_2015, grinko_iqae_2021}. The quantum advantage achieved via \acf{qae} holds only if there exists an efficient way of loading of data, namely a distribution in the asset prices, into the quantum circuit. With this purpose, \ac{qgan}s \cite{qGAN-lloyd2018, qGAN-dallaire2018} have been analyzed to address this issue \cite{qGAN-zoufal2019}.

The algorithm here proposed to solve the option pricing problem is divided in three steps. First, a circuit working on the unary basis of the asset prices is constructed. The evolution of the asset price is computed using an amplitude distributor such that the output state corresponds to a probability distribution of prices. Second, the payoff is computed using quantum gates. This steps greatly simplifies thanks to the unary representation. Finally, a \ac{qae} procedure is carried to obtain quantum advantage. This last step is common to previous approaches.

The option pricing problem suits the requirements for a unary basis approach since a great accuracy is not needed to obtain results of interests. The estimates for the number of gates indicate that the crossing point between the unary and the binary algorithm is located at least in a number of qubits rendering a good precision, $< 1\%$. This rate matches the usual precision obtained in real-world applications. It is worth mentioning that this estimation relies on machines suiting perfectly the needs of this algorithm, but real hardware architecture can lean the scale towards the unary approach. 

This chapter is organized as follows. First, the unary representation and its corresponding operations are presented and defined in Sec.~\ref{sec:unary}. Sec.~\ref{sec:background} covers the required background to present the financial problem here addressed. The application of the unary representation to finance is depicted in Sec.~\ref{sec:unary_algorithm}. A comparison between the previous binary algorithm and the unary one in this chapter is performed in Sec.~\ref{sec:comparison}, and the corresponding results are depicted in Sec.~\ref{sec:results_unary}. Final remarks can be read in Sec.~\ref{sec:conclusions_unary}.

\section{Background}\label{sec:background}

The unary algorithm here presented is constructed upon three different legs coming from different fields. They are the economical Black-Scholes model as applied for European options, the \ac{qae} procedure providing quantum advantage respect to standard Monte Carlo, and a quantum algorithm developed in the standard binary representation to compute the payoff of an European option pricing, as described in Ref. \cite{stamatopoulos_binary_2020}.

\subsection{European options and the Black-Scholes model}\label{ssec:black_scholes}
In the market of financial derivative, options are contracts signed to acquire the right to buy/sell ({\sl call/put}) some asset at a previously established price ({\sl strike}). The contract expires at a future point in time ({\sl maturity date}). The holder of the contract will only execute the call/put option if the actual price is lower/higher than the agreed strike, and thus some benefit is obtained from the trading. 

To decide whether an option is profitable at the time of the contract sign, the holder must estimate the expected payoff of the option. This quantity will depend on the evolution of the price of the asset, which follows a stochastic process. A simple, yet successful model for option pricing is the Black-Scholes model \cite{black_blackscholes_1973}, to be detailed shortly. The second step consists in computing the contract specified payoff function over the Black-Scholes distribution of prices to obtain the expected return. In particular, European options only allow to call/put the asset exactly at the maturity date. This is then the only data of interest. In contradistinction, other options in the market, for instance the Asiatic or American options rely on more sophisticated payoff functions. 

The Black-Scholes model for the evolution of an asset price is described by the stochastic equation
\begin{equation}\label{eq:BSM}
    {\rm d}S_T = S_T\, r\, {\rm d}T + S_T\, \sigma\, {\rm d}W_T, 
\end{equation}
where $T$ is the time and $S_T$ is the price at time $T$, $r$ is the interest rate, $\sigma$ is the volatility and $W_T$ describes a Brownian process. Brownian processes $W_T$ are continuous stochastic evolutions starting at $W_0=0$ and consisting of independent gaussian increments. Specifically, let $\mathcal{N}(\mu, \sigma_s)$ be a normal distribution of mean $\mu$ and standard deviation $\sigma_s$. Then, the increment related to two steps $T, S$ of the Brownian processes is $W_T - W_S \sim \mathcal{N}(0, T - S)$, for $T > S$.

The most important feature for the Black-Scholes model is that there exists a first-order-approximate analytical solution to Eq.~\eqref{eq:BSM}. The solution reads
\begin{equation}\label{eq:log_normal}
    S_T = S_0 e^{(r - \frac{\sigma^2}{2}) T} e^{\sigma W_T}\;\sim\; e^{\mathcal{N}\left(\left(r - \frac{\sigma^2}{2}\right) T, \sigma \sqrt{T}\right)},
\end{equation}
corresponding to a log-normal distribution. A more detailed description of the procedure to solve this model is outlined in App.~\ref{app:black_scholes}.

The expected return of an option is computed by integrating the payoff function over the price probability distribution. This step is usually carried by means of a Monte Carlo simulation. Depending on the option, this procedure can be extremely costly. 

For European options, the payoff function is computed as
\begin{equation}\label{eq:payoff}
f(S_T, K) = {\rm max}(0, S_T - K), 
\end{equation}
and thus the expected payoff is
\begin{equation}\label{eq:avg_payoff}
    C(S_T, K) = \int_K^\infty \left( S_T - K \right)\,dS_T,
\end{equation}
where $K$ is the agreed strike price. Since European options are executable only at the maturity date, the expected payoff can be computed only by integrating at time $T$. In contradistinction, Asiatic options return an average over time. American options allow to execute the contract at any point before the maturity date, thus the calculation of the expected payoff requires more sophisticated and costlier methods \cite{Kemna_pricing_1990}.

%{\color{blue} Our algorithm ... // Last paragraph II.A}

\subsection{Quantum Amplitude Estimation}\label{ssec:amplitude_estimation}
\ac{qae} is an inherently quantum technique designed to estimate the probability of measuring a certain outcome from a given state more efficiently as direct sampling. For a given precision, the number of function calls is quadratically reduced with respect to direct sampling  \cite{brassard_amplitude_estimation_2002, amplitude_estimation-suzuki2020}. This technique lies at the core of the quantum advantage obtained for Monte Carlo simulations \cite{montanaro_montecarlo_2015}. \ac{qae} extends the ideas developed by Grover algorithm \cite{grover}. 

The \ac{qae} procedure considers an algorithm $\mathcal{A}$ such that
\begin{equation}
    \A \ket 0_n \ket 0 = \sqrt{1 - a} \ket{\psi_0}_n\ket{0} + \sqrt{a} \ket{\psi_1}_n\ket{1}, 
\end{equation}
where the last qubit serves as a flag qubit separating {\sl good}($\ket 0$) and {\sl bad}($\ket 1$) outcomes. The states $\ket{\psi_{0,1}}_n$ can be non-orthogonal. The final state can be sampled $N$ times to obtain an estimate $\bar a$ with an accuracy
\begin{equation}
    |a - \bar a| \sim \mathcal{O}(N^{-1/2}),
\end{equation}
as dictated by the sampling error of a multinomial distribution.

For implementing \ac{qae} it is required to construct the operator 
\begin{equation}
    \Q = - \A \S_0 \A^\dagger \S_{\psi_0},
\end{equation}
where the operators $\S_0$ and $\S_{\psi_0}$ are inherited from Grover,
\begin{eqnarray}
    \S_0 & = & \mathbf{I} - 2 \ket 0_n \bra 0_n \otimes \ket 0 \bra 0 , \\ 
    \S_{\psi_0} & = & \mathbf{I} - 2 \ket{\psi_0}_n\bra{\psi_0}_n \otimes \ket 0 \bra 0.
\end{eqnarray}
The $\S_0$ operator changes the sign of the $\ket 0_n \ket 0$ state, while $\S_{\psi_0}$ takes the role of an oracle and changes the sign of all bad outcomes. 

It turns out that the eigenvalues of $\Q$ are $e^{\pm i 2 \theta_a}$, where $a = \sin^2(\theta_a / 2)$. The original algorithm makes use of the \ac{qpe} algorithm \cite{nielsen_chuang_2010}
to obtain the numerical values of these eigenvalues \cite{brassard_amplitude_estimation_2002}. The accuracy obtained with $N$ function calls is given by
\begin{equation}
 |a - \bar a| \sim \mathcal O (N^{-1})
\end{equation}
with probability at least $8/\pi^2\approx 81\%$. This approach will only be useful in the case fault tolerant computers are available. The high quality of the hardware required to perform Quantum Phase Estimation successfully prevents to use it for \ac{qae}. Further details are available in App.~\ref{app:ae_qpe}.

In order to anticipate the use of \ac{qae} techniques to the \ac{nisq} era, some approaches have emerged without the Quantum Phase Estimation requirement \cite{amplitude_estimation-suzuki2020, grinko_iqae_2021}. These examples are less resource-demanding. The key property that allows to circumvent the use of Quantum Phase Estimation is 
\begin{equation}\label{eq:q_j}
\begin{split}
    \Q^{m} \A \ket 0 = \cos\left((2 m + 1)\theta_a\right)\ket{\psi_0}_n \ket{0} + \\ + \sin\left((2 m + 1)\theta_a\right)\ket{\psi_1}_n \ket{1}.
\end{split}
\end{equation}
An integer $m$ is chosen to prepare the state here described and measure the outcome with $N$ shots, so that the value $\sin^2\left((2 m + 1)\theta_a\right)$ is estimated with a precision of $\mathcal{O}(N^{-1/2})$. The process is repeated several times with different values of $m$ defined by a set of $\{m_j\}$. Finally, all measurements are combined to extract a final estimation whose precision is bounded by $\sim N^{-1/2}M^{-1}$, with $M=\sum_{j=0}^J m_j$, where $J$ is the last index. The exact scaling of the precision depends on the choice of $\{m_j\}$. 
Further details on the iterative method are outlined in App.~\ref{app:iqae}.

\subsection{Binary algorithm}\label{ssec:binary_algorithm}

\begin{figure}
\hspace{2cm} 
\Qcircuit @C=0.85em @R=.3em @!R{
\lstick{\ket{q_0}} & \qw & \multigate{2}{\D} & \multigate{6}{\C} & \multigate{2}{\R} & \qw & \qw & \multigate{7}{\Q(\D, \C, \R)} & \ustick{\hspace{3mm} m} \qw & \qw \\
\lstick{\ket{q_1}} & \qw & \ghost{\D} & \ghost{\C} & \ghost{\R} & \qw & \qw & \ghost{\Q(\D, \C, \R)} & \qw & \qw \\
\lstick{\ket{q_2}} & \qw & \ghost{\D} & \ghost{\C} & \ghost{\R} & \qw & \qw & \ghost{\Q(\D, \C, \R)} & \qw & \qw \\
\lstick{\ket{c_0}} & \qw & \qw & \ghost{\C} & \qw & \qw & \qw & \ghost{\Q(\D, \C, \R)} & \qw & \qw \\
\lstick{\ket{c_1}} & \qw & \qw & \ghost{\C} & \qw & \qw & \qw & \ghost{\Q(\D, \C, \R)} & \qw & \qw \\
\lstick{\ket{c_2}} & \qw & \qw & \ghost{\C} & \qw & \qw & \qw & \ghost{\Q(\D, \C, \R)} & \qw & \qw \\
\lstick{\ket{b}} & \qw & \qw & \ghost{\C} & \sgate{\R}{-4} & \qw & \qw & \ghost{\Q(\D, \C, \R)} & \qw & \qw \\
\lstick{\ket{a}} & \qw & \qw & \qw & \sgate{\R}{-1} & \qw & \qw & \ghost{\Q(\D, \C, \R)} & \qw & \qw & \meter & \rstick{\rm Payoff} & 
\protect\gategroup{1}{4}{8}{5}{.7em}{--}
\protect\gategroup{1}{8}{8}{8}{1em}{(}
\protect\gategroup{1}{8}{8}{8}{1em}{)}}
\caption{Full circuit for the binary algorithm for option pricing that include all three steps, namely, the amplitude distributor $\D$, payoff estimator comprised of the comparator and payoff estimator $\C$ and $\R$ respectively, followed by components of \ac{qae}, $\Q$. The operator $\Q$ is repeated $m$ times, where $m$ depends on the \ac{qae} algorithm. The payoff is indirectly measured in the last qubit.}
\label{fig:binary_circuit}
\end{figure}
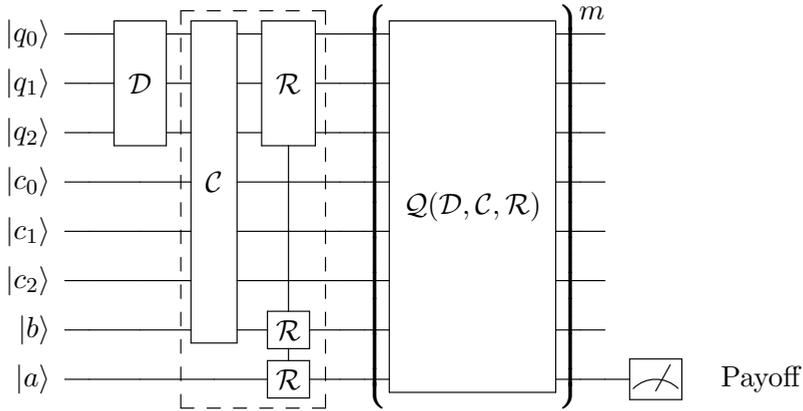

The algorithm constructed on the standard binary basis laying the groundwork for the unary algorithm here presented is introduced in Ref. \cite{stamatopoulos_binary_2020}. The algorithm is divided in three different parts. A complete scheme is depicted in Fig. \ref{fig:binary_circuit}.

{\bf Amplitude distributor:} this element encodes into the circuit the distribution of prices for a given asset with a certain interest rate and volatility. The operator representing piece is labeled as $\D$. In this algorithm, $\D$ is implemented using \ac{qgan}s \cite{qGAN-dallaire2018,qGAN-lloyd2018,qGAN-zoufal2019}. These methods demand previous knowledge on the classical solution of the Black-Scholes model from Eq.~\eqref{eq:log_normal}.

{\bf Payoff calculation:} the expected payoff is encoded into the amplitude of an ancillary qubit. For computing the payoff, it is only required to encode this information properly and then measure the ancilla qubit to retrieve the information. This piece is further subdivided in two more steps. First, a comparator $\C$ separated the prices in larger and smaller than the strike $K$. Then, a set of controlled rotations $\R$ encode the expected information into the amplitude of the ancilla qubit. 

{\bf Quantum Amplitude Estimation}: the ancilla qubit is measured using the \ac{qae} recipe to obtain an estimate of the expected payoff more efficiently than by classical methods. The operator $\Q$, which includes the $\D$ and $\R$, is applied several times. 

For further detals on the binary algorithm, the reader is referred to Refs.~\cite{stamatopoulos_binary_2020, ramos_unary_2021}.

\section{The unary representation}\label{sec:unary}

The key ingredient of the complete unary strategy is the unary represen-tation. In this representation, the Hilbert space corresponding to the computational framework is spanned by those quantum states with one qubit in the $\ket 1$ state while all other qubits remain in $\ket 0$. Thus, the general representation of a $n$-qubits quantum state in the unary basis is
\begin{multline}
\small
\ket\psi = \sum_{i=0}^{n - 1} a_i \ket{i}_n = \sum_{i=0}^{n - 1} a_i \left(\bigotimes_{j=0}^{n - 1} \ket{\delta_{i,j}} \right)  = \\ = a_{0} \ket{00\ldots 01}_n + a_{1} \ket{00\ldots 10}_n + \ldots + a_{n-2} \ket{01\ldots 00}_n + a_{n-1} \ket{10\ldots 00}_n, 
\end{multline}
where $\ket i_n$ corresponds to the $i$-th element of the unary basis and $\delta_{i,j}$ is the Kronecker delta. These states satisfy the normalization $\sum_{i=0}^{n-1}\vert \psi_i\vert^2 = 1$ as well. 

A well known example of a state in the unary representation is the $W$ state defining a three-qubit multipartite entanglement class \cite{dur_entanglement_2000}. The properties of the unary representation make it probably better suited to run on \ac{nisq} devices than the binary one. This claim will be detailed later. While it holds that exponentially more qubits are required to store equivalent quantum states, the unary approach is capable to retain more useful information and it is then much more robust against noise.

For a fixed number of qubits $n$, the unary representation only provides $n$ different states from the computational basis, as a $n$-level qudit. In this sense, adding a qubit to this system is equivalent to adding a new level to an isolated qudit. The unary representation allows for a restricted Hilbert space whose number of degrees of freedom grows linearly, unlike in the standard binary representation, where the dimensionality increases exponentially. Indeed, only $n$ out of $2^n$ states are taken into consideration, and the quantum states that belong to the unary representation are a restricted part of the total Hilbert space. 

Since the unary representation cannot expand through the whole Hilbert space, only those quantum operations compatible with this feature are applied. Such operations are entangling gates acting on the $\ket{01}, \ket{10}$ states of two qubits, namely
\begin{equation}\label{eq:U_sw_gate}
\smash{\raisebox{.75\normalbaselineskip}{
\Qcircuit @R=0.5em @C=0.3em{
& \qw & \multigate{1}{\mathcal U} & \qw \\
& \qw & \ghost{U} & \qw 
}}} = \begin{pmatrix}
1 & 0 & 0 & 0 \\ 
0 & \multicolumn{2}{c}{\multirow{2}{*}{\Huge$\mathcal U$}} & 0 \\ 
0 &   &   & 0 \\
0 & 0 & 0 & 1
\end{pmatrix}, 
\end{equation}
where $U$ is any arbitrary single-qubit operation. Implementing this operation among all possible pairs of qubits is enough to obtain arbitrary operations. In the case only gates between first neighbors are allowed, adding standard SWAP gates, namely Eq.~\eqref{eq:U_sw_gate} with $U = X$, is enough to implement any operation $U$ between any pair of qubits. This property only holds if the initial state belongs to the unary basis. Many computers start in the $\ket 0^{\otimes n}$ state, and thus adding one $X$ gate suffices to initialize the unary representation.

This problem requires gates transporting amplitude from one qubit to the neighbor one. Those gates are the equivalent to $R_y$ and $R_x$, namely partial-SWAP and partial-iSWAP gates, defined as 
\begin{figure}[h!]
\begin{eqnarray}
\label{eq:pSWAP}
\smash{\raisebox{.75\normalbaselineskip}{
\Qcircuit @R=0.7em @C=0.3em{
& \qw & \multigate{1}{\hspace{1.4cm}} & \qw \\
& \qw & \ghost{\hspace{1.4cm}} & \qw 
}}} & = & \begin{pmatrix}
1 & 0 & 0 & 0 \\ 
0 & \cos(\theta / 2) & -\sin(\theta / 2) & 0 \\ 
0 & \sin(\theta / 2)  & \cos(\theta / 2)  & 0 \\
0 & 0 & 0 & 1
\end{pmatrix}, \\ 
\label{eq:piSWAP}
\smash{\raisebox{.75\normalbaselineskip}{
\Qcircuit @R=0.7em @C=0.3em{
& \qw & \multigate{1}{\hspace{1.4cm}} & \qw \\
& \qw & \ghost{\hspace{1.4cm}} & \qw 
}}} & = &  \begin{pmatrix}
1 & 0 & 0 & 0 \\ 
0 & \cos(\theta / 2) & -i\sin(\theta / 2) & 0 \\ 
0 & -i\sin(\theta / 2)  & \cos(\theta / 2)  & 0 \\
0 & 0 & 0 & 1
\end{pmatrix},
\end{eqnarray}
\begin{textblock}{1.3}(1.4, -2.6)
{\centering 
partial-\\SWAP$(\theta)$
}
\end{textblock} 
\begin{textblock}{1.3}(1.35, -1.2)
{\centering 
partial-\\iSWAP$(\theta)$
}
\end{textblock} 
\end{figure}

Although the partial-iSWAP may seem more artificial for moving probability amplitudes from one state to other, this gate is more convenient than the standard partial-SWAP. This entangling gate comes naturally from the capacitive coupling of superconducting qubits \cite{partialiSWAP-bialczak2010, iSWAP-schuch2003}. As a matter of fact, Google's Sycamore chip in which the supremacy experiment was performed \cite{google_supremacy_2019} allows for this type of gates natively. They are also of great importance for quantum chemistry applications \cite{barkoutsos2018quantum, gard2020efficient} or combinatorial optimization \cite{hadfield2019qaoa, wang2019xy, cook2019xy}.

By correctly setting the parameters in the partial-iSWAP gates it is possible to obtain any quantum state representing a probability distribution as
\begin{equation}
\ket{\psi} = \sum_{i=0}^{n - 1} e^{i \phi_i}\sqrt{p_i} \ket{i}_n, 
\end{equation}
where $p_i$ is the probability of measuring the $i$-th state, and complex phases $e^{i\phi}$ are simply ignored. 

The unary representation resides within a restricted and small part of the Hilbert space, and all unused space can be used as a flag for the appearence of error. The post-selection mechanism provided by the unary representation is based on a high distinguishability between those states inside and outside the unary basis. Simple measurements of the output state in the $Z$ basis are enough to check if an error has occurred or not. All read-out must reflect one and only one $\ket 1$, and all other $\ket 0$s. Any failed repetition can be discarded. A trade-off between number of executions and error mitigation is obtained. The noisier a quantum computer is, the more repetitions will be ignored, in exchange of retrieving only runs where no error is detected. Notice that other error mitigation techniques are also applicable. However, the scope of this work is to focus on the native mechanism.

The properties of the unary representation make it probably better suited to run on \ac{nisq} devices than the binary one. While it holds that exponentially more qubits are required to store equivalent quantum states, the unary approach is capable to retain more useful information and it is then much more robust against noise.

\section{Unary algorithm}\label{sec:unary_algorithm}
The option pricing problem is addressed in this chapter by designing an algorithm relying on the unary representation. The economical tool used in this approach is the Black-Scholes model described in Sec.~\ref{ssec:black_scholes}\cite{black_blackscholes_1973}. The distribution of asset prices is encoded into the quantum circuit in a unary state. The structure of this algorithm follows the one in Sec.~\ref{ssec:binary_algorithm} \cite{stamatopoulos_binary_2020}, namely amplitude distributor module, payoff computation and \ac{qae} module. 

There are two main advantages of the unary representation as compared to the binary one. In terms of circuit complexity, the unary scheme allows for a significant simplification of all different pieces composing the algorithm. The unary approach brings further benefit in practice due to the native post-selection strategy that results in error mitigation. On the other hand, the unary algorithm requires more qubits than a binary one for a given precision. Both features make the unary approach adapted to run in \ac{nisq} devices.

\subsection{Description of the algorithm}
\begin{figure}[t!]
\centering
\begin{adjustwidth}{-2cm}{-1cm}
\hspace{1cm}\resizebox{.8\linewidth}{!}{
\Qcircuit @C=0.85em @R=.3em @!R{
\lstick{\ket{q_0}} & \qw & \multigate{6}{\D} & \multigate{7}{\C + \R} & \qw & \qw & \qw & \qw & \multigate{7}{(\C + \R)^\dagger} & \multigate{6}{\D^\dagger} & \qw & \qw & \qw & \qw & \qw & \qw & \multigate{6}{\D} & \multigate{7}{\C + \R} & \ustick{\hspace{2mm} m} & \qw & \meter \\
\lstick{\ket{q_1}} & \qw & \ghost{\D} & \ghost{\C + \R} & \qw & \qw & \qw & \qw & \ghost{(\C + \R)^\dagger} & \ghost{\D^\dagger} & \qw & \qw & \qw & \qw & \qw & \qw & \ghost{\A} & \ghost{\C + \R} & \qw & \qw & \meter\\
\lstick{\ket{q_2}} & \qw & \ghost{\D} & \ghost{\C + \R} & \qw & \qw & \qw & \qw & \ghost{(\C + \R)^\dagger} & \ghost{\D^\dagger} & \qw & \qw & \qw & \qw & \qw & \qw & \ghost{\A} & \ghost{\C + \R} & \qw & \qw & \meter \\
\lstick{\ket{q_3}} & \gate{X} & \ghost{\D} & \ghost{\C + \R} & \qw & \qw & \qw & \qw & \ghost{(\C + \R)^\dagger} & \ghost{\D^\dagger} & \qw & \qw & \ctrl{4} & \qw & \qw  & \qw & \ghost{\A} & \ghost{\C + \R} & \qw & \qw & \meter & \rstick{\rm Post-selection} \\
\lstick{\ket{q_4}} & \qw & \ghost{\D} & \ghost{\C + \R} & \qw & \qw & \qw & \qw & \ghost{(\C + \R)^\dagger} & \ghost{\D^\dagger} & \qw & \qw & \qw & \qw & \qw & \qw & \ghost{\A} & \ghost{\C + \R} & \qw & \qw & \meter \\
\lstick{\ket{q_5}} & \qw & \ghost{\D} & \ghost{\C + \R} & \qw & \qw & \qw & \qw & \ghost{(\C + \R)^\dagger} & \ghost{\D^\dagger} & \qw & \qw & \qw & \qw & \qw & \qw & \ghost{\A} & \ghost{\C + \R} & \qw & \qw & \meter \\
\lstick{\ket{q_6}} & \qw & \ghost{\D} & \ghost{\C + \R} & \qw & \qw & \qw & \qw & \ghost{(\C + \R)^\dagger} & \ghost{\D^\dagger} & \qw & \qw & \qw & \qw & \qw & \qw & \ghost{\A} & \ghost{\C + \R} & \qw & \qw & \meter \\
\lstick{\ket{a}} & \qw & \qw & \ghost{\C + \R} & \qw & \qw & \gate{\S_{\psi_0}} & \qw & \ghost{(\C + \R)^\dagger} & \qw & \qw & \qw & \gate{\S_0} & \qw & \qw & \qw & \qw & \ghost{\C + \R} & \qw & \qw & \meter & \rstick{\rm Payoff}\\
& & & & & & & & & & & & & & & & & \\ 
& & & & & & & & & & \Q & & & & & & &
\protect\gategroup{9}{7}{9}{18}{1em}{_\}}
\protect\gategroup{9}{6}{1}{9}{1em}{(}
\protect\gategroup{9}{18}{1}{18}{1em}{)}
\protect\gategroup{1}{21}{7}{21}{.7em}{\}}
}}
\end{adjustwidth}
    \caption{Full circuit for the option pricing algorithm in the unary representation. The gate $\D$ is the probability distributor, and $\C + \R$ represent the computation of the payoff. After applying the algorithm, the oracle $\S_{\psi_0}$, the reverse algorithm and $\S_0$ follow. The last step is applying the algorithm again. This block $\Q$ is to be repeated for Amplitude Estimation. Measurements in all qubits is a requirement for post-selection. The qubit labelled as $q_3$ is the one starting the unary representation. }
    \label{fig:full_unary}
\end{figure}
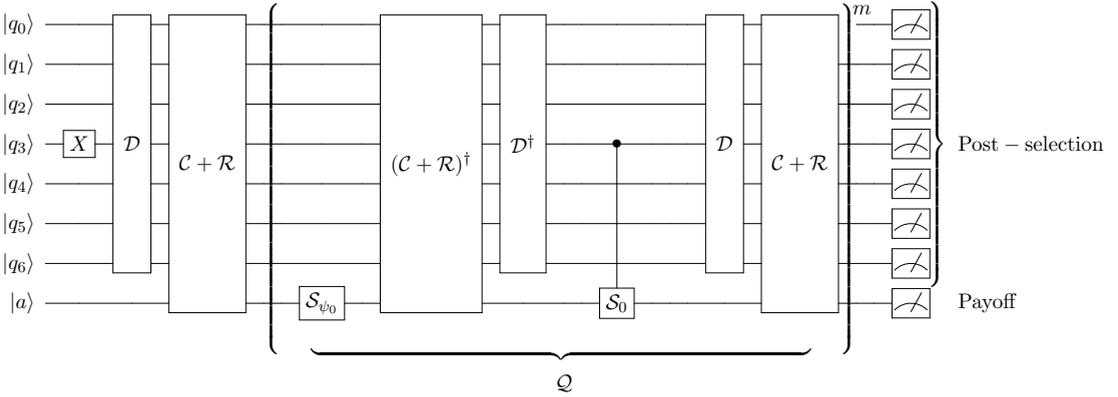

The global structure of the unary algorithm is inherited from the binary one, as outlined in Sec.~\ref{ssec:binary_algorithm}. 
A scheme for the full circuit is depicted in Fig. \ref{fig:full_unary}. In summary, the circuit is composed by one first $X$ gate that initializes the unary basis, one set of amplitude distributor $(\D)$ and payoff calculator $(\C + \R)$, and $m$ rounds of Amplitude Estimation $\Q = \A \S_0 \A^\dagger \S_{\psi_0}$. Read-out in all qubits is a requirement for post-selection to reduce errors.

\subsubsection{Amplitude distributor $\D$}
\begin{figure}[t!]
    \centering
    \includegraphics[width=.9\linewidth]{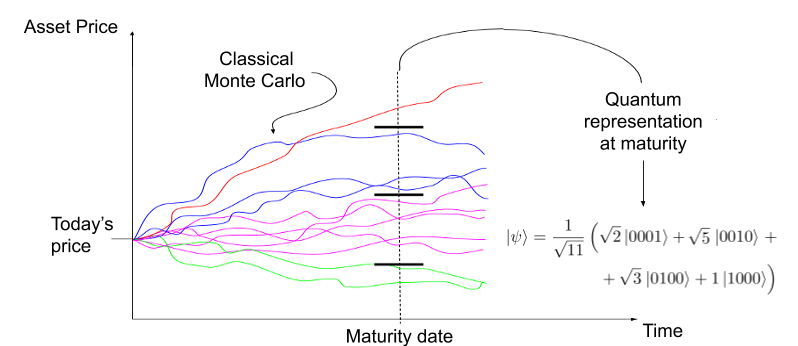}
    \caption{Scheme for the quantum representation of a given asset price at maturity date. For a given number of Monte Carlo paths, a binning scheme must be applied in such a way that the prices of the asset are separated according to its value. Different Monte Carlo paths that end up in the same bin are color coded accordingly. Each bin is mapped then to an element of the unary basis, whose coefficient is the number of Monte Carlo paths in this bin.
    The quantum representation of the asset price at maturity contains all possible Monte Carlo paths simultaneously. The precision is then bounded by the numbers of bins that can be stored on a quantum state, i. e. how many qubits are available.}
    \label{fig:MC}

\resizebox{.9\linewidth}{!}{
\Qcircuit @R=0.5em @C=0.3em{
& \lstick{\ket{0}}  & \qw & \qw & \qw & \qw & \qw & \qw & \multigate{1}{p-SWAP(\theta_1)} & \qw  & \qw\\ 
& \lstick{\ket{0}} & \qw & \qw & \qw & \qw & \qw & \qw & \ghost{p-SWAP(\theta_1)} & \qw  & \qw\\ 
&     &   &     &     &  &     &    &         &      &  \\  
&    \lstick{\vdots}        &   &  &     &  \vdots   &     &  \iddots   &         &   \vdots     &  \\            
&        &     &   &  &   &     &   &         &    &  \\     
& \lstick{\ket{0}} & \qw & \qw & \qw & \qw & \multigate{1}{p-SWAP(\theta_{n/2 - 1})} & \qw & \qw & \qw  & \qw\\ 
& \lstick{\ket{0}} & \qw & \qw & \multigate{1}{p-SWAP(\theta_{n/2})} & \qw & \ghost{p-SWAP(\theta_{n/2 - 1})} & \qw & \qw & \qw  & \qw\\ 
& \lstick{\ket{0}} & \qw & \qw & \ghost{p-SWAP(\theta_{n/2})} & \multigate{1}{p-SWAP(\theta_{n/2 + 1})} & \qw & \qw & \qw & \qw  & \qw\\ 
& \lstick{\ket{0}} & \qw & \qw & \qw & \ghost{p-SWAP(\theta_{n/2 + 1})} & \multigate{1}{p-SWAP(\theta_{n/2 + 2})} & \qw & \qw & \qw  & \qw\\ 
& \lstick{\ket{0}} & \qw & \qw & \qw & \qw & \ghost{p-SWAP(\theta_{n/2 + 2})} & \qw & \qw & \qw  & \qw\\ 
&   &    &     &     &  &     &    &         &      &  \\  
&    \lstick{\vdots}   &     &     &     &  \vdots   &     &  \ddots   &         &   \vdots     &  \\            
&    &     &     &     &   &     &   &         &    &  \\  
& \lstick{\ket{0}} & \qw & \qw & \qw & \qw & \qw & \qw & \multigate{1}{p-SWAP(\theta_{n - 1})} & \qw & \qw \\ 
& \lstick{\ket{0}} & \qw & \qw & \qw & \qw & \qw & \qw & \ghost{p-SWAP(\theta_{n - 1})} & \qw & \qw}}
\caption{Quantum circuit for loading any probability distribution in the unary representation $\D$, plus unary initalization. The circuit works as a distributor of amplitude probabilities from its middle qubit to the ones in the edges, using partial-SWAP gates that act only on nearest neighbors. Time dependence is encoded in the angles determining the gates. The first $X$ gate is needed to start the unary representation, but it does not take part in the distribution procedure. }
\label{fig:ampl_distributor}
\end{figure}

The starting point is the probability distribution of asset prices is based on the Black-Scholes solution from Eq.~\eqref{eq:log_normal}. In the unary representation, each state in the computational basis will correspond to a specified value in the price, thus the probability distribution is discretized. In particular, the qubit activated as $\ket 1$ determines the asset value. The precision is determined by the number of qubits $n$ in the circuit. The final price distribution at any time can be mapped to the unary representation by a fixed-depth quantum circuit. The probability of the asset to take a price is captured by the probability of measuring the corresponding qubit as $\ket 1$. This step takes the role of a Monte Carlo spread of asset values. See Fig.~\ref{fig:MC} for a graphical scheme on this idea.

The circuit needed to encode the asset prices into the quantum system acts as a distributor of probability amplitudes. The initial state is set as $\ket{00\ldots 010\ldots 00}_n$, i. e., the active qubit in $\ket 1$ is the middle one. To initialize the unary representation only one $X$ gate is needed. Then, coefficients in the final register encoding the asset price distribution is generated using partial-SWAP gates between the different qubits, see Eq.~\eqref{eq:pSWAP}. Every partial-SWAP gate substracts amplitude from one state and passes it to the next one. In the first step, partial-SWAP gates connect the middle qubit and its first neighbors. Subsequent steps propagate the effect to further qubits. Since the middle qubit has a probability of being measured $p_{n/2} = 1$, this mechanism distributes the whole amplitude to the rest of asset prices far from the central value. See Fig.~ \ref{fig:ampl_distributor} for a graphical description of this procedure. Note that the partial-SWAP gate can be substituted by partial-iSWAP gates, see Eq.~\eqref{eq:piSWAP}, for convenience when applied to experimental setups. The parameters needed to match the final asset price distribution can be exactly computed. A detailed procurement of those parameters is described in App.~\ref{app:amplitude_distribution}. 

Any final probability distribution in the asset prices at any time $t$ can be obtained with the Amplitude Distributor $\D$. The circuit depth is independent of time. All the necessary information, including time dependency, is carried within the set of angles defining the partial-SWAP or -iSWAP gates $\{\theta_1, \theta_2, \ldots , \theta_{n - 1}\}$. Given $n$ qubits, the depth is always $\lfloor n / 2 \rfloor + 1$. Similar ideas were exploited to describe the exact solution of the Ising model \cite{verstraete_quantum_2009,hebenstreit_compressed_2017,cerveralierta_ising_2018}.

To map a known probability distribution into the unary system, exactly $(n-1)$ parameters are required. This claim holds in the cases the final probability distribution is classically computable or remains unknown. In the first case, since the final distibution is available, the quest for quantum parameters can be addressed by solving an exact set of $n$ equations and $n-1$ variables, see App.~\ref{app:amplitude_distribution}. In case only the differential equation is known, but not its solution, other methods can attempt to solve this problem \cite{iblisdir_matrix_2007}.

\subsubsection{Payoff calculator $\C + \R$}
The circuit design to compute the expected payoff acts right after the amplitude distributor $\D$ to encode the expected return on an ancillary qubit. In the unary algorithm, this step is significantly simpler than for the binary counterpart. The procedure attempts to prepare an entangled state of the form
\begin{equation}\label{eq:final_state}
\ket{\Psi} = \sqrt{1 - a} \ket{\psi_0}_n \ket 0 + \sqrt{a}\ket{\psi_1}_n \ket 1,
\end{equation}
where $\ket{\psi_{0, 1}}_n$ are states in a superposition of the basis elements correspon-ding to asset prices below $(0)$ and above $(1)$ the strike $K$ respectively. The payoff is encoded in the amplitude $\sqrt{a}$, with $\vert a \vert \leq 1$. Notice that the expected return will in general not be bounded by $1$, thus a re-scaling must be applied to relate the quantum procedure to the exconomical values. After this step, \ac{qae} can be applied. 

\begin{figure}[t!]
    \centering
\hspace{1cm}
\Qcircuit @R=1em @C=0.6em{
& \lstick{\ket{q_0}} & \qw & \qw & \qw & \qw & \qw & \qw \\
& \lstick{\ket{q_1}} & \qw & \qw & \qw & \qw & \qw & \qw  \\
\vdots  &  &  &  \\
& \lstick{\ket{q_{k - 1}}} & \qw & \qw & \qw & \qw & \qw & \qw \\
& \lstick{\ket{q_k}} & \qw & \ctrl{5} & \qw & \qw & \qw & \qw & \rstick{{\rm Strike}\;  K} \\
& \lstick{\ket{q_{k + 1}}} & \qw & \qw & \ctrl{4} & \qw & \qw & \qw \\
\vdots  &  & \vdots & \vdots &  &  &  \\
& \lstick{\ket{q_{n - 2}}} & \qw & \qw & \qw & \ctrl{2} & \qw & \qw \\
& \lstick{\ket{q_{n - 1}}} & \qw & \qw & \qw & \qw & \ctrl{1} & \qw \\
& \lstick{\ket{a}} & \qw & \gate{R_y(\phi_{k})} & \gate{R_y(\phi_{k + 1})} & \gate{R_y(\phi_{n - 2})} & \gate{R_y(\phi_{n - 1})} & \qw & \rstick{\rm Ancilla}
\protect\gategroup{1}{6}{4}{8}{.7em}{\}}
\protect\gategroup{6}{6}{9}{8}{.7em}{\}}
}
\begin{textblock}{3}(9.5, -2.65)
No return
\end{textblock}
\begin{textblock}{3}(9.5, -1.35)
Positive \\ return
\end{textblock}
\caption{Quantum circuit that encodes the expected payoff in an ancillary qubit in the unary representation $\C + \R$. Each qubit with a mapped option price higher than the designated strike controls a c$R_y$ gate on the ancilla, where the rotation angle is a function of its contribution to the expected payoff. The comparator $\C$ is constructed through the control wires, while the $\R$ piece is performed by rotations in the last qubit.}
\label{fig:PayoffCircuit}
\end{figure}
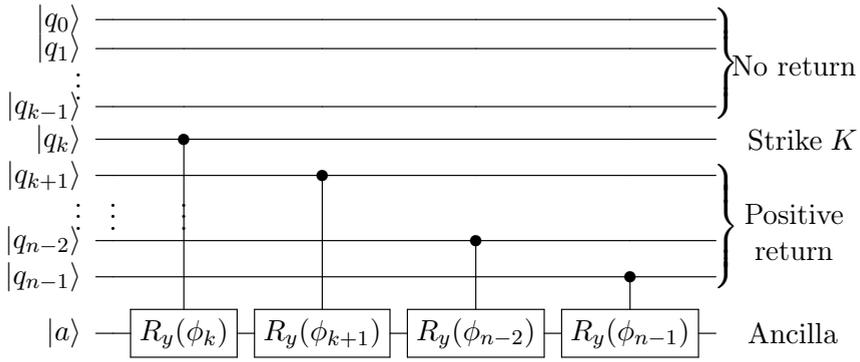

In the European option case, the most relevant point is to distinguish between those prices $S_i$ contributing to the expected return, that is prices greater than the strike $K$, and those that do not contribute. Only prices $S_i \geq K$ will have some effect on the ancillary qubit. This is acomplished by the $\C$ piece of the circuit. In the unary representation, this tasks turns out to be very simple. The computation of the expected payoff can be carried by applying single-qubit-controlled $Y$ rotations, $cR_y(\theta)$, summarized as the $\R$ operator. The control qubits are those encoding a given price $S_i$. Only those prices above the agreed strike $K$ will control applied gates to the ancillary qubit. The depth of this circuit will be $n-k$, where $k$ is the unary label of the strike $K$. See Fig.~\ref{fig:PayoffCircuit} for a graphical description of this part. 

The rotation angle for each controlled rotation depends on the price represented by the control qubit. Each price contributes differently to the expected payoff. The angle will be
\begin{equation}
\phi_i = 2 \arcsin \sqrt{\frac{S_i - K}{S_{max} - K}},
\end{equation}
where the denominator is introduced for normalization. Recall that $\vert a\vert \leq 1$. The application of the payoff calculator to a proper quantum state representing a price distribution results in
\begin{multline}\label{eq:payoff_state}
    \ket{\Psi}=\sum_{S_i\leq K}^{n-1}\sqrt{p_i}\ket{i}_n\ket{0}+\sum_{S_i>K}^{n-1}\sqrt{p_i}\cos(\phi_i/2)\ket{i}_n\ket{0}+\\
    +\sum_{S_i>K}^{n-1}\sqrt{p_i}\sqrt{\frac{S_i-K}{S_{max}-K}}\ket{i}_n\ket{1}.
\end{multline}

This state is now in the form of Eq.~\ref{eq:final_state}. The probability of measuring $\ket 1$ in the ancillary qubit is 
\begin{equation}
P(\ket a) = \sum_{S-_i > K} p_i \frac{S_i - K}{S_{max} - K}.
\end{equation}
The encoded payoff is easily recovered after measuring the probability of obtaining $\ket 1$ as the outcome. A simple multiplication times the normalization factor returns the quantity of interest. 

\subsubsection{Amplitude Estimation}
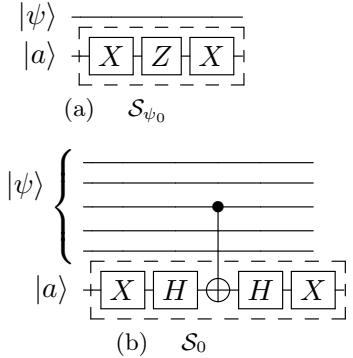
\begin{wrapfigure}{R}{.4\linewidth}
    \subfigure[\hspace{2mm} $\S_{\psi_0}$]{
    \hspace{1cm}\Qcircuit @R=0.7em @C=0.3em{
    & \lstick{\ket\psi} & \qw & \qw & \qw & \qw & \qw \\
    & \lstick{\ket a} & \qw & \gate{X} & \gate{Z} & \gate{X} & \qw \\
    \protect\gategroup{2}{4}{2}{6}{.7em}{--}
    } \vspace{2mm}    }
    \subfigure[\hspace{2mm} $\S_{0}$]{
    \hspace{1cm}\Qcircuit @R=0.7em @C=0.3em{
    & & \qw & \qw & \qw & \qw & \qw & \qw\\
    & & \qw & \qw & \qw & \qw & \qw & \qw\\
    & & \qw & \qw & \qw & \ctrl{3} & \qw & \qw\\
    & & \qw & \qw & \qw & \qw & \qw & \qw\\
    & & \qw & \qw & \qw & \qw & \qw & \qw\\
    & \lstick{\ket{a}} & \qw & \gate{X} & \gate{H} & \targ & \gate{H} & \gate{X}& \qw \\
    \protect\gategroup{6}{4}{6}{8}{.7em}{--}
    \protect\inputgroupv{1}{5}{.8em}{.8em}{\ket{\psi}}
    } \vspace{2mm}}\hspace{1cm}
    \caption{Quantum circuit representation of $\S_{\psi_0}$ (a) and $\S_0$  (b) required to perform Amplitude Estimation in the unary basis. Notice that operator $\S_0$ is much simpler in the unary representation as it does not require multi-controlled CNOT gates.}
    \label{fig:ae_circuits}
\end{wrapfigure}

Note that Eq.~\eqref{eq:payoff_state} suits the application of \ac{qae} on it. As described in Sec.~\ref{ssec:amplitude_estimation}, the key ingredient is the operator $\Q = \S_{\psi_0} \A^{\dagger} \S_{0} \A$. In this section a description on how to implement the operators $\S_{\psi_0}$ and $\S_{0}$ in the unary implementation is carried. A graphical assistance is depicted in Fig.~\ref{fig:ae_circuits}

The oracle operator $\S_{\psi_0}$ acts by identifying those elements of the quantum state corresponding to accepted outcomes. This task is already performed by the algorithm $\A$, and the information is carried by the ancilla qubit, with $\ket 1$ for accepted results. Thus, the function of this oracle can be achieved by performing local operations in the ancilla qubit. The required operation is 
\begin{equation}\label{eq:oracle}
    \S_{\psi_0} = (I^{\otimes n} \otimes (XZX)), 
\end{equation}
where the $X$ gates could even be deleted since they add a global sign. 

For the case of the operator $\S_0$ a detail that greatly simplifies this computation is remarkable .
The operator $\S_0$ is normally defined using $\ket{0}$ since most quantum algorithms start on that state, as depicted in Eq. \eqref{eq:final_state}. However, a more apt definition should instead include a generic $\ket{initial}$ state as a basis for operator $\S_0$, the state onto which the algorithm $\A$ is first applied. For the unary case, except for the first extra $X$ gate, it is possible to consider the algorithm as starting in that state of the unary basis, heavily simplifying the overall construction. That being the case, $\S_0$ can be constructed out of 2 single-qubit gates and one entangling gate. This supports a great simplification of the unary algorithm with respect to the binary one. 

With the operator $\Q$ constructed, \ac{qae} schemes can be performed. Since the unary algorithm is aimed towards \ac{nisq} devices, it is feasible to use an \ac{iqae} scheme without \ac{qpe}, as mentioned in Sec.~\ref{ssec:amplitude_estimation} and explained in detail in App. \ref{app:amplitude_estimation}. In the implementation here presented, only the options with minimal consecutive executions of $\Q$ are considered.

\subsection{Error mitigation}
The algorithms of the \ac{nisq} era need to present resilience against gate errors, noise and decoherence. The literature provides a variety of error mitigation techniques, see Refs. \cite{temme-mitigation2017, endo-mitigation2018}. Some of those techniques might find valid applications in the unary algorithms as well. However, in this unary approach the focus lies on the native error mitigation poperties described in Sec.~\ref{sec:unary}. 

Since the algorithm is designed to perform in the unary basis, all outcomes must reflect this fact. The strategy is as simple as measuring all qubits and not only the ancilla one. This step allows to ensure that the output state did not suffer errors taking it out the unary repesentation. This way, all outcomes not belonging to the unary representation are discarded. This triggers a trade-off between number of accepted samples and reduction of errors. \\

\section{Comparison to a reference algorithm}\label{sec:comparison}
The unary algorithm is conceived since the beginning as an alternative approach to the binary one with some properties that make it suit better for implementation in \ac{nisq} devices. In this section a comparison the resource demands of both algorithms in terms of circuit design and number of gates is treated. The analysis includes the proper algorithm and the \ac{qae} procedure. A detailed treatment of errors is then left for subsequent sections. 

\subsection{Gate count}\label{ssec:gate_count}
In order to properly count the number of gates required for executing the algorithm, some particular choices have been made. In practice, quantum computers make use of a native set of gates with the capability to construct any unitary with some overhead in the number of operations. The count is carried taking CNOT and partial-iSWAP gates as the native entangling gates. A chip connectivity between all qubits with theoretical common operations, see Sec.~\ref{ssec:chip_architecture} for further details, is also assumed for the sake of simplicity. The overhead of extra SWAP gates to account for non-existing connections is not considered in these calculations. In all cases, the counting for single-qubit gates is made by compiling subsequent gates into a single one. All two-qubit gates are decomposed into the native entangling gate and a number of single-qubit gates, counting for all possible overheads. This is particularly prominent in the binary case where many Toffoli gates are needed.

The unary algorithm requires $\mathcal{O}(n)$ partial-SWAP gates for accomplishing the amplitude distributor $\D$. For computing the payoff, $\mathcal{O}(\kappa n)$ controlled-rotation gates are needed, where $0 \leq \kappa \leq 1$ stands for the qubit corresponding to the strike price $K$. The results from Table~\ref{tab:gates}, left, collects the gate counting of the full circuit for both the unary and binary algorithms, as a function of the number of qubits. for CNOT and partial-iSWAP gates as the native entangling gates.

\begin{table}[t]
    \centering
    \begin{tabular}{|l|c|c|c|c|c|c|c|c|}\hline
        \multirow{2}{*}{\bf Unary} & \multicolumn{4}{c|}{CNOT} & \multicolumn{4}{c|}{partial-iSWAP} \\ \cline{2-9}
         & $\D$ & $\C + \R$ & $\S_{\psi_0}$ & $\S_0$ &$\D$ & $\C + \R$ & $\S_{\psi_0}$ & $\S_0$\\ \hline
        1-qubit gates & 2n & 2$\kappa$n & 1 & 4 & 1 & $\kappa$10n & 1 & 9\\
        2-qubit gates & 4n & 2$\kappa$n & 0 & 1 & n & $\kappa$5n & 0 & 2\\\hline
        Circuit depth & 3n & 4$\kappa$n & 1 & 5 & n/2 & 15$\kappa$n & 1 & 10 \\\hline
    \end{tabular} 
    \hfill
    \resizebox{\linewidth}{!}{
    \begin{tabular}{|l|c|c|c|c|c|c|c|c|}\hline
        \multirow{2}{*}{\bf Binary} & \multicolumn{4}{c|}{CNOT} & \multicolumn{4}{c|}{partial-iSWAP} \\ \cline{2-9}
         & $\D$ & $\C + \R$ & $\S_{\psi_0}$ & $\S_0$ & $\D$ & $\C + \R$ & $\S_{\psi_0}$ & $\S_0$ \\\hline
        1 qubit gates & 3nl  & (16+5$\kappa$)n & 1 & 20n - 23 & 8nl & (86+5$\kappa$)n & 1 & 80n - 113 \\
        2 qubit gates & nl & 14n & 0 & 12n - 18 & 2nl & 28n & 0 & 24n - 36\\\hline
        Circuit depth & nl+l & (27+2$\kappa$)n & 1 & 24n - 30 & 6nl+l & (97+2$\kappa$)n & 1 & 90n - 129\\\hline
    \end{tabular}}
    \caption{Scaling of the number of 1- and 2-qubit gates and circuit depth as a function of the number of qubits $n$ representing the asset value in unary and binary representations, for the amplitude distributor $\D$, payoff estimator $\C + \R$ and \ac{qae} operators $\S_{\psi_0}$ and $\S_0$. Ideal chips architectures are assumed. The scalings in case CNOT or partial-iSWAP gates are implemented are compared. In case the experimental device can implement both CNOT and partial-iSWAP basic gates, the total amount of gates and total depth would be reduced. For the unary circuit, the parameter $0\le \kappa\le 1$ depends on the position of the strike in the qubit register. The parameter $0\le \kappa\le 1$ characterizes the number of $1$s in the binary representation of the strike price. For the amplitude distributor, $l$ is the number of layers of the qGAN. 
    }
    \label{tab:gates}
\end{table}

The amplitude distribution module $\D$ substantially benefits from having partial-iSWAP gates as the native operation. However, this implies an overhead for the payoff computation $\C + \R$, where CNOT gates introduce a gain. In the ideal case where both partial-iSWAP interactions between first neighbors and CNOT-based connections to the ancilla qubit are available, the best possible scaling is obtained. Explicitly, the total number of gates would be $(4 \kappa+1) n+1$, and the depth of the circuit would become $(4 \kappa +\frac{1}{2})n$.

The gate count for the binary algorithm is summarized in Tab.~\ref{tab:gates}, right, in the same conditions as the unary one. The CNOT-connection turns out to be more convenient in this case. The results here provided include also the \ac{qgan} piece used for uploading the asset price distribution into the quantum circuit, thus a dependency on the number of layers emerges. However, no training cost required for \ac{qgan}s is considered.

It is worth emphasizing that the gate overhead for the unary algorithm is much lower than for the binary case. The main reason for this result is that the unary circuit does not implement any three-qubit gate, which is only decomposable in a large number of two-qubit gates. This simplification is overcome by the exponential advantage of the binary algorithm for large numbers of $n$, provided that the asset prices distribution is efficiently uploaded. 

\begin{figure}[t!]
    \centering
   	\begin{adjustwidth}{-1cm}{-2cm}
    \subfigure[\hspace{2mm} CNOT gates]{\includegraphics[width=0.3\linewidth]{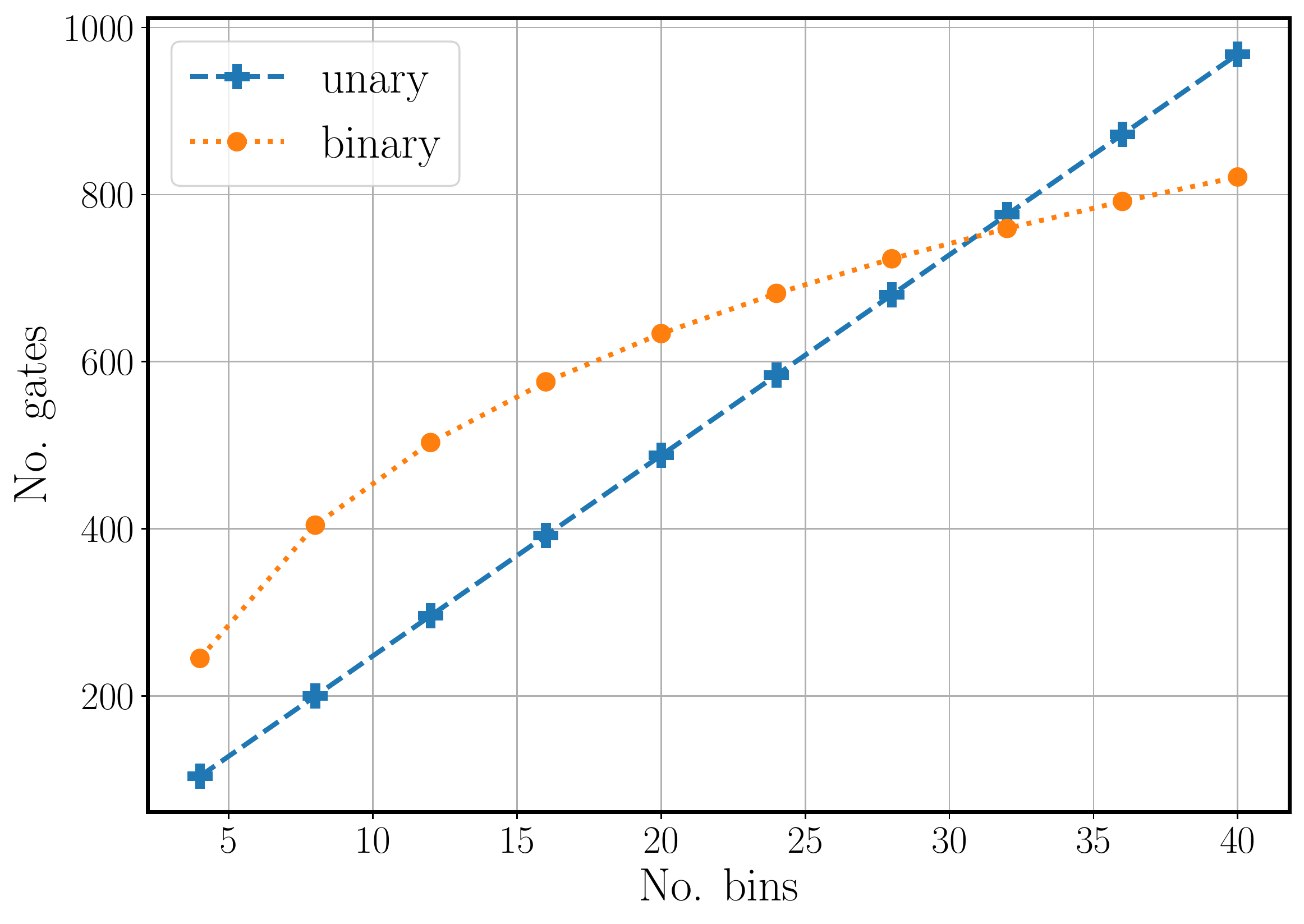}}
    \subfigure[\hspace{2mm} partial-iSWAP gates]{\includegraphics[width=0.3\linewidth]{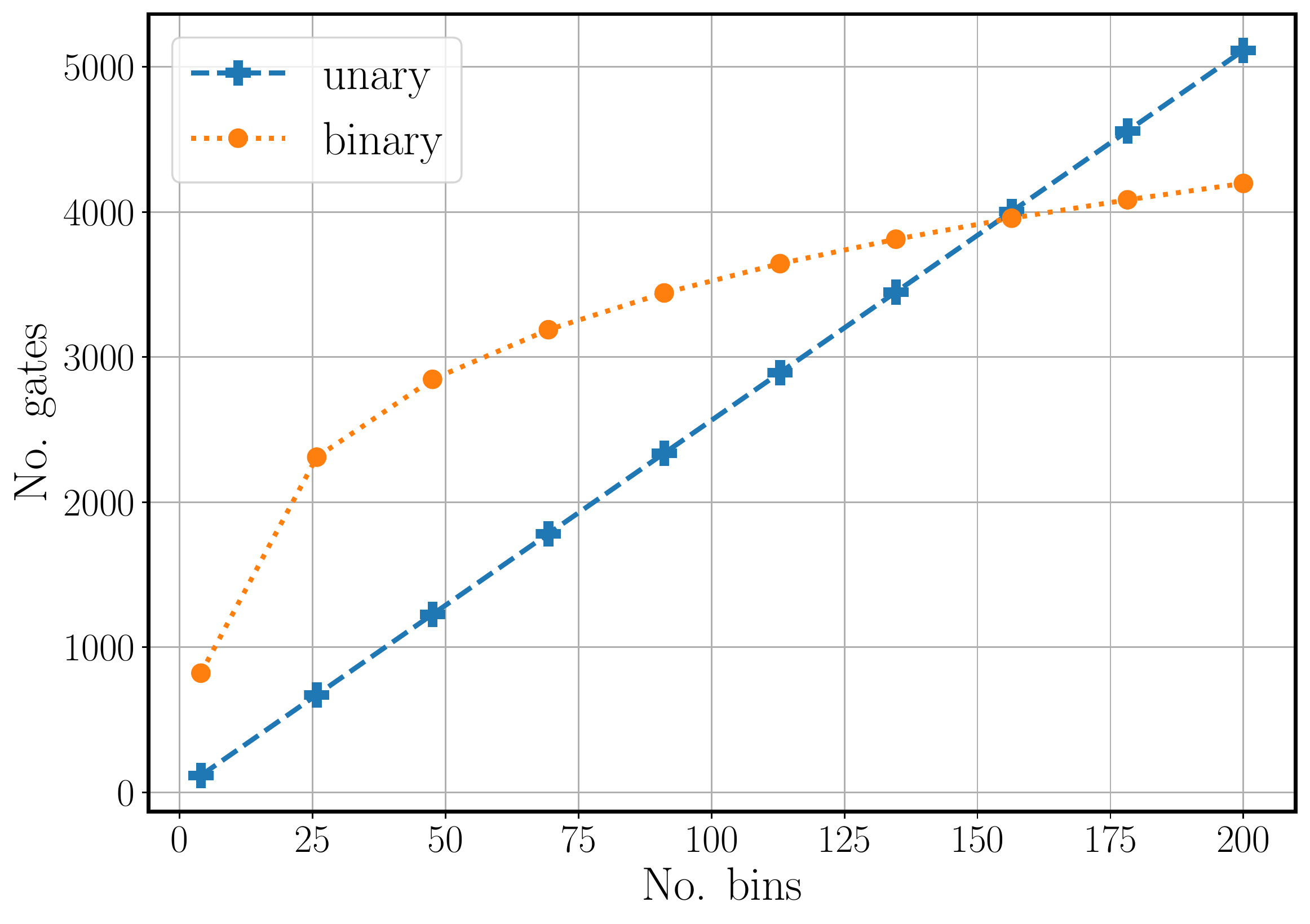}}
    \subfigure[\hspace{2mm} Best combination]{\includegraphics[width=0.3\linewidth]{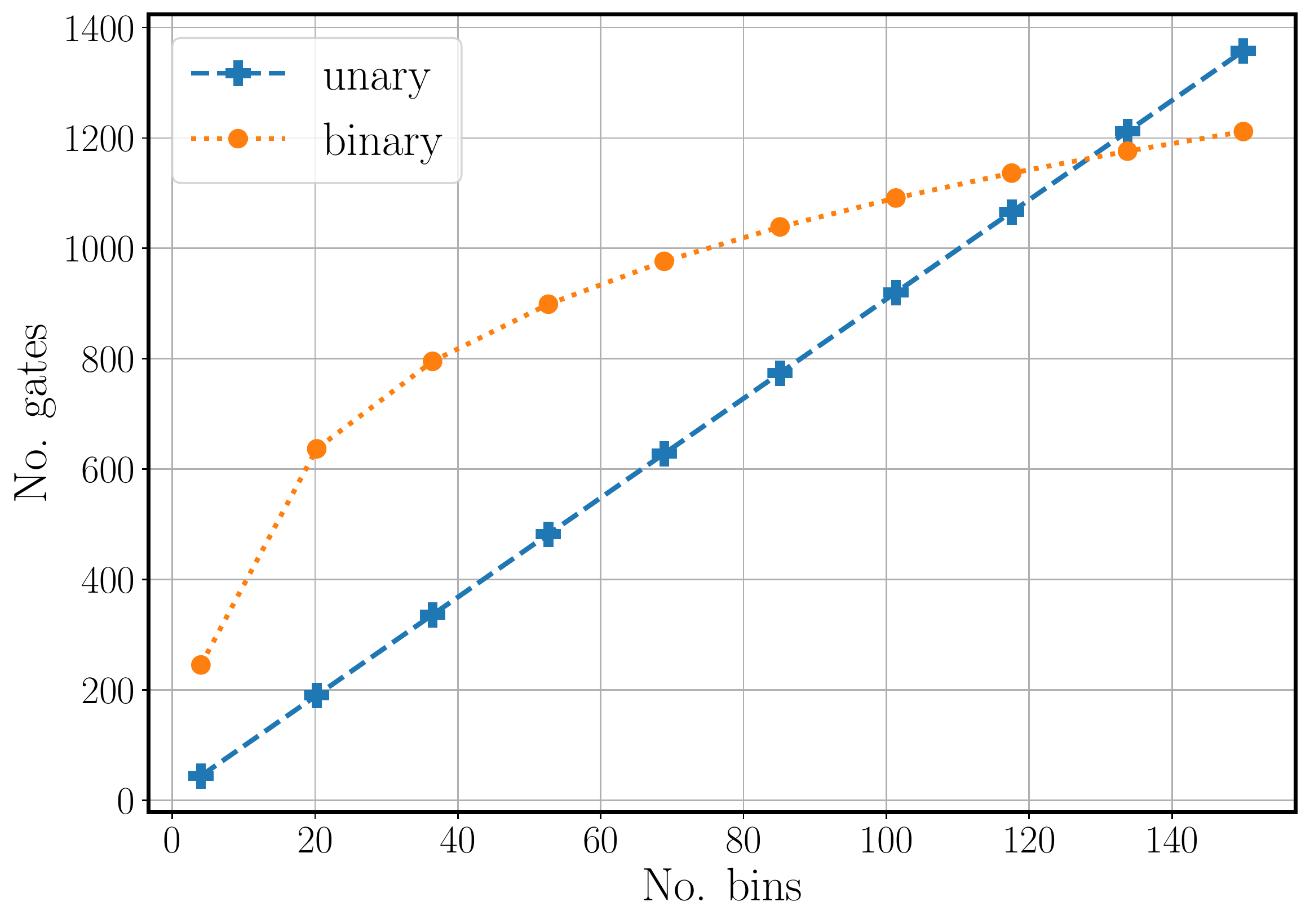}}
    \end{adjustwidth}
    \caption{Scaling of the number of gates required for the full algorithm, including a step, $m=1$, of Amplitude Estimation, with the number of bins, for different native gates: CNOT gates (a), partial-iSWAP gates (b) and the best possible combination (c), in which one is allowed both CNOT and iSWAP gates as native to the device. The scaling is calculated assuming ideal connectivity, which would largely hinder the binary implementation were that not the case.}
    \label{fig:gate_count}
\end{figure}

A complete comparison between unary and binary circuits is shown in Fig.~\ref{fig:gate_count}. In this comparison, the numbers $\kappa = 1/2$, number of controlled rotations in the unary algorithm, and $l = \log_n(2) / 2$, number of layers in the \ac{qgan} for the binary case, are fixed. The comparison is made for the same precision in the asset price representation. For a given number of $n$ bins, the unary algorithm needs $n$ qubits, thile the binary representation has enough with only $\log_2(n)$ of them, plus qubits overhead. The simple operations of the unary algorithm make this option mode convenient for a number of bins $n\sim 100$. For large values of $n$, the binary approach outperforms the unary option, as the number of gates is logarithmically lower. In case quantum resources have limited numbers and quality, as it is expected in \ac{nisq} devices, circumstances favor the unary approach to the detriment of the binary one. 

\subsection{Ideal chip architecture}\label{ssec:chip_architecture}
From the theoretical description of the algorithms described in Sec.~\ref{sec:unary_algorithm} for the unary algorithm and Ref.~\cite{stamatopoulos_binary_2020} for the binary one it is possible to infer the connectivity requirements. Those are described in this section, where the condition is that any pair of qubits with a common interaction in theory has direct connection in the chip. 

\begin{figure}[b!]
    \centering
\begin{adjustwidth}{-2cm}{-1cm}
    \subfigure[\hspace{2mm} Unary algorithm]{\includegraphics[height=.25\linewidth]{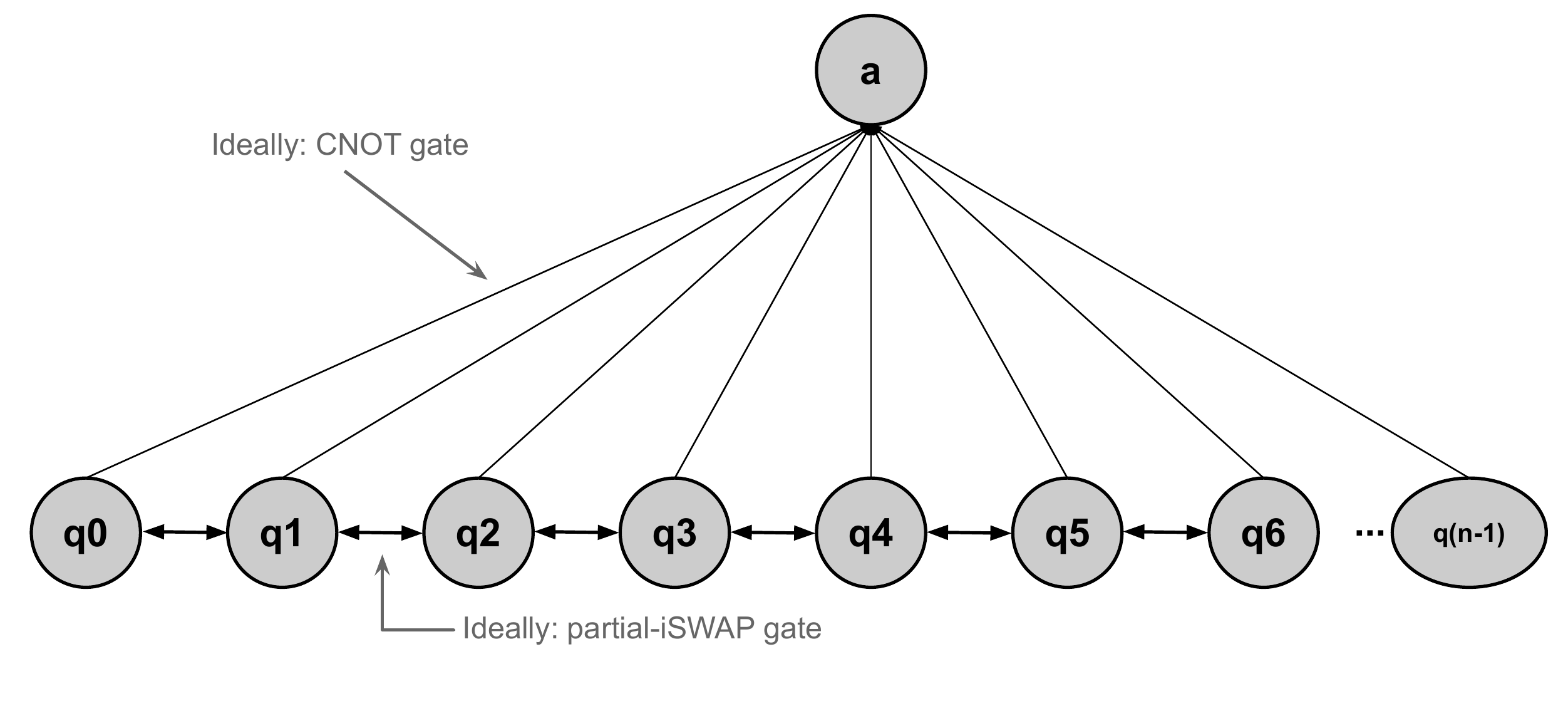}} \hfill 
      \subfigure[\hspace{2mm} Binary algorithm]{\includegraphics[height=.25\linewidth]{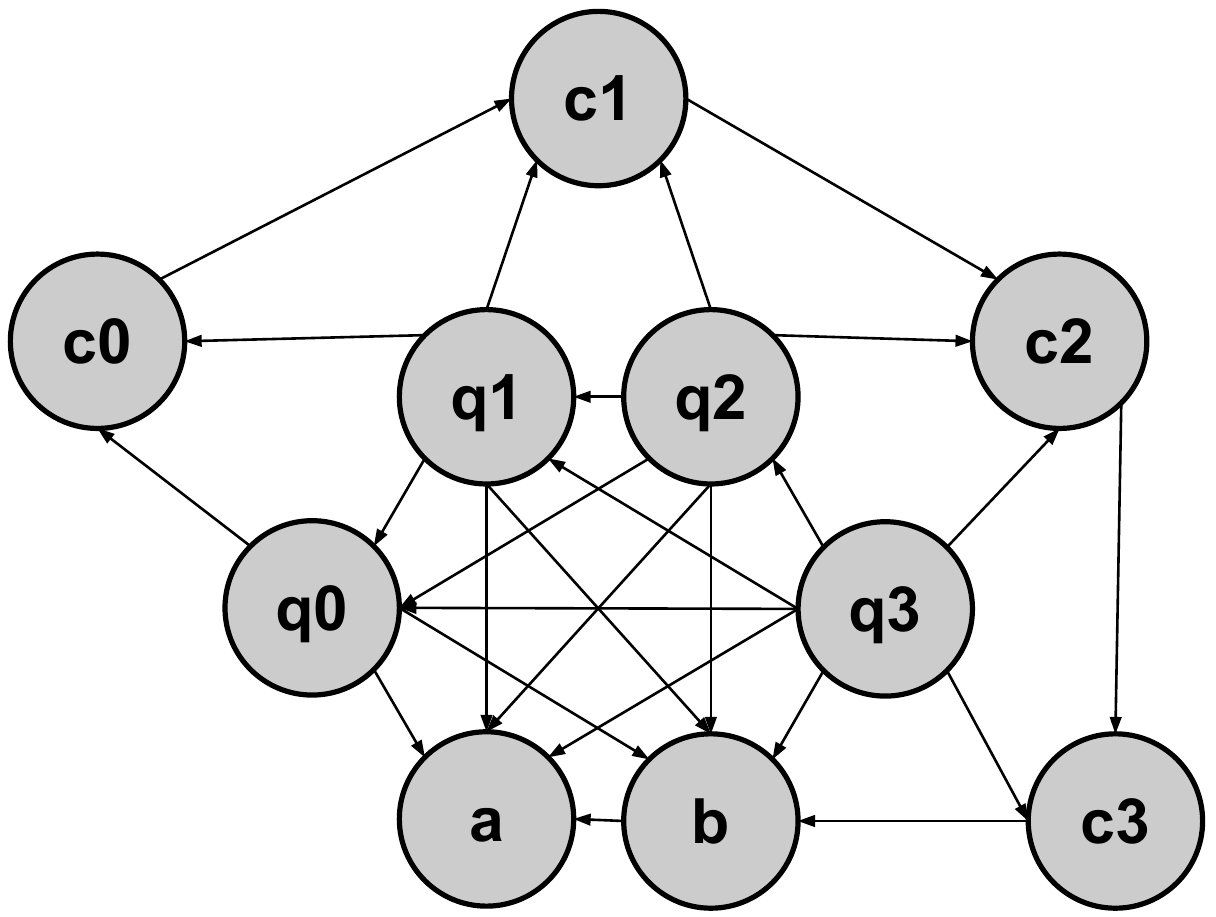}}
      \end{adjustwidth}
\caption{Ideal chip architecture to implement the unary and binary algorithms for option pricing.  a) In the unary chip, only a single ancilla qubit, labelled as \emph{a} in the image, has to be non-locally controlled by the rest of the qubits. All other interactions are first-nearest-neighbor gates. b) For the binary implementation with 4 qubits of precision, \emph{q}$_0$,\emph{q}$_1$,\emph{q}$_2$,\emph{q}$_3$, where \emph{a} and \emph{c} stand for ancillary and carrier qubit, respectively, and \emph{b} is another ancilla. The algorithm requires a number of ancillary and carrier qubits equal to the number of precision qubits plus two, 4+2 in this example. Full connectivity is needed between the precision qubits and two ancillas.}
\label{fig:chip_architecture}
\end{figure}

The unary algorithm can be performed with a very simple chip. The amplitude distribution module $\D$ only needs local interaction between first-neighbor qubits to upload the asset prices to the quantum register. Thus, the qubits can be arranged on a 1D line with two-local interactions. In addition, if this connection carries a partial-SWAP native gate, the $\D$ operations can be performed without any overhead. This realization of the quantum circuit would result in a decrease in the number of needed gates by factor of 6 in the amplitude distributor module as compared to the CNOT entangling gate. Note also that superconducting qubits allow for a natural implementation of the partial-iSWAP gate \cite{partialiSWAP-bialczak2010}. For the expected payoff, the ancillary qubit interacts with any other qubit from the quantum register, ideally relying on CNOT gates. 

On the other hand, the binary algorithm for payoff calculation needs a much more complex chip connectivity. For the sake of comparison with the simplest chip architecture presented for the unary algorithm, the most basic connectivity needed to perform the steps described for the binary scheme is displayed in Fig. \ref{fig:chip_architecture}. In the binary case, $\log_2(n)$ qubits ($q$ in the figure) are needed to store the price distribution with the same accuracy as the unary case with $n$ qubits. A total amount of $\log_2(n) + 1$ auxiliary ancilla qubits are needed (in the figure $c$ and $b$). They payoff is stored at another ancilla $a$. A total number of $2(\log_2(n) + 1)$ qubits is needed.

It is clear that the number of necessary qubits for the binary algorithm, including ancillas, scales asymptotically better than in the unary approach. Nevertheless, the need for Toffoli gates and almost full connectivity may eliminate this advantage in practical problems for NISQ devices. The simplicity of the architecture needed to implement the unary algorithm might yield an advantage over alternative algorithms as well. 

\section{Results}\label{sec:results_unary}

In this section the binary and unary algorithms for a given option pricing problem are simulated. The aim is to compare the performance of both approaches. Two main comparisons are made relating this topic. First, the performances are tested for circuits without any noise, where the only source of error comes from the sampling uncertainty at the measurement step. In a second step, both approaches are tested against increasing noise levels in order to verify the resilience to error of both unary and binary algorithms. Both steps are presented as the same process. The final target of these calculations is to discern which approach is more beneficial for \ac{nisq} computers. The calculations were carried using the {\tt qiskit}~\cite{qiskit} framework. The code is publicly available in Ref.~\cite{github_unary}.

\begin{wrapfigure}{O}{.5\linewidth}
\includegraphics[width=\linewidth]{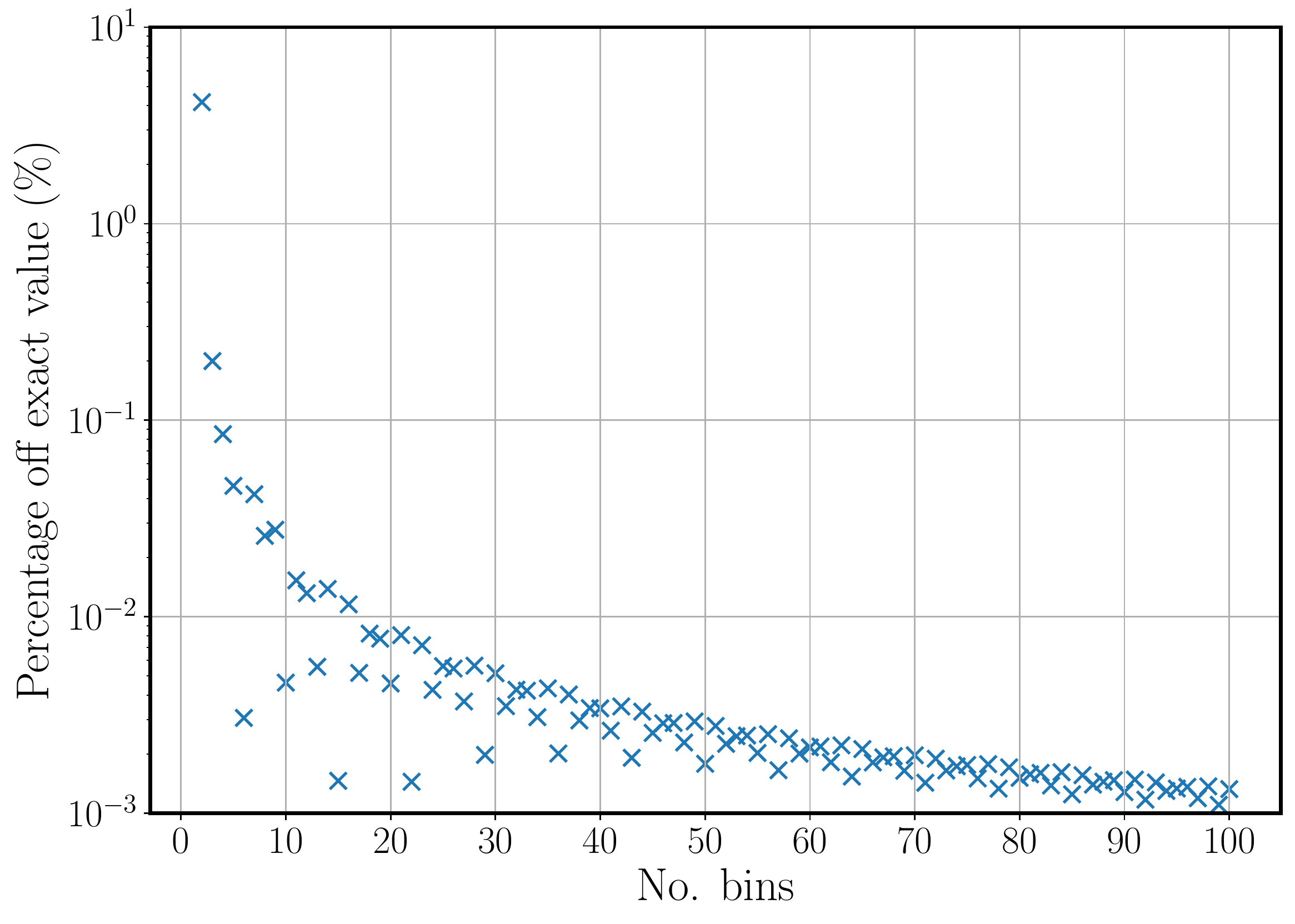}
  \caption{Percentage error from the exact value of the expected payoff, for the classical computation, as a function of the number of bins in the probability distribution. With only $\sim 50$ bins, errors for the option price below 0.5\% are already reached.}
\label{fig:bin_error_classical}
\end{wrapfigure}The financial problem addressed in this example is a standard European option with fixed properties. The asset price at $T=0$ is $S_0 = 2$, its interest rate is $r = 0.05$, and its volatility is $\sigma = 0.4$. The maturity time is $T = 0.1$ years. The agreed strike price is $K = 1.9$. The simulation of the asset price is carried taking the Black-Scholes model from Sec.~\ref{ssec:black_scholes} up to three standard deviations in the log-normal probability distribution. For the quantum circuits simulation, a 8 bins model is considered. Thus, the unary algorithm encodes the asset price distribution in $8$ qubits, and the binary one in $\log_2(8) = 3$ of them. Ideal chip structured from Sec.~\ref{ssec:chip_architecture} are considered. In terms of payoff calculation, the reference value is computed classically, with a precision of $10^4$ bins.

First of all, it is important to estimate the range of applicability of the unary and binary algorithms as a function of the number of bins $n$. From the calculations presented in Sec.~\ref{ssec:gate_count}, it is possible to conclude that the crossing point in the number of gates lies at $n \sim 100$. That is, below this threshold, the unary algorithm needs less gates than the binary one to be executed. Real-world applications require at least an accuracy $<1\%$. The question to answer here is whether this accuracy can be reached in the range of applicability of the unary or the binary algorithm. 

The error of the expected payoff as a function of the number of bins $n$ is plotted in Fig.~\ref{fig:bin_error_classical}. The error depends both in the binning $n$ and the position of the strike $K$, that is, if $K$ lies at the center or the extrema of a bin. The error will decrease in general as $n$ increases. Therefore, the results fall within a reasonable accuracy around the exact value for a sufficiently large number of bins. The results in Fig.~\ref{fig:bin_error_classical} show that $\sim 50$ bins are enough to achieve accuracies near $0.5\%$. This regimes corresponds to an advantage for the unary approach. This shows that the unary algorith can be implemented and obtain better performance than the binary one, and still return accurate results with small discretization errors. 

The noise maps used in the simulations to model the noise are simple yet descriptive. The complete model is controlled by a tunable parameters $\varepsilon$ such that it vanishes for perfect circuits, and reaches $\varepsilon = 1$ for a random execution. For single- and two-qubit gate errors, a depolarizing noise is considered. This noise is described by the transformation
\begin{equation}
\rho \rightarrow (1 - \varepsilon) \rho + \frac{\varepsilon}{d} \Tr(\rho) I,
\end{equation}
where $I$ is the identity gate of dimension $d$. The depolarizing transformation occurs after each gate, with the difference that for two-qubit gates the error grows up to $2\varepsilon$. For measurement errors, the probability of obtaining a wrong outcome is modeled as $10\varepsilon$ and symmetric, that is measuring incorrect $\ket 0$ or $\ket 1$ is equally probable. It is remarkable that no thermal relaxation nor thermal dephasing have been included into the noise models. The reason is that the execution times for the considered circuits are far below coherence times of qubits due to the shallow depth of the circuits. The execution time of a single-qubit gate is $\sim 1000$ shorter than the decoherence time. This description is adjusted to a simplified version of state-of-the-art computers~\cite{google_supremacy_2019}. 

\subsubsection{Amplitude Distribution loading - $\D$}

\begin{figure}[t!]
\begin{adjustwidth}{-2cm}{-1cm}
    \centering
    \subfigure[\hspace{2mm} Amplitude distribution $\D$\label{fig:KL}]{\includegraphics[width=.45\linewidth]{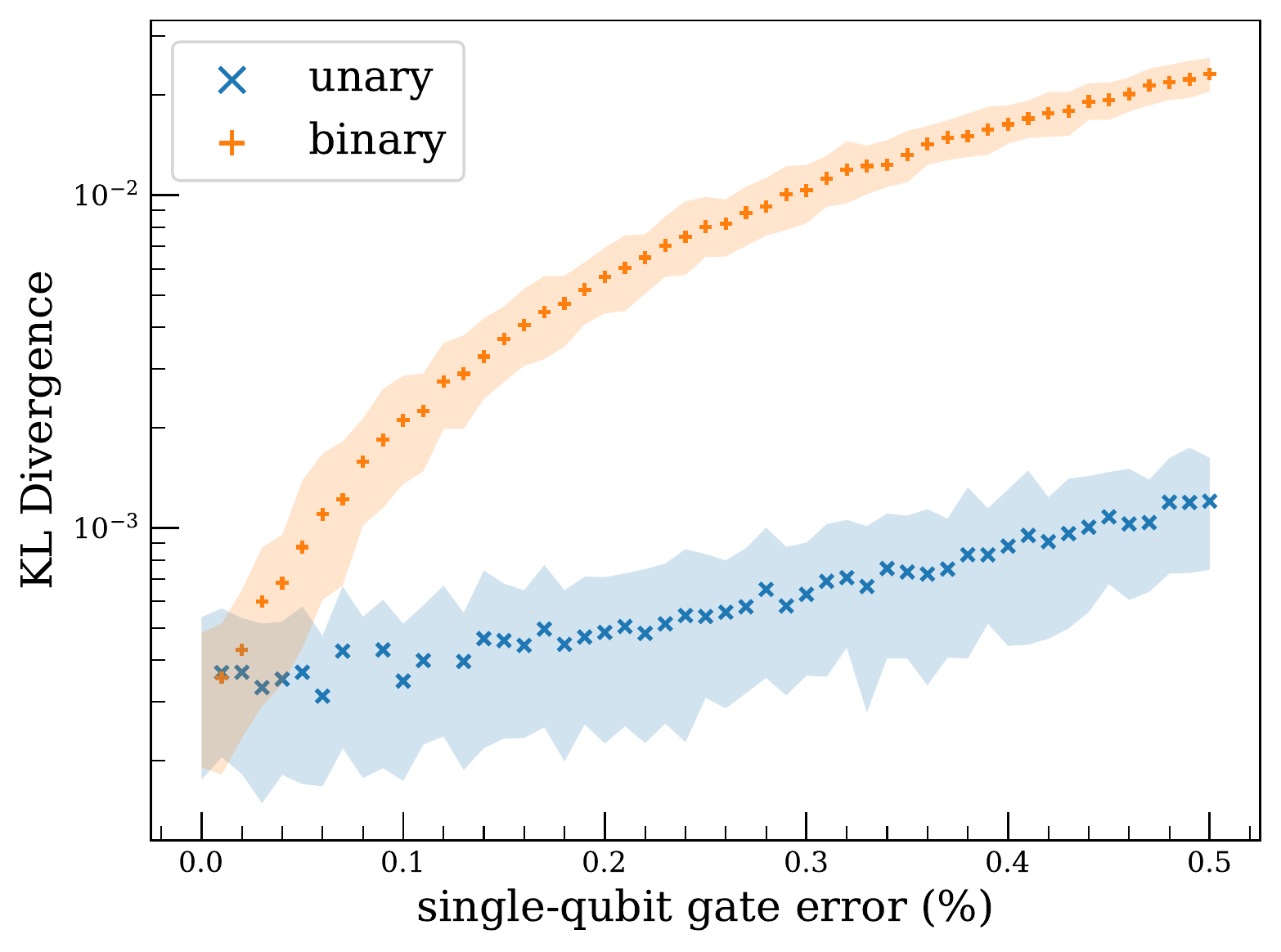}}
    \subfigure[\hspace{2mm} Expected payoff calculation $\C + \R$\label{fig:payoff}]{\includegraphics[width=.45\linewidth]{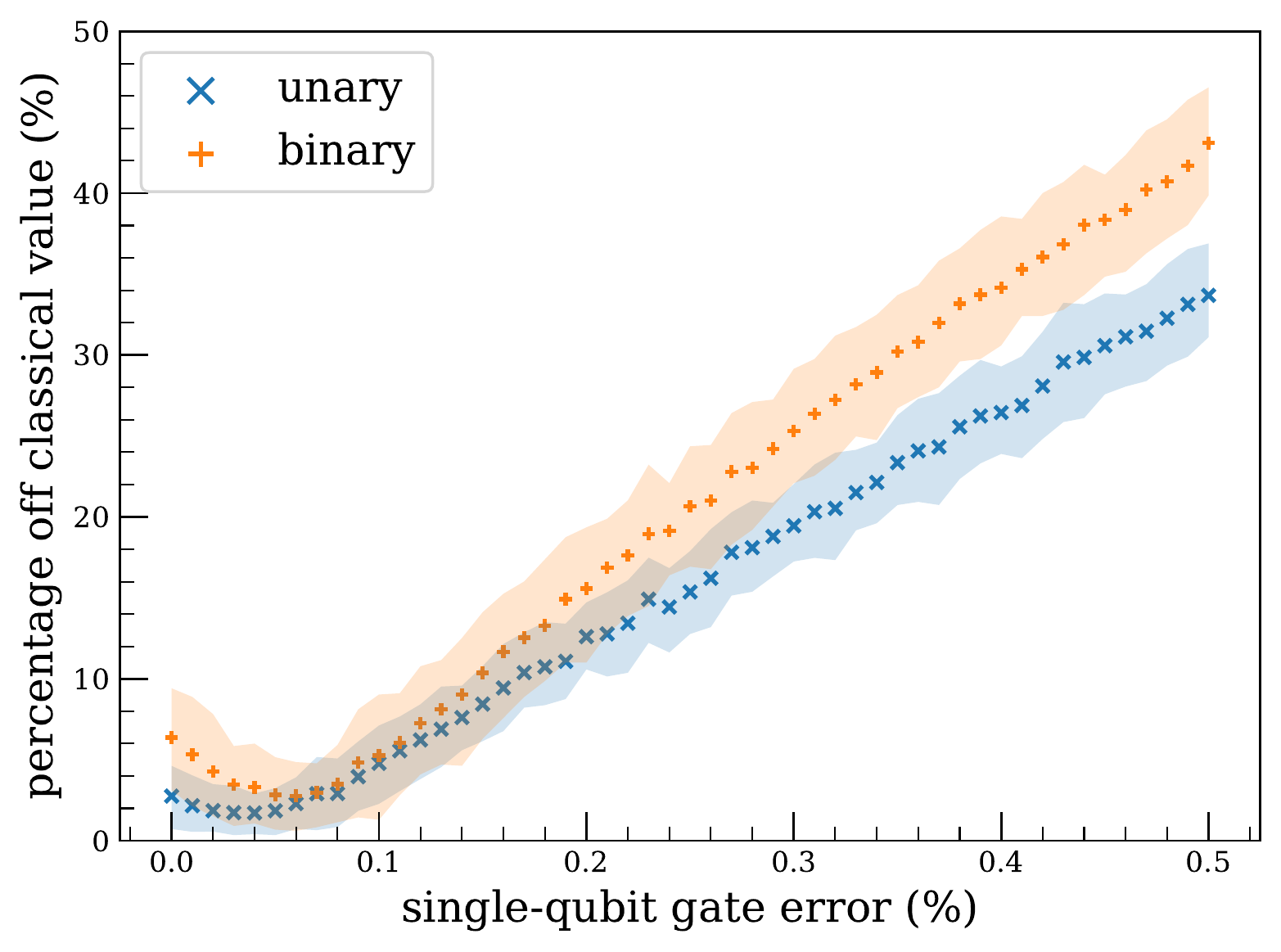}}
\end{adjustwidth}
    \caption{a) Kullback-Leibler divergence between the target probability distribution and those achieved by the quantum algorithms, b) percentage error in the payoff calculation for depolarizing and measurement errors. Calculations made for equivalent 8 unary and 3 binary qubits for depolarizing and measurement errors (only in b) ), up to 0.5\% for single-qubit gates, 1\% for two qubit gates and 5\% for read-out errors, consistent with state-of-the-art devices. Crosses stand for average results, and the shaded regions encompass the central $70\%$ of the instances. Each probability distribution is estimated using $100$ experiments with $10^4$ samples each. The shaded regions encompass the central 70\% of the instances in each case. The unary algorithm is more robust against these errors.}
    \label{fig:unary_binary_1}
\end{figure}

The capabilities of the unary and binary amplitude distribution modules are compared in Fig.~\ref{fig:KL}.
The quantity here depicted is the Kullback-Leibler divergence \cite{KL-kullback1951}. This quantity measures the distance between two probability distributions, and it vanishes when both are indistinguishable. It is clear to see that the approximation of $\D$ for the unary algorithm achieves better results than the binary distributor. For the maximum amount of noise allowed, the difference rises up to an order of magnitude.

\subsubsection{Expected Payoff - $\C + \R$}
It is shown in Fig.~\ref{fig:payoff} the average error of the expected payoff as computed with the unary and binary approaches, when compared to the classical value. The unary algorithm presents slightly better results than the binary one for all errors considered. This difference is too small to conclude that the unary approach is more convenient. Note that the deviations in the expected payoff reach a minimum for a finite value in the error parameter $\varepsilon \sim 0.005\%$. This is easy to understand by considering the theoretical expected value. Both algorithms return low approximations to the expected value, with the values of $a$ from Eq.~\eqref{eq:payoff_state} $a < 0.5$. As the noise is considered, the value of $a$ tends to its random value $a = 0.5$. Thus, there is a middle point where $a$ corresponds to an accurate approximation of the expected value. 

\subsubsection{Quantum Amplitude Estimation}
In this \ac{qae} part, results will be presented in three steps. First, instances without any noise considered are shown. The noiseless devices present convergent results within errors due to approximations. Then, an analysis of the errors and the statistical uncertainty of the expected payoff value when applying \ac{qae} with errors is performed. In this case, an extension to larger numbers of bins $n$ is carried for the unary algorithm. 

\begin{figure}[b!]
\begin{adjustwidth}{-2cm}{-1cm}
    \centering
    \subfigure[\hspace{2mm} Expected payoff, \ac{qae} for noiseless devices]{\includegraphics[width=0.5\linewidth]{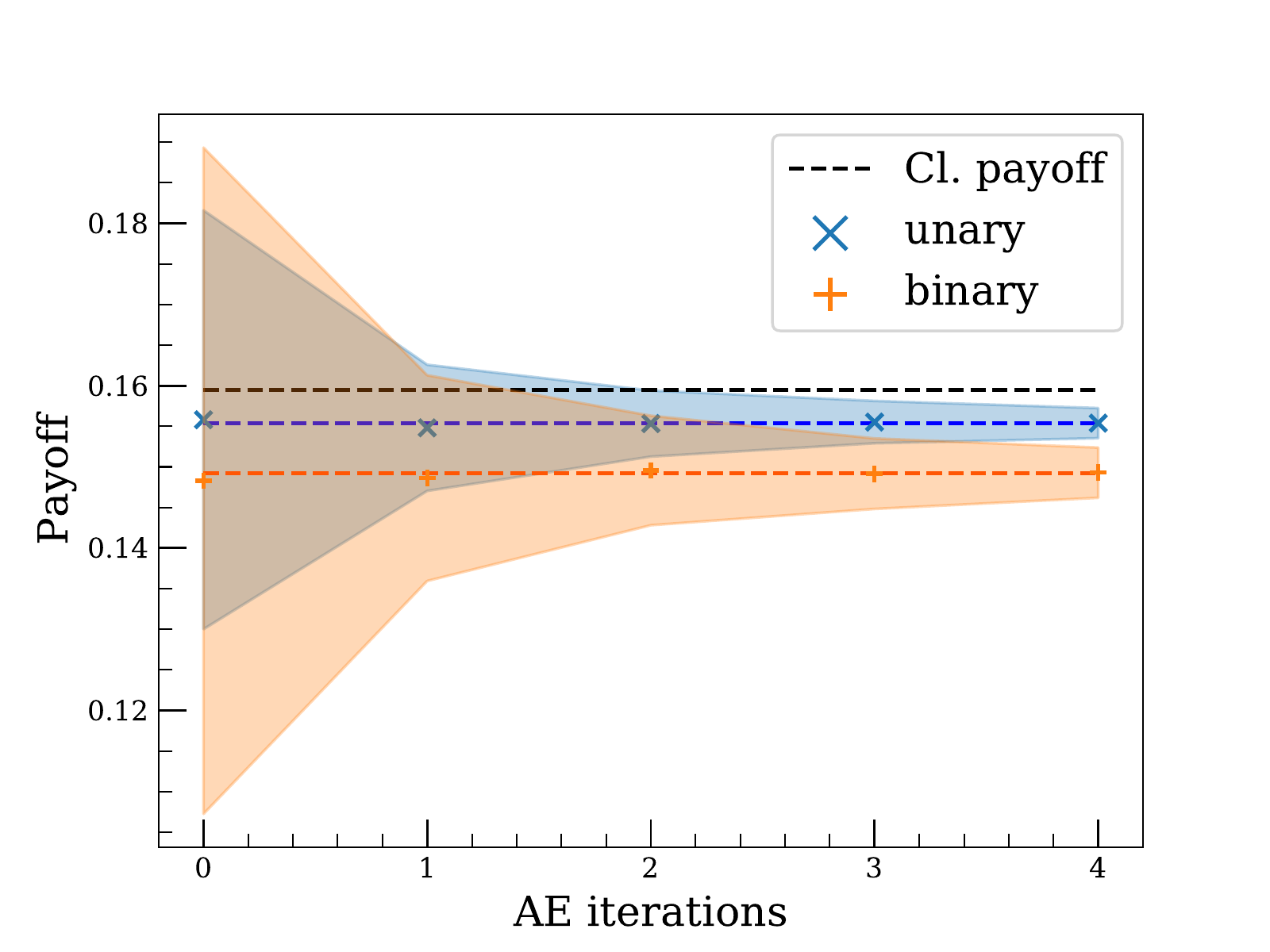}}
    \hfill 
        \subfigure[\hspace{2mm} Expected payoff, \ac{qae} for noiseless devices]{\includegraphics[width=0.45\linewidth]{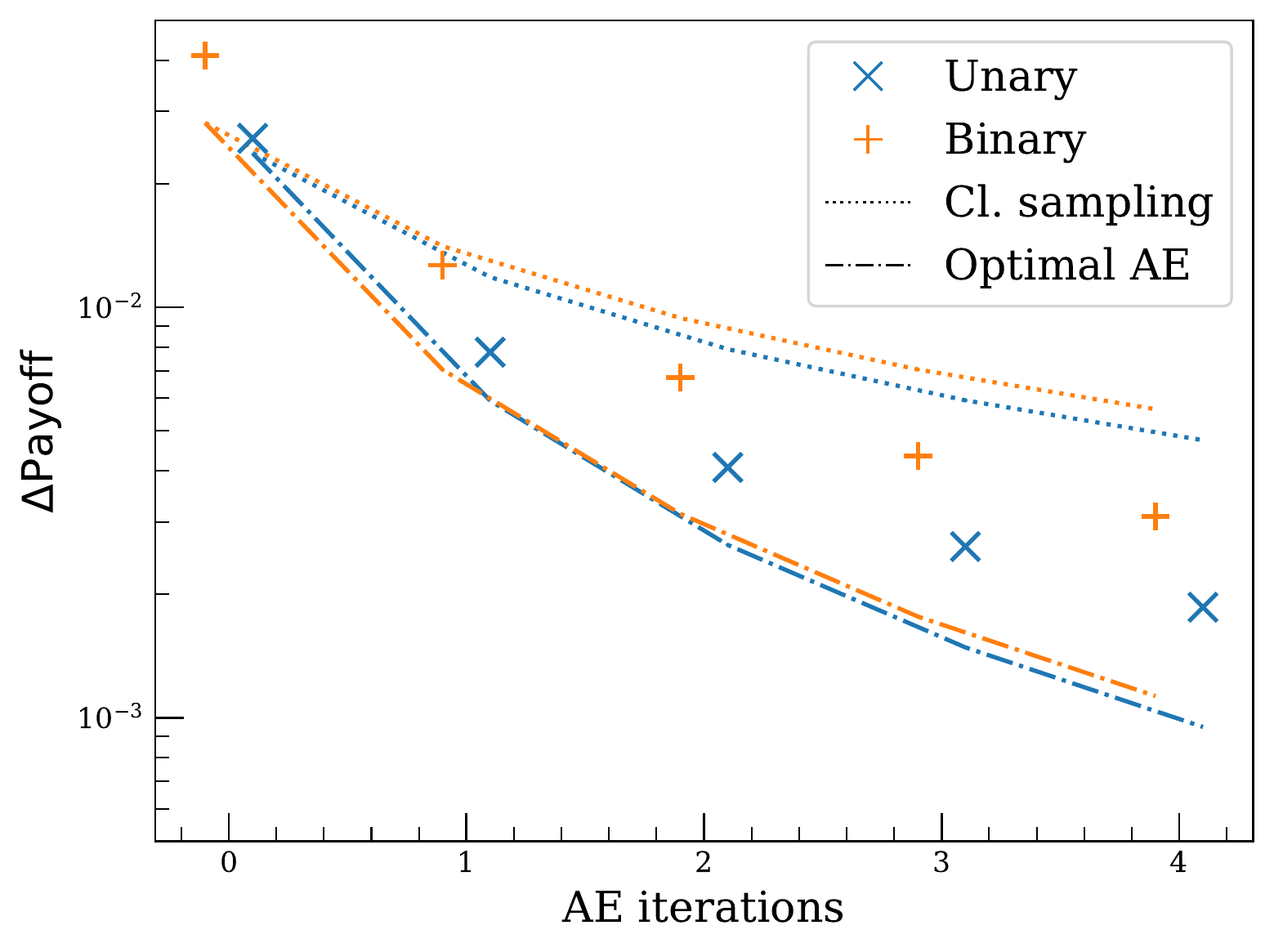}}
    \end{adjustwidth}
    \caption{a) Mean and uncertainty of the outcomes of the expected payoff, details can be found at App.~\ref{app:unary}. The dashed lines indicate the exact values. Unary and binary approaches are depicted, and convergence to the optimal values are obtained for both. Notice that these values are not the same since the outcomes of both algorithms are not equally related to the payoff due to bining. The shaded regions correspond to the statistical uncertainty. b) Statistical uncertainties in the expected payoff. The dotted lines indicate the uncertainty given by classical sampling, while the dot-dashed lines represent the optimal uncertainty provided by Amplitude Estimation. Results of the simulations lie in between. In this figure procedures with the same number of applications of the $\A$ or $\A^\dagger$ operators, for noiseless circuits, are compared.}
    \label{fig:convergence_results}
\end{figure}

Only \ac{qae} without \ac{qpe} can be performed on NISQ devices. In these simulations, a procedure based on weighted averages that consider both mean values and uncertainties is used, for a given series of \ac{qae} steps, see App. \ref{app:iqae} for further details. In our results, every instance has been repeated $100$ times. The choice of $m_j$ is linear, $m_j = j$, with $j=\{0, 1, 2, \ldots\}$, in order to control how the performance evolves. The confidence level was adjusted to $1 - \alpha = 0.95$.

\begin{figure}[t!]
    \centering
    \begin{adjustwidth}{-1cm}{-2cm}
    \subfigure[\hspace{2mm} Unary]{\includegraphics[width=.45\linewidth]{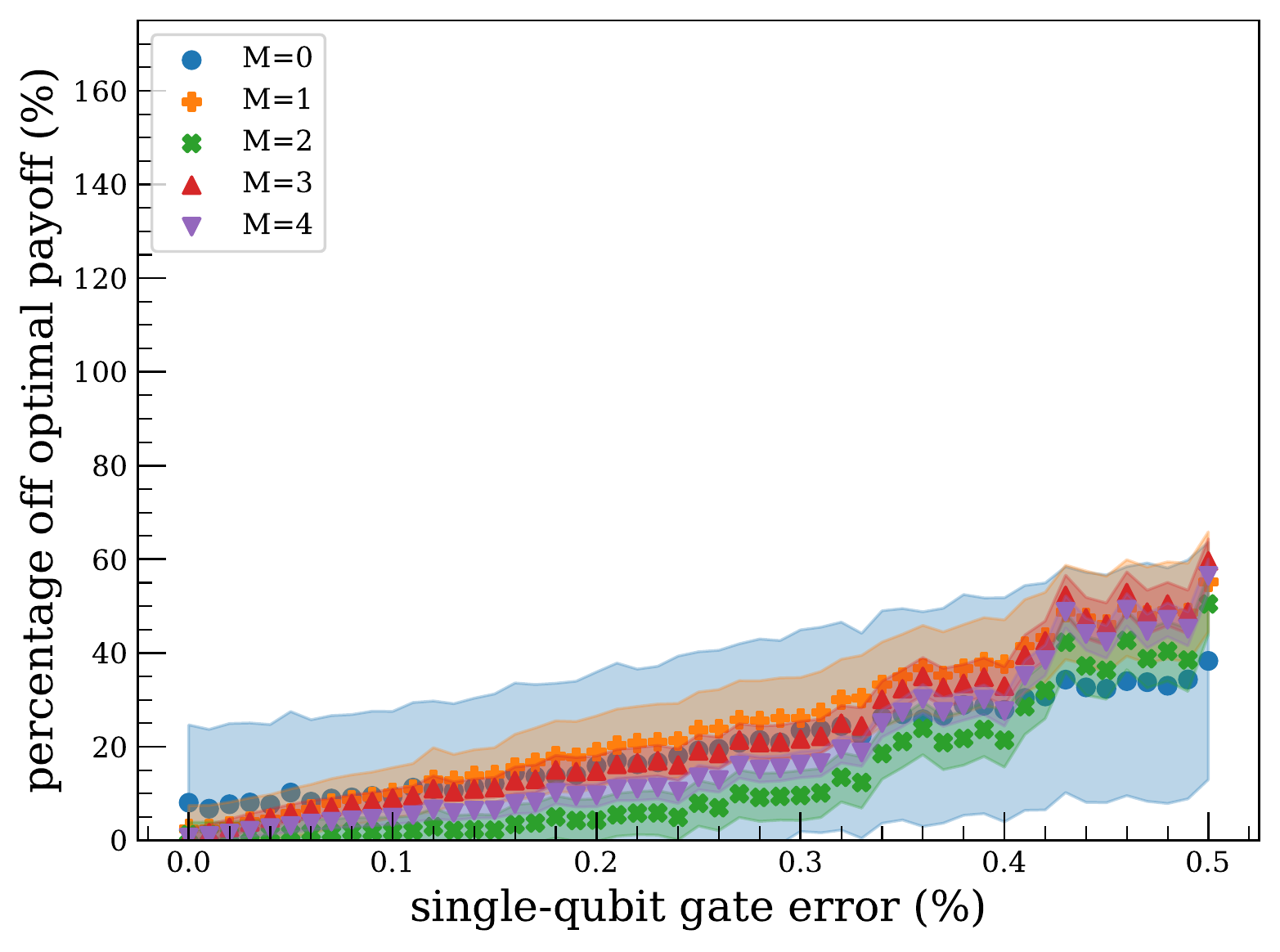}}
    \hfill \subfigure[\hspace{2mm} Binary]{\includegraphics[width=.45\linewidth]{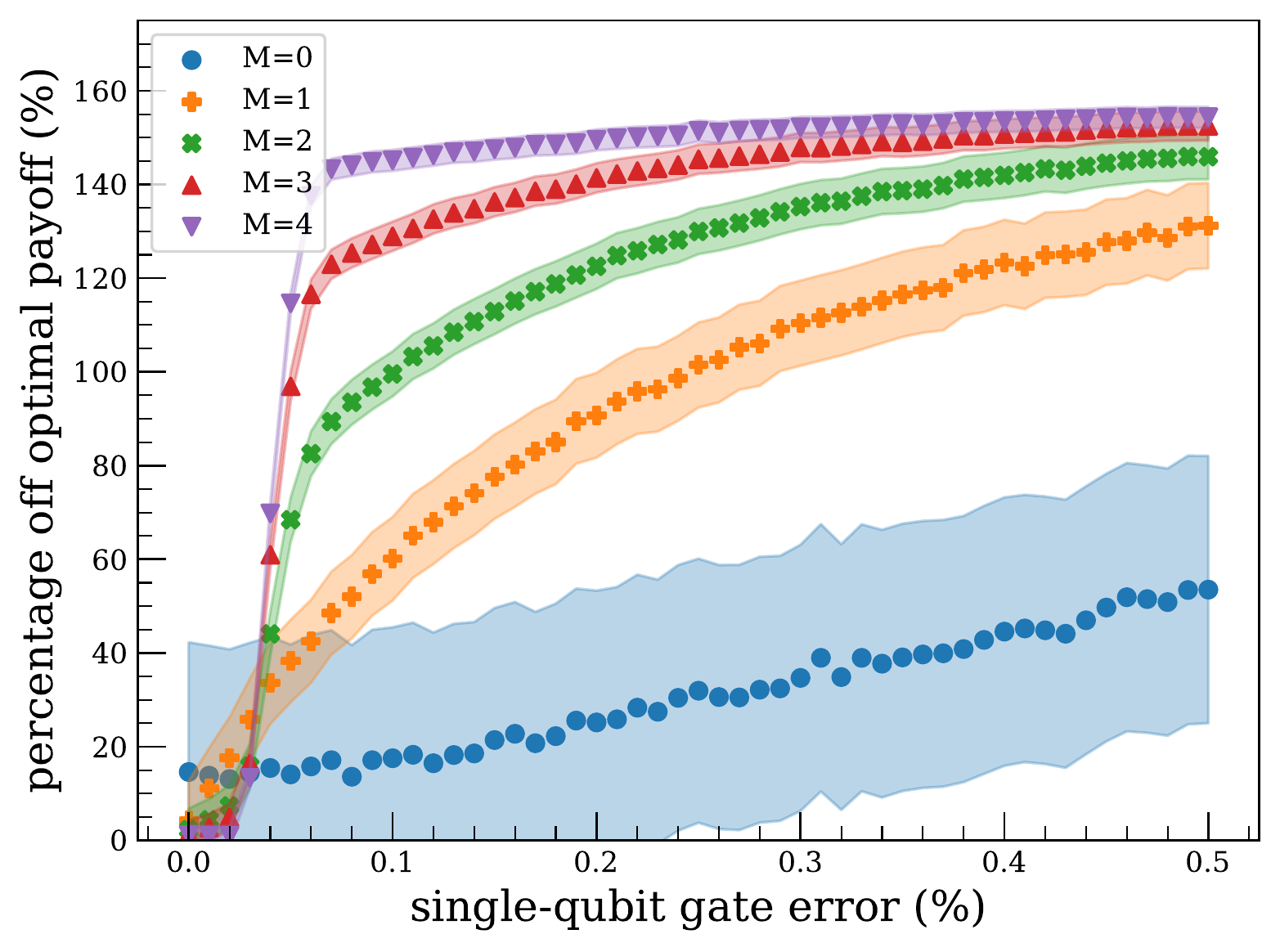}}
    \end{adjustwidth}
    \caption{Results of the errors in the expected payoff respect to the optimal value, for the unary (a) and binary (b) representation, with $M$ iterations of \ac{qae} considering depolarizing and read-out errors together. Scattering points stand for average values, while the shaded region corresponds to the statistical uncertainties. In the unary case, the expected payoff is resilient to errors, while the binary approach returns acceptable results only for $M=0$, while $M\geq 1$ rapidly saturates to a random circuit.}
   \label{fig:data} 
   \begin{adjustwidth}{-1cm}{-2cm}
\centering
    \subfigure[\hspace{2mm} Unary]{\includegraphics[width=.45\linewidth]{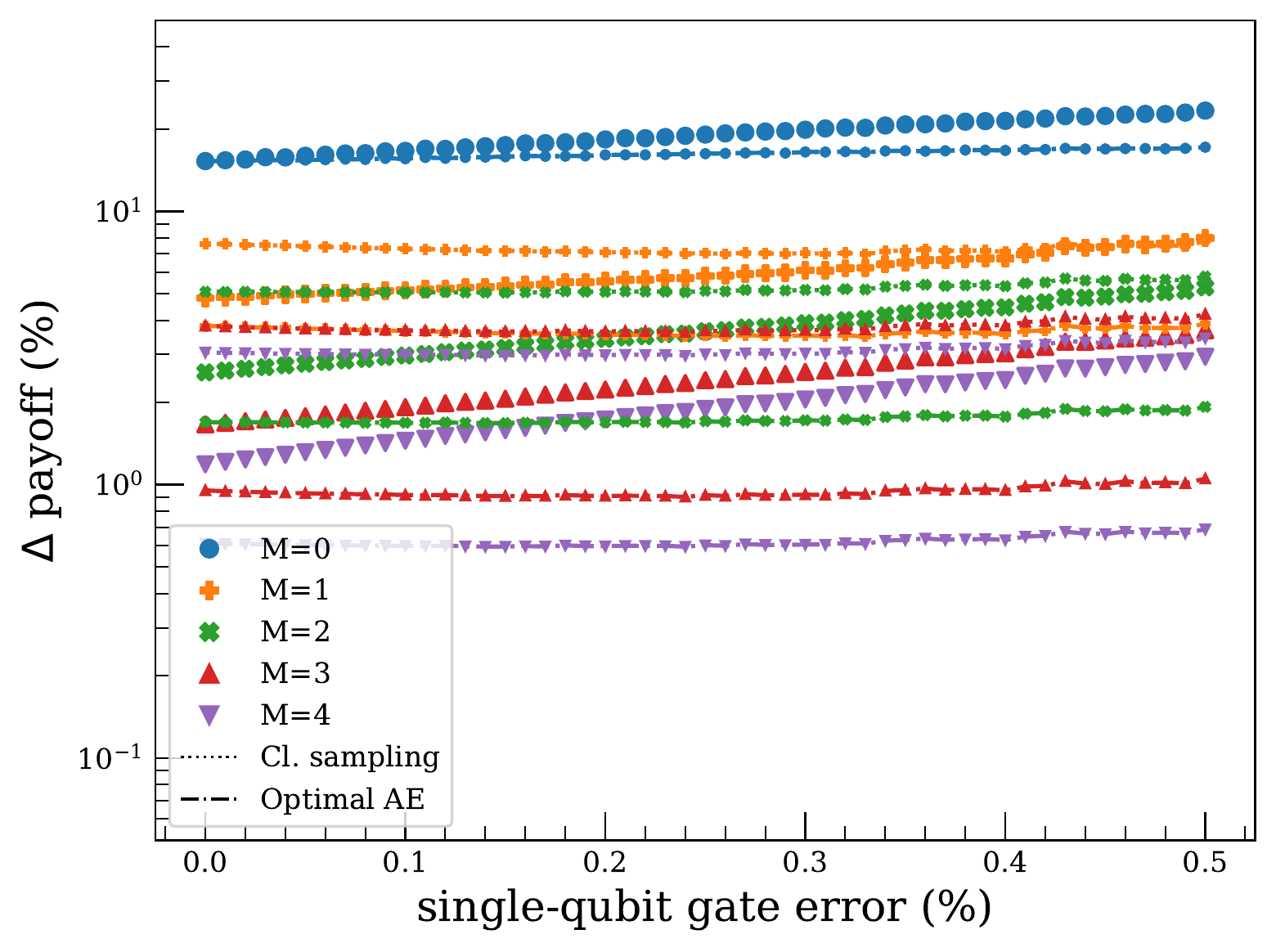}}
    \hfill \subfigure[\hspace{2mm} Binary]{\includegraphics[width=.45\linewidth]{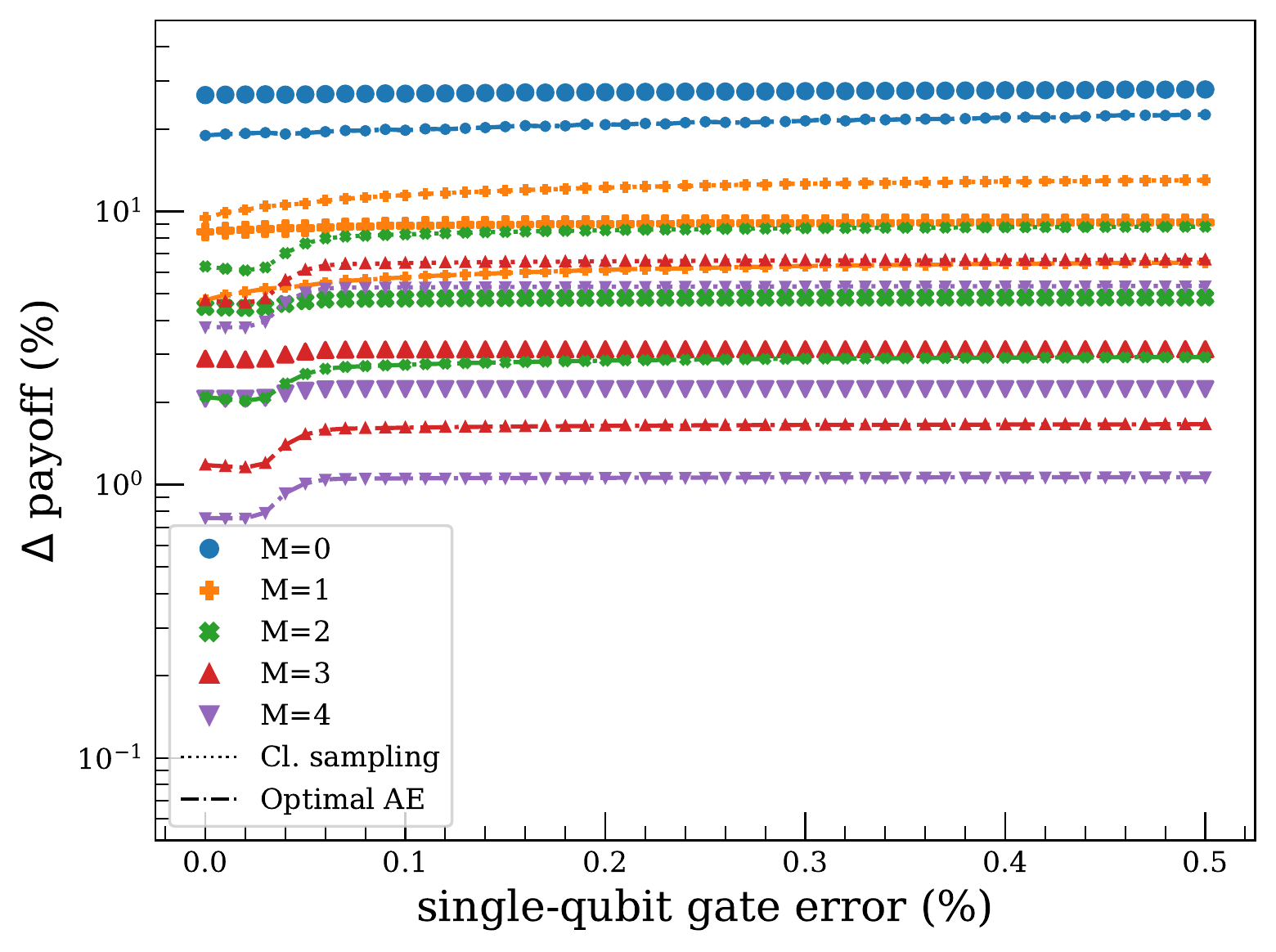}}
    \end{adjustwidth}
    \caption{Results of the sampling uncertainties of the expected payoff, same conditions as above. Scattering points represent the obtained uncertainties while dash-point lines represent theoretical bounds. For every color and symbol, the lower bound is for optimal \ac{qae}, and the upper bound is for sampling. In every case, each iteration of \ac{qae} reduces the uncertainty. For the unary case, the scattering points tend to return larger uncertainties as the errors increase, while for the binary case the uncertainties remain approximately constant. This difference is a direct consequence of the reduction of valid samples triggered by post-selection.}
    
    \label{fig:errors}
\end{figure}

Figure~\ref{fig:convergence_results} shows the increasing accuracy of the expected payoff using a \ac{qae} recipe as more iterations are utilized. These results confirm that \ac{qae} reduces the statistical uncertainty of the final results as more sophisticated circuits, that is with more iterations, are considered. 

Interesting results arise when considering the robustness against noise, in particular in this example depolarizing and read-out errors, of both unary and binary algorithms. The results in the deviation of the expected payoff with respect to the ideal case is depicted in Fig.~\ref{fig:data}. The number of \ac{qae} iterations was limited to 4 due to the computational cost of each simulation. Unary and binary algorithms show very different behaviors. First, the unary case endures the application of \ac{qae} with $M = \{0,1,2,3,4\}$ iterations when the noise levels are moderate. The errors in the expected payoff reach a $60\%$ for the maximum noise level allowed. The worsening of the results is gradual. Take, for instance $M = 2$. Results with low errors are obtained up to error rates $\varepsilon \sim 0.3\%$. Beyond this threshold, the returns become slightly more erratic. The results from the binary algorithm are totally different. When no \ac{qae} step is taken, $M=0$, the results are comparable to the unary ones, see for instance Fig.~\ref{fig:unary_binary_1}. However, the accuracy disappears completely at the first iteration $M \geq 1$ with small levels of noise $\varepsilon \sim 0.04\%$. A regimen of saturation is immediately reached in the binary case. This stationary regime corresponds to the random circuit where $a = 0.5$. The differences between both behaviors can be attributed to the post-selection regime resulting in a native mitigation of errors. The simpler structure of the unary circuit plays also a role.

It is also interesting to study the evolution of the uncertainty in the expected payoff calculation as more iterations of \ac{qae} are introduced into the circuit. This kind of errors is an exclusive consequence of the sampling uncertainty in the measurement step, which cannot be avoided. This can be observed in Fig. \ref{fig:errors}, where the obtained uncertainties are bounded between the classical sampling and the optimal \ac{qae}.

There appears a very remarkable behavior of the uncertainties in the unary approach to be noticed. The obtained uncertainties present a tendency to increase as errors get larger, unlike in the binary algorithm that does not present this feature. The reason lies in the native post-selection procedure only applicable in the unary representation. As errors become more likely to happen, the post-selection filter rejects more instances. The direct consequence is that the number of accepted shots drops for large errors, causing less certain outcomes. The joint action of this processes is that the uncertainty decreases more slowly for the unary algorithm than for the binary one. This behaviour contrasts with the error obtained in Fig. \ref{fig:data}, where the binary results reflect a poor performance.

\begin{wrapfigure}{O}{.4\linewidth}
\includegraphics[width=\linewidth]{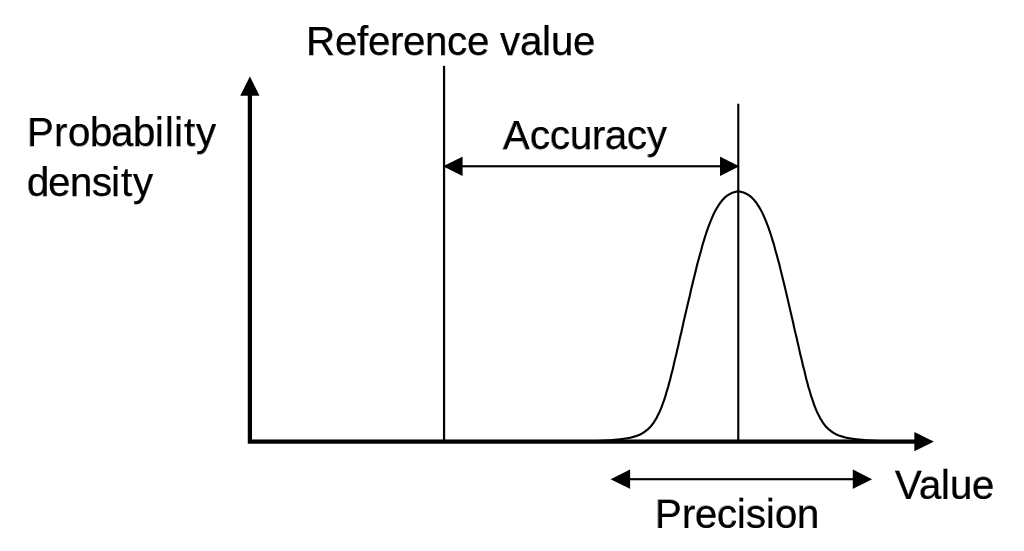}
\caption{Graphical explanation for the difference between accuracy and precision \cite{accuracy_precision}.}
\label{fig:accuracy_precision}
\end{wrapfigure}
The apparently contradictory result is related to the distinction between {\sl accuracy} and {\sl precision}. 
{\sl Accuracy} stands for how close is a measurement to the exact value of a quantity, and {\sl precision} encodes the dispersion of different measurements, see Fig.~\ref{fig:accuracy_precision}. \ac{qae} is an algorithm to increase the precision of a measurement with respect to the number of samples, but it does not provide any further information regarding the accuracy. Indeed, \ac{qae} for the binary algorithm reflects the expected tendency for the increase in precision, but comes with very poor results in accuracy. The unary algorithm grows slower in terms of precision, but maintains more accurate results. This decrease in precision might lead to losing the quantum advantage provided by \ac{qae} in the presence of significant error. In App. \ref{app:iqae} the limit of \ac{qae} iterations that can be performed given the error rates of the quantum device while still maintaining quantum advantage for the unary representation is further studied.

\begin{figure}[b!]
\centering
\begin{adjustwidth}{-1cm}{-2cm}
    \subfigure[\hspace{2mm} Errors of expected payoff]{\includegraphics[height=.35\linewidth]{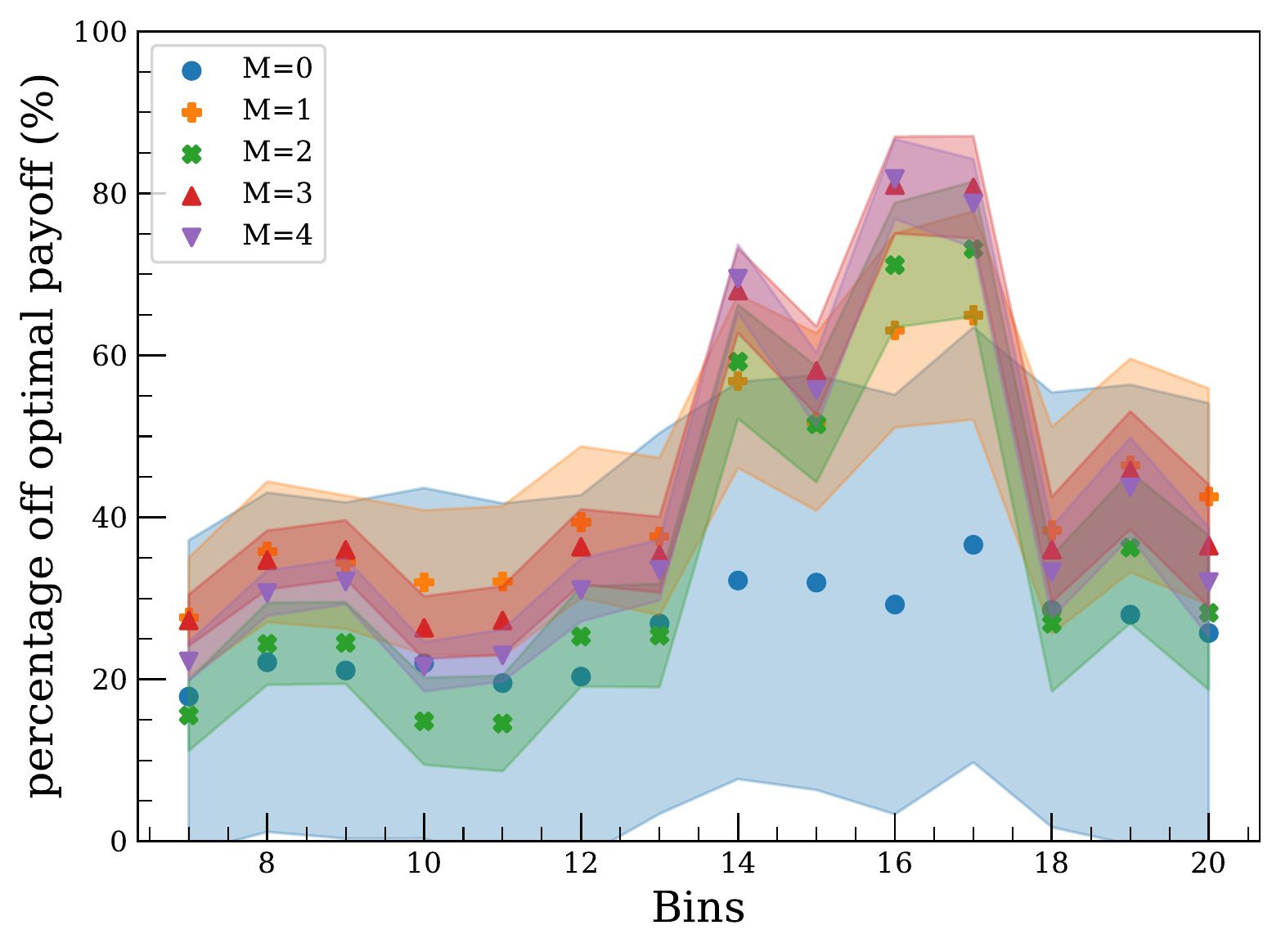}}
    \hfill
    \subfigure[\hspace{2mm} Sampling uncertainties of expected payoff \label{fig:several_bins_errors}]{\includegraphics[height=.35\linewidth]{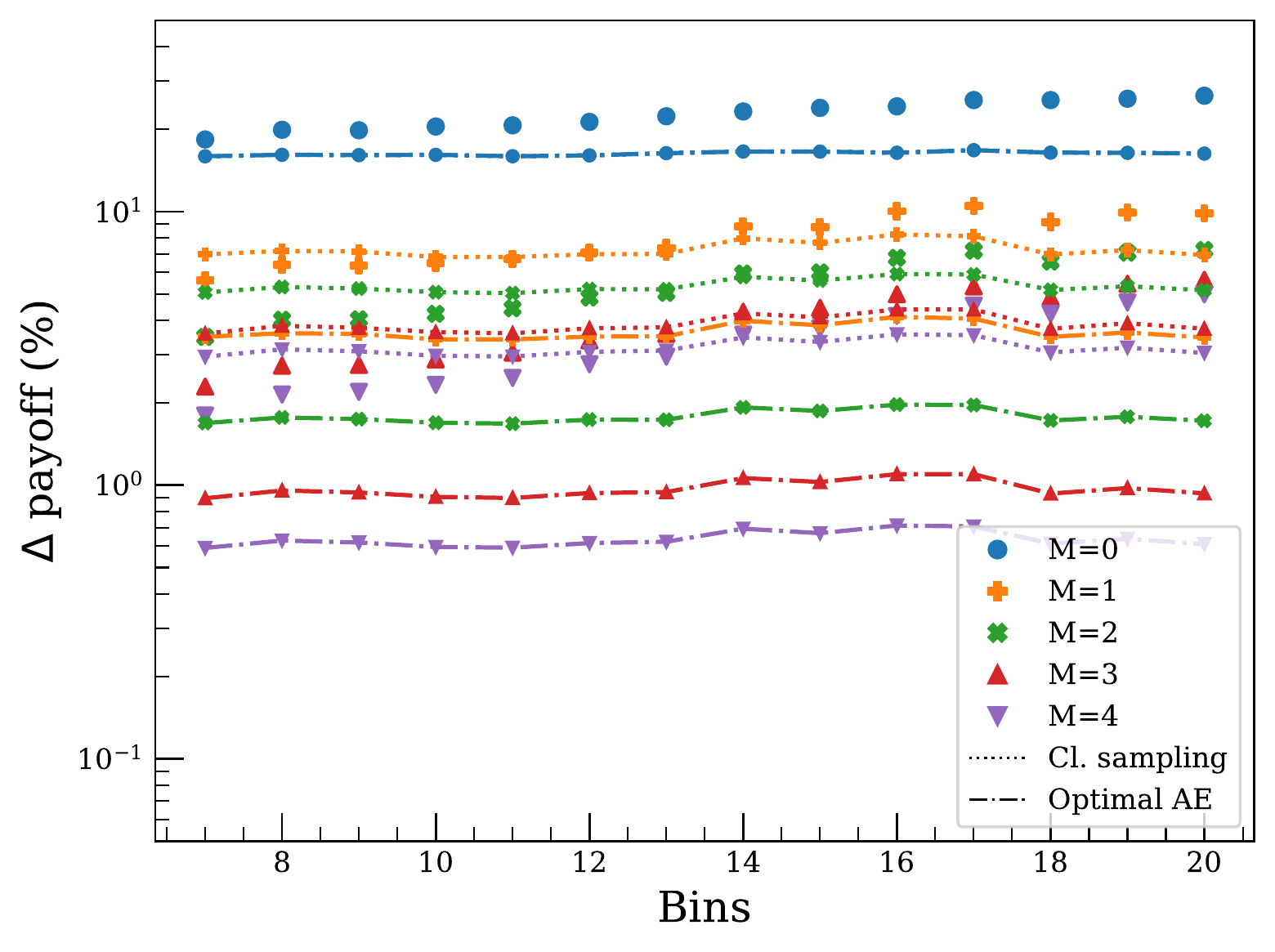}}
\end{adjustwidth}
        \caption{Results for the error and sampling uncertainties of the expected payoff for increasing number of bins for up to $M$ iterations of Amplitude Estimation for the unary approach, considering depolarizing and read-out errors together. For the error in payoff, scattering points represent the mean values obtained for the experiment, while shadowed areas include 70\% of the instances. In the sampling uncertainties, scattering points represent uncertainties obtained and dash-point lines represent theoretical bounds, where each line is accompanied with the corresponding marker. For every color and symbol, the lower bound is for optimal quantum advantage, and the upper bound is for sampling. The noise lvel is fixed to $\varepsilon = 0.3\%$. Each experiment is repeated only 10 times to reduce computational costs.} 
    \label{fig:several_bins}
\end{figure}

The results presented up to this moment support the use of \ac{qae} procedures even in \ac{nisq} devices in the unary representation. The resilience against noise shown by this approach is greater than in the binary algorithm. The noise here considered must be moderate to retrieve useful information from the calculation. However, the noise levels here considered are compatible with state-of-the-art computer~\cite{google_supremacy_2019}.

As a last step, the results of the unary algorithm are extended to larger number of bins $8 \leq n \leq 20$. Figure~\ref{fig:several_bins} show the deviation in the expected payoff and the sampling uncertainty for many different numbers $n$ with a noise level fixed at $\varepsilon = 0.3\%$ for depolarizing and measurement errors. The purpose of this calculation is to extract the behavior of the deviation in the payoff as more qubits and complex circuits are taken into account. It is clearly seen that the errors increase with the number of bins $n$, as expected. Larger systems imply more gates, and thus the errors are more likely to appear. In particular, for between 13 and 18 qubits, a larger error is observed. It is expected that this behavior encounters a completely random regime for a sufficiently large number of gates, although this regime is not observed. In contradistinction, the binary algorithm finds this situation at early stages. As in previous results, the increasing uncertainty with the number of bins $n$ reflect the more measurements rejected due to larger error probabilities. A slower convergence is the direct consequence.

\section{Conclusions}\label{sec:conclusions_unary}
This chapter has explored the strategy of encoding information in a quantum state using a low level of compression. In particular, the unary representation here presented utilizes only those state in the computational basis with only one $\ket 1$ among all qubits, while all others are $\ket 0$. This choice is not unique, but it is representative of the idea of dilluting the information across a large Hilbert space. The unary approach brings a clear advantage with respect to standard algorithms, namely the simplification achieved with the unary representation make the computation easier to execute and more resilient to noise and decoherence. It is even possible to reach quantum advantage in this regime. On the other hand, the storage capability of the quantum state is exponentially reduced. 

The range of applicability of the unary representation is a \ac{nisq} regime with few and noisy qubits, at a middle stage between state-of-the-art current and fault-tolerant computers. In case the unary representation returns a simplified circuits easier to implement in a quantum computer for small numbers of qubits, the execution is more robust against noise and more profitable. This situation cannot be maintained for large number of qubits since the exponential capabilities of standard algorithms overcome any other feature of the algorithm. 

The unary representation is tested for solving the financial problem of European option pricing. The economical problem is solved by means of the celebrated Black-Scholes models for the evolution of stochastic prices. The accuracies required in finance to solve this problem usefully are about $< 1\%$. These values are compatible with developing a unary algorithm in a regime with few qubits.
The problem is solved in three steps, namely uploading of a probability distribution of prices, computation of the expected return and an iterative \ac{qae} procedure. Each piece makes explicit use of the unary representation to simplify its implementation. 

For uploading the distribution of prices at the maturity date when the option expires, the unary representation uses a circuit where the only operations are partial-SWAP gates. The parameters controlling these gates can be found by classical means if the classical distribution is known, which is guaranteed in this problem. The circuit has a linear depth with the number of qubits. 
For the computation of the expected return, all steps can be taken with controlled-rotation gates at most. The number of operations is at most equal to the number of qubits. The quantity of interest is transported to an ancillary qubit. 
In the iterative \ac{qae} step, the required operators offer a much simpler implementation, allowing for a general reduction in the algorithmic complexity. The overall simplification comes in terms of number of gates and required connectivity to accomplich all operations. 

The unary representation permits a native post-selection method that results in a strong error mitigation. Since the unary algorithm resides in a restricted region of the Hilbert space, the outcomes must reflect this property. Any outcome with zero or more than one qubit in the $\ket 1$ state is automatically rejected. This ability reduces the acceptance of erroneous outcomes mitigating errors and increasing the performance of the quantum algorithm. In exchange, the effective number of measurements is reduced due to the active filter.  

The use of \ac{qae} pieces triggers the presence of quantum advantage by substituting Monte Carlo methods with its quantum analogues. This advantage could even be checked on state-of-the-art quantum computers. Experimentally, the advantage is exclusive of the unary algorithm since the binary one presents so erratic results that no useful quantity can be extracted. In summary, \ac{qae} on unary representation allows to maintain robust results at the price of reducing the convergence rate. Nevertheless, the attainment of quantum advantage is still feasible.

The results presented through this chapter entails that using quantum algorithms with a dilluted encoding of information may present advantages in the \ac{nisq} era with respect to the standard dense algorithms. The advantages are essentially a greater resilience to noise and a greater resilience to noise and simple implementation of operations. The advantage can only be obtained in a regime with few qubits. In some particular cases, like the one here presented, a dilluted representation can bring quantum advantage even for real-world applications.

%auto-ignore
\chapterimage{chapter_conclusions.pdf}
\chapter{Conclusions and final remarks}
\begin{adjustwidth}{4cm}{0cm}
{\sl It is better to be lucky. But I would rather 
be exact. Then when luck comes you are ready.
}

\hfill Ernest Hemingway\\
\end{adjustwidth}

This thesis covers two different strategies related to seizing quantum computers during the \ac{nisq} era, known as re-uploading and unary strategies. The aim of both strategies is to take profit of two purely quantum properties that settle the difference between quantum and classical computing: superposition (for re-uploading) and entanglement (for unary). Even though both models are related to different aspects of quantum computing, they explore different manners to treat and encode the information within a quantum computer. This freedom to choose among different approaches permits to take advantage of specific properties to overcome a variety of barriers. \\

The re-uploading strategy exploits the way to encode information in the most compressed way available. The quantum systems utilized in this case are minimalistic, starting from only one qubit, and it is shown that this approach provides enough computational power as to solve non-trivial tasks. In particular, the re-uploading strategy is a general technique to bring the fields of \acf{ml} and quantum computing together.

The tasks solved by means of the re-uploading procedure commonly involve being capable to store a function of some independent variables in several operations applied to an arbitary quantum state. The operations are driven by the independent variables and a set of tunable parameters whose optimal configuration must be found via external methods, for example classical optimizers. The differential element of the re-uploading strategy with respect to other \ac{qml} techniques is that data must be introduced sequentially and several times in the quantum system. This procedure permits to load and process data in the same step along several iterations, unlike in most \ac{qml} examples, where both steps are separate. The performance of the re-uploading method is attached both to the number of re-uploadings of the independent variables, it improves as data is introduced more times, and also to the optimality of the tunable parameters. It is demonstrated in this thesis that the re-uploading model is formally equivalent to other classical \ac{ml} models. 

Regardless of the exact implementation of the method here presented and the problems to be solved, it can be learnt from the present work that the complexity of small systems reaches high levels as compared to system of similar sizes for classical computers. One qubit, which is the fundamental unit of quantum computing, has enough room in its internal degrees of freedom as to encode arbitrarily accurate approximations of any dataset. It is demonstrated in this thesis that the re-uploading model is formally equivalent to other classical \ac{ml} models, in particular to single-layer \acf{nn}.

This fact comes from the superposition property and it is the representa-tion of the continuous nature of qubits, previous to measurement. There is an aparent contradiction with the Holevo bound, stating that one measurement of a qubit only returns one bit of information. In this case, the quantum state can be highly complex, although many measurements are required for extracting the whole description of the system. 

Another ingredient of the re-uploading scheme is the quantumly natural emergence of non-linearities, which are required for creating universal methods. The joint action of non-commutative quantum operations makes appear high-order dependencies of the independent variable. The more times the variable is introduced into the quantum system, the more possibilities of high-order terms are available, easing the process of learning arbitrary datasets. 

Therefore, the re-uploading scheme posseses two purely quantum properties that allows for a great flexibility of even single-qubit systems to solve non-trivial problems, namely superposition and non-commuting operations. However, the assistance of classical methods is still needed to manage these capabilities, that is, classical computers are needed to achieve the highest possible compression levels in a quantum computer, using the re-uploading scheme. 

From a practical perspective, it is shown that the re-uploading strategy performs successfully when facing a variety of problems on regression and classification of data. First, classical simulations of the quantum method were attempted, to pave the way towards experimental implementations. For simple problems suiting small computers, experiments on superconduc-ting and ion-trap qubits were satisfactorily performed, while problems requiring larger computers do not return meaningful results. \\ 

On the other hand, the unary strategy focuses on spreading some information across a entangled many-qubit quantum system with larger theoretical storage capabilities. That is, the information is encoded into quantum systems in a sparse manner. Thus, much of the available Hilbert space remains unused. Reducing the available space translates into a losing of asymptotical performance, but as a trade-off a great resilience against noise is obtained. Even though the computational power is limited on purpose, it is still possible to achieve quantum advantage with this procedure.

The first advantage of this scheme is that operations are simplified as compared to standard full encodings. Therefore, the collection of gates required to perform a given operation decreases both in number and in hardware requirements. This is useful in the present time since accuracies of experimental quantum gates are limited. The immediate consequence is that the cumulative error of a circuit execution is reduced.

Another advantage is that, since the Hilbert space is not crowded with information, measurements can return more information with less executions. Would information be partially lost due to errors and noise, it is possible to develop measurement strategies with mechanisms to detect errors. In particular, the example here presented exhibits a post-selection mechanism that allows to easily detect and discard corrupted outcomes.

The aim of the unary algorithm is to be useful during the first stage of the \ac{nisq} era. The trade-off between performance and resilience against noise brings an advantage for the unary algorithm, at least for those problems whose size requirements do not exceed the range of advantage. An example of usage for the unary strategy is given in this thesis, where the prominent finance problem of option pricing is solved. It is shown that \ac{nisq} computers with few qubits can exhibit advantage with with acceptable precision range in this example. \\

Both strategies are in a position to be implemented on current or near-future quantum devices, thus contributing with useful recipes to the status of \ac{nisq} computing. There are still many open research lines in both strategies to improve the performance and applicability of these methods. For example, the re-uploading strategy could benefit of efficient training procedures and specific embeddings to upload data, while the unary strategy could still explore further error mitigation and detailed differences between experimental implementations of the recipe here presented and other standard methods taken as reference.

The joint exploration of both paths covers two possible situations in the \acf{nisq} era, namely a) having few well-controlled or even logical qubits and b) having many noisy qubits. Depending on the capabilities of a given hardware and the nature of a problem to be solved,  one could choose between these two or others strategies to find the one that suits best the needs of the situation. 

The strategies here presented are examples of a potential spectrum of methods balancing the degree of compression of some information in a quantum circuit and the operational accuracy of the available hardware. As a general trend, the better control is available in the computer including circuit depth and connectivity, the denser states can be handled without a significant loss of information. 

It is of utmost importance to highlight that these methods are only available for quantum and not for classical computing. Both inherently quantum properties of superposition and entanglement are seized in the algorithmic strategies here presented. Those are hardly simulatable using classical means, in particular for quantum systems with large numbers of qubits. Therefore, this thesis aims to show that the range of possibilities that quantum computing opens up goes beyond the common belief of simultaneous processing of many numerical instances. 

This work looks for paving the way towards more sophisticated and tailored methods capable to take advantage of all quantum resources for computing, possibly by joining the effects of superposition and entanglement, and to extend its applicability to large fault-tolerant systems. The further development of experimental implementations can help in the future to bring this goal closer to reality.

\part*{Appendices}
\appendix

%auto-ignore

\chapterimage{chapter_appendix.pdf}
\chapter{Re-uploading strategy}\label{app:reuploading}

\section{Classical UAT for complex functions}\label{app:real_to_complex_uat}
The standard formulation of the UAT supports the approximation of complex function using $e^{i (\cdot)}$ as the activation function.

The approximations according to the \ac{uat} of the function are followed
\begin{equation}
    z(\vec x) = a(\vec x) + i b(\vec x), 
\end{equation}
using trigonometric functions as $\sigma(\cdot)$, 
\begin{eqnarray}
a(x) = \sum_{j=1}^N \alpha_i \cos(\vec w_j \cdot \vec x + a_j) \\
b(x) = \sum_{j=1}^N \beta_i \sin(\vec v_j \cdot \vec x + b_j).
\end{eqnarray}
Then
\begin{equation}
z(x) = \sum_{j=1}^N \alpha_i \cos(\vec w_j \cdot \vec x + a_j) + i \sum_{j=1}^N \beta_i \sin(\vec v_j \cdot \vec x + b_j),
\end{equation}

and this equation is can be rearranged as 
\begin{equation}
    z(x) = \sum_{j=1}^N \frac{\alpha_j}{2}\left( e^{i (\vec w_j \cdot \vec x + a_j)} + e^{-i ( \vec w_j \cdot \vec x + a_j)}\right) + \frac{\beta_j}{2} \left(e^{i (\vec v_j \cdot \vec x + b_j)} - e^{-i (\vec v_j \cdot \vec x + b_j)} \right),
\end{equation}

what encourages the UAT formulation for complex functions as an analogous to Eq. \eqref{eq:UAT}
\begin{equation}\label{eq:complex_UAT}
    G(\vec x) = \sum_{n=1}^N \gamma_n e^{i\delta_n} e^{i \vec u_n \cdot \vec x}.
\end{equation}

\section{Mathematical theorems for UAT}\label{app:math_theorems}

\begin{theorem}\label{th:hahn_banach}
{\bf: Hahn-Banach}~\cite{hahn_analysis_1927, banach_analysis_1929}

Set $\mathbb{K} = \mathbb{R} {\;\rm or\;} \mathbb{C}$. Let $V$ be a $\mathbb{K}-$vector space with a seminorm $p: V \rightarrow \mathbb{R}$. If $\varphi : U \rightarrow \mathbb{K}$ is a $\mathbb{K}-$linear functional on a $\mathbb{K}-$linear subspace $U\subset V$ such that
\begin{equation}
|\varphi(x)| \leq p(x) \qquad \forall x \in U,
\end{equation}
then there exists a linear extension $\psi : V \rightarrow \mathbb{K}$ of $\varphi$ to the whole space $V$ such that
\begin{eqnarray}
\psi(x) = \varphi(x) \qquad \forall x\in U \\
|\psi(x)| \leq p(x) \qquad \forall x\in V
\end{eqnarray}
\end{theorem}

\begin{theorem}\label{th:riesz}
{\bf: Riesz Representation}~\cite{riesz_analysis_1914}

Let $X$ be a locally compact Hausdorff space. For any positive linear functional $\psi$ on $C(X)$, there exists a uniruq regular Borel measure $\mu$ such that
\begin{equation}
\forall f \in C_c(X): \qquad \psi(f) = \int_X f(x) d\mu(x)
\end{equation}
\end{theorem}

\begin{theorem}\label{th:lebesgue}
{\bf: Lebesgue Bounded Convergence}~\cite{weir_analysis_1974}

Let $\lbrace f_n\rbrace$ be a sequence of complex-valued measurable functions on a measure space $(S, \Sigma, \mu)$. Suppose that $\lbrace f_n \rbrace$ converges pointwise to a function $f$ and is dominated by some integrable function $g(x)$ in the sense
\begin{equation}
|f_n(x)| \leq g(x), \qquad \int_S |g|d\mu < \infty
\end{equation}
then
\begin{equation}
\lim_{n\rightarrow \infty} \int_S f_n d\mu = \int_S f d\mu
\end{equation}
\end{theorem}

\section{Definitions of 2D functions for universality}\label{app:2D_benchmark}

\begin{figure}[h!]
\centering
    \includegraphics[width=1\linewidth]{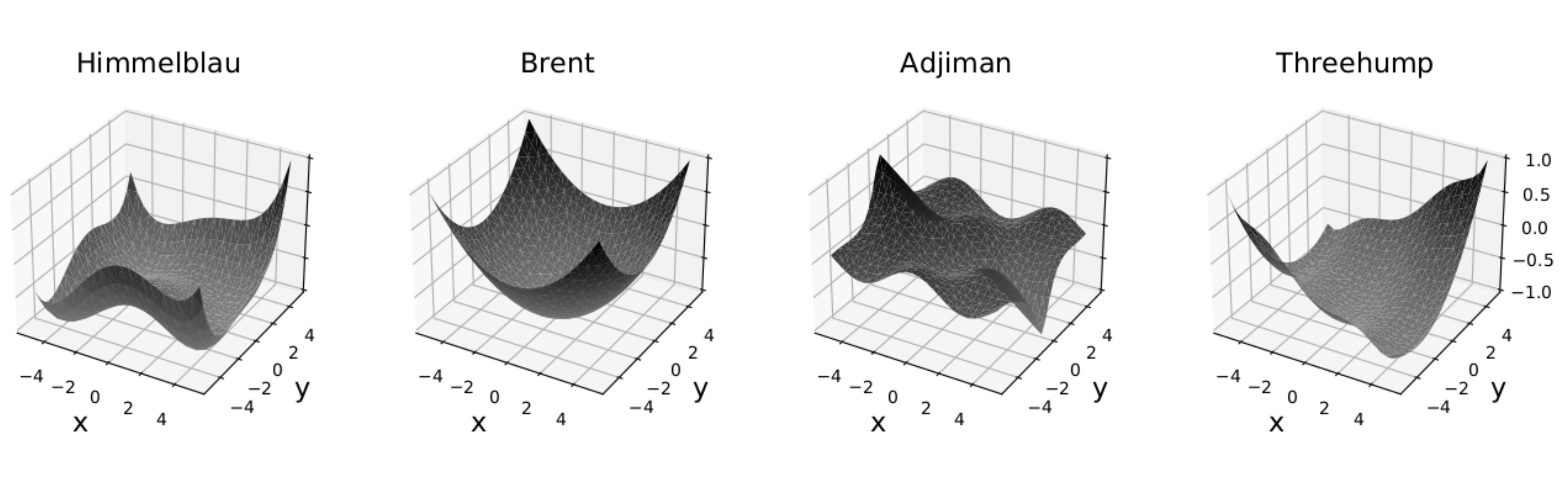}
    \caption{Graphical representation of 2-dimensional functions utilized for benchmarking. A regularization is applied to obtain $Z\in[-1,1]$.}
    \label{fig:2d_functions}
\end{figure}

The definitions used for the 2-dimensional functions \cite{2d_functions} that serve for benchmarking the proposed algorithms are defined as

\begin{eqnarray}
    {\rm Himmelblau}(x, y) = (x^2 + y - 11)^2 + (x + y^2 - 7)^2,
\\
{\rm Brent}(x, y) = \left(\frac{x}{2}\right)^2 + \left(\frac{y}{2}\right)^2 + e^{-\left(\left(\frac{x}{2} - 5\right)^2 + \left(\frac{y}{2} - 5\right)^2)\right)},
\end{eqnarray}
\begin{eqnarray}
\begin{split}
{\rm Threehump}(x, y) = 2 \left(\frac{2 x}{5}\right)^2 - 1.05 \left(\frac{2 x}{5}\right)^4 + \frac{1}{6}\left(\frac{2 x}{5}\right)^6 + \\ + \left(\frac{2 x}{5}\right) \left(\frac{2 y}{5}\right) + \left(\frac{2 y}{5}\right)^2,
\end{split}
\\
{\rm Adjiman}(x, y) = \cos(x) \sin(y) - \frac{x}{y^2 + 1},
\end{eqnarray}

where a normalization to $-1 \leq f(x, y) \leq 1$ is applied after this definition. A graphical representation of these functions is depicted in Fig. \ref{fig:2d_functions}.

\section[Superconducting experiment]{Superconducting experiment for a universal approximant}\label{app:exp_universal}

\begin{figure}[t!]
\begin{adjustwidth}{-2cm}{-1cm}
\centering
\subfigure[\hspace{1mm} $T_1$ and $T_{2E}$ fits \label{fig:coh-times} ]{\includegraphics[width=.45\linewidth]{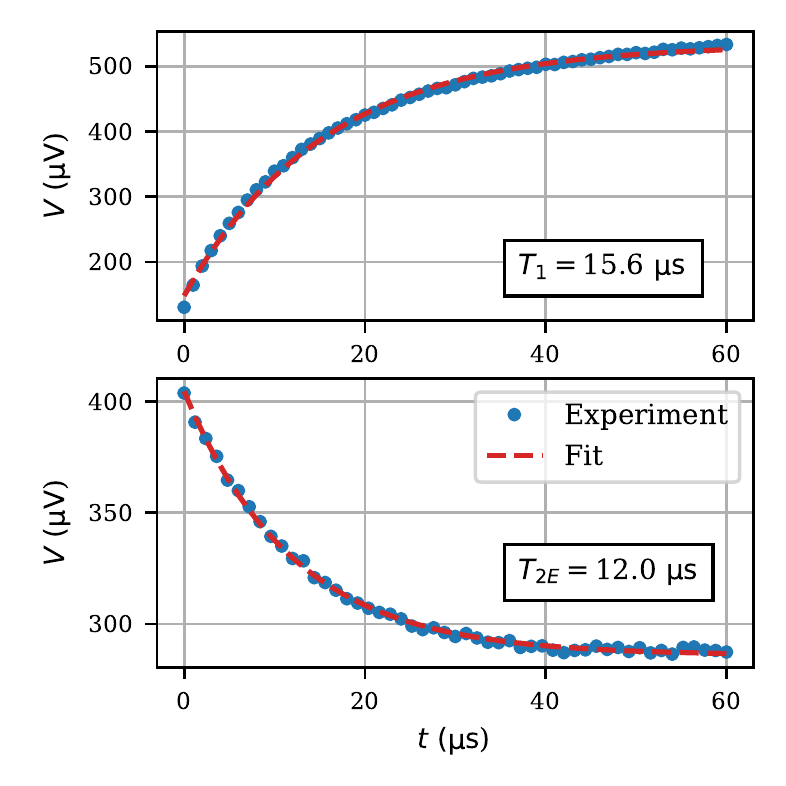}} \hfill
\subfigure[\hspace{1mm} Randomized Benchmarking \label{fig:benchmarking} ]{\includegraphics[width=.45\linewidth]{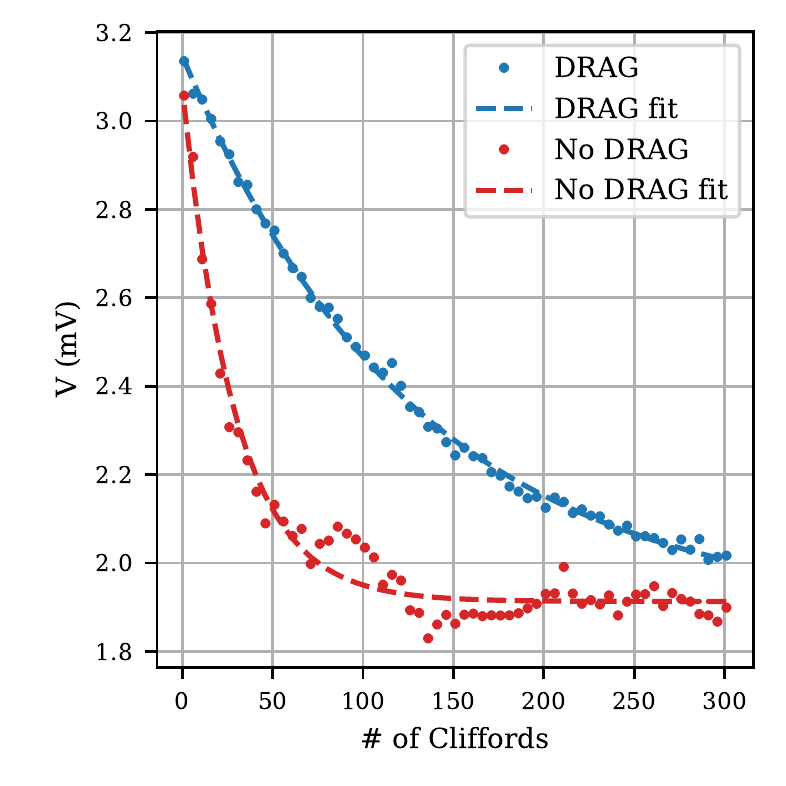}}
\end{adjustwidth}
\caption{a) $T_1$ measurement with exponential fit (top) and spin-echo measurement, $T_{2E}$, with exponential fit (bottom). b) Randomized benchmarking of the DRAG corrected pulses. The fit corresponds to the expression $Ap^n + B$, where $A$ and $B$ have dimensions of voltage, $n$ is the number of Clifford gates, and $p$ is the fidelity per gate. $\epsilon = 1-p$ is the error per gate.}
\end{figure}

The experimental implementation of the single-qubit universal approximant as detailed in Sec.~\ref{sec:benchmark} was performed in a superconducting qubit circui. The qubit is a 3D transmon geometry \cite{3d-transmon} located inside an aluminum three-dimensional cavity. The cavity bare frequency, $\omega_c = 2\pi \times 7.89$ GHz, is greatly detuned from the qubit frequency, $\omega_q = 2\pi \times 4.81 $ GHz. Hence, there is a qubit state-dependent dispersive shift on the cavity resonance, $2\vert\chi\vert = 2\pi \times 1.5 $MHz. The qubit anharmonicity is $\alpha = -2\pi \times 324$ MHz and the qubit relaxation and spin-echo decay times are, respectively, $T_1 = 15.6~\mu s$ and $T_{2E} = 12.0~\mu s$. These time scales exceed the operation times needed to implement the algorithm up to 6 layers by 2 orders of magnitude. See Fig.~\ref{fig:coh-times} for a experimental fit on these times. 

\begin{SCfigure}
\centering
\includegraphics[width=.5\linewidth]{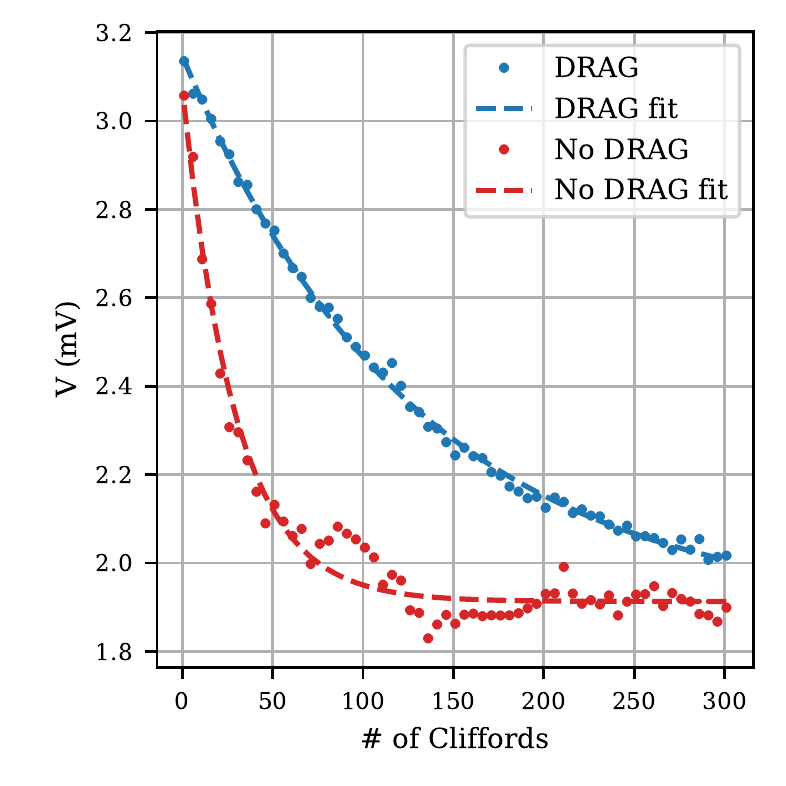}
\caption[Randomized Benchmarking]{Randomized benchmarking of the DRAG corrected pulses. The fit corresponds to the expression $Ap^n + B$, where $A$ and $B$ have dimensions of voltage, $n$ is the number of Clifford gates, and $p$ is the fidelity per gate. $\epsilon = 1-p$ is the error per gate.}
\label{fig:drag} 
\end{SCfigure}

The experiment was realized in a dilution fridge with a base temperature of approximately $20$~mK. The qubit rotation pulses were defined by an arbitrary waveform generator and then upconverted with a microwave signal generator to the gigahertz frequency range before being sent to the qubit/cavity system. The signal was low-pass filtered and attenuated by a total of 50dB before reaching the aluminum cavity. The input port of the cavity was undercoupled while the output port was overcoupled in order to maximize the readout signal amplitude. The outgoing signal was amplified by a cryogenic low noise amplifier and a second amplification stage at room temperature. The downconversion is performed with the same microwave generator as used in the upconversion of the measurement pulse, guaranteeing phase coherence in the downconversion process. The signal is read out in a digitizer, with a \ac{fpga} that demodulates and averages the results before sending the data to the main measurement computer.

\begin{figure}[b!]
\centering
\subfigure[\hspace{2mm} Full sequence]{\includegraphics[width=.45\linewidth]{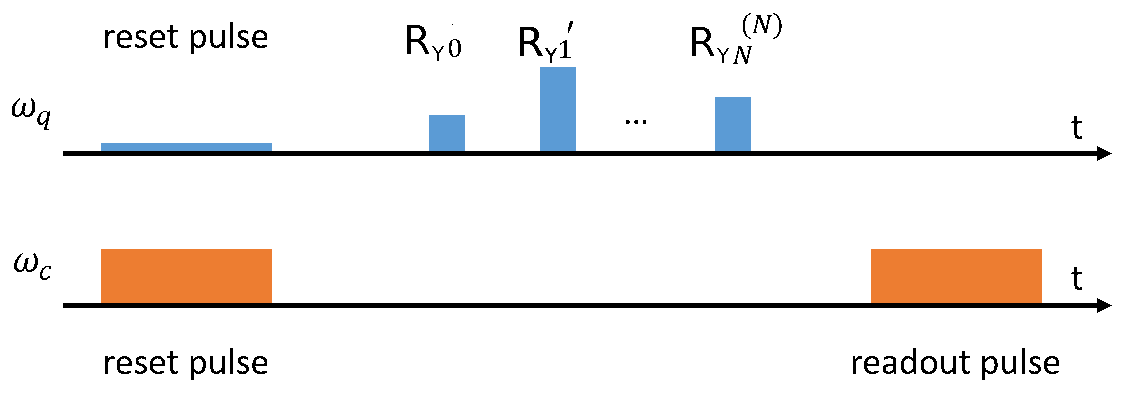}} \hfill 
\subfigure[\hspace{2mm} Pulse sequence]{\includegraphics[width=.45\linewidth]{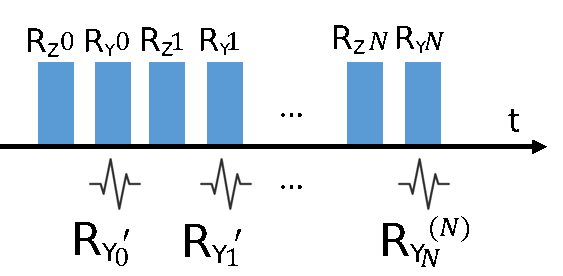}}
\caption{a) Complete pulse sequence. First, the reset protocol is performed which corresponds to two pulses at the cavity and the qubit frequencies, respectively. Note that the qubit pulse is of considerably lower amplitude than the cavity pulse. Also, both pulses have a longer duration than the qubit rotation sequence (timings not to scale). The $R_Y$ pulses are shown to have different amplitudes to determine each rotation angle. Finally, the readout corresponds to a pulse at the cavity frequency which is later read out by a digitizing card. b) Sequence performed in the experiment. Blue boxes represent actual pulses. Logical $R_Y$ and $R_Z$ rotations are explicitly shown below the blue boxes~\cite{virtual-zgates}. Note that $R_Z$ pulses do not correspond to any microwave pulse, instead subsequent pulses change rotation axis, indicated by a prime, $R_{Y,N}^{(N)}$.}
\label{fig:exp_gates} 
\end{figure}

In order to implement the gate sequences defined in Sec.~\ref{sec:benchmark}, the correspondence between logical and physical gates as shown in Fig.~\ref{fig:exp_gates} is followed. The phase of each pulse is selected at the pulse generator to modify the rotation axis, producing either $R_X$ or $R_Y$ rotations as required. The $R_Z$ rotations are, in turn, virtual~\cite{virtual-zgates}. The microwave pulses incorporate a DRAG correction~\cite{drag-a,drag-b} which leads to an error per gate $\epsilon = 0.01$ found with randomized benchmarking~\cite{randomized-benchmarking}, see Fig.~\ref{fig:benchmarking}. Randomized benchmarking measures errors in Clifford gates and not arbitrary angle rotations, which are instead used in this experiment, yet offers a reasonable estimate on the overall fidelity of the implemented gates. The gate error observed is probably limited due to a non-ideal filtering of the measurement lines in the fridge. In order to achieve better qubit state readout visibility and shorter operation times, a reset protocol is applied prior to the main sequence \cite{reset-protocol}. 

\begin{table}
    \centering
    \resizebox{\linewidth}{!}{
    \begin{tabular}{ *{13}{l} }
      \toprule
      \multicolumn{1}{c}{\textbf{Optimal}} 
      & \multicolumn{1}{c}{$p_1$}  & \multicolumn{1}{c}{$p_2$}  & \multicolumn{1}{c}{$p_3$}
      & \multicolumn{1}{c}{$p_4$}  & \multicolumn{1}{c}{$p_5$}  & \multicolumn{1}{c}{$p_6$}
      & \multicolumn{1}{c}{$p_7$}  & \multicolumn{1}{c}{$p_8$}  & \multicolumn{1}{c}{$p_9$}
      & \multicolumn{1}{c}{$p_{10}$}  & \multicolumn{1}{c}{$p_{11}$}  & \multicolumn{1}{c}{$p_{12}$}
      \\
      \textbf{parameters}
      & -2.501 & 1.685 &   1.757 & 2.105 & 3.822 & -1.788 &
 -1.507 & -4.640 & 0.430 & 1.875 & 5.038 & -1.906
      \\[0.2cm]
      \hline
      \multicolumn{1}{c}{\textbf{Rotational}} 
      & \multicolumn{2}{c}{$R_{Z,1}$}   & \multicolumn{1}{c}{$R_{Y,1}$}
      & \multicolumn{2}{c}{$R_{Z,2}$}   & \multicolumn{1}{c}{$R_{Y,2}$}
      & \multicolumn{2}{c}{$R_{Z,3}$}   & \multicolumn{1}{c}{$R_{Y,3}$}
      & \multicolumn{2}{c}{$R_{Z,4}$}   & \multicolumn{1}{c}{$R_{Y,4}$}
      \\
      \multicolumn{1}{c}{\textbf{angles$^*$}} 
      & \multicolumn{2}{c}{$p_1  + p_2 x$}   & \multicolumn{1}{c}{$p_3$}
      & \multicolumn{2}{c}{$p_4 + p_5 x$}   & \multicolumn{1}{c}{$p_6$}
      & \multicolumn{2}{c}{$p_7 + p_8 x$}   & \multicolumn{1}{c}{$p_3$}
      & \multicolumn{2}{c}{$p_{10} + p_{11} x$}   & \multicolumn{1}{c}{$p_{12}$}
      \\[0.2cm]
      \multicolumn{1}{c}{\textbf{$x=-0.5$}} 
      & \multicolumn{2}{c}{2.939}
& \multicolumn{1}{c}{1.757}
& \multicolumn{2}{c}{0.194}
& \multicolumn{1}{c}{4.495}
& \multicolumn{2}{c}{0.813}
& \multicolumn{1}{c}{0.430}
& \multicolumn{2}{c}{5.639}
& \multicolumn{1}{c}{4.377}
      \\
      \multicolumn{1}{c}{\textbf{$x=0$}} 
      & \multicolumn{2}{c}{3.782}
& \multicolumn{1}{c}{1.757}
& \multicolumn{2}{c}{2.105}
& \multicolumn{1}{c}{4.495}
& \multicolumn{2}{c}{4.776}
& \multicolumn{1}{c}{0.430}
& \multicolumn{2}{c}{1.875}
& \multicolumn{1}{c}{4.377}
      \\
      \multicolumn{1}{c}{\textbf{$x=1$}} 
      & \multicolumn{2}{c}{5.467}
& \multicolumn{1}{c}{1.757}
& \multicolumn{2}{c}{5.927}
& \multicolumn{1}{c}{4.495}
& \multicolumn{2}{c}{0.136}
& \multicolumn{1}{c}{0.430}
& \multicolumn{2}{c}{0.630}
& \multicolumn{1}{c}{4.377}
      \\[0.2cm]
      \hline
      \multicolumn{9}{l}{\small  $^*$ Angles between $0$ and $2\pi$}        
    \end{tabular}}
    \caption{Optimal parameters and angles obtained for $\relu(x)$ and 4 layers. Above the 12 parameters that define the rotational angles obtained through simulations. Below the corresponding angles of the 8 rotations for three different values of $x$. Note that $R_Y$ rotations are not $x$-dependent, hence they are equal for all three $x$ values.}
    \label{tab:rotation angles}
    
\end{table}

Figure~\ref{fig:exp_gates} shows the total pulse sequence, which includes preparation and measurement pulses in addition to the pulse sequence shown in the main text. The Y rotations are performed through microwave pulses at the qubit frequency while the Z rotations, as already stated, are phase changes in subsequent pulses. Both qubit and cavity pulses are generated at $70$~MHz and then upconverted to the gigahertz range. The qubit pulses are gaussian pulses with a total duration of $21$~ns. A proper DRAG correction is performed with a resulting error per gate of $\epsilon = 0.01$ as shown in Fig.~\ref{fig:drag}. The cavity pulse has a total length of around $2~\mu$s. The reset protocol consists of a pulse driving the qubit and a pulse driving the cavity mode, with a total duration of around $2~\mu$s. An example of the rotation angles for the $\relu(x)$ function in the 4-layer case is shown in Table~\ref{tab:rotation angles}. 

The readout consists of a cavity tone at the frequency of the cavity for the qubit in the $\ket{0}$ state. High/low transmission corresponds to the qubit being in the ground/excited state, assuming the system does not escape from the computational basis. Each data point requires around $5\cdot 10^4$ measurements in order to average out the amplifier noise. A reset protocol that drives the qubit into the ground state is implemented prior to each individual sequence. This has two benefits. The first one allows to start with a qubit state nearly polarized into the ground state. A second benefit is the reduction in the overall duration of the experiment, since the waiting time between individual measurements is not limited by the qubit relaxation time.

\section[Ion trap experiment]{Ion trap experiment for a universal classifier}\label{app:experimental_setup_classifier}

The ion-trap qubit utilized in this experiment is realized on a $^{138}$Ba$^+$ ion trapped and laser cooled in a linear blade trap. The mapping between computational and electronic states is $\ket 0 \rightarrow S_{\frac{1}{2},-\frac{1}{2}}$ and $\ket 1 \rightarrow D_{\frac{5}{2},-\frac{1}{2}}$. Both states are coupled by an electric quadrupole $E2$ transition at $1762~$nm wavelength. To manipulate the qubit an ultra-low-linewidth laser is compulsory. Laser cooling of the ion to the Doppler limit, ensures that the qubit is not influenced by the external motion. The most relevant parameters of the trapped ion in this experiment are the qubit coherence time ($5~$ms) and the Rabi $\pi$-time ($12~\mu$s) of the qubit~\cite{Yum:17,Dutta2020}. The qubit is well-characterized in terms of both its internal~\cite{Dutta2020} and external states~\cite{VanHorne2020}.

An ultra-low linewidth laser at $1762~$nm wavelength is required to control the qubit. The laser achieves an estimated linewidth of $\le 100~$Hz. Prior to performing each algorithmic cycle the qubit is first Doppler cooled to the Lamb-Dicke regime~\cite{Dutta2020} via a fast dipole transition (between S-P levels) at $493~$nm and a re-pump laser (between D-P levels) at $650~$nm. Subsequently, the qubit is initialized to the state $\ket{~0}\equiv\ket{~\text{S}_{\frac{1}{2},-\frac{1}{2}}}$ by optical pumping~\cite{Dehmelt1957}, achieved by selectively de-populating the state S$_{\frac{1}{2},+\frac{1}{2}}$. A high fidelity state initialization of $\ge 99\%$ is achieved within an optical pumping time of $\le 50\mu$s only. Once the qubit is initialized, any single qubit rotational gate is implemented by resonantly driving the qubit with full control over the laser phase, power and laser on-time. A direct digital synthesizer is used to control the phase, frequency and power of the laser as the sequential gates are applied on the qubit. The synthesizer eliminates the long term frequency drift of the synthesizer clock, thus maintaining the phase relations of the sequential gates of the classifier. As classical data is uploaded in-the-fly, the latency of uploading the synthesizer parameters play a crucial role. The latency is minimized by pre-loading the full sequence of the phase, frequency and power data to an on-chip memory of the synthesizer. The synthesizer output is then controlled by an external trigger generated from a \ac{fpga}. 

The classification algorithm requires the repetition of any pair of non-commuting rotational gates. $R_z$ and $R_y$ are chosen, instead of the active gates of the ion qubit $R_x$ and $R_y$. Active rotations are realized by the application of resonant laser at the qubit frequency. On the other hand, $R_z$ can be implemented by a combination of the other two active gates or by varying the qubit energy~\cite{McKay2017, Maslov_2017}. Both these methods are error prone as the qubit-light interaction is switched on for certain time. It is chosen instead to perform the $R_z$ by changing the laser phase without interacting with the ion~\cite{Knill2008}.The resultant error is thus limited by only the $R_y$ gate in each layer. Furthermore, every $R_z(\phi)$ gate followed by a $R_y(\theta)$ gate can be concatenated to a single gate $R(\phi, \theta)$ representing a qubit rotation, in a Bloch sphere representation, about the $(\phi,0)$ axis by angle $\theta$. The concatenation makes the effective circuit depth half of the original. 

%%%%%%%%
\begin{figure}[t!]
\begin{center}
\subfigure[\hspace{1mm} Theoretical design of the classifier circuit and its experimental implementation]{\resizebox{\linewidth}{!}{
\Qcircuit @R=0.5em @C=0.3em{
\push{\textrm{Algorithm}} & \push{\ket 0} & \qw & \gate{R_y(\theta_1)} & \qw & \gate{R_z(\theta_2)} & \qw & \gate{R_y(\theta_3)} & \qw & \gate{R_z(\theta_4)} & \qw & \gate{R_y(\theta_5)} & \qw & \push{\cdots} & & \qw & \gate{V_y} & \qw & \meter \\ 
\push{\textrm{Experiment}} & \push{\ket 0} & \qw & \gate{R(\pi / 2, \theta_1)} & \qw & \qw & \gate{R(\theta_2, \theta_3)} & \qw & \qw & \qw & \gate{R(\theta_2, \theta_3)} & \qw & \qw & \push{\cdots} & & \qw & \gate{V_y} & \qw & \meter \\
& & & & & & & & & & & & & & & & & & & & & & &
\protect\gategroup{1}{6}{3}{8}{.7em}{--}
\protect\gategroup{1}{10}{3}{12}{.7em}{--}
}
\vspace{3mm}}}
\end{center}
\subfigure[\hspace{1mm} Operational scheme for the ion-trap experiment]{\resizebox{\linewidth}{!}{
\begin{tikzpicture}
\draw (0,0.5) node  [align=left] {Doppler cool\\(DC)};
\draw[very thick,-] (0.5,0) -- (1.5,0);
\draw[very thick,-] (1.5,0) -- (1.5,1);
\draw[very thick,-] (1.5,1) -- (3.5,1);
\draw[very thick,-] (3.5,1) -- (3.5,0);
\draw[very thick,-] (3.5,0) -- (10.9,0);
\draw[very thick,-] (10.9,0) -- (10.9,1);
\draw[very thick,-] (10.9,1) -- (14.5,1);
\draw[very thick,-] (14.5,1) -- (14.5,0);

\draw (0,-0.55) node  [align=left] {Re-pump };
\draw[very thick,-] (0.5,-1.1) -- (1.5,-1.1);
\draw[very thick,-] (1.5,-1.1) -- (1.5,-0.1);
\draw[very thick,-] (1.5,-0.1) -- (4.5,-0.1);
\draw[very thick,-] (4.5,-0.1) -- (4.5,-1.1);
\draw[very thick,-] (4.5,-1.1) -- (10.9,-1.1);
\draw[very thick,-] (10.9,-1.1) -- (10.9,-0.1);
\draw[very thick,-] (10.9,-0.1) -- (14.5,-0.1);
\draw[very thick,-] (14.5,-0.1) -- (14.5,-1.1);

\draw (0,-1.6) node  [align=left] {Optical pump~(OP)};
\draw[very thick,-] (0.5,-2.2) -- (3.5,-2.2);
\draw[very thick,-] (3.5,-2.2) -- (3.5,-1.2);
\draw[very thick,-] (3.5,-1.2) -- (4.5,-1.2);
\draw[very thick,-] (4.5,-1.2) -- (4.5,-2.2);
\draw[very thick,-] (4.5,-2.2) -- (14.5,-2.2);

\draw (0.0,-2.65) node  [align=left] {Gates};
\draw[very thick,-] (0.5,-3.3) -- (4.5,-3.3);
\draw[very thick,-] (4.5,-3.3) -- (4.5,-2.3);
\draw[very thick,-] (4.5,-2.3) -- (8.9,-2.3);
\draw[very thick,-] (8.9,-2.3) -- (8.9,-3.3);
\draw[very thick,-] (5.6,-3.3) -- (5.6,-2.3);
\draw[very thick,-] (6.7,-3.3) -- (6.7,-2.3);
\draw[very thick,-] (7.8,-3.3) -- (7.8,-2.3);
\draw[very thick,-] (8.9,-3.3) -- (14.5,-3.3);
\draw (5.05,-2.8) node  [align=center] {\small $R(\phi,\theta)$};
\draw (6.15,-2.8) node  [align=center] {\small $R(\phi,\theta)$};
\draw (7.25,-2.8) node  [align=center] {$\cdots$};
\draw (8.35,-2.8) node  [align=center] {\small $R(\phi,\theta)$};

\draw (0.0,-3.7) node  [align=left] {Basis projection};
\draw[very thick,-] (0.5,-4.4) -- (8.9,-4.4);
\draw[very thick,-] (8.9,-4.4) -- (8.9,-3.4);
\draw[very thick,-] (8.9,-3.4) -- (10.9,-3.4);
\draw[very thick,-] (10.9,-4.4) -- (10.9,-3.4);
\draw[very thick,-] (10.9,-4.4) -- (14.5,-4.4);
\draw (9.9,-3.85) node  [align=center] {$V_y$};

\draw (0.0,-4.75) node  [align=left] {Reset};
\draw[very thick,-] (0.5,-5.5) -- (1.5,-5.5);
\draw[very thick,-] (1.5,-5.5) -- (1.5,-4.5);
\draw[very thick,-] (1.5,-4.5) -- (3.5,-4.5);
\draw[very thick,-] (3.5,-5.5) -- (3.5,-4.5);
\draw[very thick,-] (3.5,-5.5) -- (14.5,-5.5);

\draw (2.5,-6.2) node  [align=center] {DC\\ 300 $\mu$s};
\draw (4,-6.2) node  [align=center] {OP\\ 50 $\mu$s};
\draw (6.5,-6.2) node  [align=center] {Data uploading\\ 50-100 $\mu$s};
\draw (9.5,-6.2) node  [align=center] {Projection\\ 15$\mu$s};
\draw (12.5,-6.2) node  [align=center] {State detection\\ ~ 2 ms};
\end{tikzpicture}}}

\caption{Experimental implementation of quantum classifier circuit: (a) The algorithm to implement the quantum classifier is represented by gates in the top qubit, grouped in pairs $R_y, R_z$. In the bottom qubit, couples of gates are concatenated into one rotational gate $R(\phi,\theta)$ with modified rotation axis. For one qubit it is possible to define at most two orthogonal states, so the unitary operation $V_y$ adds any rotation to accomodate label states, see Eq.~\eqref{eq:fidelity}.(b) The time sequence used in the experiment to perform each classifier measurement.}%
  \label{fig:timeseq}
\end{figure}
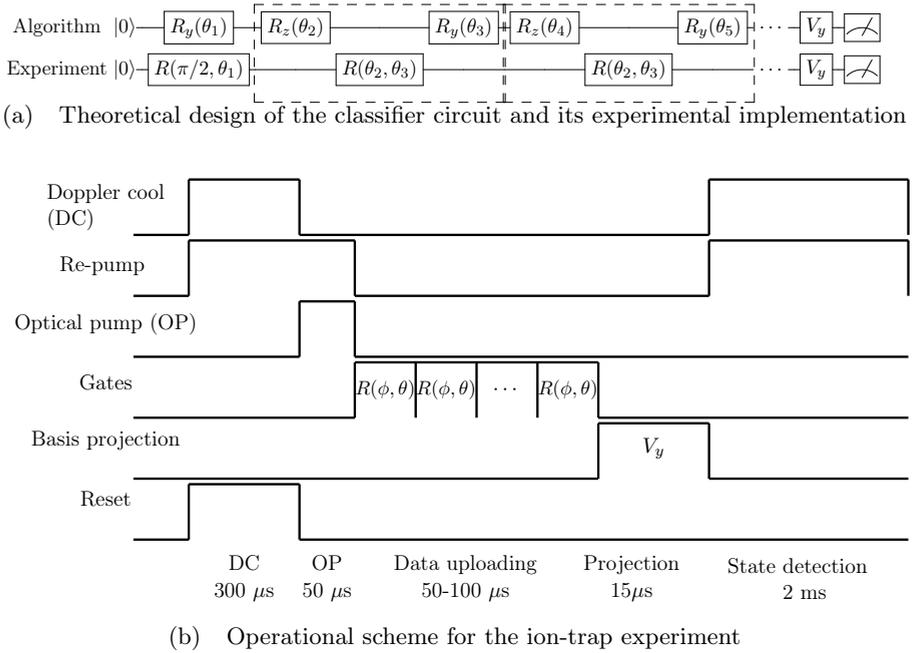

Each data point shown in Fig.~\ref{fig:exp_circle} or any of the other classifier plots, comprises of $100$ repeat experiments. Every experiment, consists of a sequence of operations performed on the qubit as shown in Fig.~\ref{fig:timeseq}. Depending on the data to be uploaded, single qubit gates are implemented by resonantly driving the qubit transition between $\ket 0, \ket 1$ with well-controlled phase and operation time, while the frequency and power of the laser are kept constant. An acousto-optic modulator controls the phase and frequency of the $1760~$nm laser implementing the rotation gates. The phase of the modulator is directly controlled by a synthesizer which supplies the radio-frequency signal to the modulator via an amplifier. The laser-qubit interaction time sets the rotation angle $\theta$. Therefore, direct and precise control over the axis and angle of rotations is available. The optimal gate to apply repeatedly on the experiment is defined as:
\begin{equation}\label{eq:matrixR}
    R(\phi, \theta) = \begin{bmatrix}
            & \cos\frac{\theta}{2}   & -ie^{-i\phi}\sin\frac{\theta}{2}  \\
            
            & -ie^{i\phi}\sin\frac{\theta}{2}  & \cos\frac{\theta}{2}.       
    \end{bmatrix}
\end{equation}

The time sequence shown in Fig.~\ref{fig:timeseq} is controlled by a \ac{fpga} with a time jitter below 1~ns. The current version of the synthesizer controlling the phase of the laser is limited by the on-chip memory to $16$ phase modulation steps thus limiting the layers to six, which is sufficient for the current discussion. Once the circuit runs over all the layers, the overlap between the final state of the qubit and a label state is measured. In case of binary classification, the final state is projected on to the label state $\ket{0}$ and the overlap is measured by observing spontaneously emitted photons while the qubit is excited by $493~$nm laser. With a photon collection time of $2~$ms, the bright state is clearly discriminated from the dark state. For any classification problem with more than two classes, two additional rotations are required before performing state discrimination. The overlap for each data point based on a threshold is designated a success (1) or failure (0) to be within a class.

\begin{figure}[t!]
    \begin{center}
            \includegraphics[width=\textwidth]{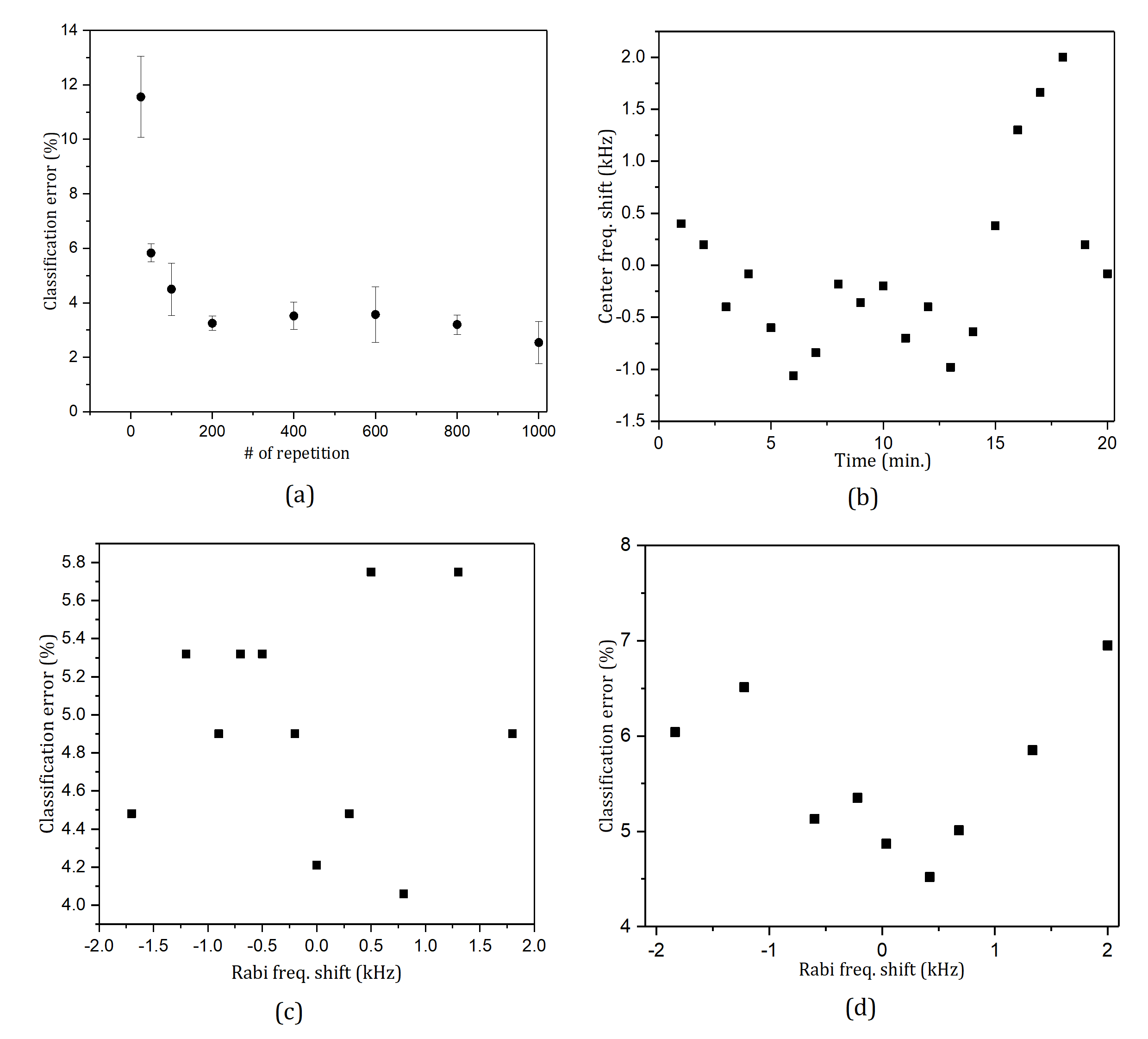}

    \end{center}
\caption{Systematic error analysis: All the results shown here are related to the binary classification of circle as in Fig.~(\ref{fig:exp_circle}). The errors are classification error. (a) The classifier error as a function of the number of repeated experiments. The error bar at each point corresponds to $1$ standard deviation of a number of repeat measurements for same number of repeated experiments under the same condition. The exact number of repeat measurements varies between $5$ and $10$. (b) Variation of the resonance frequency as a function of time. The range of Rabi frequency fluctuation within a typical experimental time of $\le 10$ minutes is about $2$ kHz. (c) Error in binary classification of a circle feature with the variation of laser frequency detuning measured in terms of the Rabi frequency. The variation in the value of classification error is  about $2\%$ within the experimental time of $\sim 10$ min. (d) The same plot as in (c) but by varying the laser power measured in terms of the Rabi frequency.}%
  \label{fig:error}
\end{figure}

The accuracy of the data re-uploading algorithm primarily relies on the the fidelity of individual single qubit rotation gate. The residual error in the gate operation is reflected in the accuracy of the classifier. In Eq.~\eqref{eq:matrixR}, the rotation angles are related to experimental values as $\phi=(\Delta/\hbar) t_{op}+\delta\phi$ and $\theta=(\Omega/\hbar) t_{op}$, where $\Delta$ is the laser detuning, $t_{op}$ is the operation time of gates, $\Omega'$ is the modified Rabi frequency and $\delta\phi$ is the relative phase of the laser with respect to the qubit. The modified Rabi frequency is $\Omega^\prime = \sqrt{\Omega_0^2+\Delta^2}$ with $\Omega_0$ denoting the resonant Rabi frequency. Furthermore, the resonant Rabi frequency is proportional to the square root of the intensity, $I_0$, at the ion position. Therefore each of the independent variables $\delta\phi$, $t_{op}$, $\Delta$ and $I_0$ contributes to the error in a rotation gate as follows: 

\begin{itemize}
    \item[$\delta \phi$]{\bf Phase:} The synthesizer controls the radio-frequency phase of the modulator which determines the relative phase of the laser. Each synthesizer is synchronized to a rubidium atomic clock which is accurate to one part in $10^{10}$ and thus contributes negligibly to the phase error. The direct digital synthesizer is however triggered by the \ac{fpga} which has time jitter below $10~$ns leading to phase noise on the qubit below $0.1\%$ for a Rabi $\pi$ time of $12~\mu$s.
   
    \item[$t_{op}$]{\bf Interaction time:} The laser-qubit interaction time is determined by the \ac{fpga}, precise up to $1~$ns. Therefore, due to the time jitter below $10~$ns, its contribution to the accuracy is below $0.1\%$. Occasional collision with the residual background gas molecule during the interaction time leads to a projection to the state $\ket{0}$, thus losing the final state information and hence error in the classification. This error becomes usually smaller with larger statistics. 
    
    \item[$\Delta$]{\bf Laser-qubit detuning:} The detuning of the laser with respect to the qubit frequency modifies the Rabi frequency. The range of Rabi frequency fluctuation within the experimental time is quantified to about $10~$min. To ensure that the accuracy is limited by systematic errors, statistical errors in the classification are measured by repeating the experiment, see Fig.~\ref{fig:error}a. The error decreases from $12\%$ to about $4\%$ for $100$ repetitions and then enters a stationary regime limited by systematic errors. The fluctuation of the laser frequency with respect to the atomic resonance is captured over a time period of $20~$min (twice the duration of an experiment) as plotted in Fig.~\ref{fig:error}b. The random variation of the Rabi frequency over time is mostly caused by the magnetic field noise as the laser frequency drift was separately measured to be $\le 5~$kHz/$24$ hrs~\cite{Yum:17}. To minimize the impact of the residual magnetic field noise, electronic levels weakly sensitive to such noise are used. In addition, the detuning also indirectly influence the modified Rabi frequency. To check its influence, the Rabi frequency is varied as shown in Fig.~\ref{fig:error}c, by varying the detuning within a $2~$kHz range (as expected from the Fig.~\ref{fig:error}b). The result shows below $5\%$ accuracy for the classifier when operating for $10~$min.
    
    \item[$I_0$]{\bf Laser intensity:} The Rabi frequency is fixed by setting the power and frequency of the laser at the start of the experiment. Any change in the Rabi frequency during the experiment leads to error in applied qubit rotation angle. %Therefore, the accuracy of the classifier, depends on the Rabi frequency fluctuations due to intensity fluctuation apart from detuning as discussed earlier. 
    The intensity is influenced by two factors, the laser power noise and the laser beam pointing error. In the experiment, the laser beam is tightly focused on the ion by a high Numerical Aperture $\sim 0.4$ in-vacuum lens. In order to capture the influence of laser power variations on the classification error, the power is varied, see Fig.~(\ref{fig:error}d). Thus it is seen that the influence of intensity noise accounts to $5\%$ error in accuracy. To avoid the influence of Rabi frequency fluctuation within the experimental time of $\sim 10~$min, the Rabi frequency is reduced from $625~$kHz to $80~$kHz such that the absolute error also reduces. This leads to an overall error of only $2\%$ on the classifier output.       
\end{itemize}

\section{Quantum circuits in NNPDF methodology}\label{app:nnpdf}

The latest implementation of the latest iteration of the NNPDF methodo-logy is described in Ref.~\cite{Carrazza:2019mzf}.
This implementation is very modular and one can seamlessly swap the \texttt{Tensorflow} based backend by any other
provider.
{\tt Qibo}, which is also partially based on \texttt{Tensorflow} can be easily integrated with the NNPDF methodology.

As previously mentioned, all results in this section corresponds to the simulation of the quantum device on classical hardware.
Such a simulation is very costly from a computational point of view which introduces a number of limitations
that need to be addressed in order to produce results in reasonable time frames.

{\bf FK reduction}: the definition of the quantum circuit depends on both the set of parameters $\theta$ and the value of
the parton momentum fraction $x$ (see Eq.~\eqref{eq:quantumcircuit}) which means the circuits needs to be simulated
once per value of $x$.
The union of all FK tables for all physical observables (following Eq.~\eqref{eq:convolution}) amounts to
several thousand values of $x$.
Since such a large number of evaluations of the quantum circuit is impracticable, a further approximation is introduced
where each partial FK table is mapped to a fixed set of 200 nodes in the x-grid.
This simplification introduces an error to the total $\chi^{2}$ of the order of $\Delta\chi^{2} = 0.14 \pm 0.01 $ when averaged
over PDF members.
This error on the cost function is however negligible for the accuracy reached in this work.

{\bf Positivity}: in the fitting basis, as defined in section~\ref{ssec:qcpdf}, the PDF cannot go negative.
Physical predictions however are computed in the flavour basis~\cite{Ball:2008by} where the rotation between basis
can make some results go negative.
However, physical observables (differential or total cross sections) cannot be. This physical constraint is included in NNPDF3.1 via fake pathological datasets.
These have not been implemented for qPDF as they correspond to a fine-tuning of the methodology which is beyond the
scope of this work.

\begin{SCfigure}
\centering
\includegraphics[width=.6\linewidth]{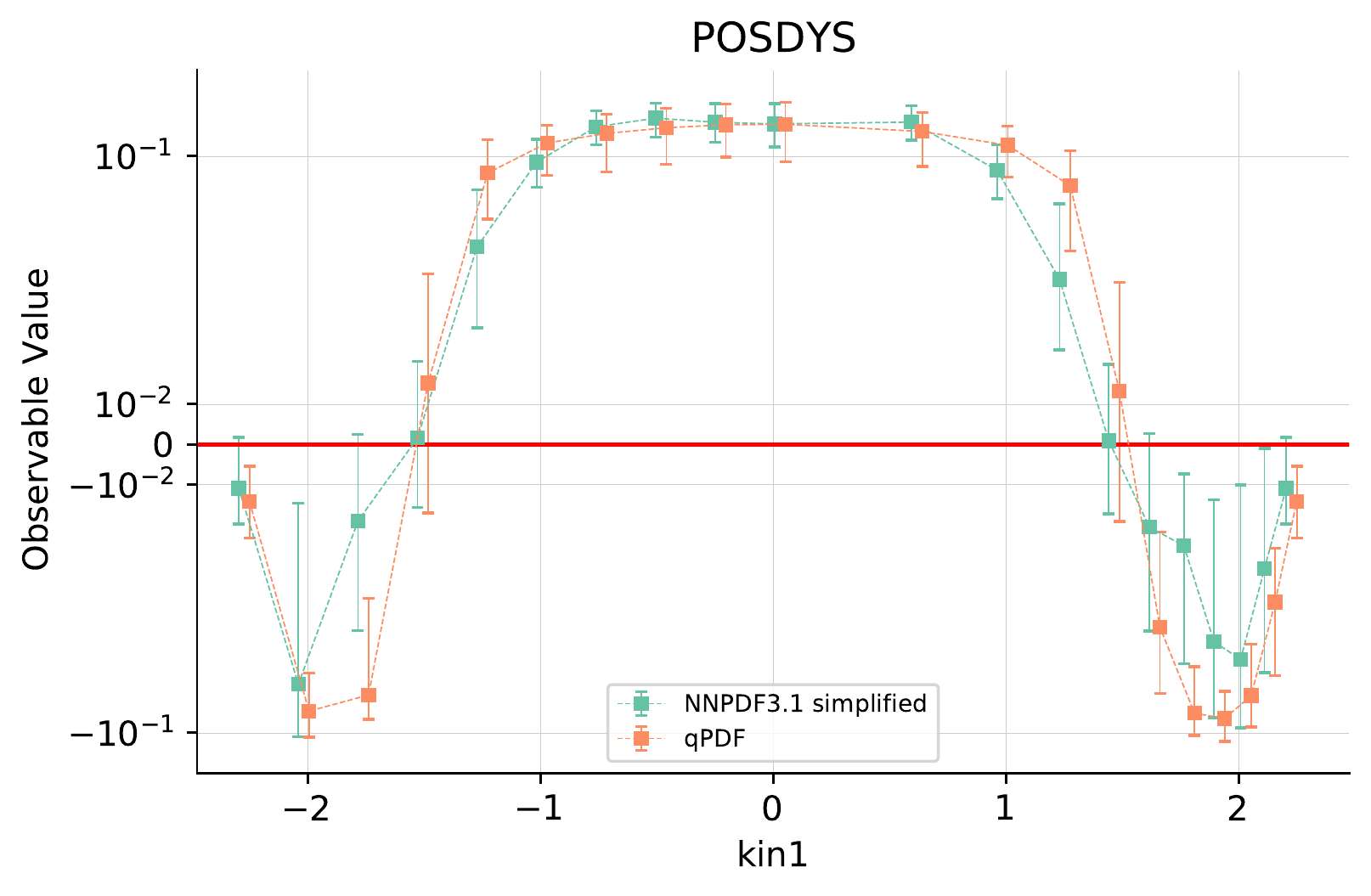} 
\caption{Predictions for a toy $s\bar{s}$ initiated Drell-Yan process with qPDF and a simplified version of NNPDF3.1
    where the positivity constraint has been removed.}\label{fig:positivity}
\end{SCfigure}

The removal of the positivity constraint from the fit introduces an unphysical distortion to the results as the PDF
could produce negative predictions for physical predictions. Such results are unphysical because they would correspond
to situations in which the probability of finding a particular phase space configuration is negative, which makes no
sense.
In Fig.~\ref{fig:positivity}, the ``negativity'' between qPDF and a version of NNPDF3.1 with the positivity constraints removed is compared.
We observe that both fits behave similarly, proving such unphysical results are a consequence of the removal of the
constraint rather than a problem in the qPDF methodology.

{\bf Momentum Sum Rule}: the PDFs as defined in Eq.~\eqref{eq:nnpdf} are normalized such that~\cite{Ball:2014uwa},
\begin{equation}
    \frac{\displaystyle\int_0^1 dx \; x\,f_{g}(x,Q_0)}{1-\displaystyle\int_0^1 dx  x f_{\Sigma}(x,Q_0)} \simeq 1,
\end{equation}
this equation is known as the momentum sum rule and it is imposed in \texttt{n3fit}
through an integration over the whole range of x which is impracticable
% \begin{align}
%     \frac{\displaystyle\int_0^1 dx \; x\,f_{g}(x,Q_0)}{1-\displaystyle\int_0^1 dx  x f_{\Sigma}(x,Q_0)} &\simeq 1;
%         \quad \displaystyle\int_0^1 dx \, f_{V}(x,Q_0) &\simeq 3; \nonumber\\
%     \displaystyle\int_0^1 dx \, f_{V_3}(x,Q_0) &\simeq 1;
%         \quad \displaystyle\int_0^1 dx \, f_{V_8}(x,Q_0) &\simeq 3;\nonumber
% \end{align}
% these rules are imposed in \texttt{n3fit} through an integration over the whole range of x which is impracticable
in this implementation for the reasons mentioned above.
Instead, in qPDF these are only checked afterwards, finding a good agreement with the expected values (despite not being
imposed at fitting time).
Indeed, for qPDF the result for the average over all replicas is $ 1.01\pm 0.01$, 
which is to be compared with the NNPDF3.1 result of $1.000 \pm 0.001$, where the constraint was imposed at fit time.

\begin{figure}[b!]
\begin{adjustwidth}{-1cm}{-2cm}
\begin{minipage}{.47\linewidth}
%\subfigure[$\chi^{2}/N$ per experiment grouping. There is a deterioration of the goodness of the
%        fit (measured by the $\chi^{2}$) for some of the experiments for the central value.
%        The goodness of the fit is very similar between the reference and
%        qPDF for most of the experiments being considered.\label{fig:expchi2}]
        {\includegraphics[width=\linewidth]{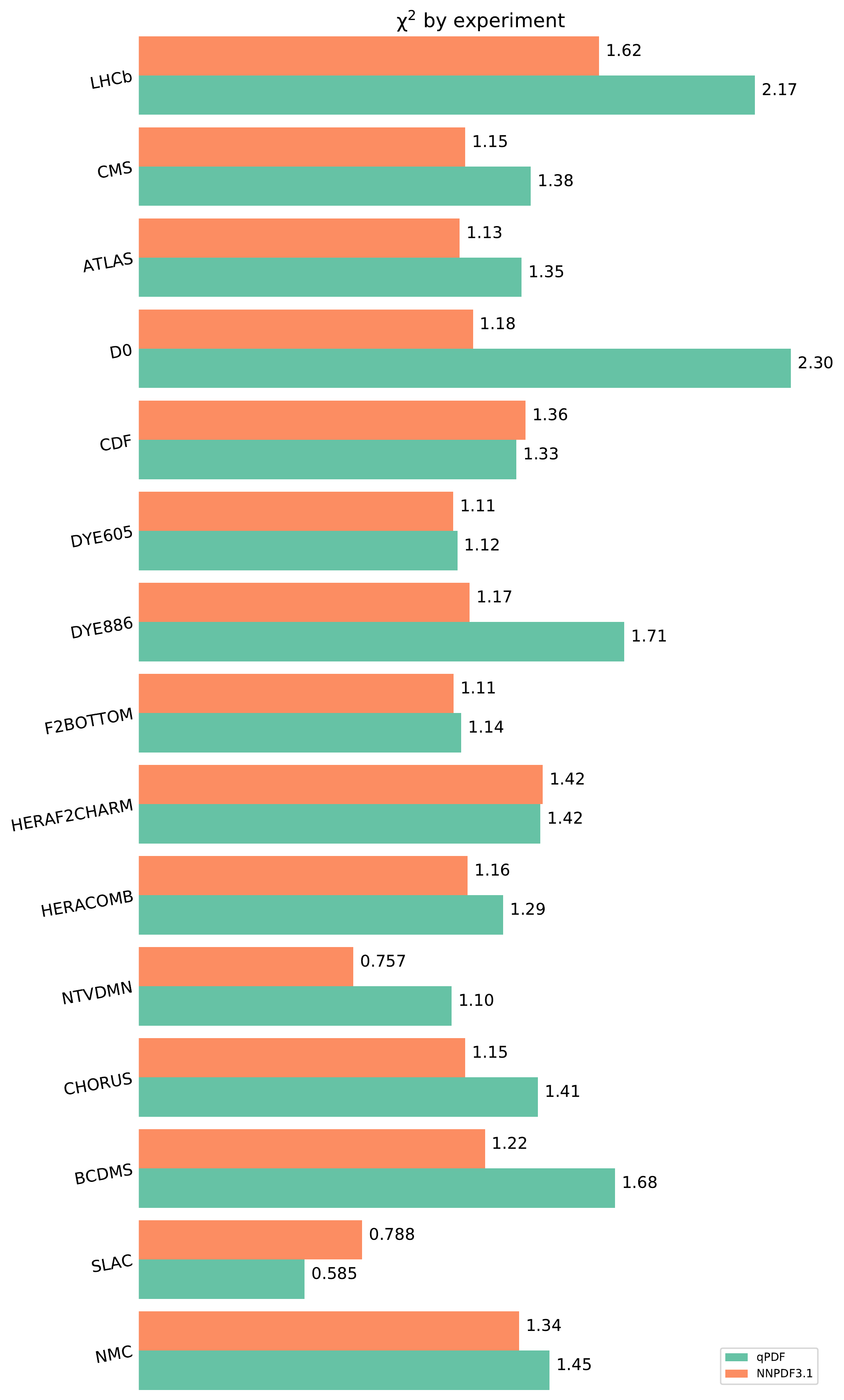}}
\end{minipage}
\hfill
\begin{minipage}{.47\linewidth}
\subfigure[Distance (as defined by Eq.~\eqref{eq:distance}) between qPDF and NNPDF3.1. When the distance is kept
    under $d(f_{i}, r_{i})=10$ the two fits are 1-$\sigma$ compatible. All partons except for $u$ and $s$ are below or
    around the 1-$\sigma$ distance for the entire range considered. Note however, by comparing to
Fig.~\ref{fig:all_flavours} that the fits for both the $u$ and $s$ quarks are compatible in the most relevant regions
for these particles.\label{fig:distance}]{\includegraphics[width=\linewidth]{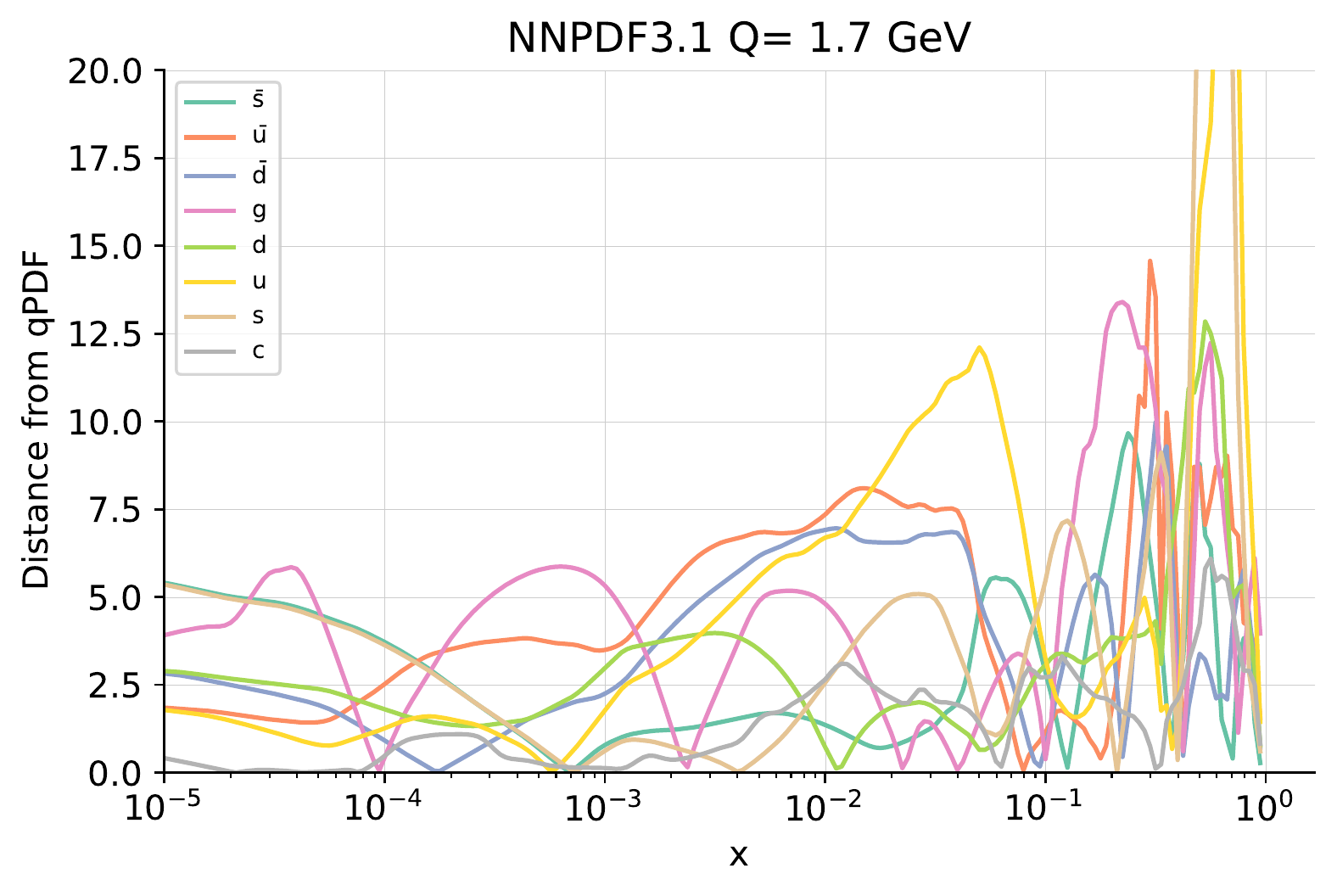}}

~\vfill

\hrule

~\vfill

\subfigure[$\chi^{2}/N$ per experiment grouping. There is a deterioration of the goodness of the
        fit (measured by the $\chi^{2}$) for some of the experiments for the central value.
        The goodness of the fit is very similar between the reference and
        qPDF for most of the experiments being considered.\label{fig:expchi2}]{\hspace{\linewidth}}
\end{minipage}
\end{adjustwidth}
\caption{Collection of extra results for the qPDF method integrated in the NNPDF methodology as extracted from LHC data. }
\begin{textblock}{10}(4.8,-2.52)
\Large $\longleftarrow$
\end{textblock}
\end{figure}

\subsubsection{Extra results}

With all ingredients implemented, it is possible to run a NNPDF3.1-like fit using the qPDF.
As a base reference for the comparison, the NNPDF3.1 NNLO fit~\cite{Ball:2017nwa} is taken,
which is the latest release by the NNPDF collaboration.
The plots comparing the NNPDF sets with qPDF are then produced using a
\texttt{reportengine}~\cite{zahari_kassabov_2019_2571601} based internal NNPDF tool. The extra result here presented complement the one showed in the main text, in particular related to Fig.~\ref{fig:fitperflavour}. 

\begin{figure}[t!]
    \centering
\begin{adjustwidth}{-2cm}{-1cm}
    \subfigure[\ Atlas jets data differential in rapidity~\cite{Aad:2014vwa}.]{
        \includegraphics[width=0.31\linewidth]{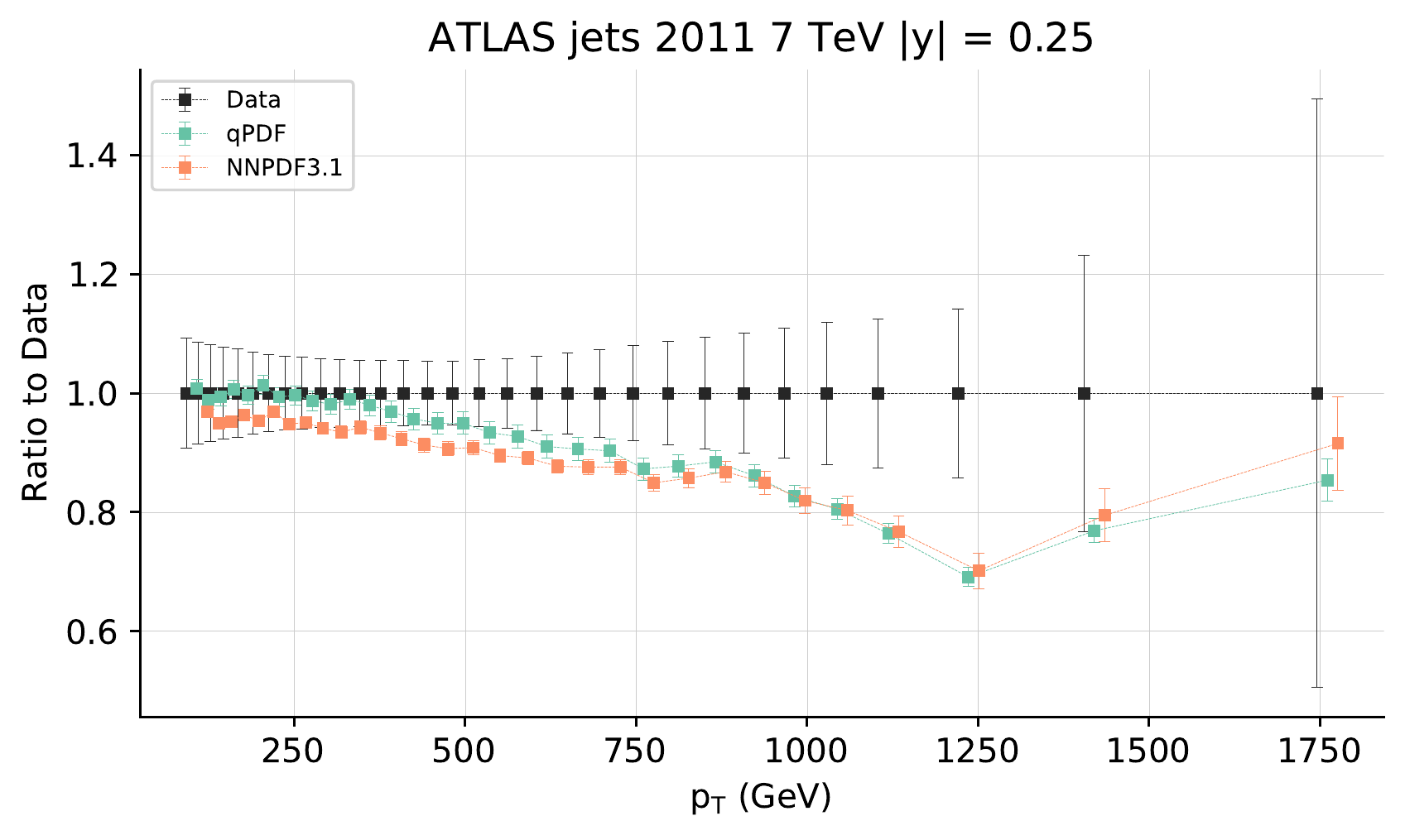}
    } \hfill
    \subfigure[\ CMS Z differential in rapidity for fixed value of $p_{T}$~\cite{Khachatryan:2015oaa}.]{
        \includegraphics[width=0.31\linewidth]{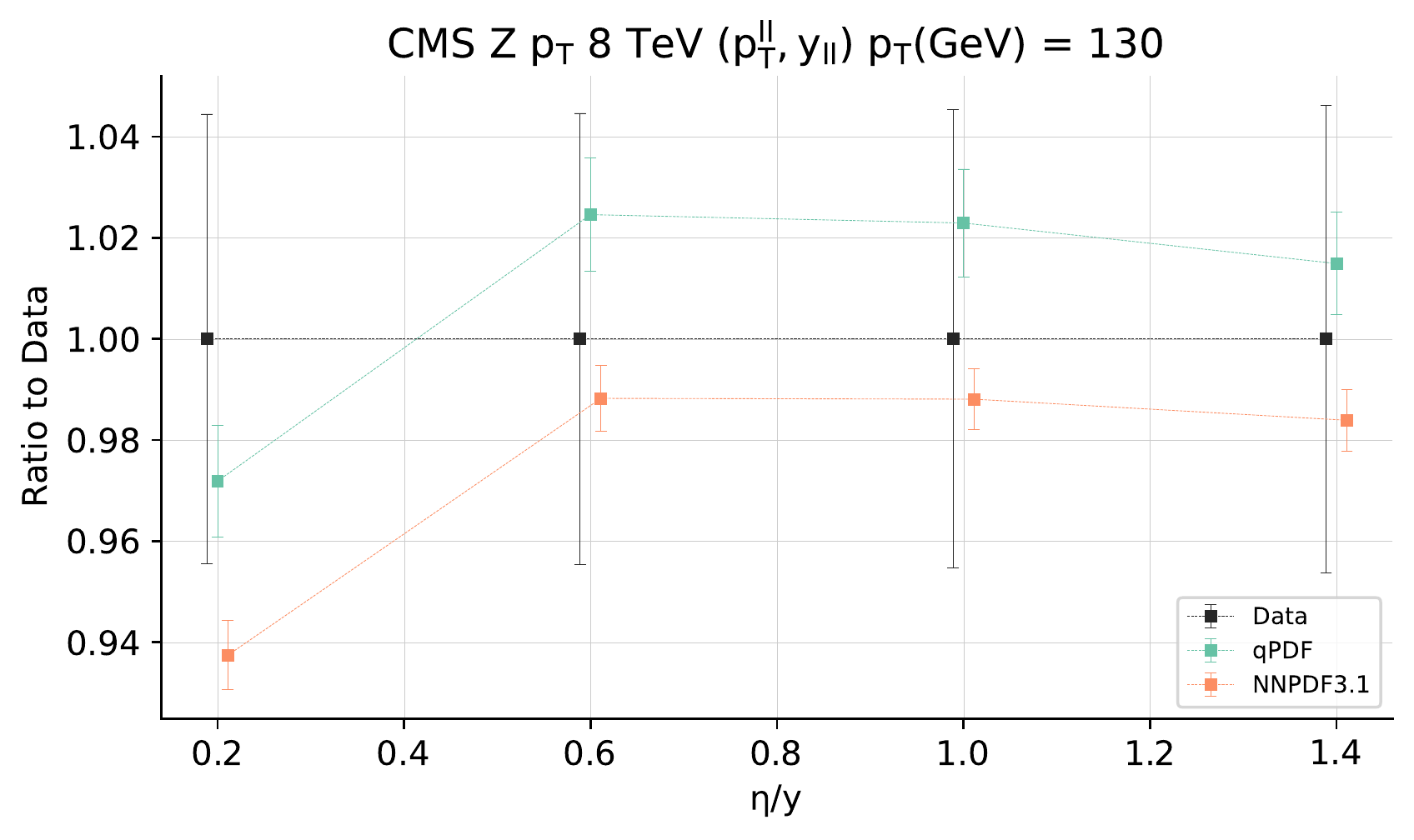}
    } \hfill
    \subfigure[\ LHCb, Z cross section differential in rapidity~\cite{Aaij:2012vn}.]{
        \includegraphics[width=0.31\linewidth]{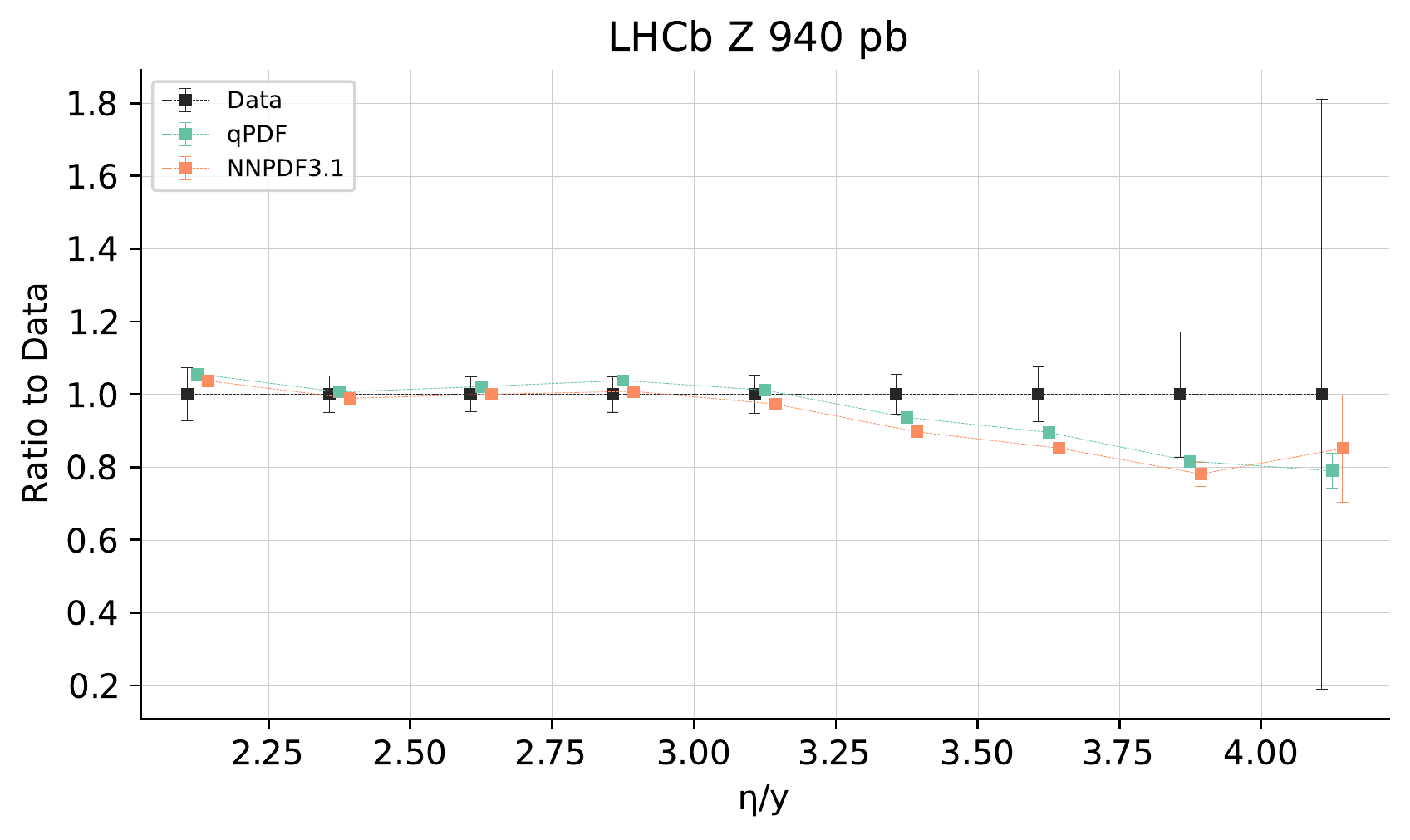}
    }
    \end{adjustwidth}
    \caption{Theoretical predictions computed with the method describe in~\cite{Bertone:2016lga} in order to compare
    the same prediction with three different PDF sets. Note that the predictions for the qPDF set is compatible with
both the experimental measurements and the released PDF set. The parton-level calculation has been performed with the
NLOjet++~\cite{Nagy:2001fj} and MCFM~\cite{Campbell:2019dru} tools.}\label{fig:expresults}
  \includegraphics[width=\linewidth]{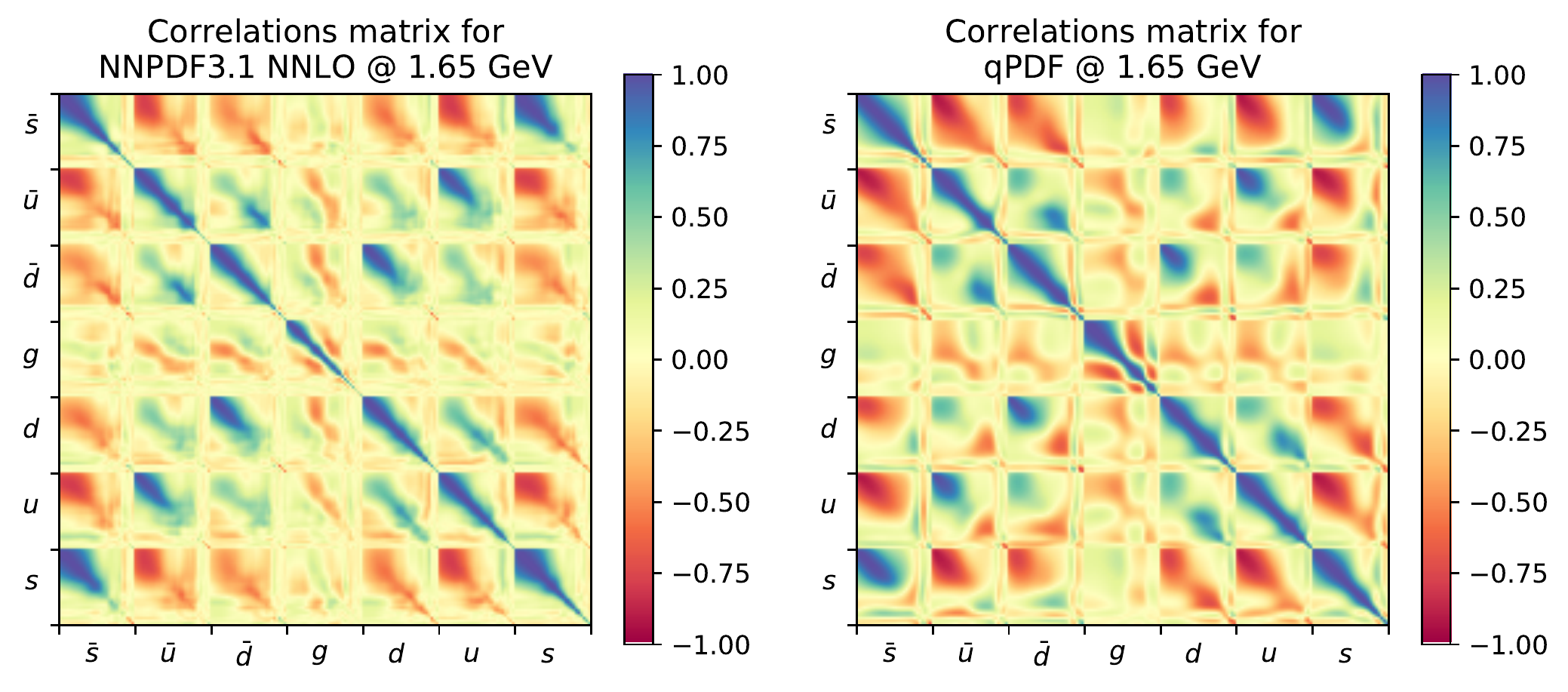}
  \caption{\label{fig:correlation}PDF correlation matrix for flavours in a grid
  of $x$ points for NNPDF3.1 NNLO (left) and the qPDF (right).}
\end{figure}

We can start by comparing the $\chi^{2}/N$ result for the datasets that have been considered in the fit,
shown in Fig.~\ref{fig:expchi2}.
One would expect a perfect fit when $\chi^{2}/N = 1$,
however this is not the case even in the reference and it is due to a combination of missing higher order corrections
(a lack of a better theory) or inconsistencies in the experimental results.
The similarity on the phenomenological results obtained by both fitting methodologies as shown in Fig.~\ref{fig:expchi2}
is well understood as well by looking at the distance plots between the qPDF~and the reference in
Fig.~\ref{fig:distance},
\begin{equation}
    d^2(f_{i}, r_{i}) = \frac{\langle f_{i} \rangle - \langle r_{i} \rangle}{\frac{1}{N_f}\sigma(f_{i})^2 +
    \frac{1}{N_{r}}\sigma(r_{i})^2},\label{eq:distance}
\end{equation}
where $i$ is the flavour being considered and $f$ and $r$ corresponds to qPDF~and the reference (NNPDF3.1)
respectively. The central value is taken over the $N$ replicas of the set, generally of the order of 100.

Indeed, for most partons the difference between both fits are under the 1-$\sigma$ level (distance equal to 10 for 100
replicas) growing up to 2-$\sigma$ for the $u$ and $s$ quarks.

In Fig.~\ref{fig:expresults} it is shown specifically a comparison between the reference NNPDF3.1 and qPDF for selected
datasets, and the LHAPDF-compatible PDF grid is provided.
The accuracy of the qPDF central value is similar to that of NNPDF3.1.
Furthermore, the error bars for the predictions of both PDF set overlap with the experimental error bars, and, in some
cases, also among themselves.

Finally, in Fig.~\ref{fig:correlation}the PDF correlations for
NNPDF3.1 and qPDF replicas using Pearson's coefficient in a fixed grid of 100
points distributed logarithmically in $x=[10^{-4},1]$ are computed.

This leads to conclude that the methodology described in this paper can be used for regression problems to unknown functional
forms such as the proton internal structure and produce results that are perfectly coherent, from a phenomenological
point of view, with the state of the art.
In addition it is believed that with adequate tuning one could achieve the same level of accuracy of the classical
approach.

We finalize this section by showing phenomenological results where the LHAPDF grids produced with this approach are used
for a full fixed order prediction.
In summary going back circle to the master equation, i.e., computing numerically Eq.~\eqref{eq:convolution}
with no approximations using state of the art tools.

\chapter{Unary strategy}\label{app:unary}
\section{Details for the Black-Scholes model}\label{app:black_scholes}

The Black-Scholes model for the evolution of an asset is based on the stochastic differential equation \cite{black_blackscholes_1973}
\begin{equation}\label{eq:ap_BSM}
    {\rm d}S_T = S_T\, r\, {\rm d}T + S_T\, \sigma\, {\rm d}W_T, 
\end{equation}
where $r$ is the interest rate, $\sigma$ is the volatility and $W_T$ describes a Brownian process.
Recall that a Brownian process $W_T$ is a continuous stochastic evolution starting at $W_0=0$ and made of independent gaussian increments. To be specific, let $\mathcal{N}(\mu, \sigma_s)$ be a normal distribution with mean $\mu$ and standard deviation $\sigma_s$. Then, the increment related to two steps of the Brownian processes is  $W_T - W_S \sim \mathcal{N}(0, T - S)$, for $T > S$.

The above differential equation can be solved analytically up to first order using Ito's lemma \cite{ItoLemma-1944}. The essential observation if that $W_T$ is treated as an independent variable with the property that $({\rm d}W_T)^2$ is of the order of ${\rm d}T$. Thus, the approximated derivative  ${\rm d}S_T$ can be written as
\begin{equation}
    {\rm d}S_T = \left( \frac{\partial S_T}{\partial T} + \frac{1}{2}\frac{\partial^2 S_T}{\partial W_T^2}\right) {\rm d}T + \frac{\partial S_T}{\partial W_T} {\rm d}W_T.
\end{equation}
By direct comparison to Eq. \eqref{eq:ap_BSM}, it is straightforward to see that
\begin{eqnarray}
    \frac{\partial S_T}{\partial W_T} = S_T\, \sigma, \\ 
    \frac{\partial S_T}{\partial T} + \frac{1}{2}\frac{\partial^2 S_T}{\partial W_T^2} = S_T\, r .
\end{eqnarray}
Using the initial condition $S_0$ at $T=0$, and the Ansatz
\begin{equation}
    S_T = S_0 \exp{(f(T) + g(W_T))},
\end{equation}
the solution for the asset price turns out to be
\begin{equation}
    S_T = S_0 e^{(r - \frac{\sigma^2}{2}) T} e^{\sigma W_T}\;\sim\; S_0 e^{\mathcal{N}\left(\left(r - \frac{\sigma^2}{2}\right) T, \sigma \sqrt{T}\right)}.
\end{equation}
This final result corresponds to a log-normal distribution.

\subsection{European Option}

An option is a contract where in its call/put form, the option holder can buy/sell an asset before a specific date or decline such a right. As a particular case, 
European options can be exercised only on the specified future date, and only depend on the price of the asset at that time. The previously agreed price that will be paid for the asset is called exercise price or \emph{strike}. The day on which the option can be exercised is called \emph{maturity date}. 

A European option payoff is defined as
\begin{equation}\label{eq:Payoff}
    f(S_T, K) = \max(0, S_T - K),
\end{equation}
where $K$ is the strike price and $T$ is the maturity date. An analytical solution exists for the payoff of this kind of options. 

The expected payoff is given by
\begin{multline}
    C(S_T, K) = {\rm average}_{S_T \geq K}\left( S_T - K \right) = \int_{d_1}^{\infty} \left(S_T - K\right) dS_T = \\ =\int_{d_1}^\infty\frac{S_0}{\sqrt{2\pi}} e^{-\frac{\left(x - \left(r - \frac{\sigma^2}{2}\right)T\right)^2}{2\sigma^2 T}} e^{\frac{-x^2}{2}} dx,
\end{multline}
yielding the analytical solution
\begin{equation}\label{eq:exp_payoff_analytical}
    C(S_T, K) = S_0 {\rm CDF}_{\mathcal N}(d_1) - K e^{-r T}{\rm CDF}_{\mathcal N}(d_2), 
\end{equation}
with 
\begin{center}
\begin{eqnarray}
    d_1 = \frac{1}{\sigma \sqrt{t}}\left( \log \frac{S_0}{K} + \left( r + \frac{\sigma^2}{2}\right) T \right) \\
    d_2 = d_1 - \sigma \sqrt{T} \\
    {\rm CDF}_{\mathcal N}(x) = \frac{1}{\sqrt{2\pi}} \int_{-\infty}^x e^{\frac{-u^2}{2}}du .
\end{eqnarray}
\end{center}

This analytical development for the European option using the Black-Scholes model cannot be extended to some exotic options, like the American (the option can be exercised at any point) or the Asian (the final option price depends on a time average of the price until the maturity date). Therefore, in a general case the expected payoff return of an option cannot be obtained analytically, but rather performing a Monte Carlo simulation where many different scenarios are taken into consideration, and a global estimation is obtained.

\section{Details for the Amplitude Distributor $\D$ in the unary basis }\label{app:amplitude_distribution}

Figure \ref{fig:ampl_distributor} is considered. In the unary basis, every qubit represents the basis element in which the qubit is $\ket{1}$. Thus, the coefficient of every element depends on as many angles as partial-SWAP gates are needed to reach its corresponding qubit. Thus, the central qubits of the circuit will depend only on 2 angles, and the number of dependencies increases one by one as the gates move to the outer part of the circuit. The very last two qubits depend on the same angles. As the procedure goes by and moves away from the middle qubit, each qubit inherits the same angle dependency than the previous ones plus an additional rotation.
Starting from the two edges, their coefficients verify the following ratios
\begin{eqnarray}
    \left\vert\frac{\psi_{0}}{\psi_{1}}\right\vert ^2 & = & \tan^2(\theta_{1} / 2) \\
    \left\vert\frac{\psi_{n-1}}{\psi_{n-2}}\right\vert ^2 & = & \tan^2(\theta_{n - 1} / 2).
\end{eqnarray}
Then $|\psi_i|^2 = p_i$, where $\lbrace p_i\rbrace$ is the target probability distribution of the asset prices at maturity. The next step corresponds to considering the qubits $1$ and $2$, as well as $n-3$, $n-2$. The relations for their coefficients must obey
\begin{eqnarray}
    \left\vert\frac{\psi_{i}}{\psi_{i+1}}\right\vert ^2 & = & \cos^2(\theta_i/2)\tan^2(\theta_{i+1} / 2) \\
    \left\vert\frac{\psi_{n-1-i}}{\psi_{n-2-i}}\right\vert ^2 & = & \cos^2(\theta_{n-i}/2)\tan^2(\theta_{n - 1 - i} / 2).
\end{eqnarray}
Then, it is straightforward to back-substitute parameters step by step until we arrive to the central qubits.
This procedure fixes all the angles for the partial-SWAP gates used in the amplitude distributor.

The exact algorithm to be followed can be also found in the provided code \cite{github_unary}.

Once the exact solution for the angles is inserted into the circuit depicted in Fig. \ref{fig:ampl_distributor}, the amplitude distributor algorithm is completed. The quantum register then reads
\begin{equation}
    \ket{\Psi}=\sum^{n-1}_{i=0}\sqrt{p_i}\ket{i}.
\end{equation}
Note that describing a probability distribution with squared amplitudes of a quantum state allows for a free phase in every coefficient of the quantum circuit. For simplicity, we will set to zero all these relative phases by only operating with real valued partial-SWAP gates.

Let us turn our attention to the gates which are needed in the above circuit.
Sharing probability between neighbor qubits can be achieved by introducing a two-qubit gate based on the SWAP and $R_y$ operations. This variant on the SWAP gate performs a partial SWAP operation, where only a piece of the amplitude is transferred from one qubit to another. This operation preserves unarity, that is the state remains as a superposition of elements of the unary basis. This partial-SWAP, can be decomposed using CNOT as the basic entangling gate as 
\begin{figure}[h!]
\begin{equation}\label{eq:ap_SWAPRy}
\smash{\raisebox{.8\normalbaselineskip}{
\Qcircuit @R=0.7em @C=0.3em{
& \qw & \multigate{1}{\hspace{1.4cm}} & \qw \\
& \qw & \ghost{\hspace{1.4cm}} & \qw 
}}} \quad = \quad
\smash{\raisebox{.8\normalbaselineskip}{
\Qcircuit @R=0.7em @C=0.3em{
& \qw & \targ & \ctrl{1} & \targ & \qw \\
& \qw & \ctrl{-1} & \gate{R_y(\theta)}& \ctrl{-1} & \qw 
}}}
\end{equation}
\begin{textblock}{1.3}(1.4, -.9)
{\centering 
partial-\\SWAP$(\theta)$
}
\end{textblock} 
\end{figure}

where the usual CNOT gate in the center of the conventional SWAP gate has been substituted by a controlled $y$-rotation, henceforth referred to as c$R_y$ gate. In turn, the c$R_y$ operation can be reworked as a combination of single-qubit gates and CNOT gates \cite{gates-barenco1995}:

\begin{equation}
\smash{\raisebox{.8\normalbaselineskip}{
\Qcircuit @R=0.7em @C=0.3em{
& \qw & \ctrl{1} & \qw \\
& \qw & \gate{R_y(\theta)} & \qw 
}}} \quad = \quad
\smash{\raisebox{.75\normalbaselineskip}{
\Qcircuit @R=0.7em @C=0.3em{
& \qw & \qw & \ctrl{1} & \qw & \ctrl{1} & \qw \\
& \qw & \gate{R_y(\theta / 2)} & \targ & \gate{R_y(- \theta / 2)}& \targ & \qw 
}}}
\end{equation}
\vspace{2mm}\\
This decomposition will come into play for the expected payoff calculation algorithm as well, albeit with angle $\phi$ in the payoff circuit.

For the purposes of this algorithm, both the CNOT and partial-iSWAP basis gates are analogous, but the direct modeling to partial-iSWAPs can economize the total number of required gates for the amplitude distributor. Partial-iSWAP gates can be used to decompose CNOT gates. More explicitly, a CNOT gate an be reproduced with two iSWAP gates, and 5 single qubit gates.

\section{\ac{qae}}\label{app:amplitude_estimation}
\ac{qae} is a general framework to estimate the probability of obtaining a certain outcome if measuring a given quantum state. This procedure allows to gain quantum advantage with respect to Monte Carlo samplings. 

\ac{qae} is in general defined from an algorithm $\A$ such that \begin{equation}
    \A \ket 0_n \ket 0 = \sqrt{1 - a} \ket{\psi_0}_n\ket{0} + \sqrt{a} \ket{\psi_1}_n\ket{1}, 
\end{equation}
where the last qubit serves as an ancilla qubit and the states $\ket{\psi_{0,1}}_n$ can be non-orthogonal. The ancilla qubit is a flag which enables to identify the states as {\sl good} ($\ket{1}$) or {\sl bad} ($\ket{0}$). The state $\A\ket 0_n \ket 0$ can be directly sampled $N$ times, and the estimate for probability of finding a good outcome will be $\bar a$, with
\begin{equation}
    |a - \bar a| \sim \mathcal{O}(N^{-1/2}),
\end{equation}
as dictated by the sampling error of a multinomial distribution.

However, \ac{qae} can improve this result. Let us first define the central operator for \ac{qae} \cite{brassard_amplitude_estimation_2002}
\begin{equation}
    \Q = - \A \S_0 \A^\dagger \S_{\psi_0},
\end{equation}
where the operators $\S_0$ and $\S_{\psi_0}$ are inherited from Grover's search algorithm \cite{grover}, being
\begin{eqnarray}
    \S_0 & = & \mathbf{I} - 2 \ket 0_n \bra 0_n \otimes \ket 0 \bra 0 , \\ 
    \S_{\psi_0} & = & \mathbf{I} - 2 \ket{\psi_0}_n\bra{\psi_0}_n \otimes \ket 0 \bra 0.
\end{eqnarray}
The $\S_0$ operator changes the sign of the $\ket 0_n \ket 0$ state, while $\S_{\psi_0}$ takes the role of an oracle and changes the sign of all bad outcomes.
\subsection[QAE with QPE]{\ac{qae} with \ac{qpe}}\label{app:ae_qpe}
The original \ac{qae} makes use of a \ac{qpe} subroutine to study the operator $\Q$, with eigenvalues $e^{\pm i 2 \theta_a}$, with $a = \sin^2(\theta_a)$~\cite{brassard_amplitude_estimation_2002}. The procedure of \ac{qpe} is then applied to extract an integer number $y \in \{0, 1, \ldots, 2^m-1\}$ such that $\bar{\theta}_a = \pi y / 2^m$ is an estimate of $\theta_a$, with $m$ the number of ancilla qubits. Recall that a \ac{qft} is required to perform \ac{qpe}.

The value of $\bar{\theta}_a$ leads to an estimate of $\bar a$, such that
\begin{equation}
  |a - \bar a| < \frac{2\pi \sqrt{a (1 - a)}}{2^m} + \frac{\pi^2}{2^{2m}} \sim \mathcal{O}\left(\frac{\pi}{2^m}\right)  
\end{equation}
with probability at least $8/\pi^2\approx 81\%$. 

The original \ac{qae} procedure requires the implementation of \ac{qpe}, which is highly resource demanding. Hence, the complexity of the circuit precludes its feasibility in the NISQ era.

\subsection[IQAE]{\ac{iqae}}\label{app:iqae}

Here, a method is presented for obtaining the most probable value of $a$ in an iterative fashion following similar methods as other \ac{qae} without \ac{qpe} algorithms. We base this procedure in the theory of confidence intervals for a binomial distribution assuming normal distributions \cite{binomial-wallis2013}.

\begin{algorithm}[t!]
\caption{\label{alg:gaussian} Algorithm for \ac{qae} based on gaussian distribution of the measurements. }
\DontPrintSemicolon
\SetKwFunction{FMain}{GaussianAmplitudeEstimation}
\SetKwFunction{FArcSin}{MultipleValuesArcsin}
\SetKwProg{Fn}{Function}{:}
\Fn{\FMain{$N_{\rm shots}$, $J$, $m_j$, $\alpha$}}
{\;
    $z \leftarrow {\rm CDF}_\mathcal{N}^{-1}(1 - \alpha / 2)$ \;
    Ensure $m_0 = 0 $ \;
    $a \leftarrow |\bra{1}\A\ket 0 |^2$ with $N_{\rm shots}$ samples\;
    $\theta_a^{(0)} \leftarrow \arcsin{\sqrt{a}}$
    $\Delta \theta_a^{(0)} = \frac{z}{2 \sqrt{N_{\rm shots}}}$\;
    \For{$j \leftarrow 1 \KwTo \; J$}
    {
        $a \leftarrow |\bra{1}Q^{m_j}A\ket 0 |^2$ with $N_{\rm shots}$ samples\;
        $\theta_{\rm array} \leftarrow $ \FArcSin{$a, m_{j-1}$} \;
        $\theta_a \leftarrow \min\left(|\theta_{\rm array} - \theta_a^{(j - 1)}|\right)$\;
        $\Delta \theta_a \leftarrow \frac{z}{2 (2 m_j + 1)\sqrt{N_{\rm shots}}}$\;
        $\theta_a^{(m_j)} \leftarrow \frac{\frac{\theta_a }{\Delta \theta_a^2} + \frac{\theta_a^{(j-1)} }{(\Delta \theta_a^{(j-1)})^2}}{\frac{1}{\Delta \theta_a^2} + \frac{1}{(\Delta \theta_a^{(j-1)})^2}}$\;
        $\Delta \theta_a^{(m_j)} \leftarrow \left(\frac{1}{\Delta \theta_a^2} + \frac{1}{(\Delta \theta_a^{(j-1)})^2}\right)^{-1/2}$\;
        $[a_j, \Delta a_j] \leftarrow [\sin^2\theta_a^{j}, \sin(2 \theta_a^{j}) \Delta \theta_a^{(j)}]$
    }
    \KwRet{$[a_j, \Delta a_j]$} }
\end{algorithm}

\begin{algorithm}[t!]
\caption{\label{alg:gaussian2} Extracting multiple values for the $\arcsin$, auxiliary function needed in Alg. \ref{alg:gaussian}.}
\DontPrintSemicolon
\SetKwFunction{FArcSin}{MultipleValuesArcsin}
\SetKwProg{Fn}{Function}{:}
\Fn{\FArcSin{$a, m$}}
{\;
    $\theta_0 \leftarrow \arcsin{\sqrt{a}}$ \tcp{The value of $\theta_0$ is bounded between $0$ and $\pi / 2$}
    The $\arcsin$ function has several solutions
    $\theta_{\rm array} \leftarrow [0] * (2 m + 1)$
    $\theta_{\rm array}[0] \leftarrow \theta_0$\;
    \For{$k \leftarrow 1 \KwTo\, m$}
    {
        $\theta_{\rm array}[2k - 1]\leftarrow k \pi - \theta_0$\;
        $\theta_{\rm array}[2k]\leftarrow k \pi + \theta_0$
    }
    $\theta_{\rm array} \leftarrow \theta_{\rm array} / (2 m + 1)$\;
    \KwRet $\theta_{\rm array}$}
\end{algorithm}

A binomial distribution with probability $a$ is considered, i. e. for every sample the chance of obtaining $1$ is $a$, while the chance of obtaining $0$ is $1 - a$. Then, if an estimate $\hat a$ of $a$ was obtained using $N$ samples, the true value of $a$ lies in the interval
\begin{equation}
    a = \hat a \pm \frac{{\rm CDF}_\mathcal{N}^{-1}(1 - \alpha/2) \sqrt{\hat a(1-\hat a)}}{2 \sqrt{N}}, 
\end{equation}
with confidence $(1 - \alpha)$.

From this result \ac{qae} can construct an iterative algorithm returning the optimal value of $a$ using \ac{qae}. A set of $m_j$ for $j={0,1,2,3,\ldots}$ is considered. For every $m_j$ the probability of obtaining $\ket 1$ is $\sin^2((2 m_j + 1) \theta_a)$, where $a = \sin^2(\theta_a)$. In the $\theta$ space, for a given $m$ the values and error of $\theta$ obtained are
\begin{equation}\label{eq:weighted}
    \theta_a = \frac{1}{2m+1}\arcsin(\sqrt{a}) \qquad \Delta \theta_a = \frac{1}{2m + 1}\frac{{\rm CDF}_\mathcal{N}^{-1}(1 - \alpha/2)}{2 \sqrt{N}}.
\end{equation}
It is important to understand two main properties of Eq. \eqref{eq:weighted}. First, there are $2m + 1$ possible values for $\theta_a$ within the interval $\theta_a \in [0, \pi / 2]$ as the $\sin^2(\cdot)$ function is $\pi$-periodical. For every new iteration it will be necessary to choose one of them.  It is very important to set $m_j = 0$ at first because this case is the only one for which $\theta_a$ corresponds to the expected value for $a$. Otherwise, several possible values of $a$ arise and it is not possible to tell which one is correct. Combining results for several values of $m_j$, it is possible to bound the uncertainty to be as small as desired.

The algorithm is based on the following statements. 
For a given collection of measurements and uncertainties $\{\theta_i, \Delta \theta_i\}$, the weighted average and uncertainty from the first $j$ terms is
\begin{equation}\label{eq:ap_exp_errors}
    \Tilde{\theta}_j = \frac{\sum_{i=0}^j \theta_i / \Delta \theta_i^2}{\sum_{i=0}^j 1 / \Delta \theta_i^2}  \qquad \Delta \Tilde{\theta}_j = \left(\sum_{i=0}^j 1 / \Delta \theta_i^2\right)^{-1/2}.
\end{equation}
Notice also that this relation is recursive, as $\Tilde{\theta}_{j+1}$ can be obtained by combining $\Tilde{\theta}_{j}$ and $\theta_{j+1}$. The same holds for uncertainties. Thus, the interpretation of this algorithm is that for every new step $j$ a new term is added to the series $\{\theta, \Delta \theta\}$. The individual uncertainties decrease as $\sim ((2 m + 1)^{-1})$, and the final global uncertainty is obtained as
\begin{equation}\label{eq:uncertainty}
    \Delta \theta = \frac{{\rm CDF}_\mathcal{N}^{-1}(1 - \alpha / 2)}{\sqrt{N}}\left( \sum_{j=0}^J (2m_j + 1)^2 \right)^{-1/2},
\end{equation}
where $J$ denotes the last iteration performed. The full recipe for the algorithm is described in Algs. \ref{alg:gaussian} and \ref{alg:gaussian2}.

In the case of a linear selection of $m_j$, i. e. $m_j = j; j=(0, 1, 2, ..., J)$, the asymptotic behavior of this uncertainty is $\Delta \theta = \mathcal{O}(N^{-1/2} M^{-3/4})$, with $M$ the sum of all $m$. For discovering it we just have to compute

\begin{equation}
    \sum_{j=0}^J (j + 1)^2 = 4 \sum_{j=0}^J j^2 + 4 \sum_{j=0}^J j + \sum_{j=0}^J 1.
\end{equation}
We now take the identities $\sum_{j=0}^J j = J(J+1)/2=M$ and $\sum_{j=0}^J j^2 = J(2J+1)(J+2)/6$. Then, it is direct to check that
\begin{equation}\label{eq:prec_lineal}
    \Delta \theta = \mathcal{O}(N^{-1/2} J^{-3/2}) = \mathcal{O}(N^{-1/2} M^{-3/4}).
\end{equation}

This behavior already surpasses the tendency of the classical sampling, but does not reach the optimal \ac{qae} with \ac{qpe}.

In the case of an exponential selection of $m_j$, i. e. $m_j = \{0\} \cup \{2^j\}; j=(0, 1, 2, ..., J)$ we can take the identities $\sum_{j=0}^J 2^j = 2^J - 1 =M$ and $\sum_{j=0}^J 2^{2j} = (2^{2J} - 1) / 3$. Then it is direct to check that
\begin{equation}\label{eq:prec_exp}
    \Delta \theta = \mathcal{O}(N^{-1/2} 2^{-J}) = \mathcal{O}(N^{-1/2} M^{-1}).
\end{equation}

\subsubsection{Extension of \ac{qae} to error-mitigation techniques}\label{app:ae_error}

The error-mitigation procedure proposed for the unary algorithm discards some of the algorithm instances to retain outcomes within the unary basis. This reduces the precision achieved in the algorithm with respect to the ones predicted in Eqs. \eqref{eq:prec_lineal} and \eqref{eq:prec_exp} in order to maintain accuracy. This section provides some lower bounds on how many \ac{qae} iterations can be done while still reaching quantum advantage. 

We will work now in the scheme where $m_j = j$. Let us assume that, in every iteration of \ac{qae}, only a fraction $\tilde p_j$ of the shots are retained. The equivalent version of Eq. \eqref{eq:uncertainty} is now
\begin{equation}
    \Delta \theta = {\rm CDF}_\mathcal{N}^{-1}(1 - \alpha / 2)\left( \sum_{j=0}^J (2m_j + 1)^2 N\tilde p_j \right)^{-1/2}.
\end{equation}

As more errors are bound to occur, $\tilde p_j$ decreases as $m_j$ increases, we can state a bound for the accuracy as
\begin{equation}
    \Delta \theta \leq \frac{{\rm CDF}_\mathcal{N}^{-1}(1 - \alpha / 2)}{\sqrt{N \tilde p_J}}\left( \sum_{j=0}^J (2m_j + 1)^2 \right)^{-1/2}, 
\end{equation}
since the precision is at least as good as the one obtained for the worst-case scenario. Comparing the trends, both in the linear and the exponential case, with the classical scaling, it is possible to see that quantum advantage is still achieved provided
\begin{equation}\label{eq:p_J}
\tilde p_J \geq M^{1 - 2\alpha},    
\end{equation}
with $\alpha = 3/4$ in the linear case and $\alpha = 1$ in the exponential case. These quantities for the linear case correspond to the dashed lines in Figs.~\ref{fig:errors} and~\ref{fig:several_bins_errors}. 

The probability of retaining a shot is at least the probability of having no errors in the circuit, considering that some double errors may lead to erroneous instances that belong even though to the unary basis. This zero-error probability in the worst case scenario, that is, at the last iteration of \ac{qae}, is written as
\begin{equation}
    p_0 = \left((1-p_e)^{a n + b}\right)^{m_J},
\end{equation}
where $p_e$ is the error of an individual gate, and $a$ and $b$ are related to the gate scaling, see Tab. \ref{tab:gates} for the details. In principle, one can expand the calculation of $p_0$ by considering different kinds of errors for different gates, but for the sake of simplicity we will focus on this analysis. Rearranging together the results for Eqs. \eqref{eq:prec_lineal}, \eqref{eq:prec_exp} and \eqref{eq:p_J} it is possible to see that quantum advantage is obtained if the individual gate errors is bounded by
\begin{equation}
    p_e < 1 - m_J^{\frac{2 - 4\alpha}{(a n + b)m_J}}.
\end{equation}

%auto-ignore
\chapterimage{qibo_logo.pdf}

\chapter{Qibo}\label{app:qibo}

\begin{adjustwidth}{8cm}{0cm}
\input{_minted-ms/19DAEC146153BF69F6D97BDF01A542EB18028B59243138779052D9B8CAF79794.pygtex}
%\begin{minted}{python}
%import qibo
%\end{minted}
\end{adjustwidth}

{\tt Qibo} is an open-source software project to write and execute both quantum circuits and adiabatic quantum computing in a user-friendly manner \cite{qibo, qibo_code}. In the global perspective, {\tt Qibo} comes as a python library to join other such as {\tt qiskit}~\cite{qiskit}, {\tt cirq}~\cite{cirq}, {\tt Forest}~\cite{forest}, {\tt Qulacs}~\cite{suzuki_qulacs_2020} or {\tt QCGPU}~\cite{qcgpu} among many others
\cite{Jones_2019,10.1007/978-3-319-27119-4_17,Steiger_2018,qsharp,zulehner2017advanced,
pednault2017paretoefficient,PhysRevLett.116.250501,
DERAEDT2007121,Fried_2018,Villalonga_2019,luo2019yaojl,bergholm2018pennylane,10.1145/3310273.3323053,10.1007/978-3-030-50433-5_35,Jones_2020,Chen_2018,
EasyChair:4050,meyerov2020simulating,moueddene2020realistic,wang2020quantum}. The effort of the {\tt Qibo} project is coordinated by TII\footnote{\href{https://tii.ae/}{Quantum Research Center, Technology Innovation Institute, Abu Dhabi, United Arab Emirates}} and Qilimanjaro\footnote{\href{http://www.qilimanjaro.tech/}{Qilimanjaro Quantum Tech, Barcelona, Spain}}.
The current status of {\tt Qibo} is that quantum algorithms can only be exactly simulated on classical hardware, but future plans forsee to extend the calculations on approximate classical methods such as the family of \ac{tn}s \cite{biamonte_tensor_2017}, and on actual quantum devices.

The structure of {\tt Qibo} is designed to be easy to use but extremely efficient to perform calculations. The final target is to make the catalogue of functionalities and the available applications increase with time. The high-level API receives the instructions from the user and allocates automatically all different ingredients through the code to be finally executed on a specific and optimized backend. The API can receive both simple instructions, such as gates and circuits, but also more ellaborated models, for example a \ac{vqe}. In this chapter a shallow review on the capabilities of {\tt Qibo} is covered. For each explanation, a short code to implement it in {\tt Qibo} is included. 

\section{Circuits}

Quantum circuits are the main paradigm for implementing a quantum computing in {\tt Qibo}. The circuits in {\tt Qibo} are an abstract object capable to implement different tools normally used in theoretical descriptions.

\subsubsection{Initialization}

The circuit object is initialized by defining the number of qubits. Automatically, an abstract object is generated to store the circuit. When executing the circuit, a vector state $\ket\psi$ with exactly $2^n$ complex coefficients is required to initialize the computation, with default $\ket\psi = \ket 0^{\otimes n}$. {\tt Qibo} also permits the use of a density matrix $\rho$, initially defined as $\rho = \ket\psi\bra\psi$, with $4^n$ complex coefficients. Density matrices imply more costly calculations both in time and in storage memory since there are more numbers to manage. However, the use of density matrices allow represent mixed states, including those affected by noise.
\input{_minted-ms/51A824172313AA5664A0C31EA2D5A6F918028B59243138779052D9B8CAF79794.pygtex}
%\begin{minted}{python}
%from qibo.models import Circuit
%qubits = 2
%c = Circuit(qubits)
%print(c)
%>>> <qibo.core.circuit.Circuit at 0x7f167347d668>
%state = c().numpy()
%print(state)
%>>> array([1.+0.j, 0.+0.j, 0.+0.j, 0.+0.j])
%\end{minted}
\subsubsection{Gates}
Quantum operations are added to the circuit subsequently to modify the state. The available operations are of three kinds
\begin{itemize}
\item Gates: Standard unitary gates $U$ are mainly used in {\tt Qibo}. These operations modify the quantum state as $U \psi$, and the density matrix as $U \rho U^\dagger$. Gates can be defined to affect one or two qubits. {\tt Qibo} allows to implement on any qubit any gate controlled by an arbitrary set of qubits. In case any gate is parameterized, the circuit incorporates a list of parameters that can be updated in subsequent executions.
\input{_minted-ms/2375952521F051EE31AE2F7242CFFB7318028B59243138779052D9B8CAF79794.pygtex}
%\begin{minted}{python}
%from qibo import gates
%import numpy as np
%qubits = 2
%c = Circuit(qubits)
%c.add(gates.X(0)) # gate X in qubit 0
%c.add(gates.H(1)) # gate H in qubit 1
%c.add(gates.RY(0, theta=np.pi / 2)) # gate RY with angle \pi / 2  in qubit 0
%c.add(gates.CNOT(0, 1)) # gate CNOT with control 0, target 1
%c.add(gates.CRY(0, 1, theta = np.pi)) # gate controlled RY with control 0, target 1, angle \pi
%c.add(gates.RY(1, theta = np.pi).controlled_by(0)) # same as above, different version
%print(c.draw())
%>>> q0: -X-RY-o-o--o--
%>>> q1: -H----X-RY-RY-
%\end{minted}
\item Channels: Quantum channels are the standard tool to introduce decoherence in a quantum circuit. Channels are defined as a set of unitary gates to be applied on a density matrix $\rho$ with different probabilities. As an additional feature, the quantum state representing the circuit is transformed automatically into a density matrix to accomodate the quantum channel.
\input{_minted-ms/E248AC2963CE65203985245166CF651918028B59243138779052D9B8CAF79794.pygtex}
%\begin{minted}{python}
%from qibo import gates
%import numpy as np
%qubits = 2
%c = Circuit(qubits)
%c.add(gates.PauliNoiseChannel(0, px = 0.05, py = 0.05, pz = 0.05)) # Pauli Noise Channel in qubit 0
%c.add(gates.ResetChannel(1, p0 = 0.05, p1 = 0.05)) # Reset Noise Channel in qubit 1
%\end{minted}
\item Measurements: the measuring step is the only chance to retrieve information from the quantum circuits. In {\tt Qibo}, measurements can be allocated at the end of the circuit to finish the computation, but also on any intermediate step. In the latter case, the circuit is modified accordingly to capture all the outcomes possibilities. 
\input{_minted-ms/32B6390478B8F7284ED651EBE035949E18028B59243138779052D9B8CAF79794.pygtex}
%\begin{minted}{python}
%qubits = 2
%c = Circuit(qubits)
%c.add(gates.X(0)) # gate X in qubit 0
%c.add(gates.H(1)) # gate H in qubit 1
%c.add(gates.M(0, collapse=True)) # measurement in qubit 0
%c.add(gates.H(0)) # gate H in qubit 0, after measurement
%c.add(gates.CNOT(0, 1)) # gate CNOT with control 0, target 1
%c.add(gates.M(0, collapse=False)) # measurement in qubit 0
%c.add(gates.M(1, collapse=False)) # measurement in qubit 1
%print(c.draw())
%>>> q0: -X-M-H-o-M-
%>>> q1: -H-----X-M-
%\end{minted}
\end{itemize}

\subsubsection{Hamiltonians}
Hamiltonians serve the purpose of measuring quantities of interest in the quantum circuit. {\tt Qibo} allows to define a hamiltonian as a full matrix by setting all components, and also from a symbolic description. In the case of quantum circuits, hamiltonians can be used to measured the expected value of quantum state, and also to compare it with the ground state. {\tt Qibo} supports corresponding tools to manage those calculations. 
\input{_minted-ms/5E017A48AEAAF7F4969283FA91D728D118028B59243138779052D9B8CAF79794.pygtex}
%\begin{minted}{python}
%from qibo.hamiltonians import Hamiltonian, SymbolicHamiltonian
%from qibo import matrices, symbols
%
%#### NUMERICAL HAMILTONIAN ####
%
%# ZZ terms
%matrix = np.kron(np.kron(matrices.Z, matrices.Z), np.kron(matrices.I, matrices.I))
%matrix += np.kron(np.kron(matrices.I, matrices.Z), np.kron(matrices.Z, matrices.I))
%matrix += np.kron(np.kron(matrices.I, matrices.I), np.kron(matrices.Z, matrices.Z))
%matrix += np.kron(np.kron(matrices.Z, matrices.I), np.kron(matrices.I, matrices.Z))
%# X terms
%matrix += np.kron(np.kron(matrices.X, matrices.I), np.kron(matrices.I, matrices.I))
%matrix += np.kron(np.kron(matrices.I, matrices.X), np.kron(matrices.I, matrices.I))
%matrix += np.kron(np.kron(matrices.I, matrices.I), np.kron(matrices.X, matrices.I))
%matrix += np.kron(np.kron(matrices.I, matrices.I), np.kron(matrices.I, matrices.X))
%# Create numerical Hamiltonian object
%numeric_ham = Hamiltonian(4, matrix)
%
%#### SYMBOLIC HAMILTONIAN ####
%# ZZ terms
%symbolic_ham = sum(symbols.Z(i) * symbols.Z(i + 1) for i in range(3))
%# periodic boundary condition term
%symbolic_ham += symbols.Z(0) * symbols.Z(3)
%# X terms
%symbolic_ham += sum(symbols.X(i) for i in range(4))
%
%# Define a Hamiltonian using the above form
%symbolic_ham = SymbolicHamiltonian(symbolic_ham)
%# This Hamiltonian is memory efficient as it does not construct the full matrix
%print(np.sum(np.abs(symbolic_ham.matrix - numeric_ham.matrix)))
%>>> 0.0
%\end{minted}

\subsubsection{Callbacks}
It is often of interest to study the change of different quantities along the execution of a given circuit. To carry this task, {\tt Qibo} implements the callbacks functionality allowing to extract those quantities when possible at any point of the circuit. Interesting and commonly used callbacks are expected values with respect to some hamiltonian, but entanglement entropy, overlap with respect to a target state or norm of the quantum state are also included.  

Callbacks can only be used in simulation backends since any measurement on an actual device would destroy the computation. This is however a useful tool to develop proof-of-concept models to scale later. 
\input{_minted-ms/9722D6AE50778586C9D7F6C79935EEE218028B59243138779052D9B8CAF79794.pygtex}
%\begin{minted}{python}
%from qibo import callbacks
%
%# create entropy callback where qubit 0 is the first subsystem
%entropy = callbacks.EntanglementEntropy([0])
%
%# initialize circuit with 2 qubits and add gates
%c = models.Circuit(2) # state is |00> (entropy = 0)
%c.add(gates.CallbackGate(entropy)) # performs entropy calculation in the initial state
%c.add(gates.H(0)) # state is |+0> (entropy = 0)
%c.add(gates.CallbackGate(entropy)) # performs entropy calculation after H
%c.add(gates.CNOT(0, 1)) # state is |00> + |11> (entropy = 1))
%c.add(gates.CallbackGate(entropy)) # performs entropy calculation after CNOT
%
%# execute the circuit using the callback
%final_state = c()
%
%print(entropy[:])
%>>> [0.0 0.0 1.0]
%\end{minted}

\subsubsection{Models}
Different circuit models play the role of pre-defined architectures with specific functionalities for solving particular problems. The prominent examples of \ac{vqe} \cite{Peruzzo_vqe_2014} and \ac{qaoa} \cite{farhi_qaoa_2014} are included, but also less known examples as the qPDF \cite{perezsalinas_proton_2021}.
\input{_minted-ms/DC87131AE5C1B94B26CB56112889DEFA18028B59243138779052D9B8CAF79794.pygtex}
%\begin{minted}{python}
%from qibo.hamiltonians import XXZ
%from qibo.models import VQE
%# create circuit ansatz for two qubits
%circuit = Circuit(2)
%circuit.add(gates.RY(0, theta=0))
%circuit.add(gates.RY(1, theta=0))
%# create XXZ Hamiltonian for two qubits
%hamiltonian = XXZ(2)
%# create VQE model for the circuit and Hamiltonian
%vqe = VQE(circuit, hamiltonian)
%# optimize using random initial variational parameters
%initial_parameters = np.random.uniform(0, 2, 2)
%result = vqe.minimize(initial_parameters)
%print('Found energy:', result[0])
%>>> Found energy: -1.9999976950940028
%\end{minted}

\subsubsection{Optimization}
Many models defined on quantum circuits depend on \ac{vqa}s where some optimization method is required. Built-in classical optimizers are available in {\tt Qibo} to be used in all methods. The optimizers are both gradient-descent \cite{nielsen_neural_2015, kingma_adam_2017}, quasi-newton \cite{l-bfgs, scipy}, or genetic \cite{cma} algorithms. 

For simulation backends there is an available option to parallelize the different pieces of optimization across different computation cores, improving the overall performance of the process. 
\input{_minted-ms/014AF5EE95E6ACAECCFDCE73253C9C6018028B59243138779052D9B8CAF79794.pygtex}
%\begin{minted}{python}
%from qibo.optimizers import optimize
%
%# create custom loss function
%# make sure the return type matches the optimizer requirements.
%def myloss(parameters, circuit):
%    circuit.set_parameters(parameters)
%    return np.square(np.sum(circuit())) # returns numpy array
%
%# create circuit ansatz for two qubits
%circuit = Circuit(2)
%circuit.add(gates.RY(0, theta=0))
%circuit.add(gates.RY(1, theta=0))
%
%# optimize using random initial variational parameters
%initial_parameters = np.random.uniform(0, 2, 2)
%best, params, extra = optimize(myloss, initial_parameters, args=(circuit))
%
%# set parameters to circuit
%circuit.set_parameters(params)
%\end{minted}

\section{Adiabatic computing}
The adiabatic computing section of {\tt Qibo} performs all necessary calculations to gradually modify a quantum state according to an adiabatic scheme \cite{farhi_adiabatic_2000}. In adiabatic computing, the ground state of a known hamiltonian is taken as the starting line. Then, a time-dependent hamiltonian is triggered in such a way that the starting one is slowly modified into another problem hamiltonian. The ground state of the problem hamiltonian encodes the desired solution to a given problem.
Then, the quantum state evolves according to the time-dependent Schrödinger equation. If the time evolution is slow enough, the overlap between the final evolved state and the ground state of the problem hamiltonian is close to $1$. 
\input{_minted-ms/E67EF8EE28A60C5897941622188A620018028B59243138779052D9B8CAF79794.pygtex}
%\begin{minted}{python}
%from qibo.hamiltonians import X, TFIM
%from qibo.models import AdiabaticEvolution
%
%h0 = X(2)
%h1 = TFIM(2, h=1)
%
%# Calculate target values (H1 ground state)
%target_state = h1.ground_state()
%target_energy = h1.eigenvalues()[0]
%
%# Check ground state
%state_energy = callbacks.Energy(h1)(target_state)
%np.testing.assert_allclose(state_energy.real, target_energy)
%
%energy = callbacks.Energy(h1)
%overlap = callbacks.Overlap(target_state)
%evolution = AdiabaticEvolution(h0, h1, lambda t: t, dt = 0.01,
%                                      callbacks=[energy, overlap])
%final_psi = evolution(final_time=1)
%print('Obtained energy:', energy[-1])
%print('Target energy:', target_energy)
%>>> Obtained energy: -2.698048119941422
%>>> Target energy: -2.82842712474619
%\end{minted}

\subsubsection{Hamiltonian}
Hamiltonians drive the time evolution in adiabatic quantum computing. {\tt Qibo} uses a Trotter implementation \cite{trotter} of the hamiltonians of interest to execute them more efficiently than if all information is conserved along the different steps.
\input{_minted-ms/A2FE20A4C74A7BCACBC009D7E50DF5E618028B59243138779052D9B8CAF79794.pygtex}
%\begin{minted}{python}
%h1 = TFIM(2, h=1, dense=False) # Add the label to avoid full matrix
%\end{minted}

\subsubsection{Solver}
Numerically solving the Schrödinger equation is far from being trivial. {\tt Qibo} incorporates different solvers to perform the evolution in several manners. Those solvers include the exponential and trotterized exponential evolution, and Runge-Kutta methods. 
\input{_minted-ms/82E8F4E1DA9530260AF23F7C0856EF6118028B59243138779052D9B8CAF79794.pygtex}
%\begin{minted}{python}
%from qibo import solvers
%
%evolution = AdiabaticEvolution(h0, h1, lambda t: t, dt = 0.01,
%                                      callbacks=[energy, overlap],
%                                       solver=solvers.RungeKutta45)
%\end{minted}

\section{Backends}
Backends are the main strength of {\tt Qibo} when competing against other similar libraries. The organization behind {\tt Qibo} plans to add quantum machines and approximate simulation methods to the available backends. However, at the time of writing this thesis, {\tt Qibo} can only be executed on exact classical simulators.
\input{_minted-ms/9F1B3A5377A52BB1C6417703B30909EA18028B59243138779052D9B8CAF79794.pygtex}
%\begin{minted}{python}
%from qibo import set_backend
%set_backend('qibojit') #Change the backend easily by selecting the name
%\end{minted}

\subsection{Classical Simulation - Hardware acceleration}

{\tt Qibo} was in origin designed to be easily used and to perform efficient computation. The most prominent feature for accomplishing this task is the capability of {\tt Qibo} to adapt its management of computation to the environment. Depending essentially on the size of the quantum circuit to be simulated, the execution can be performed on several hardware simulations:
\begin{itemize}
\item Single-thread CPU: \ac{cpu} are the main component of a classical computer. They are used as general-purpose machines. For {\tt Qibo}, this mode means that all calculation are passed to only one \ac{cpu} core. It is convenient to use this mode for small circuits, up to 15 qubits, since the memory storage does not need to be extremely large, and the overhead of using other methods do not compensate the gain. 
\item Single-thread GPU: \acf{gpu} were developed for \ac{ml} for their capability to perform linear algebra operations. This feature is now useful for simulating quantum circuits efficiently. In this mode, the \ac{cpu} passes the data to \ac{gpu} to perform the calculation. This translates into an overhead in the computation. Only for moderately large circuits this overhead is compensated by the more efficient calculation of a \ac{gpu}. The ideal range is between 15 and 30 qubits. Over that size, current \ac{gpu}s do not have enough memory to execute the calculation. 
\item Multi-thread CPU: this method is essentially useful for executing quantum circuits whose storage requirements exceed the capabilities of one \ac{cpu}. The recommended range is for more than 15 qubits. The performance does not in general overcome a \ac{gpu}.
\item Multi-thread GPU: this is the most special mode, and exclusive of {\tt Qibo}. Computational efforts are distributed accross many \ac{gpu}s with a significant overhead. This method is only efficient for simulating large circuits, over 30 qubits.
\end{itemize}

\subsubsection{QIBOJIT backend}
\ac{jit} compilation is a computational framework that optimizes the execution of a given algorithm in subsequent runs of it. In the case of {\tt Qibo}, the \ac{jit} compilation is assisted by {\tt cupy}~\cite{cupy} and {\tt numba}~\cite{numba}

For using the \ac{jit} backend, custom operators optimized for this functioning scheme were developed and implemented in {\tt Qibo}

\subsubsection{QIBOTF backend}
The first release of {\tt Qibo} was built on {\tt Tensorflow} \cite{tensorflow2015-whitepaper}. The QIBOTF backend supports custom operations to be implemented using the general {\tt Tensorflow} framework. 

This backend support hardware management as inherited from {\tt Tensorflow}. Some operations are implemented faster than in the original code, but in exchange some features, specially for automatic differentiation, were lost in the process. 

\subsubsection{TENSORFLOW and NUMPY backends}
In both cases, the calculations are performed using the {\tt einsum} method of the corresponding {\tt Tensorflow} or {\tt Numpy} package \cite{numpy}. All function-alities are taken from the parent libraries.

\section{Examples}

{\tt Qibo} looks forward to creating a community of users contributing to the development of the open-source project. As a starting line of extended collaboration, several examples were developed by the creators of {\tt Qibo}
\begin{itemize}
\item Examples inspired in the data re-uploading strategy explored in Ch.~\ref{ch:reuploading} \cite{perezsalinas_data_2020, perezsalinas_proton_2021}.
\item Examples on the unary strategy from Ch.~\ref{ch:unary} \cite{ramos_unary_2021}
\item Measuring the tangle of a three-qubit state \cite{perezsalinas_tangle_2020}.
\item The performance of a \ac{vqe} when solving a condensed matter problem as in Ref.~\cite{BravoPrieto_scaling_2020}
\item Examples of Grover's algorithm \cite{grover} for a 3SAT problem \cite{3sat} and cryptographic applications \cite{bernstein2008chacha}.
\item A quantum autoencoder \cite{BravoPrieto_autoencoders_2021}
\item The Quantum Singular Value Decomposer \cite{bravoprieto_quantum_2020}
\item Adiabatic evolutions for a exact Cover problem \cite{3sat}
\item Shor's factorization algorithm \cite{shor} and a version requiring less qubits \cite{beauregard_shor_2003}
\end{itemize}

The list of available examples is expected to grow in the short term.

\chapterimage{chapter_bibliography.pdf}
\chapter{Bibliography}

\begin{adjustwidth}{4cm}{0cm}
{\sl El Universo (que otros llaman la Biblioteca) se compone de un número indefinido, tal vez infinito, de...}

\hfill Jorge Luis Borges\\
\end{adjustwidth}

%\addcontentsline{toc}{chapter}{\textcolor{black}{Bibliography}}
%\printbibliography[heading=bibempty]
\section*{Books}
%\addcontentsline{toc}{section}{Books}

\printbibliography[heading=bibempty,type=book]
\section*{Articles}
\defbibfilter{articles}{
  type=article or
  type=inproceedings or
  type=incollection
}
%\addcontentsline{toc}{section}{Articles}
\printbibliography[heading=bibempty,filter=articles]
\section*{Preprints}
%\addcontentsline{toc}{section}{Preprints}
\printbibliography[heading=bibempty,type=misc, keyword={preprint}]
\section*{Software and others}
%\addcontentsline{toc}{section}{Software and others}
\defbibfilter{software}{
  type=online
}
\printbibliography[heading=bibempty,filter=software, keyword={software}]

@article{perezsalinas_data_2020,
  doi = {10.22331/q-2020-02-06-226},
  url = {https://doi.org/10.22331/q-2020-02-06-226},
  year = {2020},
  month = feb,
  publisher = {Verein zur Forderung des Open Access Publizierens in den Quantenwissenschaften},
  volume = {4},
  pages = {226},
  author = {Adri{\'{a}}n P{\'{e}}rez-Salinas and Alba Cervera-Lierta and Elies Gil-Fuster and Jos{\'{e}} I. Latorre},
  title = {Data re-uploading for a universal quantum classifier},
  journal = {Quantum}
}

@article{perezsalinas_qubit_2021,
  doi = {10.1103/physreva.104.012405},
  url = {https://doi.org/10.1103/physreva.104.012405},
  year = {2021},
  month = jul,
  publisher = {American Physical Society ({APS})},
  volume = {104},
  number = {1},
  author = {Adri{\'{a}}n P{\'{e}}rez-Salinas and David L{\'{o}}pez-N{\'{u}}{\~{n}}ez and Artur Garc{\'{\i}}a-S{\'{a}}ez and P. Forn-D{\'{\i}}az and Jos{\'{e}} I. Latorre},
  title = {One qubit as a universal approximant},
  journal = {Physical Review A}
}

@article{perezsalinas_proton_2021,
  doi = {10.1103/physrevd.103.034027},
  url = {https://doi.org/10.1103/physrevd.103.034027},
  year = {2021},
  month = feb,
  publisher = {American Physical Society ({APS})},
  volume = {103},
  number = {3},
  author = {Adri{\'{a}}n P{\'{e}}rez-Salinas and Juan Cruz-Martinez and Abdulla A. Alhajri and Stefano Carrazza},
  title = {Determining the proton content with a quantum computer},
  journal = {Physical Review D}
}

@article{ramos_unary_2021,
  doi = {10.1103/physreva.103.032414},
  url = {https://doi.org/10.1103/physreva.103.032414},
  year = {2021},
  month = mar,
  publisher = {American Physical Society ({APS})},
  volume = {103},
  number = {3},
  author = {Sergi Ramos-Calderer and Adri{\'{a}}n P{\'{e}}rez-Salinas and Diego Garc{\'{\i}}a-Mart{\'{\i}}n and Carlos Bravo-Prieto and Jorge Cortada and Jordi Planagum{\`{a}} and Jos{\'{e}} I. Latorre},
  title = {Quantum unary approach to option pricing},
  journal = {Physical Review A}
}

@article{stamatopoulos_binary_2020,
  doi = {10.22331/q-2020-07-06-291},
  url = {https://doi.org/10.22331/q-2020-07-06-291},
  year = {2020},
  month = jul,
  publisher = {Verein zur Forderung des Open Access Publizierens in den Quantenwissenschaften},
  volume = {4},
  pages = {291},
  author = {Nikitas Stamatopoulos and Daniel J. Egger and Yue Sun and Christa Zoufal and Raban Iten and Ning Shen and Stefan Woerner},
  title = {Option Pricing using Quantum Computers},
  journal = {Quantum}
}

@article{perezsalinas_tangle_2020,
  doi = {10.3390/e22040436},
  url = {https://doi.org/10.3390/e22040436},
  year = {2020},
  month = apr,
  publisher = {{MDPI} {AG}},
  volume = {22},
  number = {4},
  pages = {436},
  author = {Adrián Pérez-Salinas and Diego García-Martín and Carlos Bravo-Prieto and José I. Latorre},
  title = {Measuring the Tangle of Three-Qubit States},
  journal = {Entropy}
}

@misc{qibo,
      title={Qibo: a framework for quantum simulation with hardware acceleration}, 
      author={Stavros Efthymiou and Sergi Ramos-Calderer and Carlos Bravo-Prieto and Adrián Pérez-Salinas and Diego García-Martín and Artur Garcia-Saez and José Ignacio Latorre and Stefano Carrazza},
      year={2020},
      eprint={2009.01845},
      archivePrefix={arXiv},
      primaryClass={quant-ph}
}

@book{nielsen_chuang_2010, 
	place={Cambridge}, 
	title={Quantum Computation and Quantum Information: 10th Anniversary Edition},
 	DOI={10.1017/CBO9780511976667}, 
 	url = {https://doi.org/10.1017/CBO9780511976667.001},
	publisher={Cambridge University Press}, 
	author={Nielsen, Michael A. and Chuang, Isaac L.}, 
	year={2010}, isbn = {978-1107002173}}

@article{riemann_fourier_1867,
  year = {1867},
  volume = {13},
  author = {Bernhard Riemann},
  title = {Über die Darstellbarkeit einer Function durch eine trigonometrische Reihe},
  journal = {Abhandlungen der Königlichen Gesellschaft der Wissenschaften zu Göttingen}
}

@article{dirichlet_fourier_1829,
  url = {https://arxiv.org/abs/0806.1294},
  year = {1829},
  volume = {4},
  pages = {157--169},
  author = {Peter Gustav Lejeune Dirichlet},
  title = {Sur la convergence des séries trigonométriques qui servent à représenter une fonction arbitraire entre des limites données},
  journal = {Journal für die reine und angewandte Mathematik}
}

@article{carleson_fourier_1966,
author = "Carleson, Lennart",
doi = "10.1007/BF02392815",
fjournal = "Acta Mathematica",
journal = "Acta Math.",
pages = "135--157",
publisher = "Institut Mittag-Leffler",
title = "On convergence and growth of partial sums of Fourier series",
url = "https://doi.org/10.1007/BF02392815",
volume = "116",
year = "1966"
}

@online{2d_functions, 
title={BenchmarkFcns Toolbox},
author={Ardeh, Mazhar Ansari}, 
year={2016}, 
url={http://www.benchmarkfcns.xyz/}
}

@article{hahn_analysis_1927,
  title={{\"U}ber lineare Gleichungssysteme in linearen R{\"a}umen.},
  author={Hahn, Hans},
  journal={Journal f{\"u}r die reine und angewandte Mathematik},
  volume={157},
  pages={214--229},
  year={1927}
}

@article{banach_analysis_1929,
  title={Sur les fonctionnelles lin{\'e}aires II},
  author={Banach, Stefan},
  journal={Studia Mathematica},
  volume={1},
  pages={223--239},
  year={1929},
  publisher={Instytut Matematyczny Polskiej Akademii Nauk}
}

@inproceedings{riesz_analysis_1914,
  title={D{\'e}monstration nouvelle d'un th{\'e}or{\`e}me concernant les op{\'e}rations fonctionnelles lin{\'e}aires},
  author={Riesz, Fr{\'e}d{\'e}ric},
  booktitle={Annales scientifiques de l'{\'E}cole Normale Sup{\'e}rieure},
  volume={31},
  pages={9--14},
  year={1914}
}

@book{ash_analysis_1972,
  author = {Robert B. Ash},
  doi = {10.1016/c2013-0-06164-6},
  url = {https://doi.org/10.1016/c2013-0-06164-6},
  year = {1972},
  publisher = {Elsevier},
  title = {Real Analysis and Probability}, 
  isbn = {978-0120652013}
}

@book{weir_analysis_1974,
  title={General integration and measure},
  author={Weir, Alan J.},
  volume={2},
  year={1974},
  publisher={CUP Archive}, isbn = {978-0521297158}
}

@article{scikit-learn,
 title={Scikit-learn: Machine Learning in {P}ython},
 author={Pedregosa, F. and Varoquaux, G. and Gramfort, A. and Michel, V.
         and Thirion, B. and Grisel, O. and Blondel, M. and Prettenhofer, P.
         and Weiss, R. and Dubourg, V. and Vanderplas, J. and Passos, A. and
         Cournapeau, D. and Brucher, M. and Perrot, M. and Duchesnay, E.},
 journal={Journal of Machine Learning Research},
 volume={12},
 pages={2825--2830},
 year={2011}
}

@online{cma_package,
	title = {{CMA}-{ES}/pycma: r3.0.3},
	shorttitle = {{CMA}-{ES}/pycma},
	url = {https://zenodo.org/record/3764210},
	urldate = {2021-06-03},
	publisher = {Zenodo},
	author = {niko and Akimoto, Youhei and yoshihikoueno and Brockhoff, Dimo and Chan, Matthew and ARF1},
	month = apr,
	year = {2020},
	doi = {10.5281/zenodo.3764210}
}

@book{bfgs,
author = {Jorge Nocedal and Stephen J. Wright},
  doi = {10.1007/978-0-387-40065-5},
  url = {https://doi.org/10.1007/978-0-387-40065-5},
  year = {2006},
  publisher = {Springer New York},
  title = {Numerical Optimization}
}

@article{Alwall:2014hca,
  doi = {10.1007/jhep07(2014)079},
  url = {https://doi.org/10.1007/jhep07(2014)079},
  year = {2014},
  month = jul,
  publisher = {Springer Science and Business Media {LLC}},
  volume = {2014},
  number = {7},
  author = {J. Alwall and R. Frederix and S. Frixione and V. Hirschi and F. Maltoni and O. Mattelaer and H.-S. Shao and T. Stelzer and P. Torrielli and M. Zaro},
  title = {The automated computation of tree-level and next-to-leading order differential cross sections,  and their matching to parton shower simulations},
  journal = {Journal of High Energy Physics}
}

@article{Ball:2014uwa,
  doi = {10.1007/jhep04(2015)040},
  url = {https://doi.org/10.1007/jhep04(2015)040},
  year = {2015},
  month = apr,
  publisher = {Springer Science and Business Media {LLC}},
  volume = {2015},
  number = {4},
  author = {Richard D. Ball and Valerio Bertone and Stefano Carrazza and Christopher S. Deans and Luigi Del Debbio and Stefano Forte and Alberto Guffanti and Nathan P. Hartland and Jos{\'{e}} I. Latorre and Juan Rojo and Maria Ubiali},
  title = {Parton distributions for the {LHC} run {II}},
  journal = {Journal of High Energy Physics}
}

@article{Ball:2017nwa,
  doi = {10.1140/epjc/s10052-017-5199-5},
  url = {https://doi.org/10.1140/epjc/s10052-017-5199-5},
  year = {2017},
  month = oct,
  publisher = {Springer Science and Business Media {LLC}},
  volume = {77},
  number = {10},
  author = {Richard D. Ball and Valerio Bertone and Stefano Carrazza and Luigi Del Debbio and Stefano Forte and Patrick Groth-Merrild and Alberto Guffanti and Nathan P. Hartland and Zahari Kassabov and Jos{\'{e}} I. Latorre and Emanuele R. Nocera and Juan Rojo and Luca Rottoli and Emma Slade and Maria Ubiali},
  title = {Parton distributions from high-precision collider data},
  journal = {The European Physical Journal C}
}

@article{Carrazza:2019mzf,
  doi = {10.1140/epjc/s10052-019-7197-2},
  url = {https://doi.org/10.1140/epjc/s10052-019-7197-2},
  year = {2019},
  month = aug,
  publisher = {Springer Science and Business Media {LLC}},
  volume = {79},
  number = {8},
  author = {Stefano Carrazza and Juan Cruz-Martinez},
  title = {Towards a new generation of parton densities with deep learning models},
  journal = {The European Physical Journal C}
}

@article{Bertone:2016lga,
      author         = "Bertone, Valerio and Carrazza, Stefano and Hartland,
                        Nathan P.",
      title          = "{APFELgrid: a high performance tool for parton density
                        determinations}",
      journal        = "Comput. Phys. Commun.",
      volume         = "212",
      year           = "2017",
      pages          = "205-209",
      doi            = "10.1016/j.cpc.2016.10.006",
      reportNumber   = "CERN-TH-2016-103",
      SLACcitation   = "%%CITATION = ARXIV:1605.02070;%%"
}

@article{Candido:2020yat,
  doi = {10.1007/jhep11(2020)129},
  url = {https://doi.org/10.1007/jhep11(2020)129},
  year = {2020},
  month = nov,
  publisher = {Springer Science and Business Media {LLC}},
  volume = {2020},
  number = {11},
  author = {Alessandro Candido and Stefano Forte and Felix Hekhorn},
  title = {Can $\overline{MS}$ parton distributions be negative?},
  journal = {Journal of High Energy Physics}
}

@article{Carrazza:2020gss,
  doi = {10.1007/jhep12(2020)108},
  url = {https://doi.org/10.1007/jhep12(2020)108},
  year = {2020},
  month = dec,
  publisher = {Springer Science and Business Media {LLC}},
  volume = {2020},
  number = {12},
  author = {S. Carrazza and E. R. Nocera and C. Schwan and M. Zaro},
  title = {{PineAPPL}: combining {EW} and {QCD} corrections for fast evaluation of {LHC} processes},
  journal = {Journal of High Energy Physics}
}

@book{nielsen_neural_2015,
  title={Neural networks and deep learning},
  author={Nielsen, Michael A},
  volume={25},
  year={2015},
  publisher={Determination press USA},
  url = {http://neuralnetworksanddeeplearning.com/}
}

@misc{christopher_schwan_2020_3992765,
  author       = {Christopher Schwan and
                  Stefano Carrazza},
  title        = {N3PDF/pineappl: v0.3.0},
  month        = aug,
  year         = 2020,
  publisher    = {Zenodo},
  version      = {v0.3.0},
  doi          = {10.5281/zenodo.3992765},
  url          = {http://doi.org/10.5281/zenodo.3992765}
}

@incollection{cma,
  doi = {10.1007/3-540-32494-1_4},
  url = {http://doi.org/10.1007/3-540-32494-1_4},
  publisher = {Springer Berlin Heidelberg},
  pages = {75--102},
  author = {Nikolaus Hansen},
  year = {2006},
  title = {The {CMA} Evolution Strategy: A Comparing Review},
  booktitle = {Towards a New Evolutionary Computation}
}

@misc{cross2017open,
      title={Open Quantum Assembly Language},
      author={Andrew W. Cross and Lev S. Bishop and John A. Smolin and Jay M. Gambetta},
      year={2017},
      eprint={1707.03429},
      archivePrefix={arXiv},
      primaryClass={quant-ph}
}

@book{helstrom_quantum_76,
title = { Quantum detection and estimation theory / Carl W. Helstrom },
author = { Helstrom, Carl W. },
publisher = { Academic Press New York },
year = { 1976 },
isbn = { 978-0123400503 },
type = { Book },
language = { English },
subjects = { Optical communications.; Signal detection.; Estimation theory.; Quantum theory. },
url = { https://nla.gov.au/nla.cat-vn617918 }}

@article{pearson_chi_1900,
  doi = {10.1080/14786440009463897},
  url = {http://doi.org/10.1080/14786440009463897},
  year = {1900},
  month = jul,
  publisher = {Informa {UK} Limited},
  volume = {50},
  number = {302},
  pages = {157--175},
  author = {Karl Pearson},
  title = {X. On the criterion that a given system of deviations from the probable in the case of a correlated system of variables is such that it can be reasonably supposed to have arisen from random sampling},
  journal = {The London,  Edinburgh,  and Dublin Philosophical Magazine and Journal of Science}
}

@online{qibo_code,
  author       = {Stavros Efthymiou and
                  Stefano Carrazza and
                  Carlos Bravo Prieto and
                  Diego Garc\'ia-Mart\'in and
                  Sergi Ramos and
                  Adri\'an P\'erez Salinas and
                  Javier Serrano},
  title        = {Quantum-TII/qibo: Qibo 0.1.1},
  month        = oct,
  year         = 2020,
  publisher    = {Zenodo},
  version      = {v0.1.1},
  doi          = {10.5281/zenodo.4071702},
  url          = {http://doi.org/10.5281/zenodo.4071702}
}

@online{qiskit,
  doi = {10.5281/ZENODO.2562110},
  url = {http://zenodo.org/record/2562110},
  author = {Aleksandrowicz,  Gadi and others},
  title = {Qiskit: An Open-source Framework for Quantum Computing},
  publisher = {Zenodo},
  year = {2019},
  copyright = {Apache License 2.0}
}

@article{schuld_circuit_2020,
  doi = {10.1103/physreva.101.032308},
  url = {https://doi.org/10.1103/physreva.101.032308},
  year = {2020},
  month = mar,
  publisher = {American Physical Society ({APS})},
  volume = {101},
  number = {3},
  author = {Maria Schuld and Alex Bocharov and Krysta M. Svore and Nathan Wiebe},
  title = {Circuit-centric quantum classifiers},
  journal = {Physical Review A}
}

@misc{lloyd_embeddings_2020,
      title={Quantum embeddings for machine learning},
      author={Seth Lloyd and Maria Schuld and Aroosa Ijaz and Josh Izaac and Nathan Killoran},
      year={2020},
      eprint={2001.03622},
      archivePrefix={arXiv},
      primaryClass={quant-ph}
}

@article{liu_rigorous_2020,
  doi = {10.1038/s41567-021-01287-z},
  url = {https://doi.org/10.1038/s41567-021-01287-z},
  year = {2021},
  month = jul,
  publisher = {Springer Science and Business Media {LLC}},
  author = {Yunchao Liu and Srinivasan Arunachalam and Kristan Temme},
  title = {A rigorous and robust quantum speed-up in supervised machine learning},
  journal = {Nature Physics}, 
  volume = {17}, 
  pages = {1013-1017}
}

@article{rebentrost_qsvm_2014,
   title={Quantum Support Vector Machine for Big Data Classification},
   volume={113},
   ISSN={1079-7114},
   url={http://dx.doi.org/10.1103/PhysRevLett.113.130503},
   doi={10.1103/physrevlett.113.130503},
   number={13},
   journal={Physical Review Letters},
   publisher={American Physical Society (APS)},
   author={Rebentrost, Patrick and Mohseni, Masoud and Lloyd, Seth},
   year={2014},
   month={Sep},
   pages={130503}
}

@article{schuld_effect_2021,
  doi = {10.1103/physreva.103.032430},
  url = {https://doi.org/10.1103/physreva.103.032430},
  year = {2021},
  month = mar,
  publisher = {American Physical Society ({APS})},
  volume = {103},
  number = {3},
  author = {Maria Schuld and Ryan Sweke and Johannes Jakob Meyer},
  title = {Effect of data encoding on the expressive power of variational quantum-machine-learning models},
  journal = {Physical Review A}
}

@article{scipy,
  author  = {Virtanen, Pauli and Gommers, Ralf and Oliphant, Travis E. and Haberland, Matt and Reddy, Tyler and Cournapeau, David and Burovski, Evgeni and Peterson, Pearu and Weckesser, Warren and Bright, Jonathan and {van der Walt}, St{\'e}fan J. and Brett, Matthew and Wilson, Joshua and Millman, K. Jarrod and Mayorov, Nikolay and Nelson, Andrew R. J. and Jones, Eric and Kern, Robert and Larson, Eric and Carey, C J and Polat, {\.I}lhan and Feng, Yu and Moore, Eric W. and {VanderPlas}, Jake and Laxalde, Denis and Perktold, Josef and Cimrman, Robert and Henriksen, Ian and Quintero, E. A. and Harris, Charles R. and Archibald, Anne M. and Ribeiro, Ant{\^o}nio H. and Pedregosa, Fabian and {van Mulbregt}, Paul and {SciPy 1.0 Contributors}},
  title   = {{{SciPy} 1.0: Fundamental Algorithms for Scientific Computing in Python}},
  journal = {Nature Methods},
  year    = {2020},
  volume  = {17},
  pages   = {261--272},
  adsurl  = {https://rdcu.be/b08Wh},
  doi     = {10.1038/s41592-019-0686-2}
}

@article{sim_expressibility_2019,
  doi = {10.1002/qute.201900070},
  url = {https://doi.org/10.1002/qute.201900070},
  year = {2019},
  month = oct,
  publisher = {Wiley},
  volume = {2},
  number = {12},
  pages = {1900070},
  author = {Sukin Sim and Peter D. Johnson and Al{\'{a}}n Aspuru-Guzik},
  title = {Expressibility and Entangling Capability of Parameterized Quantum Circuits for Hybrid Quantum-Classical Algorithms},
  journal = {Advanced Quantum Technologies}
}

@article{nakaji_expressibility_2021,
  doi = {10.22331/q-2021-04-19-434},
  url = {https://doi.org/10.22331/q-2021-04-19-434},
  year = {2021},
  month = apr,
  publisher = {Verein zur Forderung des Open Access Publizierens in den Quantenwissenschaften},
  volume = {5},
  pages = {434},
  author = {Kouhei Nakaji and Naoki Yamamoto},
  title = {Expressibility of the alternating layered ansatz for quantum computation},
  journal = {Quantum}
}

@misc{dutta_realization_2021,
      title={Realization of an ion trap quantum classifier}, 
      author={Tarun Dutta and Adrián Pérez-Salinas and Jasper Phua Sing Cheng and José Ignacio Latorre and Manas Mukherjee},
      year={2021},
      eprint={2106.14059},
      archivePrefix={arXiv},
      primaryClass={quant-ph}
}

@article{Ball:2010de,
  doi = {10.1016/j.nuclphysb.2010.05.008},
  url = {https://doi.org/10.1016/j.nuclphysb.2010.05.008},
  year = {2010},
  month = oct,
  publisher = {Elsevier {BV}},
  volume = {838},
  number = {1-2},
  pages = {136--206},
  author = {Richard D. Ball and Luigi Del Debbio and Stefano Forte and Alberto Guffanti and Jos{\'{e}} I. Latorre and Juan Rojo and Maria Ubiali},
  title = {A first unbiased global {NLO} determination of parton distributions and their uncertainties},
  journal = {Nuclear Physics B}
}

@misc{Forte:2020yip,
    author = "Forte, Stefano and Carrazza, Stefano",
    title = "{Parton distribution functions}",
    eprint = "2008.12305",
    archivePrefix = "arXiv",
    primaryClass = "hep-ph",
    reportNumber = "TIF-UNIMI-2020-23",
    month = "8",
    year = "2020"
}

@article{l-bfgs,
  doi = {10.1137/0916069},
  url = {http://doi.org/10.1137/0916069},
  year = {1995},
  month = sep,
  publisher = {Society for Industrial and Applied Mathematics ({SIAM})},
  volume = {16},
  number = {5},
  pages = {1190--1208},
  author = {Richard H. Byrd and Peihuang Lu and Jorge Nocedal and Ciyou Zhu},
  title = {A Limited Memory Algorithm for Bound Constrained Optimization},
  journal = {{SIAM} Journal on Scientific Computing}
}

@article{bravoprieto_quantum_2020,
  doi = {10.1103/physreva.101.062310},
  url = {https://doi.org/10.1103/physreva.101.062310},
  year = {2020},
  month = jun,
  publisher = {American Physical Society ({APS})},
  volume = {101},
  number = {6},
  author = {Carlos Bravo-Prieto and Diego Garc{\'{i}}a-Mart{\'{i}}n and Jos{\'{e}} I. Latorre},
  title = {Quantum singular value decomposer},
  journal = {Physical Review A}
}

@article{cybenko_approximation_1989,
  doi = {10.1007/bf02551274},
  url = {https://doi.org/10.1007/bf02551274},
  year = {1989},
  month = dec,
  publisher = {Springer Science and Business Media {LLC}},
  volume = {2},
  number = {4},
  pages = {303--314},
  author = {G. Cybenko},
  title = {Approximation by superpositions of a sigmoidal function},
  journal = {Mathematics of Control,  Signals,  and Systems}
}

@article{eichler_new_1968,
  doi = {10.2969/jmsj/02010023},
  url = {https://doi.org/10.2969/jmsj/02010023},
  year = {1968},
  month = apr,
  publisher = {Mathematical Society of Japan (Project Euclid)},
  volume = {20},
  number = {1-2},
  author = {M. Eichler},
  title = {A new proof of the Baker-Campbell-Hausdorff formula},
  journal = {Journal of the Mathematical Society of Japan}
}

@misc{farhi_classification_2018,
      title={Classification with Quantum Neural Networks on Near Term Processors}, 
      author={Edward Farhi and Hartmut Neven},
      year={2018},
      eprint={1802.06002},
      archivePrefix={arXiv},
      primaryClass={quant-ph}
}

@incollection{hall_bakercampbellhausdorff_2015,
  doi = {10.1007/978-3-319-13467-3_5},
  url = {https://doi.org/10.1007/978-3-319-13467-3_5},
  year = {2015},
  publisher = {Springer International Publishing},
  pages = {109--137},
  author = {Brian C. Hall},
  title = {The Baker{\textendash}Campbell{\textendash}Hausdorff Formula and Its Consequences},
  booktitle = {Graduate Texts in Mathematics}
}

@article{goto_universal_2020,
  doi = {10.1103/physrevlett.127.090506},
  url = {https://doi.org/10.1103/physrevlett.127.090506},
  year = {2021},
  month = aug,
  publisher = {American Physical Society ({APS})},
  volume = {127},
  number = {9},
  author = {Takahiro Goto and Quoc Hoan Tran and Kohei Nakajima},
  title = {Universal Approximation Property of Quantum Machine Learning Models in Quantum-Enhanced Feature Spaces},
  journal = {Physical Review Letters}
}

@inproceedings{holevo_bounds_1973,
  title={Bounds for the quantity of information transmitted by a quantum communication channel},
  author={A. Holevo},
  year={1973},
  pages = {177--183}, 
  vol = {9}, 
  issue = {3}, 
  journal = {Probl. Peredachi Inf.}
}

@article{hornik_approximation_1991,
  doi = {10.1016/0893-6080(91)90009-t},
  url = {https://doi.org/10.1016/0893-6080(91)90009-t},
  year = {1991},
  publisher = {Elsevier {BV}},
  volume = {4},
  number = {2},
  pages = {251--257},
  author = {Kurt Hornik},
  title = {Approximation capabilities of multilayer feedforward networks},
  journal = {Neural Networks}
}

@article{low_hamiltonian_2019,
  doi = {10.22331/q-2019-07-12-163},
  url = {https://doi.org/10.22331/q-2019-07-12-163},
  year = {2019},
  month = jul,
  publisher = {Verein zur Forderung des Open Access Publizierens in den Quantenwissenschaften},
  volume = {3},
  pages = {163},
  author = {Guang Hao Low and Isaac L. Chuang},
  title = {Hamiltonian Simulation by Qubitization},
  journal = {Quantum}
}

@article{low_optimal_2017,
  doi = {10.1103/physrevlett.118.010501},
  url = {https://doi.org/10.1103/physrevlett.118.010501},
  year = {2017},
  month = jan,
  publisher = {American Physical Society ({APS})},
  volume = {118},
  number = {1},
  author = {Guang Hao Low and Isaac L. Chuang},
  title = {Optimal Hamiltonian Simulation by Quantum Signal Processing},
  journal = {Physical Review Letters}
}

@article{huang_power_2021,
  doi = {10.1038/s41467-021-22539-9},
  url = {https://doi.org/10.1038/s41467-021-22539-9},
  year = {2021},
  month = may,
  publisher = {Springer Science and Business Media {LLC}},
  volume = {12},
  number = {1},
  author = {Hsin-Yuan Huang and Michael Broughton and Masoud Mohseni and Ryan Babbush and Sergio Boixo and Hartmut Neven and Jarrod R. McClean},
  title = {Power of data in quantum machine learning},
  journal = {Nature Communications}
}

@article{leshno_multilayer_1993,
  doi = {10.1016/s0893-6080(05)80131-5},
  url = {https://doi.org/10.1016/s0893-6080(05)80131-5},
  year = {1993},
  month = jan,
  publisher = {Elsevier {BV}},
  volume = {6},
  number = {6},
  pages = {861--867},
  author = {Moshe Leshno and Vladimir Ya. Lin and Allan Pinkus and Shimon Schocken},
  title = {Multilayer feedforward networks with a nonpolynomial activation function can approximate any function},
  journal = {Neural Networks}
}

@misc{lloyd_quantum_2013,
      title={Quantum algorithms for supervised and unsupervised machine learning}, 
      author={Seth Lloyd and Masoud Mohseni and Patrick Rebentrost},
      year={2013},
      eprint={1307.0411},
      archivePrefix={arXiv},
      primaryClass={quant-ph}
}

@article{mitarai_quantum_2018,
  doi = {10.1103/physreva.98.032309},
  url = {https://doi.org/10.1103/physreva.98.032309},
  year = {2018},
  month = sep,
  publisher = {American Physical Society ({APS})},
  volume = {98},
  number = {3},
  author = {K. Mitarai and M. Negoro and M. Kitagawa and K. Fujii},
  title = {Quantum circuit learning},
  journal = {Physical Review A}
}

@article{preskill_quantum_2018,
  doi = {10.22331/q-2018-08-06-79},
  url = {https://doi.org/10.22331/q-2018-08-06-79},
  year = {2018},
  month = aug,
  publisher = {Verein zur Forderung des Open Access Publizierens in den Quantenwissenschaften},
  volume = {2},
  pages = {79},
  author = {John Preskill},
  title = {Quantum Computing in the {NISQ} era and beyond},
  journal = {Quantum}
}

@article{van-brunt_explicit_2018,
  doi = {10.3390/math6080135},
  url = {https://doi.org/10.3390/math6080135},
  year = {2018},
  month = aug,
  publisher = {{MDPI} {AG}},
  volume = {6},
  number = {8},
  pages = {135},
  author = {Alexander Van-Brunt and Matt Visser},
  title = {Explicit Baker{\textendash}Campbell{\textendash}Hausdorff Expansions},
  journal = {Mathematics}
}

@article{zhu_training_2019,
  doi = {10.1126/sciadv.aaw9918},
  url = {https://doi.org/10.1126/sciadv.aaw9918},
  year = {2019},
  month = oct,
  publisher = {American Association for the Advancement of Science ({AAAS})},
  volume = {5},
  number = {10},
  pages = {eaaw9918},
  author = {D. Zhu and N. M. Linke and M. Benedetti and K. A. Landsman and N. H. Nguyen and C. H. Alderete and A. Perdomo-Ortiz and N. Korda and A. Garfoot and C. Brecque and L. Egan and O. Perdomo and C. Monroe},
  title = {Training of quantum circuits on a hybrid quantum computer},
  journal = {Science Advances}
}

@article{wootters_single_1982,
  doi = {10.1038/299802a0},
  url = {https://doi.org/10.1038/299802a0},
  year = {1982},
  month = oct,
  publisher = {Springer Science and Business Media {LLC}},
  volume = {299},
  number = {5886},
  pages = {802--803},
  author = {W. K. Wootters and W. H. Zurek},
  title = {A single quantum cannot be cloned},
  journal = {Nature}
}

@article{atiyah_polyhedra_2003,
  doi = {10.1007/s00032-003-0014-1},
  url = {https://doi.org/10.1007/s00032-003-0014-1},
  year = {2003},
  month = sep,
  publisher = {Springer Science and Business Media {LLC}},
  volume = {71},
  number = {1},
  pages = {33--58},
  author = {Michael Atiyah and Paul Sutcliffe},
  title = {Polyhedra in Physics,  Chemistry and Geometry},
  journal = {Milan Journal of Mathematics}
}

@article{kang_swap_2019,
  doi = {10.1038/s41598-019-42662-4},
  url = {https://doi.org/10.1038/s41598-019-42662-4},
  year = {2019},
  month = apr,
  publisher = {Springer Science and Business Media {LLC}},
  volume = {9},
  number = {1},
  author = {Min-Sung Kang and Jino Heo and Seong-Gon Choi and Sung Moon and Sang-Wook Han},
  title = {Implementation of {SWAP} test for two unknown states in photons via cross-Kerr nonlinearities under decoherence effect},
  journal = {Scientific Reports}
}

@book{goodfellow_deep_2016,
    title={Deep Learning},
    author={Ian Goodfellow and Yoshua Bengio and Aaron Courville},
    publisher={MIT Press},
    note={\url{http://www.deeplearningbook.org}},
    year={2016}
}

@article{rumelhart_backprop_1986,
  doi = {10.1038/323533a0},
  url = {https://doi.org/10.1038/323533a0},
  year = {1986},
  month = oct,
  publisher = {Springer Science and Business Media {LLC}},
  volume = {323},
  number = {6088},
  pages = {533--536},
  author = {David E. Rumelhart and Geoffrey E. Hinton and Ronald J. Williams},
  title = {Learning representations by back-propagating errors},
  journal = {Nature}
}

@online{github_unary,
    author={Sergi Ramos-Calderer and Adri\'an P\'erez-Salinas},
    title={Quantum Finance},
    year={2020},
  url = {https://github.com/UB-Quantic/quantum-unary-option-pricing}
}

@online{github_qubit,
  author = {Adri\'an P\'erez-Salinas}, 
  title= {{Universal-Approximator}},
  year = {2021},
  url = {https://github.com/UB-Quantic/Universal-Approximator}
}

@online{github_data,
author = {Adri\'an P\'erez-Salinas},
title = {Quantum classifier with data re-uploading},
year = {2019},
url = {https://github.com/AdrianPerezSalinas/universal_qlassifier}
}

@incollection{dariano_tomography_2003,
  doi = {10.1016/s1076-5670(03)80065-4},
  url = {https://doi.org/10.1016/s1076-5670(03)80065-4},
  year = {2003},
  publisher = {Elsevier},
  pages = {205--308},
  author = {G. Mauro D'Ariano and Matteo G.A. Paris and Massimiliano F. Sacchi},
  title = {Quantum Tomography},
  booktitle = {Advances in Imaging and Electron Physics}
}

@article{future-work,
      title={Experimental optimization for single-qubit approximants}, 
      author={Adrián Pérez-Salinas and David López-Núñez and Artur García-Sáez and P. Forn-Díaz and José I. Latorre},
      year={--},
      journal={Preparation}
}

@incollection{feynman_parton_1988,
  doi = {10.1007/978-94-009-3051-3_25},
  url = {https://doi.org/10.1007/978-94-009-3051-3_25},
  year = {1988},
  publisher = {Springer Netherlands},
  pages = {289--304},
  author = {Richard P. Feynman},
  title = {The Behavior of Hadron Collisions at Extreme Energies},
  booktitle = {Special Relativity and Quantum Theory}
}

@article{Bepari2021,
  doi = {10.1103/physrevd.103.076020},
  url = {https://doi.org/10.1103/physrevd.103.076020},
  year = {2021},
  month = apr,
  publisher = {American Physical Society ({APS})},
  volume = {103},
  number = {7},
  author = {Khadeejah Bepari and Sarah Malik and Michael Spannowsky and Simon Williams},
  title = {Towards a quantum computing algorithm for helicity amplitudes and parton showers},
  journal = {Physical Review D}
}

@article{Nachman2021,
  doi = {10.1103/physrevlett.126.062001},
  url = {https://doi.org/10.1103/physrevlett.126.062001},
  year = {2021},
  month = feb,
  publisher = {American Physical Society ({APS})},
  volume = {126},
  number = {6},
  author = {Benjamin Nachman and Davide Provasoli and Wibe A. de Jong and Christian W. Bauer},
  title = {Quantum Algorithm for High Energy Physics Simulations},
  journal = {Physical Review Letters}
}

@misc{li_partonic_2021,
      title={Partonic Structure by Quantum Computing}, 
      author={Tianyin Li and Xingyu Guo and Wai Kin Lai and Xiaohui Liu and Enke Wang and Hongxi Xing and Dan-Bo Zhang and Shi-Liang Zhu},
      year={2021},
      eprint={2106.03865},
      archivePrefix={arXiv},
      primaryClass={hep-ph}
}

@article{Alexandru2019,
  doi = {10.1103/physrevd.100.114501},
  url = {https://doi.org/10.1103/physrevd.100.114501},
  year = {2019},
  month = dec,
  publisher = {American Physical Society ({APS})},
  volume = {100},
  number = {11},
  author = {Andrei Alexandru and Paulo F. Bedaque and Siddhartha Harmalkar and Henry Lamm and Scott Lawrence and Neill C. Warrington and},
  title = {Gluon field digitization for quantum computers},
  journal = {Physical Review D}
}

@article{Lamm2020,
  doi = {10.1103/physrevresearch.2.013272},
  url = {https://doi.org/10.1103/physrevresearch.2.013272},
  year = {2020},
  month = mar,
  publisher = {American Physical Society ({APS})},
  volume = {2},
  number = {1},
  author = {Henry Lamm and Scott Lawrence and Yukari Yamauchi and},
  title = {Parton physics on a quantum computer},
  journal = {Physical Review Research}
}

@article{reset-protocol,
  title = {Demonstrating a Driven Reset Protocol for a Superconducting Qubit},
  author = {Geerlings, K. and Leghtas, Z. and Pop, I. M. and Shankar, S. and Frunzio, L. and Schoelkopf, R. J. and Mirrahimi, M. and Devoret, M. H.},
  journal = {Phys. Rev. Lett.},
  volume = {110},
  issue = {12},
  pages = {120501},
  numpages = {5},
  year = {2013},
  month = {Mar},
  publisher = {American Physical Society},
  doi = {10.1103/PhysRevLett.110.120501},
  url = {https://link.aps.org/doi/10.1103/PhysRevLett.110.120501}
}

@article{randomized-benchmarking,
  title = {Scalable and Robust Randomized Benchmarking of Quantum Processes},
  author = {Magesan, Easwar and Gambetta, J. M. and Emerson, Joseph},
  journal = {Phys. Rev. Lett.},
  volume = {106},
  issue = {18},
  pages = {180504},
  numpages = {4},
  year = {2011},
  month = {May},
  publisher = {American Physical Society},
  doi = {10.1103/PhysRevLett.106.180504},
  url = {https://link.aps.org/doi/10.1103/PhysRevLett.106.180504}
}

@article{virtual-zgates,
  title = {Efficient $Z$ gates for quantum computing},
  author = {McKay, David C. and Wood, Christopher J. and Sheldon, Sarah and Chow, Jerry M. and Gambetta, Jay M.},
  journal = {Phys. Rev. A},
  volume = {96},
  issue = {2},
  pages = {022330},
  numpages = {8},
  year = {2017},
  month = {Aug},
  publisher = {American Physical Society},
  doi = {10.1103/PhysRevA.96.022330},
  url = {https://link.aps.org/doi/10.1103/PhysRevA.96.022330}
}

@article{drag-a,
  title = {Simple Pulses for Elimination of Leakage in Weakly Nonlinear Qubits},
  author = {Motzoi, F. and Gambetta, J. M. and Rebentrost, P. and Wilhelm, F. K.},
  journal = {Phys. Rev. Lett.},
  volume = {103},
  pages = {110501},
  numpages = {4},
  year = {2009},
  month = {Sep},
  publisher = {American Physical Society},
  doi = {10.1103/PhysRevLett.103.110501},
  url = {https://link.aps.org/doi/10.1103/PhysRevLett.103.110501}
}

@article{drag-b,
  title = {Optimized driving of superconducting artificial atoms for improved single-qubit gates},
  author = {Chow, J. M. and DiCarlo, L. and Gambetta, J. M. and Motzoi, F. and Frunzio, L. and Girvin, S. M. and Schoelkopf, R. J.},
  journal = {Phys. Rev. A},
  volume = {82},
  pages = {040305},
  numpages = {4},
  year = {2010},
  month = {Oct},
  publisher = {American Physical Society},
  doi = {10.1103/PhysRevA.82.040305},
  url = {https://link.aps.org/doi/10.1103/PhysRevA.82.040305}
}

@article{3d-transmon,
  title = {Observation of High Coherence in Josephson Junction Qubits Measured in a Three-Dimensional Circuit QED Architecture},
  author = {Paik, Hanhee and Schuster, D. I. and Bishop, Lev S. and Kirchmair, G. and Catelani, G. and Sears, A. P. and Johnson, B. R. and Reagor, M. J. and Frunzio, L. and Glazman, L. I. and Girvin, S. M. and Devoret, M. H. and Schoelkopf, R. J.},
  journal = {Phys. Rev. Lett.},
  volume = {107},
  pages = {240501},
  numpages = {5},
  year = {2011},
  month = {Dec},
  publisher = {American Physical Society},
  doi = {10.1103/PhysRevLett.107.240501},
  url = {https://link.aps.org/doi/10.1103/PhysRevLett.107.240501}
}

@article{Ball:2008by,
  doi = {10.1016/j.nuclphysb.2008.09.037},
  url = {https://doi.org/10.1016/j.nuclphysb.2008.09.037},
  year = {2009},
  month = mar,
  publisher = {Elsevier {BV}},
  volume = {809},
  number = {1-2},
  pages = {1--63},
  author = {Richard D. Ball and Luigi Del Debbio and Stefano Forte and Alberto Guffanti and Jos{\'{e}} I. Latorre and Andrea Piccione and Juan Rojo and Maria Ubiali},
  title = {A determination of parton distributions with faithful uncertainty estimation},
  journal = {Nuclear Physics B}
}

@online{zahari_kassabov_2019_2571601,
  author       = {Zahari Kassabov},
  title        = {{Reportengine: A framework for declarative data
                   analysis}},
  month        = feb,
  year         = 2019,
  publisher    = {Zenodo},
  version      = {v0.27},
  doi          = {10.5281/zenodo.2571601},
  url          = {http://doi.org/10.5281/zenodo.2571601}
}

@article{Aad:2014vwa,
  doi = {10.1007/jhep02(2015)153},
  url = {https://doi.org/10.1007/jhep02(2015)153},
  year = {2015},
  month = feb,
  publisher = {Springer Science and Business Media {LLC}},
  volume = {2015},
  number = {2},
  author = {G. Aad and others},
  title = {Measurement of the inclusive jet cross-section in proton-proton collisions at $s = \sqrt{7}$ {TeV} using 4.5 $fb^{-1}$ of data with the {ATLAS} detector},
  journal = {Journal of High Energy Physics}
}

@article{Khachatryan:2015oaa,
  doi = {10.1016/j.physletb.2015.07.065},
  url = {https://doi.org/10.1016/j.physletb.2015.07.065},
  year = {2015},
  month = oct,
  publisher = {Elsevier {BV}},
  volume = {749},
  pages = {187--209},
  author = {V. Khachatryan and others},
  title = {Measurement of the Z boson differential cross section in transverse momentum and rapidity in proton-proton collisions at 8 {TeV}},
  journal = {Physics Letters B}
}

@article{Aaij:2012vn,
  doi = {10.1007/jhep06(2012)058},
  url = {https://doi.org/10.1007/jhep06(2012)058},
  year = {2012},
  month = jun,
  publisher = {Springer Science and Business Media {LLC}},
  volume = {2012},
  number = {6},
  author = {R. Aaij and others},
  title = {Inclusive W and Z production in the forward region at 
		                $\sqrt{7}${TeV}},
  journal = {Journal of High Energy Physics}
}

@article{Nagy:2001fj,
  doi = {10.1103/physrevlett.88.122003},
  url = {https://doi.org/10.1103/physrevlett.88.122003},
  year = {2002},
  month = mar,
  publisher = {American Physical Society ({APS})},
  volume = {88},
  number = {12},
  author = {Zolt{\'{a}}n Nagy},
  title = {Three-Jet Cross Sections in Hadron-Hadron Collisions at Next-To-Leading Order},
  journal = {Physical Review Letters}
}

@article{Campbell:2019dru,
  doi = {10.1007/jhep12(2019)034},
  url = {https://doi.org/10.1007/jhep12(2019)034},
  year = {2019},
  month = dec,
  publisher = {Springer Science and Business Media {LLC}},
  volume = {2019},
  number = {12},
  author = {John Campbell and Tobias Neumann},
  title = {Precision phenomenology with {MCFM}},
  journal = {Journal of High Energy Physics}
}

@article{havlicek_supervised_2019,
  doi = {10.1038/s41586-019-0980-2},
  url = {https://doi.org/10.1038/s41586-019-0980-2},
  year = {2019},
  month = mar,
  publisher = {Springer Science and Business Media {LLC}},
  volume = {567},
  number = {7747},
  pages = {209--212},
  author = {Vojt{\v{e}}ch Havlí{\v{c}}ek and Antonio D. Córcoles and Kristan Temme and Aram W. Harrow and Abhinav Kandala and Jerry M. Chow and Jay M. Gambetta},
  title = {Supervised learning with quantum-enhanced feature spaces},
  journal = {Nature}
}

@misc{resch_quantum_2019,
      title={Quantum Computing: An Overview Across the System Stack}, 
      author={Salonik Resch and Ulya R. Karpuzcu},
      year={2019},
      eprint={1905.07240},
      archivePrefix={arXiv},
      primaryClass={quant-ph}
}

@article{hempel_quantum_2018,
  doi = {10.1103/physrevx.8.031022},
  url = {https://doi.org/10.1103/physrevx.8.031022},
  year = {2018},
  month = jul,
  publisher = {American Physical Society ({APS})},
  volume = {8},
  number = {3},
  author = {Cornelius Hempel and Christine Maier and Jonathan Romero and Jarrod McClean and Thomas Monz and Heng Shen and Petar Jurcevic and Ben P. Lanyon and Peter Love and Ryan Babbush and Alán Aspuru-Guzik and Rainer Blatt and Christian F. Roos},
  title = {Quantum Chemistry Calculations on a Trapped-Ion Quantum Simulator},
  journal = {Physical Review X}
}

@article{Nam_groundstate_2020,
  doi = {10.1038/s41534-020-0259-3},
  url = {https://doi.org/10.1038/s41534-020-0259-3},
  year = {2020},
  month = apr,
  publisher = {Springer Science and Business Media {LLC}},
  volume = {6},
  number = {1},
  author = {Yunseong Nam and Jwo-Sy Chen and Jungsang Kim and others},
  title = {Ground-state energy estimation of the water molecule on a trapped-ion quantum computer},
  journal = {npj Quantum Information}
}

@misc{rudolph_generation_2020,
      title={Generation of High-Resolution Handwritten Digits with an Ion-Trap Quantum Computer}, 
      author={Manuel S. Rudolph and Ntwali Bashige Toussaint and Alejandro Perdomo-Ortiz and others},
      year={2020},
      eprint={2012.03924},
      archivePrefix={arXiv},
      primaryClass={quant-ph}
}

@misc{johri_nearest_2020,
      title={Nearest Centroid Classification on a Trapped Ion Quantum Computer}, 
      author={Sonika Johri and Shantanu Debnath and Iordanis Kerenidis and others},
      year={2020},
      eprint={2012.04145},
      archivePrefix={arXiv},
      primaryClass={quant-ph}
}

@article{figgatt_complete_2017,
  doi = {10.1038/s41467-017-01904-7},
  url = {https://doi.org/10.1038/s41467-017-01904-7},
  year = {2017},
  month = dec,
  publisher = {Springer Science and Business Media {LLC}},
  volume = {8},
  number = {1},
  author = {C. Figgatt and D. Maslov and C. Monroe and others},
  title = {Complete 3-Qubit Grover search on a programmable quantum computer},
  journal = {Nature Communications}
}

@article{brown_single_2011,
  doi = {10.1103/physreva.84.030303},
  url = {https://doi.org/10.1103/physreva.84.030303},
  year = {2011},
  month = sep,
  publisher = {American Physical Society ({APS})},
  volume = {84},
  number = {3},
  author = {K. R. Brown and A. C. Wilson and Y. Colombe and C. Ospelkaus and A. M. Meier and E. Knill and D. Leibfried and D. J. Wineland},
  title = {Single-qubit-gate error below $10^{-4}$ in a trapped ion},
  journal = {Physical Review A}
}

@article{Wang_single_2021,
  doi = {10.1038/s41467-020-20330-w},
  url = {https://doi.org/10.1038/s41467-020-20330-w},
  year = {2021},
  month = jan,
  publisher = {Springer Science and Business Media {LLC}},
  volume = {12},
  number = {1},
  author = {Pengfei Wang and Chun-Yang Luan and Mu Qiao and Mark Um and Junhua Zhang and Ye Wang and Xiao Yuan and Mile Gu and Jingning Zhang and Kihwan Kim},
  title = {Single ion qubit with estimated coherence time exceeding one hour},
  journal = {Nature Communications}
}

@misc{hubregtsen_singlecomponent_2021,
      title={Single-component gradient rules for variational quantum algorithms}, 
      author={Thomas Hubregtsen and Frederik Wilde and Shozab Qasim and Jens Eisert},
      year={2021},
      eprint={2106.01388},
      archivePrefix={arXiv},
      primaryClass={quant-ph}
}

@article{sweke_stochastic_2020,
  doi = {10.22331/q-2020-08-31-314},
  url = {https://doi.org/10.22331/q-2020-08-31-314},
  year = {2020},
  month = aug,
  publisher = {Verein zur Forderung des Open Access Publizierens in den Quantenwissenschaften},
  volume = {4},
  pages = {314},
  author = {Ryan Sweke and Frederik Wilde and Johannes Meyer and Maria Schuld and Paul K. Faehrmann and Barth{\'{e}}l{\'{e}}my Meynard-Piganeau and Jens Eisert},
  title = {Stochastic gradient descent for hybrid quantum-classical optimization},
  journal = {Quantum}
}

@article{schuld_evaluating_2019,
  doi = {10.1103/physreva.99.032331},
  url = {https://doi.org/10.1103/physreva.99.032331},
  year = {2019},
  month = mar,
  publisher = {American Physical Society ({APS})},
  volume = {99},
  number = {3},
  author = {Maria Schuld and Ville Bergholm and Christian Gogolin and Josh Izaac and Nathan Killoran},
  title = {Evaluating analytic gradients on quantum hardware},
  journal = {Physical Review A}
}

@misc{kingma_adam_2017,
      title={Adam: A Method for Stochastic Optimization}, 
      author={Diederik P. Kingma and Jimmy Ba},
      year={2017},
      eprint={1412.6980},
      archivePrefix={arXiv},
      primaryClass={cs.LG}
}

@article{mccaldin_reuploading_2021,
  doi = {10.1109/access.2021.3075492},
  url = {https://doi.org/10.1109/access.2021.3075492},
  year = {2021},
  publisher = {Institute of Electrical and Electronics Engineers ({IEEE})},
  volume = {9},
  pages = {65127--65139},
  author = {Philip Easom-Mccaldin and Ahmed Bouridane and Ammar Belatreche and Richard Jiang},
  title = {On Depth,  Robustness and Performance Using the Data Re-Uploading Single-Qubit Classifier},
  journal = {{IEEE} Access}
}

@article{huembeli_characterizing_2021,
	doi = {10.1088/2058-9565/abdbc9},
	url = {https://doi.org/10.1088/2058-9565/abdbc9},
	year = 2021,
	month = {feb},
	publisher = {{IOP} Publishing},
	volume = {6},
	number = {2},
	pages = {025011},
	author = {Patrick Huembeli and Alexandre Dauphin},
	title = {Characterizing the loss landscape of variational quantum circuits},
	journal = {Quantum Science and Technology}
}

@online{reuploading_pennylane,
  author = {Shahnawaz Ahmed}, 
  title= {{Data-reuploading classifer}},
  year = 2019,
  url = {https://pennylane.ai/qml/app/tutorial_data_reuploading_classifier.html}
}

@online{reuploading_qibo,
  author = {{Qibo Team}}, 
  title= {{Data-reuploading classifer}},
  year = 2020,
  url = {https://qibo.readthedocs.io/en/stable/tutorials/reuploading_classifier/README.html}
}

@article{shor,
  doi = {10.1137/s0097539795293172},
  url = {https://doi.org/10.1137/s0097539795293172},
  year = {1997},
  month = oct,
  publisher = {Society for Industrial {\&} Applied Mathematics ({SIAM})},
  volume = {26},
  number = {5},
  pages = {1484--1509},
  author = {Peter W. Shor},
  title = {Polynomial-Time Algorithms for Prime Factorization and Discrete Logarithms on a Quantum Computer},
  journal = {{SIAM} Journal on Computing}
}

@inproceedings{grover,
  doi = {10.1145/237814.237866},
  url = {https://doi.org/10.1145/237814.237866},
  year = {1996},
  publisher = {{ACM} Press},
  author = {Lov K. Grover},
  title = {A fast quantum mechanical algorithm for database search},
  booktitle = {Proceedings of the twenty-eighth annual {ACM} symposium on Theory of computing  - {STOC} {\textquotesingle}96}, 
  jounal = {Proceedings of the twenty-eighth annual {ACM} symposium on Theory of computing  - {STOC} {\textquotesingle}96}
}

@book{gottesman_stabilizer_1997,
  title={Stabilizer codes and quantum error correction},
  author={Gottesman, Daniel},
  year={1997},
  publisher={California Institute of Technology}, 
  url = {https://arxiv.org/abs/quant-ph/9705052}, 
  eprint={quant-ph/9705052},
  archivePrefix={arXiv}
}

@article{shor_scheme_1995,
  doi = {10.1103/physreva.52.r2493},
  url = {https://doi.org/10.1103/physreva.52.r2493},
  year = {1995},
  month = oct,
  publisher = {American Physical Society ({APS})},
  volume = {52},
  number = {4},
  pages = {R2493--R2496},
  author = {Peter W. Shor},
  title = {Scheme for reducing decoherence in quantum computer memory},
  journal = {Physical Review A}
}

@article{cory_experimental_1998,
  doi = {10.1103/physrevlett.81.2152},
  url = {https://doi.org/10.1103/physrevlett.81.2152},
  year = {1998},
  month = sep,
  publisher = {American Physical Society ({APS})},
  volume = {81},
  number = {10},
  pages = {2152--2155},
  author = {D. G. Cory and M. D. Price and W. Maas and E. Knill and R. Laflamme and W. H. Zurek and T. F. Havel and S. S. Somaroo},
  title = {Experimental Quantum Error Correction},
  journal = {Physical Review Letters}
}

@article{poulin_quantum_2018,
  doi = {10.1103/physrevlett.121.010501},
  url = {https://doi.org/10.1103/physrevlett.121.010501},
  year = {2018},
  month = jul,
  publisher = {American Physical Society ({APS})},
  volume = {121},
  number = {1},
  author = {David Poulin and Alexei Kitaev and Damian S. Steiger and Matthew B. Hastings and Matthias Troyer},
  title = {Quantum Algorithm for Spectral Measurement with a Lower Gate Count},
  journal = {Physical Review Letters}
}

@article{babbush_encoding_2018,
  doi = {10.1103/physrevx.8.041015},
  url = {https://doi.org/10.1103/physrevx.8.041015},
  year = {2018},
  month = oct,
  publisher = {American Physical Society ({APS})},
  volume = {8},
  number = {4},
  author = {Ryan Babbush and Craig Gidney and Dominic W. Berry and Nathan Wiebe and Jarrod McClean and Alexandru Paler and Austin Fowler and Hartmut Neven},
  title = {Encoding Electronic Spectra in Quantum Circuits with Linear T Complexity},
  journal = {Physical Review X}
}

@article{steudtner_estimating_2020,
  doi = {10.1103/physreva.101.052329},
  url = {https://doi.org/10.1103/physreva.101.052329},
  year = {2020},
  month = may,
  publisher = {American Physical Society ({APS})},
  volume = {101},
  number = {5},
  author = {Mark Steudtner and Stephanie Wehner},
  title = {Estimating exact energies in quantum simulation without Toffoli gates},
  journal = {Physical Review A}
}

@article{qfinance-orus2019,
   title={Quantum computing for finance: Overview and prospects},
   volume={4},
   ISSN={2405-4283},
   url={http://dx.doi.org/10.1016/j.revip.2019.100028},
   DOI={10.1016/j.revip.2019.100028},
   journal={Reviews in Physics},
   publisher={Elsevier BV},
   author={Orús, Román and Mugel, Samuel and Lizaso, Enrique},
   year={2019},
   month={Nov},
   pages={100028}
}

@misc{portfolio-kerenidis2019,
    title={Quantum Algorithms for Portfolio Optimization},
    author={Iordanis Kerenidis and Anupam Prakash and Dániel Szilágyi},
    year={2019},
    eprint={1908.08040},
    archivePrefix={arXiv},
    primaryClass={math.OC}
}

@article{crashes-orus2019,
   title={Forecasting financial crashes with quantum computing},
   volume={99},
   ISSN={2469-9934},
   url={http://dx.doi.org/10.1103/PhysRevA.99.060301},
   DOI={10.1103/physreva.99.060301},
   number={6},
   journal={Physical Review A},
   publisher={American Physical Society (APS)},
   author={Orús, Román and Mugel, Samuel and Lizaso, Enrique},
   year={2019},
   month={Jun},
   pages={-}
}

@article{derivatives-martin2019,
  title={Toward pricing financial derivatives with an IBM quantum computer},
  author={Martin, Ana and Candelas, Bruno and Rodr{\'\i}guez-Rozas, {\'A}ngel and Mart{\'\i}n-Guerrero, Jos{\'e} D and Chen, Xi and Lamata, Lucas and Or{\'u}s, Rom{\'a}n and Solano, Enrique and Sanz, Mikel},
  journal={Physical Review Research},
  volume={3},
  number={1},
  pages={013167},
  year={2021},
  publisher={APS}
}

@misc{credit-egger2019,
    title={Credit Risk Analysis using Quantum Computers},
    author={Daniel J. Egger and Ricardo García Gutiérrez and Jordi Cahué Mestre and Stefan Woerner},
    year={2019},
    eprint={1907.03044},
    archivePrefix={arXiv},
    primaryClass={quant-ph}
}

@misc{portfolio-rebentrost2018,
    title={Quantum computational finance: quantum algorithm for portfolio optimization},
    author={Patrick Rebentrost and Seth Lloyd},
    year={2018},
    eprint={1811.03975},
    archivePrefix={arXiv},
    primaryClass={quant-ph}
}

@article{optimization-moll2018,
   title={Quantum optimization using variational algorithms on near-term quantum devices},
   volume={3},
   ISSN={2058-9565},
   url={http://dx.doi.org/10.1088/2058-9565/aab822},
   DOI={10.1088/2058-9565/aab822},
   number={3},
   journal={Quantum Science and Technology},
   publisher={IOP Publishing},
   author={Moll, Nikolaj and others},
   year={2018},
   month={Jun},
   pages={030503}
}

@article{optimization-rosenberg2016,
   title={Solving the Optimal Trading Trajectory Problem Using a Quantum Annealer},
   volume={10},
   ISSN={1941-0484},
   url={http://dx.doi.org/10.1109/JSTSP.2016.2574703},
   DOI={10.1109/jstsp.2016.2574703},
   number={6},
   journal={IEEE Journal of Selected Topics in Signal Processing},
   publisher={Institute of Electrical and Electronics Engineers (IEEE)},
   author={Rosenberg, Gili and Haghnegahdar, Poya and Goddard, Phil and Carr, Peter and Wu, Kesheng and de Prado, Marcos Lopez},
   year={2016},
   month={Sep},
   pages={1053–1060}
}

@article{optimization-lopez2015,
  doi = {10.2139/ssrn.2575184},
  url = {https://doi.org/10.2139/ssrn.2575184},
  year = {2015},
  publisher = {Elsevier {BV}},
  author = {Marcos Lopez de Prado},
  title = {Generalized Optimal Trading Trajectories: A Financial Quantum Computing Application},
  journal = {{SSRN} Electronic Journal}
}

@article{woerner_quantum_2019,
  doi = {10.1038/s41534-019-0130-6},
  url = {https://doi.org/10.1038/s41534-019-0130-6},
  year = {2019},
  month = feb,
  publisher = {Springer Science and Business Media {LLC}},
  volume = {5},
  number = {1},
  author = {Stefan Woerner and Daniel J. Egger},
  title = {Quantum risk analysis},
  journal = {npj Quantum Information}
}

@article{black_blackscholes_1973,
  doi = {10.1086/260062},
  url = {https://doi.org/10.1086/260062},
  year = {1973},
  month = may,
  publisher = {University of Chicago Press},
  volume = {81},
  number = {3},
  pages = {637--654},
  author = {Fischer Black and Myron Scholes},
  title = {The Pricing of Options and Corporate Liabilities},
  journal = {Journal of Political Economy}
}

@article{rebentrost_quantum_2018,
  doi = {10.1103/physreva.98.022321},
  url = {https://doi.org/10.1103/physreva.98.022321},
  year = {2018},
  month = aug,
  publisher = {American Physical Society ({APS})},
  volume = {98},
  number = {2},
  author = {Patrick Rebentrost and Brajesh Gupt and Thomas R. Bromley},
  title = {Quantum computational finance: Monte Carlo pricing of financial derivatives},
  journal = {Physical Review A}
}

@article{montanaro_montecarlo_2015,
  doi = {10.1098/rspa.2015.0301},
  url = {https://doi.org/10.1098/rspa.2015.0301},
  year = {2015},
  month = sep,
  publisher = {The Royal Society},
  volume = {471},
  number = {2181},
  pages = {20150301},
  author = {Ashley Montanaro},
  title = {Quantum speedup of Monte Carlo methods},
  journal = {Proceedings of the Royal Society A: Mathematical,  Physical and Engineering Sciences}
}

@incollection{aaronson_counting_2020,
  doi = {10.1137/1.9781611976014.5},
  url = {https://doi.org/10.1137/1.9781611976014.5},
  year = {2020},
  month = jan,
  publisher = {Society for Industrial and Applied Mathematics},
  pages = {24--32},
  author = {Scott Aaronson and Patrick Rall},
  title = {Quantum Approximate Counting,  Simplified},
  booktitle = {Symposium on Simplicity in Algorithms}
}

@article{brassard_amplitude_estimation_2002,
  doi = {10.1090/conm/305/05215},
  url = {https://doi.org/10.1090/conm/305/05215},
  year = {2002},
  publisher = {American Mathematical Society},
  pages = {53--74},
  author = {Gilles Brassard and Peter H{\o}yer and Michele Mosca and Alain Tapp},
  title = {Quantum amplitude amplification and estimation}
}

@article{grinko_iqae_2021,
  doi = {10.1038/s41534-021-00379-1},
  url = {https://doi.org/10.1038/s41534-021-00379-1},
  year = {2021},
  month = mar,
  publisher = {Springer Science and Business Media {LLC}},
  volume = {7},
  number = {1},
  author = {Dmitry Grinko and Julien Gacon and Christa Zoufal and Stefan Woerner},
  title = {Iterative quantum amplitude estimation},
  journal = {npj Quantum Information}
}

@article{qGAN-lloyd2018,
  doi = {10.1103/physrevlett.121.040502},
  url = {https://doi.org/10.1103/physrevlett.121.040502},
  year = {2018},
  month = jul,
  publisher = {American Physical Society ({APS})},
  volume = {121},
  number = {4},
  author = {Seth Lloyd and Christian Weedbrook},
  title = {Quantum Generative Adversarial Learning},
  journal = {Physical Review Letters}
}

@article{qGAN-dallaire2018,
  doi = {10.1103/physreva.98.012324},
  url = {https://doi.org/10.1103/physreva.98.012324},
  year = {2018},
  month = jul,
  publisher = {American Physical Society ({APS})},
  volume = {98},
  number = {1},
  author = {Pierre-Luc Dallaire-Demers and Nathan Killoran},
  title = {Quantum generative adversarial networks},
  journal = {Physical Review A}
}

@article{qGAN-zoufal2019,
  doi = {10.1038/s41534-019-0223-2},
  url = {https://doi.org/10.1038/s41534-019-0223-2},
  year = {2019},
  month = nov,
  publisher = {Springer Science and Business Media {LLC}},
  volume = {5},
  number = {1},
  author = {Christa Zoufal and Aur{\'{e}}lien Lucchi and Stefan Woerner},
  title = {Quantum Generative Adversarial Networks for learning and loading random distributions},
  journal = {npj Quantum Information}
}

@article{dur_entanglement_2000,
  doi = {10.1103/physreva.62.062314},
  url = {https://doi.org/10.1103/physreva.62.062314},
  year = {2000},
  month = nov,
  publisher = {American Physical Society ({APS})},
  volume = {62},
  number = {6},
  author = {W. D\"{u}r and G. Vidal and J. I. Cirac},
  title = {Three qubits can be entangled in two inequivalent ways},
  journal = {Physical Review A}
}

@article{partialiSWAP-bialczak2010,
  doi = {10.1038/nphys1639},
  url = {https://doi.org/10.1038/nphys1639},
  year = {2010},
  month = apr,
  publisher = {Springer Science and Business Media {LLC}},
  volume = {6},
  number = {6},
  pages = {409--413},
  author = {R. C. Bialczak and M. Ansmann and M. Hofheinz and E. Lucero and M. Neeley and A. D. O'Connell and D. Sank and H. Wang and J. Wenner and M. Steffen and A. N. Cleland and J. M. Martinis},
  title = {Quantum process tomography of a universal entangling gate implemented with Josephson phase qubits},
  journal = {Nature Physics}
}

@article{iSWAP-schuch2003,
   title={Natural two-qubit gate for quantum computation using theXYinteraction},
   volume={67},
   ISSN={1094-1622},
   url={http://dx.doi.org/10.1103/PhysRevA.67.032301},
   DOI={10.1103/physreva.67.032301},
   number={3},
   journal={Physical Review A},
   publisher={American Physical Society (APS)},
   author={Schuch, Norbert and Siewert, Jens},
   year={2003},
   month={Mar},
   pages={-}
}

@article{google_supremacy_2019,
   title={Quantum supremacy using a programmable superconducting processor},
   volume={574},
   ISSN={1476-4687},
   url={http://dx.doi.org/10.1038/s41586-019-1666-5},
   DOI={10.1038/s41586-019-1666-5},
   number={7779},
   journal={Nature},
   publisher={Springer Science and Business Media LLC},
   author={Arute, Frank and others},
   year={2019},
   month={Oct},
   pages={505–510}
}

@article{gard2020efficient,
   title={Efficient symmetry-preserving state preparation circuits for the variational quantum eigensolver algorithm},
   volume={6},
   ISSN={2056-6387},
   url={http://dx.doi.org/10.1038/s41534-019-0240-1},
   DOI={10.1038/s41534-019-0240-1},
   number={1},
   journal={npj Quantum Information},
   publisher={Springer Science and Business Media LLC},
   author={Gard, Bryan T. and Zhu, Linghua and Barron, George S. and Mayhall, Nicholas J. and Economou, Sophia E. and Barnes, Edwin},
   year={2020},
   month={Jan},
   pages={-}
}

@article{barkoutsos2018quantum,
   title={Quantum algorithms for electronic structure calculations: Particle-hole Hamiltonian and optimized wave-function expansions},
   volume={98},
   ISSN={2469-9934},
   url={http://dx.doi.org/10.1103/PhysRevA.98.022322},
   DOI={10.1103/physreva.98.022322},
   number={2},
   journal={Physical Review A},
   publisher={American Physical Society (APS)},
   author={Barkoutsos, Panagiotis Kl. and others},
   year={2018},
   month={Aug},
   pages={-}
}

@article{wang2019xy,
   title={XY
 mixers: Analytical and numerical results for the quantum alternating operator ansatz},
   volume={101},
   ISSN={2469-9934},
   url={http://dx.doi.org/10.1103/PhysRevA.101.012320},
   DOI={10.1103/physreva.101.012320},
   number={1},
   journal={Physical Review A},
   publisher={American Physical Society (APS)},
   author={Wang, Zhihui and Rubin, Nicholas C. and Dominy, Jason M. and Rieffel, Eleanor G.},
   year={2020},
   month={Jan},
   pages={-}
}

@article{hadfield2019qaoa,
   title={From the Quantum Approximate Optimization Algorithm to a Quantum Alternating Operator Ansatz},
   volume={12},
   ISSN={1999-4893},
   url={http://dx.doi.org/10.3390/a12020034},
   DOI={10.3390/a12020034},
   number={2},
   journal={Algorithms},
   publisher={MDPI AG},
   author={Hadfield, Stuart and Wang, Zhihui and O’Gorman, Bryan and Rieffel, Eleanor and Venturelli, Davide and Biswas, Rupak},
   year={2019},
   month={Feb},
   pages={34}
}

@inproceedings{cook2019xy,
  doi = {10.1109/qce49297.2020.00021},
  url = {https://doi.org/10.1109/qce49297.2020.00021},
  year = {2020},
  month = oct,
  publisher = {{IEEE}},
  author = {Jeremy Cook and Stephan Eidenbenz and Andreas Bartschi},
  title = {The Quantum Alternating Operator Ansatz on Maximum k-Vertex Cover},
  booktitle = {2020 {IEEE} International Conference on Quantum Computing and Engineering ({QCE})}
}

@article{Kemna_pricing_1990,
  doi = {10.1016/0378-4266(90)90039-5},
  url = {https://doi.org/10.1016/0378-4266(90)90039-5},
  year = {1990},
  month = mar,
  publisher = {Elsevier {BV}},
  volume = {14},
  number = {1},
  pages = {113--129},
  author = {A.G.Z. Kemna and A.C.F. Vorst},
  title = {A pricing method for options based on average asset values},
  journal = {Journal of Banking {\&} Finance}
}

@article{amplitude_estimation-suzuki2020,
   title={Amplitude estimation without phase estimation},
   volume={19},
   ISSN={1573-1332},
   number={2},
   journal={Quantum Information Processing},
   publisher={Springer Science and Business Media LLC},
   author={Suzuki, Yohichi and Uno, Shumpei and Raymond, Rudy and Tanaka, Tomoki and Onodera, Tamiya and Yamamoto, Naoki},
   year={2020},
   month={Jan},
   url={http://dx.doi.org/10.1007/s11128-019-2565-2},
   DOI={10.1007/s11128-019-2565-2},
   pages={-}
}

@article{verstraete_quantum_2009,
   title={Quantum circuits for strongly correlated quantum systems},
   volume={79},
   ISSN={1094-1622},
   url={http://dx.doi.org/10.1103/PhysRevA.79.032316},
   DOI={10.1103/physreva.79.032316},
   number={3},
   journal={Physical Review A},
   publisher={American Physical Society (APS)},
   author={Verstraete, Frank and Cirac, J. Ignacio and Latorre, José I.},
   year={2009},
   month={Mar},
   pages={-}
}

@article{hebenstreit_compressed_2017,
  doi = {10.1103/physreva.95.052339},
  url = {https://doi.org/10.1103/physreva.95.052339},
  year = {2017},
  month = may,
  publisher = {American Physical Society ({APS})},
  volume = {95},
  number = {5},
  author = {M. Hebenstreit and D. Alsina and J. I. Latorre and B. Kraus},
  title = {Compressed quantum computation using a remote five-qubit quantum computer},
  journal = {Physical Review A},
   pages={-}
}

@article{cerveralierta_ising_2018,
   title={Exact Ising model simulation on a quantum computer},
   volume={2},
   ISSN={2521-327X},
   url={http://dx.doi.org/10.22331/q-2018-12-21-114},
   DOI={10.22331/q-2018-12-21-114},
   journal={Quantum},
   publisher={Verein zur Forderung des Open Access Publizierens in den Quantenwissenschaften},
   author={Cervera-Lierta, Alba},
   year={2018},
   month={Dec},
   pages={114}
}

@article{iblisdir_matrix_2007,
  doi = {10.1103/physrevb.75.104305},
  url = {https://doi.org/10.1103/physrevb.75.104305},
  year = {2007},
  month = mar,
  publisher = {American Physical Society ({APS})},
  volume = {75},
  number = {10},
  author = {S. Iblisdir and R. Or{\'{u}}s and J. I. Latorre},
  title = {Matrix product states algorithms and continuous systems},
  journal = {Physical Review B}
}

@article{KL-kullback1951,
  year = {1951},
  month = mar,
  publisher = {Institute of Mathematical Statistics},
  volume = {22},
  number = {1},
  pages = {79--86},
  author = {S. Kullback and R. A. Leibler},
  title = {On Information and Sufficiency},
  journal = {The Annals of Mathematical Statistics},
  doi = {10.1214/aoms/1177729694},
  url = {https://doi.org/10.1214/aoms/1177729694}
  }

@online{accuracy_precision,
author = {Pekaje},
title = {Accuracy and precision},
year = {2007},
url = {https://commons.wikimedia.org/w/index.php?curid=1862863},
OPTurldate = {19th July 2021}
}

@article{temme-mitigation2017,
  doi = {10.1103/physrevlett.119.180509},
  url = {https://doi.org/10.1103/physrevlett.119.180509},
  year = {2017},
  month = nov,
  publisher = {American Physical Society ({APS})},
  volume = {119},
  number = {18},
  author = {Kristan Temme and Sergey Bravyi and Jay M. Gambetta},
  title = {Error Mitigation for Short-Depth Quantum Circuits},
  journal = {Physical Review Letters}
}

@article{endo-mitigation2018,
  doi = {10.1103/physrevx.8.031027},
  url = {https://doi.org/10.1103/physrevx.8.031027},
  year = {2018},
  month = jul,
  publisher = {American Physical Society ({APS})},
  volume = {8},
  number = {3},
  author = {Suguru Endo and Simon C. Benjamin and Ying Li},
  title = {Practical Quantum Error Mitigation for Near-Future Applications},
  journal = {Physical Review X}
}

@article{ItoLemma-1944,
title = {Stochastic integral},
author = {Itô, Kiyosi},
journal = {Proceedings of the Imperial Academy},
volume = {20},
number = {8},
pages = {519-524},
year = {1944},
publisher = {The Japan Academy},
}

@article{binomial-wallis2013,
  year = {2013},
  month = {Jul},
  publisher = {Informa {UK} Limited},
  volume = {20},
  number = {3},
  pages = {178--208},
  author = {Sean Wallis},
  title = {Binomial Confidence Intervals and Contingency Tests: Mathematical Fundamentals and the Evaluation of Alternative Methods},
  journal = {Journal of Quantitative Linguistics},
  DOI = {10.1080/09296174.2013.799918},
  url = {https://doi.org/10.1080/09296174.2013.799918},
  }

@book{mitchell_machine_1997,
  title={Machine learning},
  author={Mitchell, Tom M and others},
  publisher={McGraw Hill},
  isbn={978-0070428077},
  year={1997}, 
  url={http://www.cs.cmu.edu/~tom/mlbook.html}
}

@book{michalski_machine_2013,
  title={Machine Learning: An Artificial Intelligence Approach},
  author={Michalski, R.S. and Carbonell, J.G. and Mitchell, T.M.},
  isbn={978-3662124055},
  series={Symbolic Computation},
  url={https://books.google.es/books?id=-eqpCAAAQBAJ},
  year={2013},
  publisher={Springer Berlin Heidelberg}
}

@article{deng_mnist_2012,
  title={The mnist database of handwritten digit images for machine learning research},
  author={Deng, Li},
  journal={IEEE Signal Processing Magazine},
  volume={29},
  number={6},
  pages={141--142},
  year={2012},
  publisher={IEEE}
}

@article{chess,
  doi = {10.1016/s0004-3702(01)00129-1},
  url = {https://doi.org/10.1016/s0004-3702(01)00129-1},
  year = {2002},
  month = jan,
  publisher = {Elsevier {BV}},
  volume = {134},
  number = {1-2},
  pages = {57--83},
  author = {Murray Campbell and A.Joseph Hoane and Feng-hsiung Hsu},
  title = {Deep Blue},
  journal = {Artificial Intelligence}
}

@article{go,
  doi = {10.1038/nature16961},
  url = {https://doi.org/10.1038/nature16961},
  year = {2016},
  month = jan,
  publisher = {Springer Science and Business Media {LLC}},
  volume = {529},
  number = {7587},
  pages = {484--489},
  author = {David Silver and Aja Huang and Chris J. Maddison and Arthur Guez and Laurent Sifre and George van den Driessche and Julian Schrittwieser and Ioannis Antonoglou and Veda Panneershelvam and Marc Lanctot and Sander Dieleman and Dominik Grewe and John Nham and Nal Kalchbrenner and Ilya Sutskever and Timothy Lillicrap and Madeleine Leach and Koray Kavukcuoglu and Thore Graepel and Demis Hassabis},
  title = {Mastering the game of Go with deep neural networks and tree search},
  journal = {Nature}
}

@article{alpha_fold,
  doi = {10.1038/s41586-021-03819-2},
  url = {https://doi.org/10.1038/s41586-021-03819-2},
  year = {2021},
  month = jul,
  publisher = {Springer Science and Business Media {LLC}},
  author = {John Jumper and Richard Evans and Alexander Pritzel and Tim Green and Michael Figurnov and Olaf Ronneberger and Kathryn Tunyasuvunakool and Russ Bates and Augustin {\v{Z}}{\'{\i}}dek and Anna Potapenko and Alex Bridgland and Clemens Meyer and Simon A. A. Kohl and Andrew J. Ballard and Andrew Cowie and Bernardino Romera-Paredes and Stanislav Nikolov and Rishub Jain and Jonas Adler and Trevor Back and Stig Petersen and David Reiman and Ellen Clancy and Michal Zielinski and Martin Steinegger and Michalina Pacholska and Tamas Berghammer and Sebastian Bodenstein and David Silver and Oriol Vinyals and Andrew W. Senior and Koray Kavukcuoglu and Pushmeet Kohli and Demis Hassabis},
  title = {Highly accurate protein structure prediction with {AlphaFold}},
  journal = {Nature}
}

@book{russell_artificial_2010,
 author = {Russell, Stuart},
 title = {Artificial intelligence: a modern approach},
 publisher = {Prentice Hall},
 year = {2010},
 address = {Upper Saddle River, New Jersey},
 isbn = {978-0136042594}
 }

@book{duda_pattern_2012,
  title={Pattern Classification},
  author={Duda, R.O. and Hart, P.E. and Stork, D.G.},
  isbn={978-1118586006},
  url={https://books.google.es/books?id=Br33IRC3PkQC},
  year={2012},
  publisher={Wiley}
}

@article{EstivillCastro_clustering_2002,
  doi = {10.1145/568574.568575},
  url = {https://doi.org/10.1145/568574.568575},
  year = {2002},
  month = jun,
  publisher = {Association for Computing Machinery ({ACM})},
  volume = {4},
  number = {1},
  pages = {65--75},
  author = {Vladimir Estivill-Castro},
  title = {Why so many clustering algorithms},
  journal = {{ACM} {SIGKDD} Explorations Newsletter}
}

@article{Kaelbling_reinforcement_1996,
  doi = {10.1613/jair.301},
  url = {https://doi.org/10.1613/jair.301},
  year = {1996},
  month = may,
  publisher = {{AI} Access Foundation},
  volume = {4},
  pages = {237--285},
  author = {L. P. Kaelbling and M. L. Littman and A. W. Moore},
  title = {Reinforcement Learning:  A Survey},
  journal = {Journal of Artificial Intelligence Research}
}

@inproceedings{Schmidhuber_ageneral_1998,
    author = {Jürgen Schmidhuber},
    title = {A General Method for Incremental Self-Improvement and Multi-Agent Learning},
    booktitle = {Evolutionary Computation: Theory and Applications. Scientific Publ. Co., Singapore. In},
    year = {1998},
    pages = {81--123}
}

@article{carleo_ml_2019,
  doi = {10.1103/revmodphys.91.045002},
  url = {https://doi.org/10.1103/revmodphys.91.045002},
  year = {2019},
  month = dec,
  publisher = {American Physical Society ({APS})},
  volume = {91},
  number = {4},
  author = {Giuseppe Carleo and Ignacio Cirac and Kyle Cranmer and Laurent Daudet and Maria Schuld and Naftali Tishby and Leslie Vogt-Maranto and Lenka Zdeborov{\'{a}}},
  title = {Machine learning and the physical sciences},
  journal = {Reviews of Modern Physics}
}

@book{hoffmann_nn_1992,
  doi = {10.1007/978-3-322-83200-9},
  url = {https://doi.org/10.1007/978-3-322-83200-9},
  isbn ={ 978-3322832009},
  year = {1992},
  publisher = {Vieweg$+$Teubner Verlag},
  author = {Norbert Hoffmann},
  title = {Simulation Neuronaler Netze}
}

@article{Hammer2005,
  doi = {10.1162/0899766053491878},
  url = {https://doi.org/10.1162/0899766053491878},
  year = {2005},
  month = may,
  publisher = {{MIT} Press - Journals},
  volume = {17},
  number = {5},
  pages = {1109--1159},
  author = {Barbara Hammer and Alessio Micheli and Alessandro Sperduti},
  title = {Universal Approximation Capability of Cascade Correlation for Structures},
  journal = {Neural Computation}
}

@article{Kramer1991,
  doi = {10.1002/aic.690370209},
  url = {https://doi.org/10.1002/aic.690370209},
  year = {1991},
  month = feb,
  publisher = {Wiley},
  volume = {37},
  number = {2},
  pages = {233--243},
  author = {Mark A. Kramer},
  title = {Nonlinear principal component analysis using autoassociative neural networks},
  journal = {{AIChE} Journal}
}

@article{Kingma_autoencoders_2019,
  doi = {10.1561/2200000056},
  url = {https://doi.org/10.1561/2200000056},
  year = {2019},
  publisher = {Now Publishers},
  volume = {12},
  number = {4},
  pages = {307--392},
  author = {Diederik P. Kingma and Max Welling},
  title = {An Introduction to Variational Autoencoders},
  journal = {Foundations and Trends{\textregistered} in Machine Learning}
}

@misc{li2015constructing,
      title={Constructing Long Short-Term Memory based Deep Recurrent Neural Networks for Large Vocabulary Speech Recognition}, 
      author={Xiangang Li and Xihong Wu},
      year={2015},
      eprint={1410.4281},
      archivePrefix={arXiv},
      primaryClass={cs.CL}
}

@article{Valueva2020,
  doi = {10.1016/j.matcom.2020.04.031},
  url = {https://doi.org/10.1016/j.matcom.2020.04.031},
  year = {2020},
  month = nov,
  publisher = {Elsevier {BV}},
  volume = {177},
  pages = {232--243},
  author = {M.V. Valueva and N.N. Nagornov and P.A. Lyakhov and G.V. Valuev and N.I. Chervyakov},
  title = {Application of the residue number system to reduce hardware costs of the convolutional neural network implementation},
  journal = {Mathematics and Computers in Simulation}
}

@article{ackley_learning_1985,
  doi = {10.1207/s15516709cog0901_7},
  url = {https://doi.org/10.1207/s15516709cog0901_7},
  year = {1985},
  month = jan,
  publisher = {Wiley},
  volume = {9},
  number = {1},
  pages = {147--169},
  author = {David H. Ackley and Geoffrey E. Hinton and Terrence J. Sejnowski},
  title = {A Learning Algorithm for Boltzmann Machiness},
  journal = {Cognitive Science}
}

@book{du_neural_2019,
  title={Neural Networks and Statistical Learning},
  author={Du, K.L. and Swamy, M.N.S.},
  isbn={978-1447174523},
  url={https://books.google.ru/books?id=IUmvDwAAQBAJ},
  year={2019},
  publisher={Springer London}
}

@article{Graves2009,
  doi = {10.1109/tpami.2008.137},
  url = {https://doi.org/10.1109/tpami.2008.137},
  year = {2009},
  month = may,
  publisher = {Institute of Electrical and Electronics Engineers ({IEEE})},
  volume = {31},
  number = {5},
  pages = {855--868},
  author = {A. Graves and M. Liwicki and S. Fernandez and R. Bertolami and H. Bunke and J. Schmidhuber},
  title = {A Novel Connectionist System for Unconstrained Handwriting Recognition},
  journal = {{IEEE} Transactions on Pattern Analysis and Machine Intelligence}
}

@book{pardalos_non-convex_2017,
 author = {Pardalos, P. M.},
 title = {Non-convex multi-objective optimization},
 publisher = {Springer International Publishing Imprint Springer},
 year = {2017},
 address = {Cham},
 isbn = {978-3319610054}
 }

@book{jain_non-convex_2017,
 author = {Jain, Prateek},
 title = {Non-convex optimization for machine learning},
 publisher = {Now Publishers},
 year = {2017},
 address = {Hanover, Massachusetts},
 isbn = {978-1680833683}
 }

@article{Nash_tnc_1984,
  doi = {10.1137/0721052},
  url = {https://doi.org/10.1137/0721052},
  year = {1984},
  month = aug,
  publisher = {Society for Industrial {\&} Applied Mathematics ({SIAM})},
  volume = {21},
  number = {4},
  pages = {770--788},
  author = {Stephen G. Nash},
  title = {Newton-Type Minimization via the Lanczos Method},
  journal = {{SIAM} Journal on Numerical Analysis}
}

@article{nelder_mead_1965,
  title={A Simplex Method for Function Minimization},
  author={J. Nelder and R. Mead},
  journal={Comput. J.},
  year={1965},
  volume={7},
  pages={308-313}
}

@article{powell_1964,
  doi = {10.1093/comjnl/7.2.155},
  url = {https://doi.org/10.1093/comjnl/7.2.155},
  year = {1964},
  month = feb,
  publisher = {Oxford University Press ({OUP})},
  volume = {7},
  number = {2},
  pages = {155--162},
  author = {M. J. D. Powell},
  title = {An efficient method for finding the minimum of a function of several variables without calculating derivatives},
  journal = {The Computer Journal}
}

@article{Qian1999,
  doi = {10.1016/s0893-6080(98)00116-6},
  url = {https://doi.org/10.1016/s0893-6080(98)00116-6},
  year = {1999},
  month = jan,
  publisher = {Elsevier {BV}},
  volume = {12},
  number = {1},
  pages = {145--151},
  author = {Ning Qian},
  title = {On the momentum term in gradient descent learning algorithms},
  journal = {Neural Networks}
}

@inproceedings{nesterov_method_1983,
  title={A method for unconstrained convex minimization problem with the rate of convergence O (1/k\^{} 2)},
  author={Nesterov, Yurii},
  booktitle={Doklady an ussr},
  volume={269},
  pages={543--547},
  year={1983}
}

@misc{zeiler_adadelta_2012,
      title={ADADELTA: An Adaptive Learning Rate Method}, 
      author={Matthew D. Zeiler},
      year={2012},
      eprint={1212.5701},
      archivePrefix={arXiv},
      primaryClass={cs.LG}
}

@book{sutskever2013training,
  title={Training recurrent neural networks},
  author={Sutskever, Ilya},
  year={2013},
  publisher={University of Toronto, Canada}, 
  isbn = {978-0499220660}, 
  url = {https://dl.acm.org/doi/book/10.5555/2604780}, 
  doi = {10.5555/2604780}
}

@book{spall_stochastic_2005,
  title={Introduction to stochastic search and optimization: estimation, simulation, and control},
  author={Spall, James C},
  volume={65},
  year={2005},
  publisher={John Wiley \& Sons}
}

@article{spall_spsa_1998,
  title={An overview of the simultaneous perturbation method for efficient optimization},
  author={Spall, James C},
  journal={Johns Hopkins apl technical digest},
  volume={19},
  number={4},
  pages={482--492},
  year={1998}
}

@online{cuda,
  author={NVIDIA and Vingelmann, P\'eter and Fitzek, Frank H.P.},
  title={CUDA, release: 10.2.89},
  year={2020},
  url={https://developer.nvidia.com/cuda-toolkit}
}

@article{Cortes_svc_1995,
  doi = {10.1007/bf00994018},
  url = {https://doi.org/10.1007/bf00994018},
  year = {1995},
  month = sep,
  publisher = {Springer Science and Business Media {LLC}},
  volume = {20},
  number = {3},
  pages = {273--297},
  author = {Corinna Cortes and Vladimir Vapnik},
  title = {Support-vector networks},
  journal = {Machine Learning}
}

@article{benhur_svc_2002,
    author = {Ben-Hur, Asa and Horn, David and Siegelmann, Hava T. and Vapnik, Vladimir},
    title = {Support Vector Clustering},
    year = {2002},
    publisher = {JMLR.org},
    volume = {2},
    issn = {1532-4435},
    doi = {10.5555/944790.944807},
    url = {https://dl.acm.org/doi/10.5555/944790.944807},
    month = mar,
    pages = {125--137}, 
    journal={The Journal of Machine Learning Research}
    }

@book{press_numerical_1986,
author = {Press, William H. and Flannery, Brian P. and Teukolsky, Saul A. and Vetterling, William T.},
title = {Numerical Recipes: The Art of Scientific Computing},
year = {1986},
isbn = {978-0521308119},
publisher = {Cambridge University Press},
address = {USA}, 
url = {https://dl.acm.org/doi/book/10.5555/6771}, doi = {10.5555/6771}
}

@misc{cerezo_variational_2020,
      title={Variational Quantum Algorithms}, 
      author={M. Cerezo and Andrew Arrasmith and Ryan Babbush and Simon C. Benjamin and Suguru Endo and Keisuke Fujii and Jarrod R. McClean and Kosuke Mitarai and Xiao Yuan and Lukasz Cincio and Patrick J. Coles},
      year={2020},
      eprint={2012.09265},
      archivePrefix={arXiv},
      primaryClass={quant-ph}
}

@misc{bharti_noisy_2021,
      title={Noisy intermediate-scale quantum (NISQ) algorithms}, 
      author={Kishor Bharti and Alba Cervera-Lierta and Thi Ha Kyaw and Tobias Haug and Sumner Alperin-Lea and Abhinav Anand and Matthias Degroote and Hermanni Heimonen and Jakob S. Kottmann and Tim Menke and Wai-Keong Mok and Sukin Sim and Leong-Chuan Kwek and Alán Aspuru-Guzik},
      year={2021},
      eprint={2101.08448},
      archivePrefix={arXiv},
      primaryClass={quant-ph}
}

@article{Vidal_mps_2004,
   title={Efficient Simulation of One-Dimensional Quantum Many-Body Systems},
   volume={93},
   ISSN={1079-7114},
   url={http://dx.doi.org/10.1103/PhysRevLett.93.040502},
   DOI={10.1103/physrevlett.93.040502},
   number={4},
   journal={Physical Review Letters},
   publisher={American Physical Society (APS)},
   author={Vidal, Guifré},
   year={2004},
   month={Jul}
}

@article{orus_tn_2014,
  doi = {10.1016/j.aop.2014.06.013},
  url = {https://doi.org/10.1016/j.aop.2014.06.013},
  year = {2014},
  month = oct,
  publisher = {Elsevier {BV}},
  volume = {349},
  pages = {117--158},
  author = {Román Orús},
  title = {A practical introduction to tensor networks: Matrix product states and projected entangled pair states},
  journal = {Annals of Physics}
}

@misc{biamonte_tensor_2017,
      title={Tensor Networks in a Nutshell}, 
      author={Jacob Biamonte and Ville Bergholm},
      year={2017},
      eprint={1708.00006},
      archivePrefix={arXiv},
      primaryClass={quant-ph}
}

@article{verstraete_tn_2008,
   title={Matrix product states, projected entangled pair states, and variational renormalization group methods for quantum spin systems},
   volume={57},
   ISSN={1460-6976},
   url={http://dx.doi.org/10.1080/14789940801912366},
   DOI={10.1080/14789940801912366},
   number={2},
   journal={Advances in Physics},
   publisher={Informa UK Limited},
   author={Verstraete, F. and Murg, V. and Cirac, J.I.},
   year={2008},
   month={Mar},
   pages={143–224}
}

@article{Vidal_mera_2008,
   title={Class of Quantum Many-Body States That Can Be Efficiently Simulated},
   volume={101},
   ISSN={1079-7114},
   url={http://dx.doi.org/10.1103/PhysRevLett.101.110501},
   DOI={10.1103/physrevlett.101.110501},
   number={11},
   journal={Physical Review Letters},
   publisher={American Physical Society (APS)},
   author={Vidal, G.},
   year={2008},
   month={Sep}
}

@article{Vidal_entanglement_2007,
   title={Entanglement Renormalization},
   volume={99},
   ISSN={1079-7114},
   url={http://dx.doi.org/10.1103/PhysRevLett.99.220405},
   DOI={10.1103/physrevlett.99.220405},
   number={22},
   journal={Physical Review Letters},
   publisher={American Physical Society (APS)},
   author={Vidal, G.},
   year={2007},
   month={Nov}
}

@article{orus_tn2_2014,
   title={Advances on tensor network theory: symmetries, fermions, entanglement, and holography},
   volume={87},
   ISSN={1434-6036},
   url={http://dx.doi.org/10.1140/epjb/e2014-50502-9},
   DOI={10.1140/epjb/e2014-50502-9},
   number={11},
   journal={The European Physical Journal B},
   publisher={Springer Science and Business Media LLC},
   author={Orús, Román},
   year={2014},
   month={Nov}
}

@article{Cheng_supervised_2021,
   title={Supervised learning with projected entangled pair states},
   volume={103},
   ISSN={2469-9969},
   url={http://dx.doi.org/10.1103/PhysRevB.103.125117},
   DOI={10.1103/physrevb.103.125117},
   number={12},
   journal={Physical Review B},
   publisher={American Physical Society (APS)},
   author={Cheng, Song and Wang, Lei and Zhang, Pan},
   year={2021},
   month={Mar}
}

@misc{convy_mutual_2021,
      title={Mutual Information Scaling for Tensor Network Machine Learning}, 
      author={Ian Convy and William Huggins and Haoran Liao and K. Birgitta Whaley},
      year={2021},
      eprint={2103.00105},
      archivePrefix={arXiv},
      primaryClass={quant-ph}
}

@misc{torlai_quantum_2020,
      title={Quantum process tomography with unsupervised learning and tensor networks}, 
      author={Giacomo Torlai and Christopher J. Wood and Atithi Acharya and Giuseppe Carleo and Juan Carrasquilla and Leandro Aolita},
      year={2020},
      eprint={2006.02424},
      archivePrefix={arXiv},
      primaryClass={quant-ph}
}

@misc{wang_anomaly_2020,
      title={Anomaly Detection with Tensor Networks}, 
      author={Jinhui Wang and Chase Roberts and Guifre Vidal and Stefan Leichenauer},
      year={2020},
      eprint={2006.02516},
      archivePrefix={arXiv},
      primaryClass={cs.LG}
}

@misc{martyn_entanglement_2020,
      title={Entanglement and Tensor Networks for Supervised Image Classification}, 
      author={John Martyn and Guifre Vidal and Chase Roberts and Stefan Leichenauer},
      year={2020},
      eprint={2007.06082},
      archivePrefix={arXiv},
      primaryClass={quant-ph}
}

@misc{stoudenmire_supervised_2017,
      title={Supervised Learning with Quantum-Inspired Tensor Networks}, 
      author={E. Miles Stoudenmire and David J. Schwab},
      year={2017},
      eprint={1605.05775},
      archivePrefix={arXiv},
      primaryClass={stat.ML}
}

@misc{reyes_multiscale_2020,
      title={A Multi-Scale Tensor Network Architecture for Classification and Regression}, 
      author={Justin Reyes and Miles Stoudenmire},
      year={2020},
      eprint={2001.08286},
      archivePrefix={arXiv},
      primaryClass={stat.ML}
}

@misc{huang_provably_2021,
      title={Provably efficient machine learning for quantum many-body problems}, 
      author={Hsin-Yuan Huang and Richard Kueng and Giacomo Torlai and Victor V. Albert and John Preskill},
      year={2021},
      eprint={2106.12627},
      archivePrefix={arXiv},
      primaryClass={quant-ph}
}

@article{Peruzzo_vqe_2014,
  doi = {10.1038/ncomms5213},
  url = {https://doi.org/10.1038/ncomms5213},
  year = {2014},
  month = jul,
  publisher = {Springer Science and Business Media {LLC}},
  volume = {5},
  number = {1},
  author = {Alberto Peruzzo and Jarrod McClean and Peter Shadbolt and Man-Hong Yung and Xiao-Qi Zhou and Peter J. Love and Al{\'{a}}n Aspuru-Guzik and Jeremy L. O'Brien},
  title = {A variational eigenvalue solver on a photonic quantum processor},
  journal = {Nature Communications}
}

@misc{farhi_qaoa_2014,
      title={A Quantum Approximate Optimization Algorithm}, 
      author={Edward Farhi and Jeffrey Goldstone and Sam Gutmann},
      year={2014},
      eprint={1411.4028},
      archivePrefix={arXiv},
      primaryClass={quant-ph}
}

@article{christopher_solovay_2006,
    author = {Dawson, Christopher M. and Nielsen, Michael A.},
    title = {The Solovay-Kitaev Algorithm},
    year = {2006},
    issue_date = {January 2006},
    publisher = {Rinton Press, Incorporated},
    address = {Paramus, NJ},
    volume = {6},
    number = {1},
    issn = {1533-7146},
    journal = {Quantum Info. Comput.},
    month = jan,
    pages = {81–95},
    numpages = {15},
    doi={10.5555/2011679.2011685}, url={https://dl.acm.org/doi/10.5555/2011679.2011685}
    }

@misc{bittel_training_2021,
      title={Training variational quantum algorithms is NP-hard -- even for logarithmically many qubits and free fermionic systems}, 
      author={Lennart Bittel and Martin Kliesch},
      year={2021},
      eprint={2101.07267},
      archivePrefix={arXiv},
      primaryClass={quant-ph}
}

@article{McClean_bp_2018,
   title={Barren plateaus in quantum neural network training landscapes},
   volume={9},
   ISSN={2041-1723},
   url={http://dx.doi.org/10.1038/s41467-018-07090-4},
   DOI={10.1038/s41467-018-07090-4},
   number={1},
   journal={Nature Communications},
   publisher={Springer Science and Business Media LLC},
   author={McClean, Jarrod R. and Boixo, Sergio and Smelyanskiy, Vadim N. and Babbush, Ryan and Neven, Hartmut},
   year={2018},
   month={Nov}
}

@misc{dunjko_machine_2017,
      title={Machine learning \& artificial intelligence in the quantum domain}, 
      author={Vedran Dunjko and Hans J. Briegel},
      year={2017},
      eprint={1709.02779},
      archivePrefix={arXiv},
      primaryClass={quant-ph}
}

@article{PerdomoOrtiz_opportunities_2018,
  doi = {10.1088/2058-9565/aab859},
  url = {https://doi.org/10.1088/2058-9565/aab859},
  year = {2018},
  month = jun,
  publisher = {{IOP} Publishing},
  volume = {3},
  number = {3},
  pages = {030502},
  author = {Alejandro Perdomo-Ortiz and Marcello Benedetti and John Realpe-G{\'{o}}mez and Rupak Biswas},
  title = {Opportunities and challenges for quantum-assisted machine learning in near-term quantum computers},
  journal = {Quantum Science and Technology}
}

@misc{kübler_inductive_2021,
      title={The Inductive Bias of Quantum Kernels}, 
      author={Jonas M. Kübler and Simon Buchholz and Bernhard Schölkopf},
      year={2021},
      eprint={2106.03747},
      archivePrefix={arXiv},
      primaryClass={quant-ph}
}

@article{Lloyd_qpca_2014,
   title={Quantum principal component analysis},
   volume={10},
   ISSN={1745-2481},
   url={http://dx.doi.org/10.1038/nphys3029},
   DOI={10.1038/nphys3029},
   number={9},
   journal={Nature Physics},
   publisher={Springer Science and Business Media LLC},
   author={Lloyd, Seth and Mohseni, Masoud and Rebentrost, Patrick},
   year={2014},
   month={Jul},
   pages={631–633}
}

@misc{schuld_supervised_2021,
      title={Supervised quantum machine learning models are kernel methods}, 
      author={Maria Schuld},
      year={2021},
      eprint={2101.11020},
      archivePrefix={arXiv},
      primaryClass={quant-ph}
}

@article{GilVidal_redundancy_2020,
  doi = {10.3389/fphy.2020.00297},
  url = {https://doi.org/10.3389/fphy.2020.00297},
  year = {2020},
  month = aug,
  publisher = {Frontiers Media {SA}},
  volume = {8},
  author = {Francisco Javier Gil Vidal and Dirk Oliver Theis},
  title = {Input Redundancy for Parameterized Quantum Circuits},
  journal = {Frontiers in Physics}
}

@article{Meyer_variational_2021,
  doi = {10.1038/s41534-021-00425-y},
  url = {https://doi.org/10.1038/s41534-021-00425-y},
  year = {2021},
  month = jun,
  publisher = {Springer Science and Business Media {LLC}},
  volume = {7},
  number = {1},
  author = {Johannes Jakob Meyer and Johannes Borregaard and Jens Eisert},
  title = {A variational toolbox for quantum multi-parameter estimation},
  journal = {npj Quantum Information}
}

@article{Amari_natural_1998,
  doi = {10.1162/089976698300017746},
  url = {https://doi.org/10.1162/089976698300017746},
  year = {1998},
  month = feb,
  publisher = {{MIT} Press - Journals},
  volume = {10},
  number = {2},
  pages = {251--276},
  author = {Shun-ichi Amari},
  title = {Natural Gradient Works Efficiently in Learning},
  journal = {Neural Computation}
}

@article{fisher_mathematical_1922,
  author = {Ronald Fisher},
  doi = {10.1098/rsta.1922.0009},
  url = {https://doi.org/10.1098/rsta.1922.0009},
  year = {1922},
  month = jan,
  publisher = {The Royal Society},
  volume = {222},
  number = {594-604},
  pages = {309--368},
  title = {On the mathematical foundations of theoretical statistics},
  journal = {Philosophical Transactions of the Royal Society of London. Series A,  Containing Papers of a Mathematical or Physical Character}
}

@article{Savage_fisher_1976,
  doi = {10.1214/aos/1176343456},
  url = {https://doi.org/10.1214/aos/1176343456},
  year = {1976},
  month = may,
  publisher = {Institute of Mathematical Statistics},
  volume = {4},
  number = {3},
  author = {Leonard J. Savage},
  title = {On Rereading R. A. Fisher},
  journal = {The Annals of Statistics}
}

@article{Harrow_gradient_2021,
  doi = {10.1103/physrevlett.126.140502},
  url = {https://doi.org/10.1103/physrevlett.126.140502},
  year = {2021},
  month = apr,
  publisher = {American Physical Society ({APS})},
  volume = {126},
  number = {14},
  author = {Aram W. Harrow and John C. Napp},
  title = {Low-Depth Gradient Measurements Can Improve Convergence in Variational Hybrid Quantum-Classical Algorithms},
  journal = {Physical Review Letters}
}

@article{Stokes_quantum_2020,
  doi = {10.22331/q-2020-05-25-269},
  url = {https://doi.org/10.22331/q-2020-05-25-269},
  year = {2020},
  month = may,
  publisher = {Verein zur Forderung des Open Access Publizierens in den Quantenwissenschaften},
  volume = {4},
  pages = {269},
  author = {James Stokes and Josh Izaac and Nathan Killoran and Giuseppe Carleo},
  title = {Quantum Natural Gradient},
  journal = {Quantum}
}

@misc{meyer_fisher_2021,
      title={Fisher Information in Noisy Intermediate-Scale Quantum Applications}, 
      author={Johannes Jakob Meyer},
      year={2021},
      eprint={2103.15191},
      archivePrefix={arXiv},
      primaryClass={quant-ph}
}

@misc{beckey_variational_2020,
      title={Variational Quantum Algorithm for Estimating the Quantum Fisher Information}, 
      author={Jacob L. Beckey and M. Cerezo and Akira Sone and Patrick J. Coles},
      year={2020},
      eprint={2010.10488},
      archivePrefix={arXiv},
      primaryClass={quant-ph}
}

@misc{gacon_simultaneous_2021,
      title={Simultaneous Perturbation Stochastic Approximation of the Quantum Fisher Information}, 
      author={Julien Gacon and Christa Zoufal and Giuseppe Carleo and Stefan Woerner},
      year={2021},
      eprint={2103.09232},
      archivePrefix={arXiv},
      primaryClass={quant-ph}
}

@article{Cerezo_fisher_2021,
  doi = {10.1088/2058-9565/abfbef},
  url = {https://doi.org/10.1088/2058-9565/abfbef},
  year = {2021},
  month = jun,
  publisher = {{IOP} Publishing},
  volume = {6},
  number = {3},
  pages = {035008},
  author = {M Cerezo and Akira Sone and Jacob L Beckey and Patrick J Coles},
  title = {Sub-quantum Fisher information},
  journal = {Quantum Science and Technology}
}

@article{Uvarov_bp_2021,
   title={On barren plateaus and cost function locality in variational quantum algorithms},
   volume={54},
   ISSN={1751-8121},
   url={http://dx.doi.org/10.1088/1751-8121/abfac7},
   DOI={10.1088/1751-8121/abfac7},
   number={24},
   journal={Journal of Physics A: Mathematical and Theoretical},
   publisher={IOP Publishing},
   author={Uvarov, A V and Biamonte, J D},
   year={2021},
   month={May},
   pages={245301}
}

@article{Uvarov_variational_2020,
  doi = {10.1103/physrevb.102.075104},
  url = {https://doi.org/10.1103/physrevb.102.075104},
  year = {2020},
  month = aug,
  publisher = {American Physical Society ({APS})},
  volume = {102},
  number = {7},
  author = {Alexey Uvarov and Jacob D. Biamonte and Dmitry Yudin},
  title = {Variational quantum eigensolver for frustrated quantum systems},
  journal = {Physical Review B}
}

@article{BravoPrieto_scaling_2020,
   title={Scaling of variational quantum circuit depth for condensed matter systems},
   volume={4},
   ISSN={2521-327X},
   url={http://dx.doi.org/10.22331/q-2020-05-28-272},
   DOI={10.22331/q-2020-05-28-272},
   journal={Quantum},
   publisher={Verein zur Forderung des Open Access Publizierens in den Quantenwissenschaften},
   author={Bravo-Prieto, Carlos and Lumbreras-Zarapico, Josep and Tagliacozzo, Luca and Latorre, José I.},
   year={2020},
   month={May},
   pages={272}
}

@misc{bravoprieto_variational_2020,
      title={Variational Quantum Linear Solver}, 
      author={Carlos Bravo-Prieto and Ryan LaRose and M. Cerezo and Yigit Subasi and Lukasz Cincio and Patrick J. Coles},
      year={2020},
      eprint={1909.05820},
      archivePrefix={arXiv},
      primaryClass={quant-ph}
}

@misc{holmes_connecting_2021,
      title={Connecting ansatz expressibility to gradient magnitudes and barren plateaus}, 
      author={Zoë Holmes and Kunal Sharma and M. Cerezo and Patrick J. Coles},
      year={2021},
      eprint={2101.02138},
      archivePrefix={arXiv},
      primaryClass={quant-ph}
}

@misc{arrasmith_effect_2020,
      title={Effect of barren plateaus on gradient-free optimization}, 
      author={Andrew Arrasmith and M. Cerezo and Piotr Czarnik and Lukasz Cincio and Patrick J. Coles},
      year={2020},
      eprint={2011.12245},
      archivePrefix={arXiv},
      primaryClass={quant-ph}
}

@article{Cerezo_bp_2021,
  doi = {10.1038/s41467-021-21728-w},
  url = {https://doi.org/10.1038/s41467-021-21728-w},
  year = {2021},
  month = mar,
  publisher = {Springer Science and Business Media {LLC}},
  volume = {12},
  number = {1},
  author = {M. Cerezo and Akira Sone and Tyler Volkoff and Lukasz Cincio and Patrick J. Coles},
  title = {Cost function dependent barren plateaus in shallow parametrized quantum circuits},
  journal = {Nature Communications}
}

@misc{wang_noiseinduced_2021,
      title={Noise-Induced Barren Plateaus in Variational Quantum Algorithms}, 
      author={Samson Wang and Enrico Fontana and M. Cerezo and Kunal Sharma and Akira Sone and Lukasz Cincio and Patrick J. Coles},
      year={2021},
      eprint={2007.14384},
      archivePrefix={arXiv},
      primaryClass={quant-ph}
}

@online{cirq,
  doi = {10.5281/ZENODO.4062499},
  url = {https://zenodo.org/record/4062499},
  author = {{Cirq Developers}},
  title = {Cirq},
  publisher = {Zenodo},
  year = {2021},
  copyright = {Apache License 2.0}
}

@misc{forest,
    title={A Practical Quantum Instruction Set Architecture},
    author={Robert S. Smith and Michael J. Curtis and William J. Zeng},
    year={2016},
    eprint={1608.03355},
    archivePrefix={arXiv},
    primaryClass={quant-ph}
}

@misc{suzuki_qulacs_2020,
      title={Qulacs: a fast and versatile quantum circuit simulator for research purpose}, 
      author={Yasunari Suzuki and Yoshiaki Kawase and Yuya Masumura and Yuria Hiraga and Masahiro Nakadai and Jiabao Chen and Ken M. Nakanishi and Kosuke Mitarai and Ryosuke Imai and Shiro Tamiya and Takahiro Yamamoto and Tennin Yan and Toru Kawakubo and Yuya O. Nakagawa and Yohei Ibe and Youyuan Zhang and Hirotsugu Yamashita and Hikaru Yoshimura and Akihiro Hayashi and Keisuke Fujii},
      year={2020},
      eprint={2011.13524},
      archivePrefix={arXiv},
      primaryClass={quant-ph}
}

@misc{qcgpu,
      title={Simulating Quantum Computers Using OpenCL}, 
      author={Adam Kelly},
      year={2018},
      eprint={1805.00988},
      archivePrefix={arXiv},
      primaryClass={quant-ph}
}

@article{gates-barenco1995,
   title={Elementary gates for quantum computation},
   volume={52},
   ISSN={1094-1622},
   url={http://dx.doi.org/10.1103/PhysRevA.52.3457},
   DOI={10.1103/physreva.52.3457},
   number={5},
   journal={Physical Review A},
   publisher={American Physical Society (APS)},
   author={Barenco, Adriano and Bennett, Charles H. and Cleve, Richard and DiVincenzo, David P. and Margolus, Norman and Shor, Peter and Sleator, Tycho and Smolin, John A. and Weinfurter, Harald},
   year={1995},
   month={Nov},
   pages={3457–3467}
}

@article{Yum:17,
  doi = {10.1364/josab.34.001632},
  url = {https://doi.org/10.1364/josab.34.001632},
  year = {2017},
  month = jul,
  publisher = {The Optical Society},
  volume = {34},
  number = {8},
  pages = {1632},
  author = {Dahyun Yum and Debashis De Munshi and Tarun Dutta and Manas Mukherjee},
  title = {Optical barium ion qubit},
  journal = {Journal of the Optical Society of America B}
}

@article{Dutta2020,
  doi = {10.1038/s41534-019-0234-z},
  url = {https://doi.org/10.1038/s41534-019-0234-z},
  year = {2020},
  month = jan,
  publisher = {Springer Science and Business Media {LLC}},
  volume = {6},
  number = {1},
  author = {T. Dutta and M. Mukherjee},
  title = {A single atom noise probe operating beyond the Heisenberg limit},
  journal = {npj Quantum Information}
}

@article{VanHorne2020,
  doi = {10.1038/s41534-020-0264-6},
  url = {https://doi.org/10.1038/s41534-020-0264-6},
  year = {2020},
  month = may,
  publisher = {Springer Science and Business Media {LLC}},
  volume = {6},
  number = {1},
  author = {Noah Van Horne and Dahyun Yum and Tarun Dutta and Peter H\"{a}nggi and Jiangbin Gong and Dario Poletti and Manas Mukherjee},
  title = {Single-atom energy-conversion device with a quantum load},
  journal = {npj Quantum Information}
}

@article{Dehmelt1957,
  doi = {10.1103/physrev.105.1487},
  url = {https://doi.org/10.1103/physrev.105.1487},
  year = {1957},
  month = mar,
  publisher = {American Physical Society ({APS})},
  volume = {105},
  number = {5},
  pages = {1487--1489},
  author = {H. G. Dehmelt},
  title = {Slow Spin Relaxation of Optically Polarized Sodium Atoms},
  journal = {Physical Review}
}

@article{Maslov_2017,
  doi = {10.1088/1367-2630/aa5e47},
  url = {https://doi.org/10.1088/1367-2630/aa5e47},
  year = {2017},
  month = feb,
  publisher = {{IOP} Publishing},
  volume = {19},
  number = {2},
  pages = {023035},
  author = {Dmitri Maslov},
  title = {Basic circuit compilation techniques for an ion-trap quantum machine},
  journal = {New Journal of Physics}
}

@article{McKay2017,
  doi = {10.1103/physreva.96.022330},
  url = {https://doi.org/10.1103/physreva.96.022330},
  year = {2017},
  month = aug,
  publisher = {American Physical Society ({APS})},
  volume = {96},
  number = {2},
  author = {David C. McKay and Christopher J. Wood and Sarah Sheldon and Jerry M. Chow and Jay M. Gambetta},
  title = {Efficient 
		Z
		 gates for quantum computing},
  journal = {Physical Review A}
}

@article{Knill2008,
  doi = {10.1103/physreva.77.012307},
  url = {https://doi.org/10.1103/physreva.77.012307},
  year = {2008},
  month = jan,
  publisher = {American Physical Society ({APS})},
  volume = {77},
  number = {1},
  author = {E. Knill and D. Leibfried and R. Reichle and J. Britton and R. B. Blakestad and J. D. Jost and C. Langer and R. Ozeri and S. Seidelin and D. J. Wineland},
  title = {Randomized benchmarking of quantum gates},
  journal = {Physical Review A}
}

@article{Ackley_bm_1985,
  doi = {10.1207/s15516709cog0901_7},
  url = {https://doi.org/10.1207/s15516709cog0901_7},
  year = {1985},
  month = jan,
  publisher = {Wiley},
  volume = {9},
  number = {1},
  pages = {147--169},
  author = {David H. Ackley and Geoffrey E. Hinton and Terrence J. Sejnowski},
  title = {A Learning Algorithm for Boltzmann Machines},
  journal = {Cognitive Science}
}

@article{Zoufal_bm_2021,
   title={Variational quantum Boltzmann machines},
   volume={3},
   ISSN={2524-4914},
   url={http://dx.doi.org/10.1007/s42484-020-00033-7},
   DOI={10.1007/s42484-020-00033-7},
   number={1},
   journal={Quantum Machine Intelligence},
   publisher={Springer Science and Business Media LLC},
   author={Zoufal, Christa and Lucchi, Aurélien and Woerner, Stefan},
   year={2021},
   month={Feb}
}

@misc{shingu_bm_2020,
      title={Boltzmann machine learning with a variational quantum algorithm}, 
      author={Yuta Shingu and Yuya Seki and Shohei Watabe and Suguru Endo and Yuichiro Matsuzaki and Shiro Kawabata and Tetsuro Nikuni and Hideaki Hakoshima},
      year={2020},
      eprint={2007.00876},
      archivePrefix={arXiv},
      primaryClass={quant-ph}
}

@article{Dixit_training_2021,
   title={Training Restricted Boltzmann Machines With a D-Wave Quantum Annealer},
   volume={9},
   ISSN={2296-424X},
   url={http://dx.doi.org/10.3389/fphy.2021.589626},
   DOI={10.3389/fphy.2021.589626},
   journal={Frontiers in Physics},
   publisher={Frontiers Media SA},
   author={Dixit, Vivek and Selvarajan, Raja and Alam, Muhammad A. and Humble, Travis S. and Kais, Sabre},
   year={2021},
   month={Jun}
}

@article{Romero_autoencoders_2017,
  doi = {10.1088/2058-9565/aa8072},
  url = {https://doi.org/10.1088/2058-9565/aa8072},
  year = {2017},
  month = aug,
  publisher = {{IOP} Publishing},
  volume = {2},
  number = {4},
  pages = {045001},
  author = {Jonathan Romero and Jonathan P Olson and Alan Aspuru-Guzik},
  title = {Quantum autoencoders for efficient compression of quantum data},
  journal = {Quantum Science and Technology}
}

@article{Wan_generalisation_2017,
  doi = {10.1038/s41534-017-0032-4},
  url = {https://doi.org/10.1038/s41534-017-0032-4},
  year = {2017},
  month = sep,
  publisher = {Springer Science and Business Media {LLC}},
  volume = {3},
  number = {1},
  author = {Kwok Ho Wan and Oscar Dahlsten and Hl{\'{e}}r Kristj{\'{a}}nsson and Robert Gardner and M. S. Kim},
  title = {Quantum generalisation of feedforward neural networks},
  journal = {npj Quantum Information}
}

@misc{verdon_universal_2018,
      title={A Universal Training Algorithm for Quantum Deep Learning}, 
      author={Guillaume Verdon and Jason Pye and Michael Broughton},
      year={2018},
      eprint={1806.09729},
      archivePrefix={arXiv},
      primaryClass={quant-ph}
}

@article{Pepper_autoencoder_2019,
  doi = {10.1103/physrevlett.122.060501},
  url = {https://doi.org/10.1103/physrevlett.122.060501},
  year = {2019},
  month = feb,
  publisher = {American Physical Society ({APS})},
  volume = {122},
  number = {6},
  author = {Alex Pepper and Nora Tischler and Geoff J. Pryde},
  title = {Experimental Realization of a Quantum Autoencoder: The Compression of Qutrits via Machine Learning},
  journal = {Physical Review Letters}
}

@article{BravoPrieto_autoencoders_2021,
   title={Quantum autoencoders with enhanced data encoding},
   volume={2},
   ISSN={2632-2153},
   url={http://dx.doi.org/10.1088/2632-2153/ac0616},
   DOI={10.1088/2632-2153/ac0616},
   number={3},
   journal={Machine Learning: Science and Technology},
   publisher={IOP Publishing},
   author={Bravo-Prieto, Carlos},
   year={2021},
   month={Jul},
   pages={035028}
}

@article{Du_expressive_2020,
  doi = {10.1103/physrevresearch.2.033125},
  url = {https://doi.org/10.1103/physrevresearch.2.033125},
  year = {2020},
  month = jul,
  publisher = {American Physical Society ({APS})},
  volume = {2},
  number = {3},
  author = {Yuxuan Du and Min-Hsiu Hsieh and Tongliang Liu and Dacheng Tao},
  title = {Expressive power of parametrized quantum circuits},
  journal = {Physical Review Research}
}

@misc{verdon_quantum_2019,
      title={A quantum algorithm to train neural networks using low-depth circuits}, 
      author={Guillaume Verdon and Michael Broughton and Jacob Biamonte},
      year={2019},
      eprint={1712.05304},
      archivePrefix={arXiv},
      primaryClass={quant-ph}
}

@article{Liu_differentiable_2018,
  doi = {10.1103/physreva.98.062324},
  url = {https://doi.org/10.1103/physreva.98.062324},
  year = {2018},
  month = dec,
  publisher = {American Physical Society ({APS})},
  volume = {98},
  number = {6},
  author = {Jin-Guo Liu and Lei Wang},
  title = {Differentiable learning of quantum circuit Born machines},
  journal = {Physical Review A}
}

@article{Coyle_born_2020,
  doi = {10.1038/s41534-020-00288-9},
  url = {https://doi.org/10.1038/s41534-020-00288-9},
  year = {2020},
  month = jul,
  publisher = {Springer Science and Business Media {LLC}},
  volume = {6},
  number = {1},
  author = {Brian Coyle and Daniel Mills and Vincent Danos and Elham Kashefi},
  title = {The Born supremacy: quantum advantage and training of an Ising Born machine},
  journal = {npj Quantum Information}
}

@misc{romero_variational_2019,
      title={Variational quantum generators: Generative adversarial quantum machine learning for continuous distributions}, 
      author={Jonathan Romero and Alan Aspuru-Guzik},
      year={2019},
      eprint={1901.00848},
      archivePrefix={arXiv},
      primaryClass={quant-ph}
}

@misc{altaisky_quantum_2001,
  title={Quantum neural network},
  author={Altaisky, M. V.},
  eprint={quant-ph/0107012},
  year={2001},
  archivePrefix={arXiv},
  primaryClass={quant-ph}
}

@article{Torrontegui_unitary_2019,
   title={Unitary quantum perceptron as efficient universal approximator},
   volume={125},
   ISSN={1286-4854},
   url={http://dx.doi.org/10.1209/0295-5075/125/30004},
   DOI={10.1209/0295-5075/125/30004},
   number={3},
   journal={EPL (Europhysics Letters)},
   publisher={IOP Publishing},
   author={Torrontegui, E. and García-Ripoll, J. J.},
   year={2019},
   month={Mar},
   pages={30004}
}

@article{Cong_quantum_2019,
  doi = {10.1038/s41567-019-0648-8},
  url = {https://doi.org/10.1038/s41567-019-0648-8},
  year = {2019},
  month = aug,
  publisher = {Springer Science and Business Media {LLC}},
  volume = {15},
  number = {12},
  pages = {1273--1278},
  author = {Iris Cong and Soonwon Choi and Mikhail D. Lukin},
  title = {Quantum convolutional neural networks},
  journal = {Nature Physics}
}

@inproceedings{franken_explorations_2020,
  title={Explorations in Quantum Neural Networks with Intermediate Measurements.},
  author={Franken, Lukas and Georgiev, Bogdan},
  booktitle={ESANN},
  pages={297--302},
  year={2020}
}

@misc{zhang_trainability_2020,
      title={Toward Trainability of Quantum Neural Networks}, 
      author={Kaining Zhang and Min-Hsiu Hsieh and Liu Liu and Dacheng Tao},
      year={2020},
      eprint={2011.06258},
      archivePrefix={arXiv},
      primaryClass={quant-ph}
}

@misc{pesah_absence_2020,
      title={Absence of Barren Plateaus in Quantum Convolutional Neural Networks}, 
      author={Arthur Pesah and M. Cerezo and Samson Wang and Tyler Volkoff and Andrew T. Sornborger and Patrick J. Coles},
      year={2020},
      eprint={2011.02966},
      archivePrefix={arXiv},
      primaryClass={quant-ph}
}

@inproceedings{Dunjko_advances_2017,
  doi = {10.1109/smc.2017.8122616},
  url = {https://doi.org/10.1109/smc.2017.8122616},
  year = {2017},
  month = oct,
  publisher = {{IEEE}},
  author = {Vedran Dunjko and Jacob M. Taylor and Hans J. Briegel},
  title = {Advances in quantum reinforcement learning},
  booktitle = {2017 {IEEE} International Conference on Systems,  Man,  and Cybernetics ({SMC})}
}

@article{Dunjko_quantum_2016,
  doi = {10.1103/physrevlett.117.130501},
  url = {https://doi.org/10.1103/physrevlett.117.130501},
  year = {2016},
  month = sep,
  publisher = {American Physical Society ({APS})},
  volume = {117},
  number = {13},
  author = {Vedran Dunjko and Jacob M. Taylor and Hans J. Briegel},
  title = {Quantum-Enhanced Machine Learning},
  journal = {Physical Review Letters}
}

@article{Dong_quantum_2008,
  doi = {10.1109/tsmcb.2008.925743},
  url = {https://doi.org/10.1109/tsmcb.2008.925743},
  year = {2008},
  month = oct,
  publisher = {Institute of Electrical and Electronics Engineers ({IEEE})},
  volume = {38},
  number = {5},
  pages = {1207--1220},
  author = {Daoyi Dong and  Chunlin Chen and  Hanxiong Li and  Tzyh-Jong Tarn},
  title = {Quantum Reinforcement Learning},
  journal = {{IEEE} Transactions on Systems,  Man,  and Cybernetics,  Part B (Cybernetics)}
}

@article{Jerbi_quantum_2021,
   title={Quantum Enhancements for Deep Reinforcement Learning in Large Spaces},
   volume={2},
   ISSN={2691-3399},
   url={http://dx.doi.org/10.1103/PRXQuantum.2.010328},
   DOI={10.1103/prxquantum.2.010328},
   number={1},
   journal={PRX Quantum},
   publisher={American Physical Society (APS)},
   author={Jerbi, Sofiene and Trenkwalder, Lea M. and Poulsen Nautrup, Hendrik and Briegel, Hans J. and Dunjko, Vedran},
   year={2021},
   month={Feb}
}

@inproceedings{lockwood_reinforcement_2020,
  title={Reinforcement learning with quantum variational circuit},
  author={Lockwood, Owen and Si, Mei},
  booktitle={Proceedings of the AAAI Conference on Artificial Intelligence and Interactive Digital Entertainment},
  volume={16},
  number={1},
  pages={245--251},
  year={2020}
}

@article{Chen_variational_2020,
  doi = {10.1109/access.2020.3010470},
  url = {https://doi.org/10.1109/access.2020.3010470},
  year = {2020},
  publisher = {Institute of Electrical and Electronics Engineers ({IEEE})},
  volume = {8},
  pages = {141007--141024},
  author = {Samuel Yen-Chi Chen and Chao-Han Huck Yang and Jun Qi and Pin-Yu Chen and Xiaoli Ma and Hsi-Sheng Goan},
  title = {Variational Quantum Circuits for Deep Reinforcement Learning},
  journal = {{IEEE} Access}
}

@inproceedings{lockwood_atari_2021,
  title={Playing Atari with Hybrid Quantum-Classical Reinforcement Learning},
  author={Lockwood, Owen and Si, Mei},
  booktitle={NeurIPS 2020 Workshop on Pre-registration in Machine Learning},
  pages={285--301},
  year={2021},
  organization={PMLR}
}

@misc{crawford_reinforcement_2019,
      title={Reinforcement Learning Using Quantum Boltzmann Machines}, 
      author={Daniel Crawford and Anna Levit and Navid Ghadermarzy and Jaspreet S. Oberoi and Pooya Ronagh},
      year={2019},
      eprint={1612.05695},
      archivePrefix={arXiv},
      primaryClass={quant-ph}
}

@article{Lamata_superconducting_2017,
  doi = {10.1038/s41598-017-01711-6},
  url = {https://doi.org/10.1038/s41598-017-01711-6},
  year = {2017},
  month = may,
  publisher = {Springer Science and Business Media {LLC}},
  volume = {7},
  number = {1},
  author = {Lucas Lamata},
  title = {Basic protocols in quantum reinforcement learning with superconducting circuits},
  journal = {Scientific Reports}
}

@article{cardenaslopez_multiqubit_2018,
  doi = {10.1371/journal.pone.0200455},
  url = {https://doi.org/10.1371/journal.pone.0200455},
  year = {2018},
  month = jul,
  publisher = {Public Library of Science ({PLoS})},
  volume = {13},
  number = {7},
  pages = {e0200455},
  author = {F. A. C{\'{a}}rdenas-L{\'{o}}pez and L. Lamata and J. C. Retamal and E. Solano},
  title = {Multiqubit and multilevel quantum reinforcement learning with quantum technologies},
  journal = {{PLOS} {ONE}}
}

@article{Yu_reconstruction_2019,
  doi = {10.1002/qute.201800074},
  url = {https://doi.org/10.1002/qute.201800074},
  year = {2019},
  month = mar,
  publisher = {Wiley},
  volume = {2},
  number = {7-8},
  pages = {1800074},
  author = {Shang Yu and Francisco Albarr{\'{a}}n-Arriagada and Juan Carlos Retamal and Yi-Tao Wang and Wei Liu and Zhi-Jin Ke and Yu Meng and Zhi-Peng Li and Jian-Shun Tang and Enrique Solano and Lucas Lamata and Chuan-Feng Li and Guang-Can Guo},
  title = {Reconstruction of a Photonic Qubit State with Reinforcement Learning},
  journal = {Advanced Quantum Technologies}
}

@article{hhl,
  doi = {10.1103/physrevlett.103.150502},
  url = {https://doi.org/10.1103/physrevlett.103.150502},
  year = {2009},
  month = oct,
  publisher = {American Physical Society ({APS})},
  volume = {103},
  number = {15},
  author = {Aram W. Harrow and Avinatan Hassidim and Seth Lloyd},
  title = {Quantum Algorithm for Linear Systems of Equations},
  journal = {Physical Review Letters}
}

@article{Zhao_bayesian_2019,
  doi = {10.1007/s42484-019-00004-7},
  url = {https://doi.org/10.1007/s42484-019-00004-7},
  year = {2019},
  month = may,
  publisher = {Springer Science and Business Media {LLC}},
  volume = {1},
  number = {1-2},
  pages = {41--51},
  author = {Zhikuan Zhao and Alejandro Pozas-Kerstjens and Patrick Rebentrost and Peter Wittek},
  title = {Bayesian deep learning on a quantum computer},
  journal = {Quantum Machine Intelligence}
}

@misc{schiffer_adiabatic_2021,
      title={Adiabatic Spectroscopy and a Variational Quantum Adiabatic Algorithm}, 
      author={Benjamin F. Schiffer and Jordi Tura and J. Ignacio Cirac},
      year={2021},
      eprint={2103.01226},
      archivePrefix={arXiv},
      primaryClass={quant-ph}
}

@misc{garciasaez_aavqe_2018,
      title={Addressing hard classical problems with Adiabatically Assisted Variational Quantum Eigensolvers}, 
      author={A. Garcia-Saez and J. I. Latorre},
      year={2018},
      eprint={1806.02287},
      archivePrefix={arXiv},
      primaryClass={quant-ph}
}

@article{Giovannetti_qram_2008,
  doi = {10.1103/physrevlett.100.160501},
  url = {https://doi.org/10.1103/physrevlett.100.160501},
  year = {2008},
  month = apr,
  publisher = {American Physical Society ({APS})},
  volume = {100},
  number = {16},
  author = {Vittorio Giovannetti and Seth Lloyd and Lorenzo Maccone},
  title = {Quantum Random Access Memory},
  journal = {Physical Review Letters}
}

@article{Gisin_communication_2007,
  doi = {10.1038/nphoton.2007.22},
  url = {https://doi.org/10.1038/nphoton.2007.22},
  year = {2007},
  month = mar,
  publisher = {Springer Science and Business Media {LLC}},
  volume = {1},
  number = {3},
  pages = {165--171},
  author = {Nicolas Gisin and Rob Thew},
  title = {Quantum communication},
  journal = {Nature Photonics}
}

@article{Bennett_cryptography_1992,
  doi = {10.1007/bf00191318},
  url = {https://doi.org/10.1007/bf00191318},
  year = {1992},
  month = jan,
  publisher = {Springer Science and Business Media {LLC}},
  volume = {5},
  number = {1},
  pages = {3--28},
  author = {Charles H. Bennett and Fran{\c{c}}ois Bessette and Gilles Brassard and Louis Salvail and John Smolin},
  title = {Experimental quantum cryptography},
  journal = {Journal of Cryptology}
}

@article{quantummaterials_2016,
  doi = {10.1038/nphys3668},
  url = {https://doi.org/10.1038/nphys3668},
  year = {2016},
  month = feb,
  publisher = {Springer Science and Business Media {LLC}},
  volume = {12},
  number = {2},
  pages = {105--105},
  title = {The rise of quantum materials},
  journal = {Nature Physics}, 
  author = {{Nature Editorial}}
}

@article{Feynman_simulating_1982,
  doi = {10.1007/bf02650179},
  url = {https://doi.org/10.1007/bf02650179},
  year = {1982},
  month = jun,
  publisher = {Springer Science and Business Media {LLC}},
  volume = {21},
  number = {6-7},
  pages = {467--488},
  author = {Richard P. Feynman},
  title = {Simulating physics with computers},
  journal = {International Journal of Theoretical Physics}
}

@misc{preskill_feynman_2021,
      title={Quantum computing 40 years later}, 
      author={John Preskill},
      year={2021},
      eprint={2106.10522},
      archivePrefix={arXiv},
      primaryClass={quant-ph}
}

@article{Turing_computable_1938,
  doi = {10.1112/plms/s2-43.6.544},
  url = {https://doi.org/10.1112/plms/s2-43.6.544},
  year = {1938},
  publisher = {Wiley},
  volume = {s2-43},
  number = {1},
  pages = {544--546},
  author = {A. M. Turing},
  title = {On Computable Numbers,  with an Application to the Entscheidungsproblem. A Correction},
  journal = {Proceedings of the London Mathematical Society}
}

@article{deutsch_quantum_1985,
  title={Quantum theory, the Church--Turing principle and the universal quantum computer},
  author={Deutsch, David},
  journal={Proceedings of the Royal Society of London. A. Mathematical and Physical Sciences},
  volume={400},
  number={1818},
  pages={97--117},
  year={1985},
  publisher={The Royal Society London}
}

@article{bernstein_quantum_1997,
  title={Quantum complexity theory},
  author={Bernstein, Ethan and Vazirani, Umesh},
  journal={SIAM Journal on computing},
  volume={26},
  number={5},
  pages={1411--1473},
  year={1997},
  publisher={SIAM}
}

@book{arora_computational_2009,
  title={Computational complexity: a modern approach},
  author={Arora, Sanjeev and Barak, Boaz},
  year={2009},
  publisher={Cambridge University Press}, 
  isbn = {978-0521424264}
}

@article{Vazirani_quantum_2002,
  doi = {10.1090/psapm/058/1922899},
  url = {https://doi.org/10.1090/psapm/058/1922899},
  year = {2002},
  publisher = {American Mathematical Society},
  pages = {193--217},
  author = {Umesh V. Vazirani},
  title = {A survey of quantum complexity theory}
}

@article{moore_cramming_1965,
  doi = {10.1109/n-ssc.2006.4785860},
  url = {https://doi.org/10.1109/n-ssc.2006.4785860},
  year = {1965},
  month = apr,
  volume = {38},
  number = {8},
  pages = {114},
  author = {Gordon E. Moore},
  title = {Cramming more components onto integrated circuits},
  journal = {Electronics}
}

@article{Haroche_quantum_1996,
  doi = {10.1063/1.881512},
  url = {https://doi.org/10.1063/1.881512},
  year = {1996},
  month = aug,
  publisher = {{AIP} Publishing},
  volume = {49},
  number = {8},
  pages = {51--52},
  author = {Serge Haroche and Jean-Michel Raimond},
  title = {Quantum Computing: Dream or Nightmare?},
  journal = {Physics Today}
}
\section*{Acronyms}
%\addcontentsline{toc}{section}{Acronyms}
\printacronyms[heading=none]

\cleardoublepage

\newpage
\thispagestyle{empty}

\newpage

\begingroup
\begin{tikzpicture}[remember picture,overlay]
\coordinate [below=10cm] (midpoint) at (current page.north);
\node at (current page.north west)
{\begin{tikzpicture}[remember picture,overlay]
\node[anchor=north west,inner sep=0pt] at (0,0) {\includegraphics[height=\paperheight]{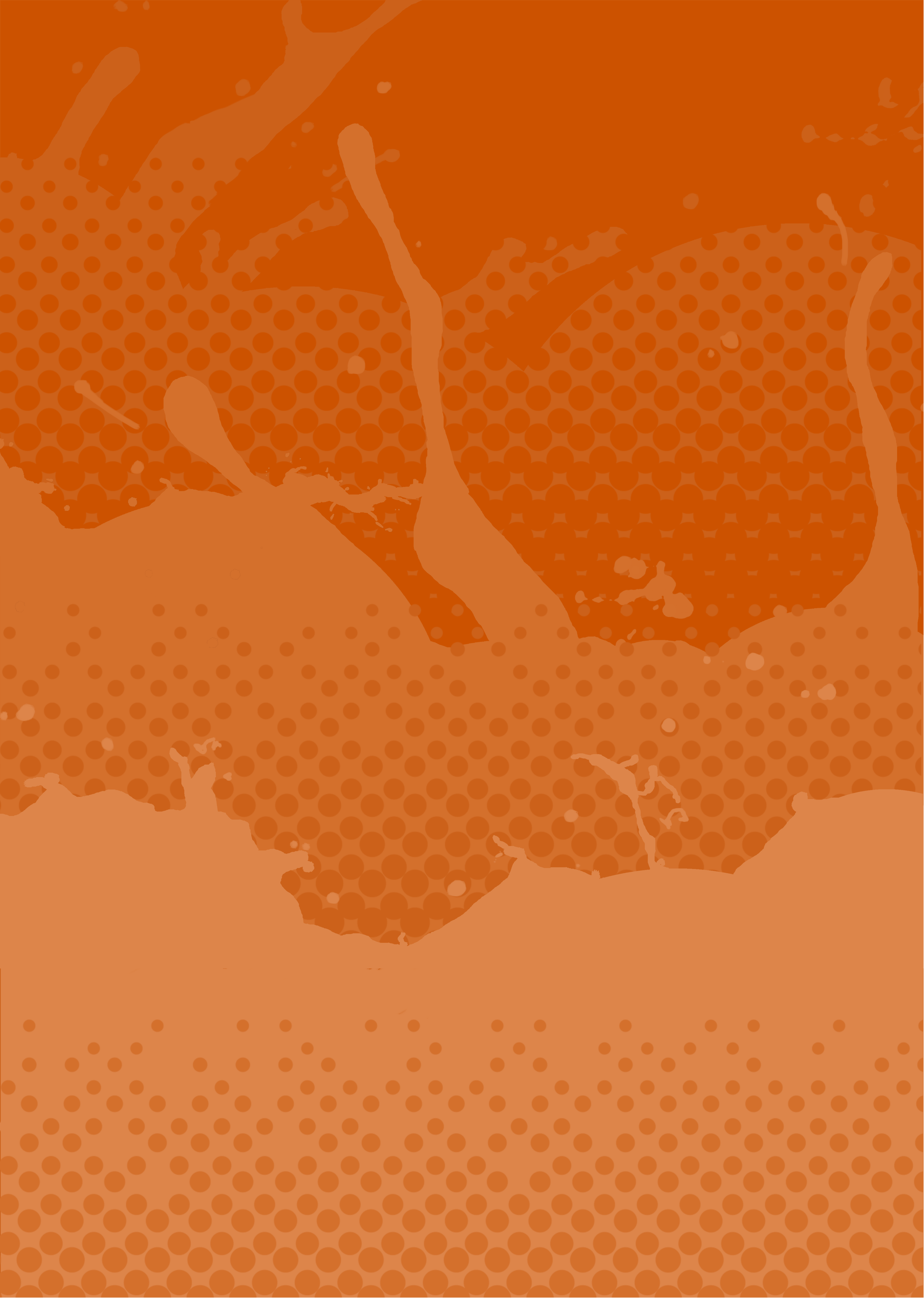}}; % Background image
\draw[anchor=north west] (midpoint) node at (2, -3) [fill=white,fill opacity=0,text opacity=1,inner sep=0cm]{\large\centering\bfseries\parbox[c][][t]{13cm}{Quantum computing is a nascent technology with prospects to have a huge impact in the world. Its current status, however, only counts on small and noisy quantum computers whose performance is limited. In this thesis, two different strategies are explored to take advantage of inherently quantum properties and propose recipes to seize quantum computing since its advent. First, the re-uploading strategy is a variational algorithm related to machine learning. It consists in introducing data several times along a computation accompanied by tunable parameters. This process permits the circuit to learn and mimic any behavior. This capability emerges naturally from the quantum properties of the circuit. Second, the unary strategy aims to reduce the density of information stored in a quantum circuit to increase its resilience against noise. This trade-off between performance and robustness brings an advantage for noisy devices, where small but meaningful quantum speed-ups can be found. 
\\[15pt]}  % Author name
};
\draw[anchor=north west] (midpoint) node at (-1, -17) [fill=blue,fill opacity=0,text opacity=1,inner sep=1cm]{\Huge\centering\bfseries\parbox[c][][t]{\paperwidth}{\centering \includegraphics[width=10cm]{logo_ub} \\ \includegraphics[width=10cm]{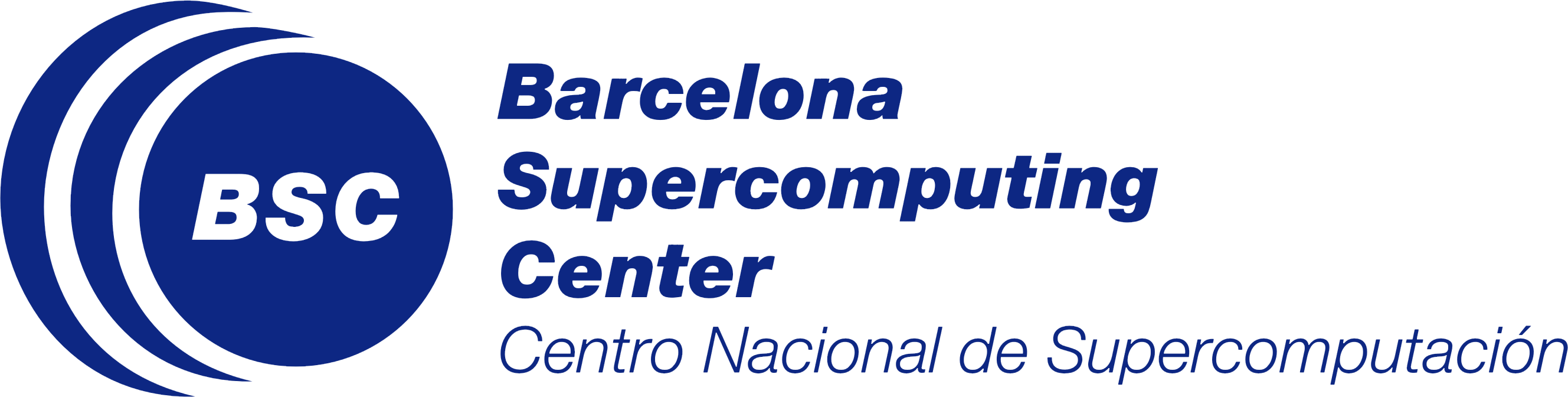}}  % logo 
};
\end{tikzpicture}};
\end{tikzpicture}

\endgroup 

\end{document}